\documentclass[a4paper,12pt,twoside]{book}

\fontfamily{lmr}\selectfont
\usepackage[greek,english]{babel}
\usepackage[utf8x]{inputenc}
\usepackage{hyperref}
\usepackage{libertine}
\usepackage{graphicx}
\usepackage{booktabs}
\usepackage{caption}
\usepackage[sort]{cite}
\usepackage[list=true,listformat=simple]{subcaption}
\usepackage{comment}
\usepackage{glossaries}
\usepackage[usenames,dvipsnames,svgnames,table]{xcolor}
\usepackage{cgiannoula-phd-ntua}
\usepackage{pdfpages}

\usepackage{fancyhdr}
\usepackage[Sonny]{fncychap}
\ChNameVar{\Large}
\ChNumVar{\Huge}
\ChTitleVar{\Huge}

\usepackage{algorithm}% http://ctan.org/pkg/algorithms
\usepackage{multirow}% http://ctan.org/pkg/algorithmicx
\usepackage{color}
\usepackage{amsmath}
\usepackage{amsfonts}

\usepackage{titlesec}
\newcommand\YUGE{\fontsize{48}{60}\selectfont}
\newcommand\YUGEN{\fontsize{36}{60}\selectfont}
\titleformat{\chapter}[display]
{\bfseries\Huge}
{\filleft\MakeUppercase{\chaptertitlename} \YUGE\thechapter}
{8ex}
{\titlerule\vspace{4ex}\filleft\YUGEN}
[\vspace{4ex}\titlerule]

\usepackage{fancyhdr}            % Permits header customization. See header section below.
\fancypagestyle{plain}{
    \fancyhead{}
    \renewcommand{\headrulewidth}{0pt}
    \cfoot{\thepage}
}
\renewcommand{\headrulewidth}{0pt}
\fancyheadoffset[RE,LO]{-0.0\textwidth}
\newcommand{\ColorTM}{\emph{ColorTM}}
\newcommand{\BalColorTM}{\emph{BalColorTM}}

\usepackage{amsmath}
\usepackage{listings}
\definecolor{gray}{rgb}{0.5,0.5,0.5}
\lstdefinestyle{mystyle}{
    keywordstyle=\color{blue}\textbf,
    numberstyle=\footnotesize\color{codegray},
    basicstyle=\footnotesize\fontfamily{txtt}\selectfont,
    breakatwhitespace=false,
    breaklines=true,
    captionpos=b,
    keepspaces=true,
    numbers=left,
    numbersep=5pt,
    showspaces=false,
    showstringspaces=false,
    showtabs=false,
    tabsize=2,
    xleftmargin=1.8em,
}
\usepackage{array}
\newcolumntype{L}[1]{>{\raggedright\let\newline\\\arraybackslash\hspace{0pt}}m{#1}}
\newcolumntype{C}[1]{>{\centering\let\newline\\\arraybackslash\hspace{0pt}}m{#1}}
\newcolumntype{R}[1]{>{\raggedleft\let\newline\\\arraybackslash\hspace{0pt}}m{#1}}
\newcommand{\smartpq}{\emph{SmartPQ}}
\newcommand{\numa}{NUMA-aware}

\newcommand{\notnuma}{NUMA-oblivious}
\newcommand{\insrt}{\emph{insert}} 
\newcommand{\delete}{\emph{deleteMin}}
\newcommand{\nuddle}{\emph{Nuddle}}
\newcommand{\algomode}{algorithmic}
\newcommand{\ffwd}{\emph{ffwd}}

\usepackage{paralist}

\definecolor{codegray}{rgb}{0.5,0.5,0.5}
\lstdefinestyle{mystyle2}{
    keywordstyle=\color{blue}\textbf,
    numberstyle=\footnotesize\color{codegray},
    basicstyle=\footnotesize\fontfamily{txtt}\selectfont,
    breakatwhitespace=false,
    breaklines=true,
    captionpos=b,
    keepspaces=true,
    numbers=left,
    numbersep=5pt,
    showspaces=false,
    showstringspaces=false,
    showtabs=false,
    tabsize=2,
    firstnumber=last,
    xleftmargin=1.8em,
}

\definecolor{blue(pigment)}{rgb}{0.2, 0.2, 0.7}
\definecolor{dgreen}{rgb}{0.0, 0.8, 0.0}
\usepackage[usenames,dvipsnames,svgnames,table]{xcolor}
\usepackage{titlesec}
\usepackage{dblfloatfix}
\definecolor{ocre}{RGB}{243,102,25}
\definecolor{amber}{rgb}{1.0, 0.49, 0.0}
\definecolor{darkbyzantium}{rgb}{0.36, 0.22, 0.33}
\definecolor{darkseagreen}{rgb}{0.56, 0.74, 0.56}
\definecolor{darkspringgreen}{rgb}{0.09, 0.45, 0.27}
\definecolor{dollarbill}{rgb}{0.52, 0.73, 0.4}
\definecolor{MidnightBlue}{rgb}{0.1, 0.1, 0.44}

\definecolor{darkgreen}{RGB}{0,50,0}
\definecolor{olivegreen}{RGB}{68,85,37}
\definecolor{darkred}{RGB}{190,0,0}
\definecolor{darkblue}{RGB}{0,0,190}
\definecolor{darkgreen}{RGB}{0,50,0}
\definecolor{olivegreen}{RGB}{68,85,37}
\definecolor{darkred}{RGB}{190,0,0}
\definecolor{darkblue}{RGB}{0,0,190}
\definecolor{mygreen}{RGB}{2,108,69}

\definecolor{bostonuniversityred}{rgb}{0.8, 0.0, 0.0}
\newcommand\fix[1]{\noindent{\color{black}{#1}}} 
\newcommand{\SynCron}{\emph{SynCron}}
\newcommand{\myEngine}{Synchronization Engine}
\newcommand{\myEngineShort}{SE}

\newcommand{\masterSE}{\emph{Master SE}}

\newcommand{\myTableShort}{ST}
\newcommand{\mySyncVar}{syncronVar}
\newcommand{\hier}{\emph{Hier}}
\newcommand{\naive}{\emph{Central}}
\newcommand{\ideal}{\emph{Ideal}}
\newcommand{\mpsync}{message-passing}
\newcommand{\Mpsync}{Message-passing}

\definecolor{blue(pigment)}{rgb}{0.2, 0.2, 0.7}

\usepackage{tikz}
\newcommand*\circled[1]{\tikz[baseline=(char.base)]{\node[shape=circle,fill,inner sep=0.5pt] (char) {\textbf{\textcolor{white}{#1}}};}}
\newcommand*\rectangled[1]{\tikz[baseline=(char.base)]{\node[shape=rectangle,fill,inner sep=1pt, minimum width=0.34cm, rounded corners] (char) {\textbf{\textcolor{white}{#1}}};}}

\newcommand{\SparseP}{\emph{SparseP}}
\newcommand{\spmv}{SpMV}
\newcommand{\equallySized}{\emph{equally-sized}}
\newcommand{\equallyWidth}{\emph{equally-wide}}
\newcommand{\variableSized}{\emph{variable-sized}}
\newcommand{\dpuActive}{2528 DPUs}

\usepackage{multirow}
\usepackage{colortbl}
\usepackage{adjustbox}
\usepackage{setspace}
\usepackage{array,multirow}
\usepackage{makecell, tabularx}
\usepackage{color,soul}
\usepackage[most]{tcolorbox}
\usepackage{rotating, graphicx}
\usepackage{algpseudocode}
\usepackage{titlesec}
\usepackage{afterpage}

\newcommand\blankpage{%
    \null
    \thispagestyle{empty}%
    \addtocounter{page}{-1}%
    \newpage}
    
\newcommand\blankpagecgiannou{%
    \null
    \thispagestyle{empty}%
    \addtocounter{page}{-1}%
    \centering 
    \vspace{260pt}
    \emph{\selectlanguage{english}To my loving parents, Maria and Christoforos.\selectlanguage{greek}\\
    Στους αγαπημένους μου γονείς, Μαρία και Χριστόφορο.}
    \newpage}
\tcbset{
    left=8pt,
    right=8pt,
    enhanced,
    halign=center,
    colback=gray!16,
    colframe=black,
    boxrule=0.8pt,
    width=\dimexpr\textwidth\relax,
}

\algrenewcommand\algorithmiccomment[2][\normalsize]{{#1\hfill\(\triangleright\) #2}}
\usepackage{geometry} \title{Accelerating Irregular Applications \\ via Efficient Synchronization \\ \vspace{8pt} and Data Access Techniques}

\author{CHRISTINA GIANNOULA}
\date{September 2022}

\place{Athens}
\yearr{2022}
\authornominative{CHRISTINA GIANNOULA}
\authortitle{Doctor of Electrical and Computer Engineer, NTUA}
\advisors{Georgios Goumas \\ Nectarios Koziris\\  Onur Mutlu}

\committeeone{Georgios Goumas \\  Associate Professor NTUA}
\committeetwo{Nectarios Koziris\\ Professor NTUA}
\committeethree{Onur Mutlu \\ Professor ETH Zurich}
\committeefour{Dionisios Pnevmatikatos \\ Professor NTUA}
\committeefive{Stefanos Kaxiras \\ Professor Uppsala University}
\committeesix{
Dimitris Gizopoulos \\ Professor University of Athens}
\committeeseven{Vasileios Papaefstathiou \\  Assistant Professor University of Crete}
		
\submitdate{Accepted by the seven-member committee on 27/09/2022.}

\algnewcommand{\LineComment}[1]{\Statex \(\triangleright\) #1}

\algrenewcommand{\algorithmicindent}{0.8em}

\newcommand{\eat}[1]{}

\usepackage{setspace}
\onehalfspace

\fontfamily{lmr}\selectfont

\begin{document}
\maketitle
%\includepdf[pages={1}]{Doctoral_Dissertation-CGiannoula_signed.pdf}
\selectlanguage{english}
\afterpage{\blankpagecgiannou}
\afterpage{\blankpage}
\chapter*{Acknowledgments}

This doctoral thesis is the culmination of five and half years of hard work throughout my PhD studies. My PhD journey was a significant source of learning and growth for me both professionally and personally. There have been many who have supported me and contributed in different ways. These acknowledgments comprise a brief and humble attempt to thank their invaluable contributions.

First and foremost, I wholeheartedly thank my advisors, Prof. Georgios Goumas, Prof. Nectarios Koziris, and Prof. Onur Mutlu. I am very grateful to Prof. Georgios Goumas for helping me to find and follow a very interesting research direction for me, supporting and advising me with great patience and tolerance, motivating me to work on the fields of high-performance computing and computer architecture and providing me a very comfortable, safe and stimulating environment to grow. I also thank him for being open to my collaborations with researchers from other institutions. I am extremely grateful to Prof. Onur Mutlu for his generous guidance, resources and opportunities which constitute the key to my professional growth, success and  research achievements. I thank him for giving me the invaluable opportunity to work with him and his research group, providing rigorous feedback to my paper submissions and talks, teaching me how to think critically, write comprehensively, and perform impactful research. His motivation for top-notch research and his passion for excellence were a constant source of inspiration and have significantly shaped my research mindset and personality. I am grateful to Prof. Nectarios Koziris for giving the opportunity to be a member of his research laboratory, inspiring me with his incredible passion for teaching and working in the field of computer architecture, as well as his continuous support and the encouraging environment he has provided. My advisors' influence and constant encouragement provided real-life lessons and shaped my personality as a researcher, scientist and engineer.

I thank my committee members, Dionisios Pnevmatikatos, Stefanos Kaxiras, Dimitris Gizopoulos, and Vasileios Papaefstathiou for supervising this thesis. Their feedback and suggestions were valuable to improving my doctoral thesis and its constituent works.

I am grateful to Nandita Vijaykumar for being a great mentor and encouraging me to become a strong and independent researcher. During my visit at the SAFARI research group of ETH Zurich, Nandita helped me to stay motivated, taught me how to find the right research problem to work on and how to perform quality research. I also wholeheartedly thank Athena Elafrou, Foteini Strati, Ivan Fernandez and Thomas Lagos for being my closest collaborators and friends throughout PhD studies, the many long hours of brainstorming and our stimulating discussions. I am very grateful for their endless support, valuable feedback, kindness, positivity, and confidence in myself, as well as our invaluable synergy and friendship.

Furthermore, I thank all the CSLab group members for being great colleagues, supporting my research and enabling a productive working environment. I want to especially thank Kostis Nikas, Vasileios Karakostas, Nikela Papadopoulou and Dimitris Siakavaras for providing significant intellectual and technical support in my research contributions and willingly sharing their expertise with me. I am grateful to all the students and mentees with who I worked closely: Foteini Strati, Athanasios Peppas, and Thrasyvoulos-Fivos Iliadis. Their work significantly helped me in completing my PhD thesis.

I am immensely grateful to all the members of the SAFARI research group for creating a rich, stimulating and highly motivating research environment. During my visit at the SAFARI research group, I realized that innovative research can vastly benefit from close collaboration among the group members. The valuable advice of the SAFARI group members, consisting of concrete guidelines and useful methodologies, helped me to work effectively and efficiently and stay focused in tackling my research obstacles. I want to especially thank Juan Gomez-Luna, Lois Orosa, Konstantinos Kanellopoulos and Nika Mansouri-Ghiasi for their rigorous feedback and criticism on my progress and research, their friendship, their technical and intellectual suggestions, as well as for generously sharing their deep knowledge on the field of computer architecture with me. %I am tremendously lucky to have interacted and worked with the SAFARI research group.

I gratefully acknowledge financial support from my PhD scholarships. Specifically, it was a great honor for me to receive a PhD Fellowship (October 2017 - March 2020) from the General Secretariat for Research and Technology (GSRT) and the Hellenic Foundation for Research and Innovation (HFRI) and a PhD award (September 2021 - October 2022) funded by the Foundation for Education and European Culture (IPEP).

I am immensely grateful to my friends for their support, companionship and patience. I want to particularly thank Katerina Bogiatzoglou, Konstantina Kada, Vicky Routsi, Orestis Alpos, Stamatios Kourkoutas, Stamatios Anoustis, Artemis Zografou, Marina Gourgioti, Katerina Tsesmeli, Isidora Tourni, Athina Kyriakou, Foteini Strati, and Thomas Lagos for our endless conversations, countless laughs, fun nights out and beautiful trips and excursions that were the best discharge from the hard work that I did during my PhD studies.

Last but not least, I am tremendously blessed and would like to express my profound gratitude to my parents, Maria and Christoforos, and my sister, Chara, for their unconditional love and endless encouragement throughout my PhD journey, as well as their valuable support to pursue my academic dreams. I thank my mother for continuously supporting each and every step of this journey. I thank my father for always believing in myself and for helping me with his optimism to pursue my dreams. I thank my sister for her valuable support and patience throughout my PhD studies. This dissertation would not be possible without them. I will be forever grateful to my loving family for the dedication, support, patience, love and opportunities they have given me.

%\selectlanguage{greek}
%\input{sections/abstract_gr}
\selectlanguage{english}
\chapter*{Abstract}
Irregular applications comprise an increasingly important workload domain for many fields, including bioinformatics, chemistry, graph analytics, physics, social sciences and machine learning. Therefore, achieving high performance and energy efficiency in the execution of emerging irregular applications is of vital importance. While there is abundant research on accelerating irregular applications, in this thesis, we identify two critical challenges. First, irregular applications are hard to scale to a high number of parallel threads due to high synchronization overheads. Second, irregular applications have complex memory access patterns and exhibit low operational intensity, and thus they are bottlenecked by expensive data access costs.

This doctoral thesis studies the root causes of inefficiency of irregular applications in modern computing systems, and aims to fundamentally address such inefficiencies, by 1) proposing low-overhead synchronization techniques among parallel threads in cooperation with 2) well-crafted data access policies. Our approach leads to high system performance and energy efficiency on the execution of irregular applications in modern computing platforms, \fix{both processor-centric CPU systems and memory-centric Processing-In-Memory (PIM) systems.}

We make \fix{four major} contributions to accelerating irregular applications in different contexts including CPU and Near-Data-Processing (NDP) (or Processing-In-Memory (PIM)) systems. \fix{First, we design \ColorTM{}, a novel parallel graph coloring algorithm for CPU systems that trades off using synchronization with lower data access costs. \ColorTM{} proposes an efficient data management technique co-designed with a speculative synchronization scheme implemented on Hardware Transactional Memory, and significantly outperforms prior state-of-the-art graph coloring algorithms across a wide range of real-world graphs. Second, we propose \smartpq, an adaptive priority queue that achieves high performance under all various contention scenarios in Non-Uniform Memory Access (NUMA) CPU systems. \smartpq{} tunes itself by dynamically switching between a \notnuma{} and a \numa{} \algomode{} mode, thus providing low data access costs in high contention scenarios, and high levels of parallelism in low contention scenarios. Our evaluations show that \smartpq{} achieves the highest throughput over prior state-of-the-art \numa{} and \notnuma{} concurrent priority queues under various contention scenarios and even when contention varies during runtime. Third, we introduce \SynCron, the first practical and lightweight hardware synchronization mechanism tailored for NDP systems. \SynCron{} minimizes synchronization overheads in NDP systems by (i) adding low-cost hardware support near memory for synchronization acceleration, (ii) directly buffering the synchronization variables in a specialized cache memory structure, (ii) implementing a hierarchical message-passing communication scheme, and (iv) integrating a hardware-only overflow management scheme to avoid performance degradation when hardware resources for synchronization tracking are exceeded. We demonstrate that \SynCron{} outperforms prior state-of-the-art approaches both in performance and energy consumption using a wide range of irregular applications, and has low hardware area and power overheads. Fourth, we design \SparseP, the first library for high-performance Sparse Matrix Vector Multiplication (\spmv) on real PIM systems. \SparseP{} is publicly-available and includes a wide range of data partitioning, load balancing, compression and synchronization techniques to accelerate this irregular kernel in current and future PIM systems. We also extensively characterize the widely used \spmv{} kernel on a real PIM architecture, and provide recommendations for software, system and hardware designers of future PIM systems.}

Overall, we demonstrate that the execution of irregular applications in CPU and NDP/PIM architectures can be significantly accelerated by co-designing lightweight synchronization approaches along with well-crafted data access policies. \fix{Specifically, we show that efficient synchronization and data access techniques can provide high amount of parallelism, low-overhead inter-thread communication and low data access and data movement costs in emerging irregular applications, thus significantly improving system performance and system energy. This doctoral thesis also bridges the gap between processor-centric CPU systems and memory-centric PIM systems in the critically-important area of irregular applications.} We hope that this dissertation inspires future work in co-designing software algorithms with \fix{cutting-edge} computing platforms to significantly accelerate emerging irregular applications.

\vspace{8pt}

\noindent\textbf{Keywords:} Irregular Applications, Synchronization, Efficient Data Access Techniques, Multicore Systems, Processing-In-Memory Architectures
%\selectlanguage{english}
\tableofcontents
\listoffigures
\listoftables

\pagestyle{fancy}
\fancyhead[RE,LO]{\chaptertitlename \hspace{2mm}\thechapter\chaptermark}
\fancyhead[LE,RO]{\thepage}
\fancyheadoffset[RE,LO]{-0.0\textwidth}
\renewcommand{\headrulewidth}{1pt}
\cfoot{}

%\selectlanguage{greek}
%\input{sections/abstract_gr_extended}
\selectlanguage{english}
\chapter{Introduction}\label{IntroChapter}
%\markboth{Introduction}{Introduction}

Irregular applications such as graph processing, data analytics, sparse linear algebra and dynamic pointer-chasing constitute an \fix{important part of software systems we rely on}. These applications lie at the heart of many important \fix{workloads} including deep neural networks~\cite{Han2016Deep,parashar2017scnn,Boroumand2018Google,gao2017tetris,Kim2016Neurocube,Liu2018Processing}, bioinformatics~\cite{Cali2020GenASM,Cali2022SeGram,Kim2017GrimFilter}, databases~\cite{Drumond2017mondrian,Boroumand2019Conda}, data analytics~\cite{Gilbert2006High,Kepner2015GraphsMA,ahn2015scalable,Nai2017GraphPIM,Zhang2018GraphP,Youwei2019GraphQ,Ahn2015PIMenabled,Boroumand2019Conda,boroumand2017lazypim}, large-scale simulations~\cite{Canning1996Tight,Zhang2021Gamma,che2009rodinia,Dongarra2016High,dongarra1996sparse,Ruder2016Overview}, medical imaging~\cite{Elafrou2018SparseX,che2009rodinia,Elafrou2019Conflict}, economic modeling~\cite{Elafrou2018SparseX,che2009rodinia,Elafrou2019Conflict}, and scientific applications~\cite{Elafrou2019Conflict,Dongarra2016High,che2009rodinia,dongarra1996sparse,Ruder2016Overview,Elafrou2018SparseX}. Therefore, optimizing and accelerating irregular applications is of vital importance, and thus a large corpus of research proposes either software designs~\cite{Calciu2017Black,ffwd,Dhulipala2017Julienne,ligra,Welsh1967Upper,Lu2015Balanced,Jones1993Parallel,Deveci2016Parallel,Tas2017Greed,Gebremedhin2000Scalable,Boman2005Scalable,Catalyurek2012GraphColoring,Rokos2015Fast,Giannoula2018Combining,Hasenplaugh2014Ordering,Maciej2020GC,sagonas,sundell,adaptivepq,Wimmer_Martin,rihani,Brodal,Zhang,Sanders1998Randomized,Sagonas2016TheCA,Rab2020NUMA,hotspot,fraser,fomitchev,herlihy,herlihy_art,dick,pim_cds,Pugh1990SkipLists,choe2019concurrent,lotan_shavit,linden_jonsson,spraylist,flatcombining1,flatcombining2,Calciu2013Message,Klaftenegger2014Delegation,Lozi2012Remote,Petrovic2015Performance,Suleman2009Accelerating,blackbox,google,athena,ppopp2018,cluster2018,benatia,Gronquist2021Deep,Memeti2019Using,KusumNG2016Safe,Michie1968Memo,Meng2019APattern,Dhulipala2020APattern,Sedaghati2015Automatic,Benatia2016SparseMF,Pengfei2020LISA,smart,Elafrou2018SparseX,Buluc2011Reduced,Elafrou2017PerformanceAA,Kjolstad2017Taco,Merrill2016Merge,Willcock2006Accelerating,Williams2007Optimization,Namashivayam2021Variable,Tang2015Optimizing,Elafrou2019Conflict,Vuduc2005oski,Elafrou2017PerformanceXeon,Rong2016Sparso,Xie2018CVR,Xiao2021CASpMV,Hou2017Auto,Pinar1999Improving,Liu2013Efficient,Mellor2004Optimizing,Oliker2002Effects,Vuduc2005Fast,Toledo1997Improving,Temam1992Characterizing,Aktemur2018ASM,Zhao2020Exploring,Bolz2003Sparse,Hong2018Efficient,Liu2014AnEfficient,Wu2010Efficient,Guo2014APerformance,su2012ClSpMV,Steinberger2017Globally,Shengen2014YaSpMV,Bell99Implementing,Choi2010Model,Pichel2012Optimization,Sun2011Optimizintg,Vazquez2011New,Yang2011Fast,Elafrou2019BASMAT,Filippone2017Sparse,Lee2008Adaptive,Bisseling2005Communication,Bylina2014Performance,Page2018Scalability,Kayaaslan2015Semi,Liu2018Towards,Catalyurek1999Hypergraph,Vastenhouw2005Two,Nastea1996Load,Pelt2014Medium,Grandjean2012Optimal,Boman2013Scalable} or hardware mechanisms~\cite{Zhang2021Gamma,Mukkara2019PHI,Abeydeera2020Chronos,Giannoula2022SparsePPomacs,Giannoula2022SparsePSigmetrics,ahn2015scalable,Gomez2022Benchmarking,Giannoula2021SynCron,devaux2019,Mutlu2019Processing,Ghose2019Workload,mutlu2020modern,Gomez2021Analysis,Gomez2021Benchmarking,Stone1970Logic,Kautz1969Cellular,shaw1981non,kogge1994,gokhale1995processing,patterson1997case,oskin1998active,kang1999flexram,Mai:2000:SMM:339647.339673,Draper:2002:ADP:514191.514197,aga.hpca17,Eckert2018Neural,Fujiki2019Duality,kang.icassp14,seshadri.micro17,seshadri.arxiv16,Seshadri:2015:ANDOR,Seshadri2013RowClone,Angizi2019GraphiDe,kim.hpca18,kim.hpca19,gao2020computedram,chang.hpca16,Xin2020ELP2IM,li.micro17,deng.dac2018,hajinazarsimdram,Rezaei2020NoM,Wang2020Figaro,Ali2020InMemory,Li2016Pinatubo,angizi2018pima,angizi2018cmp,angizi2019dna,levy.microelec14,kvatinsky.tcasii14,Shafiee2016ISAAC,kvatinsky.iccd11,kvatinsky.tvlsi14,gaillardon2016plim,Bhattacharjee2017ReVAMP,Hamdioui2015Memristor,xie2015fast,hamdioui2017myth,Yu2018Memristive,fernandez2020natsa,Cali2020GenASM,kim.bmc18,Ahn2015PIMenabled,boroumand2017lazypim,Boroumand2019Conda,Chi2016PRIME,Farmahini2015NDA,Gao2015Practical,Gao2016HRL,Gu2016Leveraging,Guo2014APerformance,hashemi2016accelerating,Hsieh2016accelerating,Kim2017GrimFilter,LeBeane2015Data,Nai2017GraphPIM,Drumond2017mondrian,Dai2018GraphH,Zhang2018GraphP,Youwei2019GraphQ,ferreira2021pluto,Olgun2021QuacTrng,Singh2020NEROAN,asghari-moghaddam.micro16,BabarinsaI15,kim2016bounding,morad.taco15,pattnaik.pact16,zhang.hpdc14,Zhu2013Accelerating,denzler2021casper,boroumand2021polynesia,boroumand2021icde,singh2021fpga,singh2021accelerating,herruzo2021enabling,yavits2021giraf,asgarifafnir,seshadri.bookchapter17,diab2022high,diab2022hicomb,fujiki2018memory,zha2020hyper,Saugata2018Enabling,Mutlu2013Memory,Mutlu2014Research,Mutlu2019InDram,SESHADRI2017107,impica,DBLP:conf/isca/AkinFH15,huang2020heterogeneous,santos2017operand,wen2017rebooting,besta2021sisa,lloyd2015memory,elliott1999computational,zheng2016tcam,landgraf2021combining,rodrigues2016scattergather,lloyd2018dse,lloyd2017keyvalue,gokhale2015rearr,jacob2016compiling,sura2015data,nair2015evolution,xi2020memory,Singh2019Napel,pugsley2014ndc,gao2017tetris,Nair2015Active,Balasubramonian2014Near,Asgari2020Alrescha,Hegde2019ExTensor,Zhang2016CambriconX,Pal2018OuterSpace,Nurvitadhi2016Hardware,Eric2020sigma,Zhou2018CambriconS,Mishra2017Fine,zhang2021asplos,zhang2020sparch,parashar2017scnn,Hwang2020Centaur,qin2020sigma,Sadi2019Efficient,Fowers2014AHigh,Grigoras2015Accelerating,Lin2010Design,Umuroglu2014Anenergy,Kanellopoulos2019SMASH,Mukkara2018Exploiting,Nurvitadhi2015Sparse,Cali2022SeGram} to accelerate the execution of such applications.

In this dissertation, we identify three important characteristics of irregular applications that \fix{critically} affect their performance. First, irregular applications exhibit \emph{inherent imbalance} as a result of the real-world input data sets given. Specifically, the concrete pieces involved in the underlying data structures and program data of irregular applications are \emph{not} of equal size. For example, the matrices involved in linear algebra kernels are very sparse, i.e., the vast majority of elements are zeros~\cite{Kanellopoulos2019SMASH,Elafrou2018SparseX,Elafrou2017PerformanceAA,YouTubeGraph,FacebookGraph,Goumas2008Understanding,White97Improving,Helal2021ALTO,Pelt2014Medium,Goumas2009Performance,Elafrou2019BASMAT}, and in most real-world matrices the number of non-zero elements per row shows high disparity and imbalance across the rows of the matrix~\cite{Barabasi2009Scale}. Similarly, the real-world graphs involved in graph processing workloads typically have a power-law distribution, i.e., only a \emph{few} vertices have a very \emph{high} adjacency degree, while the vast majority of the remaining vertices of the graph have a very low adjacency degree~\cite{Giannoula2022SparsePSigmetrics,Giannoula2022SparsePPomacs,Tang2015Optimizing}, \fix{which causes} high disparity and imbalance in the number of edges across vertices. Therefore, \emph{naively} parallelizing such workloads to a large number of threads in modern computing platforms can incur 1) high load imbalance across parallel threads, and 2) high disparity in the amount of computation \fix{versus} memory accesses executed across parallel threads. Second, irregular applications exhibit \emph{random memory access patterns}, i.e., the memory accesses performed are neither sequential nor strided, and they are \emph{input-driven} dependent. Such complex memory access patterns are very hard to predict. Therefore, irregular applications \fix{exhibit} complex data dependencies, poor spatial and temporal data locality, and high data movement overheads to transfer data between memory and processors of commodity computing systems. Third, most irregular applications have \emph{low operational intensity}, i.e., the amount of useful arithmetic operations performed by the processors \fix{compared to} the amount of data necessary to perform these operations is very low. \fix{As a result, irregular applications are memory-bound kernels. They can be significantly bottlenecked by the memory subsystem, incurring high latency costs and excessive memory bandwidth consumption to access data through memory.}

As such, irregular applications constitute an important and emerging workload domain. However, at the same time, it is very challenging to achieve high performance and energy efficiency in the execution of such workloads in modern computing systems \fix{due to the large memory and communication bottlenecks.} Overall, irregular applications have several important inherent characteristics that necessitate new approaches both in the software, i.e., re-designing parallel algorithms, and the hardware level, i.e., re-designing key components of modern computing architectures, to achieve high system performance, \fix{and cooperatively between the software and the hardware.}

\vspace{-10pt}
\section{Motivation: Excessive Synchronization and High Memory Intensity Degrade the Execution of Irregular Applications}
\vspace{-2pt}

Modern computing systems and state-of-the-art parallel algorithms have two important implications that render the efficient execution of irregular applications a significantly challenging task.

\begin{figure}[H]
    %\vspace{4pt}
    \centering
    \includegraphics[width=0.98\linewidth]{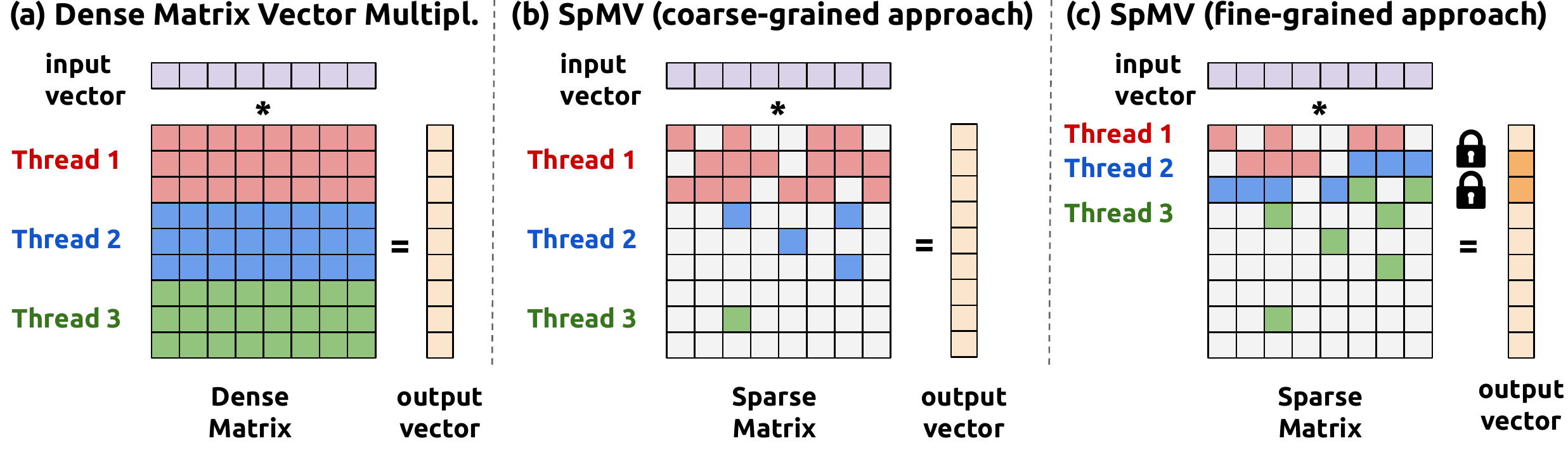}
    \caption{(a) Dense Matrix Vector Multiplication using three parallel threads. (b) Sparse Matrix Vector Multiplication with a coarse-grained parallelization strategy among three parallel threads. (c) Sparse Matrix Vector Multiplication with a fine-grained parallelization strategy among three parallel threads. The colored cells of each matrix represent non-zero elements.}
    \label{fig:intro-sync-example}
    \vspace{-10pt}
\end{figure}

\noindent\textbf{Implication 1: Excessive Synchronization.}
To achieve high system performance in a multithreaded execution context, load balance among parallel threads is \fix{critical}.  \fix{Therefore, software engineers employ} a \emph{fine-grained} parallelization strategy among parallel threads in irregular applications due to the inherent imbalance exhibited in the input data sets involved.  For example, Figure~\ref{fig:intro-sync-example} compares a \emph{regular} Dense Matrix Vector Multiplication (DEMV) with an \emph{irregular} Sparse Matrix Vector Multiplication (\spmv{}). In the DEMV execution, \fix{a coarse-grained parallelization strategy (Figure~\ref{fig:intro-sync-example}a), in which the rows of the matrix are equally distributed across parallel threads, can easily achieve high load balance.} However, using a coarse-grained parallelization strategy to parallelize the irregular \spmv{} kernel (Figure~\ref{fig:intro-sync-example}b) results in significantly high load imbalance among parallel threads, due to the high disparity in the number of non-zero elements processed across parallel threads, causing \fix{large} performance overheads. Thus, a fine-grained parallelization strategy among parallel threads is necessary, e.g., Figure~\ref{fig:intro-sync-example}c. Unfortunately, this approach however results in excessive and frequent synchronization among parallel threads. In the example of the irregular \spmv{} kernel, \fix{with a fine-grained parallelization strategy, parallel threads that process non-zero elements of the \emph{same} row of the matrix (Figure~\ref{fig:intro-sync-example}c), use synchronization primitives (e.g., locks, mutexes) to ensure atomicity and correctness, when performing write updates on the \emph{same} element of the output vector.} Therefore, a large amount of processor cycles is spent on communication and synchronization with significant performance overheads~\cite{David2013Everything,Goumas2008Understanding,Elafrou2019Conflict,Macieh2017pushpull,ligra}.

At the application level, existing fine-grained parallel algorithms (e.g.,~\cite{lotan_shavit,linden_jonsson,sagonas,sundell,spraylist,adaptivepq,Wimmer_Martin,rihani,Brodal,Zhang,Sanders1998Randomized,Sagonas2016TheCA,Rab2020NUMA}) typically \emph{lack} well-tuned synchronization implementations~\cite{Roghanchi2017ffwd,Calciu2017Black}, and/or do \emph{not} \fix{achieve high system performance under all \emph{various} contention scenarios.} Recent works~\cite{David2013Everything,Antic2016Locking,Roghanchi2017ffwd,Calciu2017Black} demonstrate that (i) naive synchronization schemes used in irregular applications can cause high memory traffic with significant latency access costs, and (ii) the best-performing synchronization scheme varies depending on the \fix{levels of contention among parallel threads and the characteristics of the underlying hardware platform.} At the architecture level, even though numerous hardware synchronization mechanisms have been proposed~\cite{Liang2015MISAR,Zhu2007SSB,abell2011glocks,sampson2006exploiting,abellan2010g,oh2011tlsync,Sergi2016WiSync,akgul2001system,Vallejo2010Architectural,Leiserson1992CM5,Kaxiras2010SARC,Choi2011DeNovo,Kaxiras2013ANew,Sung2014DeNovoND,Lebeck1995Dynamic,Ros2012Complexity,Laudon1997SGI,Alverson1990Tera,gottlieb1998NYU}, most of them incur high hardware cost to be implemented in commodity systems, require important cross-stack modifications and/or have narrow programming interfaces, \fix{and thus they are} hard to adopt.

\noindent\textbf{Implication 2: High Memory Intensity.} 
Irregular applications involve random memory access patterns, have low operational intensity and are primarily bottlenecked by the memory subsystem~\cite{Gomez2021Benchmarking,Gomez2022Benchmarking,Elafrou2017PerformanceAA,Elafrou2018SparseX,Goumas2008Understanding,Goumas2009Performance,Elafrou2019Conflict,ahn2015scalable,Nai2017GraphPIM,Zhang2018GraphP,Youwei2019GraphQ,Ahn2015PIMenabled,Boroumand2019Conda,boroumand2017lazypim,Boroumand2018Google}. Thus, irregular applications incur high memory intensity with significant data access costs, and a large fraction of the application's execution time is spent on data accesses and/or waiting for data to be transferred between memory and processors. Things become even worse \fix{with the large growth in input data set sizes as well as} intermediate data used and generated during runtime. Therefore, irregular applications need to process increasingly large \fix{volumes} of data (input data sets with tens or hundreds of GBs memory footprints~\cite{Satish2014Navigating,Dhulipala2017Julienne}), and need to effectively handle the high data demand.

\begin{figure}[H]
    \centering
    \includegraphics[width=0.98\linewidth]{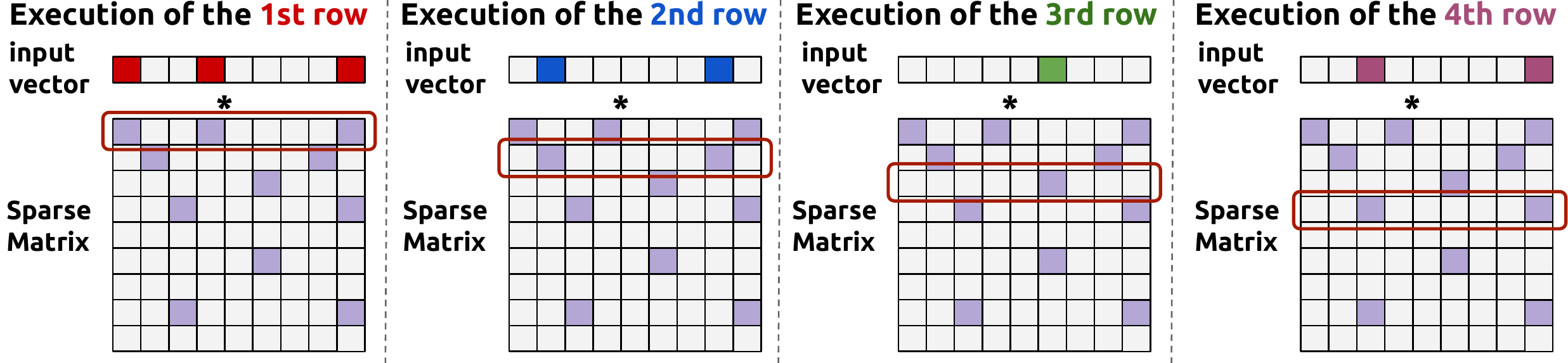}
    \vspace{-4pt}
    \caption{An example \spmv{} execution on the first four rows of a sparse 9$\times$9 matrix with only 10 non-zero elements. The execution steps are performed at a row granularity. The colored cells of the matrix with purple color represent non-zero elements, and the colored cells of the input vector represent the elements of the input vector that are processed/accessed at each execution step.}
    \label{fig:intro-mem-example}
    \vspace{-12pt}
\end{figure}

We \fix{demonstrate the aforementioned critical performance implication with an example, i.e., the \spmv{} kernel execution.} The \spmv{} kernel performs $\mathcal{O}(NNZ)$ operations on $\mathcal{O}(N+NNZ)$ amount of data (assuming a square matrix), where $NNZ$ is the number of  non-zero elements of the input matrix and $N$ is the number of columns of the input matrix (equal to the number of elements of the input vector). However, the real-world matrices involved are very sparse~\cite{Kanellopoulos2019SMASH,Elafrou2018SparseX,Elafrou2017PerformanceAA,YouTubeGraph,FacebookGraph,Goumas2008Understanding,White97Improving,Helal2021ALTO,Pelt2014Medium}. For instance, the matrices that represent Facebook’s and YouTube’s network connectivity contain \emph{only} 0.0003\%~\cite{YouTubeGraph,Kanellopoulos2019SMASH} and 2.31\%~\cite{FacebookGraph,Kanellopoulos2019SMASH} non-zero elements, respectively. Figure~\ref{fig:intro-mem-example} presents an example \spmv{} execution on the first four rows of a sparse 9$\times$9 matrix with only 10 non-zero elements, i.e., having $\sim$0.17 operational intensity when assuming single precision non-zero elements (i.e., integers). As shown in Figure~\ref{fig:intro-mem-example}, the accesses to the elements of the input vector are random and irregular, and they depend on the sparsity pattern of the matrix that is given as input. The data accesses to the elements of the input vector are very hard to predict, since they are affected by the particular characteristics of the input matrix, and are typically performed using the main memory of commodity systems, \fix{which often has high latency and low bandwidth~\cite{ghoseibm2019,Mutlu2014Research}.} Thus, \spmv{} execution is highly limited by the irregular data accesses to the elements of the input vector and the data movement costs of accessing the elements of the input vector, which cause significant performance overheads in the total execution time~\cite{Elafrou2018SparseX,Shengen2014YaSpMV,Elafrou2017PerformanceAA,Goumas2009Performance,Karakasis2009Performance,Vuduc2005Fast,im2004sparsity,Vuduc2003PhD,Vuduc2002Performance,White97Improving,Elafrou2017PerformanceXeon,Elafrou2019BASMAT,Kanellopoulos2019SMASH}.

Two recent works~\cite{Boroumand2018Google,Boroumand2021Google} explain that the energy overheads of data movement across the memory hierarchy of commodity systems can be significantly higher than that of computation in irregular applications. \fix{First, Boroumand et al.~\cite{Boroumand2018Google} show that moving data
between memory and processors causes more than 60\% of the total system energy efficiency in several irregular Google's consumer workloads. Second, Boroumand et al.~\cite{Boroumand2021Google} demonstrate that the commercial Google Edge TPU unit~\cite{GoogleTPU} spends 50.3\% of its total energy on off-chip memory accesses across a wide range of irregular machine learning applications, including convolutional neural networks, transducers and recurrent neural networks. Multiple other works (e.g.,~\cite{Oliveira2021Damov,Yan2015Characterizing}) provide analysis of data movement bottlenecks in a variety of  irregular workloads. Therefore, we conclude that the high memory intensity of irregular applications causes significant bottlenecks and high overheads both in performance and energy consumption.}

At the application level, many parallel applications and software packages do \emph{not} handle data well (e.g.,~\cite{Gebremedhin2000Scalable,Boman2005Scalable,Catalyurek2012GraphColoring,Rokos2015Fast,Lu2015Balanced,lotan_shavit,linden_jonsson,sagonas,sundell,spraylist}), and/or do \emph{not} adapt their parallelization strategies to the particular characteristics of the input data given. Recent works~\cite{Vijaykumar2018Locality,Vijaykumar2018Xmem,Luo2014Characterizing,Koppula2019Eden,hajinazar2020virtual,Vijaykumar2022MetaSys} highlight that different pieces of program data have different performance characteristics (latency/bandwidth/parallelism sensitivity), and inherent properties. Consequently, \emph{data-oblivious} policies, i.e., \fix{policies that} are designed without taking into consideration the properties of the application data they handle, result in lost performance optimization opportunities, which could be achieved by exploiting data properties. Similarly, at the architecture level, existing hardware mechanisms (e.g.,~\cite{Siakavaras2017Combining,Brown2016Investigating,Herlihy1993Transactional,Yoo2013Performance}) are designed without considering modern applications’ memory access patterns and overwhelming data demand, \fix{and as such, they cause} frequent data movement across the entire system and significant data access costs.

\vspace{-12pt}
\section{Our Approach: Efficient Synchronization and Data Access Techniques for Irregular Applications}
\vspace{-4pt}
In this dissertation, we study a wide range of irregular applications, including graph processing, data analytics, pointer-chasing and sparse linear algebra, and explore their performance implications on two modern computing platforms: (i) the \emph{processor-centric Non-Uniform Memory Access (NUMA) CPU} architectures, and (ii) the \emph{memory-centric Processing-In-Memory (PIM)} (or Near-Data-Processing (NDP)) architectures. The NUMA CPU architectures constitute the dominant hardware platform in today's computing systems, and have been significantly optimized over \fix{decades} to integrate general-purpose cores with high computation capability. The PIM/NDP architectures have been recently  commercialized~\cite{upmem,devaux2019,Gomez2021Benchmarking,Gomez2022Benchmarking,Gomez2021Analysis,Mutlu2019Processing,Hadi2016Chameleon,Lee2021HardwareAA,Kwon2021Function}, and represent a promising \fix{disruptive} paradigm to alleviate the costs of data movement across the memory hierarchy. PIM/NDP architectures equip memory chips with a large number of low-area and low-power cores with relatively low computation capability, and alleviate data movement overheads by performing computation close to where the application data resides. Therefore, PIM/NDP architectures provide high levels of parallelism and very large memory bandwidth.

We \fix{posit} that, moving forward, both hardware mechanisms and parallel algorithms need to consider the synchronization needs and memory access patterns of irregular applications as the two major priorities to significantly improve system performance and system energy efficiency, when employing hundreds or thousands of parallel threads. In particular, modern software and hardware designs for irregular applications should provide two \fix{major types of optimization approaches: (1) efficient synchronization, and (2) efficient data access techniques.}

\noindent\textbf{Efficient Synchronization Techniques.} Modern computing platforms need to support low-cost and practical hardware synchronization mechanisms, \fix{while} parallel algorithmic designs need to provide fine-grained communication and adaptive synchronization approaches among parallel threads to significantly accelerate the execution of irregular applications. Lightweight synchronization techniques are highly effective at execution of irregular applications, since they improve system performance and energy efficiency by (1) mitigating coherence/communication traffic overheads caused when synchronizing hundreds or thousands of parallel threads, and (2) exposing high levels of fine-grained parallelism thanks to enabling low-cost communication and coordination among parallel threads. \fix{We demonstrate the benefits of efficient synchronization in four different contexts. First, we design a \emph{speculative} synchronization scheme for the widely used graph coloring kernel~\cite{Welsh1967Upper} (Chapter~\ref{ColorTMChapter}), which \emph{speculatively} performs most computations and data accesses \emph{outside} the critical section, and thus effectively minimizes synchronization costs and provides high levels of parallelism on the graph coloring kernel by enabling short critical sections with small memory footprints. Second, we propose an \emph{adaptive} concurrent priority queue (Chapter~\ref{SmartPQChapter}), which dynamically tunes its parallelization strategy between two algorithmic modes (a \numa{} and a \notnuma{} mode) \emph{without} using barrier synchronization, and thus achieving \emph{negligible} synchronization costs upon transitions. Third, we introduce a low-cost and practical hardware synchronization mechanism tailored for NDP architectures (Chapter~\ref{SynCronChapter}), which significantly improves system performance and system energy efficiency in a wide variety of irregular parallel applications including pointer chasing, graph processing, and time series analysis. Fourth, we implement three synchronization schemes among parallel threads of a multithreaded PIM core in the \spmv{} kernel (Chapter~\ref{SparsePChapter}), and show that the \emph{lock-free} synchronization scheme provides significant performance benefits over the \emph{lock-based} synchronization schemes in a real PIM system, by providing higher amount of parallelism among parallel threads.}

\noindent\textbf{Efficient Data Access Techniques.} Modern computing systems need to eliminate data movement overheads, \fix{while} parallel algorithmic designs need to provide well-crafted data distribution and data-aware parallelization strategies (by exploiting properties of data), as well as adaptive cache and memory management techniques (by leveraging characteristics of the underlying hardware), \fix{in order to minimize data access costs in the execution of irregular applications.} \textit{Data-aware} parallel algorithms and \emph{memory-centric} architectures can significantly improve performance and energy efficiency in the execution of irregular applications by (1) reducing data access and communication costs, and (2) better leveraging the available memory bandwidth of the underlying hardware to increase the efficiency of the \fix{application execution}. \fix{We demonstrate the benefits of efficient data access techniques in four different contexts. First, we propose an \emph{eager} coloring conflict policy to detect and resolve coloring inconsistencies arised among parallel threads in the graph coloring kernel~\cite{Welsh1967Upper} (Chapter~\ref{ColorTMChapter}), which effectively reduces data access costs by accessing conflicted vertices \emph{immediately} using the low-cost  on-chip caches of multicore CPU platforms. Second, we design (i) a concurrent priority queue (Chapter~\ref{SmartPQChapter}) having a parallelization strategy that is aware of the non-uniform distribution (\numa{}) of the underlying data structure in a NUMA CPU architecture, and thus achieves higher performance benefits (by minimizing data access costs) in high-contention scenarios over state-of-the-art concurrent priority queues~\cite{spraylist,lotan_shavit}, which are oblivious to the non-uniform distribution (\notnuma{}) of the underlying data structure in a NUMA CPU architecture, and we also propose (ii) an \emph{adaptive} concurrent priority queue (Chapter~\ref{SmartPQChapter}), which \emph{dynamically} tunes its parallelization strategy between a \numa{} and a \notnuma{} \algomode{} mode depending on the contention levels of the current workload, and thus provides high throughput in NUMA CPU systems under all various contention scenarios (by reducing data access costs and achieving high amount of parallelism). Third, we add a specialized low-cost cache memory structure inside synchronization accelerators for NDP systems (Chapter~\ref{SynCronChapter}) to \emph{directly} buffer synchronization variables, and thus minimize latency accosts costs by avoiding costly memory accesses for synchronization. Fourth, we introduce various data partitioning techniques to efficiently map the irregular \spmv{} execution kernel on near-bank PIM systems~\cite{upmem,devaux2019,Lee2021HardwareAA} (Chapter~\ref{SparsePChapter}), and show that the best-performing \spmv{} execution on a near-bank PIM system~\cite{upmem,devaux2019} (Chapter~\ref{SparsePChapter}) is achieved using intelligent \emph{data-aware} algorithmic designs that (i) trade off computation for lower data movement overheads, and (ii) select their parallelization strategy and data partitioning policy based on the particular sparsity pattern of the input matrix, i.e., by exploiting properties of the input data. We also observe that the memory-bound \spmv{} kernel on a state-of-the-art \emph{memory-centric} PIM system achieves a much higher fraction of the machine’s peak performance compared to that on state-of-the-art \emph{processor-centric} CPU and GPU systems (Chapter~\ref{SparsePChapter}).}

\vspace{-10pt}
\subsection{Thesis Statement}

In this dissertation, we propose parallelization techniques and algorithmic designs, along with hardware mechanisms that enable lightweight synchronization and \fix{low-cost data accesses} in emerging irregular applications running in processor-centric CPU and memory-centric PIM/NDP architectures. \fix{Specifically, we propose (1) a novel parallel algorithm that minimizes synchronization and data access costs in the graph coloring kernel execution on CPU systems, (2) an adaptive concurrent priority queue that provides high amount of parallelism and minimizes memory traffic in NUMA CPU systems, (3) an end-to-end hardware mechanism that enables low-cost synchronization in NDP systems, and (4) several efficient algorithmic designs that provide low synchronization and data transfer costs in the \spmv{} kernel execution on real near-bank PIM systems.}

This dissertation, hence, provides \fix{substantial} evidence for the following thesis statement:\\
\vspace{-16pt}

\begin{tcolorbox}[width=0.96\columnwidth,colback=white,colframe=white,halign=left]
\fix{\textbf{\textit{Low-overhead synchronization approaches in cooperation with well-crafted data access techniques can significantly improve performance and energy efficiency of important data-intensive irregular applications.}}}
\end{tcolorbox}
%data-intensive

\vspace{-18pt}
\section{Overview of Our Research}
\vspace{-4pt}
We propose four new approaches to accelerate irregular applications in CPU and PIM/NDP architectures via efficient synchronization and data access techniques, which we briefly describe next. We also put our contributions in the context of relevant prior work and provide detailed discussions of and comparisons to prior work in Chapters~\ref{ColorTMChapter}-~\ref{SparsePChapter}.

\vspace{-8pt}
\subsection{\ColorTM~\cite{Giannoula2018Combining,colortmGithub,Giannoula2022ColorTM}: High-Performance and Balanced Parallel Graph Coloring on Multicore CPU Platforms (Chapter~\ref{ColorTMChapter})}

Graph coloring is an important graph processing kernel, and it is widely used in many real-world end-applications including the conflicting job scheduling~\cite{Welsh1967Upper,Marx2004Graph,Arkin1987Scheduling,Marx2004GRAPHCP,Ramaswami1989Distributed}, register allocation~\cite{Chaitin1981Register,Chaitin1982Register,Briggs1994Improvements,Chen2018Register,Cohen2010Processor} and sparse linear algebra~\cite{Coleman1983Estimation,Saad1994SparseKit,Jones1993Efficient,Gebremedhin2005WhatColor}. The total runtime of the graph coloring kernel typically adds to the overall parallel overhead of the real-world end-application, and thus a high-performance parallel graph coloring algorithm for modern multicore platforms is necessary. Prior works~\cite{Catalyurek2012GraphColoring,Gebremedhin2000Scalable,Rokos2015Fast,Boman2005Scalable,Lu2015Balanced} that parallelize the graph coloring kernel are still inefficient (as we demonstrate in Chapter~\ref{ColorTMChapter}), because they detect and resolve the coloring inconsistencies arised among parallel threads with a \emph{lazy} approach: they detect and resolve the coloring inconsistencies much later in the runtime compared to the time that the coloring inconsistencies appeared. As a result, prior approaches access the conflicted vertices of the graph multiple times, mainly using the expensive last levels of the memory hierarchy (e.g., main memory) of commodity multicore platforms, thus incurring high data access costs.

To this end, we design \ColorTM{}~\cite{Giannoula2018Combining,colortmGithub,Giannoula2022ColorTM}, a high-performance parallel graph coloring algorithm for multicore platforms. \ColorTM{} is designed to provide both low synchronization overheads and low data access costs via two key techniques. First, we introduce an \emph{eager} conflict detection and resolution approach of the coloring inconsistencies arised among parallel threads: coloring inconsistencies are immediately detected and resolved at the time they appear. This way in \ColorTM{}, the conflicted vertices are accessed multiple times, using the low-cost lower levels of the memory hierarchy of multicore platforms, thus achieving low data access costs. Second, we design a \emph{speculative} computation and synchronization scheme, in which parallel threads speculatively perform computations and data accesses \emph{outside} the critical section to enable short critical sections with small memory footprints. This key technique provides high levels of parallelism and low synchronization costs by executing multiple small and short critical sections in parallel. Next, we extend our algorithmic design to propose a \emph{balanced} parallel graph coloring algorithm, named \BalColorTM{}~\cite{colortmGithub}, in which the vertices of the graph are \emph{almost equally} distributed across the color classes produced. Enabling color classes with almost equal sizes can provide  high hardware resource utilization and high load balance among parallel threads in the real-world end-applications of graph coloring.

We evaluate \ColorTM{} and \BalColorTM{} on a modern multicore platform using a wide variety of large real-world graphs with diverse characteristics. In Chapter~\ref{ColorTMChapter}, we show that \ColorTM{} and \BalColorTM{} significantly outperform prior state-of-the-art graph coloring algorithms~\cite{Catalyurek2012GraphColoring,Gebremedhin2000Scalable,Rokos2015Fast,Boman2005Scalable,Lu2015Balanced}, while also achieving high coloring quality. We also demonstrate that \ColorTM{} and \BalColorTM{} can provide significant performance improvements in real-world end-applications, e.g., Community Detection~\cite{FORTUNATO201075}. \ColorTM{} and \BalColorTM{} are freely and publicly available~\cite{colortmGithub} at {\href{https://github.com/cgiannoula/ColorTM}{\textcolor{blue}{github.com/cgiannoula/ColorTM}}} to enable further research on the graph coloring kernel in modern multicore systems.

\vspace{-10pt}
\subsection{\smartpq~\cite{Strati2019AnAdaptive}: An Adaptive Concurrent Priority Queue for NUMA CPU Architectures (Chapter~\ref{SmartPQChapter})}

Concurrent priority queues lie at the heart of many important applications including graph processing~\cite{Kolmogorov,Lasalle,Thorup,CLRS,Prim} and discrete event simulations~\cite{Tang,Marotta,Xu}. Prior works~\cite{lotan_shavit,linden_jonsson,sagonas,sundell,Wimmer_Martin,rihani,Brodal,Zhang,spraylist,Heidarshenas2020Snug,blackbox,ffwd} have designed concurrent priority queues for modern NUMA architectures. These implementations for concurrent priority queues are typically of two types: (i) \notnuma{} concurrent priority queues~\cite{lotan_shavit,linden_jonsson,sagonas,sundell,Wimmer_Martin,rihani,Brodal,Zhang,spraylist,Heidarshenas2020Snug}, in which the parallelization strategy implemented is \emph{oblivious} to the non-uniform memory access costs of the underlying memory subsystem, and (ii) \numa{} concurrent priority queues~\cite{blackbox,ffwd}, in which the parallelization strategy implemented takes into consideration the non-uniform memory access costs of the underlying memory subsystem. We examine prior state-of-the-art concurrent priority queues~\cite{spraylist,lotan_shavit,ffwd} on a NUMA CPU system using a wide variety of contention scenarios, and find that none of the prior state-of-the-art concurrent priority queues performs best across all various contention scenarios. Specifically, \notnuma{} concurrent priority queues provide high levels of parallelism, low data access costs, and high performance in \insrt-dominated scenarios, which typically exhibit low-contention, since parallel threads may work on different parts of the underlying data structure. In contrast, \notnuma{} concurrent priority queues cause high data movement traffic in the memory subsystem of a NUMA architecture, and incur significant performance slowdowns over the \numa{} concurrent priority queues in \delete{}-dominated workloads, which exhibit very high contention, since parallel threads highly compete to remove the first few elements of the underlying data structure.

To this end, we propose \smartpq{}~\cite{Strati2019AnAdaptive}, an adaptive concurrent priority queue for NUMA architectures that achieves the highest performance in all different contention scenarios, and even when the contention of the workload varies over time. \smartpq{} is designed to provide high levels of parallelism and low data access and data movement costs under all various contention scenarios. To achieve this, \smartpq{} \emph{dynamically} adapts itself over time between a \notnuma{} and a \numa{} \algomode{} mode depending on the contention levels of the workload. \smartpq{} integrates (i) \textit{NUMA Node Delegation} (\nuddle{}), a \emph{generic} framework to wrap \emph{any} arbitrary \notnuma{} concurrent data structure, and transform it into its \numa{} counterpart, and (ii) a simple decision tree \emph{classifier}, which predicts the best-performing \algomode{} mode (between a \notnuma{} and a \numa{} \algomode{} mode) given the current contention levels of the workload. This way \smartpq{} uses the \numa{} \nuddle{} priority queue in \delete{}-dominated workloads, and switches to \emph{directly} using the \nuddle{}'s underlying \notnuma{} concurrent priority queue in \insrt-dominated scenarios, thus enabling low data access costs in all various contention scenarios.

We evaluate \smartpq{} in a modern NUMA CPU system using a wide range of contention scenarios, and also using synthetic benchmarks that \emph{dynamically} vary the contention of the workload over time. In Chapter~\ref{SmartPQChapter}, we demonstrate that \smartpq{} achieves the highest performance over prior state-of-the-art \notnuma{} and \numa{} concurrent priority queues~\cite{spraylist,lotan_shavit,ffwd} in \emph{all various} contention scenarios and at \emph{any} point in time with 87.9\% success rate.

\vspace{-10pt}
\subsection{\SynCron~\cite{Giannoula2021SynCron}: Efficient Synchronization Support for NDP Architectures (Chapter~\ref{SynCronChapter})}

NDP architectures~\cite{mutlu-isca2008,mutlu2020modern,Mutlu2019Processing,ahn2015scalable,Ahn2015PIMenabled,Balasubramonian2014Near} alleviate the expensive data movement between processors and memory by performing computation close to where the application data resides. Typical NDP architectures support several NDP units connected to each other, with each unit comprising multiple NDP cores close to memory~\cite{Kim2013memory,Youwei2019GraphQ,Tsai2018Adaptive,ahn2015scalable,Hsieh2016TOM,Boroumand2018Google,Zhang2018GraphP}. Therefore, NDP architectures provide high levels of parallelism, low memory access latency, and large aggregate memory bandwidth, thereby being a very good fit to accelerate irregular applications such as genome analysis~\cite{Kim2017GrimFilter,Cali2020GenASM}, graph processing~\cite{ahn2015scalable,Nai2017GraphPIM,Zhang2018GraphP,Youwei2019GraphQ,Ahn2015PIMenabled,Boroumand2019Conda,boroumand2017lazypim}, databases~\cite{Drumond2017mondrian,Boroumand2019Conda}, pointer-chasing workloads~\cite{liu2017concurrent,choe2019concurrent,Hsieh2016accelerating,hashemi2016accelerating}, and sparse neural networks~\cite{Boroumand2018Google,gao2017tetris,Kim2016Neurocube,Liu2018Processing}. However, to fully leverage the benefits of NDP architectures for these irregular parallel applications, an effective synchronization solution for NDP systems is necessary.

Numerous prior works~\cite{abell2011glocks,sampson2006exploiting,abellan2010g,oh2011tlsync,Sergi2016WiSync,akgul2001system,Zhu2007SSB,Vallejo2010Architectural,Liang2015MISAR,yilmazer2013hql,eltantawy2018warp,kessler1993crayTA,scott1996synchronization,Laudon1997SGI,zhang2004highly,gottlieb1998NYU,Alverson1990Tera,Jordan1983Performance,Smith1978Pipelined,Dally1992TheMessage,Keckler1998Exploiting,Goodman1989EfficientSP,Lenoski1992Dash,Leiserson1992CM5} propose synchronization solutions for \emph{processor-centric} CPU, GPU and Massively Parallel Processing (MPP) systems. However, these synchronization solutions are not efficient or suitable for the \emph{memory-centric} NDP systems (Chapter~\ref{SynCronChapter}), which are fundamentally different from commodity \emph{processor-centric} systems. First, synchronization approaches for CPU systems are typically implemented upon the underlying hardware cache coherence protocols, but most NDP systems do \emph{not} support hardware cache coherence (e.g.,~\cite{Tsai2018Adaptive,Ghose2019Workload,ahn2015scalable,Zhang2018GraphP,Youwei2019GraphQ}). Second, synchronization in GPUs and MPPs is implemented in dedicated hardware atomic units, known as \emph{remote atomics}. However, synchronization using remote atomics has been shown to be inefficient, since it causes high global traffic and hotspots~\cite{Wang2019Fast,li2015fine,yilmazer2013hql,eltantawy2018warp,Mukkara2019PHI}. Finally, prior hardware synchronization mechanisms~\cite{abell2011glocks,abellan2010g,oh2011tlsync,Zhu2007SSB,Vallejo2010Architectural,Liang2015MISAR,Leiserson1992CM5,Sergi2016WiSync} tailored for commodity processor-centric systems are not suitable for memory-centric NDP systems, because they would either incur high hardware costs to be implemented in large-scale NDP systems (e.g., ~\cite{abell2011glocks,abellan2010g,oh2011tlsync,Sergi2016WiSync}) or cause excessive network traffic across the NDP units of the system with significant performance overheads upon contention (e.g., ~\cite{Zhu2007SSB,Vallejo2010Architectural,Liang2015MISAR,Leiserson1992CM5}).

To this end, we design \SynCron{}~\cite{Giannoula2021SynCron}, a low-overhead hardware synchronization mechanism tailored for memory-centric NDP architectures. \SynCron{} consists of four key techniques. First, we offload synchronization among NDP cores to dedicated low-cost hardware units to avoid the need for complex coherence protocols and expensive atomic operations. Second, we directly buffer the synchronization variables in a specialized cache memory structure to avoid costly memory accesses for synchronization. Third, we coordinate synchronization among NDP cores of several NDP units via a hierarchical \mpsync{} scheme to minimize synchronization traffic across NDP units of the system under high-contention scenarios. Fourth, when applications with frequent synchronization oversubscribe the hardware synchronization resources, we use an efficient and programmer-transparent overflow management scheme to avoid costly fallback solutions and minimize overheads.

In Chapter~\ref{SynCronChapter}, we demonstrate that \SynCron{} achieves the goals of performance, cost, programming ease, and generality by covering a wide range of synchronization primitives. In addition, we show that \SynCron{} significantly improves system performance and energy efficiency across a wide range of irregular applications including pointer-chasing, graph applications, and time series analysis, while also has low area and power overheads to be integrated into the compute die of NDP units.

\vspace{-10pt}
\subsection{\SparseP~\cite{SparsePLibrary,giannoula2022sigmetrics,Giannoula2022SparseParXiv,Giannoula2022SparsePExarXiv,Giannoula2022SparsePPomacs,Giannoula2022SparsePISBLSI}: Towards Efficient Sparse Matrix Vector Multiplication on Real PIM Architectures (Chapter~\ref{SparsePChapter})}

The \spmv{} kernel has been characterized as one of the most thoroughly studied scientific computation kernels~\cite{Goumas2008Understanding,Elafrou2018SparseX}, and is a fundamental linear algebra kernel for important applications from the scientific computing, machine learning, graph analytics and high performance computing domains. In commodity processor-centric systems like CPU and GPU systems, \spmv{}  is a memory-bandwidth-bound kernel for the majority of real sparse matrices, and is bottlenecked by data movement between memory and processors~\cite{Gomez2021Benchmarking,Elafrou2018SparseX,Elafrou2017PerformanceAA,Elafrou2019Conflict,Xie2021SpaceA,Goumas2009Performance,Pal2018OuterSpace,Gomez2021Analysis,Vuduc2002Performance,Vuduc2003PhD,Vuduc2005Fast,Karakasis2009Performance,dongarra1996sparse,im2004sparsity,Liu2018Towards,Kourtis2011CSX,Goumas2008Understanding,Kourtis2008Optimizing,Elafrou2017PerformanceXeon}. To alleviate the data movement bottleneck, several manufacturers have already started to commercialize near-bank PIM architectures~\cite{Lee2021HardwareAA,upmem,Hadi2016Chameleon,devaux2019,Kwon2021Function,Gomez2021Analysis,Gomez2021Benchmarking,Gu2020iPIM,Cho2020Near,Cho2021Accelerating,Kumar2020Parallel,Nag2021OrderLight,Park2021Trim,Sadredini2021Sunder,Gu2021DLUX,Aga2019coml,Shin2018MCDRAM,Cho2020MCDRAM,Yazdanbakhsh2018InDRAM,Alves2015Opportunities,li2017drisa}, after decades of research efforts. \textit{Near-bank} PIM designs tightly couple a PIM core with each DRAM bank, exploiting bank-level parallelism to expose high on-chip memory bandwidth of standard DRAM to processors. Two \textit{real} near-bank PIM architectures are Samsung's FIMDRAM~\cite{Lee2021HardwareAA,Kwon2021Function} and the UPMEM PIM system~\cite{upmem2018,devaux2019,Gomez2021Analysis,Gomez2021Benchmarking}.

Recent works leverage near-bank PIM architectures to provide high performance and energy benefits on bioinformatics~\cite{lavenier2020Variant,Gomez2021Benchmarking,Gomez2021Analysis,Lavenier2016DNA}, skyline
computation~\cite{Zois2018Massively}, compression~\cite{Nider2020Processing} and neural network~\cite{Lee2021HardwareAA,Gomez2021Benchmarking,Gomez2021Analysis,Gu2020iPIM,Cho2021Accelerating} kernels. A recent study~\cite{Gomez2021Analysis,Gomez2021Benchmarking} provides PrIM benchmarks~\cite{PrIMLibrary}, which are a collection of 16 kernels for evaluating near-bank PIM architectures, like the UPMEM PIM system. Similarly, a recent work~\cite{Gomez2022Machine} implements several machine learning kernels, i.e., linear regression, logistic regression,
decision tree, k-means clustering, on the UPMEM PIM system to understand the potential of a modern general-purpose PIM architecture to accelerate machine learning training. However, there is \emph{no} prior work to efficiently map the \spmv{} execution kernel on near-bank PIM systems, and thoroughly study the widely used, memory-bound \spmv{} kernel on a real PIM system.

To this end, we design efficient \spmv{} algorithms tailored for current and future real PIM systems, which are publicly available in the \SparseP{} software package~\cite{SparsePLibrary,giannoula2022sigmetrics,Giannoula2022SparseParXiv,Giannoula2022SparsePExarXiv,Giannoula2022SparsePPomacs,Giannoula2022SparsePISBLSI}. The \SparseP{} software package includes 25 \spmv{} kernels for real PIM systems, which are designed to provide high levels of parallelism, low synchronization costs, low data movement overheads, as well as to effectively leverage the immense memory bandwidth supported in near-bank PIM architectures. Specifically, \SparseP{} supports (1) the four most popular compressed matrix formats, (2) a wide range of data types, (3) two types of well-crafted data partitioning techniques of the sparse matrix to DRAM banks of PIM-enabled memory modules, (4) various load balancing schemes across thousands of PIM cores, (5) various load balancing schemes across several threads of a multithreaded PIM core, and (6) three synchronization approaches among threads within multithreaded PIM core.

We conduct an extensive and comprehensive study of \SparseP{} kernels on the memory-centric UPMEM PIM system~\cite{Gomez2022Benchmarking,Gomez2021Analysis,devaux2019,Gomez2021Benchmarking}, the first publicly-available real-world PIM architecture. In Chapter~\ref{SparsePChapter}, we analyze the \spmv{} execution (1) using one single multithreaded PIM core, (2) using thousands of PIM cores, and (3) comparing its performance and energy consumption with that achieved on processor-centric CPU and GPU systems. Based on our rigorous experimental results and observations, we provide programming recommendations for software designers and suggestions for hardware and system designers of future PIM systems. Our \SparseP{} software package is freely and publicly available at {\href{https://github.com/CMU-SAFARI/SparseP}{\textcolor{blue}{https://github.com/CMU-SAFARI/SparseP}}} to enable further research on the irregular \spmv{} kernel in current and future PIM systems.

\vspace{-14pt}
\section{Contributions}
\vspace{-2pt}

This dissertation explores lightweight synchronization approaches in cooperation with efficient data access techniques to accelerate irregular applications both in processor-centric CPU systems and memory-centric NDP/PIM systems. \fix{This doctoral thesis aims to bridge the gap between processor-centric CPU systems and memory-centric PIM systems in the critically-important area of irregular applications.} Based on our rigorous experimental results and observations, we provide programming recommendations for software designers and suggestions for hardware and system designers of CPU and NDP/PIM systems in Chapters~\ref{ColorTMChapter}-~\ref{SparsePChapter}. In summary, this dissertation makes the following major contributions:

\begin{itemize}
\setlength\itemsep{-2pt}
\vspace{-8pt}
    \item We introduce \ColorTM{}, a novel algorithmic design to accelerate the widely used irregular graph coloring kernel on modern multicore CPU platforms. We introduce a \emph{speculative} synchronization and computation approach on graph coloring to mitigate synchronization overheads. We propose an \emph{eager} detection and resolution policy of the coloring inconsistencies arised among parallel threads to minimize data access costs. We extend our algorithmic design to present \BalColorTM{}, an efficient \emph{balanced} graph coloring kernel, which produces color classes with almost equal sizes. We demonstrate the effectiveness of \ColorTM{} and \BalColorTM{} at significantly outperforming prior state-of-the-art parallel graph coloring algorithms, and providing high performance benefits on a real-world end-application using a wide variety of large real-world graphs with diverse characteristics. Chapter~\ref{ColorTMChapter} describes \ColorTM{} and \BalColorTM{} and their evaluations in detail.
    
    \item We propose \smartpq{}, an adaptive concurrent priority queue for NUMA CPU architectures. We introduce \nuddle{}, a generic technique to obtain efficient \numa{} concurrent data structures by wrapping any arbitrary \notnuma{} concurrent data structure. We design the adaptive \smartpq{} that uses the \numa{} \nuddle{} concurrent priority queue under high-contention scenarios, and switches to \emph{directly} using the \nuddle's underlying \notnuma{} concurrent priority queue under low-contention scenarios. This way \smartpq{} provides high levels of parallelism, low data access costs, and significant performance benefits in modern NUMA CPU systems under all various contention scenarios, and even when the contention of the workload varies over time. We show the effectiveness of \smartpq{} at providing significant performance benefits over prior state-of-the-art \numa{} and \notnuma{} concurrent priority queues under various contention scenarios. Chapter~\ref{SmartPQChapter} describes \smartpq{} and its evaluations in detail.
    
    \item We introduce \SynCron, the first end-to-end hardware synchronization mechanism for NDP architectures. \SynCron{} provides low-overhead synchronization in the execution of irregular applications on NDP systems, has low hardware costs, supports many synchronization primitives, and implements an easy-to-use high-level synchronization interface. We design low-cost synchronization units that coordinate synchronization across NDP cores, and directly buffer synchronization variables in a specialized cache memory to avoid costly memory accesses to them. We integrate an efficient hierarchical \mpsync{} synchronization scheme, and  hardware-only programmer-transparent overflow management to mitigate performance overheads when hardware resources are exceeded. We demonstrate the effectiveness of \SynCron{} at significantly improving system performance and system energy efficiency using a wide range of irregular parallel applications, including pointer-chasing, graph processing, and time series analysis, and under various contention scenarios. Chapter~\ref{SynCronChapter} describes \SynCron{} and its evaluations in detail.

    \item We propose \SparseP{}, the first open-source software package of 25 efficient \spmv{} kernels tailored for real near-bank PIM architectures. We support several well-crafted data partitioning techniques of the sparse matrix to PIM-enabled memory and various load balancing schemes across PIM cores and across threads of a multithreaded PIM core to trade off computation balance across PIM cores for lower data transfer costs to PIM-enabled memory. We include three different synchronization approaches among several threads within a multithreaded PIM core to minimize synchronization overheads and achieve high levels of parallelism. We perform the first comprehensive study of the widely used irregular \spmv{} kernel on the UPMEM PIM architecture, the first real commercial PIM architecture, using various compressed matrix storage formats, many data types, and 26 sparse matrices with diverse sparsity patterns. We demonstrate that the \spmv{} execution on the memory-centric UPMEM PIM system with 2528 PIM cores achieves a much higher fraction of the machine's peak performance compared to that on the state-of-the-art processor-centric CPU and GPU systems, and also provides high energy efficiency. Chapter~\ref{SparsePChapter} describes \SparseP{} and its evaluations in detail.
    
\end{itemize}

\section{Outline}
\vspace{-2pt}

This dissertation is organized into 8 chapters. %Chapter~\ref{GrekAbstractChapter} presents the extended summary of this dissertation in Greek. 
Chapter~\ref{IntroChapter} describes and motivates the thesis statement of this dissertation, and also briefly describes the research contributions of this dissertation. Chapter~\ref{ColorTMChapter} introduces \ColorTM{}, a new algorithmic design to accelerate the irregular graph coloring kernel in modern CPU architectures, and presents its experimental study on a modern multicore platform. Chapter~\ref{SmartPQChapter} introduces \smartpq{}, an adaptive concurrent priority queue that achieves high performance in NUMA CPU architectures under all various contention scenarios, and presents its respective evaluations. Chapter~\ref{SynCronChapter} introduces \SynCron{}, an end-to-end hardware mechanism to support efficient and low-cost synchronization in NDP systems, and presents its evaluations across a wide variety of irregular applications including graph processing, pointer-chasing and time series analysis. Chapter~\ref{SparsePChapter} introduces \SparseP{},  the first open-source library of 25 algorithms to efficiently execute the irregular \spmv{} kernel on real PIM architectures, and presents a comprehensive experimental study of these \spmv{} kernels on the first real commercial PIM architecture. Chapter~\ref{FutureWorkChapter} presents future research directions and concluding remarks of this dissertation. Chapter~\ref{OtherWorksChapter} presents several other research works of the author of this dissertation. Chapter~\ref{AppendixChapter} presents additional experimental results and descriptions for the \SparseP{} contribution (Chapter~\ref{SparsePChapter}). %Chapter~\ref{GlossaryChapter} contains a dictionary table of the main keyword terms from Greek to English.

%ColorTM
\lstset{style=mystyle}
\chapter{\ColorTM{}}\label{ColorTMChapter}
\section{Overview}

Graph coloring assigns colors to the vertices of a graph such that any two adjacent vertices have different colors. Graph coloring kernel is widely used in many important real-world applications including the conflicting job scheduling~\cite{Welsh1967Upper,Marx2004Graph,Arkin1987Scheduling,Marx2004GRAPHCP,Ramaswami1989Distributed}, register allocation~\cite{Chaitin1981Register,Chaitin1982Register,Briggs1994Improvements,Chen2018Register,Cohen2010Processor}, sparse linear algebra~\cite{Coleman1983Estimation,Saad1994SparseKit,Jones1993Efficient,Gebremedhin2005WhatColor}, machine learning (e.g., to select non-similar samples that form an effective training set), and chromatic scheduling of graph processing applications~\cite{Kaler2016Executing,Kaler2014Executing}. For instance, the chromatic scheduling execution is as follows: given the vertex coloring of a graph, chromatic scheduling performs $N$ steps that are executed \emph{serially}, where $N$ is the number of colors used to color the graph, and at each step the vertices assigned to the \emph{same} color are processed \emph{in parallel}, i.e., representing independent tasks that are executed concurrently. In addition, it is of vital importance that programmers manage the registers of modern CPUs effectively, and thus compilers~\cite{Chen2018Register,Cohen2010Processor} optimize the register allocation problem via graph coloring: compilers construct undirected graphs, named register inference graphs (RIGs), with vertices representing the variables used in the source code and edges between vertices representing variables that are simultaneously active at some point in the program execution, and then compilers leverage the graph coloring kernel to identify independent variables that can be allocated on the same registers, i.e., if there \emph{no} edge in the RIG connecting the associated vertices of the variables.

To achieve high system performance in the aforementioned real-world scenarios, software designers need to improve three key aspects of the graph coloring kernel. First, they need to minimize the number of colors used to color the input graph. For example, in the chromatic scheduling scheme minimizing the number of colors used to color the graph reduces the number of the sequential steps performed in the multithreaded end-application. However, minimizing the number of colors in graph coloring is an NP-hard problem~\cite{Garey1974Some}, and thus prior works~\cite{Maciej2020GC,Hasenplaugh2014Ordering,Coleman1983Estimation,Arkin1987Scheduling,Marx2004GRAPHCP,brelaz1979,Matula1983Smallest,Karp1985Fast,Luby1985,Goldberg1987Parallel} introduce ordering heuristics that generate effective graph colorings with a relatively small number of colors. Second, given that the execution time of the graph coloring kernel adds to the overall parallel overhead of the real-world end-application, software engineers need to design high-performance graph coloring algorithms for modern multicore computing systems. Third, an effective graph coloring for a real-world end-application necessitates a balanced distribution of the vertices across the color classes, i.e., the sizes of the color classes to be almost the same. Producing color classes with high skew in their sizes, i.e., high disparity in the number of vertices distributed across color classes, typically causes load imbalance and low resource utilization in real-world end-application. For example, in the register allocation scenario high disparity in the sizes of the color classes  results to a large number of registers needed (high financial costs), equal to the size of the largest color class produced, while a \emph{large} portion of the registers remains idle (unused) for a \emph{long} time during the program execution (i.e., in time periods corresponding to many color classes with small sizes), thus causing low hardware resource utilization. Therefore, software designers need to propose \emph{balanced} and \emph{fast} graph coloring algorithms for commodity computing systems. Our \textbf{goal} in this work is to improve the last two key aspects of the graph coloring kernel by introducing high-performance and balanced multithreaded graph coloring algorithms for modern multicore platforms.

With a straightforward parallelization of graph coloring, coloring conflicts may arise when two parallel threads assign the same color to adjacent vertices. To deal with this problematic case, recent works~\cite{Catalyurek2012GraphColoring,Gebremedhin2000Scalable,Rokos2015Fast,Boman2005Scalable,Lu2015Balanced} perform two additional phases: a conflict detection phase, which iterates over the vertices of the graph to detect coloring inconsistencies between adjacent vertices, and a conflict resolution phase, which iterates over the detected conflicted vertices to resolve the coloring inconsistencies via recoloring. Nevertheless, these prior works~\cite{Catalyurek2012GraphColoring,Gebremedhin2000Scalable,Rokos2015Fast,Boman2005Scalable,Lu2015Balanced} are still inefficient, as we demonstrate in Section~\ref{eval}, because they need to traverse the \emph{whole} graph at least two times (one for coloring the vertices and one for detecting coloring conflicts), and also detect and resolve coloring conflicts with a \emph{lazy} approach, i.e., much later in the runtime compared to the time that the coloring conflicts appeared. As a result, prior approaches access the conflicted vertices of the graph multiple times, however mainly using the expensive last levels of the memory hierarchy (e.g., main memory) of commodity multicore platforms, thus incurring high data access costs.

In this work, we present \ColorTM{}~\cite{colortmGithub}, a high-performance graph coloring algorithm for multicore platforms. \ColorTM{} is designed to provide low synchronization and data access costs. Our algorithm proposes (a) an \emph{eager} conflict detection and resolution approach, i.e., immediately detecting and resolving coloring inconsistencies when they arise, such that to minimize data access costs by accessing conflicted vertices immediately using the low-cost lower levels of the memory hierarchy of multicore platforms, and (b) a speculative computation and synchronization scheme, i.e., leveraging  Hardware Transactional Memory (HTM) and speculatively performing computations and data accesses outside the critical section, such that to provide high levels of parallelism and low synchronization costs by executing multiple small and short critical sections in parallel. Specifically, \ColorTM{} consists of three steps: for each vertex on the graph, it (i) speculatively finds a candidate legal color by recording the colors of the adjacent vertices, (ii) validates and updates the color of the current vertex by checking the colors of the critical adjacent vertices within an HTM transaction to detect potential coloring conflicts, and (iii) eagerly repeats steps (i) and (ii) for the current vertex multiple times until a valid coloring is found.

However, \ColorTM{} does not provide any guarantee on the sizes of the color classes relative to each other. As we demonstrate in our evaluation (Section~\ref{eval}), the color classes produced by \ColorTM{} for a real-world graphs have high disparity in the number of vertices across them, thus causing load imbalance and low resource utilization in real-world end-applications. Therefore, we extend our algorithmic design to propose a \emph{balanced} graph coloring algorithm, named \BalColorTM{}~\cite{colortmGithub}.  \BalColorTM{} achieves high system performance and produces highly balanced color classes, i.e., having almost the same number of vertices across color classes, targeting to provide high hardware resource utilization and load balance in the real-world end-applications of graph coloring.

We evaluate \ColorTM{} and \BalColorTM{} on a dual socket Intel Haswell server using a wide variety of large real-world graphs with diverse characteristics. \ColorTM{} improves performance by 12.98$\times$ on average using 56 parallel threads compared to state-of-the-art graph coloring algorithms, while providing similar coloring quality. \BalColorTM{} outperforms prior state-of-the-art balanced graph coloring algorithms by 1.78$\times$ on average using 56 parallel threads, and provides the best color balancing quality over prior schemes (See Section~\ref{eval}). Finally, we study the effectiveness of our proposed \ColorTM{} and \BalColorTM{} in parallelizing a widely used real-world end-application, i.e., Community Detection~\cite{FORTUNATO201075}, and demonstrate that our proposed algorithmic designs can provide significant performance improvements in real-world scenarios. \ColorTM{} and \BalColorTM{} are publicly available~\cite{colortmGithub} at {\href{https://github.com/cgiannoula/ColorTM}{\textcolor{blue}{github.com/cgiannoula/ColorTM}}}.

The main \textbf{contributions} of this work are:
\begin{itemize}%[$\textbf{--}$]
\vspace{-8pt}
\setlength\itemsep{-2pt}
    \item We design high-performance and balanced graph coloring algorithms, named \ColorTM{} and \BalColorTM{}, for modern multicore platforms.
    \item We leverage HTM to efficiently detect coloring inconsistencies between adjacent vertices (processed by different parallel threads) with low synchronization costs. We propose an eager conflict resolution approach to efficiently resolve coloring inconsistencies in multithreaded executions by minimizing data access costs.
    \item We evaluate \ColorTM{} and \BalColorTM{} using a wide variety of large real-world graphs and demonstrate that they provide significant performance improvements over prior state-of-the-art graph coloring algorithms. Our proposed algorithmic designs significantly improve performance in multithreaded executions of real-world end-applications.
\end{itemize}

\section{Prior Graph Coloring Algorithms}

In this section, we describe prior state-of-the-art graph coloring algorithms~\cite{Catalyurek2012GraphColoring,Gebremedhin2000Scalable,Rokos2015Fast,Boman2005Scalable,Lu2015Balanced}. Section~\ref{sec:greedy} presents the sequential graph coloring algorithm. Section~\ref{sec:unbalanced} describes prior parallel (no guarantee on the sizes of color classes) graph coloring algorithms proposed in the literature, while Section~\ref{sec:balanced} presents prior balanced (color classes are highly balanced) graph coloring algorithms proposed in the literature.

\subsection{The Greedy Algorithm} \label{sec:greedy}

Figure~\ref{alg:greedy} presents the sequential graph coloring algorithm, called \emph{Greedy}~\cite{Welsh1967Upper}. Consider an undirected graph $G = (V, E)$, and the neighborhood $N(v)$ of a vertex $v \in V$ defined as $N(v) = \{u \in V : (v,u) \in E\}$. For each vertex $v$ of the graph, Greedy records the colors of $v's$ adjacent vertices in a forbidden set of colors, and  assigns the minimum legal color to the vertex $v$ based on the forbidden set of colors.

%putting basicstyle=\scriptsize it becomes smaller
\begin{figure}[H]
\begin{lstlisting}[language=C]
(*\bfseries Input:*) Graph G=(V,E)
Let N(v) be the adjacent vertices of the vertex v
for each (*$v \in V$*) do
 forbidColors = (*$\emptyset$*)
 for each (*$u \in N(v)$*) do
  forbidColors = forbidColors (*$\cup$*) {(*$u$*).color}
 (*$v$*).color = minLegalColor(forbidColors)
\end{lstlisting}
\vspace{-6pt}
\caption{The Greedy algorithm.}
\label{alg:greedy}
\end{figure}

The Greedy approach produces at most $\Delta+1$ colors~\cite{Welsh1967Upper}, where $\Delta$ is the degree of the graph $G$. The degree of the graph is defined as $\Delta$ = $max_{v \in V}\{deg(v)\}$, where $deg(v)$ is the degree of a vertex $v$, which is the number of its adjacent vertices $deg(v) =\lvert N(v) \rvert$. However, finding the minimum number of colors to color a graph $G$ is an NP-hard problem~\cite{Mitchem1976OnVA}. In this work, we have experimented with the \emph{first-fit} ordering heuristic~\cite{Welsh1967Upper}, in which the vertices of the graph are processed and colored in the order they appear in the input graph representation $G$, since this heuristic can provide high coloring quality based on prior works~\cite{Hasenplaugh2014Ordering,Lovsz1989AnOG,Welsh1967Upper}. 
%In the \emph{first-fit} ordering heuristic, the vertices of the graph are processed and colored in the order they appear in the input graph representation $G$. 
We leave the exploration of other ordering heuristics for future work.

\subsection{Prior Parallel Graph Coloring Algorithms} \label{sec:unbalanced}

To parallelize the graph coloring problem, the vertices of the graph are distributed among parallel threads. However, due to crossing edges, the coloring subproblems assigned to parallel threads are not independent, and the parallel algorithm may terminate with an invalid coloring. Specifically, a race condition arises when two parallel threads assign the same color to adjacent vertices. The algorithm implies that when a parallel thread updates the color of a vertex, the forbidden set of colors of the adjacent vertices has not been changed. Thus, the nature of this algorithm imposes that the reads to the colors of the adjacent vertices of a vertex $v$ have to be executed \emph{atomically} with the write-update to the color of the vertex $v$.

\subsubsection{The SeqSolve Algorithm}

Figure~\ref{alg:seqsolve} presents the parallel graph coloring algorithm proposed by Gebremedhin et al.~\cite{Gebremedhin2000Scalable}, henceforth referred to as SeqSolve. This algorithm consists of three steps: (i) multiple parallel threads iterate over the whole graph and speculatively color the vertices of the graph with \emph{no} synchronization (lines 3-6), (ii) multiple parallel threads iterate over the whole graph and detect coloring inconsistencies that appeared in the (i) step (lines 7-13), and (iii) only \emph{one} single thread resolves the detected coloring inconsistencies by re-coloring the conflicted vertices (lines 14-16). Since the (iii) step is executed by only a single thread, no coloring inconsistencies appear after this step. Note that when a coloring conflict arises between two adjacent vertices, only \emph{one} of the involved adjacent vertices needs to be re-colored, e.g., using a simple order heuristic among the vertices (line 11).

\begin{figure}[H]
\begin{lstlisting}[language=C]
(*\bfseries Input:*) Graph G=(V,E)
Let (*$N(v)$*) be the adjacent vertices of the vertex (*$v$*)
(*\textcolor{mygreen}{// Speculative Coloring - Step (i)}*)
for each (*$v \in V$*) do (*\textcolor{blue}{in parallel}*)
  Assign the minimum legal color to the vertex (*$v$*)
(*\textcolor{bostonuniversityred}{\_\_barrier\_\_}*)
(*\textcolor{mygreen}{// Detection of Coloring Inconsistencies - Step (ii)}*)
R = (*$\emptyset$*) (*\textcolor{mygreen}{// Set of Conflicted Vertices}*)
for each (*$v \in V$*) do (*\textcolor{blue}{in parallel}*)
  for each (*$u \in N(v)$*) do
    if (((*$v$*).color == (*$u$*).color) && ((*$v$*) < (*$u$*)))
      R = R (*$\cup$*) v
(*\textcolor{bostonuniversityred}{\_\_barrier\_\_}*)
(*\textcolor{mygreen}{// Sequential Resolution of Coloring Conflicts - Step (iii)}*)
for each (*$v \in R$*) do
  Assign the minimum legal color to the vertex (*$v$*)
\end{lstlisting}
\vspace{-4pt}
\caption{The SeqSolve algorithm.}
\label{alg:seqsolve}
\end{figure}

In the SeqSolve algorithm, we make three key observations. First, if the number of coloring conflicts arised in a multithreaded execution is low, the algorithm might scale well~\cite{Gebremedhin2000Scalable}. However, as the number of parallel threads increases and the graph becomes denser, i.e., the vertices of the graph have a large number of adjacent vertices, many more coloring conflicts arise in multithreaded executions. In such scenarios, a large number of coloring inconsistencies is resolved sequentially, i.e., by only \emph{one} single thread, thus achieving limited parallelism. Second, we note SeqSolve traverses the whole graph at least two times, i.e., step (i) and step (ii). Assuming large real-world graphs that do not typically fit in the Last Level Cache (LLC) of contemporary multicore platforms, the whole graph is traversed twice using the main memory, thus incurring high data access costs. Third, we observe that SeqSolve detects and resolves the coloring conflicts \emph{lazily}, i.e., much later in the runtime compared to the time that the coloring conflicts appears. Specifically, a coloring inconsistency in a vertex $v$ might appear in step (i). However, SeqSolve detects the coloring inconsistency in vertex $v$ in step (ii), i.e., after \emph{first} coloring \emph{all} the remaining vertices of the graph. Similarly, SeqSolve resolves the coloring inconsistency of the vertex $v$ in step (iii), i.e., after \emph{first} detecting if coloring inconsistencies exist in \emph{all} the remaining vertices of the graph (step (ii)). As a result, many conflicted vertices are accessed multiple times in the runtime, however with a lazy approach, i.e., accessing them through the expensive last levels of the memory hierarchy of commodity platforms, thus incurring high data access costs.

\subsubsection{The IterSolve Algorithm}

Figure~\ref{alg:itersolve} presents the parallel graph coloring algorithm proposed by Boman et al.~\cite{Boman2005Scalable,Catalyurek2012GraphColoring}, henceforth referred to as IterSolve. This algorithm consists of two repeated steps: (i) multiple parallel threads iterate over the uncolored vertices of the graph and speculatively color the uncolored vertices of the graph with \emph{no} synchronization (lines 5-8), (ii) multiple parallel threads iterate over the recently-colored vertices of the graph and detect coloring inconsistencies appeared in the (i) step (lines 9-15). The steps (i) and (ii) are iteratively repeated until there are \emph{no} coloring inconsistencies in any adjacent vertices of the graph.

\begin{figure}[H]
\begin{lstlisting}[language=C]
(*\bfseries Input:*) Graph G=(V,E)
Let (*$N(v)$*) be the adjacent vertices of the vertex (*$v$*)
(*$U = V$*)
while (*$U \neq \emptyset$*) 
  (*\textcolor{mygreen}{// Speculative Coloring - Step (i)}*)
  for each (*$v \in U$*) do (*\textcolor{blue}{in parallel}*)
    Assign the minimum legal color to the vertex (*$v$*)
  (*\textcolor{bostonuniversityred}{\_\_barrier\_\_}*)
  (*\textcolor{mygreen}{// Detection of Coloring Inconsistencies - Step (ii)}*)
  R = (*$\emptyset$*) (*\textcolor{mygreen}{ // Set of Conflicted Vertices}*)
  for each (*$v \in U$*) do (*\textcolor{blue}{in parallel}*)
    for each (*$u \in N(v)$*) do
      if (((*$v$*).color == (*$u$*).color) && ((*$v$*) < (*$u$*)))
        R = R (*$\cup$*) v
 (*\textcolor{bostonuniversityred}{\_\_barrier\_\_}*)
  (*$U = R$*)
\end{lstlisting}
\vspace{-4pt}
\caption{The IterSolve algorithm.}
\label{alg:itersolve}
\end{figure}

In the IterSolve algorithm, we make four key observations. First, the programmer needs to explicitly define forward progress in the source code, so that the IterSolve algorithm terminates. Specifically, to ensure forward progress when a coloring inconsistency appears between two adjacent vertices, the programmer needs to \emph{explicitly} define only \emph{one} of them to be re-colored (line 13), e.g., based on the vertices' ids. Otherwise, the two adjacent vertices may \emph{always} obtain the same color, if they are always being processed by different threads. Second, similarly to SeqSolve, IterSolve traverses the whole graph at least two times (steps (i) and (ii)), i.e., in the first iteration of the while loop in line 4, where the set $U$ is equal to the set $V$ (line 3). In the first iteration of the while loop, the whole large real-world graph is accessed through the main memory twice, thus incurring high data access costs. Third, similarly to SeqSolve, IterSolve detects and resolves the coloring conflicts \emph{lazily}. Specifically, a coloring inconsistency in a vertex $v$ might appear in step (i) (line 7), it is detected in step (ii) (line 13), i.e., after \emph{first} coloring \emph{all} the remaining uncolored vertices of the graph. Moreover, IterSolve resolves the coloring inconsistency of a vertex $v$ in step (i) (with re-coloring), i.e., after \emph{first} detecting if coloring inconsistencies exist in \emph{all} the remaining recently-colored vertices of the graph (step (ii)). Thus, IterSolve incurs high data access costs on the many conflicted vertices, which are accessed multiple times in the runtime with \emph{lazy} approach, through the last levels of the memory hierarchy of commodity platforms. Fourth, the iterative process of resolving coloring conflicts may introduce new conflicts, and thus, IterSove might need additional iterations to fix them. This scenario may happen when adjacent vertices are assigned to the \emph{same} thread and incur coloring inconsistencies, they will be assigned and processed by \emph{different} parallel threads in the next iteration. The authors of the original IterSolve papers~\cite{Boman2005Scalable,Catalyurek2012GraphColoring} empirically observe that a few iterations of IterSolve are needed to produce a valid coloring. However, the authors used \emph{synthetic}  and \emph{not} real-world graphs in their evaluation. In addition, the more iterations are needed, the more \emph{lazy} traversals on the conflicted vertices of the graph are performed, which can significantly degrade performance.

\subsubsection{The IterSolveR Algorithm}

Figure~\ref{alg:itersolveR} presents the parallel graph coloring algorithm proposed by Rokos et al.~\cite{Rokos2015Fast}, henceforth referred to as IterSolveR. Rokos et al. observed that the IterSolve algorithm (Figure~\ref{alg:itersolve}) can be improved by merging the steps (i) and (ii) into a single \emph{detect-and-re-color} step, thus eliminating one of the two barrier synchronizations of IterSolve (lines 8 and 15 in Figure~\ref{alg:itersolve}). When a coloring inconsistency on a vertex $v$ is found, the vertex $v$ can be immediately re-colored (line 18 in Figure~\ref{alg:itersolveR}). However, the new re-coloring on the vertex $v$ may \emph{again} introduce a coloring inconsistency in multithreaded executions, since re-colorings are performed \emph{concurrently} by multiple parallel threads (line 11). Therefore, the vertex $v$ is marked as \emph{recently-re-colored} vertex (line 19), and needs to be  re-validated in the next iteration of IterSolveR. Overall, IterSolverR (Figure~\ref{alg:itersolveR}) first speculatively colors all the vertices of the graph and marks them as \emph{recenlty-colored} vertices (lines 3-
6). Then, it executes one single repeated step (lines 8-21): multiple parallel threads iterate over the recently-colored vertices of the graph, and detect if coloring inconsistencies have appeared, which in that case are immediately resolved via re-coloring. This step is repeated until there are no \emph{recently-re-colored} vertices: in one single iteration of this step, there are \emph{no} coloring inconsistencies detected in any adjacent vertices of the graph.

\begin{figure}[t]
\begin{lstlisting}[language=C]
(*\bfseries Input:*) Graph G=(V,E)
Let (*$N(v)$*) be the adjacent vertices of the vertex (*$v$*)
(*\textcolor{mygreen}{// Speculative Coloring (Step 0)}*)
for each (*$v \in V$*) do (*\textcolor{blue}{in parallel}*)
    Assign the minimum legal color to the vertex (*$v$*)
(*\textcolor{bostonuniversityred}{\_\_barrier\_\_}*)
(*$U = V$*) (*\textcolor{mygreen}{ // Mark all Vertices as Recently-Colored}*)
while (*$U \neq \emptyset$*) 
  R = (*$\emptyset$*) (*\textcolor{mygreen}{// Set of Recently-Colored Vertices}*)
  (*\textcolor{mygreen}{// Conflict Detection and Resolution (Step (i))}*)
  for each (*$v \in U$*) do (*\textcolor{blue}{in parallel}*)
    (*\textcolor{blue}{bool}*) conflict-detected = false
    for each (*$u \in N(v)$*) do
      if (((*$v$*).color == (*$u$*).color) && ((*$v$*) < (*$u$*)))
        conflict-detected = true
        break
    if (conflict-detected == true)
      Assign the minimum legal color to vertex (*$v$*)
      R = R (*$\cup$*) v (*\textcolor{mygreen}{ // Mark vertex v as Recently-Colored}*)
  (*\textcolor{bostonuniversityred}{\_\_barrier\_\_}*)
  (*$U = R$*)
\end{lstlisting}
\vspace{-4pt}
\caption{The IterSolveR algorithm.}
\label{alg:itersolveR}
\end{figure}

In the IterSolveR algorithm, even though one barrier synchronization is eliminated compared to IterSolve, we observe that IterSolveR still traverses the whole graph at least twice: (i) in Step 0 (lines 4-5), and (ii) in the first iteration of the while loop in line 8, where the set $U$ is equal to the set $V$ (line 7), including all the vertices of the graph. Thus, IterSolveR traverses the large real-world graph twice through the main memory, incurring high data access costs. In addition, we find that similarly to SeqSolve and IterSolve, the IterSolveR algorithm also detects the coloring inconsistencies \emph{lazily}. Specifically, a coloring inconsistency on a vertex $v$ might appear in the re-coloring process of lines 17-19, since the re-coloring process is \emph{concurrently} executed on multiple conflicted vertices by multiple parallel threads. However, re-coloring inconsistencies of lines 17-19 are detected in the next iteration of the step (i) in lines 13-16, i.e., after \emph{first} processing \emph{all} the remaining vertices of the set $U$ (line 11). Therefore, as we demonstrate in our evaluation (Section~\ref{eval:unbalanced}), IterSolveR is still inefficient, incurring high data access costs on multiple conflicted vertices which are accessed multiple times in the runtime with a \emph{lazy} approach.

\subsection{Prior Balanced Graph Coloring Algorithms} \label{sec:balanced}

To provide a balanced coloring on a graph in which the color classes produced include almost the same number of vertices, a two-step process is typically used: (i) obtain an initial graph coloring using a balanced-oblivious algorithm (e.g., Section~\ref{sec:unbalanced}), and (ii) obtain a balanced graph coloring using a balanced-aware (henceforth referred to as balanced for simplicity) graph coloring algorithm, as we describe next. Specifically, given a graph $G = (V, E)$, we can assume that the number of colors produced by the initial coloring step (i) is $C$. A strictly balanced graph coloring results in the size of \emph{each} color class being $b = V / C$.\footnote{Please note that in our work we make the following assumption: in a real-world end-application, the vertices of the graph represent sub-tasks that have almost equal load/weights of computation. If the vertices of the input graph have different load/weights of computation, a pre-processing step needs to be applied in the original graph: vertices with large computation weights/load are split into multiple smaller vertices, each of them has one weight/load unit of computation.} Therefore, we refer to the color classes whose sizes are greater than $b$ as \emph{over-full} classes, and those whose sizes are less than $b$ as \emph{under-full} classes. Balanced graph coloring algorithms leverage the quantity of $b$, which can be extracted by first executing an initial balanced-oblivious graph coloring on the graph, in order to provide balanced color classes on a graph.

\subsubsection{The Color-Centric (CLU) Algorithm}

Figure~\ref{alg:CLU} presents the color-centric balanced graph coloring algorithm proposed by Lu et al.~\cite{Lu2015Balanced}, henceforth referred to as CLU. In this scheme, vertices belonging in the \emph{same} color class are processed concurrently, and a \emph{subset} of vertices from each over-full color class is moved to under-full color classes in order to achieve high color balance. Only vertices belonging in over-full color classes are considered for re-coloring, while graph coloring balance is achieved \emph{without} increasing the number of color classes produced by the initial graph coloring.

\begin{figure}[H]
\vspace{-12pt}
\begin{lstlisting}[language=C]
(*\bfseries Input:*) Graph G=(V,E)
Obtain an initial coloring on G
Let (*$C$*) be the number of colors produced
Let (*$b = V / C$*) be the perfect balance
Let (*$Q$*) be the set of vertices of the over-full color classes
for each (*$c \in Q$*) do (*\textcolor{mygreen}{// Process the Over-Full Color Classes}*)
  Let (*$R(c)$*) be the set of vertices with color (*$c$*)
  for each (*$v \in R(c)$*) do (*\textcolor{blue}{in parallel}*)
    if (the size of the color class (*$c$*) <= (*$b$*))
      continue (*\textcolor{mygreen}{// Color Class is Balanced}*)
    Let (*$k$*) be the index of the minimum under-full color class that is permissible to vertex (*$v$*)
    if ((*$k$*) exists) (*\textcolor{mygreen}{// Re-Coloring}*)
      (*$v$*).color = k
      (*\emph{Atomically}*) decrease the size of the color class (*$c$*) 
      (*\emph{Atomically}*) increase the size of the color class (*$k$*) 
  (*\textcolor{bostonuniversityred}{\_\_barrier\_\_}*)
\end{lstlisting}
\vspace{-4pt}
\caption{The CLU algorithm.}
\label{alg:CLU}
\vspace{-14pt}
\end{figure}

The CLU algorithm (Figure~\ref{alg:CLU}) processes the over-full color classes sequentially (lines 6 and 16), while vertices belonging at the same over-full color class are processed concurrently (line 8). CLU iterates over each vertex $v$ of an over-full color class, and finds the minimum color of an under-full color class that is permissible to be assigned at the vertex $v$ (line 11). If such a color exists, the vertex $v$ is re-colored with a color of an under-full color class (lines 12-15). The CLU algorithm iterates over the vertices of each over-full color class until that particular over-full class becomes balanced at a certain point in the execution, i.e., until when its size becomes smaller or equal to $b$ (lines 9-10). Then, the vertices belonging on that color class are no longer considered for re-coloring (line 10). Thus, this algorithm terminates when either vertex-balance across color classes is achieved or vertex-balance across color classes is no longer available, i.e., there are no more permissible re-colorings for any vertex belonging in an over-full color class.

In the CLU algorithm, we make two key observations. First, parallel threads always process vertices of the \emph{same} color, thus no coloring inconsistencies are produced: since vertices had the same color in the initial coloring, they are \emph{not} adjacent vertices, and thus they can be re-colored with the same color of an under-full color class without violating correctness. This way CLU requires \emph{only} one iteration over the vertices of all the over-full color classes. Second, the parallel performance of CLU depends on the number of the over-full color classes produced in the initial coloring. CLU requires $F$ steps, where $F$ is the number of over-full color classes produced in the initial coloring. At \emph{each} of these steps, i.e., for each over-full color class on the initial coloring, CLU introduces a barrier synchronization among parallel threads (line 16). This way it increases the synchronization costs, which might significantly degrade scalability in multithreaded executions.

\subsubsection{The Vertex-Centric (VFF) Algorithm}

Figure~\ref{alg:VFF} presents the vertex-centric balanced graph coloring algorithm proposed by Lu et al.~\cite{Lu2015Balanced}, henceforth referred to as VFF. The VFF algorithm is the balanced graph coloring counterpart of the IterSolve algorithm (Figure~\ref{alg:itersolve}). In this scheme, vertices from \emph{different} color classes are processed concurrently by parallel threads. Thus, in contrast to CLU, VFF introduces coloring inconsistencies. However, similarly to CLU, in VFF only vertices belonging in over-full color classes are considered for re-coloring, i.e., to be moved to under-full color classes, while graph coloring balance is also achieved \emph{without} increasing the number of color classes produced by the initial graph coloring.

\begin{figure}[H]
\begin{lstlisting}[language=C]
(*\bfseries Input:*) Graph G=(V,E)
Let (*$N(v)$*) be the adjacent vertices of the vertex (*$v$*)
Obtain an initial coloring on G
Let (*$C$*) be the number of colors produced
Let (*$b = V / C$*) be the perfect balance
Let (*$Q$*) be the set of vertices of the over-full color classes
while (*$Q \neq \emptyset$*) (*\textcolor{mygreen}{// Process the Over-Full Color Classes}*)
  (*\textcolor{mygreen}{// Speculative Re-Coloring - Step (i)}*)
  for each (*$v \in Q$*) do (*\textcolor{blue}{in parallel}*) 
    Let (*$c$*) be the current color of the vertex (*$v$*)
    if ((c != -1) && (the size of the color class (*$c$*) <= (*$b$*)))
      continue(*\textcolor{mygreen}{// Color Class is Balanced}*)
    Let (*$k$*) be the index of the minimum under-full color class that is permissible to vertex (*$v$*)
    if ((*$k$*) exists)(*\textcolor{mygreen}{// Re-Coloring}*)
      (*$v$*).color = k
      (*\emph{Atomically}*) decrease the size of the color class (*$c$*) 
      (*\emph{Atomically}*) increase the size of the color class (*$k$*)  
  (*\textcolor{bostonuniversityred}{\_\_barrier\_\_}*)
  (*\textcolor{mygreen}{// Detection of Coloring Inconsistencies - Step (ii)}*)
  R = (*$\emptyset$*) (*\textcolor{mygreen}{ // Conflicted Vertices of Over-Full Color Classes}*)
  for each (*$v \in Q$*) do (*\textcolor{blue}{in parallel}*)
    for each (*$u \in N(v)$*) do
      if (((*$v$*).color == (*$u$*).color) && ((*$v$*) < (*$u$*)))
        R = R (*$\cup$*) v
        (*$v$*).color = -1
 (*\textcolor{bostonuniversityred}{\_\_barrier\_\_}*)
  (*$Q = R$*)
\end{lstlisting}
\vspace{-4pt}
\caption{The VFF algorithm.}
\label{alg:VFF}
\vspace{-8pt}
\end{figure}

Similarly to IterSolve, VFF (Figure~\ref{alg:VFF}) consists of two repeated steps: (i) multiple parallel threads iterate over vertices of over-full color classes and speculatively re-color them with permissible colors of under-full color classes, if possible (lines 8-18), and (ii) multiple parallel threads iterate over the recently re-colored vertices and detect coloring inconsistencies that appeared in the (i) step (lines 19-26). Similarly to CLU, VFF iterates over the vertices of an over-full color class until that particular over-full class becomes balanced at a certain point in the execution, i.e., until when its size becomes smaller or equal to $b$ (lines 11-12). Then, the vertices belonging on that particular color class are no longer considered for re-coloring (line 12). The steps (i) and (ii) are iteratively repeated until there are \emph{no} coloring inconsistencies in any adjacent vertices of the graph, and the algorithm terminates when either vertex-balance across color classes is achieved or vertex-balance across color classes is no longer available, i.e., there are no more permissible re-colorings for any vertex belonging in an over-full color class.

Since VFF is the balanced graph coloring counterpart of IterSolve, we report similar key observations for them. First, VFF detects and resolves the coloring conflicts \emph{lazily}. Specifically, a coloring inconsistency in a vertex $v$ might appear in step (i), while it is detected in step (ii), i.e., after \emph{first} iterating over \emph{all} the remaining vertices of over-full color classes. Moreover, VFF resolves the coloring inconsistency in a vertex $v$ in step (i) (re-coloring), i.e., after \emph{first} detecting if coloring inconsistencies exist in \emph{all} the remaining recently re-colored vertices (in step (ii) of the previous iteration). Thus, VFF incurs high data access costs due to accessing multiple conflicted vertices in the runtime through the last levels of the memory hierarchy of commodity platforms.  Second, the iterative process of resolving coloring conflicts may introduce new conflicts, and thus, VFF might need additional iterations to fix them. This scenario may happen when adjacent vertices are assigned to the \emph{same} thread and incur coloring inconsistencies, they will be assigned and processed by \emph{different} parallel threads in the next iteration. Note that the more iterations are needed, the more \emph{lazy} traversals on the conflicted vertices of the graph are performed, which might significantly degrade performance.

\subsubsection{The Recoloring Algorithm}

Figure~\ref{alg:Recoloring} presents the re-coloring balanced graph coloring algorithm proposed by Lu et al.~\cite{Lu2015Balanced}, henceforth referred to as Recoloring. Recoloring is similar to the VFF (Figure~\ref{alg:VFF}) and IterSolve (Figure~\ref{alg:itersolve}) schemes. The key idea of this algorithm is that after performing an initial graph coloring with $C$ colors, \emph{all} the vertices of the graph are re-colored, having an additional condition on the color selection in order to achieve better vertex balance across color classes compared to that produced by the initial graph coloring. Specifically, Recoloring leverages the perfect balance $b = V / C$ known from the initial graph coloring, and keeps track the sizes of the color classes during the execution in order to improve vertex balance across color classes as follows: each vertex is re-colored using the minimum permissible color $k$ such that the size of the color class $k$ is less than $b$.

Similarly to IterSolve and VFF, Recoloring (Figure~\ref{alg:Recoloring}) consists of two repeated steps: (i) multiple parallel threads iterate over all the vertices of the graph and speculatively re-color them with a new permissible color $k$, that satisfies the condition that the size of the color class $k$ is less than $b$ (lines 12-17), and (ii) multiple parallel threads iterate over the recently re-colored vertices and detect coloring inconsistencies that appeared in the (i) step (lines 18-25). The steps (i) and (ii) are iteratively repeated until there are \emph{no} coloring inconsistencies in any adjacent vertices of the graph. In contrast to VFF and CLU, Recoloring does not guarantee that the graph color balance achieved uses the same number of colors with the initial graph coloring. To avoid producing a large number of color classes, the Recoloring scheme~\cite{Lu2015Balanced} (Figure~\ref{alg:Recoloring}) re-colors the vertices of the graph with the following order: assuming that the vertices of the graph are ordered such that the vertices of the same color class are listed consecutively (line 6), Recoloring iterates over the vertex sets of the color classes in the reverse order compared to that produced in the initial graph coloring, i.e., starting from the vertices assigned to the color class with the largest index (See line 8). The rationale behind this heuristic is that the vertices that are "difficult" to color, i.e., in the initial graph coloring they are assigned to a color class with large index, will be processed \emph{early}, thus aiming to produce a small number of color classes. For more details, we refer the reader to~\cite{Lu2015Balanced}.

\vspace{-10pt}
\begin{figure}[H]
\begin{lstlisting}[language=C]
(*\bfseries Input:*) Graph G=(V,E)
Let (*$N(v)$*) be the adjacent vertices of the vertex (*$v$*)
Obtain an initial coloring on G 
Let (*$C$*) be the number of colors produced
Let (*$b = V / C$*) be the perfect balance
Let (*$K(j)$*) be the set of vertices (*$u$*) with color j 
(*\textcolor{mygreen}{// $K(j) = \{u \in V, u.color = j \} $ }*)
Construct the order set (*$W=\{ K(C), K(C-1), ..., K(1), K(0)\}$*)
Initialize the sizes of the (*$C$*) color classes to 0
(*$Q = W$*)
while (*$Q \neq \emptyset$*) (*\textcolor{mygreen}{// Re-Color the Whole Graph}*)
  (*\textcolor{mygreen}{// Speculative Coloring - Step (i)}*)
  for each (*$v \in Q$*) do (*\textcolor{blue}{in parallel}*) 
    Let (*$k$*) be the minimum color that is permissible to the vertex (*$v$*) such that the size of the color class (*$k$*) is less than (*$b$*) (*\textcolor{mygreen}{// Balanced Color Classes}*)
    (*$v$*).color = k
    (*\emph{Atomically}*) increase the size of the color class (*$k$*) 
  (*\textcolor{bostonuniversityred}{\_\_barrier\_\_}*)
  (*\textcolor{mygreen}{// Detection of Coloring Inconsistencies - Step (ii)}*)
  R = (*$\emptyset$*) (*\textcolor{mygreen}{ // Set of Conflicted Vertices}*)   
  for each (*$v \in Q$*) do (*\textcolor{blue}{in parallel}*)
    for each (*$u \in N(v)$*) do
      if (((*$v$*).color == (*$u$*).color) && ((*$v$*) < (*$u$*)))
        (*\emph{Atomically}*) decrease the size of the color class (*$v$*).color 
        R = R (*$\cup$*) v
  (*\textcolor{bostonuniversityred}{\_\_barrier\_\_}*)
  (*$Q = R$*)
\end{lstlisting}
\vspace{-4pt}
\caption{The Recoloring algorithm.}
\label{alg:Recoloring}
\vspace{-14pt}
\end{figure}

In Recoloring, we make three key observations. First, Recoloring traverses the whole graph, i.e., it re-colors \emph{all} the vertices of the graph, while CLU and VFF re-color only a \emph{subset} of the vertices of over-full color classes. As a result, Recoloring performs a much larger number of computations and memory accesses compared to VFF and CLU. Second, similarly to IterSolve and VFF, Recoloring detects and resolves coloring inconsistencies with a \emph{lazy} approach, thus incurring high data access costs. Recoloring may also introduce new conflicts, thus resulting in additional iterations to fix them. Third, even though Recoloring employs a different vertex ordering heuristic to re-color vertices compared to that used in the initial graph coloring (vertices are colored with the order they appear in the input graph), there is \emph{no} guarantee on the number of color classes that will be produced. As we demonstrate in our evaluation (Section~\ref{eval:balanced}), Recoloring might significantly increase the number of color classes produced compared to that produced in the initial graph coloring.

\section{\ColorTM: Overview}

Our proposed algorithmic design is a high-performance graph coloring algorithm for multicore platforms. \ColorTM{} provides low synchronization and data access costs by relying on two key techniques, that we describe in detail next.

\subsection{Speculative Computation and Synchronization}

As already discussed, the graph coloring kernel implies that the reads to the colors of the adjacent vertices of a vertex $v$ have to be executed atomically with the write-update to the color of the vertex $v$. Figure~\ref{alg:naive} presents a straightforward parallelization scheme of the graph coloring problem. A naive parallelization approach would be to distributed the vertices of the graph across parallel threads, and for each vertex to include within a critical section the whole block of code that computes and assigns a permissible color to that vertex. However, this approach results in large critical sections with large data access footprints and long duration, and significantly limits the amount of parallelism and the scalability to a large number of threads. 

\vspace{-4pt}
\begin{figure}[H]
\begin{lstlisting}[language=C]
(*\bfseries Input:*) Graph G=(V,E)
for each (*$v \in V$*) do (*\textcolor{blue}{in parallel}*)
  (*\textcolor{mygreen}{// Atomic Coloring Step (i)}*)
  (*\textcolor{bostonuniversityred}{begin\_critical\_section}*)
  Compute and assign the minimum legal color to the vertex (*$v$*)
  (*\textcolor{bostonuniversityred}{end\_critical\_section}*)
\end{lstlisting}
\vspace{-4pt}
\caption{A Naive Approach.}
\label{alg:naive}
\vspace{-8pt}
\end{figure}

We observe that it is not necessary to include inside the critical section (i) the computations performed to find a permissible color for a vertex $v$, and (ii) the accesses to \emph{all} the adjacent vertices of the vertex $v$. Figure~\ref{alg:colortm-overview} presents an overview of \ColorTM. For each vertex $v$, we design \ColorTM{} to implement a speculative computation scheme through two sub-steps: (i) speculatively compute a permissible color $k$ for the vertex $v$ (line 5) without using synchronization and track the set of critical adjacent vertices (line 6), i.e., a subset of $v$'s adjacent vertices that can cause coloring inconsistencies with the vertex $v$ (See Section~\ref{sec:critical-vertices} for more details), and (ii) execute a critical section (using synchronization) that validates the speculative color $k$ computed in step (i) over the colors of the critical adjacent vertices (lines 8-9) and assigns the color $k$ to the vertex $v$, if the validation succeeds (lines 10-14). With the proposed speculative computation scheme, we provide small critical sections, i.e., having small data access footprints and short duration, thus achieving high amount of parallelism and high scalability to  a large number of threads.

\begin{figure}[t]
\begin{lstlisting}[language=C]
(*\bfseries Input:*) Graph G=(V,E)
for each (*$v \in V$*) do (*\textcolor{blue}{in parallel}*)
  RETRY:
  (*\textcolor{mygreen}{// Speculative Computation}*)
  Compute a speculative minimum color (*$k$*) that is permissible to the vertex (*$v$*)
  Keep track the critical adjacent vertices of the vertex (*$v$*)
  (*\textcolor{bostonuniversityred}{begin\_critical\_section}*)
  (*\textcolor{mygreen}{// Validate Coloring}*)
  Compare (*$k$*) with the colors of the critical adjacent vertices 
  if (no coloring conflict)
    (*v*).color = (*$k$*)
    (*\textcolor{bostonuniversityred}{end\_critical\_section}*)
  else
    (*\textcolor{bostonuniversityred}{end\_critical\_section}*)
    goto RETRY (*\textcolor{mygreen}{// Eager Resolution}*)
\end{lstlisting}
\vspace{-4pt}
\caption{\ColorTM: Overview.}
\label{alg:colortm-overview}
\vspace{-12pt}
\end{figure}

In addition, we leverage Hardware Transactional Memory (HTM) to implement synchronization on critical sections (lines 7, 12, and 14 of Figure~\ref{alg:colortm-overview}). HTM enables a speculative synchronization mechanism: multiple critical sections of parallel threads are executed concurrently with an optimistic approach that they will not cause any data inconsistency,  even though their data access sets might overlap. In contrast, fine-grained locking with software-based locks (e.g., provided by the pthread library) constitutes a more conservative synchronization approach:  multiple critical sections of parallel threads are executed concurrently, \emph{only} if their data access sets do \emph{not} overlap. Therefore, HTM can enable a higher number of critical sections to be executed in parallel compared to that enabled with the fine-grained locking scheme. We provide more details in Section~\ref{sec:htm-vs-lock}. With the speculative synchronization approach of HTM, \ColorTM{} further minimizes synchronization costs and provides high amount of parallelism.

\subsection{Eager Coloring Conflict Detection and Resolution}

We design \ColorTM{} to detect and resolve coloring inconsistencies \emph{eagerly}, i.e., immediately detecting and resolving coloring inconsistencies  at the time that the coloring conflicts appear. This way, the conflicted vertices are accessed multiple times, however within a short time during runtime. Therefore, application data corresponding to conflicted vertices can remain and be located in the first levels of the memory hierarchy of commodity platforms (i.e., in the low-cost cache memories), thus enabling \ColorTM{} to improve performance by achieving low data access costs.

In Figure~\ref{alg:colortm-overview}, parallel threads concurrently compute speculative colors for multiple vertices of the graph (lines 4-6), and at that time coloring inconsistencies may appear. Then, parallel threads \emph{immediately} detect possible coloring conflict inconsistencies for the current vertices using synchronization (lines 7-14). This way, parallel threads detect conflicts by accessing application data with low access latencies, since the data accessed in lines 7-14 has just been accessed within a short time, i.e., in lines 4-6. Next, if coloring conflicts arise (line 13), parallel threads \emph{immediately} resolve the coloring conflicts by directly retrying to find new colors for the current vertices (\texttt{goto RETRY} inline 15) (without proceeding to process new vertices). This way, parallel threads resolve conflict inconsistencies by accessing application data with low access latencies, since the data accessed in lines 4-6 after the execution of \texttt{goto RETRY} has just been accessed within a short time, i.e., in lines 7-14 of the previous iteration.

In \ColorTM{}, we highlight two important key design choices. First, \ColorTM{} executes only \emph{one} single parallel step (line 2). In contrast to prior state-of-the-art parallel graph coloring algorithms~\cite{Gebremedhin2000Scalable,Boman2005Scalable,Catalyurek2012GraphColoring,Rokos2015Fast,Lu2015Balanced}, \ColorTM{} \emph{completely} avoids barrier synchronization among parallel threads: multiple parallel threads repeatedly iterate over each vertex of the graph until a valid coloring is found. By completely avoiding barrier synchronization, \ColorTM{} can provide high scability. Second, \ColorTM{} does not perform re-colorings to vertices: once a vertex is assigned a permissible color, it will \emph{not} be re-colored again during the runtime. This way, colored vertices will \emph{not} introduce coloring inconsistencies with vertices that will be processed next. Prior \emph{lazy} iterative graph coloring schemes including IterSolve, IterSolveR, VFF and Recoloring do \emph{not} use data synchronization when they assign permissible colors to vertices. This way, many vertices are re-colored multiple times with different colors during runtime, and thus new additional coloring inconsistencies might be introduced due to re-colorings. Instead, \ColorTM{} employs HTM synchronization (lines 7, 12 and 14 of Figure~\ref{alg:colortm-overview}) when it assigns permissible colors to vertices (line 11 of Figure~\ref{alg:colortm-overview}). This way, vertices are assigned only \emph{one} final color during the runtime, thus avoiding introducing new coloring inconsistencies due to re-colorings.

\section{\ColorTM: Detailed Design}

\ColorTM{}~\cite{Giannoula2018Combining} is a high-performance graph coloring algorithm that leverages HTM to implement synchronization among parallel threads, and performs speculative computations outside the critical section in order to minimize the memory footprint and computations executed inside the critical section. In the section, we describe the detailed design and correctness of \ColorTM{}. We also extend our proposed design to introduce a new balanced graph coloring algorithm, named \BalColorTM{}, which  evenly distributes the vertices of the graph across color classes.

\subsection{Speculative Synchronization via HTM}\label{sec:htm-vs-lock}

\ColorTM{} leverages HTM to implement synchronization among parallel threads instead of using fine-grained locking. As already discussed, HTM is a more optimistic synchronization approach and can provide higher levels of parallelism compared to the fine-grained locking scheme. Specifically, multiple critical sections with \emph{overlapped} data access sets can be executed in parallel with HTM, while they need to be executed sequentially with fine-grained locking. 

\begin{figure}[H]
    \centering
    \includegraphics[width=0.78\linewidth]{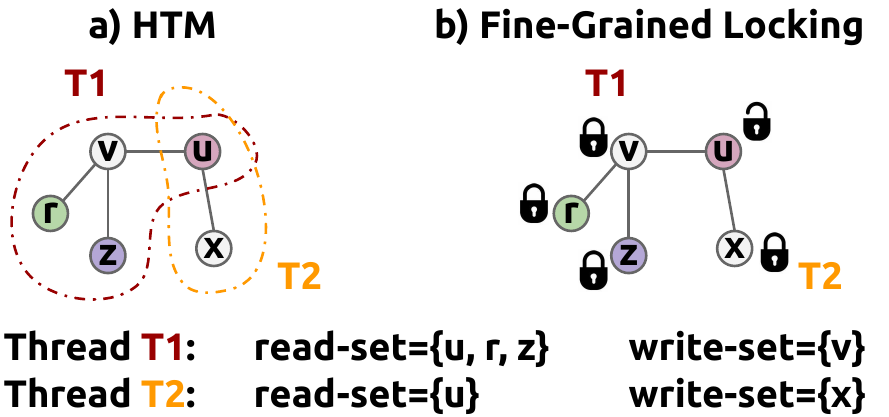}
    \vspace{-2pt}
    \caption{An example execution scenario in which threads $T1$ and $T2$ attempt to concurrently find colors for the vertices $v$ and $x$, respectively, using a) HTM and b) fine-grained locking for synchronization. The white circles represent uncolored vertices, and the colorful circles represent vertices that have already obtained a color.}
    \label{fig:htm_vs_lock}
    \vspace{-12pt}
\end{figure}

Figure~\ref{fig:htm_vs_lock} provides an example of the aforementioned scenario in graph coloring. Consider the scenario where thread $T1$ attempts to assign a color to the vertex $v$, and  thread $T2$ attempts to assign a color to the vertex $x$. Thread $T1$ needs to \emph{atomically} read the colors of the adjacent vertices of the vertex $v$, i.e., $u,r,z$ vertices, and write the corresponding color to the vertex $v$. Similarly, Thread $T2$ needs to \emph{atomically} read the colors of the adjacent vertices of the vertex $x$, i.e., $u$ vertex, and write the corresponding color to the vertex $x$. With HTM (Figure~\ref{fig:htm_vs_lock}a), $T1$'s and $T2$'s transactions can be executed and committed concurrently: neither the write-set of $T1$'s transaction does \emph{not} conflict with the read-set of $T2$'s transaction, nor the write-set of $T2$'s transaction does \emph{not} conflict with the read-set of $T1$'s transaction. Therefore, even though $T1$'s and $T2$'s critical sections have \emph{overlapped} data access sets, i.e., both of them include the color of the vertex $u$ in their read-sets, they can be executed concurrently with HTM. In contrast, with fine-grained locking, $T1$'s and $T2$'s critical sections are executed \emph{sequentially} (Figure~\ref{fig:htm_vs_lock}b): threads $T1$ and $T2$ compete to acquire the \emph{same} lock, i.e., the lock associated with the vertex $u$, in order to execute their critical sections. Thus, only \emph{one} of threads $T1$ and $T2$ will acquire the lock, and will proceed. Given that $T1$'s and $T2$'s critical sections have \emph{overlapped} data access sets, i.e., both of them include the color of the vertex $u$ in their read-sets, they will be executed sequentially when using the fine-grained locking scheme for synchronization. As a result, we conclude that in graph coloring HTM can provide higher levels of parallelism compared to fine-grained locking.

%\vspace{-10pt}

To this end, \ColorTM{} employs HTM to deal with race conditions that arise when parallel threads concurrently process adjacent vertices. HTM can detect and resolve coloring inconsistencies among parallel threads as follows:
\begin{compactitem}[$\textbf{--}$]
\item \textbf{HTM can detect coloring conflicts:} HTM detects coloring conflicts that arise due to crossing edges. For a vertex $v$ to be colored, we enclose within the transaction (i) the memory location that stores the color of the current vertex $v$ (the transaction's write-set), and (ii) the memory locations that store the colors of the critical adjacent vertices of the vertex $v$ (the transaction's read-set). When parallel threads  attempt to concurrently update-write the colors of adjacent vertices using different transactions, the HTM mechanism detects read-write conflicts across the running transactions: a running transaction attempts to write the read-set of another running transaction. Figure~\ref{fig:htm_detection} provides an example scenario on how HTM detects coloring inconsistencies among two parallel threads. When the thread $T1$ attempts to color the vertex $v$ using HTM, the corresponding running transaction includes the memory location of the color of the vertex $v$ in its write-set, and the memory locations of the colors of the $v$'s adjacent vertices, i.e., $u$, $r$ and $z$ vertices, in its read-set. Similarly, when the thread $T2$ attempts to color the vertex $u$ using HTM, the corresponding running transaction includes the memory location of the color of the vertex $u$ in its write-set, and the memory locations of the colors of the $u$'s adjacent vertices, i.e., $v$ and $x$ vertices, in its read-set. When $T1's$ and $T2's$ transactions are executed concurrently, HTM detects a read-write conflict either on the color of the vertex $v$ or the color of the vertex $u$: either $T1's$ transaction attempts to write the read-set of $T2's$ transaction or $T2's$ transaction attempts to write the read-set of $T1's$ transaction. Therefore, one of the two running transactions will be aborted by the HTM mechanism, and the other one will be committed.

%\vspace{-14pt}
\begin{figure}[H]
    \centering
    \includegraphics[width=0.74\linewidth]{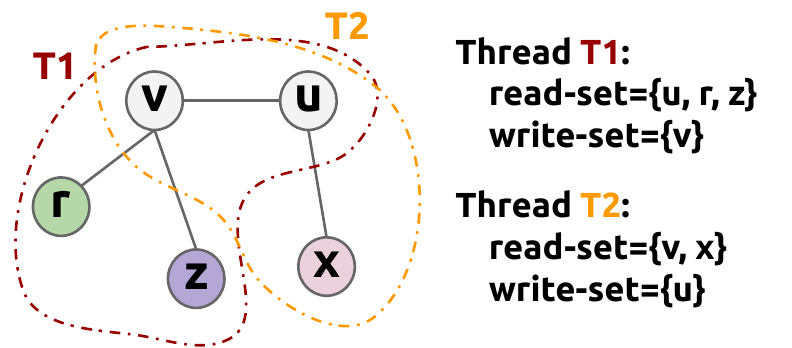}
    \vspace{-2pt}
    \caption{An example execution scenario in which threads $T1$ and $T2$ attempt to concurrently update the colors of the vertices $v$ and $u$ respectively, using two different transactions, and the HTM mechanism detects read-write conflicts to their data sets. The white circles represent uncolored vertices, and the colorful circles represent vertices that have already obtained a color.}
    \label{fig:htm_detection}
    \vspace{-8pt}
\end{figure}

\item \textbf{HTM can resolve coloring conflicts:} In case of $n$ conflicting running transactions (read-write conflicts explained in Figure~\ref{fig:htm_detection}), the HTM mechanism aborts $n-1$ running transactions and commits only one of them. In prior graph coloring schemes such as SeqSolve (line 11 of Figure~\ref{alg:seqsolve}), IterSolve (line 13 of Figure~\ref{alg:itersolve}), VFF (line 23 of Figure~\ref{alg:VFF}) and Recoloring (line 22 of Figure~\ref{alg:Recoloring}), the programmer explicitly defines a coloring conflict resolution policy among conflicted vertices to guarantee forward progress, i.e., the programmer explicitly defines which of the conflicted vertices will be re-colored next. In contrast, in \ColorTM{} when coloring conflicts arise among multiple running transactions, the programmer does \emph{not} need to explicitly define a conflict resolution policy: the HTM mechanism itself commits one of the multiple conflicted transactions and aborts the remaining running transactions. Thus, the conflict resolution policy implemented in the underlying hardware mechanism of HTM determines which vertices will continue to be processed for coloring.
\end{compactitem}

However, currently available HTM systems~\cite{Herlihy1993Transactional,Yoo2013Performance,Cain2013Robust,Wang2012Evaluation} are best-effort HTM implementations that do \emph{not} guarantee forward progress: a transaction may always fail to commit and thus, a non-transactional execution path (\emph{fallback path}) needs to be implemented. The most common fallback path is to implement a coarse-grained locking solution: each transaction can be retried up to a predefined number of times (pre-determined threshold), and if it exceeds this threshold, it fall backs to the acquisition of global lock, which allows only one single thread to execute its critical section. To implement this, the global lock is added to the transactions’ read sets: inside the transaction the thread always reads the value of the global lock variable. During the multithreaded execution, when the transaction of a parallel thread exceeds the predefined threshold of retries, the parallel thread acquires the global lock by writing to the value of the global lock variable, and then the concurrent running transactions of the remaining threads are aborted (read-write conflict) and wait until the global lock is released.

\subsection{Critical Adjacent Vertices}\label{sec:critical-vertices}

\ColorTM{} implements a speculative computation approach to achieve high performance. Specifically, for each vertex $v$, all necessary computations to find a permissible color $k$ are performed outside the critical section (line 5 in Figure~\ref{alg:colortm-overview}) such that avoid unnecessary computations inside the critical sections. Within the critical section, \ColorTM{} \emph{only} validates the speculative color $k$ (line 9 in Figure~\ref{alg:colortm-overview}) by comparing it with the colors of the adjacent vertices of vertex $v$. However, the speculative color $k$ for a vertex $v$ does \emph{not} need to be validated with the colors of \emph{all} the adjacent vertices of vertex $v$: we observe that some adjacent vertices can be omitted from the validation process of the critical section, because they do not cause \emph{any} coloring inconsistency with the vertex $v$. Specifically, we can omit from the validation step performed within the critical section the following adjacent vertices of vertex $v$:

\begin{inparaenum}
    \noindent\item \textbf{The adjacent vertices that are assigned to be processed by the same thread with the vertex $\mathbf{v}$.} Given that the vertices of the graph are distributed across multiple threads, coloring conflicts cannot arise between adjacent vertices that are assigned to the \emph{same} parallel thread. Therefore, we omit from the validation step of the critical section the adjacent vertices assigned to the same thread as the current vertex $v$.
    \\
    \noindent\item \textbf{The adjacent vertices that have already obtained a color.} As already explained, \ColorTM{} does \emph{not} perform re-colorings to the vertices of the graph: once a vertex is assigned a permissible color within the critical section (using synchronization), it will \emph{not} be re-colored again during runtime. Multiple parallel threads repeatedly iterate over a vertex until a valid coloring is found, which is assigned to it using data synchronization, and then proceed to the remaining vertices. Therefore, in \ColorTM{} coloring conflicts do \emph{not} arise between adjacent vertices that have already obtained a color: the colors assigned to adjacent vertices are taken into consideration in the computations performed outside the critical section (line 5 in Figure~\ref{alg:colortm-overview}) to find a speculative color for the current vertex, and will not be modified when the critical section is executed (lines 7-15 in Figure~\ref{alg:colortm-overview}), since \ColorTM{} does \emph{not} perform re-colorings. Therefore, adjacent vertices of a vertex $v$ that have already obtained a color when the speculative coloring computation step (line 5 in Figure~\ref{alg:colortm-overview}) is executed, do not cause any coloring inconsistency when critical section is executed (lines 7-15 in Figure~\ref{alg:colortm-overview}). Hence, we can safely omit from the validation step of the critical section the adjacent vertices that have already been assigned a color.
\end{inparaenum}
 
\begin{figure}[t]
    %\vspace{-12pt}
    \centering
    \includegraphics[width=0.74\linewidth]{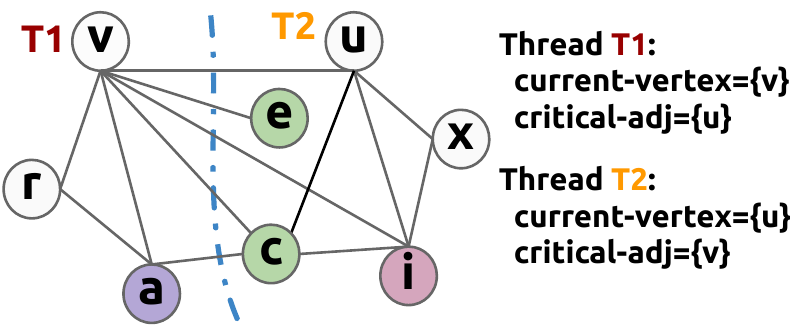}
    \vspace{-2pt}
    \caption{An example execution scenario in which the graph is partitioned across two parallel threads. The white circles represent uncolored vertices, and the colorful circles represent vertices that have already obtained a color. When the threads $T1$ and $T2$ attempt to color the vertices $v$ and $u$, respectively, the critical adjacent vertices that need to be validated within the critical section (HTM) are \emph{only} the vertices $u$ and $v$, respectively.}
    \label{fig:critical_vertices}
    \vspace{-8pt}
\end{figure}

Figure~\ref{fig:critical_vertices} presents an example execution scenario of a graph partitioned across two parallel threads $T1$ and $T2$. In Figure~\ref{fig:critical_vertices}, the white vertices represent uncolored vertices and the colorful vertices represent vertices that have already obtained a color during runtime. In this scenario, threads $T1$ and $T2$ attempt to color the vertices $v$ and $u$, respectively. According to our described optimizations, the adjacent vertices that need to be validated inside the critical sections (via HTM) of the vertices $v$ and $u$ are \emph{only} the vertices $u$ and $v$, respectively.
 
Overall, for the current vertex $v$ to be colored, the necessary adjacent vertices that need to be validated inside the critical section, referred to as \emph{critical} adjacent vertices, are the \emph{uncolored} adjacent vertices assigned to different parallel threads compared to the thread to which the vertex $v$ is assigned to. By accessing inside the critical section \emph{only} a few data needed to ensure correctness, \ColorTM{} provides short critical sections and small transaction footprints, and achieves high levels of parallelism and low synchronization costs, i.e., low abort ratio in hardware transactions of HTM (See Figure~\ref{coloring-abort-ratio}). Note that having large transactions footprints in HTM transactions can cause three important problems: (i) if the transaction read- and write-sets are large, the available hardware buffers of HTM may be oversubscribed (hardware overflow), and in that case the HTM mechanism will abort the running transactions due to capacity aborts, (ii) if the duration of a running transaction is long (e.g., due to expensive data accesses), the running transactions may be aborted due to a time interrupt (when the duration of a transaction exceeds the time scheduling quantum, the OS scheduler schedules out the software thread from the hardware thread and the transaction is aborted), and (iii) the longer the transactions last and the larger their data sets are, the greater the probability that running transactions are aborted due to (read-write) data conflicts among them.

\subsection{Implementation Details}

Figure~\ref{alg:colortm-detail} presents  \ColorTM{} in detail. \ColorTM{} distributes the vertices of the graph across multiple threads, which color the vertices of the graph through one single parallel step (lines 4-29): multiple parallel threads repeatedly iterate over each vertex of the graph until a valid coloring on each vertex is performed.

%\vspace{-14pt}
\begin{figure}[t]
\begin{lstlisting}[language=C]
(*\bfseries Input:*) Graph G=(V,E)
Let (*$N(v)$*) be the adjacent vertices of the vertex (*$v$*)
Let tid be the unique id of each parallel thread
for each (*$v \in V$*) do (*\textcolor{blue}{in parallel}*)
  RETRY:
  (*\textcolor{mygreen}{// Speculative Computation}*)
  R = (*$\emptyset$*) (*\textcolor{mygreen}{ // Track Forbidden Colors}*)
  C = (*$\emptyset$*) (*\textcolor{mygreen}{ // Track Critical Adjacent Vertices}*)
  for each (*$u \in N(v)$*) do
    R = R (*$\cup$*) (*$u$*).color
    if ((hasColor((*$u$*)) == false) && (get_threadID((*$u$*)) != tid))
      C = C (*$\cup$*) (*$u$*) (*\textcolor{mygreen}{ // Critical Adjacent Vertices Are the Uncolored Vertices Assigned to Another Thread}*)
  (*$k$*) = compute_speculative_color(R)  (*\textcolor{mygreen}{ // Compute a Speculative Color $k$ for the Vertex $v$}*)
  (*\textcolor{mygreen}{// Validate Coloring}*)
  if (C == (*$\emptyset$*)) (*\textcolor{mygreen}{ // Skip the Validation Step, If There Are No Critical Adjacent Vertices}*)
    (*$v$*).color = (*$k$*)
  else
    (*\textcolor{bostonuniversityred}{begin\_transaction}*)
    (*\textcolor{blue}{bool}*) valid = true
    for each (*$u \in C$*) do (*\textcolor{mygreen}{ //  Validate the Colors of the Critical Adjacent Vertices Over the Speculative Color}*)
      if ((*$u$*).color == (*$k$*)) 
        valid = false
        break
    if (valid == true) (*\textcolor{mygreen}{ //  If the Validation Succeeded, Assign the Speculative Color to the Vertex $v$}*)
      (*$v$*).color = (*$k$*)
      (*\textcolor{bostonuniversityred}{end\_transaction}*)
    else (*\textcolor{mygreen}{ //  If the Validation Failed, Immediately Retry to Find a New Color for the Vertex $v$}*)
      (*\textcolor{bostonuniversityred}{end\_transaction}*)
      goto RETRY (*\textcolor{mygreen}{// Eager Resolution}*)
\end{lstlisting}
\caption{The \ColorTM{} algorithm.}
\label{alg:colortm-detail}
\vspace{-8pt}
\end{figure}

For each vertex $v$, there are two sub-steps. In the first sub-step (lines 6-13), the parallel thread keeps track (i) the forbidden set of colors assigned to the adjacent vertices of the vertex $v$ (line 10), and (ii) the critical adjacent vertices of the vertex $v$ (lines 11-12), which are the uncolored adjacent vertices assigned to different parallel threads (line 11), and then computes a speculative color $k$ that is permissible for the vertex $v$ using the \texttt{compute\_speculative\_color()} function (line 13). In the second sub-step (lines 14-29), the parallel thread validates and assigns (if allowed) the speculative color $k$ to the vertex $v$ using data synchronization via HTM (lines 18-29). Specifically, the colors of the critical adjacent vertices are compared to the speculative color $k$ within a hardware transaction (lines 20-23) to ensure that the color $k$ is still permissible to be assigned to the vertex $v$. If the validation succeeds (line 24), the color $k$ is assigned to the vertex $v$ within the same transaction (line 25) to ensure correctness: recall that the reads on the colors of the critical adjacent vertices need to be executed \emph{atomically} with the write-update on the color of the vertex $v$. Instead, if the validation step fails due to a coloring inconsistency appearing during runtime (line 27), the parallel thread \emph{repeatedly} and \emph{eagerly} retries to find a new permissible color for the current vertex $v$ (line 29). Note that if there are \emph{no} critical adjacent vertices to be validated (line 15), the speculative color $k$ is directly assigned to the vertex $v$ \emph{without} using synchronization (line 16).

Note that in the second sub-step (lines 14-29), \ColorTM{} does \emph{not} check if the colors of the critical adjacent vertices have not been modified since the first sub-step (lines 6-13). Instead, the validation of the second sub-step \emph{only} checks that the colors of the critical adjacent vertices are different from the speculative color $k$ computed in the first sub-step (line 13). In the meantime, different parallel threads may have just assigned new colors to critical adjacent vertices, which however are different from the color $k$, and thus causing \emph{no} coloring inconsistencies. In that scenario, the validation of the second sub-step succeeds. This way, \ColorTM{} provides high levels of parallelism: multiple parallel threads that have just assigned \emph{different} colors than the color $k$ to critical adjacent vertices of the vertex $v$ will \emph{not} cause any validation failure in the critical section of the vertex $v$, and the corresponding running transaction will be safely committed.

\subsection{Progress and Correctness}

We clarify in detail how \ColorTM{} resolves the race conditions that may arise during runtime. There are two race conditions that may cause coloring inconsistencies in multithreaded executions. First, while a parallel thread computes a speculative color $k$ for the vertex $v$ (lines 9-13 of Figure~\ref{alg:colortm-detail}), different parallel threads may have just assigned the color $k$ to one or more adjacent vertices of the vertex $v$. In that scenario, the validation step of lines 20-23 of Figure~\ref{alg:colortm-detail} fails (line 22, 27), since the speculative color $k$ has been assigned to at least one critical adjacent vertex (line 21). Then, the corresponding parallel thread will retry to find a new permissible color for the vertex $v$ (line 29). Second, a race condition arises when $n$ parallel threads (assuming $n$ > 1) attempt to write-update the same color $k$ to $n$ adjacent vertices (fully connected adjacent vertices) within $n$ different running transactions. In that scenario, the HTM mechanism detects read-write data conflicts on running transactions, because one (or more) running transaction attempts to write to the read-sets of another running transactions. Recall that the colors of the critical adjacent vertices are included in the read-set of each running transaction (lines 21 of Figure~\ref{alg:colortm-detail}). Then, the HTM mechanism aborts $n-1$ running transactions, and commits only \emph{one} of them. When the aborted $n-1$ transactions retry (each transaction can retry up to a predefined number of times), the validation step of lines 20-23 fails (lines 27 of Figure~\ref{alg:colortm-detail}), since at that time the $n-1$ parallel threads observe that there is one critical adjacent vertex that has just been assigned to the color $k$ (the committed transaction). Afterwards, since the validation failed, the $n-1$ parallel threads will retry to find new permissible colors for their current vertices (lines 27-29 of Figure~\ref{alg:colortm-detail}).

Finally, we clarify that \ColorTM{} provides forward progress and eventually terminates: each parallel thread retries to find a new permissible color for a current vertex $v$ (line 29 of Figure~\ref{alg:colortm-detail}) up to a limited number of retries. Specifically, a parallel thread retries to find a new color for a vertex $v$, when the validation step of lines 20-23 of Figure~\ref{alg:colortm-detail} fails. However, for each vertex $v$ the validation step can fail up to a bounded number of times: the validation step fails when one (or more) critical adjacent vertex has been assigned to the same color $k'$ with the speculative color $k$ computed for the vertex $v$. Therefore, in the worst case, the validation step might fail up to $deg(v)$ times, where $deg(v)$ is the adjacency degree of the vertex $v$. When all $v$'s adjacent vertices have obtained a color, there are \emph{no} critical adjacent vertices to be validated (line 15 of Figure~\ref{alg:colortm-detail}), and thus, the speculative color $k$ is directly assigned to the vertex $v$ (line 16 of Figure~\ref{alg:colortm-detail}), and the validation step is omitted. As a result, each parallel thread retries to find a color for each vertex $v$ of the graph at most $deg(v)$ times. However, in our evaluation, we find that the validation step fails only for \emph{a few} times: across all our evaluated large real-world graphs (Table~\ref{matrix-info}) and using a large number of parallel threads (up to 56 threads) the validation step failures are less than 0.01\%. Overall, we conclude that \ColorTM{} \emph{correctly} handles all the race conditions that may arise in multithreaded executions of the graph coloring kernel, and \emph{effectively}  terminates with a valid coloring.

\begin{figure}[!ht]
\begin{lstlisting}[language=C]
(*\bfseries Input:*) Graph G=(V,E)
Let (*$N(v)$*) be the adjacent vertices of the vertex (*$v$*)
Obtain an initial coloring on G
Let (*$C$*) be the number of colors produced
Let (*$b = V / C$*) be the perfect balance
Let (*$Q$*) be the set of vertices of the over-full color classes
for each (*$v \in Q$*) do (*\textcolor{blue}{in parallel}*)
  Let (*$c$*) be the current color of the vertex (*$v$*)
  if (the size of the color class (*$c$*) <= (*$b$*))
      continue(*\textcolor{mygreen}{// Color Class is Balanced}*)
  RETRY:
  (*\textcolor{mygreen}{// Speculative Computation}*)
  R = (*$\emptyset$*) (*\textcolor{mygreen}{ // Track Forbidden Colors}*)
  C = (*$\emptyset$*) (*\textcolor{mygreen}{ // Track Critical Adjacent Vertices}*)
  for each (*$u \in N(v)$*) do
    R = R (*$\cup$*) (*$u$*).color
    if ((isOverFull((*$u$*).color) == true) && (get_threadID((*$u$*)) != tid))
      C = C (*$\cup$*) (*$u$*) (*\textcolor{mygreen}{ // Critical Adjacent Vertices Are the Vertices of Over-Full Color Classes That Are Assigned to Another Thread}*)
  (*$k$*) = compute_speculative_color(R)   
  Let (*$k$*) be the index of the minimum under-full color class that is permissible to the vertex (*$v$*)
  if ((*$k$*) exists) (*\textcolor{mygreen}{// Validate Coloring}*)
    if (C == (*$\emptyset$*)) (*\textcolor{mygreen}{ // Skip the Validation Step, If There Are No Critical Adjacent Vertices}*)
      (*$v$*).color = (*$k$*)
      (*\emph{Atomically}*) decrease the size of the color class (*$c$*) 
      (*\emph{Atomically}*) increase the size of the color class (*$k$*)  
    else
      (*\textcolor{bostonuniversityred}{begin\_transaction}*)
      (*\textcolor{blue}{bool}*) valid = true
      for each (*$u \in C$*) do (*\textcolor{mygreen}{ //  Validate the Colors of the Critical Adjacent Vertices Over the Speculative Color}*)
        if ((*$u$*).color == (*$k$*)) 
          valid = false
          break
      if (valid == true) (*\textcolor{mygreen}{ //  If the Validation Succeeded, Set the Speculative Color to the Vertex $v$}*)
        (*$v$*).color = (*$k$*)
        (*\textcolor{bostonuniversityred}{end\_transaction}*)
        (*\emph{Atomically}*) decrease the size of the color class (*$c$*) 
        (*\emph{Atomically}*) increase the size of the color class (*$k$*)
      else (*\textcolor{mygreen}{ //  If the Validation Failed, Immediately Retry to Find a New Color for the Vertex $v$}*)
        (*\textcolor{bostonuniversityred}{end\_transaction}*)
        goto RETRY (*\textcolor{mygreen}{// Eager Resolution}*)
  else
    continue
\end{lstlisting}
\vspace{-2pt}
\caption{The \BalColorTM{} algorithm.}
\label{alg:balcolortm}
\end{figure}

\subsection{The \BalColorTM{} Algorithm}

Figure~\ref{alg:balcolortm} presents the \emph{balanced} counterpart of \ColorTM{}, named as \BalColorTM{}. Similarly to CLU and VFF, in \BalColorTM{} (i) only the vertices of the over-full color classes are considered for re-coloring, i.e., to be moved from over-full to under-full color classes in order to achieve high vertex-balance across color classes, and (ii) graph coloring balance is achieved \emph{without increasing} the number of color classes produced by the initial graph coloring (e.g., using \ColorTM).

Similarly to \ColorTM, \BalColorTM{} (Figure~\ref{alg:balcolortm}) has one single parallel step (lines 7-42): multiple parallel threads repeatedly iterate over each vertex of the over-full color classes until either a valid re-coloring to an under-full class is performed, or there is no permissible re-coloring for this vertex to an under-full color class (line 42). For each vertex of an over-full color class $c$, there are two sub-steps. In the first sub-step (lines 8-20), the parallel thread keeps track the forbidden set of colors assigned to the adjacent vertices of the vertex $v$ (line 16), and the set of the critical adjacent vertices (lines 17-18) of the vertex $v$. In \BalColorTM{}, note that the \emph{critical} adjacent vertices of a vertex $v$ (line 17) are the adjacent vertices that (i) belong to \emph{over-full} color classes (recall that the vertices assigned under-full color classes are \emph{not} considered to be re-colored/moved, and thus they do not cause any coloring inconsistency during runtime), and (ii) are assigned to different threads compared to the parallel thread in which the vertex $v$ is assigned to. Then, the parallel thread speculatively computes a color $k$ of an under-full color class that is permissible to be assigned to the vertex $v$ (lines 19-20). If a permissible color $k$ exists (without increasing the number of color classes produced by the initial graph coloring), the parallel thread attempts to assign the speculative color $k$ to the vertex $v$ in the second sub-step (lines 21-42). If there is \emph{no} permissible color $k$ of an under-full color class (line 41), the parallel threads continue to process the next vertices (line 42). In the second sub-step, if there are critical adjacent vertices that need to be validated, the parallel thread validates the speculative color $k$ over the colors of the critical adjacent vertices within an HTM transaction (lines 27-39). If the validation succeeds (line 33), the parallel thread moves the vertex $v$ from the color class $c$ to the color class $k$ by re-coloring it (line 34), and atomically updates the sizes of the color classes $c$ and $k$ (lines 36-37) accordingly. If the validation step fails due to a coloring inconsistency appearing during runtime (line 38), the parallel thread \emph{eagerly} retries to find a new permissible color of an under-full color class for the vertex $v$ (line 40). Finally, note that \BalColorTM{} iterates over the vertices of each over-full color class until that particular over-full class becomes balanced at a certain point in the execution (lines 9-10), i.e., until the size of the particular color class becomes smaller or equal to $b=V/C$. Then, the vertices belonging to that color class are no longer considered for re-coloring (line 10). Overall, \BalColorTM{} terminates when either vertex-balance across color classes is achieved or vertex-balance across color classes is no longer available, i.e., there are no more permissible re-colorings for any vertex belonging to an over-full color class.

Similarly to \ColorTM, \BalColorTM{} \emph{completely} avoids barrier synchronization, since it includes only one single parallel step, thus minimizing synchronization costs. Moreover, it also integrates an \emph{eager} approach to detect and resolve coloring conflicts appearing during runtime among parallel threads, that concurrently move vertices from over-full to under-full color classes. With the eager coloring policy, \BalColorTM{} provides high performance by minimizing access latency costs to application data. Finally, \BalColorTM{} effectively implements short critical sections (short running transactions with small transaction footprints) by (i) speculatively performing the computations to find permissible colors for the vertices of the over-full color classes \emph{outside} the critical section (lines 9-13), and (ii) accessing inside the critical sections \emph{only} the necessary data to ensure correctness, i.e., for each vertex $v$ \BalColorTM{} \emph{only} accesses the colors of a small subset of $v$'s adjacent vertices (critical adjacent vertices). Via short running transactions, \BalColorTM{} achieves low synchronization costs and provides high amount of parallelism.

\section{Evaluation Methodology}
\vspace{-6pt}
We conduct our evaluation using a 2-socket Intel Haswell server with an Intel Xeon E5-2697 v3 processor with 28 physical cores and 56 hardware threads. The processor runs at 2.6 GHz and each physical core has its own L1 and L2 caches of sizes 32 KB and 256 KB, respectively. Each socket includes a shared 35 MB L3 cache. We statically pin each software thread to a hardware thread, and enable hyperthreading only on 56-thread executions, unless otherwise stated. In our evaluation (Section~\ref{eval}), the numbers reported are averaged across 5 runs of each experiment.

Table~\ref{matrix-info} shows the characteristics of the large real-world graphs used in our evaluation. We select 18 representative graphs from the Suite Matrix Collection that vary in vertex and graph degrees, and are used in different application domains. For each graph, Table~\ref{matrix-info} presents the number of vertices (\#vertices), the number of edges (\#edges), the maximum (\textbf{$deg_{max}$}) degree, the average (\textbf{$deg_{avg}$}) degree and the standard deviation of the vertices' degrees (\textbf{$deg_{std}$}), and the last column of this table shows the ratio of the standard deviation of the vertices' degrees to the average degree (\textbf{$\frac{deg_{std}}{deg_{avg}}$}).

\begin{table}[ht]
\centering
\resizebox{\columnwidth}{!}{
\begin{tabular}{|l||r|r|r|r|r|r|}
    \hline
    \textbf{Graph Name} & \textbf{\#Vertices} &\textbf{\#Edges} & $ \mathbf{deg_{max}}$ &$\mathbf{deg_{avg}}$ & $\mathbf{deg_{std}}$ &$\mathbf{\frac{deg_{std}}{deg_{avg}}}$ \\
    \hline \hline
    
    Queen\_4147 (\textbf{qun}) & 4147110 & 329499284 & 81 & 79.45 & 6.34 & 0.080 \\ \hline
    
    Geo\_1438 (\textbf{geo}) & 1437960 & 63156690 & 57 & 43.92 & 4.39 & 0.100 \\ \hline
    
    Flan\_1565 (\textbf{fln}) & 1564794 & 117406044 & 81 & 75.03 & 11.43 & 0.152 \\ \hline
    
    Bump\_2911 (\textbf{bum}) & 2911419 & 127729899 & 195 & 43.87 & 6.96 & 0.159 \\ \hline
    
    Serena (\textbf{ser}) & 1391349 & 64531701 & 249 & 46.38 & 9.24 & 0.199 \\ \hline
    
    delaunay\_n24 (\textbf{del}) & 16777216 & 100663202 & 26 & 5.99 & 1.34 & 0.222 \\ \hline
    
    rgg\_n\_2\_23\_s0 (\textbf{rgg}) & 8388608 & 127002786 & 40 & 15.14 & 3.89 & 0.257 \\ \hline
    
    kmer\_A2a (\textbf{kmr}) & 170728175 & 360585172 & 40 & 2.11 & 0.57 & 0.267 \\ \hline
    
    cage15 (\textbf{cag}) & 5154859 & 99199551 & 47 & 19.24 & 5.73 & 0.298 \\ \hline
    
    road\_usa (\textbf{usa}) & 23947347 & 57708624 & 9 & 2.41 & 0.93 & 0.386 \\ \hline
    
    dielFilterV3real (\textbf{dlf}) & 1102824 & 89306020 & 270 & 80.98 & 36.56 & 0.451 \\ \hline
    
    audikw\_1 (\textbf{aud}) & 943695 & 77651847 & 345 & 82.29 & 42.44 & 0.516 \\ \hline
    
    vas\_stokes\_2M (\textbf{vas}) & 2146677 & 65129037 & 637 & 30.34 & 37.18 & 1.226 \\ \hline
    
    stokes (\textbf{stk}) & 11449533 & 349321980 & 720 & 30.51 & 41.44 & 1.358 \\ \hline
    
    uk-2002 (\textbf{uk}) & 18520486 & 298113762 & 2450 & 16.10 & 27.53 & 1.710 \\ \hline
    
    soc-LiveJournal1 (\textbf{soc}) & 4847571 & 68993773 & 20293 & 14.23 & 36.08 & 2.535 \\ \hline
    
    arabic-2005 (\textbf{arb}) & 22744080 & 639999458 & 9905 & 28.14 & 78.84 & 2.802 \\ \hline
    
    FullChip (\textbf{fch}) & 2987012 & 26621990 & 2312481 & 8.91 & 1806.80 & 202.725 \\ \hline

    %\hline
\end{tabular}
}
\vspace{0pt}
\caption{Large Real-World Graph Dataset.}
\label{matrix-info}
\vspace{-6pt}
\end{table}
\section{Evaluation}\label{eval}
\vspace{-8pt}
This section evaluates the proposed \ColorTM{} and \BalColorTM{} algorithms. First, we compare the coloring quality and the performance %and the execution behavior of \ColorTM{} 
over prior state-of-the-art graph coloring algorithms, as well as the execution behavior of \ColorTM{} (Section~\ref{eval:unbalanced}). Second, we compare the color balancing quality and the performance of \BalColorTM{} over prior state-of-the-art balanced graph coloring algorithms, as well as the execution behavior of \BalColorTM{} (Section~\ref{eval:balanced}). Finally, we evaluate the performance of  %PageRank~\cite{page1999pagerank} and 
Community Detection~\cite{FORTUNATO201075} by parallelizing it using \ColorTM{} and \BalColorTM{} (Section~\ref{real-apps}) via chromatic scheduling.

\subsection{Analysis of Parallel Graph Coloring Algorithms}\label{eval:unbalanced}
\vspace{-4pt}
We compare the following parallel graph coloring implementations:
\begin{compactitem}
\item The sequential Greedy algorithm presented in Figure~\ref{alg:greedy}.
\item The SeqSolve algorithm presented in Figure~\ref{alg:seqsolve}.
\item The IterSolve algorithm presented in Figure~\ref{alg:itersolve}.
\item The IterSolveR algorithm presented in Figure~\ref{alg:itersolveR}.
\item A variant of our proposed algorithm (Figure~\ref{alg:colortm-detail}) that uses fine-grained locking instead of HTM, henceforth referred to as ColorLock. Specifically, each vertex of the graph is associated with a software-based lock. In the beginning of the critical section (line 18 in Figure~\ref{alg:colortm-detail}), parallel threads acquire the corresponding locks of both the current vertex $v$ and the \emph{critical} adjacent vertices of the vertex $v$. Then, when the critical section ends (lines 26 and 28 in Figure~\ref{alg:colortm-detail}), parallel threads release the acquired locks. To avoid deadlocks, we impose a global order  when acquiring/releasing locks based on the vertices' id: parallel threads acquire/release locks of multiple vertices starting from the lock associated with the vertex with the smallest vertex id, iterating via an increasing order of the vertices' ids, and finishing to the lock associated with the vertex with the highest vertex id.
\item Our proposed \ColorTM{} algorithm (Figure~\ref{alg:colortm-detail}) that leverages HTM. Each transaction can retry up to 50 times, before resorting to a non-transactional fallback path. The non-transactional path is a coarse-grained locking solution for the critical section (lines 18-28 in Figure~\ref{alg:colortm-detail}).
\end{compactitem}
For a fair comparison, in all graph coloring schemes we color the vertices in the order they appear in the input graph representation (first-fit ordering heuristic~\cite{Hasenplaugh2014Ordering}).

\subsubsection{Analysis of the Coloring Quality}

Table~\ref{coloring-quality} compares the coloring quality of all parallel graph coloring implementations in single-threaded and multithreaded executions.

\begin{table}[H]
\centering
\resizebox{0.66\columnwidth}{!}{
\begin{tabular}{|l||c|c|c|c|}
    \hline
    \textbf{Coloring} & \textbf{1} & \textbf{14} & \textbf{28} & \textbf{56} \\
    \textbf{Scheme} & \textbf{thread} & \textbf{threads} & \textbf{threads} & \textbf{threads} \\
    \hline \hline
    
    Greedy & 42.58 & - & - & -  \\ \hline
    SeqSolve & 42.58 & 42.34 & 42.33 & 42.18  \\ \hline
    IterSolve & 42.58 & 44.05 & 43.94 & 44.04  \\ \hline
    IterSolveR & 42.58 & 43.61 & 43.88 & 44.58  \\ \hline
    ColorLock & 42.58 & 45.75 & 45.67 & 46.14  \\ \hline
    \textbf{ColorTM} & 42.58 & 46.20 & 45.77 & 46.28 \\ \hline
    %\hline
\end{tabular}
}
\vspace{2pt}
\caption{The geometric mean on the number of colors produced across all large real-world graphs (lower is better) for each parallel graph coloring implementation using one core (1 thread), all cores of one socket (14 threads), all cores of two sockets (28 threads), and the maximum hardware thread capacity of our machine with hyperthreading enabled (56 threads). }
\label{coloring-quality}
\vspace{-14pt}
\end{table}

We make two key observations. First, there is low variability on the number of colors used across the different graph coloring schemes. The parallel graph coloring schemes provide similar graph coloring quality, because the number of colors produced is primarily determined by the order in which the vertices are colored~\cite{Maciej2020GC,Hasenplaugh2014Ordering}. In this work, we use the \emph{first-fit} ordering heuristic in all schemes, i.e., coloring the vertices in the order they appear in the input graph representation, and we leave the experimentation of other ordering heuristics for future work. Second, we find that in most schemes the coloring quality becomes slightly worse as the number of threads increases. As the number of threads increases, the number of coloring conflicts that arise during runtime typically increases, and thus parallel threads might resolve coloring inconsistencies by introducing a few additional color classes. The SeqSolve scheme does not typically increase the number of colors used in multithreaded executions, because the coloring inconsistencies are resolved using one single thread. Overall, we conclude that since all graph coloring schemes employ the same ordering heuristic, they provide similar coloring quality.

\begin{figure}[!ht]
\begin{minipage}{1.0\columnwidth}
\centering
\includegraphics[width=\columnwidth]{ 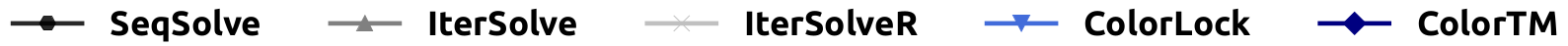}
\end{minipage}
\begin{minipage}{1.0\columnwidth}
\centering
\includegraphics[scale=0.04]{ 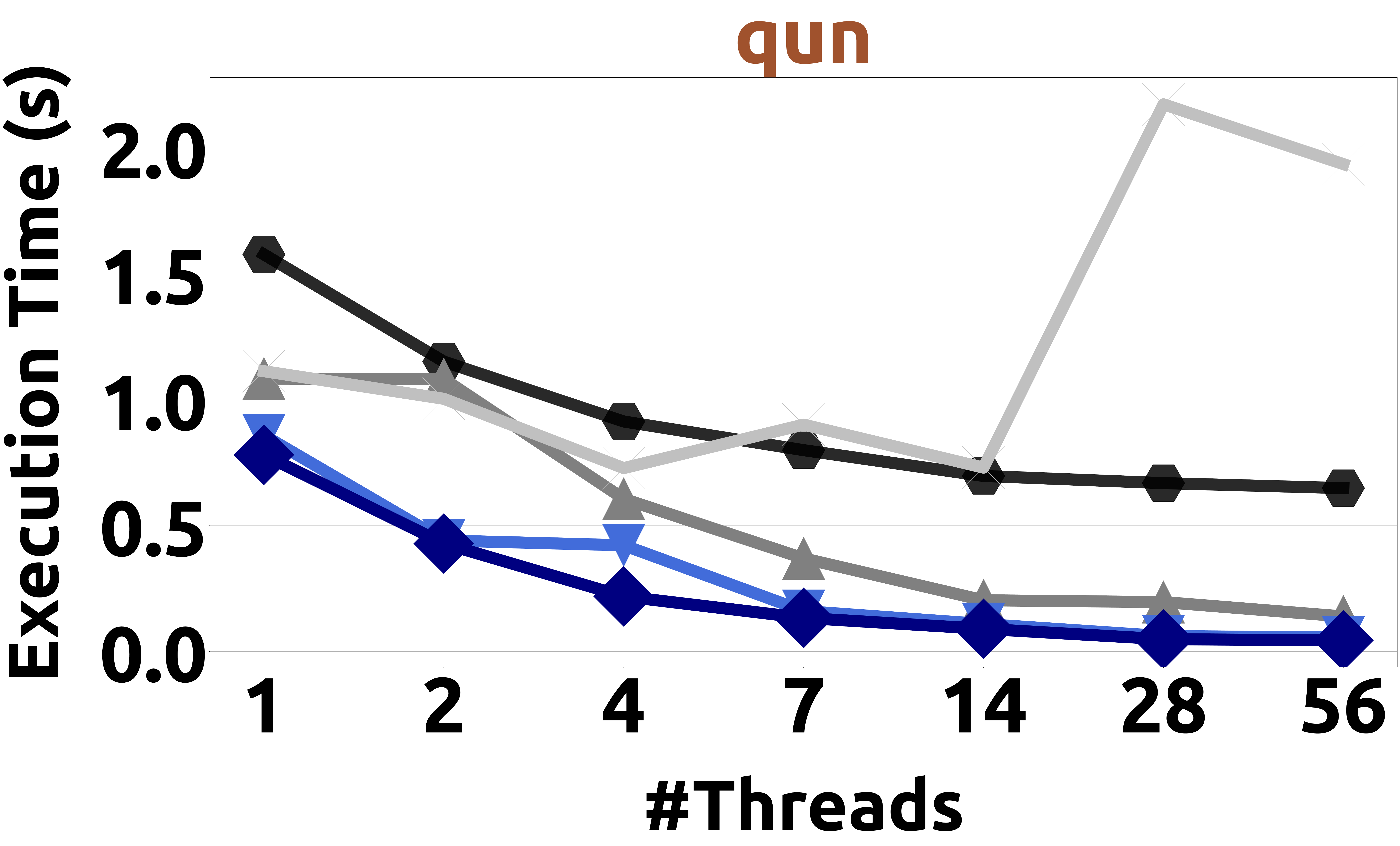}
\includegraphics[scale=0.04]{ 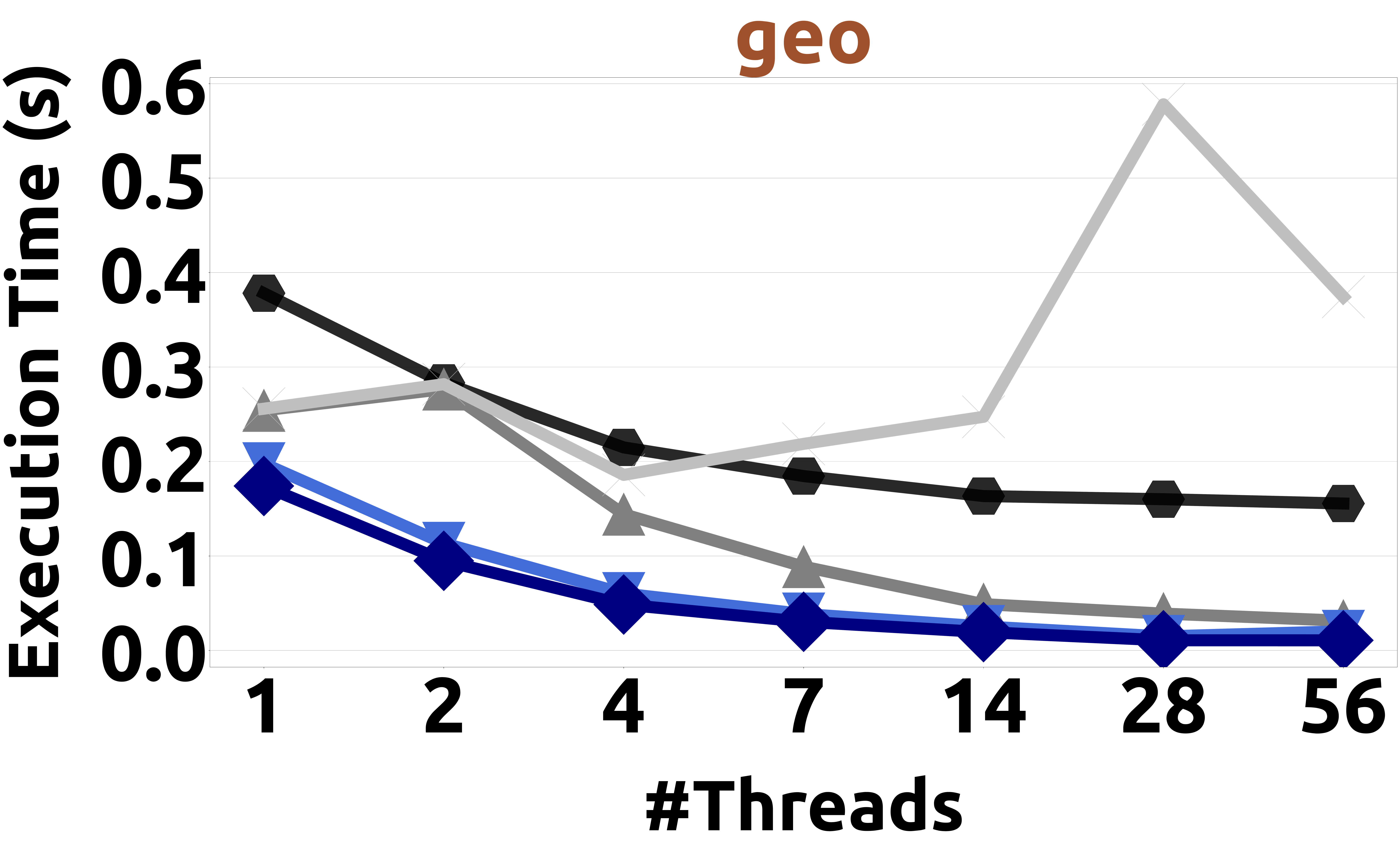}
\includegraphics[scale=0.04]{ 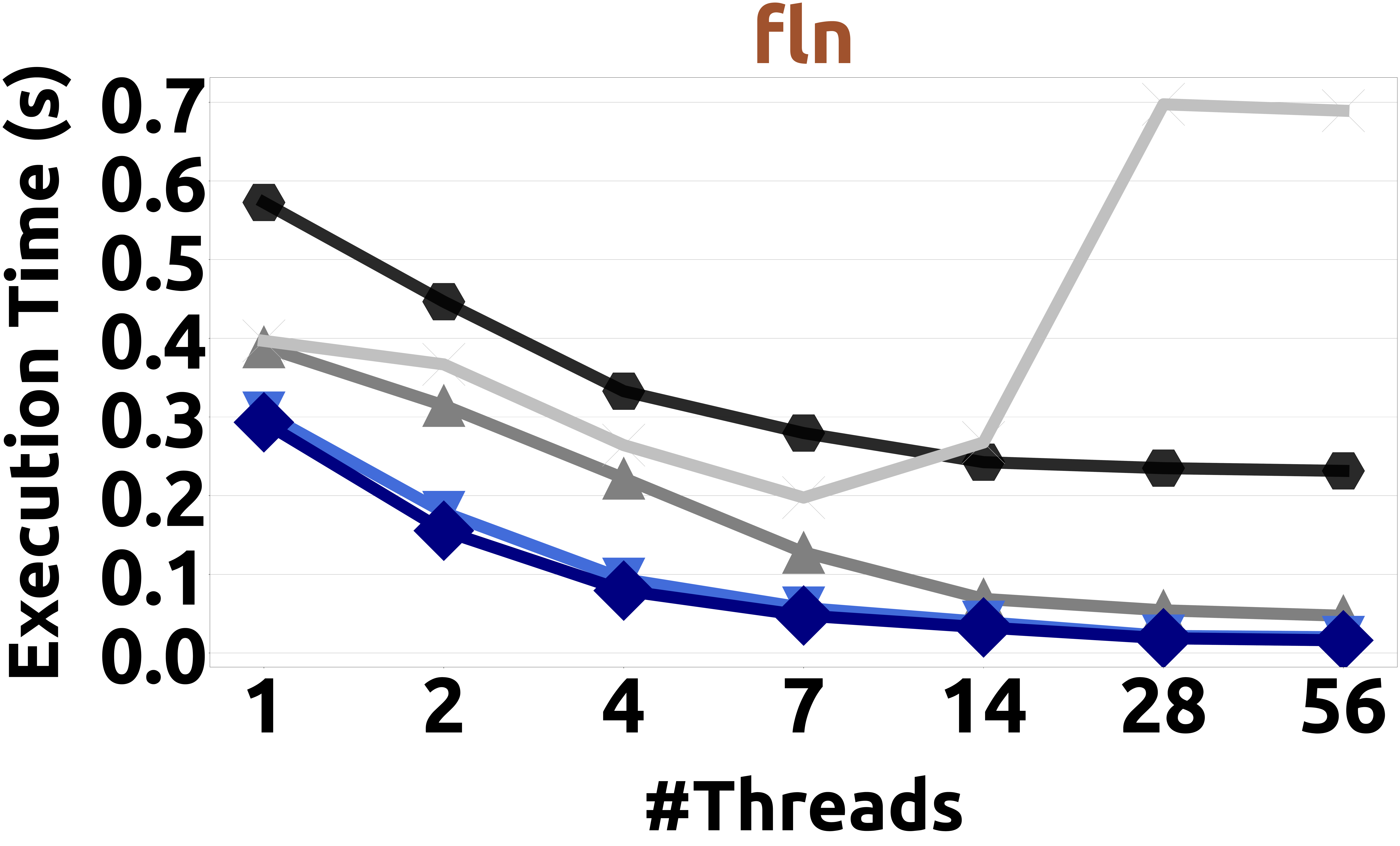}
\end{minipage}
\begin{minipage}{1.0\textwidth}
\centering
\includegraphics[scale=0.04]{ 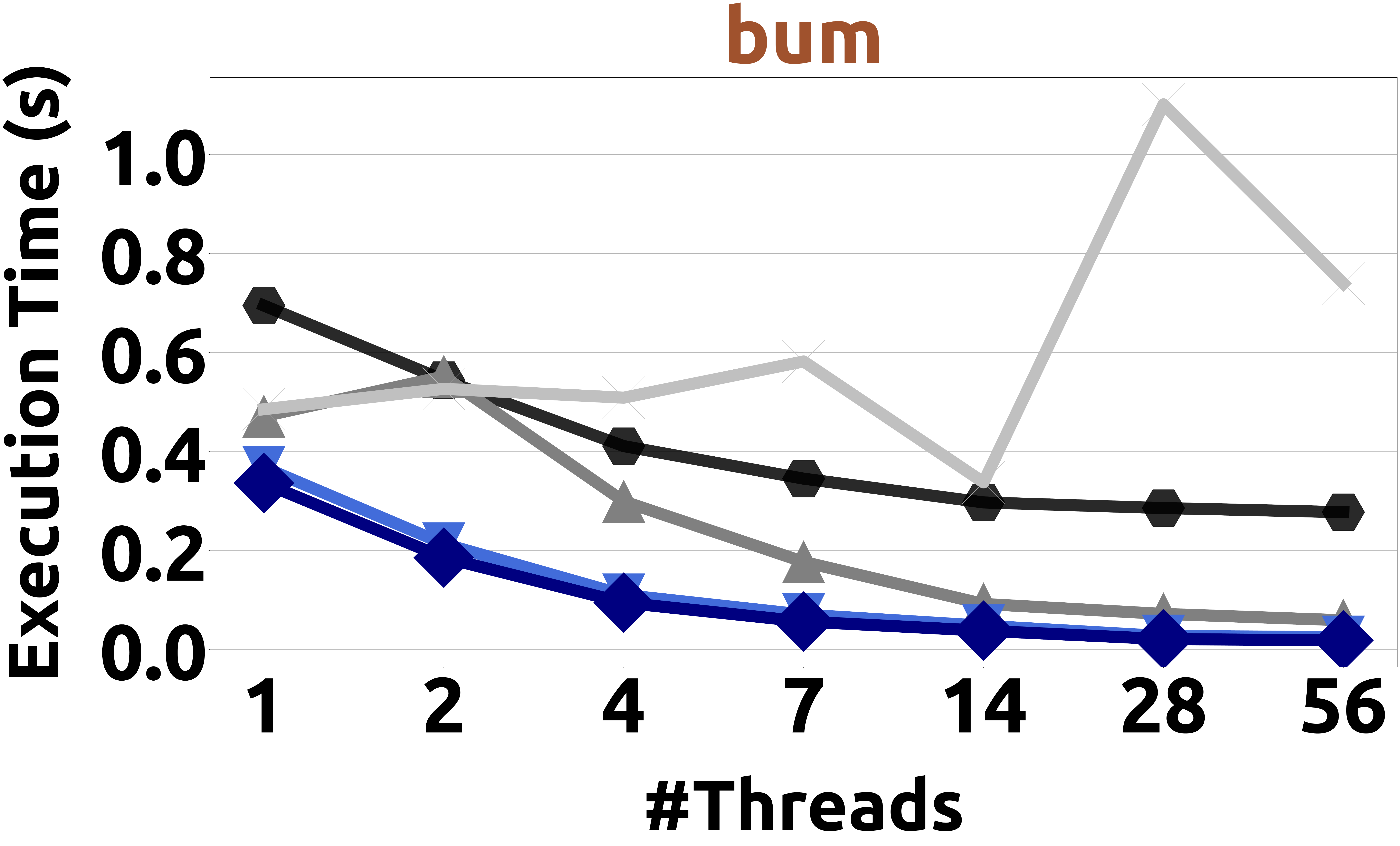}
\includegraphics[scale=0.04]{ 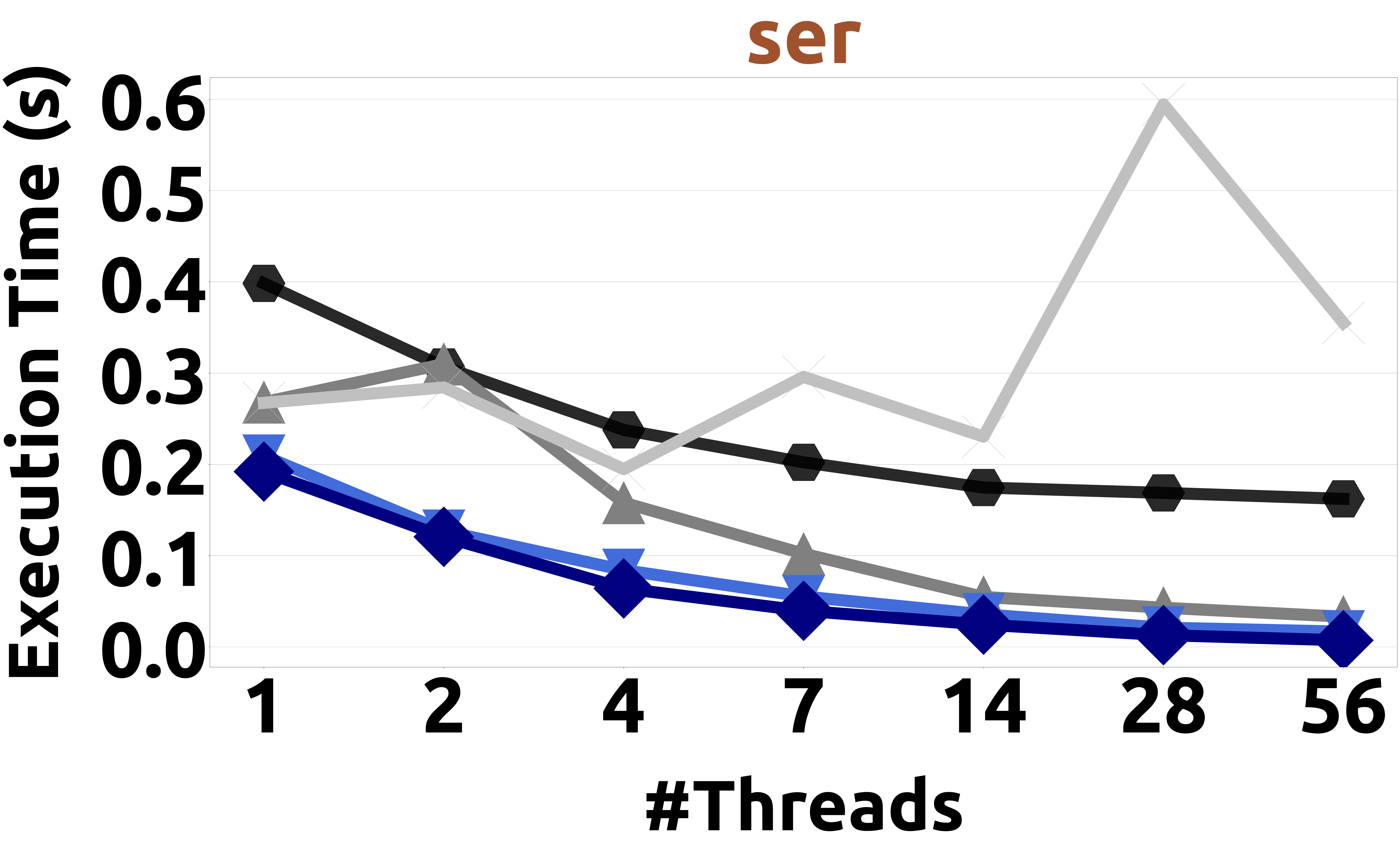}
\includegraphics[scale=0.04]{ 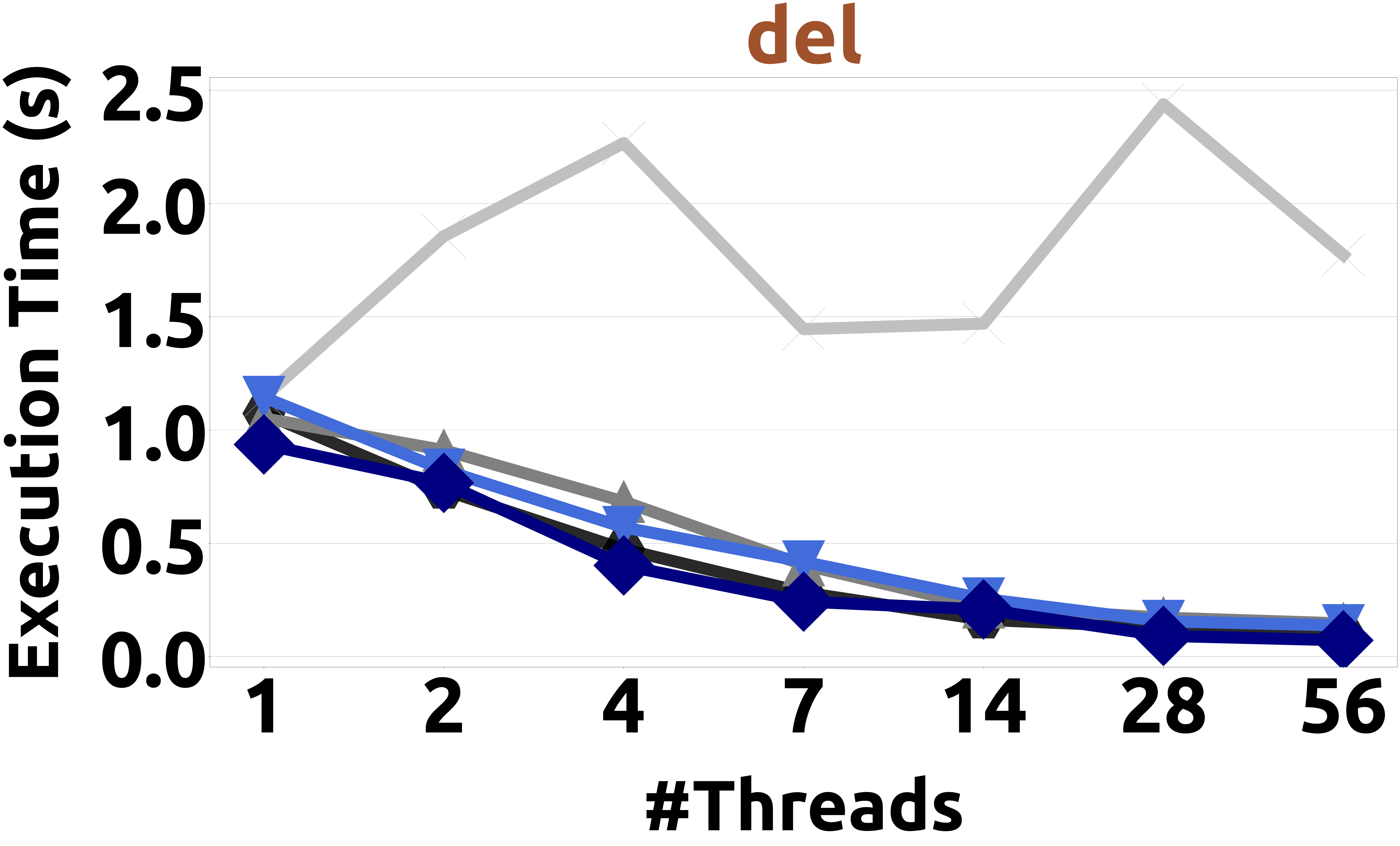}
\end{minipage}
\begin{minipage}{1.0\columnwidth}
\centering
\includegraphics[scale=0.0394]{ 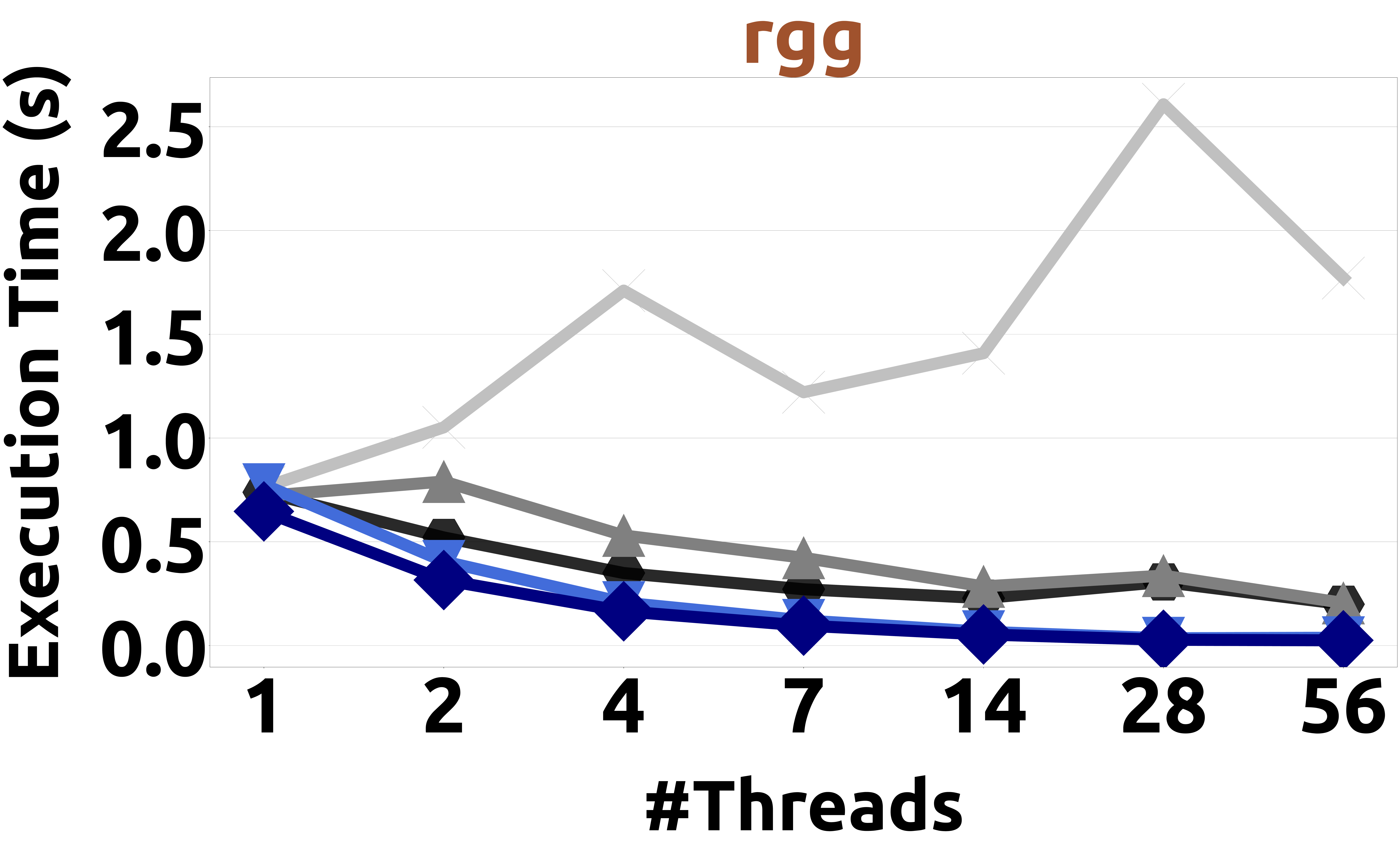}
\includegraphics[scale=0.0394]{ 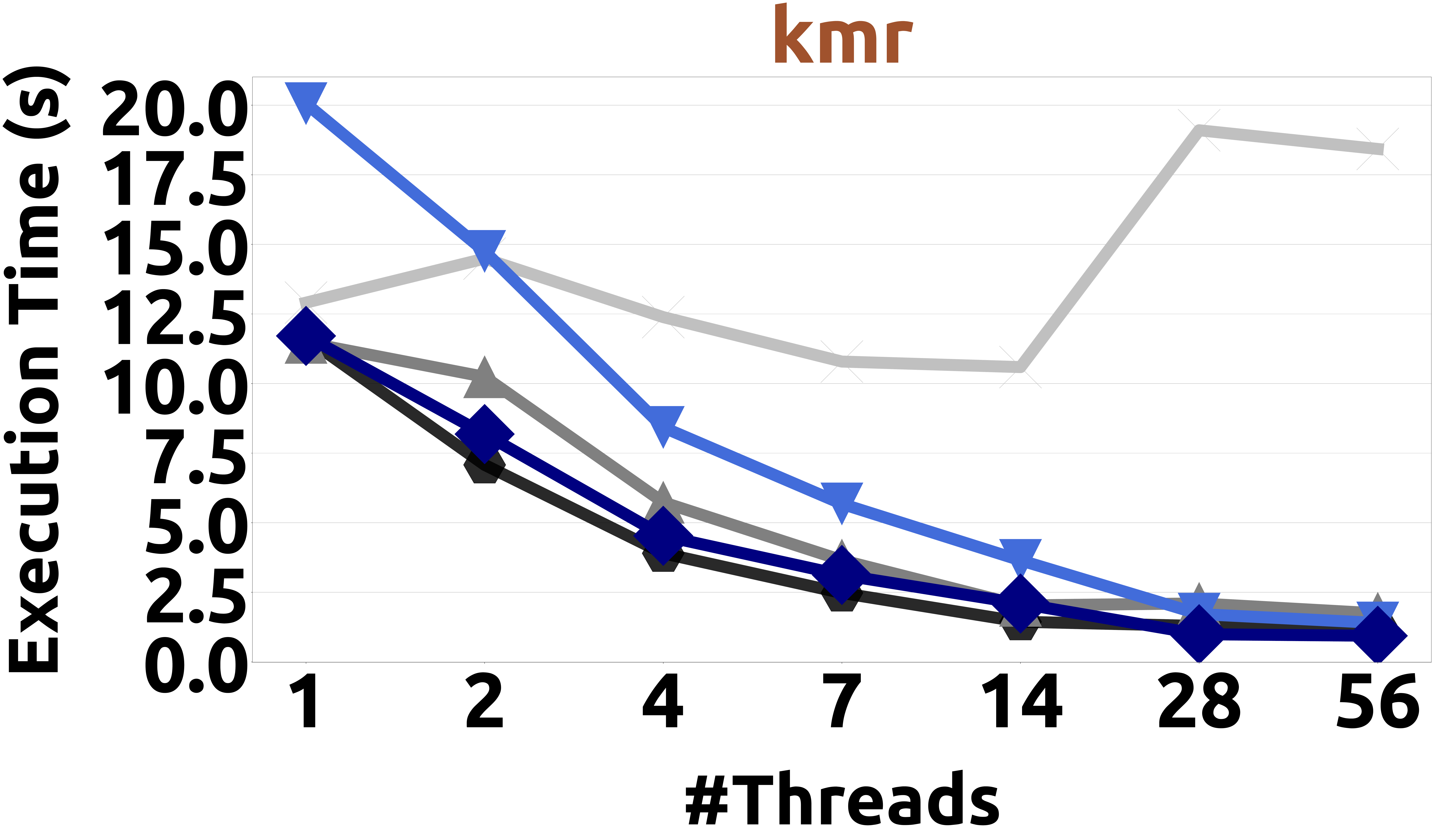}
\includegraphics[scale=0.0394]{ 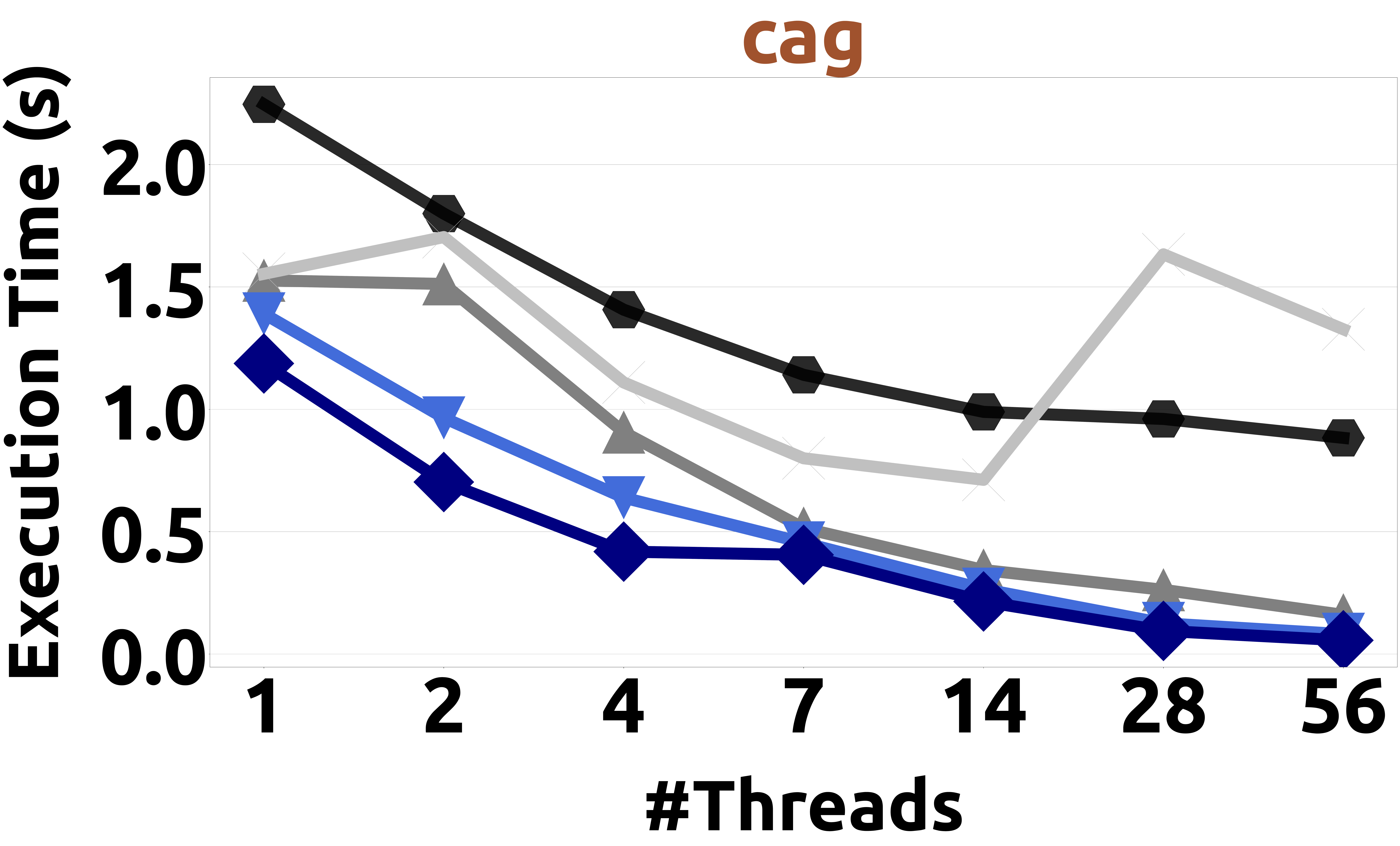}
\end{minipage}
\begin{minipage}{1.0\columnwidth}
\centering
\includegraphics[scale=0.04]{ 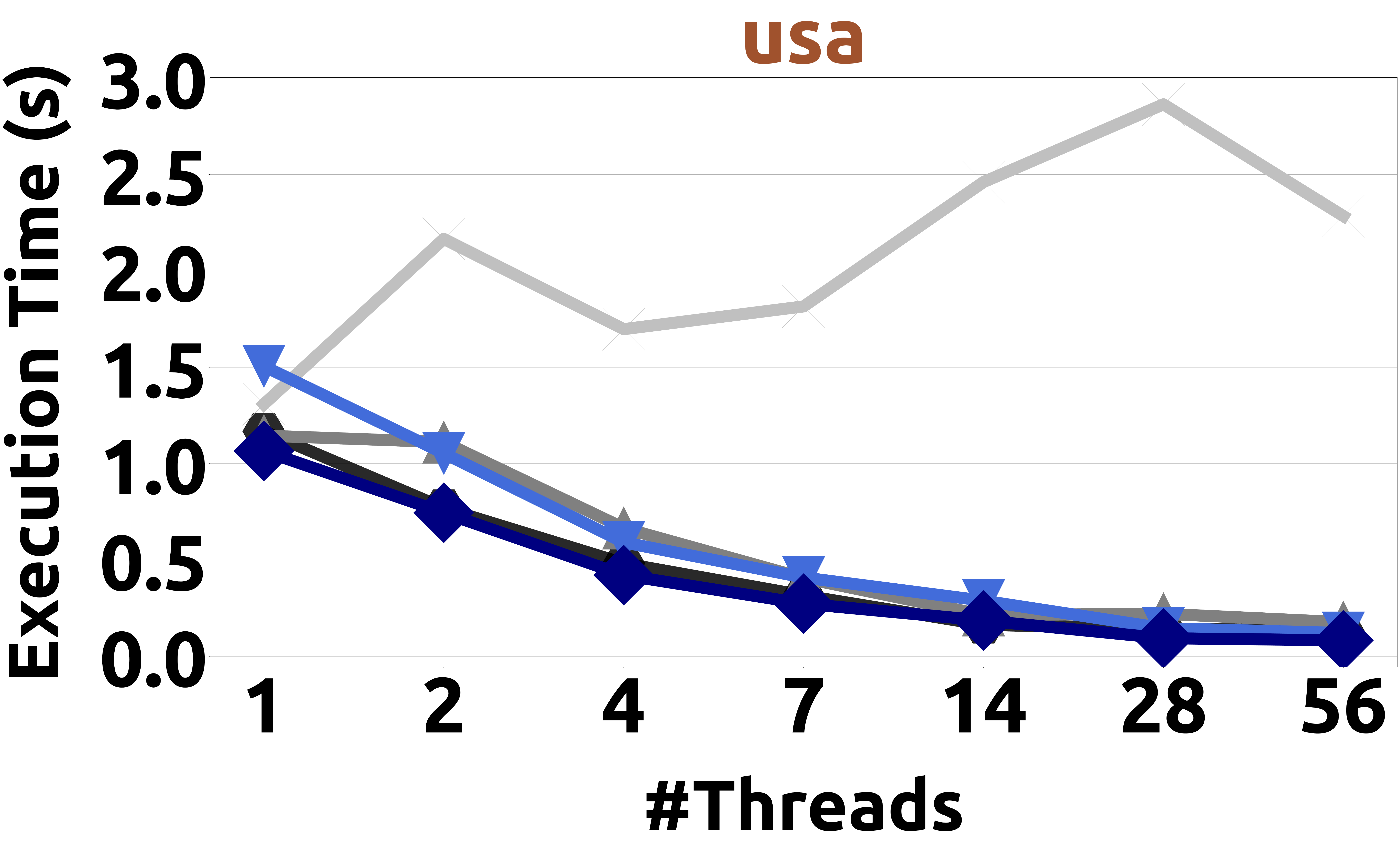}
\includegraphics[scale=0.04]{ 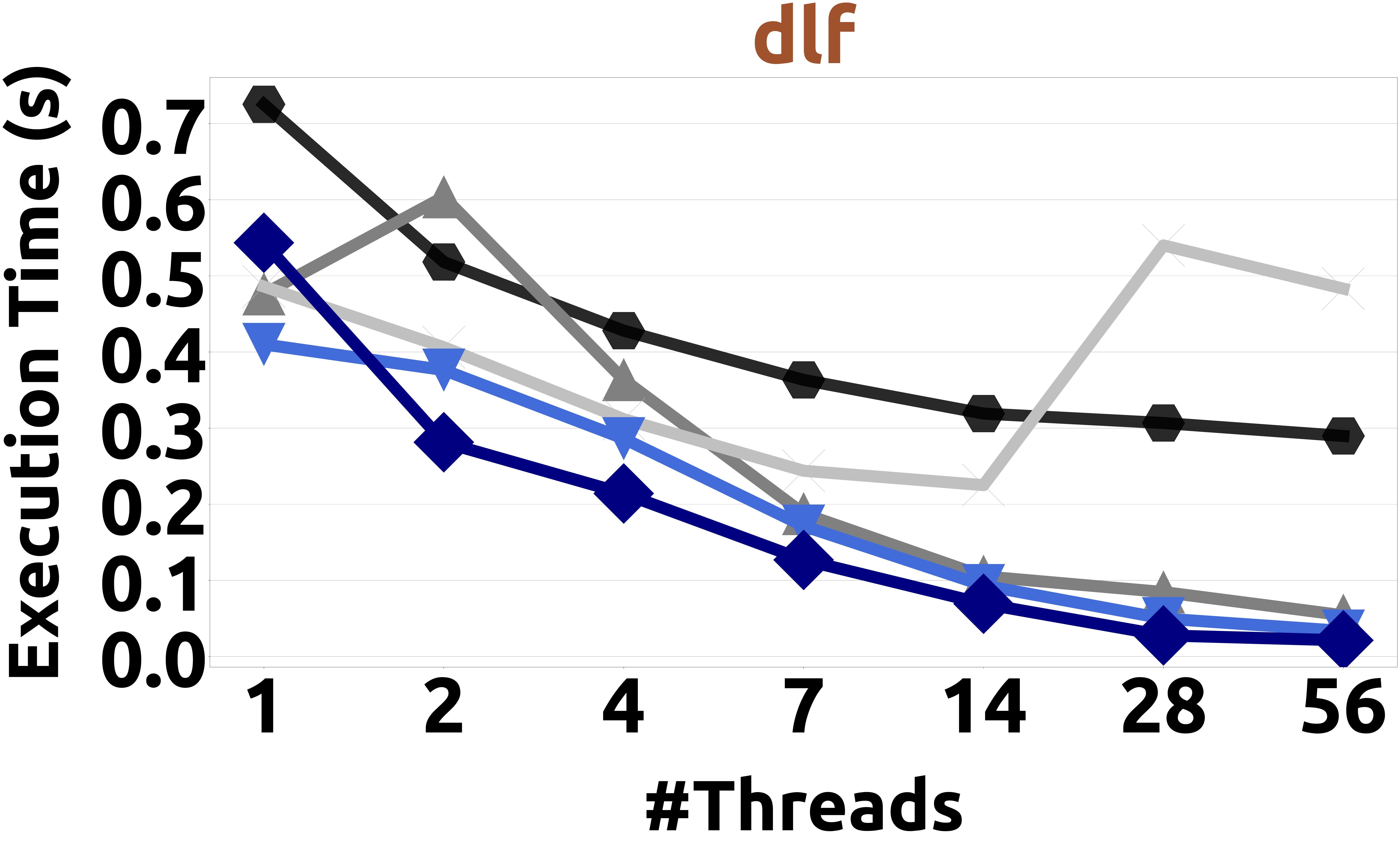}
\includegraphics[scale=0.04]{ 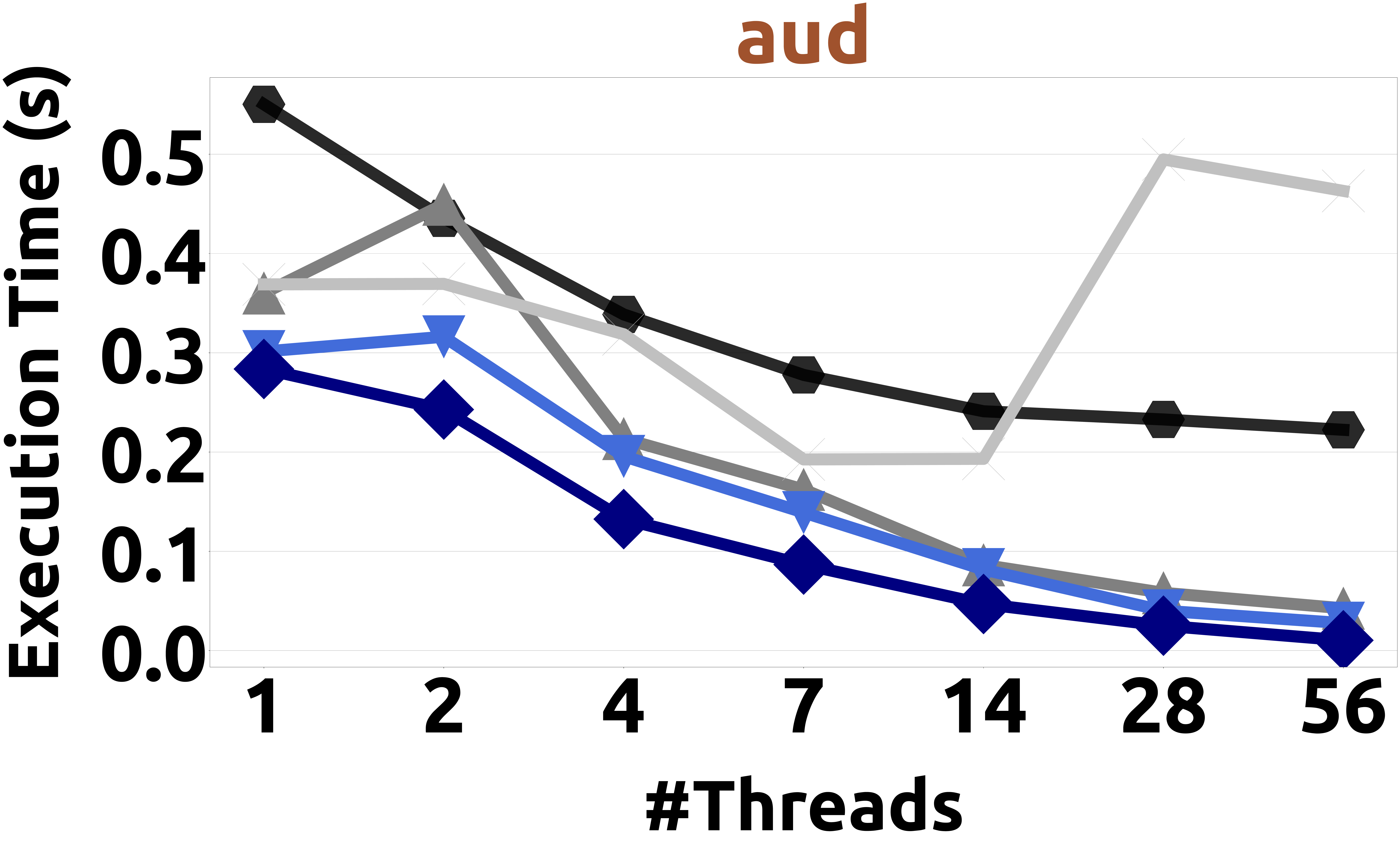}
\end{minipage}
\begin{minipage}{1.0\columnwidth}
\centering
\includegraphics[scale=0.04]{ 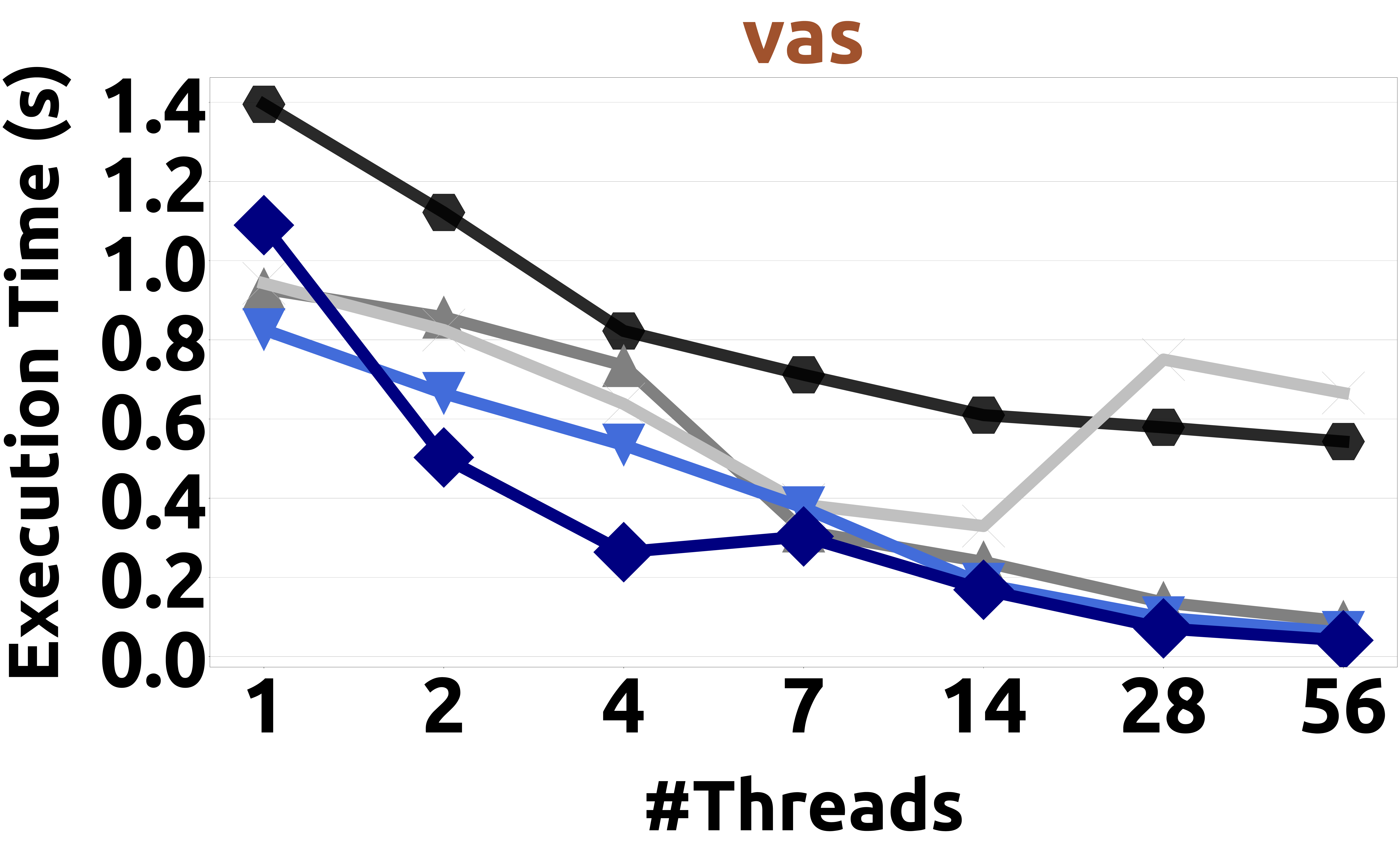}
\includegraphics[scale=0.04]{ 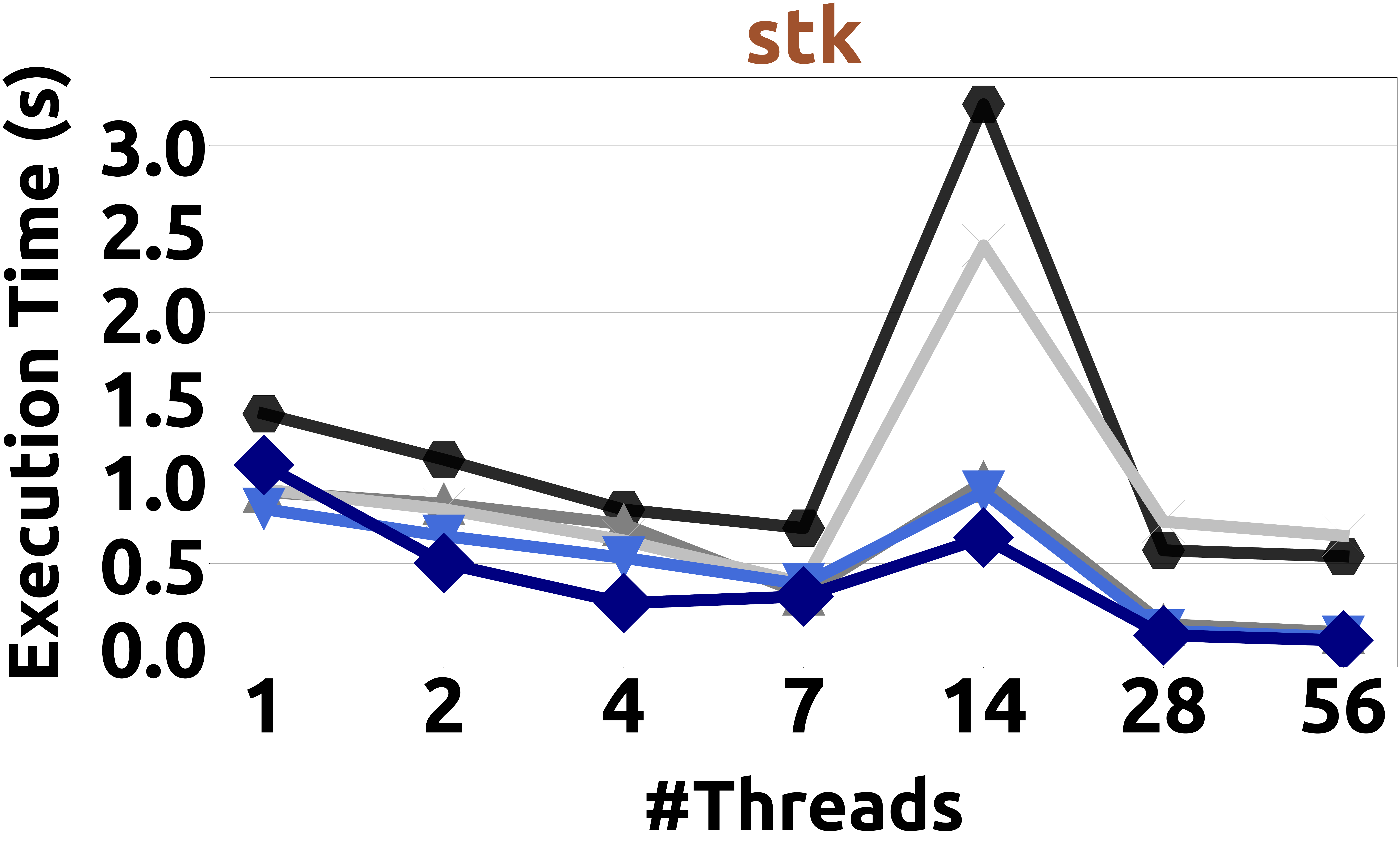}
\includegraphics[scale=0.04]{ 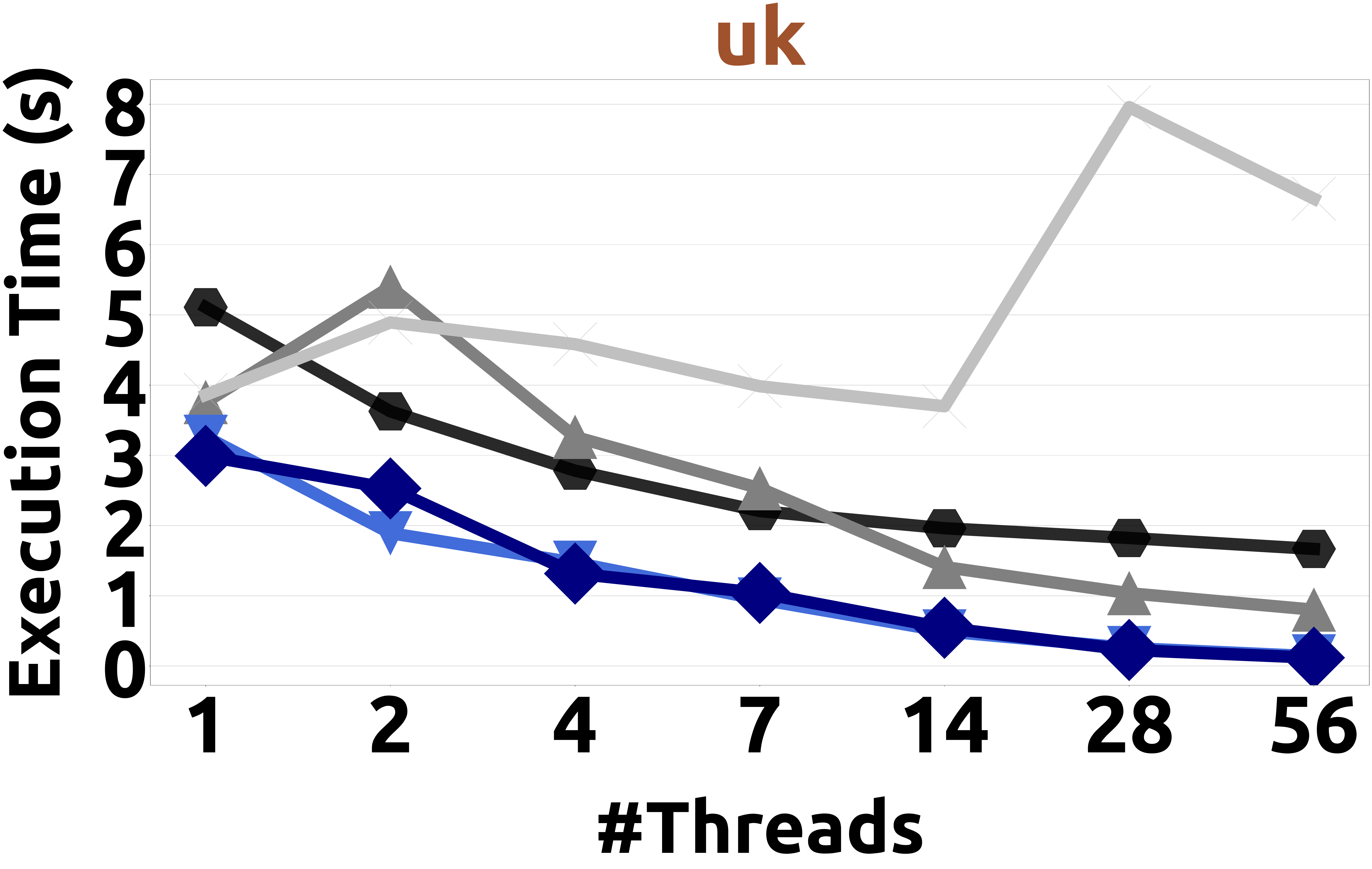}
\end{minipage}
\begin{minipage}{1.0\textwidth}
\centering
\includegraphics[scale=0.04]{ 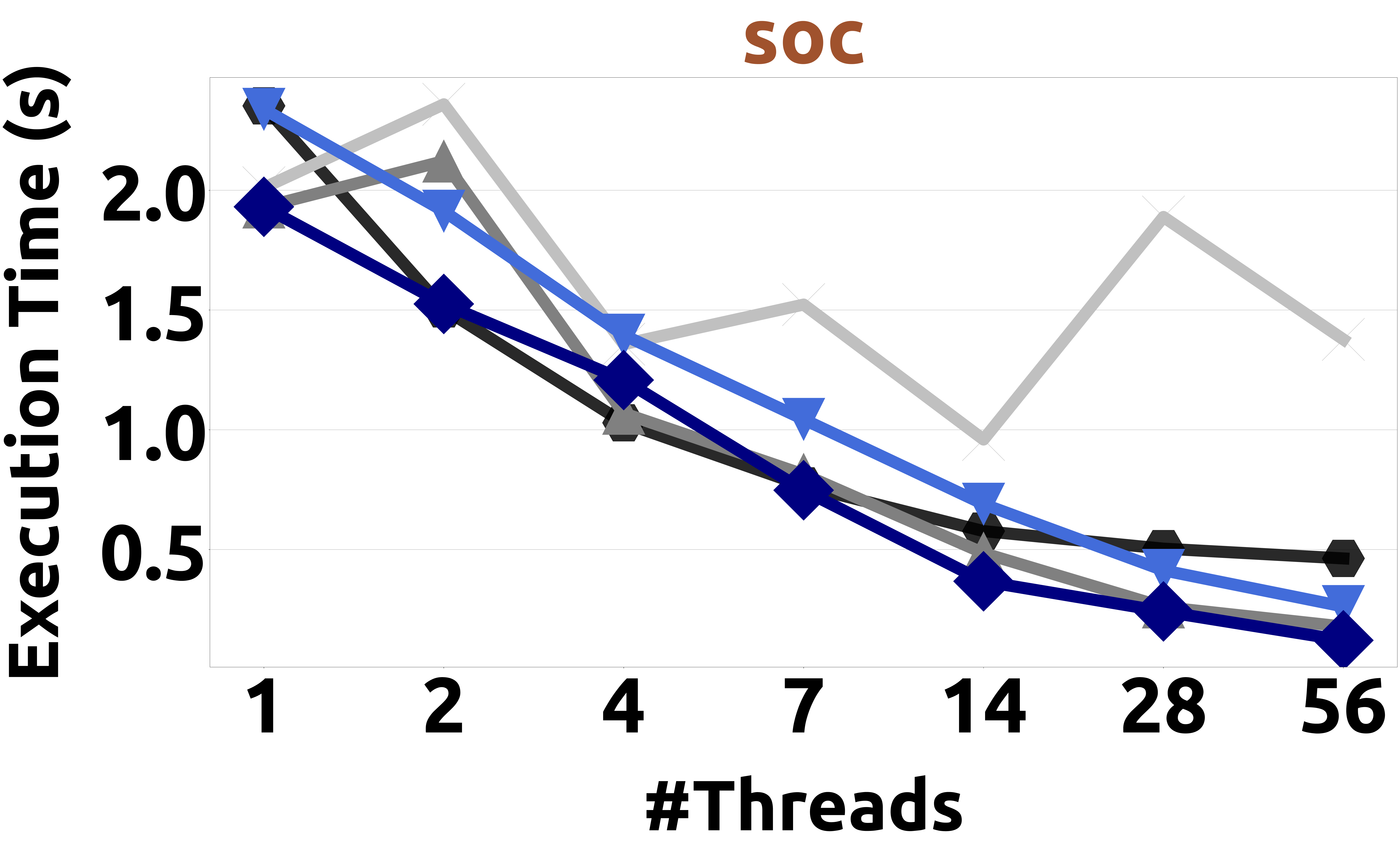}
\includegraphics[scale=0.04]{ 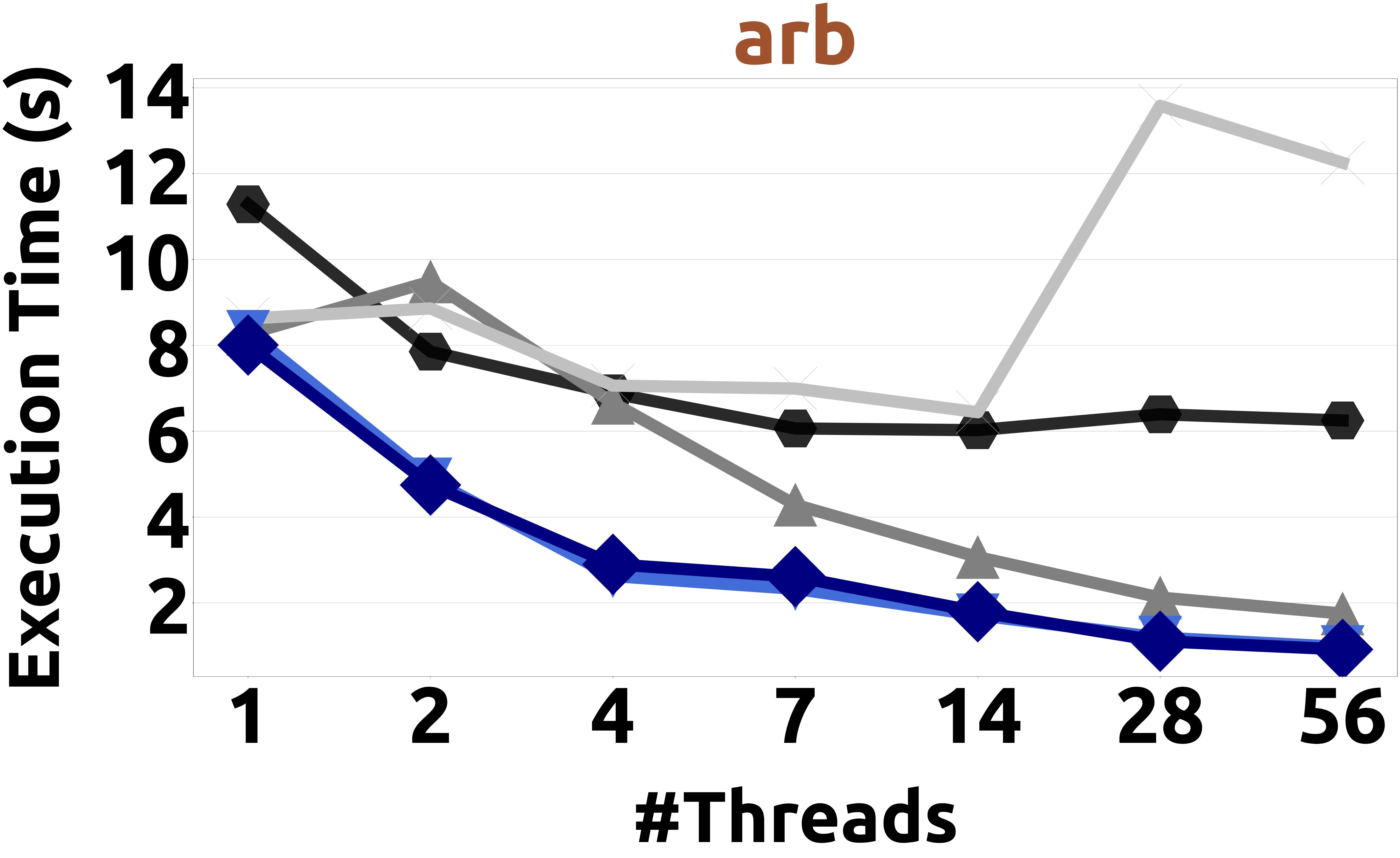}
\includegraphics[scale=0.04]{ 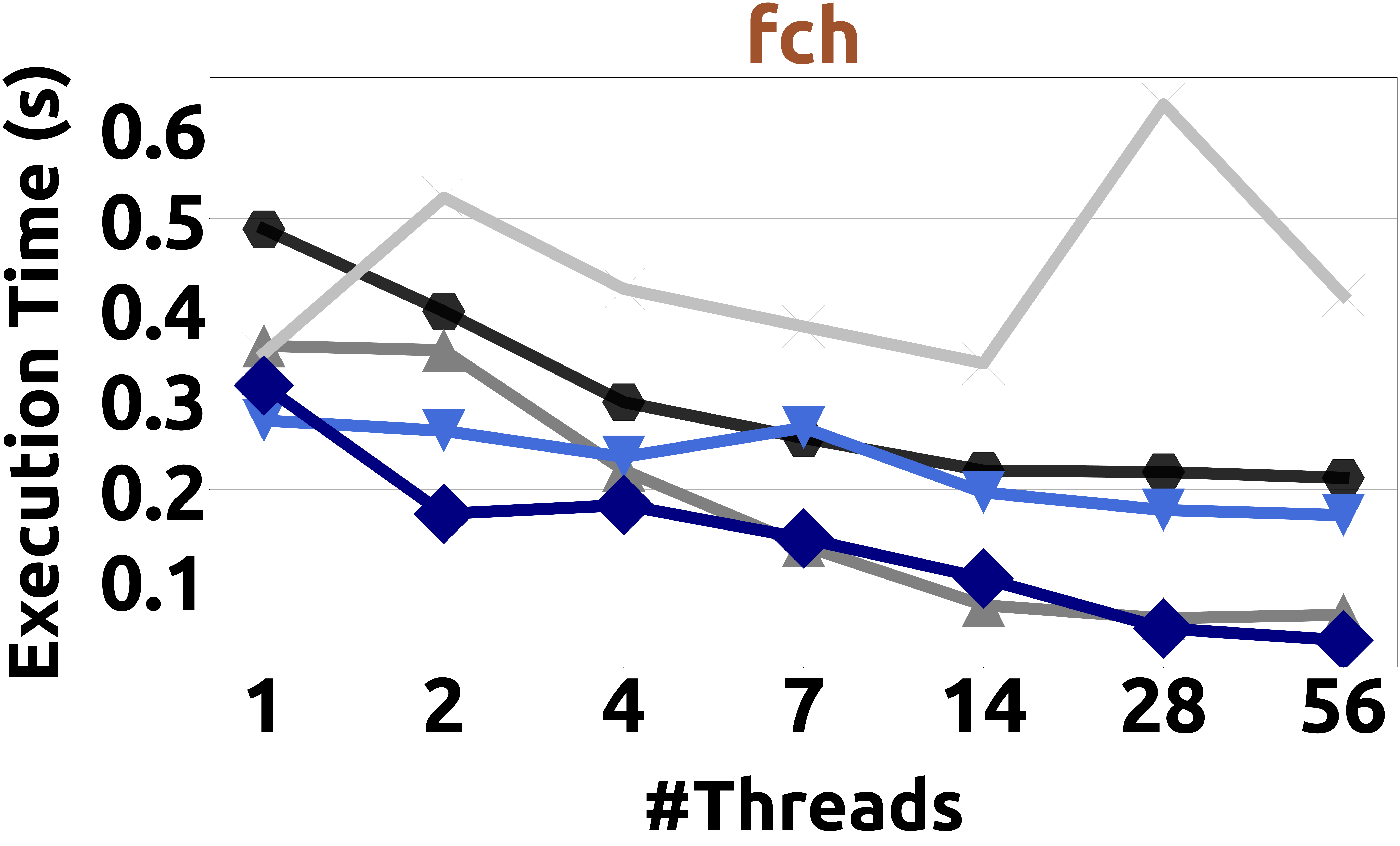}
\end{minipage}
\vspace{0pt}
\caption{Scalability achieved by all parallel graph coloring implementations in large real-world graphs. }
\label{coloring-scalability}
\vspace{-18pt}
\end{figure}

\subsubsection{Performance Comparison}

Figure~\ref{coloring-scalability} evaluates the scalability achieved by all parallel graph coloring implementations in our large real-world graphs, when increasing the number of threads from 1 to 56, i.e., the maximum available hardware thread capacity of our machine.

We draw three findings. First, \ColorTM{}  and ColorLock achieve the lowest execution time across all schemes in single-threaded executions. Using one single thread, \ColorTM{} and ColorLock on average outperform SeqSolve by 1.55$\times$ and 1.42$\times$, respectively, and they on average outperform IterSolve by 1.17$\times$ and 1.06$\times$, respectively. With only one thread, \ColorTM{} and ColorLock have identical executions to the sequential Greedy algorithm (Figure~\ref{alg:greedy}): thanks to the optimizations proposed in Section~\ref{sec:critical-vertices}, the list of critical adjacent vertices that need to be validated inside the critical section is empty, and thus \ColorTM{} and ColorLock \emph{completely} eliminate using synchronization (either HTM of fine-grained locking). Second, we find that IterSolveR exhibits the lowest scalability across all schemes. IterSolveR merges two parallel for-loops into a single parallel for-loop in order to eliminate one of the two barriers used in IterSolve. Even though IterSolveR reduces the barrier synchronization costs, it increases the load imbalance among parallel threads, thus causing significant performance overheads. Third, we observe that the scalability of SeqSolve, IterSolve, and IterSolveR is highly affected by the NUMA effect, i.e., the non-uniform memory access latencies to the application data. For example, when increasing the number of threads from 7 to 14 (only one NUMA socket is used) the performance of SeqSolve, IterSolve, IterSolveR, ColorLock and \ColorTM{} improves by 1.24$\times$, 1.75$\times$, 1.06$\times$, 1.62$\times$ and 1.65$\times$, respectively, averaged across all large graphs. However, when increasing the number of threads from 14 to 28, i.e., using both NUMA sockets of our machine, the performance of SeqSolve and IterSolve \emph{only} improves by 1.03$\times$ and 1.26$\times$, respectively, while the performance of and IterSolveR  decreases by 2.13$\times$, averaged across all large graphs. In contrast, when increasing the number of threads from 14 to 28, the performance of ColorLock and \ColorTM{} significantly improves by 1.77$\times$ and 1.97$\times$, respectively, averaged across all graphs. This is because our proposed algorithmic design implemented in ColorLock and \ColorTM{} leverages better the deep memory hierarchy of commodity multicore platforms thanks to its \emph{eager} conflict detection and resolution policy, thus achieving lower data access costs. Overall, we conclude that our proposed algorithmic design achieves the best scalability in modern multicore platforms.

\begin{figure}[H]
\begin{minipage}{1.0\columnwidth}
\centering
\includegraphics[width=\columnwidth]{ 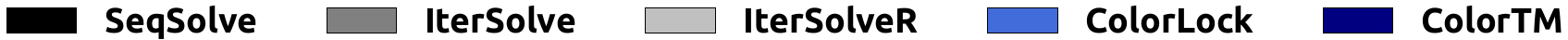}
\end{minipage}
\begin{minipage}{1.0\columnwidth}
\centering
\includegraphics[width=\columnwidth]{ sections/ColorTM/results/speedup_14.pdf}
\end{minipage}
\begin{minipage}{1.0\columnwidth}
\centering
\includegraphics[width=\columnwidth]{ sections/ColorTM/results/speedup_28.pdf}
\end{minipage}
\begin{minipage}{1.0\columnwidth}
\centering
\includegraphics[width=\columnwidth]{ sections/ColorTM/results/speedup_56.pdf}
\end{minipage}
\vspace{-1pt}
\caption{Speedup achieved by all parallel graph coloring implementations over the sequential Greedy scheme in large real-world graphs using all cores of one socket (14 threads), all cores of two sockets (28 threads), and the maximum hardware thread capacity of our machine with hyperthreading enabled (56 threads). }
\label{coloring-speedup}
\vspace{-6pt}
\end{figure}

Figure~\ref{coloring-speedup} compares the speedup achieved by all schemes over the sequential Greedy scheme, when varying the number of hardware threads used in all large real-world graphs.

We make two key observations. First, all parallel graph coloring schemes achieve lower speedup in very irregular graphs including the \texttt{soc}, \texttt{arb} and \texttt{fch} graphs, compared to all the remaining real-world graphs. In very irregular graphs, the number of edges per vertex significantly vary across vertices~\cite{Giannoula2022SparsePSigmetrics,Giannoula2022SparsePPomacs,Tang2015Optimizing}: typically only a few vertices have a much larger number of edges over the vast majority of the remaining vertices of the graph. Therefore, in irregular graphs parallel threads typically cause more coloring inconsistencies than regular graphs, which are resolved during runtime, increasing the execution time. Second, we find that \ColorTM{} achieves significant performance improvements over all the prior state-of-the-art graph coloring schemes. \ColorTM{} outperforms SeqSolve, IterSolve, and IterSolveR by 3.43$\times$, 1.71$\times$ and 5.83$\times$ respectively, when using 14 threads, and by 8.46$\times$, 2.84$\times$ and 27.66$\times$ respectively, when using the maximum hardware thread capacity of our machine (56 threads). This is because SeqSolve, IterSolve, and IterSolveR traverse \emph{all} the vertices of the graph at least twice, and employ a \emph{lazy} conflict resolution policy, thus incurring high data access costs. Instead, \ColorTM{} traverses more than once \emph{only} the conflicted vertices, and resolves coloring inconsistencies with an \emph{eager} approach, thus better leveraging the deep memory hierarchy of multicore platforms and reducing data access costs. In addition, \ColorTM{} outperforms ColorLock by 1.34$\times$ and 1.67$\times$ when using 14 and 56 threads, respectively. As explained, HTM is a speculative hardware-based synchronization mechanism, and thus \ColorTM{} provides high performance improvements over ColorLock thanks to significantly minimizing data access and synchronization costs. Note that in the fine-grained locking approach of ColorLock, for each adjacent vertex accessed inside the critical section, the parallel thread needs to acquire and release the corresponding software-based lock, thus performing additional memory accesses in the memory hierarchy for accessing the lock variable. Overall, we conclude that \ColorTM{} significantly outperforms all prior state-of-the-art parallel graph coloring algorithms across a wide variety of large real-world graphs.

\begin{figure}[H]
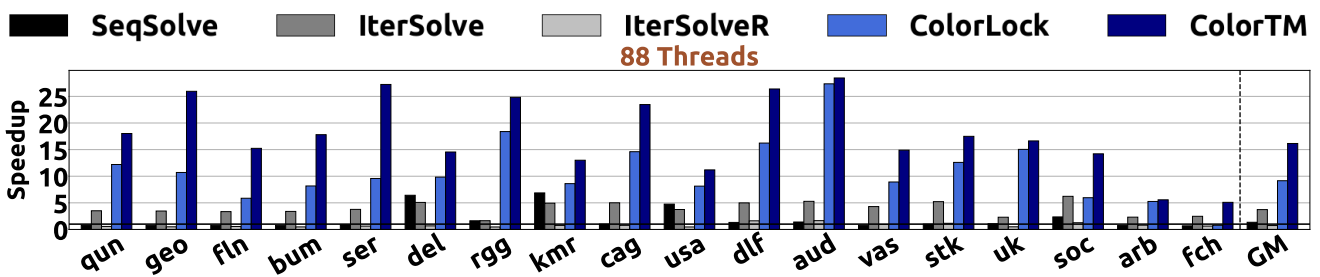

\begin{minipage}{1.0\columnwidth}
\centering
\includegraphics[width=\columnwidth]{ sections/ColorTM/results/legend_speedup.png}
\end{minipage}
\begin{minipage}{1.0\columnwidth}
\centering
\includegraphics[width=\columnwidth]{ sections/ColorTM/results/speedup_88.pdf}
\end{minipage}
\vspace{-1pt}
\caption{Speedup achieved by all parallel graph coloring implementations over the sequential Greedy scheme in large real-world graphs using the maximum hardware thread capacity of an Intel Broadwell server with hyperthreading enabled (88 threads).}
\label{coloring-speedup-broady}
\vspace{-10pt}
\end{figure}

To confirm the performance benefits of \ColorTM{} across multiple computing platforms, we evaluate all schemes on a 2-socket Intel Broadwell server with an Intel Xeon E5-2699 v4 processor at 2.2 GHz having 44 physical cores and 88 hardware threads. Figure~\ref{coloring-speedup-broady} compares the speedup achieved by all schemes over the sequential Greedy scheme in all large real-world graphs using 88 threads, i.e., the maximum hardware thread capacity of the Intel Broadwell server. We find that \ColorTM{} provides significant performance benefits over prior state-of-the-art graph coloring algorithms, achieving 11.98$\times$, 4.33$\times$ and 22.06$\times$ better performance over SeqSolve, IterSolve, and IterSolveR, respectively.

\subsubsection{Analysis of \ColorTM{}  Execution}\label{eval:aborts}

\begin{figure}[H]
\vspace{-4pt}
\begin{minipage}{1.0\textwidth}
\centering
\includegraphics[width=\columnwidth]{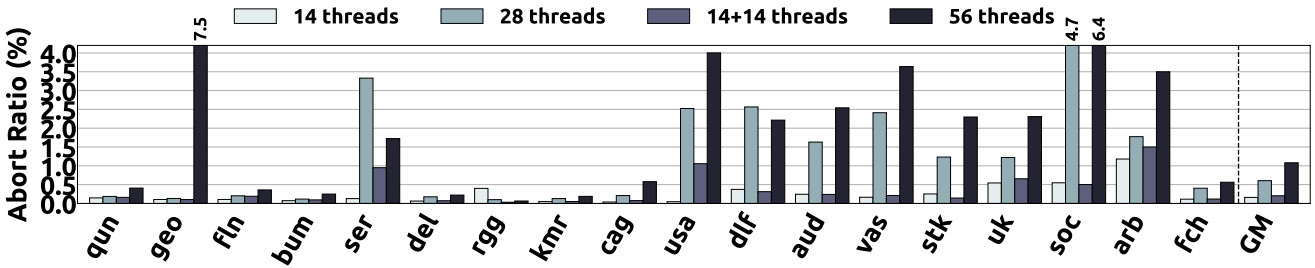}
\end{minipage}
\vspace{-1pt}
\caption{Abort ratio exhibited by \ColorTM{} in all large real-world graphs. }
\label{coloring-abort-ratio}
\vspace{-14pt}
\end{figure}

We further analyze the HTM-related execution behavior of our proposed \ColorTM{} and \BalColorTM{} algorithms. Figure~\ref{coloring-abort-ratio} presents the abort ratio of \ColorTM{}, i.e., the number of transactional aborts divided by the number of attempted transactions, in all real-world graphs, as the number of threads increases. In the 14-thread execution, we pin all thread on one single NUMA socket. In the 28-thread execution, we pin threads on both NUMA sockets of our machine with hyperthreading disabled. In the (14+14)-thread execution, we pin all 28 threads on the same \emph{single} socket with hyperthreading enabled. In the 56-thread execution, we use the maximum hardware thread capacity of our machine.

We make three key observations. First, we find that the abort ratio becomes high in real-world graphs which have high maximum degree and high standard deviation of the vertices' degrees, e.g., \texttt{dlf}, \texttt{aud}, \texttt{vas}, \texttt{stk}, \texttt{uk}, \texttt{soc} and \texttt{arb} graphs. In graphs with high vertex degree, the transaction data access footprint is large and parallel threads compete for the same adjacent vertices with a high probability, thus causing aborts in HTM. Second, we observe that when using both sockets of our machine, the transactional aborts in \ColorTM{} significantly increase due to the NUMA effect. Specifically, averaged across all graphs the (14+14)-thread execution of \ColorTM{} exhibits 2.97$\times$ lower abort ratio compared to the 28-thread execution of \ColorTM{}. Due to the NUMA effect, the memory accesses to the application data are very expensive. As a result, the duration of the transactions increases, thus increasing the probability of conflict aborts among running transactions (See more details in the next experiment). Third, we observe that \ColorTM{} exhibits a very low abort ratio. \ColorTM{} has \emph{only} 1.08\% abort ratio on average across all real-world graphs, when using the maximum hardware thread capacity (56 threads) of our machine. Our proposed \emph{speculative} algorithmic design effectively reduces the amount of computations and data accesses performed inside the critical section (inside the HTM transaction), thus effectively decreasing the transaction's footprint and duration. As a result, \ColorTM{} provides high amount of parallelism and low interference among parallel threads. We conclude that \ColorTM{} has low synchronization and interference costs among a large number of parallel threads, even in real-world graphs with high vertex degree.

\begin{figure}[H]
\vspace{-2pt}
\begin{minipage}{1.0\textwidth}
\centering
\includegraphics[width=0.8\columnwidth]{ 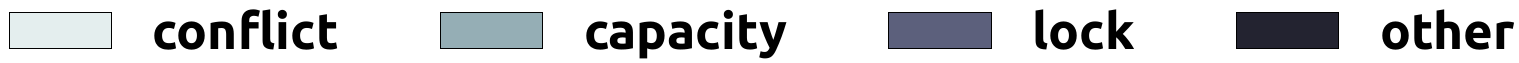}
\end{minipage}
\begin{minipage}{1.0\textwidth}
\centering
\includegraphics[scale=0.035]{ 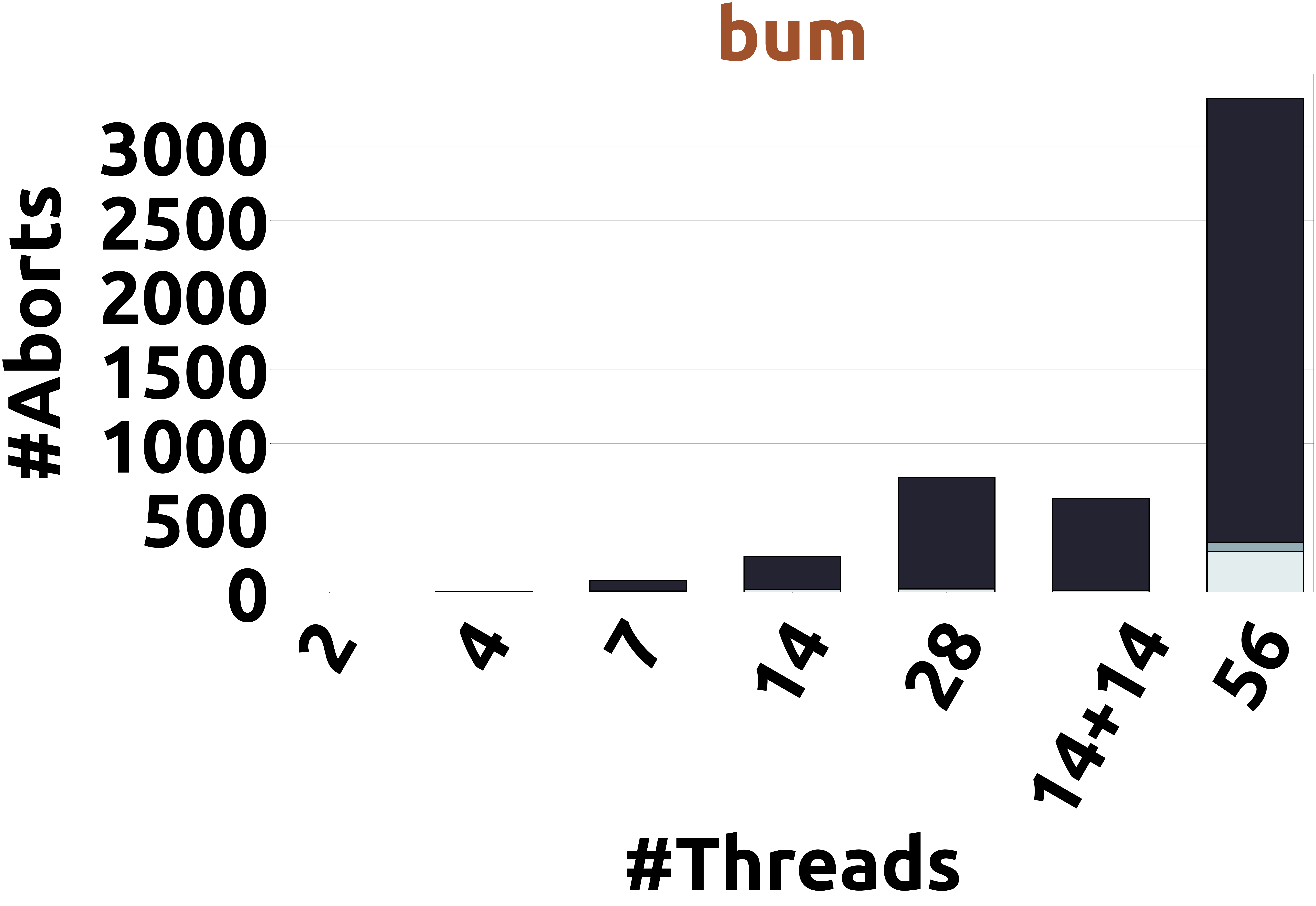}
\includegraphics[scale=0.035]{ 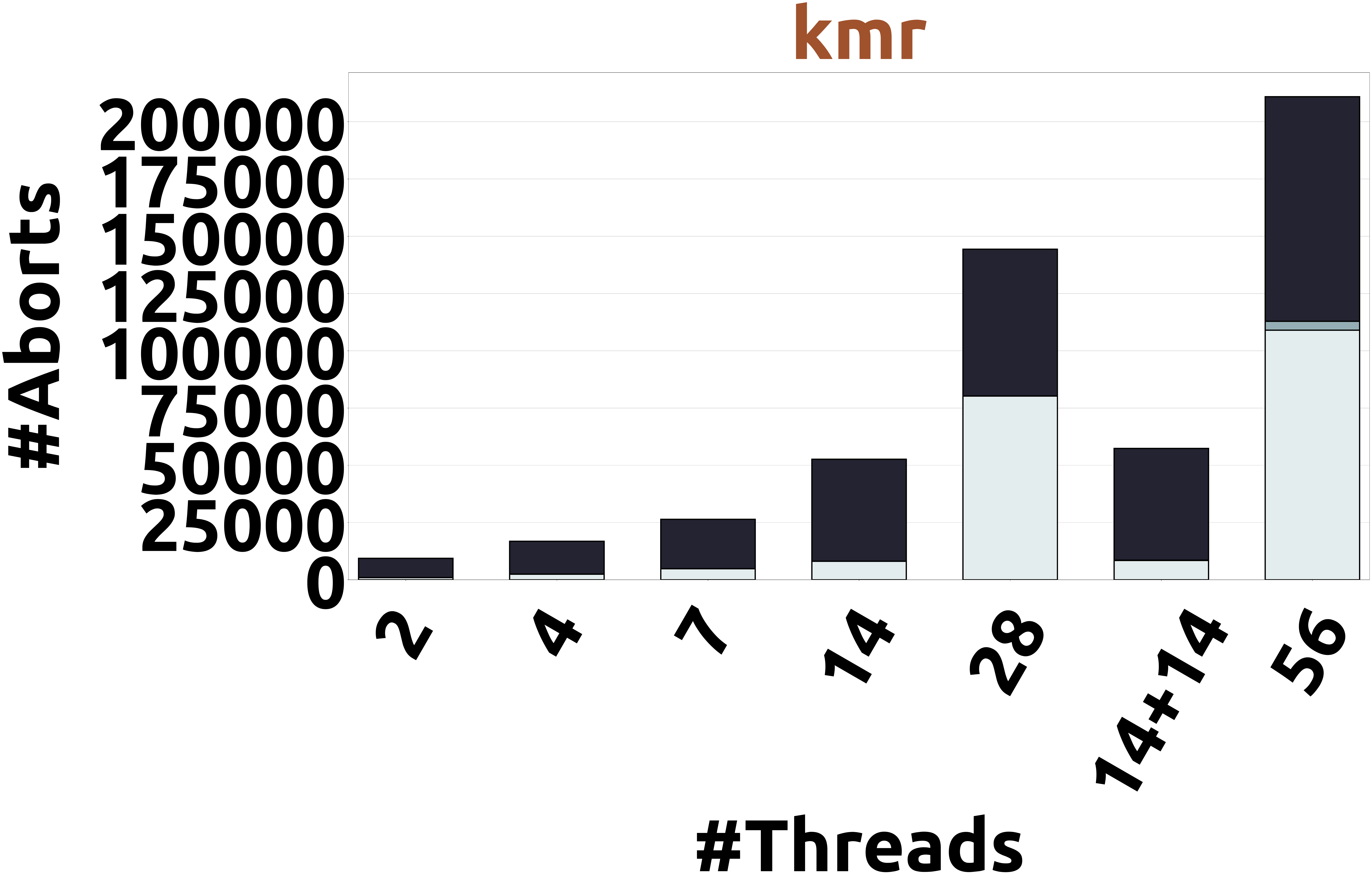}
\includegraphics[scale=0.035]{ 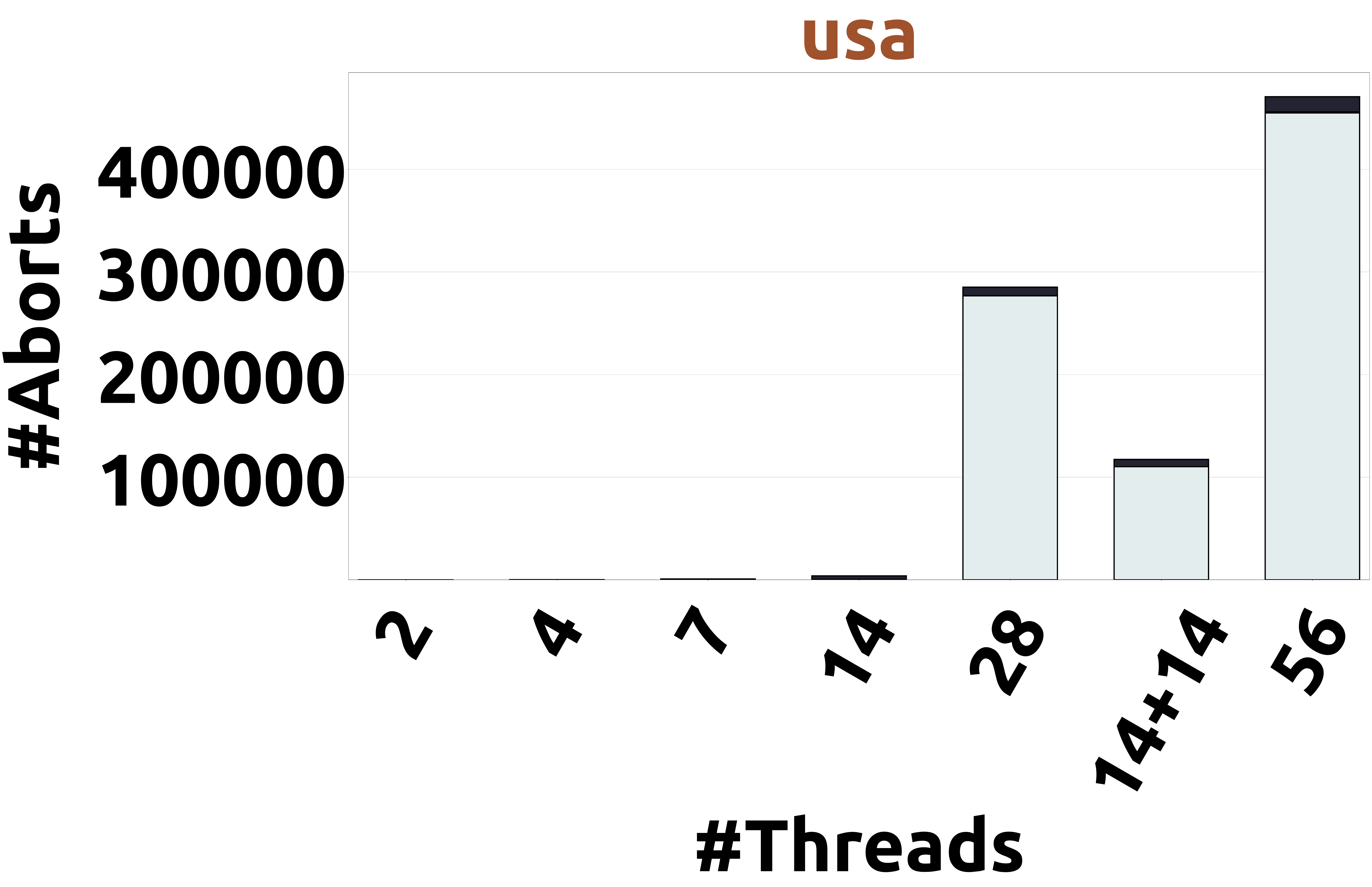}
\end{minipage}
\begin{minipage}{1.0\textwidth}
\centering
\includegraphics[scale=0.035]{ 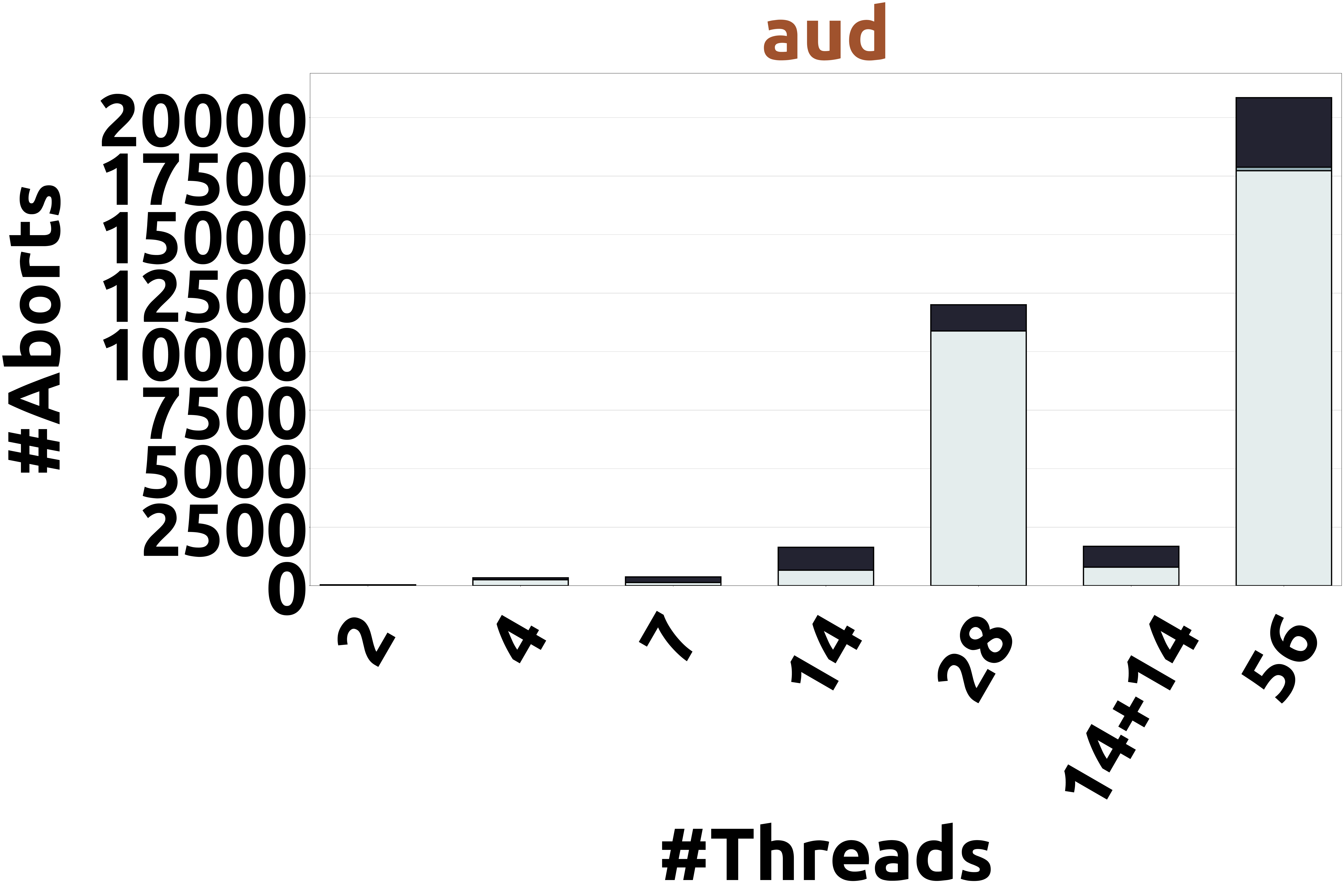}
\includegraphics[scale=0.035]{ 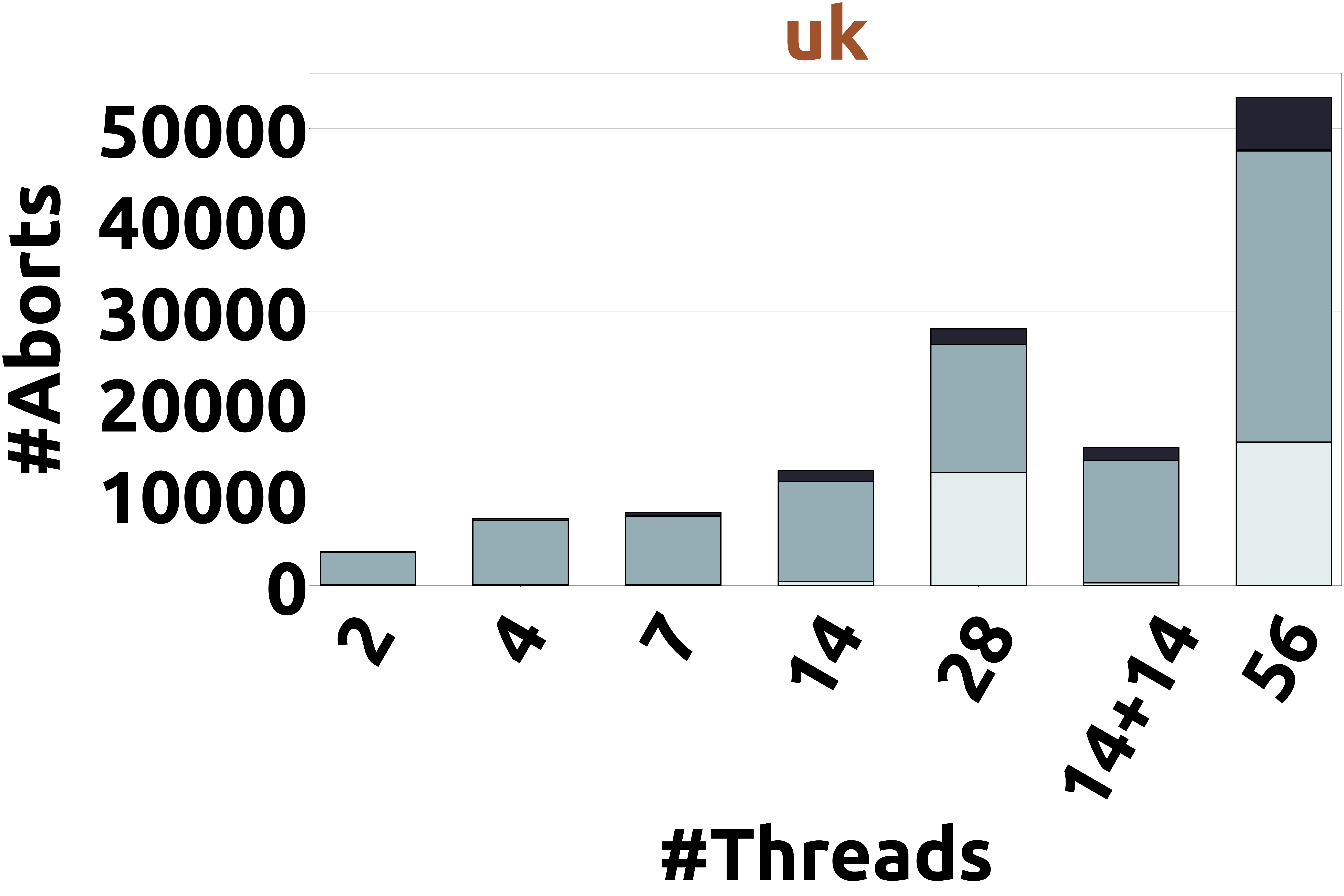}
\includegraphics[scale=0.035]{ 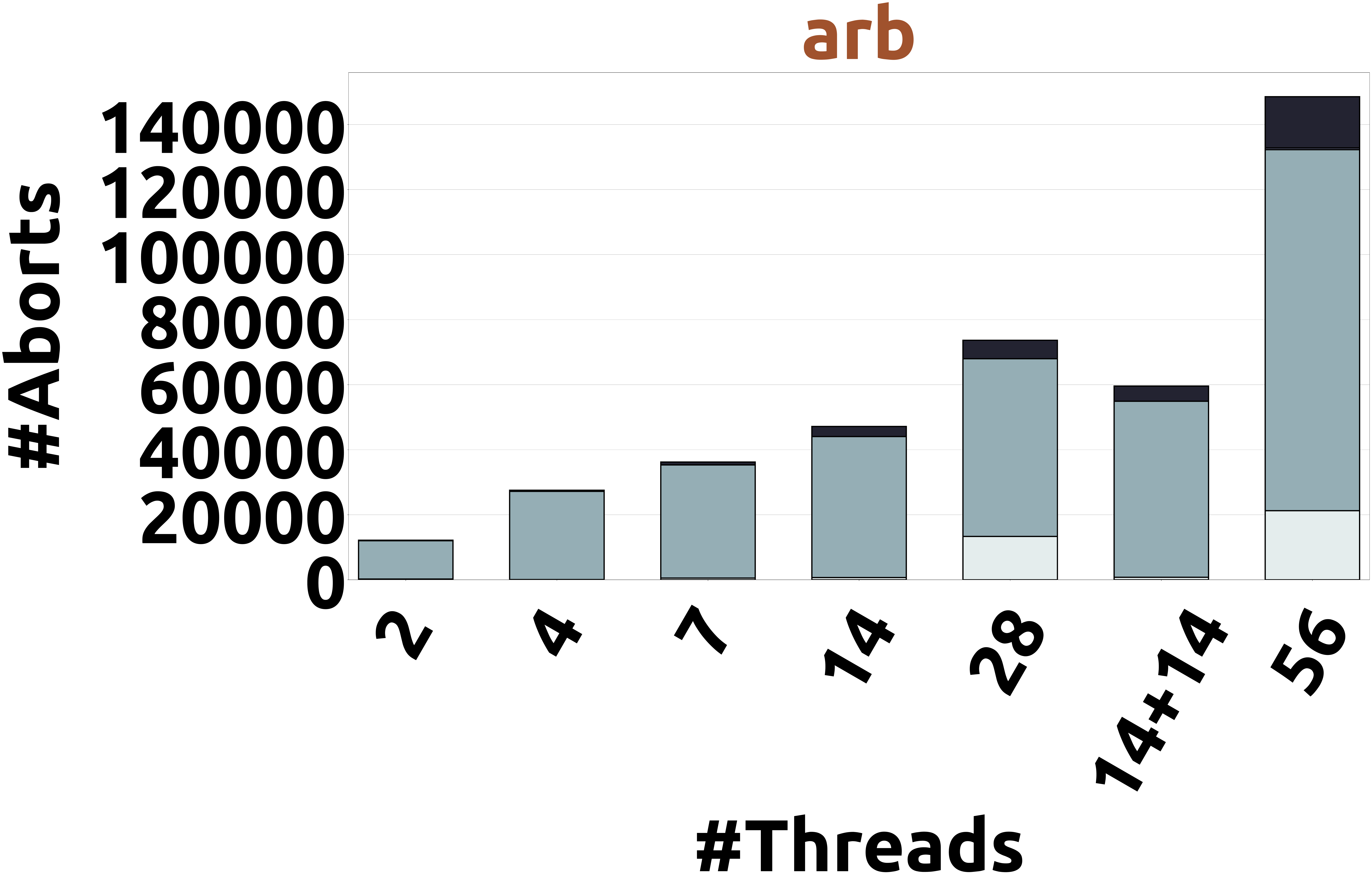}
\end{minipage}
\vspace{-2pt}
\caption{Breakdown of different types of aborts exhibited by \ColorTM{} in real-world graphs.}
\label{coloring-abort-breakdown}
\vspace{-10pt}
\end{figure}

Figure~\ref{coloring-abort-breakdown} presents the breakdown of different types of aborts exhibited by \ColorTM{} in a representative subset of real-world graphs. We break down the transactional aborts into four types: (i) \emph{conflict} aborts: they appear when a running transaction executed by a parallel thread attempts to write the read-set of another running transaction executed by a different thread, (ii) \emph{capacity} aborts: they appear when the memory footprint of a running transaction exceeds the size of the hardware transactional buffers, (iii) \emph{lock} aborts: current HTM implementations~\cite{Herlihy1993Transactional,Yoo2013Performance,Cain2013Robust,Wang2012Evaluation} provide no guarantee that any transaction will eventually commit inside the transactional path, and thus the programmer provides an alternative non-transactional fallback path, i.e., falling back to the acquisition of coarse-grained lock that allows only a single thread to enter the critical section, and forces aborts to the transactions of all the remaining threads \footnote{To achieve this, the lock is added to each transaction’s read set, so that when the
lock is acquired by a thread (write to the lock variable), the remaining threads are aborted and wait until the lock is released.}, and (iv) \emph{other} aborts: they appear when a transaction fails due to other reasons such as cache line evictions, interrupts and/or when the duration of a transaction exceeds the scheduling quantum and the OS scheduler schedules out the software thread from the hardware thread, aborting the transaction. Note that since the fallback path lock is just a variable in the source code, some conflict aborts are caused by the writes in this lock variable. Thus, a part of the lock aborts is counted as conflict aborts in our measurements.

We draw three findings. First, we find that the \emph{conflict} aborts significantly increase across all graphs when using both sockets of our machine due to the NUMA effect. For example, the number of conflicts aborts in the 28-thread executions is 3.32$\times$ higher compared to that in the 14-thread executions. As already mentioned, the NUMA effect significantly increases the duration of the running transactions, and thus the probability of causing conflict aborts among running transactions is high. Second, as number of threads increases, e.g., when comparing the 56-thread execution over the 28-thread execution, the number of conflict aborts increases by 1.05$\times$. This is because partitioning the graph to a higher number of threads results in a higher number of crossing edges among parallel threads, which in turn results in a larger list of critical adjacent vertices that is validated inside the HTM transactions. Therefore, the transaction footprint increases, thus increasing the probability of causing conflict aborts. Third, we find that in graphs with very high maximum degree, e.g., \texttt{uk} and \texttt{arb} graphs, the capacity aborts constitute a large portion of total aborts. In such graphs, the data access footprint of the transactions is large, resulting to a high probability of exceeding the hardware buffers. Overall, our analysis demonstrates that current HTM implementations are severely limited by the NUMA effect~\cite{Brown2016Investigating}, and incur high performance costs when using more than one NUMA socket on the machine. To this end, we recommend hardware designers to improve the HTM implementations in NUMA machines, and suggest software designers to propose intelligent algorithmic schemes and data partitioning approaches that minimize the expensive memory accesses to remote NUMA sockets inside the HTM transactions.

\subsection{Analysis of Balanced Graph Coloring Algorithms}\label{eval:balanced}

We compare the following balanced graph coloring implementations:
\begin{compactitem}
\item The CLU algorithm presented in Figure~\ref{alg:CLU}.
\item The VFF algorithm presented in Figure~\ref{alg:VFF}.
\item The Recoloring algorithm presented in Figure~\ref{alg:Recoloring}.
\item Our proposed \BalColorTM{} algorithm (Figure~\ref{alg:balcolortm}) that leverages HTM. Each transaction is retried up to 50 times, before resorting to a non-transactional fallback path. The non-transactional path is a coarse-grained lock scheme for the critical section (lines 27-39 in Figure~\ref{alg:balcolortm}).
\end{compactitem}

For a fair comparison, in all graph coloring schemes we color the vertices in the order they appear in the color classes produced by the initial coloring.

\subsubsection{Analysis of Color Balancing Quality}
Table~\ref{balancing-color-quality} compares the quality of balance in the color class sizes produced by the balanced-oblivious \ColorTM{} and all our evaluated balanced graph coloring implementations. Similarly to~\cite{Lu2015Balanced}, we evaluate the color balancing quality using the relative standard deviation of the color class sizes expressed in \%, which is defined as the ratio of the standard deviation of the color class sizes to the average color class size. The closer the value of this metric is to 0.00, the better is the color balance. For the \ColorTM{} and Recoloring schemes, we also include in parentheses the number of color classes produced. As already explained in Section~\ref{sec:balanced}, the CLU, VFF, and \BalColorTM{} schemes produce the same number of color classes with the initial coloring. In this experiment, we evaluate all algorithms using the maximum hardware thread capacity of our machine, i.e., 56 threads, in order to evaluate the color balancing quality of all schemes using the maximum available parallelism provided by the underlying hardware platform.

%\vspace{-12pt}
\begin{table}[t]
\centering
\resizebox{0.96\columnwidth}{!}{
\begin{tabular}{|l||r r|r|r|r r|r|}
    \hline
    \textbf{Input} & \multicolumn{2}{c|}{\textbf{Initial Coloring}} &
    \multicolumn{5}{c|}{\textbf{Balanced Graph Coloring Schemes}}\\
    \textbf{Graph} & \multicolumn{2}{c|}{\textbf{\ColorTM{}}} & \textbf{CLU} & \textbf{VFF} & \multicolumn{2}{c|}{\textbf{Recoloring}} & \textbf{\BalColorTM{}} \\
    \hline \hline
    
    \textbf{qun} & 63.62 & (48) & 0.212 & 1.669 & 14.739 & (48) & 0.009\\ \hline
    
    \textbf{geo} & 70.28  & (36) & 0.321 & 0.635 &  17.664 &  (34) & 0.020\\ \hline
    
    \textbf{fln} &  65.42&  (45) & 0.576 & 0.611 &  20.384 &  (51)  & 0.044\\ \hline
    
    \textbf{bum} &  64.32 & (36) &  0.179 & 0.647 & 17.950 &  (33)  & 0.009 \\ \hline
    
    \textbf{ser} & 73.64 & (39) & 0.405 & 0.751 &  16.651 & (38)  & 0.024\\ \hline
    
    \textbf{del} & 100.06 &  (9) & 0.002 & 0.013 & 35.136 &  (10) &  0.001\\ \hline
    
    \textbf{rgg} &  115.30 & (22) & 0.018 & 3.783 &  21.799 &  (23)  & 0.003\\ \hline
    
    \textbf{kmr} & 189.79 & (11) & 0.0003 & 0.0002 &  31.492 & (12) &  0.0004\\ \hline
    
    \textbf{cag} &  122.89 & (19) & 0.014 & 0.649 &  34.197 & (20)  & 0.005\\ \hline
    
    \textbf{usa} &  105.09 & (5) & 0.001 &  0.024 & 0.0005 & (5)  & 0.0005\\ \hline
    
    \textbf{dlf} &   57.95 & (54) & 2.58 & 2.53 &  22.551 &  (57) & 3.01\\ \hline
    
    \textbf{aud} &  84.02 & (60) & 5.243 & 2.780 &  19.498 & (54) & 3.575\\ \hline
    
    \textbf{vas} & 144.18 &  (38) & 0.084 & 18.527 &  25.373 &  (34) &  0.016\\ \hline
    
    \textbf{stk} & 141.41 & (35) & 0.016 & 17.684 &  25.375 & (34)   & 0.003 \\ \hline
    
    \textbf{uk} &  1882.66 & (944) & 0.437 & 0.237 & 65.994 &  (1355) & 1.732\\ \hline
    
    \textbf{soc} & 945.35 & (324) & 1.136 &  1.466 & 58.190 &  (459) & 1.886\\ \hline
    
    \textbf{arb} & 3351.79 & (3248) & 0.681 & 1.499 & 68.521 & (4772)  &  3.410 \\ \hline
    
    \textbf{fch} &   125.70 &  (9) &  0.012 & 0.271 & 33.854 & (10)  & 0.451\\ \hline

    %\hline
\end{tabular}
}
\vspace{2pt}
\caption{Color balancing quality achieved by \ColorTM{} and all balanced graph coloring implementations in the large real-world graphs. We present the relative standard deviation (in \%) on the sizes of the color classes obtained by each scheme (lower is better). In \ColorTM{} and Recoloring, we provide inside the parentheses the number of color classes produced. The CLU, VFF and \BalColorTM{} produce the same number of color classes with the initial coloring scheme.}
\label{balancing-color-quality}
\vspace{-4pt}
\end{table}

We draw three findings from Table~\ref{balancing-color-quality}. First, we observe that the balanced-oblivious \ColorTM{} scheme incurs very high disparity in the sizes of the color classes produced. Specifically, the color balancing quality of \ColorTM{} is 1887.01$\times$, 287.70$\times$, 10.32$\times$, and 4266.03$\times$ worse than that of CLU, VFF, Recoloring and \BalColorTM{}, respectively. Second, even though Recoloring is effective over \ColorTM{} by providing better color balancing quality, its color balancing quality is the worst compared to all the remaining balanced graph coloring schemes. In addition, in highly irregular graphs (graphs with high maximum degree and high standard deviation in the vertices' degrees) such as \texttt{uk}, \texttt{soc} and \texttt{arb}, Recoloring significantly increases the number of color classes produced over the initial coloring. Recoloring re-colors the vertices of the graph with a different order compared to that used in the initial graph coloring scheme, which in turn may introduce new additional color classes. Third, we find that \BalColorTM{} provides the best color balancing quality compared to all prior state-of-the-art balanced graph coloring schemes. Specifically, the color balancing quality of \BalColorTM{} is 2.26$\times$, 14.82$\times$ and 413.31$\times$ better compared to that of CLU, VFF and Recoloring, respectively. Overall, we conclude that our proposed \BalColorTM{} provides the best color balancing quality over prior state-of-the-art schemes in all large real-world graphs.

%\vspace{-12pt}
\begin{figure}[!ht]
\begin{minipage}{1.0\textwidth}
\centering
\includegraphics[width=\columnwidth]{ 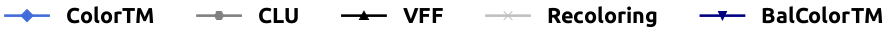}
\end{minipage}
\begin{minipage}{1.0\textwidth}
\centering
\includegraphics[scale=0.04]{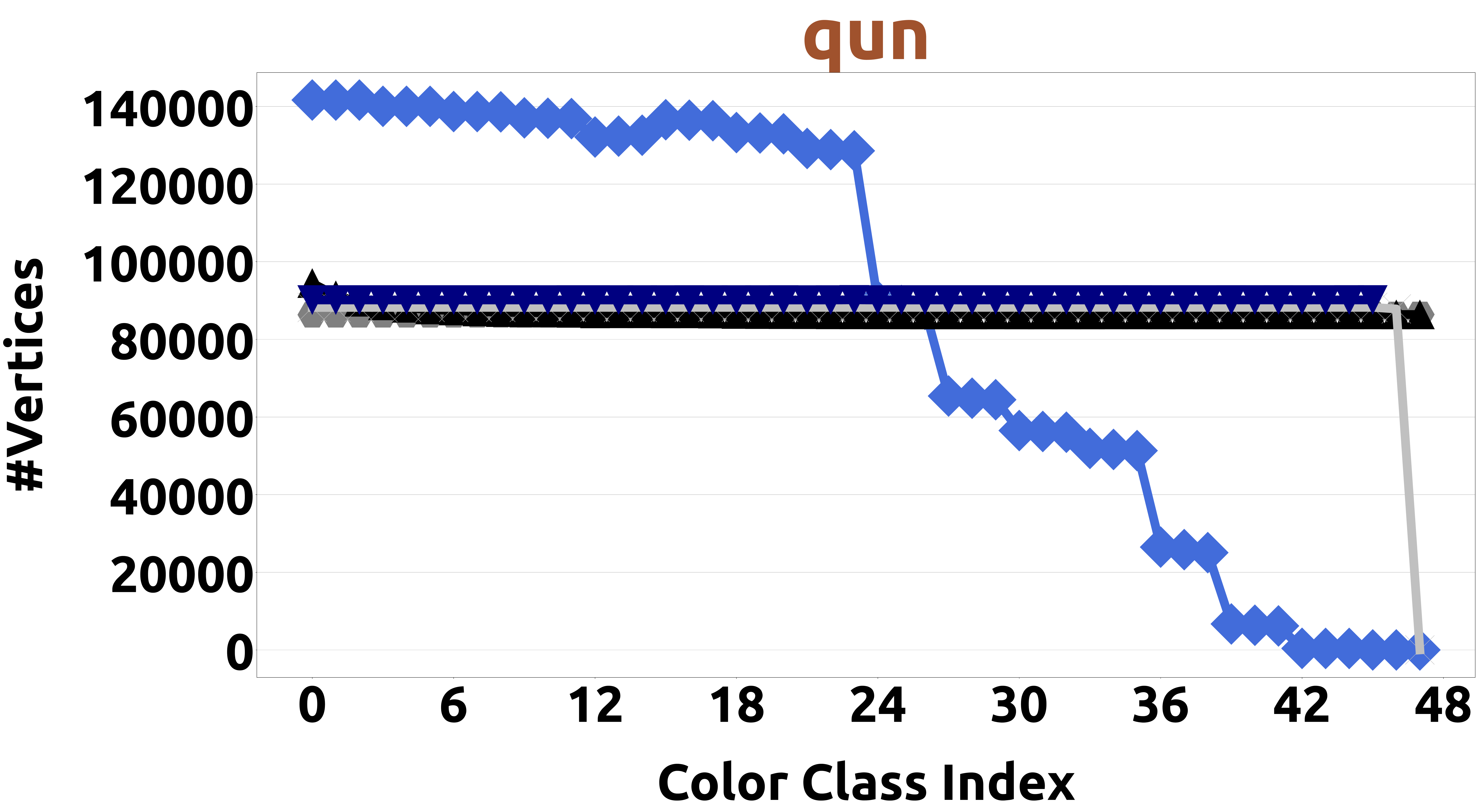}
\includegraphics[scale=0.04]{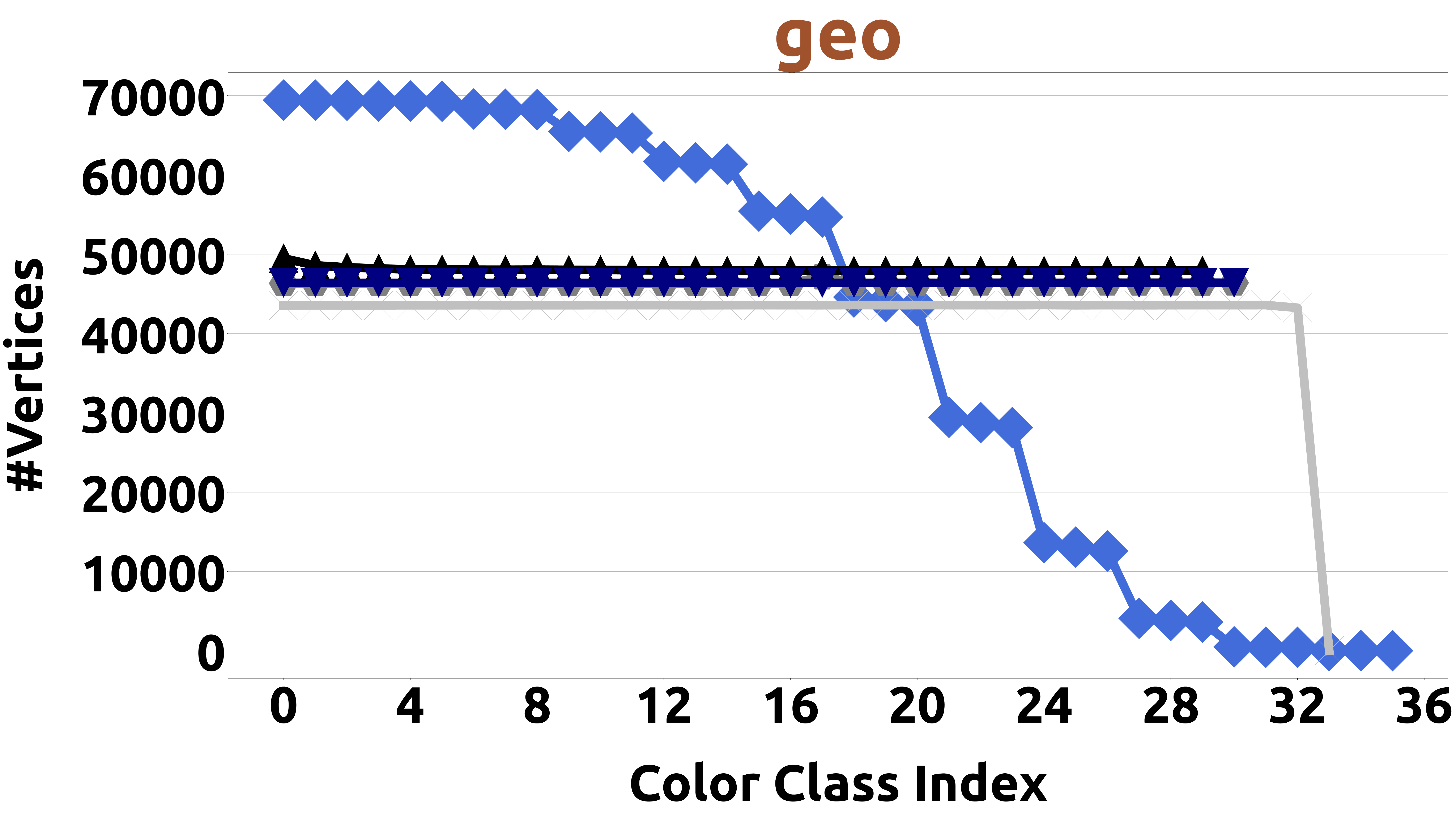}
\includegraphics[scale=0.04]{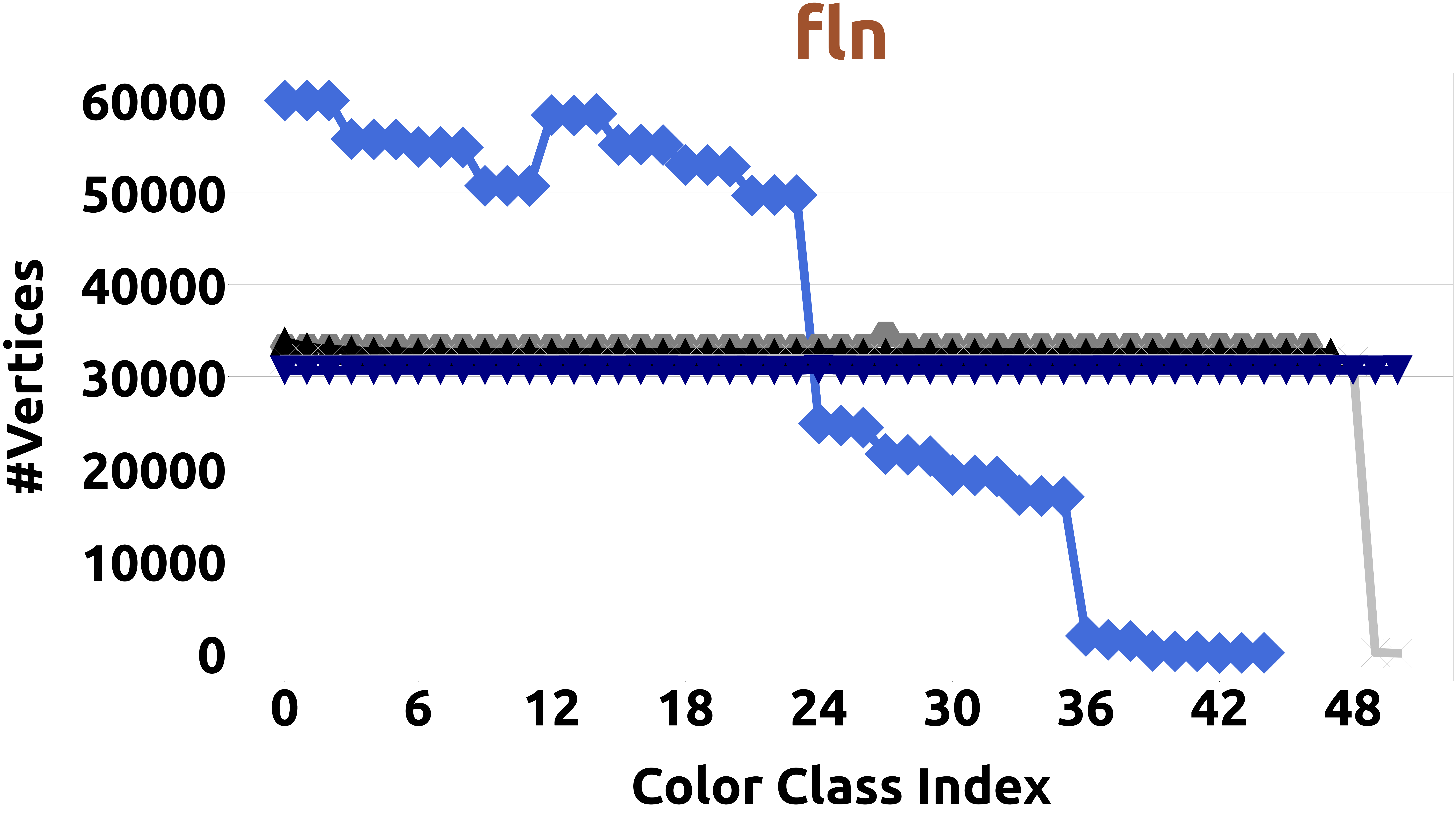}
\end{minipage}
\begin{minipage}{1.0\textwidth}
\centering
\includegraphics[scale=0.041]{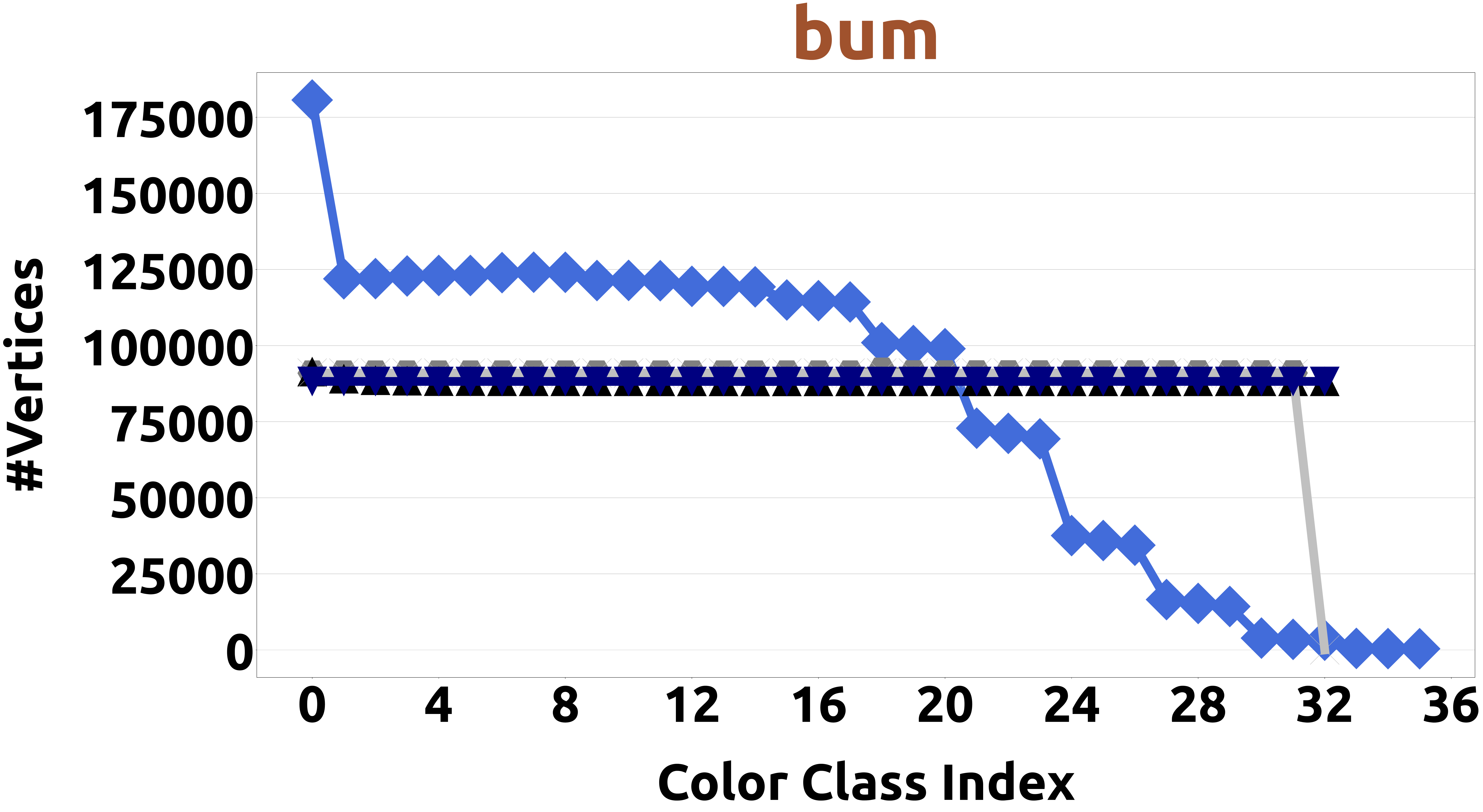}
\includegraphics[scale=0.041]{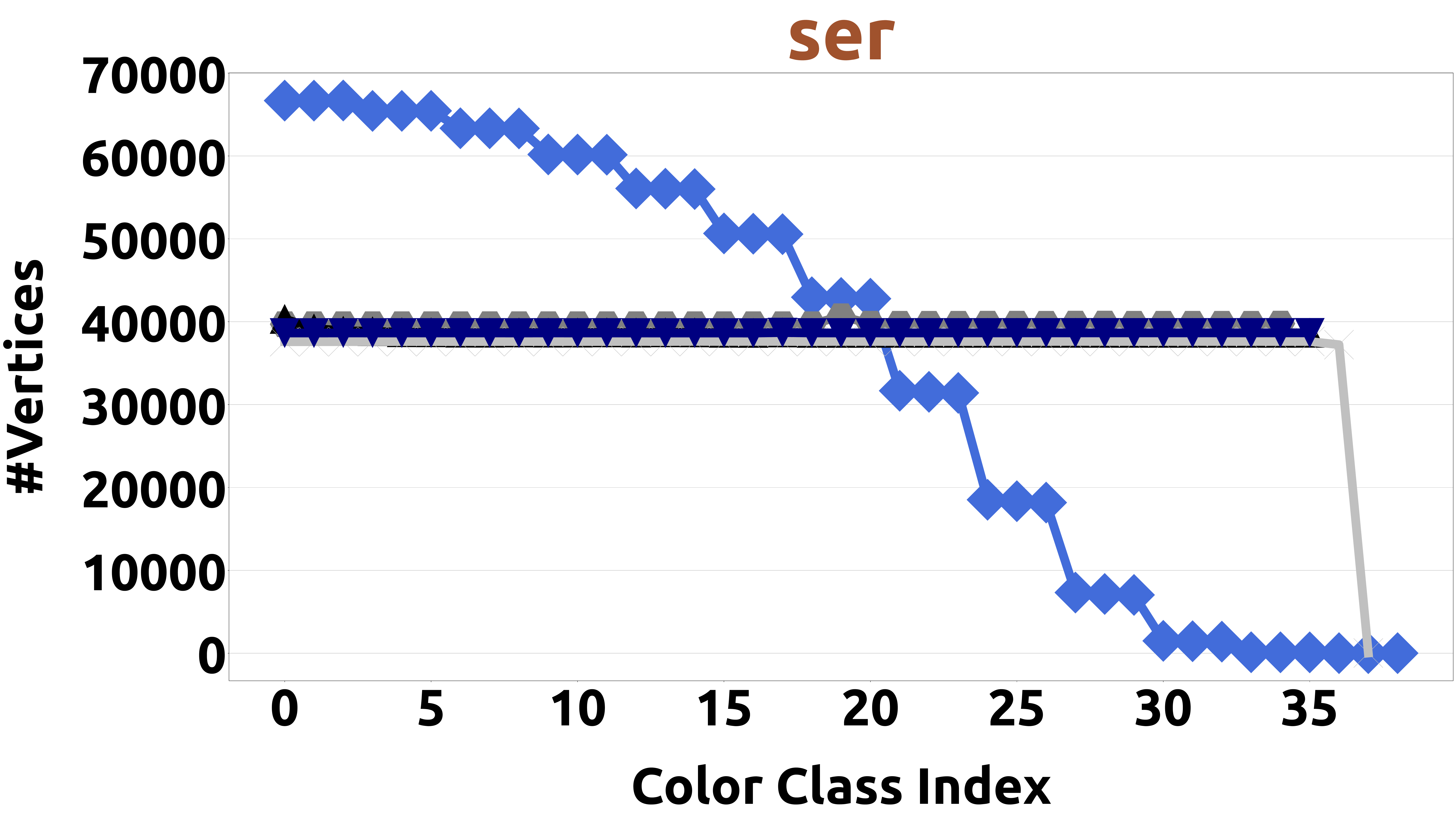}
\includegraphics[scale=0.041]{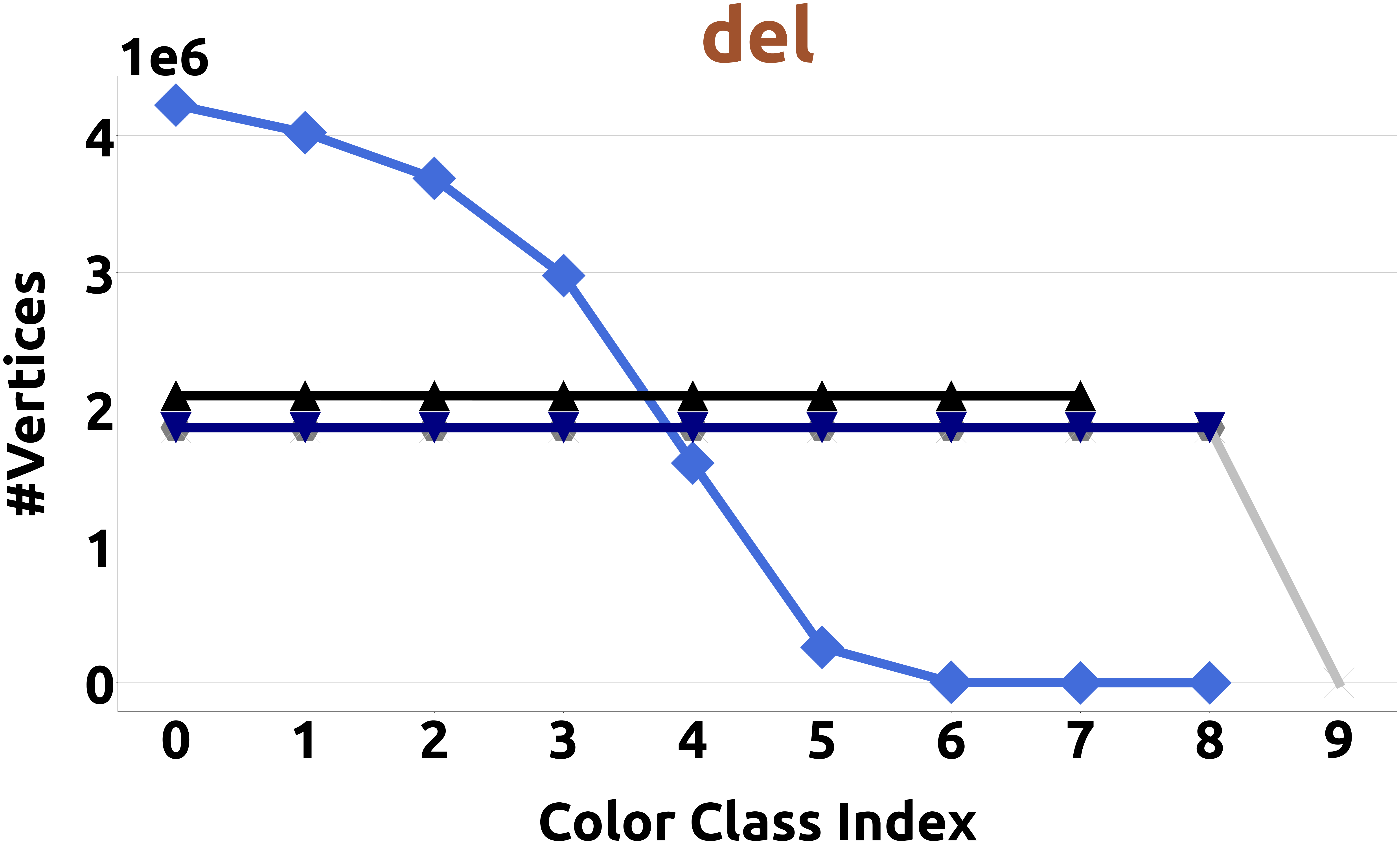}
\end{minipage}
\begin{minipage}{1.0\textwidth}
\centering
\includegraphics[scale=0.0414]{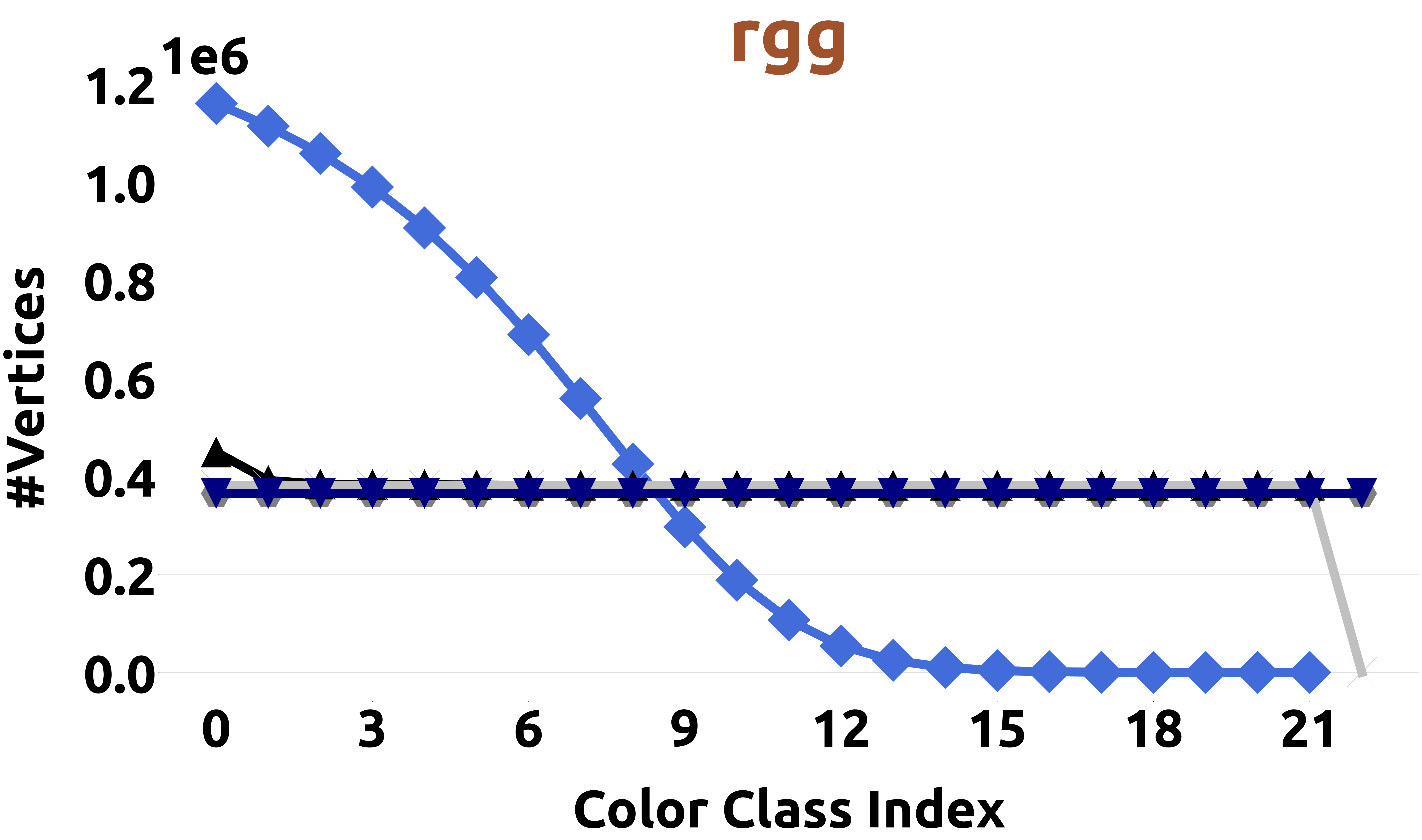}
\includegraphics[scale=0.0414]{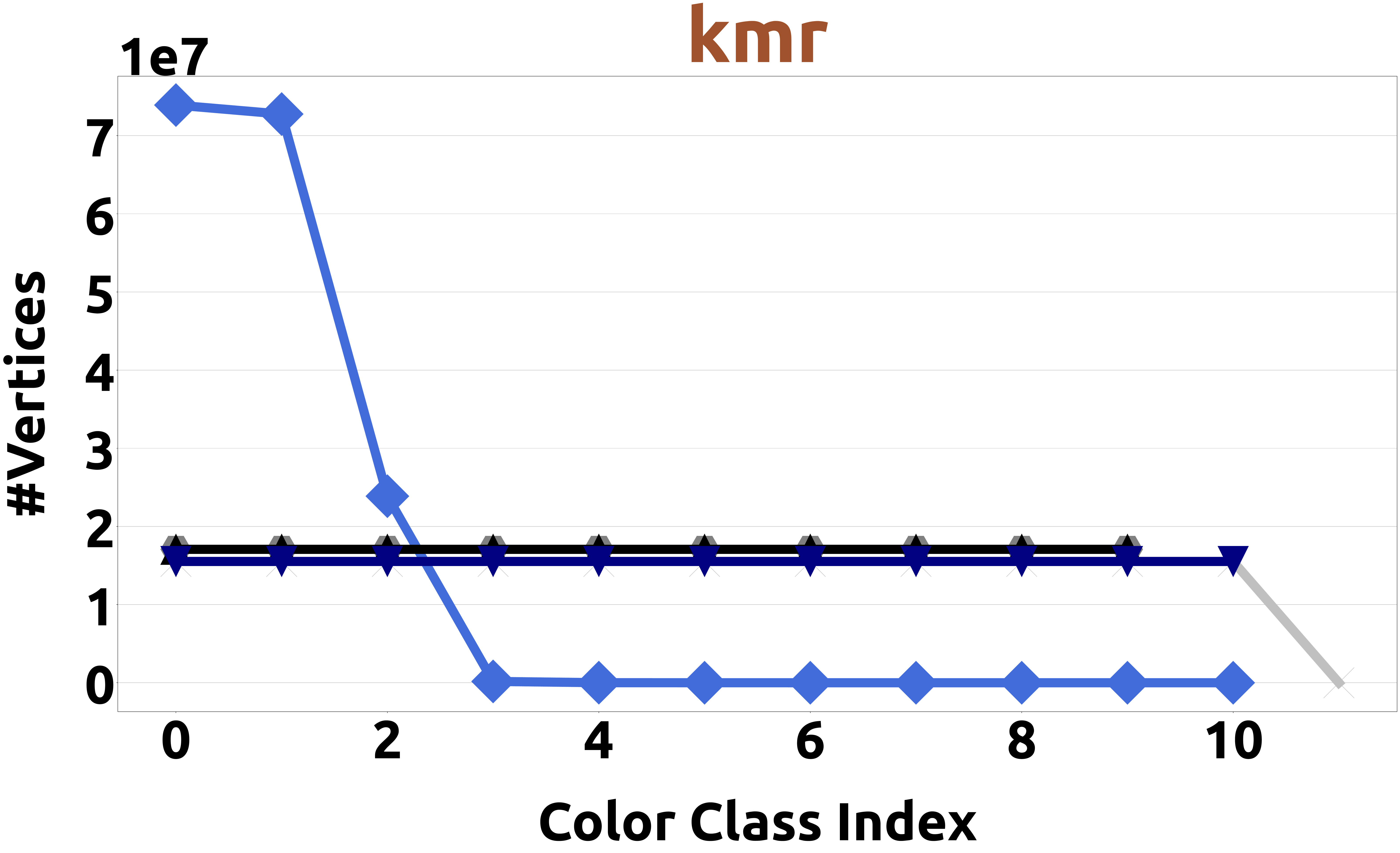}
\includegraphics[scale=0.0414]{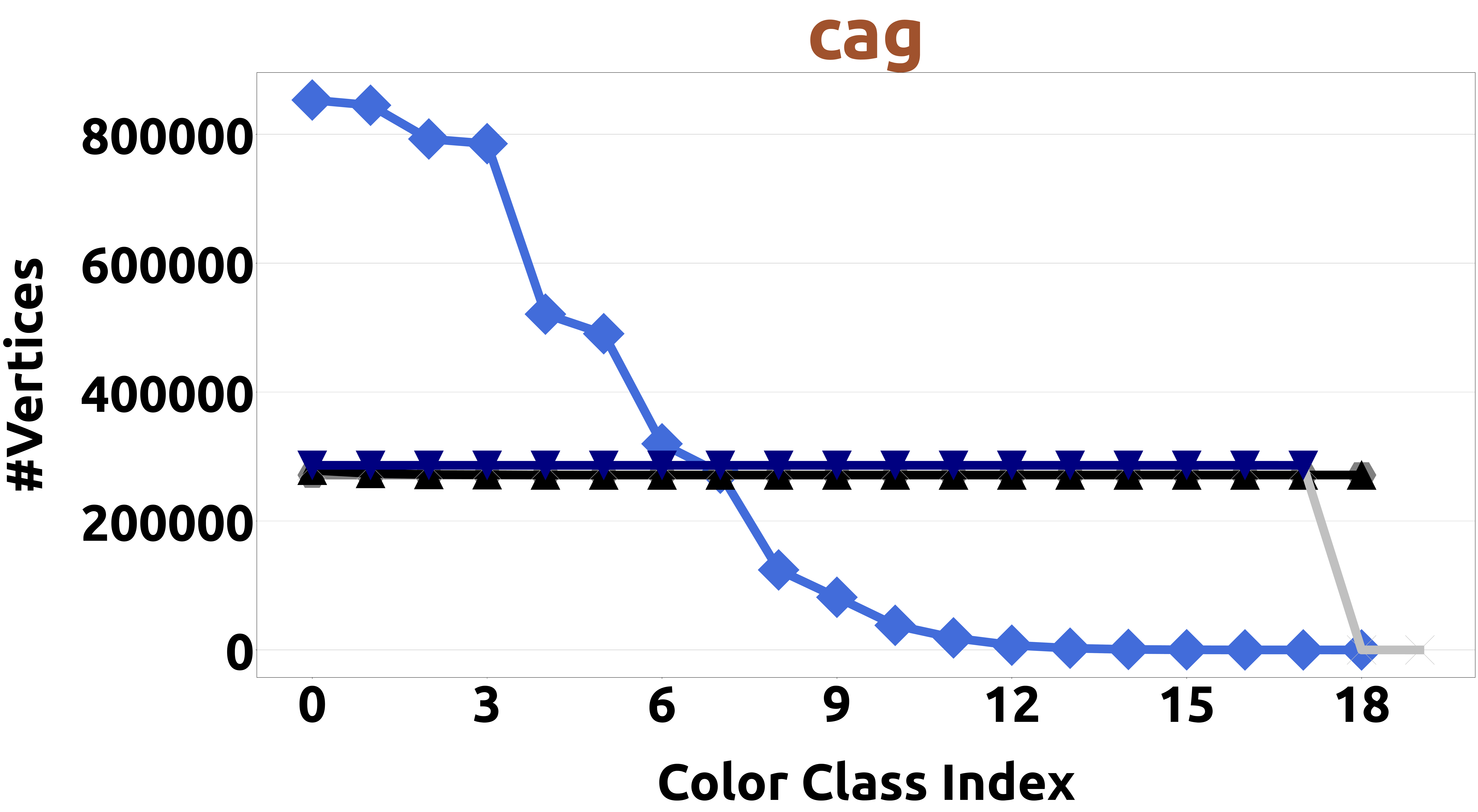}
\end{minipage}
\begin{minipage}{1.0\textwidth}
\centering
\includegraphics[scale=0.041]{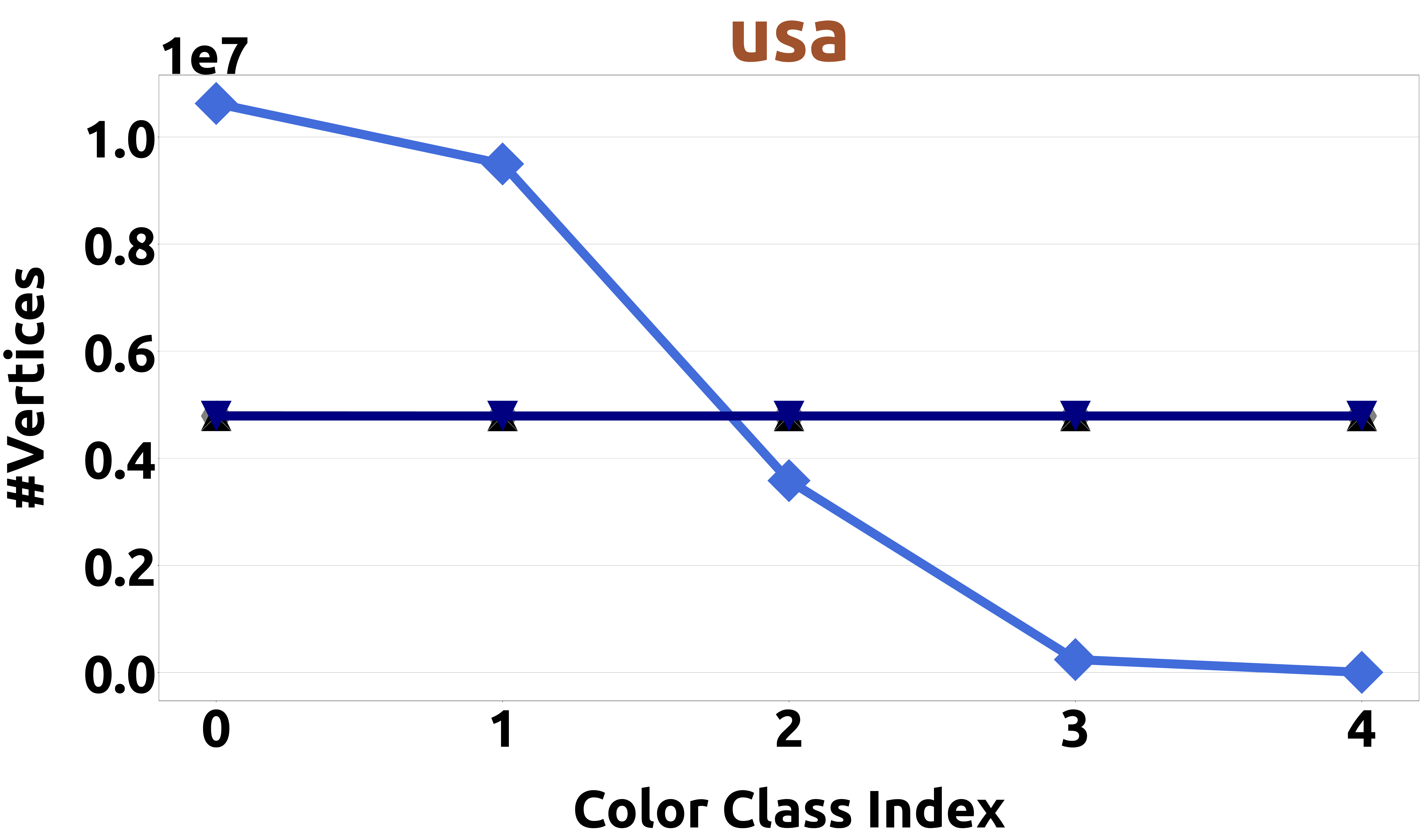}
\includegraphics[scale=0.041]{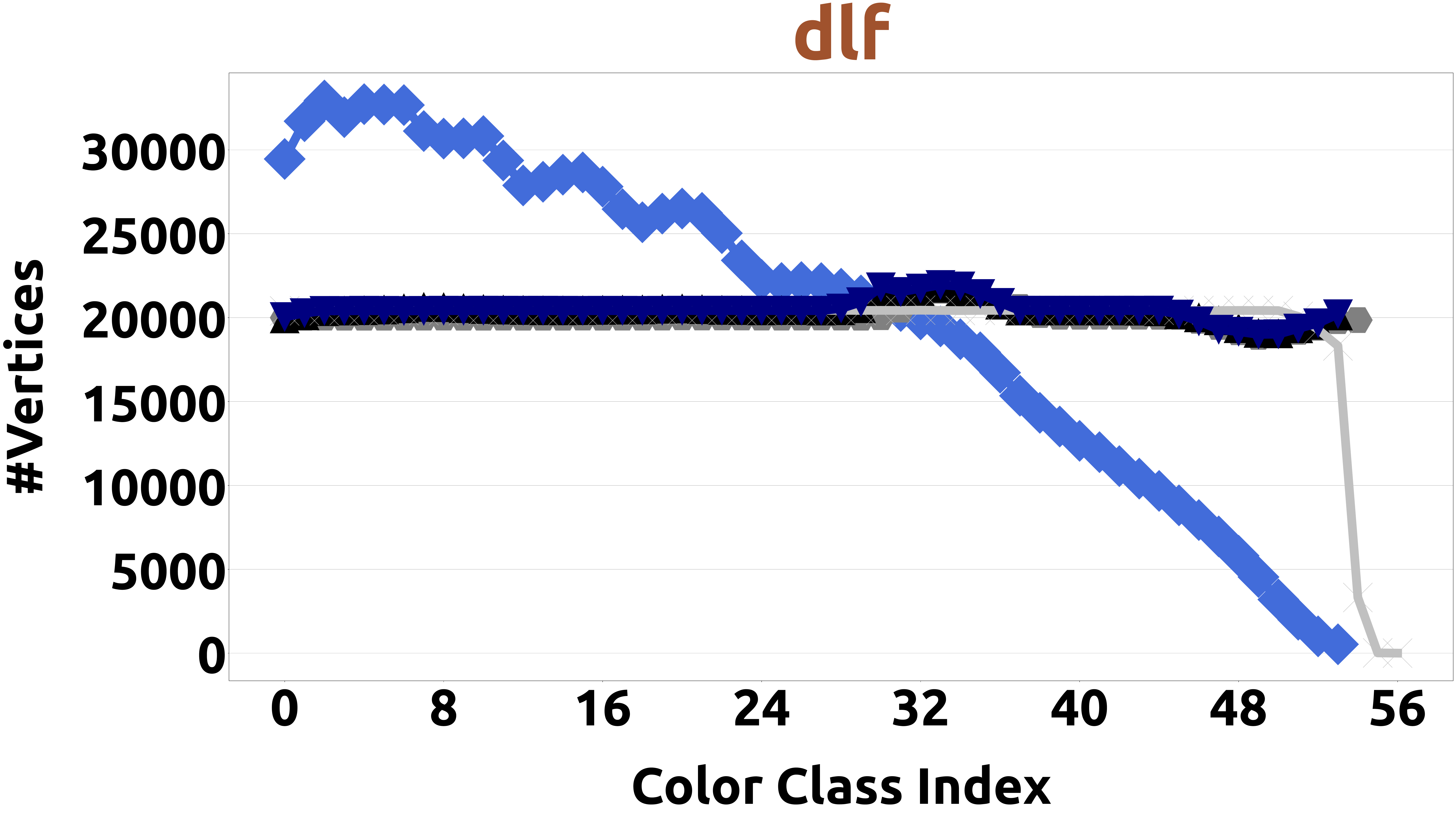}
\includegraphics[scale=0.041]{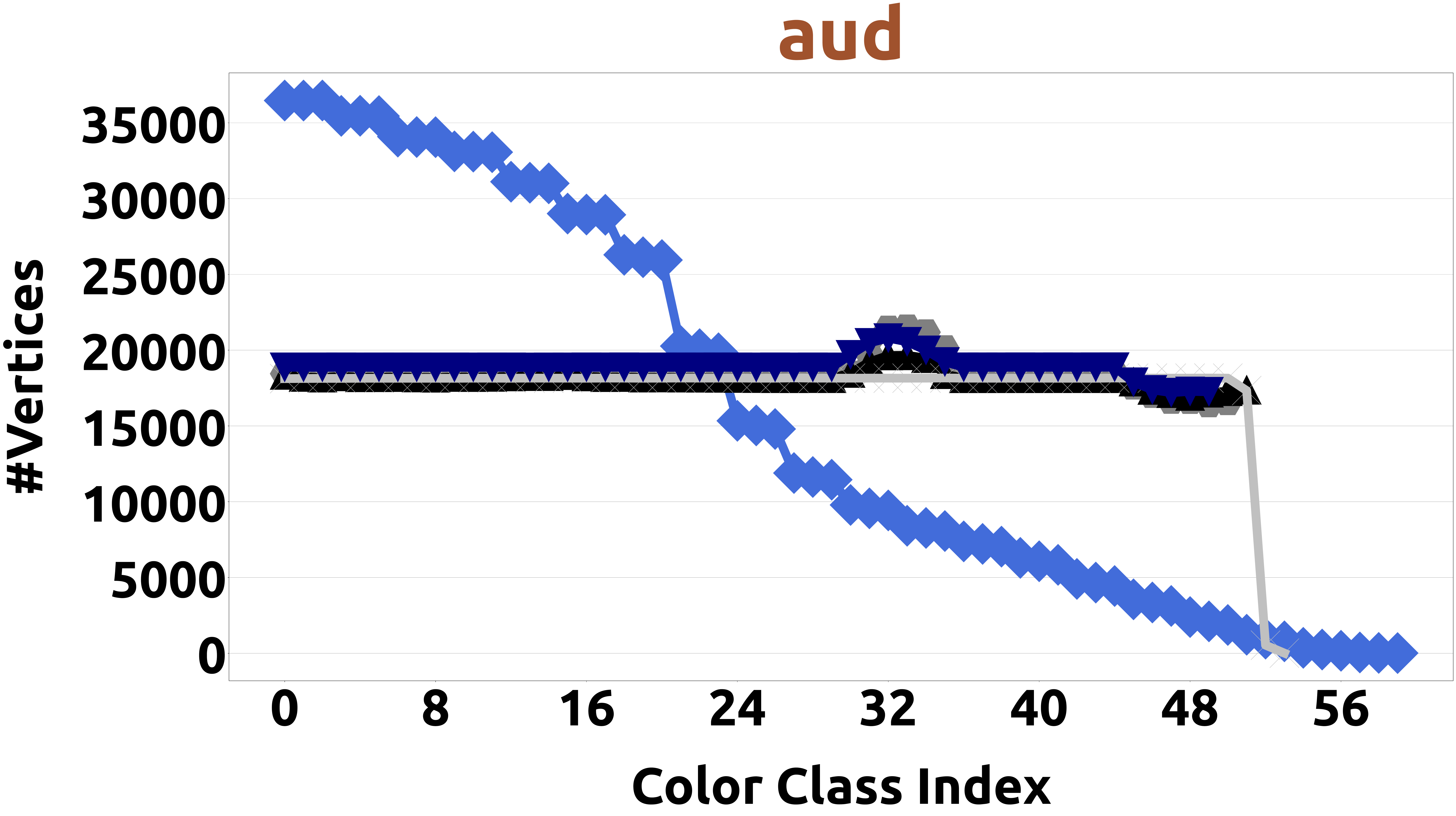}
\end{minipage}
\begin{minipage}{1.0\textwidth}
\centering
\includegraphics[scale=0.0414]{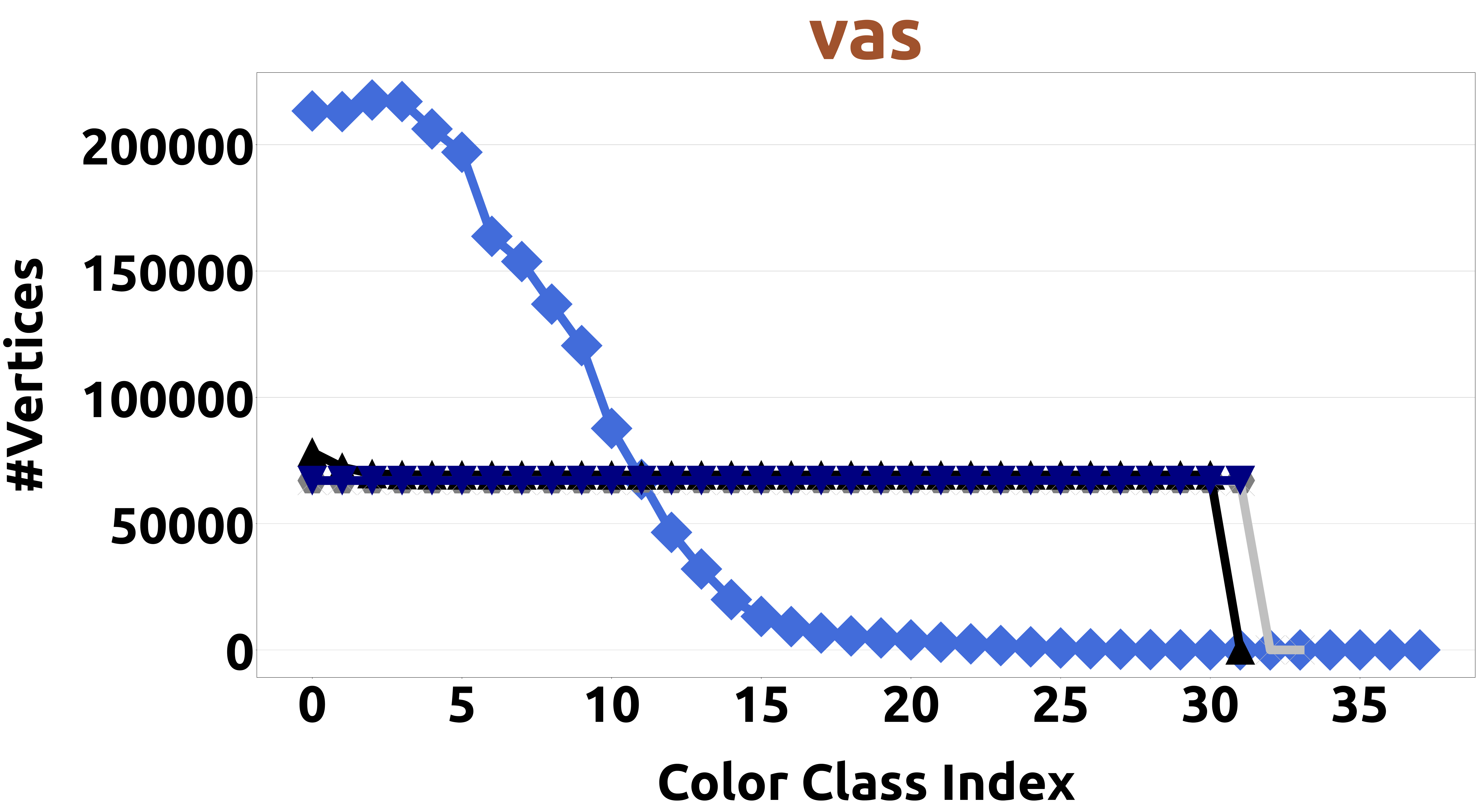}
\includegraphics[scale=0.0414]{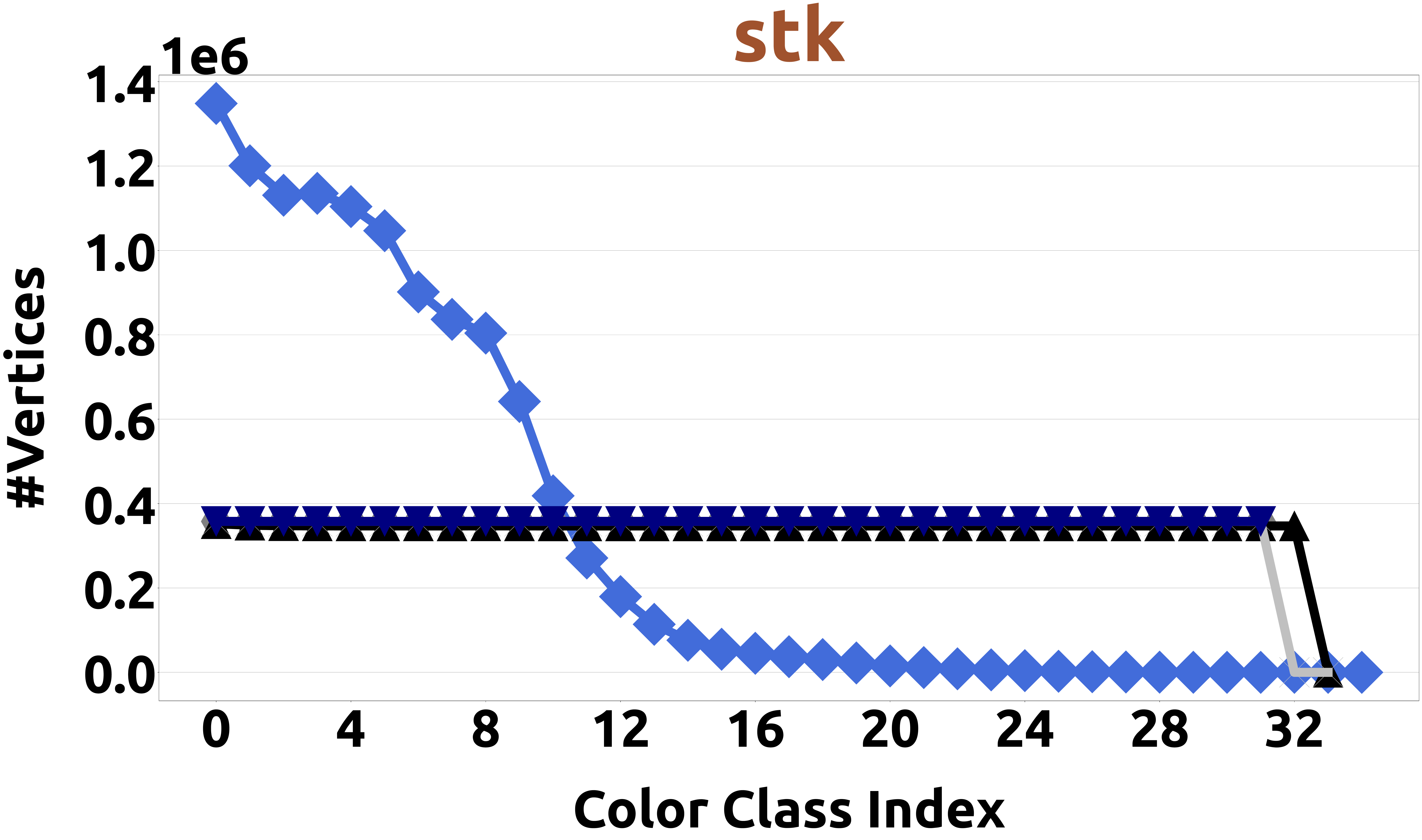}
\includegraphics[scale=0.0414]{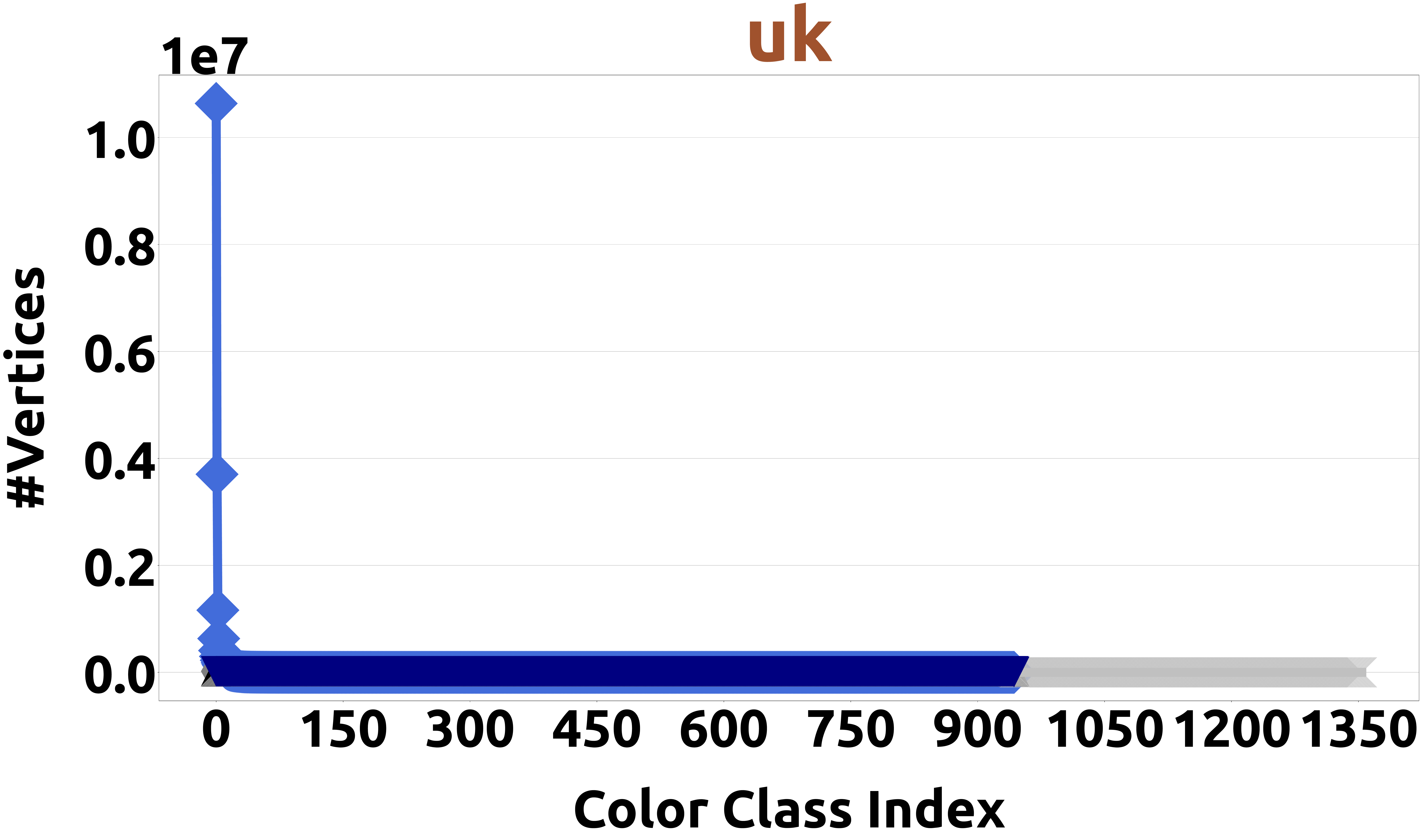}
\end{minipage}
\begin{minipage}{1.0\textwidth}
\centering
\includegraphics[scale=0.042]{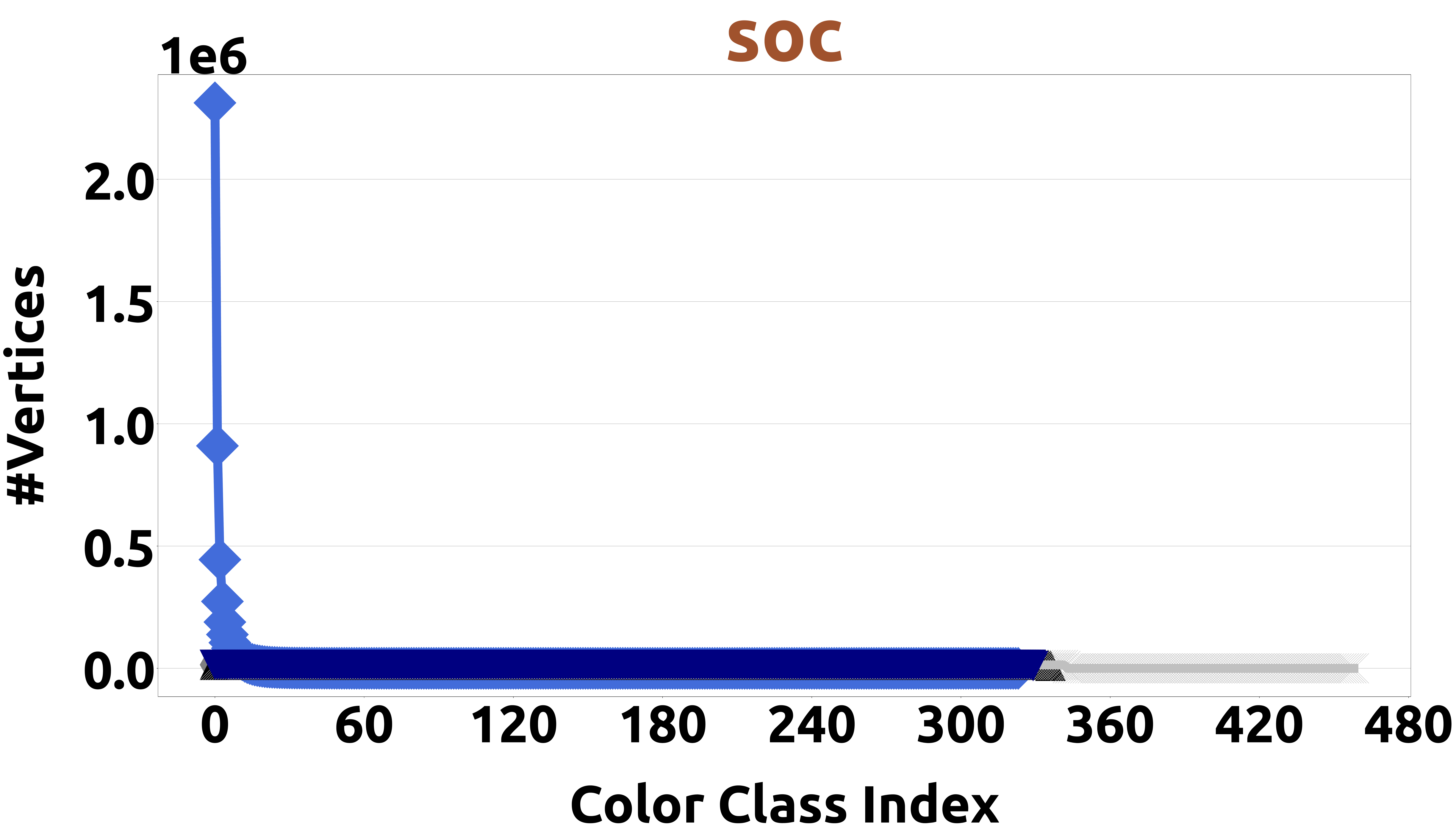}
\includegraphics[scale=0.042]{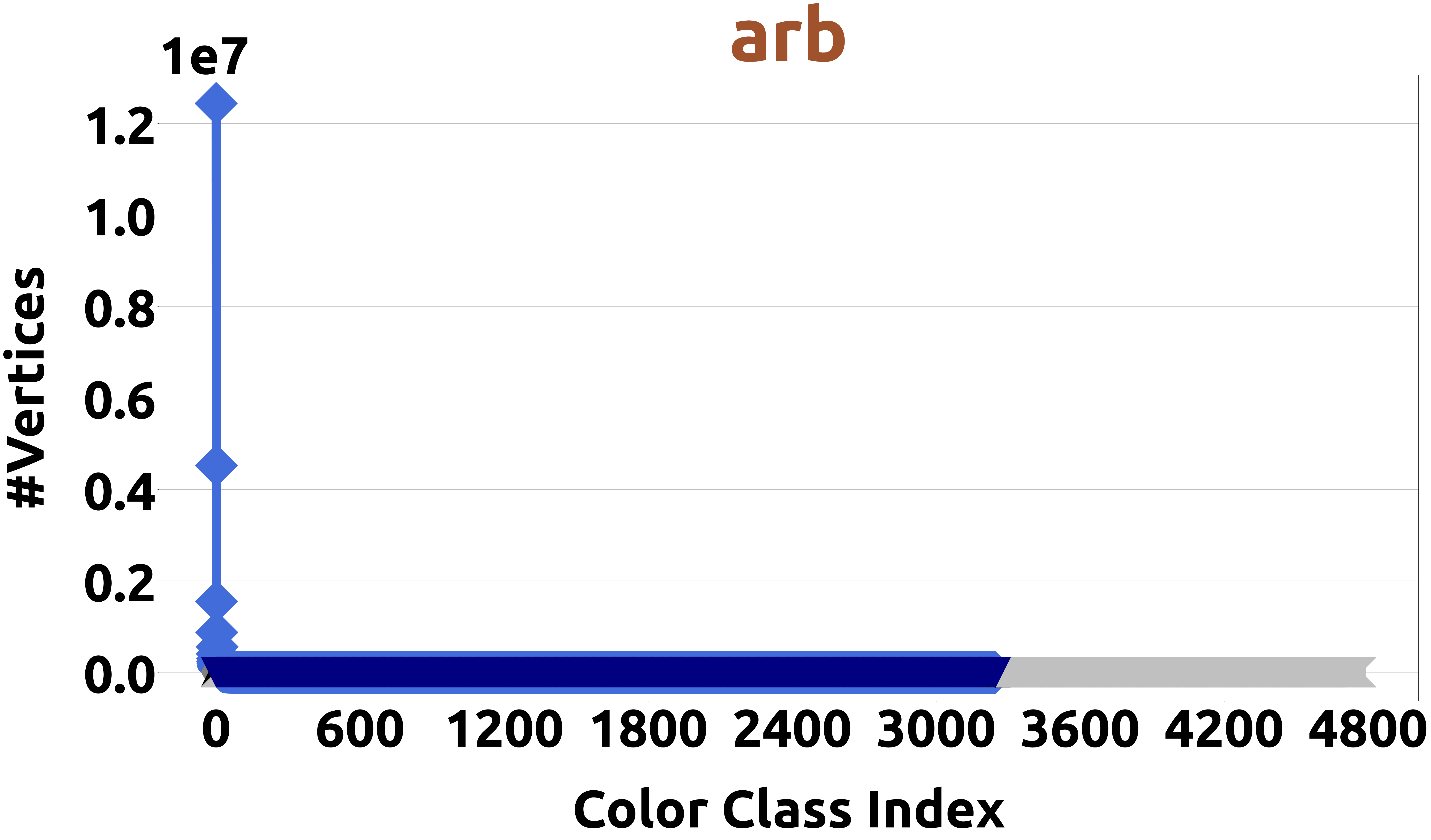}
\includegraphics[scale=0.042]{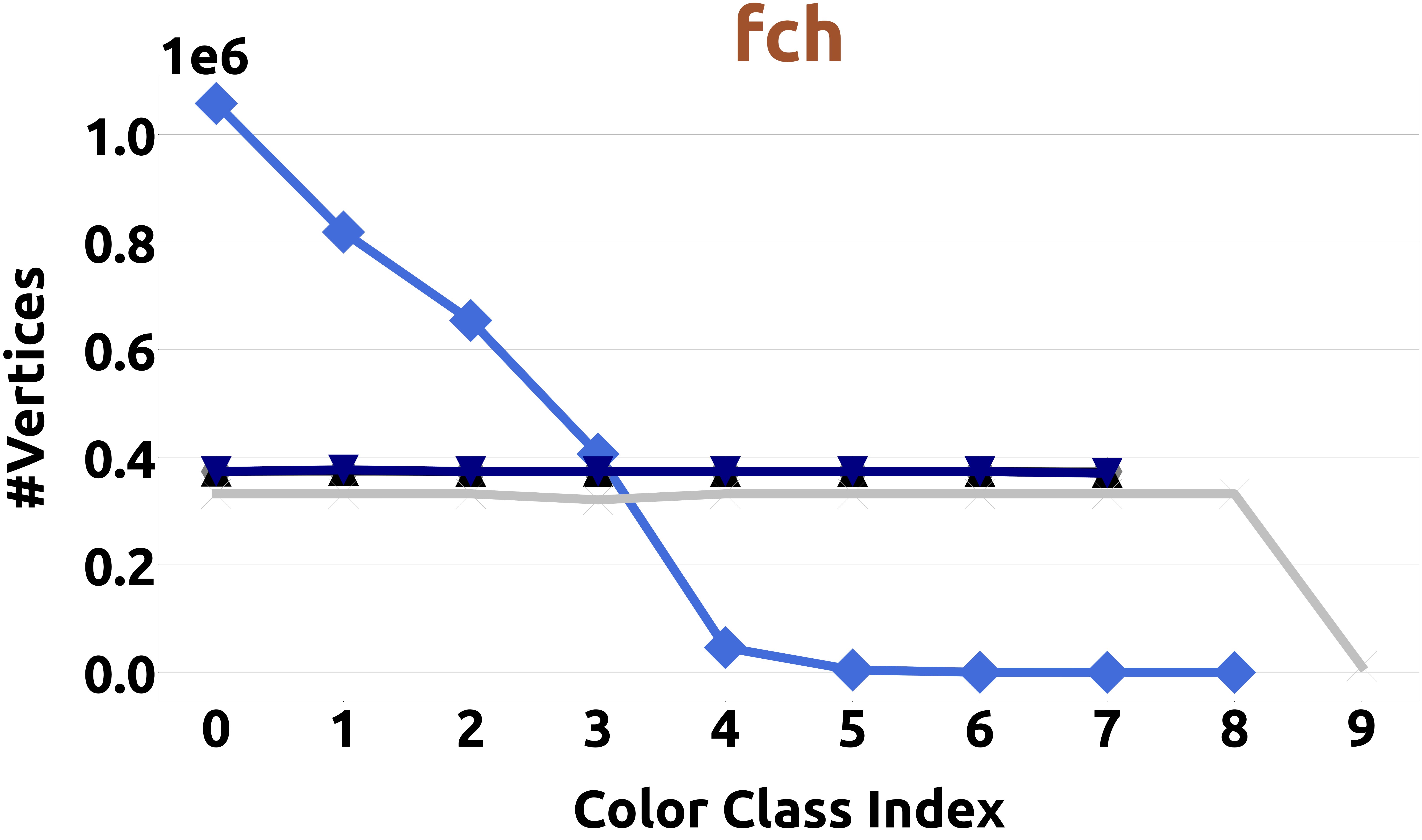}
\end{minipage}
\vspace{-1pt}
\caption{Distribution of color class sizes produced by \ColorTM{} and all our evaluated balanced graph coloring schemes. Note that small color class sizes result in reduced parallelism in the real-world end-application.}
\label{balancing-distribution}
\vspace{-10pt}
\end{figure}

To better illustrate the effect of balancing the vertices across color classes, we present in Figure~\ref{balancing-distribution} the sizes of all the color classes produced by \ColorTM{}, CLU, VFF, Recoloring and \BalColorTM{} for a representative subset of our evaluated real-world graphs. The \texttt{uk}, \texttt{soc} and \texttt{arb} graphs are web social networks~\cite{Boldi2004Web} with a highly power-law distribution~\cite{Giannoula2022SparsePSigmetrics,Giannoula2022SparsePPomacs,Tang2015Optimizing}: only a \emph{few} vertices have a very \emph{high} degree, while the vast majority of the remaining vertices of the graph has very low degree. In such graphs, \ColorTM{} inserts the vast majority of the vertices in the first few color classes, and the remaining \emph{few} vertices are assigned to different separate color classes. Moreover, as explained, Recoloring introduces a large number of \emph{new} additional color classes in such real-world graphs.

\subsubsection{Performance Comparison}
\begin{figure}[t]
\begin{minipage}{1.0\textwidth}
\centering
\includegraphics[width=0.8\columnwidth]{ 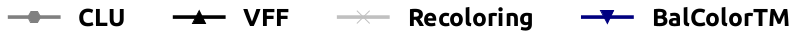}
\end{minipage}
\begin{minipage}{1.0\textwidth}
\centering
\includegraphics[scale=0.040]{ 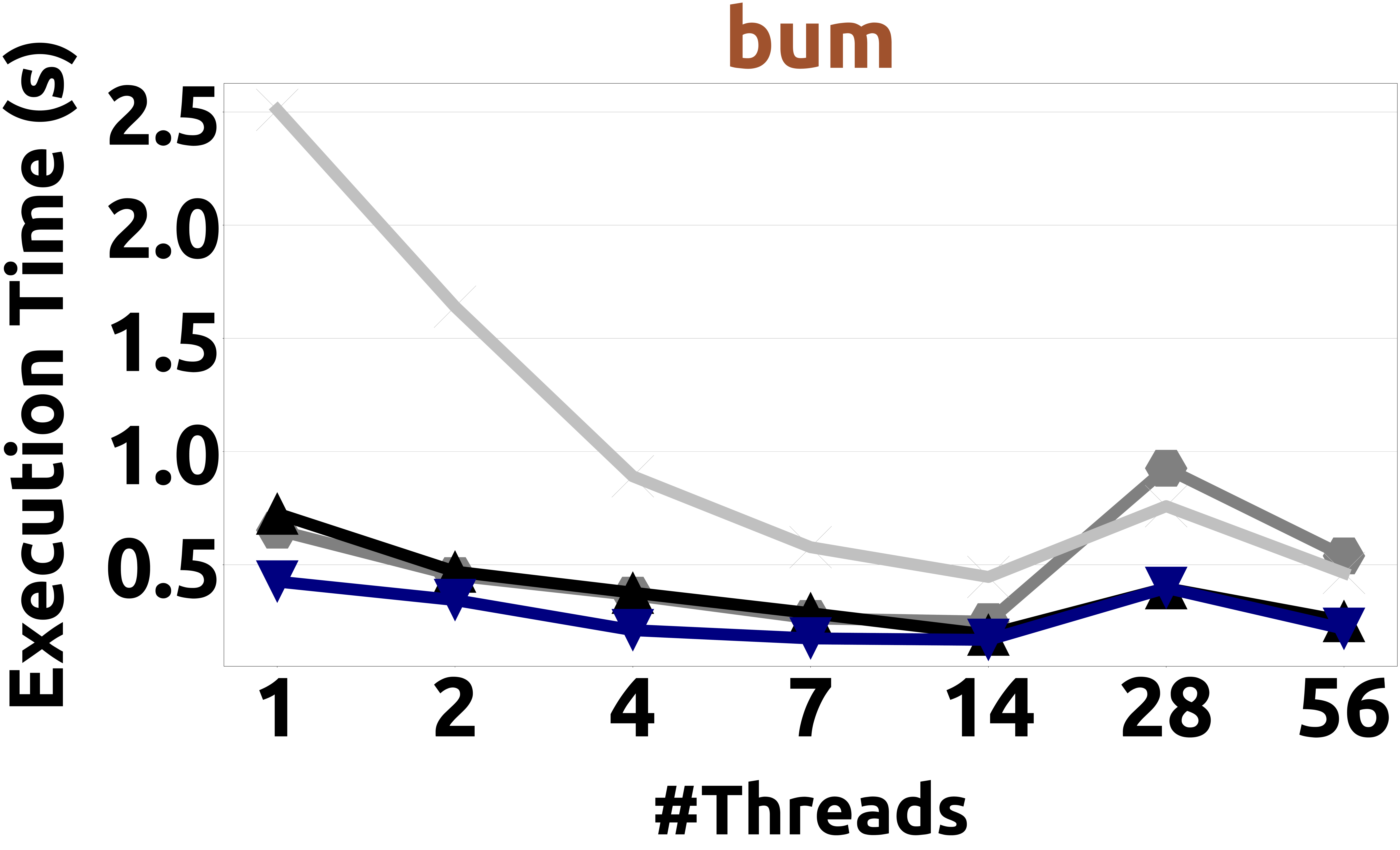}
\includegraphics[scale=0.040]{ 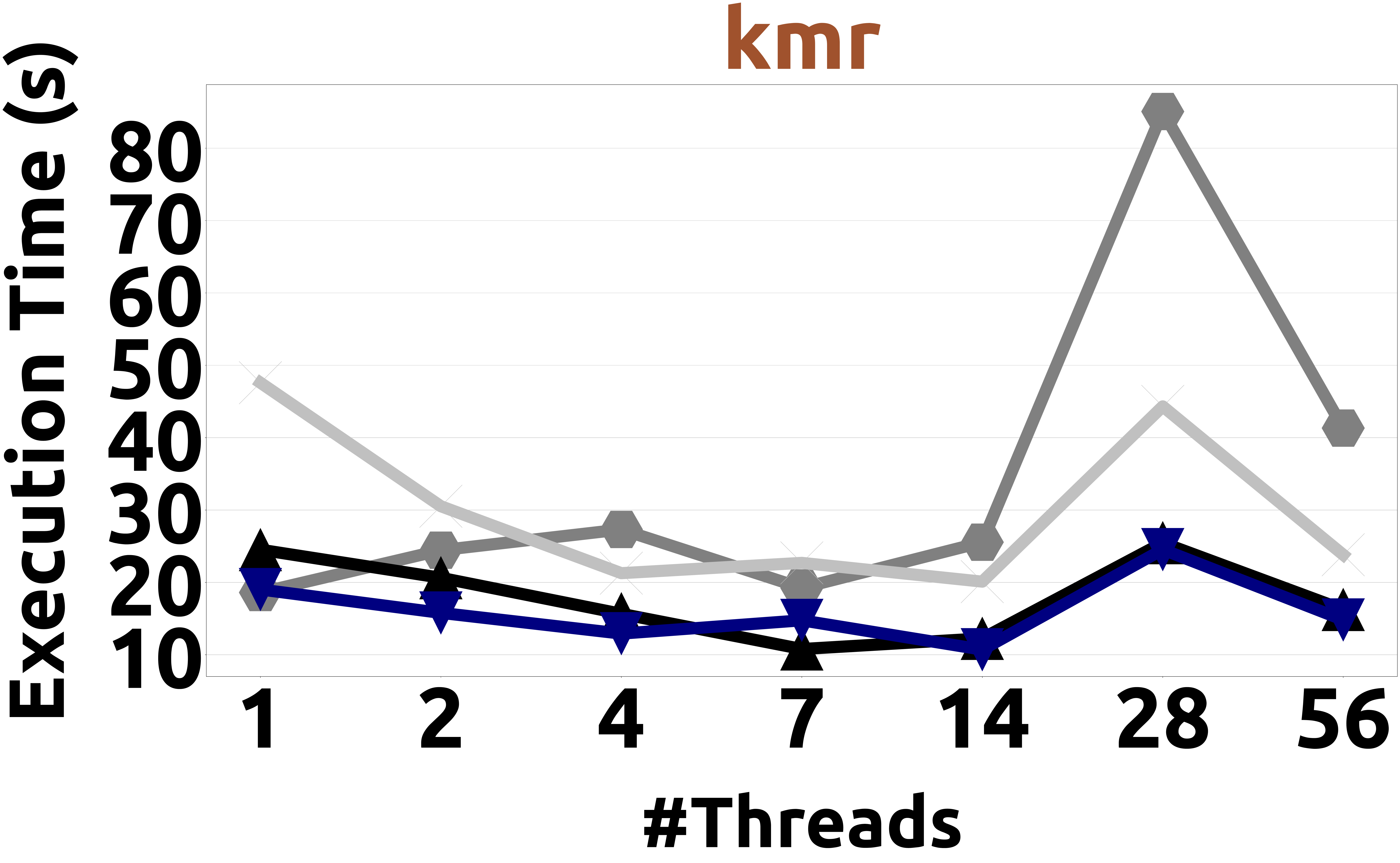}
\includegraphics[scale=0.040]{ 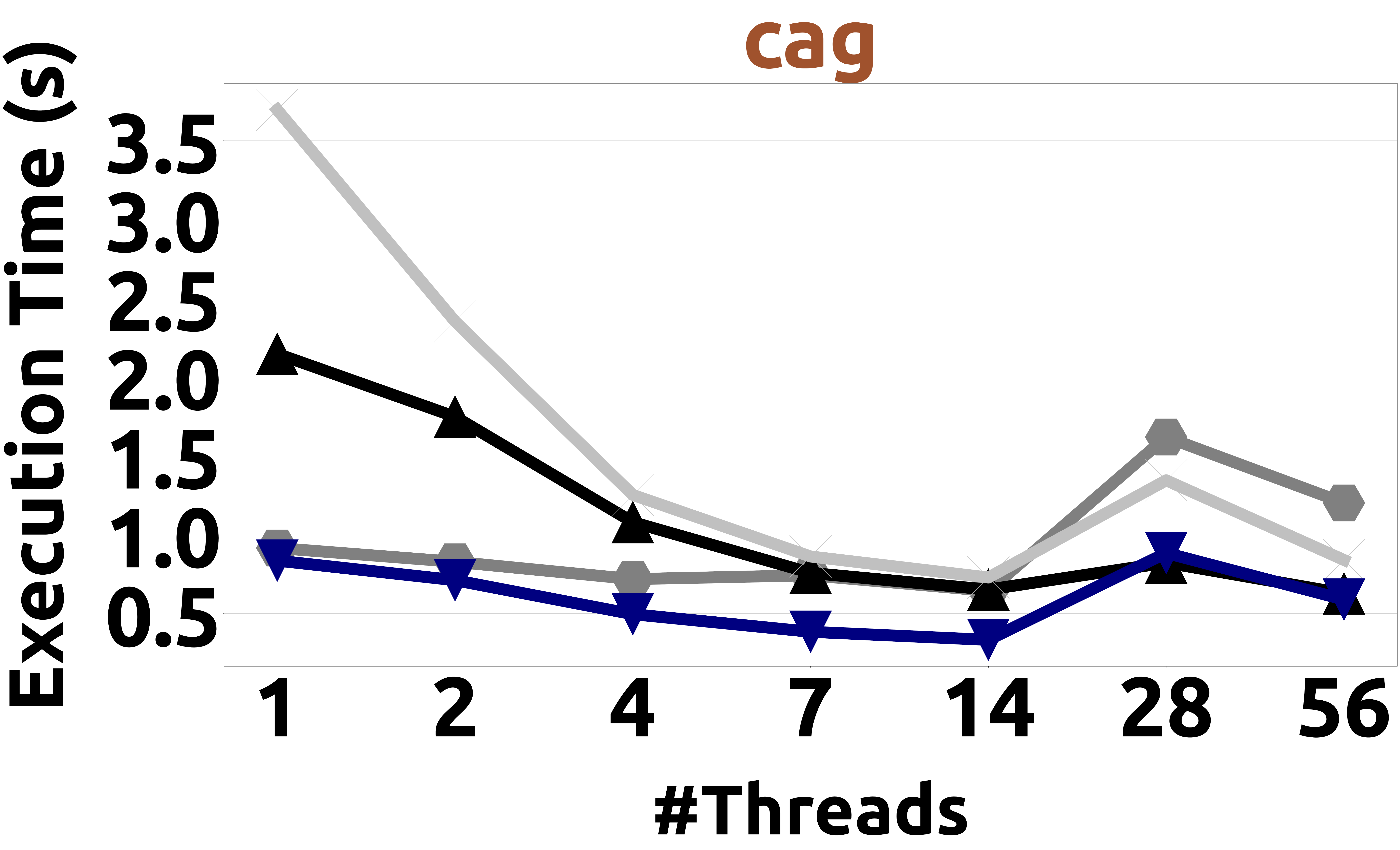}
\end{minipage}
\begin{minipage}{1.0\textwidth}
\centering
\includegraphics[scale=0.040]{ 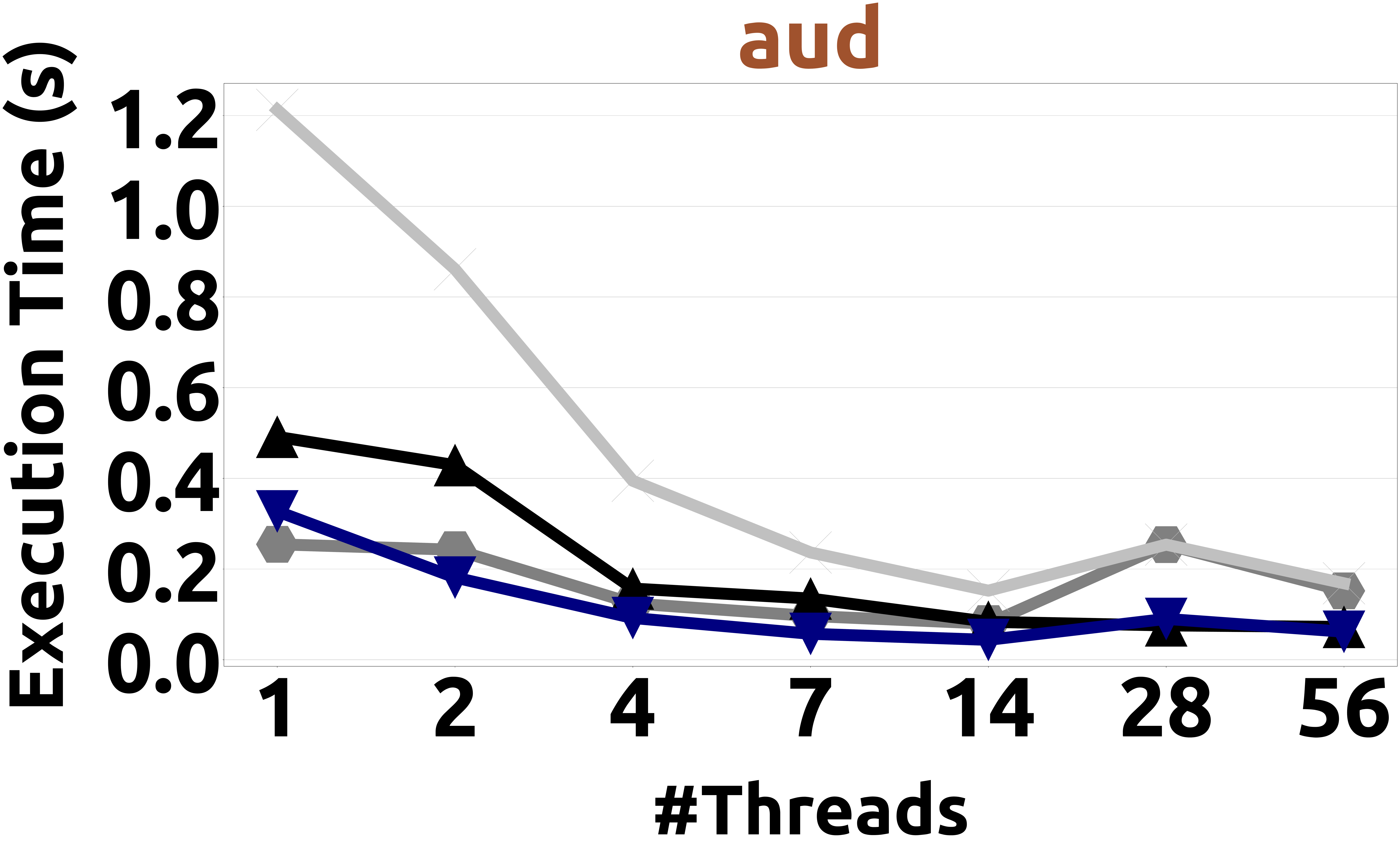}
\includegraphics[scale=0.040]{ 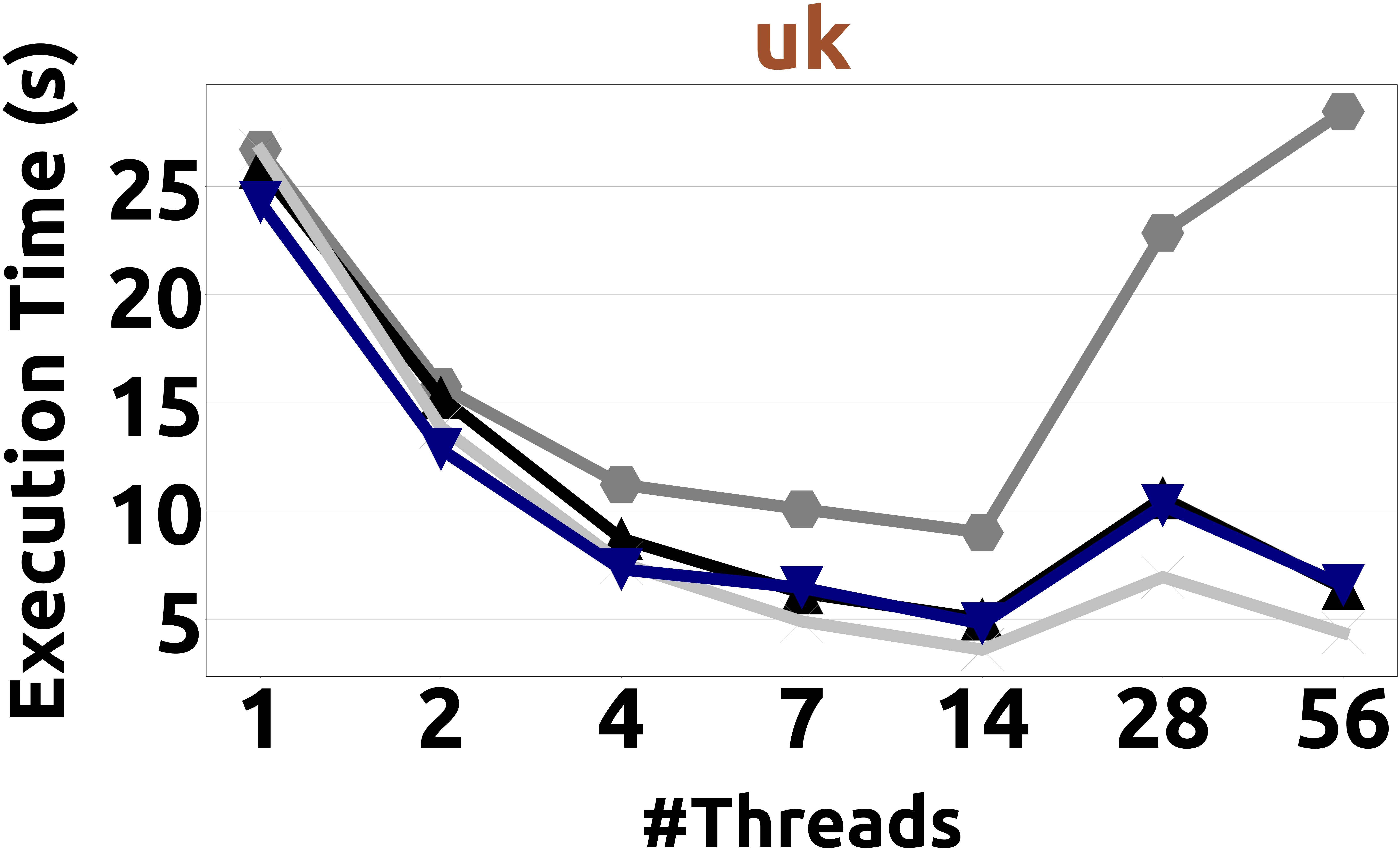}
\includegraphics[scale=0.040]{ 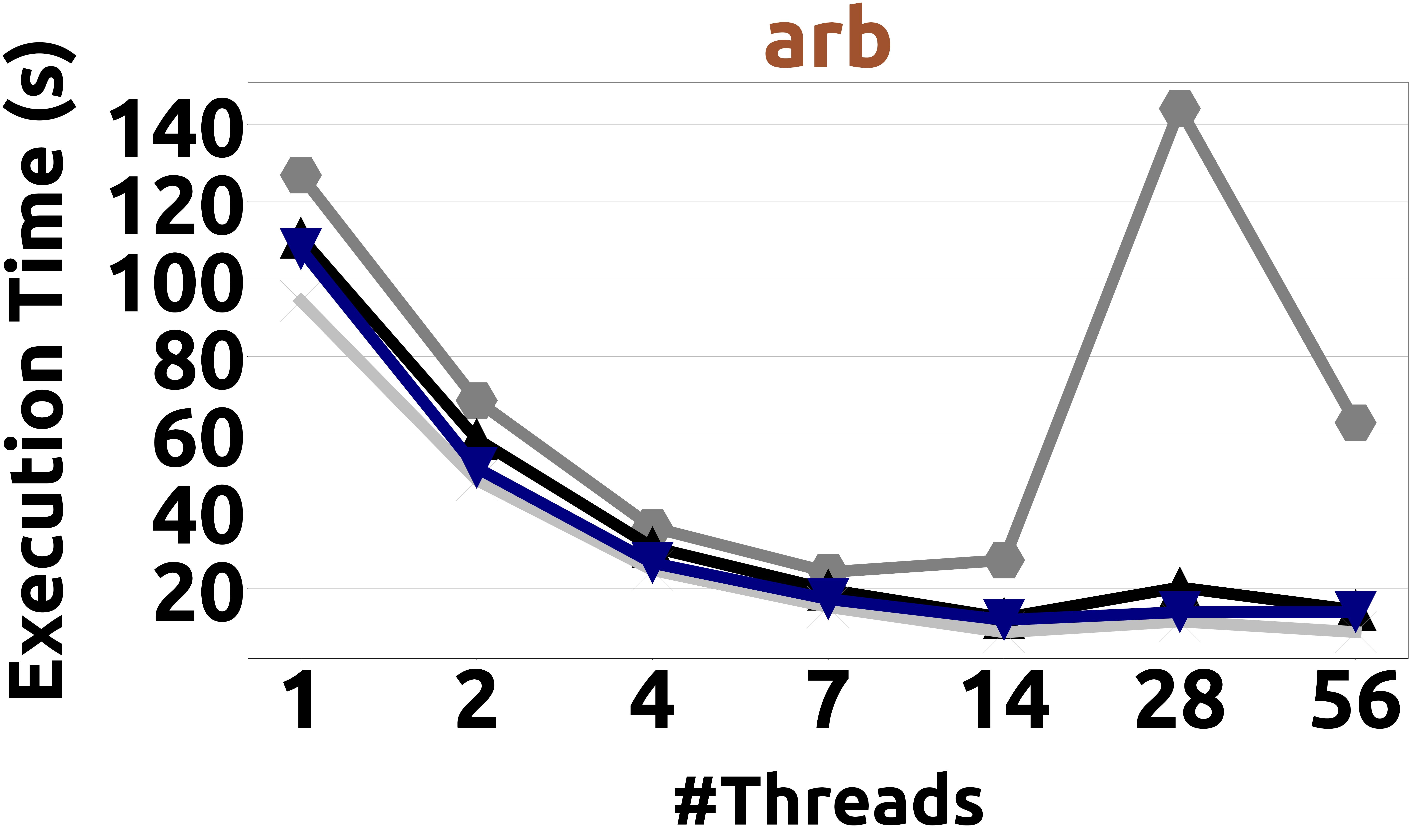}
\end{minipage}
\vspace{-2pt}
\caption{Scalability achieved by all balanced graph coloring implementations in large real-world graphs.}
\label{balancing-scalability}
\vspace{-18pt}
\end{figure}

Figure~\ref{balancing-scalability} evaluates the scalability achieved by all balanced graph coloring implementations in a representative subset of our evaluated large real-world graphs, as the number of threads increases from 1 to 56, i.e., up to the maximum available hardware thread capacity of our machine. We present the execution time of \emph{only} the kernel that balances the vertices across color classes (excluding the execution time of the initial graph coloring).

%\vspace{-14pt}

We draw three findings. First, we observe that Recoloring achieves the worst performance over all balanced graph coloring schemes. Even in the single-threaded executions, Recoloring performs by 3.21$\times$, 2.26$\times$ and 3.69$\times$ worse than CLU, VFF and \BalColorTM{}, respectively, because it executes a much larger amount of computation, memory accesses and synchronization. Recall that Recoloring processes and re-colors \emph{all} the vertices of the graph, while the remaining balanced graph coloring schemes re-color only a \emph{subset} of the vertices of the graph. Note that in \texttt{uk} and \texttt{arb} graphs, all balanced graph coloring schemes need to re-color a \emph{large} portion of the graph's vertices, thus performing closely to each other. Second, we find that the scalability of all schemes is affected by the NUMA effect, however \BalColorTM{} on average scales well even when using all available hardware threads and both NUMA sockets of our machine. When increasing the number of threads from 28 to 56, the performance of \BalColorTM{} improves by 1.55$\times$ averaged across all large graphs. Third, we find that in contrast to the graph coloring kernel, in many real-world graphs the performance of the balanced graph coloring kernel scales up to 14 threads, and degrades when using 56 threads. This is because the balanced graph coloring kernel has a lower amount of parallelism (a small subset of the vertices of the graph are re-colored by parallel threads) than the graph coloring kernel. Thus, our analysis demonstrates that when a kernel has low levels of parallelism, the best performance is achieved using a smaller number of parallel threads than the available hardware threads on the multicore platform. To this end, we suggest software designers of real-world end-applications to \emph{on-the-fly} adjust the number of parallel threads used to parallelize each different sub-kernel of the end-application based on the parallelization needs of each particular sub-kernel.
%We conclude that our proposed balanced graph coloring algorithm achieves high scalability in contemporary multicore platforms.

\begin{figure}[t]
\begin{minipage}{1.0\textwidth}
\centering
\includegraphics[width=0.9\columnwidth]{ 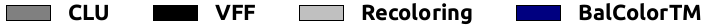}
\end{minipage}
\begin{minipage}{1.0\textwidth}
\centering
\includegraphics[width=\columnwidth]{ sections/ColorTM/results/bspeedup_14.pdf}
\end{minipage}
\begin{minipage}{1.0\textwidth}
\centering
\includegraphics[width=\columnwidth]{ sections/ColorTM/results/bspeedup_28.pdf}
\end{minipage}
\begin{minipage}{1.0\textwidth}
\centering
\includegraphics[width=\columnwidth]{ sections/ColorTM/results/bspeedup_56.pdf}
\end{minipage}
\vspace{-1pt}
\caption{Speedup achieved by all balanced graph coloring implementations over the CLU scheme in large real-world graphs using all cores of one socket (14 threads), all cores of two sockets (28 threads), and the maximum hardware thread capacity of our machine with hyperthreading enabled (56 threads). }
\label{balancing-speedup}
\vspace{-14pt}
\end{figure}

Figure~\ref{balancing-speedup} compares the speedup achieved by all balanced graph coloring schemes normalized to the CLU scheme in all large real-world graphs. We compare the actual kernel time that balances the vertices across color classes.

%\vspace{-14pt}

We observe that \BalColorTM{} outperforms all prior state-of-the-art balanced graph coloring schemes across all various large real-world graphs with a large number of parallel threads used. \BalColorTM{} outperforms CLU, VFF and Recoloring by on average 1.89$\times$, 1.33$\times$ and 2.06$\times$ respectively, when using 14 threads. Moreover, \BalColorTM{} outperforms CLU, VFF and Recoloring by on average 2.61$\times$, 1.05$\times$ and 1.68$\times$ respectively, when using 56 threads, i.e., the maximum hardware thread capacity of our machine. Overall, \BalColorTM{} performs best over all prior schemes in all large real-world graphs. Therefore, considering the fact that \BalColorTM{} also provides the best color balancing quality over prior schemes, we conclude that our proposed algorithmic design is a highly efficient and effective parallel graph coloring algorithm for modern mutlicore platforms.

\begin{figure}[H]
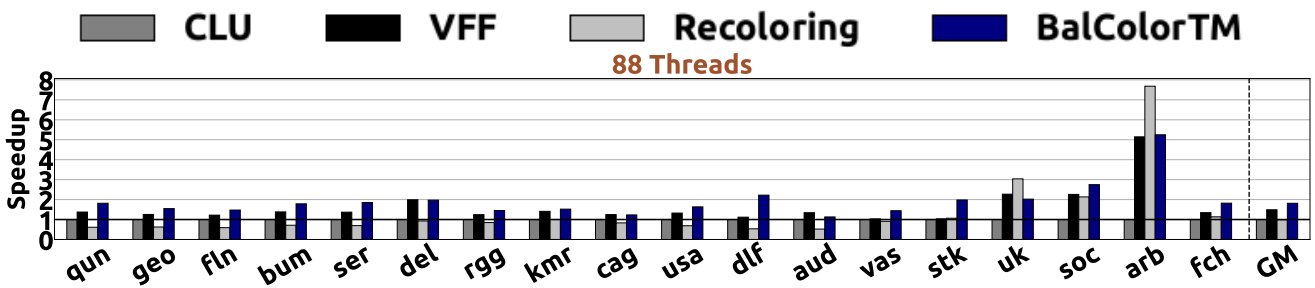

\vspace{-10pt}
\begin{minipage}{1.0\textwidth}
\centering
\includegraphics[width=0.9\columnwidth]{ sections/ColorTM/results/legend_bspeedup.png}
\end{minipage}
\begin{minipage}{1.0\textwidth}
\centering
\includegraphics[width=\columnwidth]{ sections/ColorTM/results/bspeedup_88.pdf}
\end{minipage}
\vspace{-1pt}
\caption{Speedup achieved by all balanced graph coloring implementations over the CLU scheme in large real-world graphs using the maximum hardware thread capacity of an Intel Broadwell server with hyperthreading enabled (88 threads).}
\label{balancing-speedup-broady}
%\vspace{-4pt}
\end{figure}

To confirm the performance benefits of \BalColorTM{} across multiple computing platforms, we evaluate all schemes on a 2-socket Intel Broadwell server with an Intel Xeon E5-2699 v4 processor at 2.2 GHz having 44 physical cores and 88 hardware threads. Figure~\ref{balancing-speedup-broady} compares the speedup achieved by all balanced graph coloring schemes normalized to the CLU scheme in all large real-world graphs using 88 threads, i.e., the maximum hardware thread capacity of the Intel Broadwell server. We find that \BalColorTM{} provides significant performance benefits over prior state-of-the-art graph coloring algorithms, achieving 1.82$\times$, 1.22$\times$ and 1.84$\times$ better performance over CLU, VFF, and Recoloring, respectively.

%\vspace{-14pt}

\subsubsection{Analysis of \BalColorTM{}  Execution}

\begin{figure}[H]
\vspace{-8pt}
\begin{minipage}{1.0\textwidth}
\centering
\includegraphics[width=\textwidth]{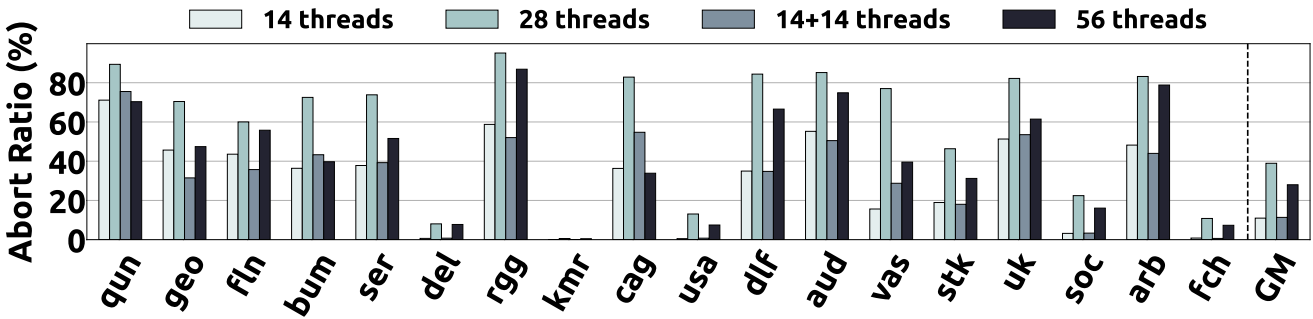}
\end{minipage}
\vspace{-2pt}
\caption{Abort ratio exhibited by \BalColorTM{} in all large real-world graphs. }
\label{balancing-abort-ratio}
\vspace{-14pt}
\end{figure}

Figure~\ref{balancing-abort-ratio} presents the abort ratio of \BalColorTM{}, i.e., the number of transactional aborts divided by the number of attempted transactions, in all real-world graphs, as the number of threads increases. In the 14-thread execution, we pin all thread on one single socket. In the 28-thread execution, we pin threads on both NUMA sockets of our machine with hyperthreading disabled. In the (14+14)-thread execution, we pin all 28 threads on the same single socket with hyperthreading enabled. In the 56-thread execution, we use the maximum hardware thread capacity of our machine.

We make two key observations. First, we observe that \BalColorTM{} on average incurs higher abort ratio over \ColorTM{}, reaching up to 80\% abort ratio in some multithreaded executions. Specifically, \BalColorTM{} incurs 68.55$\times$, 64.35$\times$, 55.83$\times$ and 25.91$\times$ higher abort ratio (averaged across all real-world graphs) over \ColorTM{}, when using 14, 28, (14+14), and 56 threads, respectively. This is because \BalColorTM{} processes and re-colors a much smaller number of vertices (a small subset of the vertices of the graph) compared to \ColorTM{}, which instead processes and colors \emph{all} the vertices of the graph. As a result, parallel threads compete for the same data and memory locations with a much higher probability in \BalColorTM{} compared to \ColorTM{}, thus incurring higher abort ratio and synchronization costs. Second, we find that in \emph{all} real-world graphs the vast majority of transactional aborts are \emph{conflict} aborts. Specifically, the portion of conflict aborts is more than 95\% in all real-world graphs for all multithreaded executions. Typically, the lower parallelization needs a parallel kernel has, the higher data contention among parallel threads it incurs. Overall, our analysis demonstrates that using a high number of parallel threads results in high contention on shared data due to low amount of parallelism of the balanced graph coloring kernel. The aforementioned high contention causes high synchronization overheads. To this end, we recommend software designers of real-world end-applications to design adaptive parallelization schemes that trade off the amount of parallelism provided for lower synchronization costs.

\subsection{Analysis of a Real-World Scenario}~\label{real-apps}
\vspace{-14pt}

In this section, we study the performance benefits of our proposed graph coloring schemes, i.e., \ColorTM{} and \BalColorTM{}, when parallelizing a widely used real-world end-application, i.e., Community Detection, via chromatic scheduling. Specifically, we compare the following parallel implementations to execute the Community Detection application:
\begin{compactitem}
\item The parallelization scheme for the Louvain method~\cite{Lu2015Parallel,Chavarria2014Scaling,Blondel2008Fast} provided by Grappolo suite~\cite{grappoloGithub}, henceforth referred to as SimpleCD, in which the vertices are processed as they appear in the input graph representation. The algorithm consists of multiple iterations. First, each vertex is placed in a community of its own. Then, multiple iterations are performed until a convergence criterion is met. Within each iteration, all vertices are processed concurrently by multiple parallel threads, and a greedy decision is made to decide whether each vertex should be moved to a different community (selected from one of its adjacent vertices) or should remain in its current community, targeting to maximize the net modularity gain. For more details, we refer the reader to~\cite{Lu2015Parallel,Ghosh2018Distributed,Naim2017Community,Halappanavar2017Scalable}.
\item The chromatic scheduling parallelization approach using \ColorTM{} to color the vertices of the graph, henceforth referred to as \ColorTM CD, in which the vertices are processed in the order they are distributed in the color classes. The end-to-end Community Detection execution can be broken down in two steps: (i) the time to color the vertices of the graph with \ColorTM{}, and (ii) the time to classify the vertices of the graph into communities via chromatic scheduling parallelization approach. The (ii) step processes the color classes produced by the (i) step sequentially, and all vertices of the same color class are processed in parallel.
\item The chromatic scheduling parallelization approach using \ColorTM{} to color the vertices of the graph and \BalColorTM{} to balance the vertices across color classes produced, henceforth referred to as \BalColorTM CD, in which the vertices are processed in the order they are distributed in the color classes. The end-to-end Community Detection execution can be broken down in three steps: (i) the time to color the vertices of the graph with \ColorTM{}, (ii) the time to balance the vertices of the graph across color classes, and (iii) the time to classify the vertices of the graph into communities via chromatic scheduling parallelization approach. The (iii) step processes the color classes produced by the (ii) step sequentially, and all vertices of the same color class are processed in parallel.
\end{compactitem}

\begin{figure}[t]
\begin{minipage}{1.0\textwidth}
\centering
\includegraphics[width=0.8\columnwidth]{ 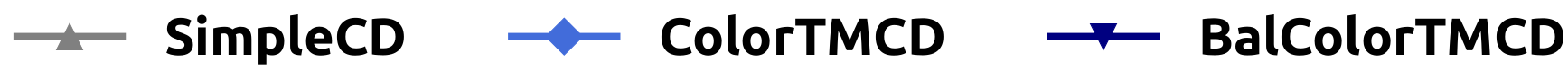}
\end{minipage}
\begin{minipage}{1.0\textwidth}
\centering
\includegraphics[scale=0.038]{ 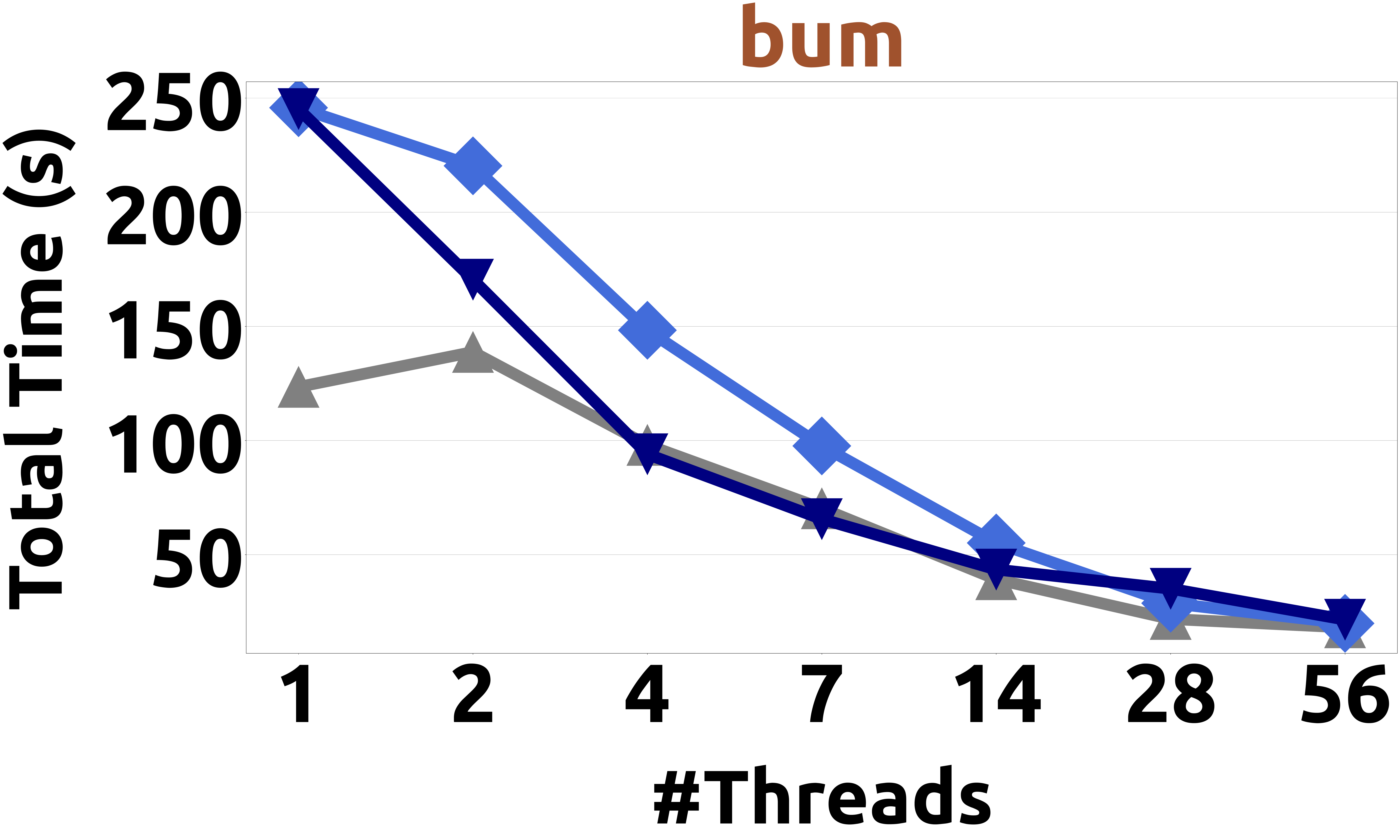}
\includegraphics[scale=0.038]{ 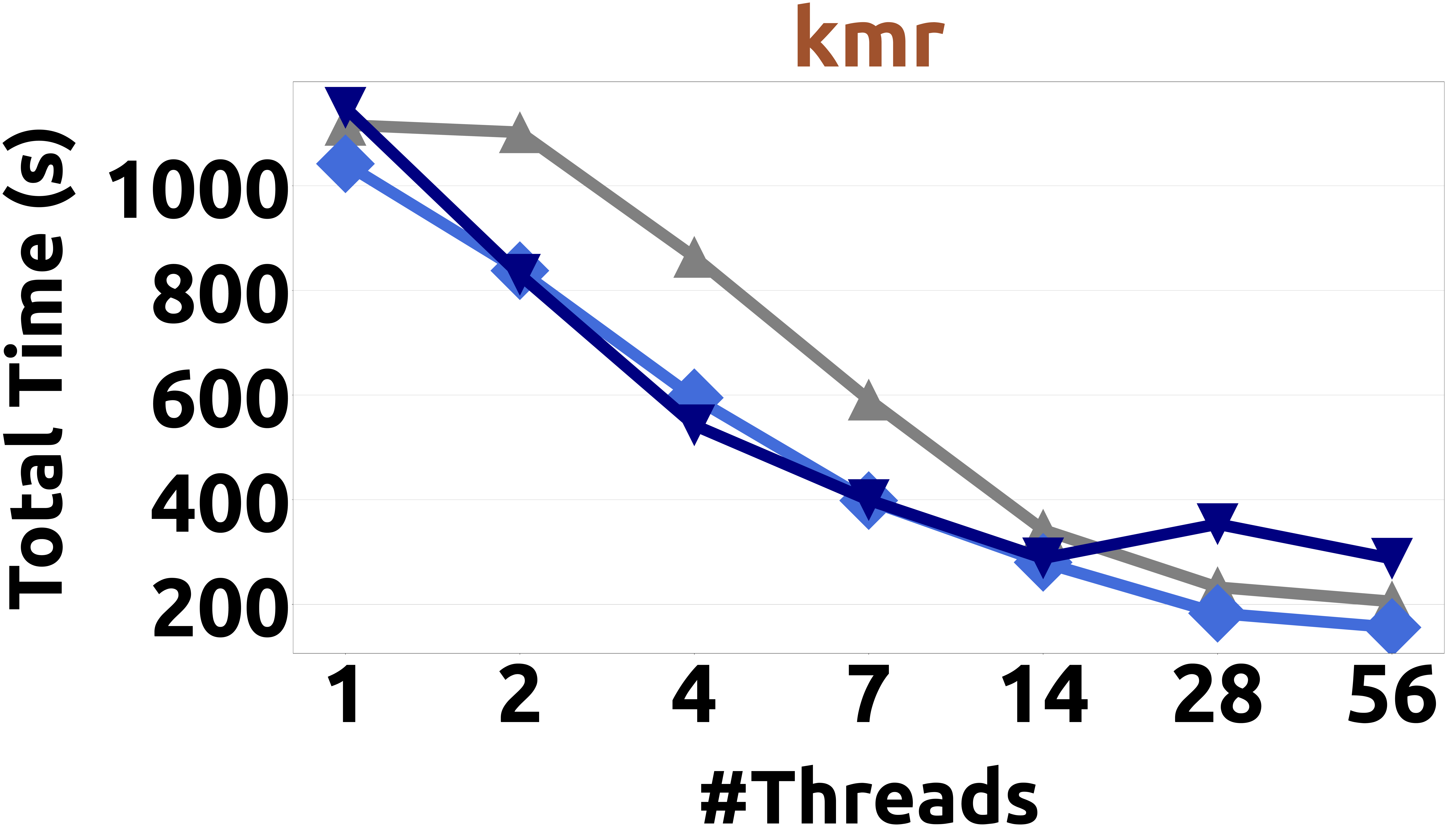}
\includegraphics[scale=0.038]{ 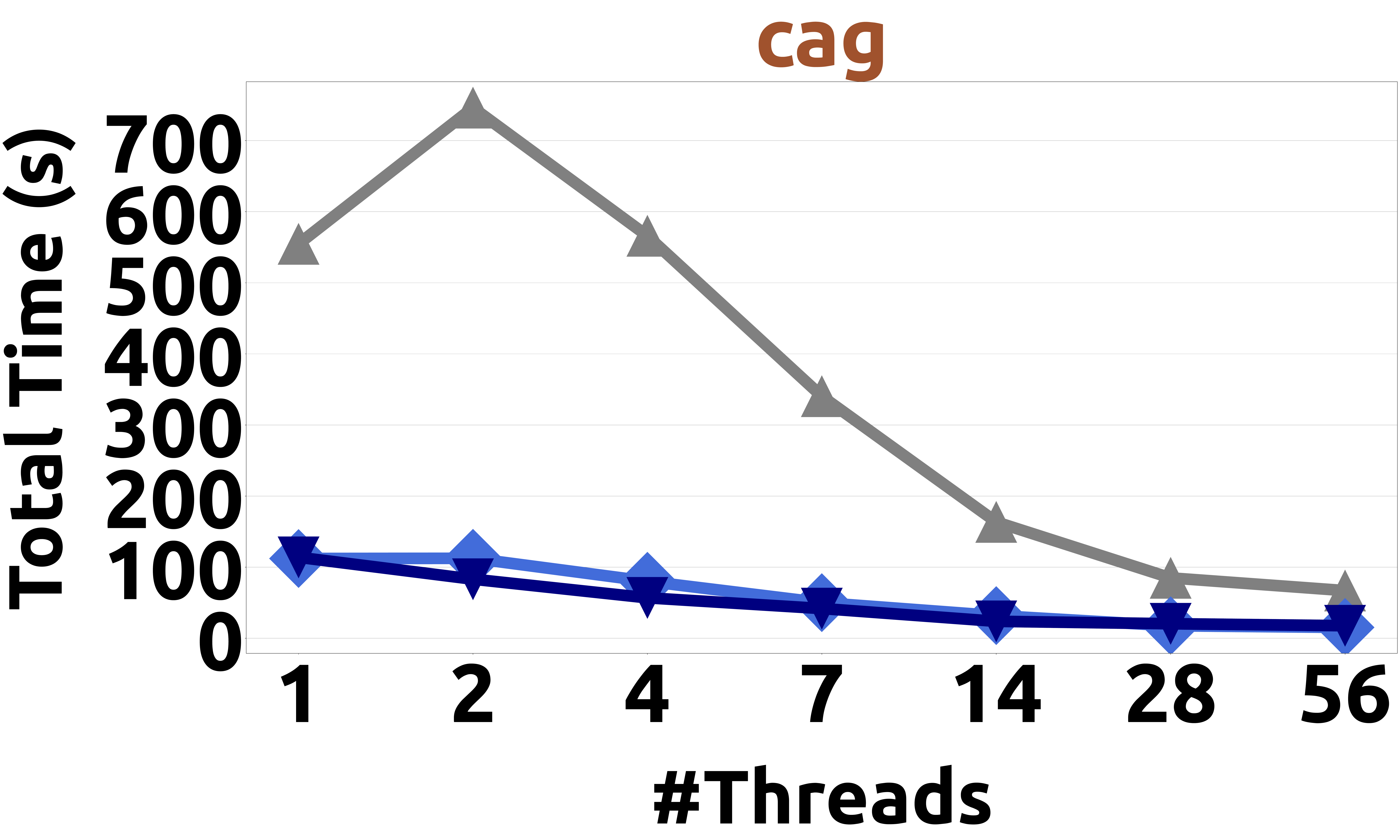}
\end{minipage}
\begin{minipage}{1.0\textwidth}
\centering
\includegraphics[scale=0.038]{ 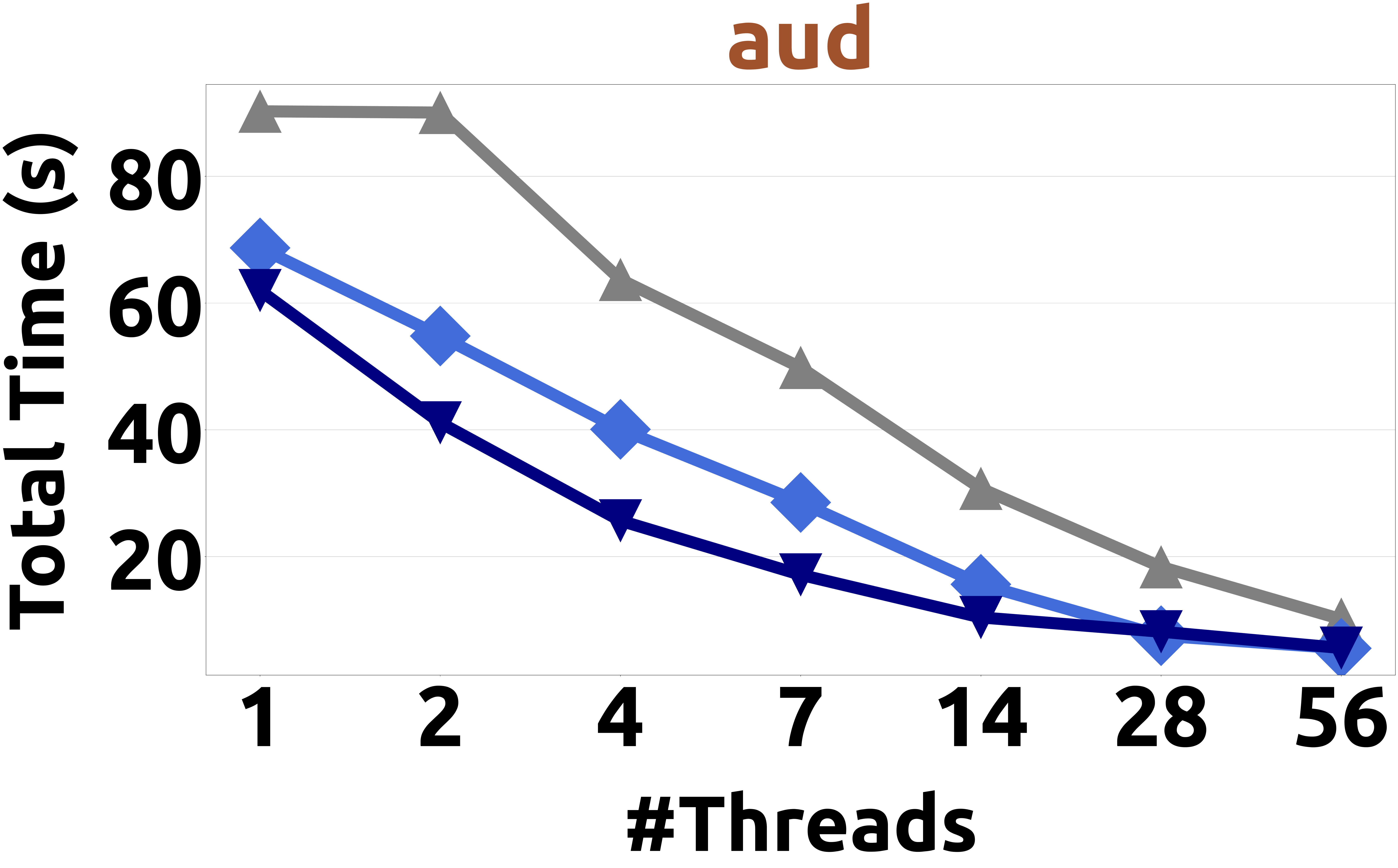}
\includegraphics[scale=0.038]{ 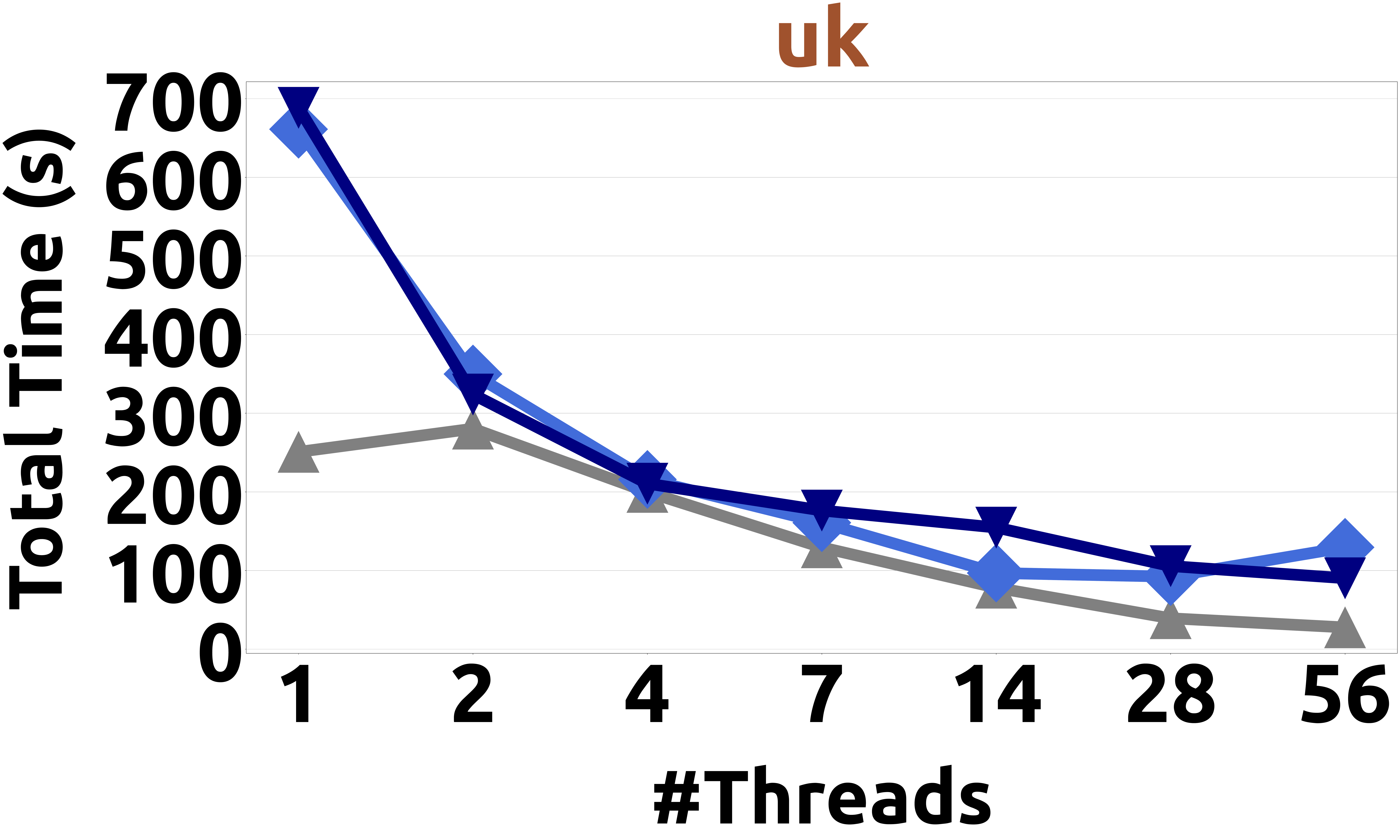}
\includegraphics[scale=0.038]{ 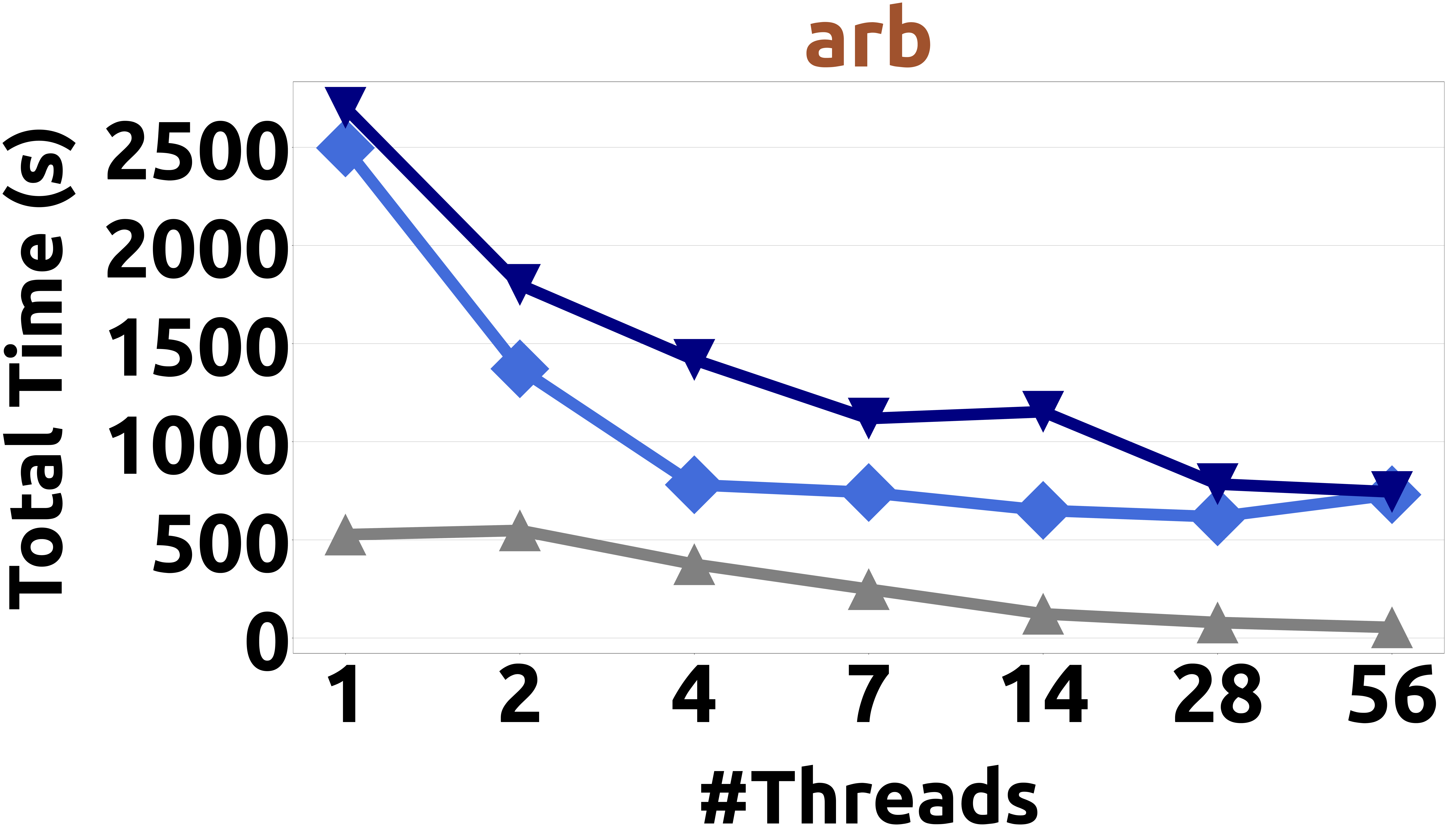}
\end{minipage}
\vspace{-2pt}
\caption{Scalability of the end-to-end Community Detection execution achieved by (i) the Grappolo~\cite{grappoloGithub} parallelization approach of the Louvain method (SimplCD) and (ii) the chromatic scheduling parallelization approach with \ColorTM{} (\ColorTM CD) and (iii) the chromatic scheduling parallelization approach with both \ColorTM{} and \BalColorTM{} (\BalColorTM CD) in large real-world graphs.}
\label{community-scalability}
\vspace{-14pt}
\end{figure}

Figure~\ref{community-scalability} evaluates the scalability of all the end-to-end Community Detection parallel implementations in a representative subset of large real-world graphs, as the number of parallel threads increases. We present the \emph{total} end-to-end execution time, i.e., in \ColorTM CD we account for the time to color the vertices of the graph (coloring step), and in \BalColorTM CD we account for the time to color the vertices of the graph (coloring step), and the time to balance the vertices across color classes (balancing step).

We draw two findings. First, we find that \ColorTM CD and \BalColorTM CD scale well in large real-world graphs. For example, when increasing the number of threads from 1 to 56, \ColorTM CD improves performance by 12.34$\times$ and 3.44$\times$ in \texttt{bum} and \texttt{arb} graphs, respectively. Similarly, when increasing the number of threads from 1 to 56, \BalColorTM CD improves performance by 11.38$\times$ and 3.63$\times$ in \texttt{bum} and \texttt{arb} graphs, respectively. However, we observe that in \texttt{uk} and \texttt{arb} graphs, SimpleCD outperforms both \ColorTM CD and \BalColorTM CD. In these two graphs, \ColorTM{} and \BalColorTM{} produce the largest number of color classes  compared to all the remaining real-world graphs (See Table~\ref{balancing-color-quality}), i.e., they produce 944 and 3248 colors for the \texttt{uk} and \texttt{arb} graphs, respectively. As a result, in \texttt{uk} and \texttt{arb} graphs the chromatic scheduling parallelization approach of \ColorTM CD and \BalColorTM CD executes 944 and 3248 times of \emph{barrier} synchronization among parallel threads, respectively, thus incurring higher synchronization costs over SimpleCD. Second, the scalability of \BalColorTM CD is affected more by the NUMA effect compared to that of \ColorTM CD. Specifically, when increasing the number of threads from 14 to 28, the performance of \ColorTM CD improves by 1.63$\times$ averaged across all real-world graphs, while the performance of \BalColorTM CD only improves by 1.22$\times$. Similarly, when increasing the number of threads from 14 to 56, the performance of \ColorTM CD improves by 1.98$\times$, while the performance of \BalColorTM CD improves by 1.50$\times$. We find that even though balancing the sizes of color classes provides higher load balance across parallel threads of real-world end-applications, it might because more remote expensive memory accesses across NUMA sockets of modern multicore machines.

\begin{figure}[t]
\begin{minipage}{1.0\textwidth}
\centering
\includegraphics[width=0.88\columnwidth]{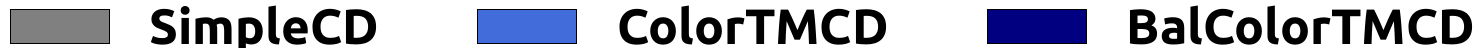}
\end{minipage}
\begin{minipage}{1.0\textwidth}
\centering
\includegraphics[width=\textwidth]{sections/ColorTM/results/speedup_communitydetection_only14.pdf}
\end{minipage}
\begin{minipage}{1.0\textwidth}
\centering
\includegraphics[width=\textwidth]{sections/ColorTM/results/speedup_communitydetection_only28.pdf}
\end{minipage}
\begin{minipage}{1.0\textwidth}
\centering
\includegraphics[width=\textwidth]{sections/ColorTM/results/speedup_communitydetection_only56.pdf}
\end{minipage}
\vspace{-2pt}
\caption{Speedup of the actual kernel of the Community Detection execution achieved by (i) SimpleCD (D), (ii) \ColorTM CD (C) and (iii) \BalColorTM CD (B)  in large real-world graphs using all cores of one socket (14 threads), all cores of two sockets (28 threads), and the maximum hardware thread capacity of our machine with hyperthreading enabled (56 threads). }
\label{community-speedup-kernel}
\vspace{-14pt}
\end{figure}

Figure~\ref{community-speedup-kernel} shows the actual kernel time (without accounting for performance overheads introduced by the coloring and balancing steps) of Community Detection by comparing the speedup of \ColorTM CD and \BalColorTM CD over SimpleCD in all our evaluated large real-world graphs.

%\vspace{-10pt}

We draw two key findings. First, \BalColorTM{} can on average outperform \ColorTM{}, when considering only the actual kernel time of Community Detection, by providing better load balance among parallel threads. When only the actual kernel time of Community Detection is considered (excluding the performance overheads introduced by the coloring and balancing steps), \BalColorTM CD on average outperforms \ColorTM CD by 1.27$\times$, 1.01$\times$ and 1.12$\times$ when using 14, 28, and 56 threads, respectively. Second, parallelizing the Community Detection using \ColorTM{}  and \BalColorTM{} provides significant performance speedups over SimpleCD, the state-of-the-art paralellization approach of Louvain method of Community Detection~\cite{Lu2015Parallel,Chavarria2014Scaling,Blondel2008Fast,grappoloGithub}. Specifically, \ColorTM CD improves the performance of the actual kernel time of Community Detection compared to SimpleCD by 1.40$\times$, 1.34$\times$, and 1.20$\times$, when using 14, 28, and 56 threads, respectively. In addition, \BalColorTM CD improves the performance of the actual kernel time of Community Detection compared to SimpleCD by 1.77$\times$, 1.34$\times$, and 1.34$\times$, when using 14, 28, and 56 threads, respectively. We conclude that our proposed graph coloring algorithmic designs can provide high performance benefits in real-world end-applications which are parallelized using coloring.

\begin{figure}[t]
\begin{minipage}{1.0\textwidth}
\centering
\includegraphics[width=0.88\columnwidth]{ 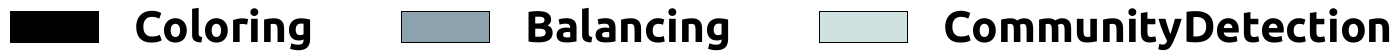}
\end{minipage}
\begin{minipage}{1.0\textwidth}
\centering
\includegraphics[width=\textwidth]{ sections/ColorTM/results/speedup_communitydetection_breakdown14.pdf}
\end{minipage}
\begin{minipage}{1.0\textwidth}
\centering
\includegraphics[width=\textwidth]{ sections/ColorTM/results/speedup_communitydetection_breakdown28.pdf}
\end{minipage}
\begin{minipage}{1.0\textwidth}
\centering
\includegraphics[width=\textwidth]{ sections/ColorTM/results/speedup_communitydetection_breakdown56.pdf}
\end{minipage}
\vspace{-2pt}
\caption{Speedup breakdown of the end-to-end Community Detection execution achieved by (i) SimpleCD (D), (ii) \ColorTM CD (C) and (iii) \BalColorTM CD (B)  in large real-world graphs using all cores of one socket (14 threads), all cores of two sockets (28 threads), and the maximum hardware thread capacity of our machine with hyperthreading enabled (56 threads). }
\label{community-speedup}
\vspace{-10pt}
\end{figure}

Figure~\ref{community-speedup} presents the speedup breakdown of \ColorTM CD and \BalColorTM CD over SimpleCD in all our evaluated large real-world graphs. The performance is broken down in three steps: (i) the coloring step to color the vertices of the graph (\textbf{Coloring}), (ii) the balancing step to balance the vertices across color classes (\textbf{Balancing}), and (iii) the actual Community Detection kernel time (\textbf{CommunityDetection}).

We make two key observations. First, \BalColorTM CD on average outperforms \ColorTM CD when using up to 14 threads (using one single NUMA socket). When considering the end-to-end execution including the performance overheads introduced by the coloring and balancing steps, \BalColorTM CD outperforms \ColorTM CD by 1.19$\times$ when using 14 threads, while it performs on average 1.18$\times$ and 1.10$\times$ worse over \ColorTM CD, when using 28 and 56 threads, respectively. We find that the performance overhead introduced in the balancing step of \BalColorTM CD is not compensated in the runtime of the actual kernel time of Community Detection when using both NUMA sockets of our machine. Second, we observe that both \ColorTM CD and \BalColorTM CD can provide high performance in Community Detection. \ColorTM CD on average outperforms SimpleCD by 1.38$\times$, 1.33$\times$ and 1.19$\times$, when using 14, 28 and 56 threads, respectively. \BalColorTM CD on average outperforms SimpleCD by 1.64$\times$, 1.10$\times$ and 1.08$\times$, when using 14, 28 and 56 threads, respectively. In addition, we observe that \BalColorTM CD provides significant performance speedups over Simple CD in many graphs such as \texttt{fln}, \texttt{del}, \texttt{cag}, \texttt{aud}, \texttt{soc} and \texttt{fch}, reaching up to 10.36$\times$ with 56 threads. Overall, we conclude that our proposed parallel graph coloring algorithms can provide significant performance improvements in real-world end-applications, e.g., parallelizing Community Detection with chromatic scheduling, across a wide variety of input data sets with diverse characteristics.

\section{Recommendations} 

This section presents our key takeaways in the form of recommendations for software and hardware designers.

\noindent\textbf{Recommendation \#1.} \textit{Optimize the Hardware Transactional Memory implementation on NUMA multicore systems.} Figures~\ref{coloring-abort-ratio} and ~\ref{balancing-abort-ratio} demonstrate the number of transactional aborts significantly increases when using both NUMA sockets of our machine. Accessing data of remote NUMA sockets within HTM transactions increases the duration of the transactions, thus potentially causing transactional aborts: long-running HTM transactions increase the probability of incurring read-write conflicts among them, while they might suffer from time interrupt aborts when the OS scheduler schedules out the software threads from the hardware threads. Overall, we find that current HTM implementations are severely limited by the NUMA effect~\cite{Brown2016Investigating}, which degrades the benefits of HTM on synchronization among parallel threads. To this end, we suggest that hardware designers of multicore systems provide a NUMA-aware HTM implementation for modern multicore systems.

\noindent\textbf{Recommendation \#2.} \textit{Design intelligent data partitioning techniques of real-world graphs across NUMA sockets of modern systems.} Figure~\ref{coloring-abort-breakdown} shows that the number of \emph{conflicts} (read-write) aborts among running HTM transactions significantly increases when using both sockets of our evaluated machine. This is because expensive accesses to remote data increase the duration of the HTM transactions, and thus the probability of causing conflicts aborts among long-running transactions becomes very high. Thus, we conclude that the performance of parallel algorithms might significantly degrade when accessing application data from remote NUMA sockets within the critical section. Therefore, we recommend that software designers of parallel graph processing kernels design effective data partitioning techniques of real-world graphs across NUMA sockets of modern systems to minimize contention and synchronization overheads among parallel threads.

\noindent\textbf{Recommendation \#3.}
\textit{Design adaptive parallel applications that on-the-fly adjust the number of parallel threads used to parallelize their sub-kernels based on the parallelization needs of each particular sub-kernel.} Figure~\ref{coloring-scalability} shows that all parallel graph coloring schemes scale up to 56 threads, i.e., all available hardware threads of our machine. However, Figure~\ref{balancing-scalability} shows that  balanced graph coloring schemes typically scale up 14 threads, thus achieving the best performance with 14 parallel threads, while their performance degrades when using all available hardware threads of our machine (56 threads). The graph coloring kernel has high parallelization needs, since all the vertices of the large real-world graph need to be processed (colored) by parallel threads. Instead, the balance coloring kernel has lower parallelization needs, since typically a small subset of the vertices of the graph need to be processed (re-colored) by parallel threads. We demonstrate in Section~\ref{real-apps} that the execution times of the graph coloring and balance coloring kernels add to the overall execution time of the real-world end-application. Thus, we conclude that to achieve high system performance in the end-to-end execution of real-world applications, we need to dynamically tune the number of parallel threads used to parallelize the sub-kernels of the end-applications depending on the parallelization needs of each particular sub-kernel. To this end, we recommend that software designers provide adaptive parallel applications that on-the-fly adjust the number of parallel threads used to parallelize each sub-kernel of the end-applications based on the parallelization needs of the particular sub-kernel.
\section{Related Work}

A handful of prior works~\cite{Welsh1967Upper,Gebremedhin2000Scalable,Boman2005Scalable,Catalyurek2012GraphColoring,Rokos2015Fast,Lu2015Balanced,Tas2017Greed,Maciej2020GC,Hasenplaugh2014Ordering,Jones1993Parallel,Deveci2016Parallel,Tas2017Greed,Giannoula2018Combining} has examined the graph coloring kernel in modern multicore platforms. Welsh and Powell~\cite{Welsh1967Upper} propose the original sequential Greedy algorithm that colors the vertices of the graph using the \emph{first-fit} heuristic. %, i.e., for each uncolored vertex selecting the minimum available color to provide graph coloring with a small number of colors. 
Recent prior works~\cite{Gebremedhin2000Scalable,Boman2005Scalable,Catalyurek2012GraphColoring,Rokos2015Fast} parallelize Greedy by proposing the SeqSolve, IterSolve and IterSolveR schemes described in Section~\ref{sec:unbalanced}. We compare \ColorTM{} with these prior schemes in Section~\ref{eval:unbalanced}, and demonstrate that our proposed \ColorTM{} outperforms these state-of-the-art schemes across a wide variety of real-world graphs. Jones and Plassmann~\cite{Jones1993Parallel} design an algorithm, named JP, that colors the vertices of the graph by identifying independent sets of vertices: in each iteration, the algorithm finds and selects an independent set of vertices that can be colored concurrently. However, JP is a recursive algorithm that typically runs longer than the original Greedy~\cite{Giannoula2018Combining,Hasenplaugh2014Ordering,Maciej2020GC}, since it performs more computations and needs more synchronization points, i.e., parallel threads need to synchronize at each iteration of processing independent sets of vertices. Moreover, the original paper~\cite{Jones1993Parallel} shows that JP provides good performance mostly in $\mathcal{O}(1)$-degree graphs. In contrast, our work efficiently parallelizes the original and widely used Greedy algorithm for graph coloring, and our proposed parallel algorithms achieve significant performance improvements across a wide variety of real-world graphs and using a large number of parallel threads. 

Deveci et al.~\cite{Deveci2016Parallel} present an edge-centric parallelization scheme for graph coloring which is better suited for GPUs. \ColorTM{} and \BalColorTM{} can be straightforwardly extended to color the vertices of a graph by equally distributing the edges of the graph among parallel threads. We leave the exploration of edge-centric graph coloring schemes for future work. Future work also comprises the experimentation of the graph coloring kernel on multicore computing platforms such as modern GPUs~\cite{Grosset2011Evaluating,Osama2019Graph,Chen2017Efficient,Che2015Graph} and Processing-In-Memory systems~\cite{Giannoula2022SparsePPomacs,Giannoula2022SparsePSigmetrics,Giannoula2021SynCron,fernandez2020natsa,Gomez2021Analysis,Gomez2022Benchmarking,Gao2015Practical,ahn2015scalable,Nai2017GraphPIM,Youwei2019GraphQ}. Maciej et al.~\cite{Maciej2020GC} and Hasenplaugh et al.~\cite{Hasenplaugh2014Ordering} propose new vertex \emph{ordering} heuristics for graph coloring. Ordering heuristics define the order in which Greedy colors the vertices of the graph in order to improve the coloring quality by minimizing the number of colors used. Instead, our work aims to improve system performance by proposing efficient parallelization schemes. For a fair comparison, we employ the \emph{first-fit} ordering heuristic (the vertices of the graph are colored in the order they appear in the input graph representation) in all parallel algorithms evaluated in Sections~\ref{eval:unbalanced} and ~\ref{eval:unbalanced}. \ColorTM{} and \BalColorTM{} can support various ordering heuristics~\cite{Maciej2020GC,Hasenplaugh2014Ordering,Coleman1983Estimation,Arkin1987Scheduling,Marx2004GRAPHCP,brelaz1979,Matula1983Smallest,Karp1985Fast,Luby1985,Goldberg1987Parallel,Alabandi2020Increasing} by assigning the vertices of the graph to parallel threads with a particular order. We leave the evaluation of various vertex ordering heuristics for future work.

Lu et al.~\cite{Lu2015Balanced} design \emph{balanced} graph coloring algorithms to efficiently balance the vertices across the color classes. We compare \BalColorTM{} with their proposed algorithms, i.e., CLU, VFF, Recoloring, in Section~\ref{sec:balanced}, and demonstrate that our proposed \BalColorTM{} scheme on average performs best across all large real-world graphs. Tas et al.~\cite{Tas2017Greed} propose balanced graph coloring algorithms for bitpartie graphs, i.e., graphs whose vertices can be divided into two disjoint and independent sets $U$ and $V$, and every edge $(u,v)$ either connects a vertex from $U$ to $V$ or a vertex from $V$ to $U$. In contrast, \ColorTM{} and \BalColorTM{} are designed to be general, and efficiently color any \emph{arbitrary} real-world graph using a large number of parallel threads. In addition, Tas et al.~\cite{Tas2017Greed} also explore the distance-2 graph coloring kernel on multicore architectures, in which any two vertices $u$ and $v$ with an edge-distance at most 2 are assigned with different colors. Instead, our work efficiently parallelizes the distance-1 graph coloring kernel on multicore platforms, in which any two adjacent vertices of the graph connected with a \emph{direct} edge are assigned with different colors. Finally, prior works propose algorithms for edge coloring~\cite{Holyer1981}, dynamic or streaming coloring~\cite{Sallinen2016High,Yuan2017,Bossek2019Runtime,Barba2019Dynamic,bhattacharya2018dynamic,solomon2020improved}, k-distance coloring~\cite{bozda2010Distributed,Bozdag2005APD} and sequential exact coloring~\cite{Jinkun2017Reduction,Verma2015,Hebrard2019Hybrid}. All these works are not closely related to our work, since we focus on designing high-performance parallel algorithms for the distance-1 vertex graph coloring kernel.
\section{Summary}

In this work, we explore the graph coloring kernel on multicore platforms, and propose \ColorTM{} and \BalColorTM{}, two novel algorithmic designs for high performance and balanced graph coloring on modern computing platforms. \ColorTM{} and \BalColorTM{} achieve high system performance through two key techniques: (i) \emph{eager} conflict detection and resolution of the coloring inconsistencies that arise when adjacent vertices are concurrently processed by different parallel threads, and (ii) \emph{speculative} computation and synchronization among parallel threads by leveraging Hardware Transactional Memory. Via the eager coloring conflict detection and resolution policy, \ColorTM{} and \BalColorTM{} effectively leverage the deep memory hierarchy of modern multicore  platforms and minimize access costs to application data. Via the speculative computation and synchronization approach, \ColorTM{} and \BalColorTM{} minimize synchronization costs among parallel threads and provide high amount of parallelism. Our evaluations demonstrate that our proposed parallel graph coloring algorithms outperform prior state-of-the-art approaches across a wide range of large real-world graphs. \ColorTM{} and \BalColorTM{} can also provide significant performance improvements in real-world scenarios. We conclude that \ColorTM{} and \BalColorTM{} are highly efficient graph coloring algorithms for modern multicore systems, and hope that this work encourages further studies of the graph coloring kernel in modern computing platforms.

%SmartPQ
\lstset{style=mystyle2}
\chapter{\smartpq{}}\label{SmartPQChapter}
\section{Overview}

Concurrent data structures are widely used in the software stack, i.e., kernel, libraries and applications. Prior works~\cite{ffwd,blackbox,numask,adaptivepq} discuss the need for efficient and scalable concurrent data structures for commodity Non-Uniform Memory Access (NUMA) architectures. Pointer chasing data structures such as linked lists, skip lists and search trees have inherently low-contention, since their operations need to de-reference a non-constant number of pointers before completing. Recent works~\cite{blackbox,ascylib,Siakavaras2021RCUHTM} have shown that lock-free algorithms ~\cite{Harris-ll,Michael-ll, fraser, Ellen-bst, Howley-bst, Natarajan-bst} of such data structures can scale to hundreds of threads. On the other hand, data structures such as queues and stacks typically incur high-contention, when accessed by many threads. In these data structures, concurrent threads compete for the \emph{same} memory locations, incurring excessive traffic and non-uniform memory accesses between nodes of a NUMA system.

In this work, we focus on priority queues, which are widely used in a variety of applications, including task scheduling in real-time and computing systems~\cite{Xu}, discrete event simulations~\cite{Tang,Marotta} and graph applications~\cite{Kolmogorov,Lasalle,Thorup}, e.g., Single Source Shortest Path~\cite{CLRS} and Minimum Spanning Tree~\cite{Prim}. Similarly to skip-lists and search trees, in \insrt{} operation, concurrent priority queues typically have high levels of parallelism and low-contention, since threads may work on different parts of the data structure. Therefore, concurrent \notnuma{} implementations~\cite{lotan_shavit, linden_jonsson, sagonas, sundell, Wimmer_Martin, rihani, Brodal, Zhang} can scale up to a high number of threads. In contrast, in \delete{} operation, \emph{all} threads compete for deleting the highest-priority element of the queue, thus competing for the \emph{same} memory locations (similarly to queues and stacks), and creating a contention spot. In \delete{}-dominated workloads, concurrent priority queues typically incur high-contention and low parallelism. To achieve higher parallelism, \emph{relaxed} priority queues have been proposed in the literature~\cite{spraylist,Heidarshenas2020Snug}, in which \delete{} operation returns an element among the \emph{first few} (high-priority) elements of the priority queue. However, such \notnuma{} implementations are still inefficient in NUMA architectures, as we demonstrate in Section~\ref{sec:experimental}. Therefore, to improve performance in NUMA systems, \numa{} implementations have been proposed~\cite{blackbox,ffwd}.

\begin{figure}[t]
    \centering
    \includegraphics[scale=0.23]{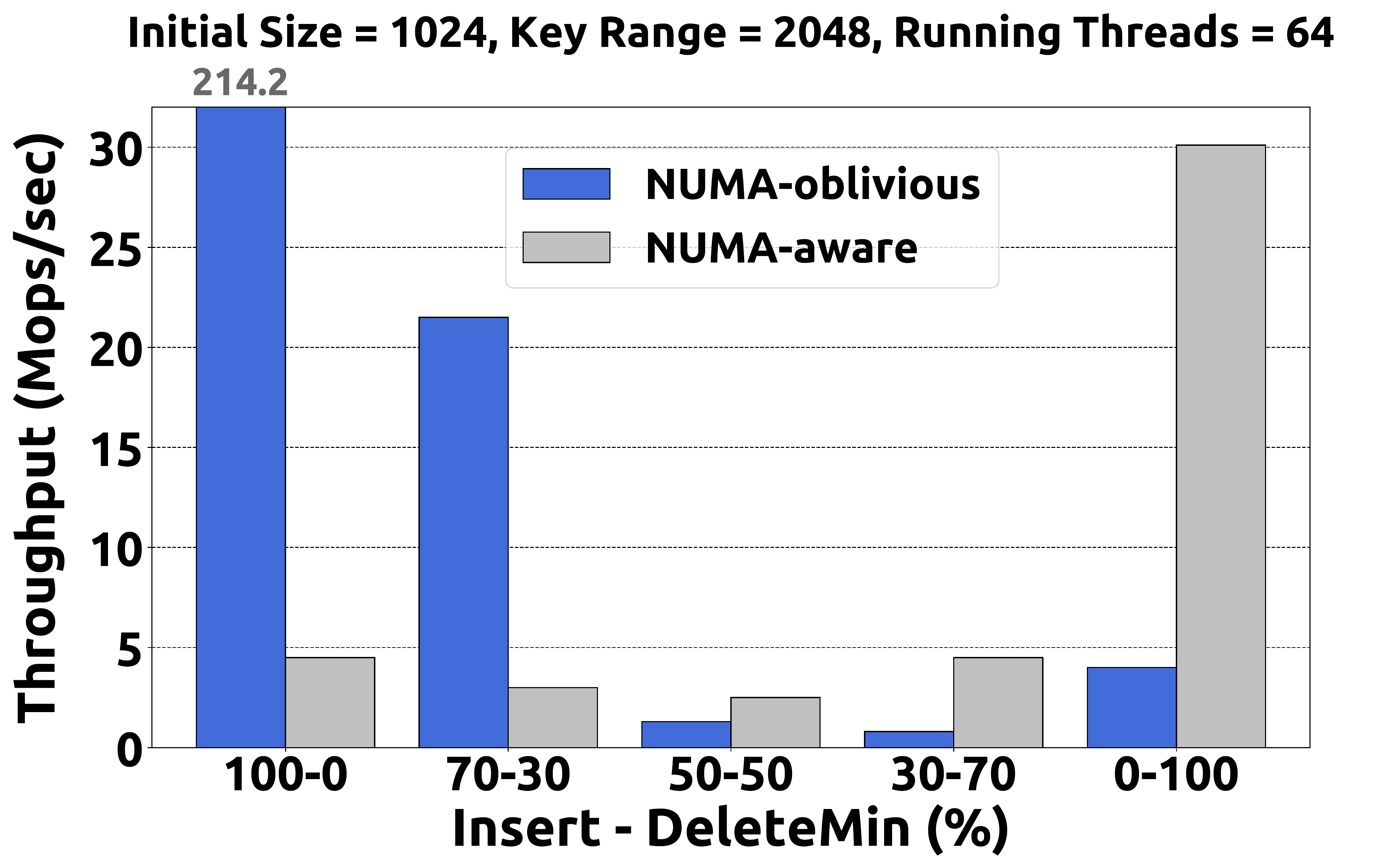}
    \caption{Throughput achieved by a \notnuma{}~\cite{spraylist,herlihy} and a \numa{}~\cite{ffwd} priority queue, both initialized with 1024 keys. We use 64 threads that perform a mix of \insrt{} and \delete{} operations in parallel, and the key range is set to 2048 keys. We use \emph{all} NUMA nodes of a 4-node NUMA system, the characteristics of which are presented in Section~\ref{sec:experimental}. }
    \label{fig:motivation}
    \vspace{-8pt}
\end{figure}

We examine \numa{} and \notnuma{} concurrent priority queues with a wide variety of contention scenarios in NUMA architectures, and find that the performance of a priority queue implementation is becoming increasingly dependent on both the contention levels of the workload and the underlying computing platform. This is illustrated in Figure~\ref{fig:motivation}, which shows the throughput achieved by a \notnuma{} and a \numa{} priority queue using a 4-node NUMA system. Even though in a \insrt-dominated scenario, e.g., when having 100\% \insrt{} operations, the \notnuma{} implementation achieves significant performance gains over the \numa{} one, when contention increases, i.e., the percentage of \delete{} operations increases, the \notnuma{} implementation incurs non-negligible performance slowdowns over the \numa{} priority queue. We conclude that none of the priority queues performs best across \emph{all} contention workloads.

Our \textbf{goal} in this work is to design a concurrent priority queue that (i) achieves \emph{the highest performance under all various contention scenarios}, and (ii) performs best even when the contention of the workload \emph{varies} over time.

To this end, our contribution is twofold. First, we introduce \textit{NUMA Node Delegation} (\nuddle{}), a generic technique to obtain \numa{} data structures, by effectively transforming \emph{any} concurrent \notnuma{} data structure into the corresponding \numa{} implementation. In other words, \nuddle{} is a framework to wrap any \emph{concurrent} \notnuma{} data structure and transform it into an efficient \numa{} one. \nuddle{} extends \ffwd~\cite{ffwd} by enabling multiple server threads, instead of only one, to execute operations in parallel on behalf of client threads. In contrast to \ffwd, which aims to provide single threaded data structure performance, \nuddle{} targets data structures which are able to scale up to a number of threads such as priority queues.

Second, we propose \smartpq, an adaptive concurrent priority queue that achieves the \emph{highest} performance under \emph{all} contention workloads and \emph{dynamically} adapts itself over time between a \notnuma{} and a \numa{} \algomode{} mode. \smartpq{} integrates (i) \nuddle{} to \emph{efficiently} switch between the two \algomode{} modes with \emph{very low} overhead, and (ii) a simple decision tree \emph{classifier}, which predicts the best-performing \algomode{} mode given the expected contention levels of a workload.

\begin{figure}[t]
    \centering
    \includegraphics[scale=0.78]{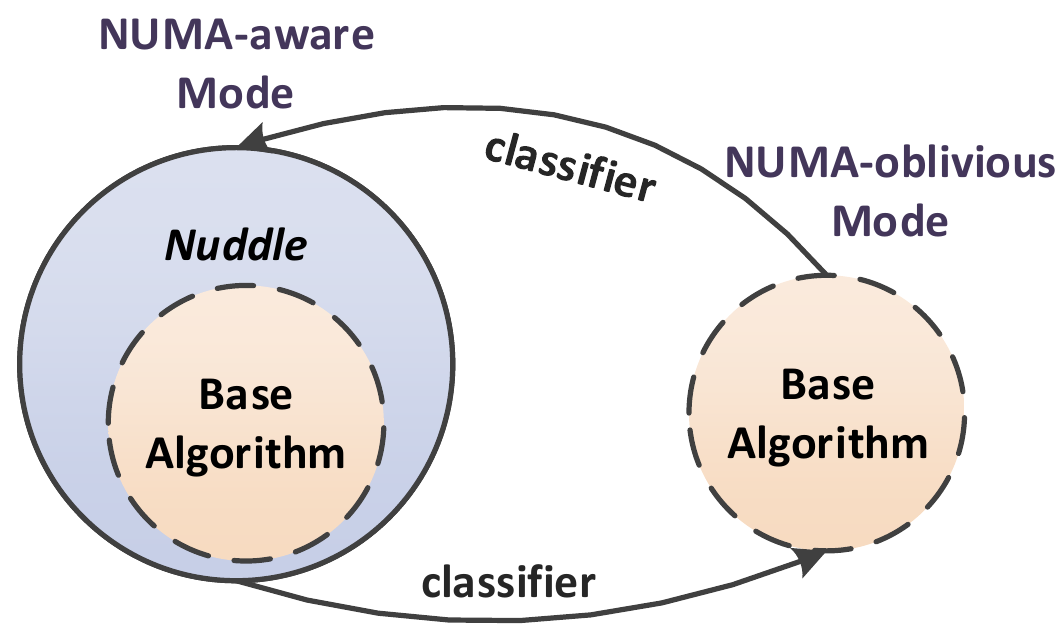}
    \caption{High-level overview of \smartpq. \smartpq{} dynamically adapts its algorithm to the contention levels of the workload based on the prediction of a simple classifier.}
\label{fig:smartpq}
\vspace{-10pt}
\end{figure}

Figure~\ref{fig:smartpq} presents  an overview of \smartpq, where we use the term \textit{base algorithm} to denote \emph{any} arbitrary concurrent \notnuma{} data structure. \smartpq{} relies on three key ideas. First, client threads can execute operations using either \nuddle{} (\numa{} mode) or its underlying \notnuma{} base algorithm (\notnuma{} mode). Second, \smartpq{} incorporates a decision-making mechanism to decide upon transitions between the two modes. Third, \smartpq{} exploits the fact that the actual underlying implementation of \nuddle{} is a \emph{concurrent} \notnuma{} data structure. Client threads in both \algomode{} modes access the data structure in the same way, i.e.,  with no actual change in the way data is accessed. Therefore, \smartpq{} switches from one mode to another with \emph{no} synchronization points between transitions.

We evaluate a wide range of contention scenarios and compare \nuddle{} and \smartpq{} with state-of-the-art \notnuma{}~\cite{spraylist,lotan_shavit} and \numa{}~\cite{ffwd} concurrent priority queues. We also evaluate \smartpq{} using synthetic benchmarks that \emph{dynamically} vary their contention workload over time. Our evaluation shows that \smartpq{} adapts between its two \algomode{} modes with negligible performance overheads, and achieves the highest performance in \emph{all} contention workloads and at \emph{any} point in time with 87.9\% success rate.

The main \textbf{contributions} of this work are:
\begin{itemize}%[$\textbf{--}$]
\vspace{-10pt}
\setlength\itemsep{-2pt}
    \item We propose \nuddle{}, a generic technique to obtain \numa{} concurrent data structures.
    \item We design a simple classifier to predict the best-performing implementation among \notnuma{} and \numa{} priority queues given the contention levels of a workload.
    \item We propose \smartpq, an adaptive concurrent priority queue that achieves the highest performance, even when contention varies over time.
    \item We evaluate \nuddle{} and \smartpq{} with a wide variety of contention scenarios, and demonstrate that \smartpq{} performs best over prior state-of-the-art concurrent priority queues.
\end{itemize}

\section{NUMA Node Delegation (\nuddle)}
\label{sec:nuddle}

\subsection{Overview}
NUMA Node Delegation (\nuddle) is a generic technique to obtain \numa{} data structures by automatically transforming \emph{any} concurrent \notnuma{} data structure into an efficient \numa{} implementation. \nuddle{} extends \ffwd~\cite{ffwd}, a client-server software mechanism which is based on the delegation technique~\cite{Calciu2013Message,Klaftenegger2014Delegation,Lozi2012Remote,Petrovic2015Performance,Suleman2009Accelerating}.

\begin{figure}[t]
    \centering
    \hspace{-8pt}\includegraphics[scale=0.54]{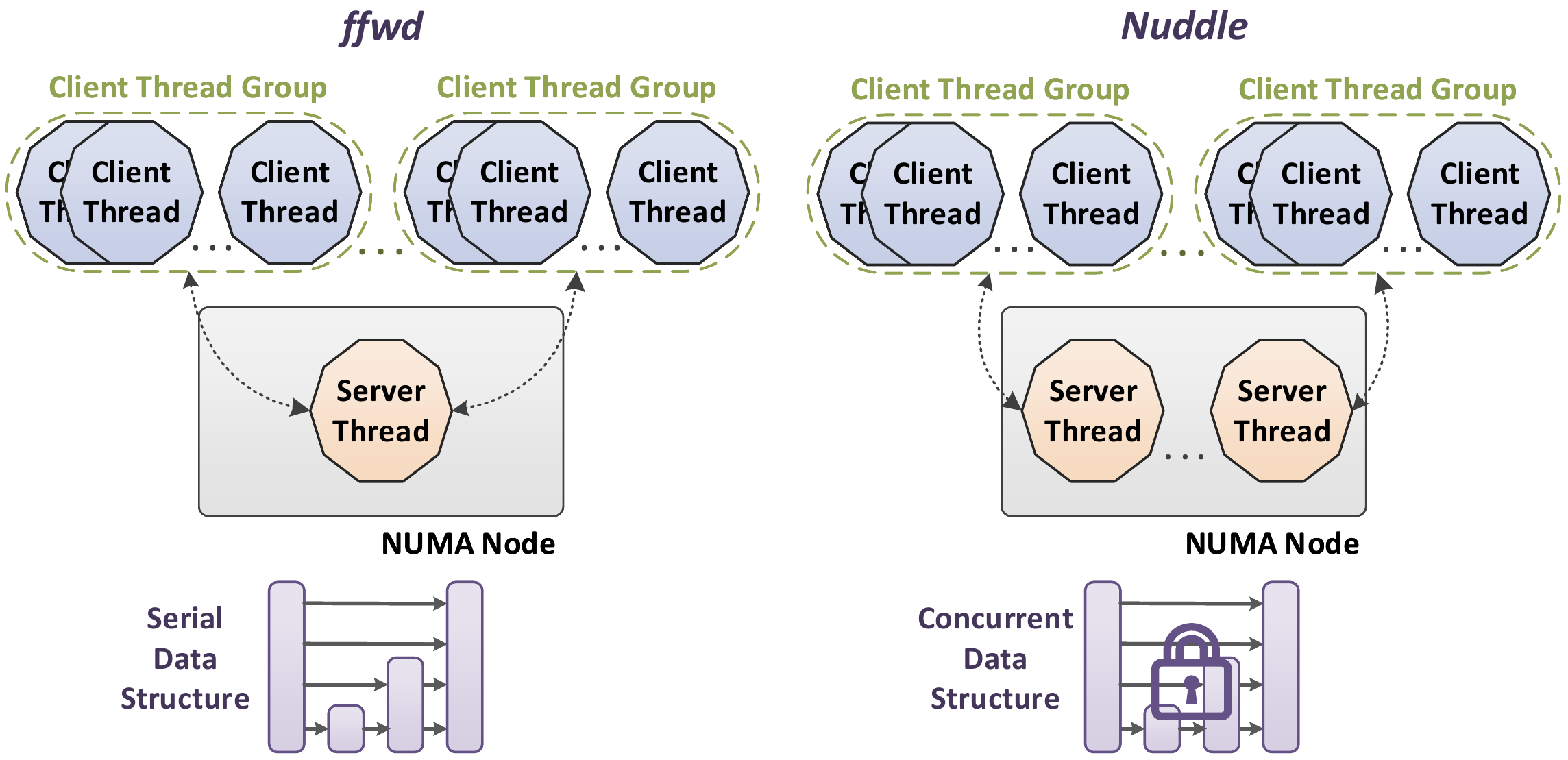}
    \caption{High-level design of \ffwd~\cite{ffwd} and \nuddle{}. \nuddle{} locates \emph{all} server threads at the \emph{same} NUMA node to design a \numa{} scheme, and associates each of them to multiple client thread groups. \nuddle{} uses the communication protocol proposed in \ffwd~\cite{ffwd}.}
    \label{fig:nuddle}
    \vspace{-10pt}
\end{figure}

Figure~\ref{fig:nuddle} left shows the high-level overview of \ffwd, which has three key design characteristics. First, \emph{all} operations performed by multiple client threads are \emph{delegated} to one \emph{single} dedicated thread, called \textit{server} thread.  The server thread performs operations in the data structure on behalf of its client threads. This way, the data structure remains in the memory hierarchy of a \emph{single} NUMA node, avoiding non-uniform memory accesses to remote data. Second, \ffwd{} eliminates the need for synchronization, since the shared data structure is no longer accessed by multiple threads: only a single server thread directly modifies the data structure, and therefore, \ffwd{} uses a \emph{serial asynchronized} implementation of the underlying data structure. Third, \ffwd{} provides an efficient communication protocol between the server thread and client threads that minimizes cache coherence overheads. Specifically, \ffwd{} reserves dedicated cache lines to exchange request and response messages between the client threads and sever thread. Multiple client threads are grouped together to minimize the response messages from the server thread: one response cache line is shared among multiple client threads belonging to the same client thread group. For more details, we refer the reader to the original paper~\cite{ffwd}.

Figure~\ref{fig:nuddle} right presents the high-level overview of \nuddle, which is based on three key ideas. First, \nuddle{} deploys \emph{multiple} servers to perform operations on behalf of multiple client threads. Specifically, client threads are grouped in client thread groups, and each sever thread serves multiple client thread groups. This way, multiple server threads \emph{concurrently} perform operations on the data structure, achieving high levels of parallelism up to a number of server threads. Second, \nuddle{} locates all server threads to the \emph{same} NUMA node to keep the data structure in the memory hierarchy of one \emph{single} NUMA node, and propose a \numa{} approach. Client threads can be located at any  NUMA node. Third, since multiple servers can \emph{concurrently} update the \emph{shared} data structure, \nuddle{} uses a \emph{concurrent} \notnuma{} implementation (i.e., which includes synchronization primitives when accessing the shared data) of the underlying data structure to ensure correctness. Third, \nuddle{} employs the same client-server communication protocol with \ffwd to carefully manage memory accesses and minimize cache coherence traffic and latency.

\ffwd{} targets inherently serial data structures, whose concurrent performance cannot be better than that of \emph{single} threaded performance. In contrast, \nuddle{} targets data structures that can scale up to a number of concurrent threads. Priority queue is a typical example of such a data structure. In \insrt{} operation, priority queue can scale up to multiple threads, which can \emph{concurrently} update the shared data. In contrast, \delete{} operation is inherently serial: at each time only \emph{one} thread can update the shared data, since \emph{all} threads compete for the highest-priority element of the queue. However, as we mentioned, in relaxed priority queues (e.g., SprayList~\cite{spraylist}), even \delete{} operation can be parallelized to some extent.

\subsection{Implementation Details}

Figures~\ref{alg:structs},~\ref{alg:init} and~\ref{alg:functions} present the code of a priority queue implementation using \nuddle{}. We denote with red color the core operations of the \emph{base algorithm}, which is used as the underlying concurrent \notnuma{} implementation of \nuddle. Note that even though in this work we focus on priority queues, \nuddle{} is a \emph{generic} framework for any type of concurrent data structure.

\textbf{Helper Structures.} \nuddle{} includes three \emph{helper} structures (Figure~\ref{alg:structs}), which are needed for client-server communication. First, the main structure of \nuddle, called  {\fontfamily{lmtt}\selectfont \textit{struct nuddle\_pq}}, wraps the \emph{base algorithm} ({\fontfamily{lmtt}\selectfont \textit{nm\_oblv\_set}}), and includes a few additional fields, which are used to associate client  thread groups to server threads in the initialization step. Second, each client thread has its own {\fontfamily{lmtt}\selectfont \textit{struct client}} structure with a dedicated request and a dedicated response cache line. The request cache line is exclusively written by the client thread and read by the associated server thread, while the response cache line is exclusively written by the server thread and read by all client threads that belong to the same client thread group. Third, each server thread has its own {\fontfamily{lmtt}\selectfont \textit{struct server}} structure that includes an array of requests ({\fontfamily{lmtt}\selectfont \textit{my\_clients}}), each of them is shared with a client thread, and an array of responses ({\fontfamily{lmtt}\selectfont \textit{my\_responses}}), each of them is shared with all client threads of the \emph{same} client thread group.

%putting basicstyle=\scriptsize it becomes smaller
\begin{figure}[t]
\begin{lstlisting}[language=C]
#define cache_line_size 128
typedef char cache_line[cache_line_size];

struct nuddle_pq {              
  (*\textcolor{darkred}{nm\_oblv\_set}*) *base_pq;      
  int servers, groups, clnt_per_group; 
  int server_cnt, clients_cnt, group_cnt;   
  cache_line *requests[groups][clnt_per_group]; 
  cache_line *responses[groups];     
  lock *global_lock;                
};

struct client {
  cache_line *request, *response;   
  int clnt_pos;
};

struct server {        
  (*\textcolor{darkred}{nm\_oblv\_set}*) *base_pq;
  cache_line *my_clients[], *my_responses[];
  int my_groups, clnt_per_group;  
};
\end{lstlisting}
\vspace{-2pt}
\caption{Helper structures of \nuddle{}.}
\label{alg:structs}
\vspace{-10pt}
\end{figure}

\textbf{Initialization Step.} Figure~\ref{alg:init} describes the initialization functions of \nuddle. {\fontfamily{lmtt}\selectfont \textit{initPQ()}} initializes (i) the underlying data structure using the corresponding function of the \emph{base algorithm} (line 25), and (ii) the additional fields  of {\fontfamily{lmtt}\selectfont \textit{struct nuddle\_pq}}. For this function, programmers need to specify the number of server threads and the maximum number of client threads to properly allocate cache lines needed for communication among them. Programmers also specify the size of the client thread group (line 27), which is typically 7 or 15, if the cache line is 64 or 128 bytes, respectively. As explained in \ffwd~\cite{ffwd}, assuming 8-byte return values, a dedicated 64-byte (or 128-byte) response cache line can be shared between up to 7 (or 15) client threads, because it also has to include one additional toggle bit for each client  thread. After initializing {\fontfamily{lmtt}\selectfont \textit{struct nuddle\_pq}}, each running thread calls either {\fontfamily{lmtt}\selectfont \textit{initClient()}} or {\fontfamily{lmtt}\selectfont \textit{initServer()}} depending on its role. Each thread initializes its own helper structure ({\fontfamily{lmtt}\selectfont \textit{struct client}} or {\fontfamily{lmtt}\selectfont \textit{struct server}}) with request and response cache lines of the corresponding shared arrays of {\fontfamily{lmtt}\selectfont \textit{struct nuddle\_pq}}. Server threads undertake client thread groups with a round-robin fashion, such that the load associated with client threads is balanced among them. In function {\fontfamily{lmtt}\selectfont \textit{initServer()}}, it is the programmer's responsibility to properly pin software server threads to hardware contexts (line 56), such that  server threads are located in the \emph{same} NUMA node, and the programmer fully benefits from the \nuddle{} technique. Moreover, given that client threads of the \emph{same} client thread group share the same response cache line, the programmer could pin client threads of the \emph{same} client thread group to hardware contexts of the \emph{same} NUMA node to minimize cache coherence overheads. Finally, since the request and response arrays of {\fontfamily{lmtt}\selectfont \textit{struct nuddle\_pq}} are \emph{shared} between all threads, a global lock is used when updating them to ensure mutual exclusion.

\begin{figure}[!ht]
\begin{lstlisting}[language=C]
struct nuddle_pq *initPQ(int servers, int max_clients) {
  struct nuddle_pq *pq = allocate_nuddle_pq();
  (*\textcolor{darkred}{\_\_base\_init(pq->base\_pq);}*)
  pq->servers = servers;
  pq->clnt_per_group = client_group(cache_line_size);
  pq->groups = (max_clients +
    pq->clnt_per_group-1) / pq->clnt_per_group;
  pq->server_cnt = 0;
  pq->client_cnt = 0;
  pq->group_cnt = 0;
  pq->requests = malloc(groups * clnt_per_group);
  pq->responses = malloc(groups);
  init_lock(pq->global_lock);
  return pq;
}

struct client *initClient(struct nuddle_pq *pq) {
  struct client *cl = allocate_client();
  acquire_lock(pq->global_lock);
  cl->request = &(pq->requests[group_cnt][clients_cnt]);
  cl->response = &(pq->responses[group_cnt]);
  cl->pos = pq->client_cnt;
  pq->client_cnt++;
  if (pq->client_cnt % pq->clnt_per_group == 0) {
    pq->clients_cnt = 0;
    pq->group_cnt++;
  }
  release_lock(pq->global_lock);
  return cl;
}

struct server *initServer(struct nuddle_pq *pq, int core)
{
  set_affinity(core);
  struct server *srv = allocate_server();
  srv->base_pq = pq->base_pq;
  srv->my_groups = 0;
  srv->clnt_per_group = pq->clnt_per_group;
  acquire_lock(pq->global_lock);
  int j = 0;
  for(i = 0; i < pq->groups; i++)
    if(i % pq->servers == pq->server_cnt) {
      srv->my_clients[j] = pq->requests[i][0..gr_clnt];
      srv->my_responses[j++] = pq->responses[i];
      srv->my_groups++;
     }
  pq->server_cnt++;
  release_lock(pq->global_lock);
  return srv;
}
\end{lstlisting}
\vspace{-4pt}
\caption{Initialization functions of \nuddle{}.}
\label{alg:init}
\vspace{-12pt}
\end{figure}

\textbf{Main API.} Figure~\ref{alg:functions} shows the core functions of \nuddle, where we omit the corresponding functions for \delete{} operation, since they are very similar to that of \insrt{} operation. Both \insrt{} and \delete{} operations of \nuddle{} have similar API with the classic API of prior state-of-the-art priority queue implementations~\cite{spraylist,lotan_shavit,sagonas,linden_jonsson}. However, we separate the corresponding functions for client threads and server threads. A client thread writes its request to a dedicated request cache line (line 75) and then waits for the server thread's response. In contrast, a server thread directly executes operations in the data structure using the core functions of the \emph{base algorithm} (line 82). Moreover, a server thread can serve client threads using the {\fontfamily{lmtt}\selectfont\textit{serve\_requests()}} function. A server thread iterates over its own client thread groups and executes the requested operations in the data structure. The server thread buffers individual return values for clients to a local cache line ({\fontfamily{lmtt}\selectfont \textit{resp}} in lines 92 and 94) until it finishes processing all requests for the current client thread group. Then, it writes all responses to the \emph{shared} response cache line of that client thread group (line 96), and proceeds to its next client thread group.

\begin{figure}[H]
\begin{lstlisting}[language=C]
int insert_client(struct client *cl, int key, (*\textcolor{blue}{\textbf{\scriptsize{int64\_t}}}*) value)
{
  cl->request = write_req("insert", key, value);
  while (cl->response[cl->pos] == 0) ;
  return cl->response[cl->pos];
}

int insert_server(struct server *srv, int key, (*\textcolor{blue}{\textbf{\scriptsize{int64\_t}}}*) value)
{
  return (*\textcolor{darkred}{\_\_base\_insrt}*)(srv->base_pq, key, value);
}

void serve_requests(struct server *srv) {
  for(i = 0; i < srv->mygroups; i++) {
    cache_line resp;
    for(j = 0; j < srv->clnt_per_group; j++) {
      key = srv->my_clients[i][j].key;
      value = srv->my_clients[i][j].value;
      if (srv->my_clients[i][j].op == "insert")
        resp[j] = (*\textcolor{darkred}{\_\_base\_insrt}*)(srv->base_pq, key, value);  
      else if (srv->my_clients[i][j].op == "deleteMin")
        resp[j] = (*\textcolor{darkred}{\_\_base\_delMin}*)(srv->base_pq);
    }
    srv->my_responses[i] = resp;
  }
}
\end{lstlisting}
\vspace{-2pt}
\caption{Functions used by server threads and client threads to perform operations using \nuddle{}.}
\label{alg:functions}
\vspace{-10pt}
\end{figure}

\section{\smartpq} \label{SmartPQbl}

We propose \smartpq, an adaptive concurrent priority queue which tunes itself by \emph{dynamically} switching between \notnuma{} and \numa{} \algomode{} modes, in order to perform best in \emph{all} contention workloads and at \emph{any} point in time, even when contention varies over time.

Designing an adaptive priority queue involves addressing two major challenges: (i) how to switch from one \algomode{} mode to the other with \emph{low overhead}, and (ii) \emph{when} to switch from one \algomode{} mode to the other.

To address the first challenge, we exploit the fact that the actual underlying implementation of \nuddle{} is a \emph{concurrent} \notnuma{} implementation. We select \nuddle{}, as the \numa{} \algomode{} mode of \smartpq{}, and its underlying \emph{base algorithm}, as the \notnuma{} \algomode{} mode of \smartpq. Threads can perform operations in the data structure using either \nuddle{} or its underlying \emph{base algorithm}, with \emph{no actual change} in the way data is accessed. As a result, \smartpq{} can switch between the two \algomode{} modes \emph{without} needing a synchronization point between transitions, and without violating correctness.

To address the second challenge, we design a simple decision tree classifier (Section~\ref{sec:classifier}), and train it to select the best-performing \algomode{} mode between \nuddle{}, as the \numa{} \algomode{} mode of \smartpq, and its underlying \emph{base algorithm}, as the \notnuma{} mode of \smartpq. Finally, we add a \emph{lightweight} decision-making mechanism in \smartpq{} (Section~\ref{sec:smartpq_details}) to dynamically tune itself over time between the two \algomode{} modes. We describe more details in next sections.

\subsection{Selecting the Algorithmic Mode}

\subsubsection{The Need for a Machine Learning Approach}

\begin{figure}[H]
%\centering
    \centering\captionsetup[subfloat]{labelfont=bf}
  \begin{subfigure}[h]{0.48\textwidth}
    \centering
    \includegraphics[scale=0.2]{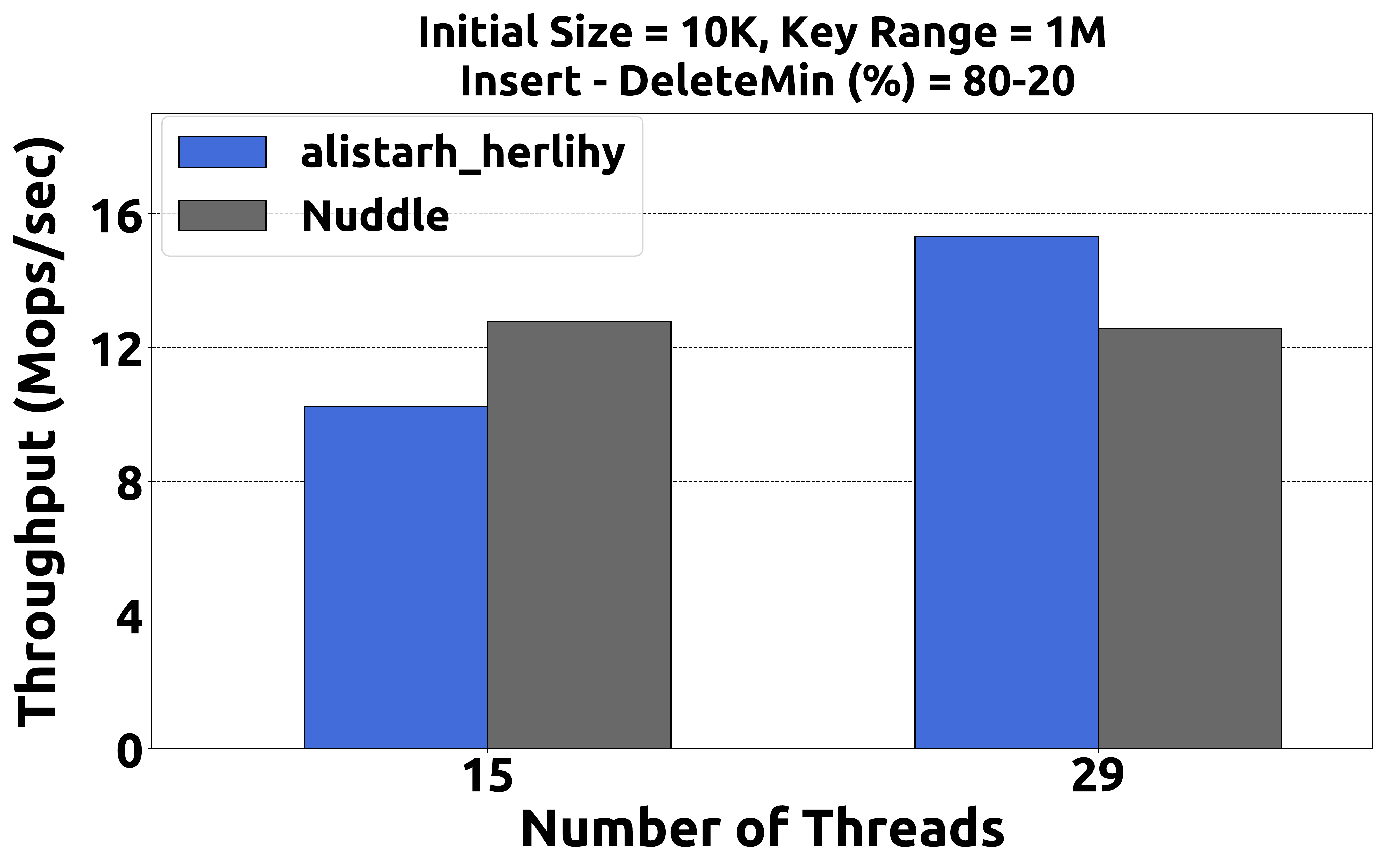}
    \vspace{-6pt}
    \caption{Varying the number of threads.} 
    %\label{fig:coherence} 
 \end{subfigure}
  ~
  \begin{subfigure}[h]{0.48\textwidth}
   \centering
    \includegraphics[scale=0.2]{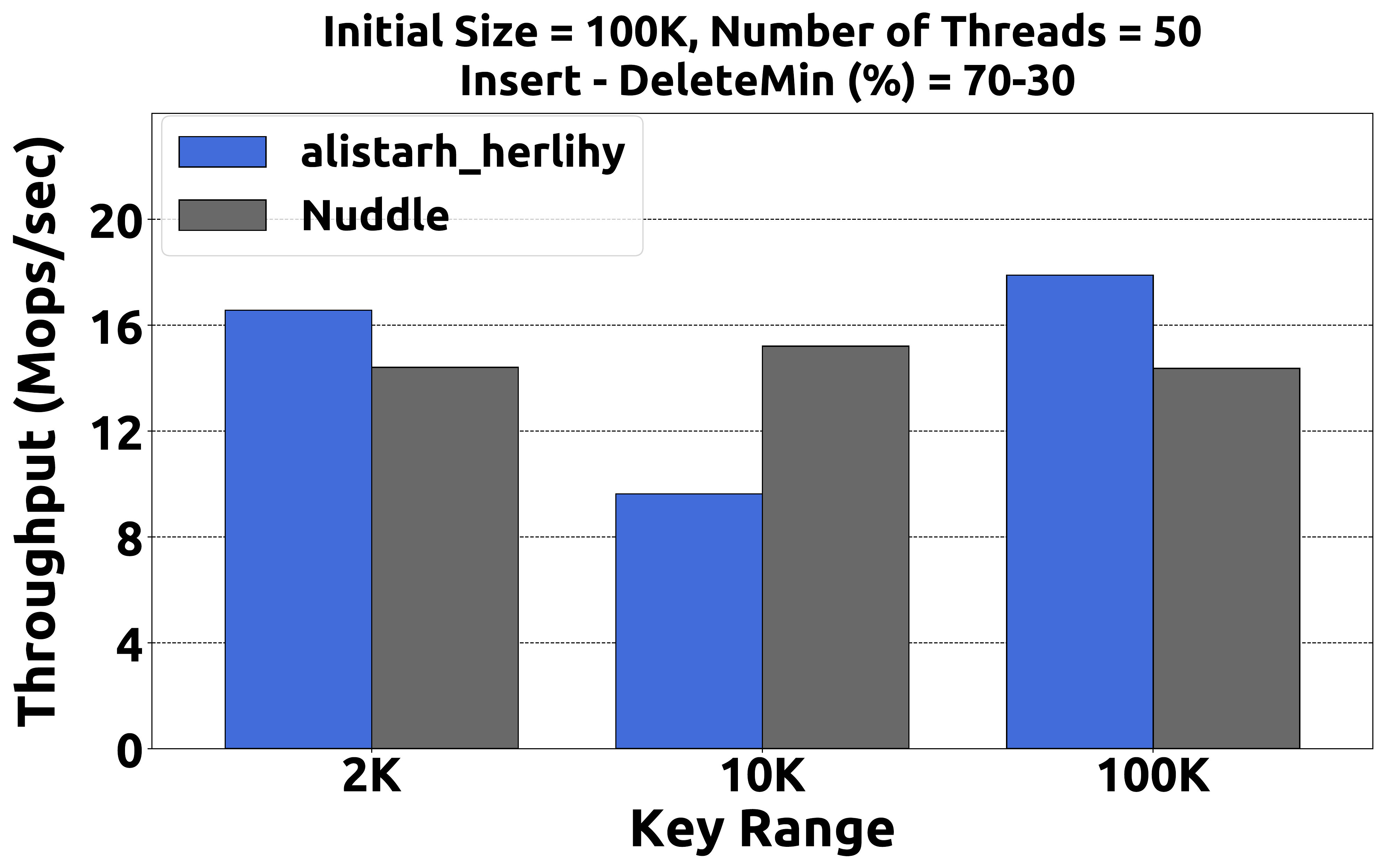}
    \vspace{-6pt}
    \caption{Varying the key range.}
    %\label{fig:numaness}
  \end{subfigure} 
  %\vspace{-6pt}
  \caption{Throughput achieved by \nuddle{} (using 8 server threads) and its underlying \notnuma{} \emph{base algorithm}, i.e., \emph{alistarh\_herlihy}~\cite{spraylist,herlihy}, when we vary (a) the number of threads that perform operations in the shared data structure, and (b) the key range of the workload.}
  \label{fig:nuddle_spraylist}
  \vspace{-8pt}
\end{figure}

Selecting the best-performing algorithmic mode can be solved in various ways. For instance, one could take an empirical exhaustive approach: measure the throughput achieved by the two algorithmic modes for all various contention scenarios on the target NUMA system, and then use the \algomode{} mode that achieves the highest throughput on future runs of the same contention workload on the target NUMA system. Even though this is the most accurate method, it (i) incurs \emph{substantial} overhead and effort to sweep over \emph{all} various contention workloads, and (ii) would need a large amount of memory to store the best-performing \algomode{} mode for all various scenarios. Furthermore, it is not trivial to construct a statistical model to predict the best-performing \algomode{} mode, since the performance of an algorithm is also affected by the characteristics of the underlying computing platform. Figure~\ref{fig:nuddle_spraylist} summarizes these observations by comparing \nuddle{}  with its underlying \emph{base algorithm} in a 4-node NUMA system. For the \emph{base algorithm}, we use \emph{alistarh\_herlihy} priority queue~\cite{spraylist,herlihy}, since this is the \notnuma{} implementation that achieves the highest performance, according to our evaluation (Section~\ref{sec:experimental}).

Figure~\ref{fig:nuddle_spraylist}a demonstrates that the best-performing \algomode{} mode depends on multiple parameters, such as the number of threads that perform operations in the shared data structure, the size of the data structure, the operation workload, i.e., the percentage of \insrt/\delete{} operations. Specifically, when the number of threads increases, we may expect that the performance of the \notnuma{} \emph{alistarh\_herlihy} degrades due to higher contention. In contrast, with 80\% \insrt{} operations when increasing the number of threads to 29, \emph{alistarh\_herlihy} outperforms \nuddle. This is because the size of the priority queue and the range of keys used in the workload are relatively large, while the percentage of \delete{} operations is low. In this scenario, threads may not compete for the same elements, working on different parts of the data structure, and thus, the \notnuma{} \emph{alistarh\_herlihy} achieves higher throughput compared to the \numa{} \nuddle.

Figure~\ref{fig:nuddle_spraylist}b demonstrates that the best-performing \algomode{} mode cannot be straightforwardly predicted, and also depends on the characteristics of underlying hardware~\cite{Strati2019AnAdaptive}. In \insrt-dominated workloads, as the key range increases, threads may update different parts of the shared data structure. We might, thus, expect that after a certain point of increasing the key range, the \notnuma{} \emph{alistarh\_herlihy} will always outperform \nuddle, since the contention decreases. However, we note that, even though the performance of \nuddle{} remains constant, as expected, the performance of \emph{alistarh\_herlihy} highly varies as the key range increases due to the hyperthreading effect. When using more than 32 threads, hyperthreading is enabled in our NUMA system (Section~\ref{sec:experimental}). The hyperthreading pair of threads shares the L1 and L2 caches, and thus, these threads may either thrash or benefit from each other depending on the characteristics of L1 and L2 caches (e.g., size, eviction policy), and the elements accessed in each operation.

Considering the aforementioned non-straightforward behavior, we resort to a machine learning approach as the basis of our prediction mechanism.

\subsubsection{Decision Tree Classifier}\label{sec:classifier}
We formulate the selection of the \algomode{} mode as a classification problem, and leverage supervised learning techniques to train a simple classifier to predict the best-performing \algomode{} mode for each contention workload. For our classifier, we select decision trees, since they are commonly used in classification models for multithreaded workloads~\cite{athena,Doddipalli2012Ensemble,Polat2009ANovel,Meng2019APattern,Dhulipala2020APattern,Sloan2012Algorithmic,Sedaghati2015Automatic,Benatia2016SparseMF}, and incur low training and inference overhead. Moreover, they are easy to interpret and thus, be incorporated to our proposed priority queue (Section~\ref{sec:smartpq_details}). We generate the decision tree classifier using the scikit-learn machine learning toolkit~\cite{scikit}.

\textbf{\textit{1) Class Definition:}} 
We define the following classes: (a) the \textbf{\notnuma} class that stands for the \notnuma{} \algomode{} mode, (b) the \textbf{\numa{}} class that stands for the \numa{} \algomode{} mode, and (c) the \textbf{\textit{neutral}} class that stands for a tie, meaning that either a \numa{} or a \notnuma{} implementation can be selected, since they achieve similar performance. We include a neutral class for two reasons: (i) when using \emph{only one} socket of a NUMA system, \numa{} implementations deliver similar throughput with \notnuma{} implementations, and (ii) in an adaptive data structure, which \emph{dynamically} switches between the two \algomode{} modes, we want to configure a transition from one \algomode{} mode to another to occur when the difference in their throughput is relatively high, i.e., greater than a certain threshold. Otherwise, the adaptive data structure might continuously oscillate between the two modes, without delivering significant performance improvements or even causing performance degradation.

\textbf{\textit{2) Extracted Features}}: Table~\ref{table:features} explains the four features  of the contention workload which are used in our classifier  targeting priority queues. We assume that the contention workload is known a priori, and thus, we can easily extract the features needed for classification. Section~\ref{sec:discussion} discusses how to on-the-fly extract these features.

\begin{table}[t]
\centering
%\ra{0.9}
\resizebox{0.7\linewidth}{!}{
\begin{tabular}{l l} 
 \toprule
 Feature & Definition  \\ [0.5ex] \midrule \midrule
% \hline\hline
 \multirow{2}{*}{\shortstack[l]{\#Threads}} & The number of active threads \\
  & that perform operations in the data structure \\
 Size & The current size of the priority queue\\
% key\_range & The range of keys used in requested \\
% & operations\\
Key\_range & The range of keys used in the workload \\
 \multirow{1}{*}{\shortstack[l]{\% \insrt /\delete}} & The percentage of \insrt /\delete{} operations \\
 \bottomrule
\end{tabular}
}
\caption{The features of the contention workload which are used for classification.}
\label{table:features}
\vspace{-8pt}
\end{table}

\textbf{\textit{3) Generation of Training Data:}}
To train our classifier, we develop microbenchmarks, in which threads repeatedly execute random operations on the priority queue for 5 seconds. We select \nuddle{}, as the \numa{} implementation, and \emph{alistarh\_herlihy}, as its underlying \notnuma{} implementation, since this is the best-performing \notnuma{} priority queue (Section~\ref{sec:experimental}). We use a variety of values for the features needed for classification (Table~\ref{table:features}). Our training data set consists of 5525 different contention workloads. Finally, we pin software threads to hardware contexts of the evaluated NUMA system in a round-robin fashion, and thus, the classifier is trained with this thread placement. We leave the exploration of the thread placement policy for future work.

\textbf{\textit{4) Labeling of Training Data:}}
Regarding the labeling of our training data set, we set the threshold for tie between the two \algomode{} modes to an empirical value of 1.5 Million operations per second. When the difference in throughput between the two \algomode{} modes is less than this threshold, the \textit{neutral} class is selected as label. Otherwise, we select the class that corresponds to the \algomode{} mode that achieves the highest throughput.

The final decision tree classifier has only 180 nodes, half of which are leaves. It has a very low depth of 8, that is the length of the longest path in the tree, and thus, a \emph{very low} traversal cost (2-4 ms in our evaluated NUMA system).

\subsection{Implementation Details} \label{sec:smartpq_details}

\begin{figure}{t}
\hspace{8pt}
\begin{subfigure}{1\linewidth}
\begin{lstlisting}[language=C]
struct smartpq {              
  (*\textcolor{darkred}{nm\_oblv\_set}*) *base_pq;      
  int servers, groups, clnt_per_group; 
  int server_cnt, clients_cnt, group_cnt;   
  cache_line *requests[groups][clnt_per_group]; 
  cache_line *responses[groups];     
  lock *global_lock;    
  (*\textcolor{mygreen}{int *algo;}*) (*\textcolor{mygreen}{// 1: NUMA-oblivious (default), 2: NUMA-aware}*)
};

struct client {
  (*\textcolor{darkred}{nm\_oblv\_set}*) (*\textcolor{mygreen}{*base\_pq;}*)
  (*\textcolor{mygreen}{int *algo;}*)  
  cache_line *request, *response;   
  int clnt_pos;             
};

struct client *initClient(struct smartpq *pq) { 
  ... lines 40-49 of Fig. 5 ...
  (*\textcolor{mygreen}{cl->base\_pq = pq->base\_pq;}*)
  (*\textcolor{mygreen}{cl->algo = \&(pq->algo);}*)
  release_lock(pq->global_lock);  
  return cl; 
}

int insert_client(struct client *cl, int key, float value) {
  (*\textcolor{mygreen}{if(*(cl->algo) == 1) \{ }*)       
    return (*\textcolor{darkred}{\_\_base\_insert(cl->base\_pq,key,value);}*)
  (*\textcolor{mygreen}{\} else \{ // *(cl->algo) == 2}*)
    ... lines 75-77 of Fig. 6 ...
  (*\textcolor{mygreen}{\}}*)                         
}   

void serve_requests(struct server *srv) {
  (*\textcolor{mygreen}{if(*(srv->algo) == 2)\{}*) 
    for(i = 0; i < srv->mygroups; i++) {
      cache_line resp;
      for(j = 0; j < srv->clnt_per_group; j++) {
        key = srv->my_clients[i][j].key;
        value = srv->my_clients[i][j].value;
        if (srv->my_clients[i][j].op == "insert")
          resp[j] = (*\textcolor{darkred}{\_\_base\_insrt}*)(srv->base_pq, key, value);  
        else if (srv->my_clients[i][j].op == "deleteMin")
          resp[j] = (*\textcolor{darkred}{\_\_base\_delMin}*)(srv->base_pq);
      }
      srv->my_responses[i] = resp;
    }
  (*\textcolor{mygreen}{\} else}*)
    (*\textcolor{mygreen}{return;}*)
}

void decisionTree(struct server struct client *str, int nthreads,  (* \\ *)             int size, int key_range,  double insert\_deleteMin) {
  (*\textcolor{mygreen}{int algo = 0;}*)
  (*\textcolor{mygreen}{... code for decision tree classifier ...}*)
  (*\textcolor{mygreen}{if (algo != 0) //  0: neutral}*)  
    (*\textcolor{mygreen}{*(str->algo) = algo;}*)
}
\end{lstlisting}
\end{subfigure}
\vspace{-6pt}
\caption{The modified code of \nuddle{} with the decision-making mechanism to implement \smartpq{}.}
\label{alg:smartpq}
\end{figure}

Figure~\ref{alg:smartpq} presents the modified code of \nuddle{} adding the decision-making mechanism (using green color) to implement \smartpq.
We extend the main structure of \nuddle{}, renamed to {\fontfamily{lmtt}\selectfont \textit{struct smartpq}}, by adding an additional field, called {\fontfamily{lmtt}\selectfont \textit{algo}}, to keep track the current \algomode{} mode, (either \notnuma{} or \numa{}). Similarly, {\fontfamily{lmtt}\selectfont \textit{struct client}} and {\fontfamily{lmtt}\selectfont \textit{struct server}} structures are extended with an additional {\fontfamily{lmtt}\selectfont \textit{algo}} field (e.g., line 111), which is a pointer to the {\fontfamily{lmtt}\selectfont \textit{algo}} field of {\fontfamily{lmtt}\selectfont \textit{struct smartpq}}. Each active thread initializes this pointer either in {\fontfamily{lmtt}\selectfont \textit{initClient()}} or {\fontfamily{lmtt}\selectfont \textit{initServer()}} depending on its role (e.g., line 119). This way, all threads share the same \algomode{} mode at \emph{any} point in time. In {\fontfamily{lmtt}\selectfont \textit{struct client}}, we also add a pointer to the shared data structure (line 110), which is used by client threads to \emph{directly} perform operations in the data structure in case of \notnuma{} mode. Specifically, we modify the core functions of client threads, i.e., {\fontfamily{lmtt}\selectfont \textit{insert\_client()}} and {\fontfamily{lmtt}\selectfont \textit{deleteMin\_client()}}, such that client threads either directly execute their operations in the data structure (e.g., line 126), or delegate them to server threads (e.g., line 127-128), with respect to the current \algomode{} mode. In contrast, the core functions of server threads do not need any modification. Finally, we wrap the code of {\fontfamily{lmtt}\selectfont \textit{serve\_requests}} function, i.e., the lines 86-97 of Figure~\ref{alg:functions}, with an if/else statement on the {\fontfamily{lmtt}\selectfont \textit{algo}} field (lines 133, 146 in Fig.~\ref{alg:smartpq}), such that server threads poll at client threads' requests only in \numa{} mode. In \notnuma{} mode, {\fontfamily{lmtt}\selectfont \textit{serve\_requests}} function returns without doing nothing. This way, programmers do not need to take care of calls on this function in their code, when the \notnuma{} mode is selected.

The {\fontfamily{lmtt}\selectfont \textit{decisionTree()}} function describes the interface with our proposed decision tree classifier, where the input arguments are associated with its features. In frequent time lapses, one or more threads may call this function to check if a transition to another \algomode{} mode is needed. If this is the case, the {\fontfamily{lmtt}\selectfont \textit{algo}} field of {\fontfamily{lmtt}\selectfont \textit{struct smartpq}} is updated (line 154 in Fig.~\ref{alg:smartpq}), and \smartpq{} switches \algomode{} mode, i.e., all active threads start executing their operations using the new \algomode{} mode. If the classifier predicts the neutral class (line 153), the {\fontfamily{lmtt}\selectfont \textit{algo}} field is not updated, and thus \smartpq{} remains at the currently selected \algomode{} mode.

\section{Experimental Evaluation}
\label{sec:experimental}

In our experimental evaluation, we use a 4-socket Intel Sandy Bridge-EP server equipped with 8-core Intel Xeon CPU E5-4620 processors providing a total of 32 physical cores and 64 hardware contexts. The processor runs at 2.2GHz and each physical core has its own L1 and L2 cache of sizes 64KB and 256KB, respectively. A 16MB L3 cache is shared by all cores in a NUMA socket and the RAM is 256GB. We use GCC 4.9.2 with -O3 optimization flag enabled to compile all implementations.

%\vspace{-pt}

Our evaluation includes the following concurrent priority queue implementations:
\begin{compactitem}[$\textbf{--}$]
\item \textit{alistarh\_fraser}~\cite{fraser,spraylist}: A \notnuma, relaxed priority queue~\cite{spraylist} based on Fraser's skip-list~\cite{fraser} available at ASCYLIB library~\cite{ascylib}.
\item \textit{alistarh\_herlihy}~\cite{herlihy,spraylist}: A \notnuma, relaxed priority queue~\cite{spraylist} based on Herlihy's skip-list~\cite{herlihy} available at ASCYLIB library~\cite{ascylib}.
\item \textit{lotan\_shavit}~\cite{lotan_shavit}: A \notnuma{} priority queue available at ASCYLIB library~\cite{ascylib}.
\item \ffwd~\cite{ffwd}: A \numa{} priority queue based on the delegation technique~\cite{Calciu2013Message,Klaftenegger2014Delegation,Lozi2012Remote,Petrovic2015Performance,Suleman2009Accelerating}, which includes \emph{only} one server thread to perform operations on behalf of \emph{all} client threads.
\item \nuddle: Our proposed \numa{} priority queue, which uses \emph{alistarh\_herlihy} as the underlying \emph{base algorithm}.
\item \smartpq: Our proposed adaptive priority queue, which uses \nuddle{} as the \numa{} mode, and \emph{alistarh\_herlihy} as the \notnuma{} \emph{base algorithm}.
\end{compactitem}

We evaluate the concurrent priority queue implementations in the following way:
\begin{compactitem}[$\textbf{--}$]
\item Each execution lasts 5 seconds, during which each thread performs randomly chosen operations. We also tried longer durations and got similar results.
\item Between consecutive operations in the data structure each thread executes a delay loop of 25 pause instructions. This delay is intentionally added in our benchmarks to better simulate a real-life application, where operations in the data structure are intermingled with other instructions in the application.
\item At the beginning of each run, the priority queue is initialized with elements the number of which is described at each figure.
\item Each software thread is pinned to a hardware context. Hyperthreading is enabled when using more than 32 software threads. When exceeding the number of available hardware contexts of the system, we oversubscribe software threads to hardware contexts.
\item We pin the first 8 threads to the first NUMA node, and consecutive client thread groups of 7 client threads each, to NUMA nodes in a round-robin fashion. 
\item In \notnuma{} implementations, any allocation needed in the operation is executed on demand, and memory affinity is determined by the first touch policy.
\item In \numa{} implementations, since our NUMA system has 64-byte cache lines, the response cache line is shared between up to 7 client threads, using 8-byte return values.
\item In \nuddle, the first 8 threads represent server threads. Server threads repeatedly execute the  {\fontfamily{lmtt}\selectfont \textit{serve\_requests}} function, and then a randomly chosen operation until time is up.
\item We have disabled the automatic Linux Balancing \cite{numa-balancing} to get consistent and stable results.
\item All reported results are the average of 10 independent executions with no significant variance.
\end{compactitem}

\subsection{Throughput of \nuddle}\label{sec:simbple_bench}
\begin{figure}[t]
\centering
\includegraphics[width=\linewidth]{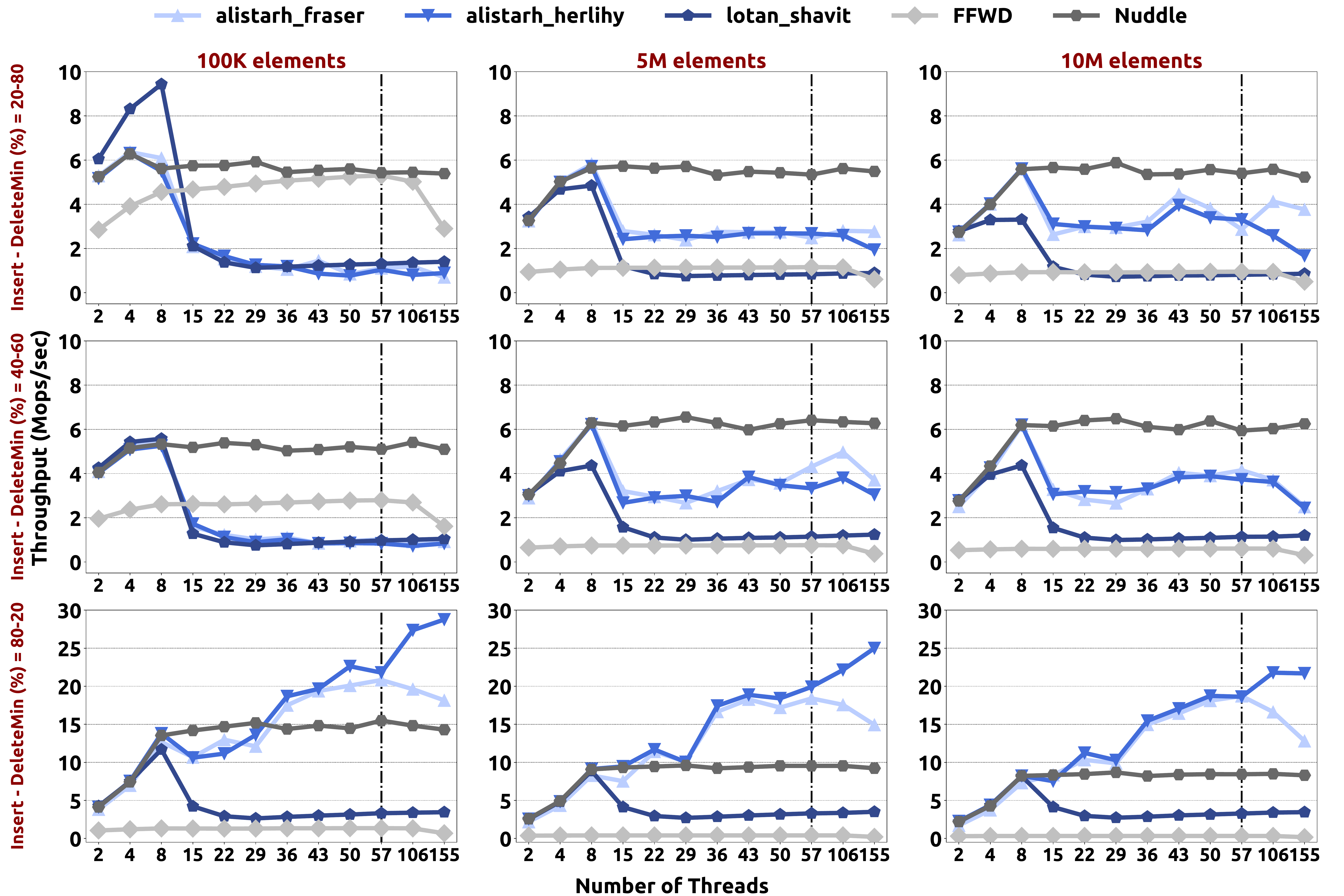}
\caption{Throughput of concurrent priority queue implementations. The columns show different priority queue sizes using the key range of double the elements of each size. The rows show different operation workloads. The vertical line in each plot shows the point after which we oversubscribe software threads to hardware contexts.}
\label{fig:simple_bench}
\vspace{-12pt}
\end{figure}

Figure~\ref{fig:simple_bench} presents the throughput achieved by concurrent priority queue implementations for various sizes and operation workloads. \numa{} priority queue implementations, i.e., \ffwd{} and \nuddle, achieve high throughput in \delete-dominated workloads: \nuddle{} performs best in \emph{all} \delete-dominated workloads, while \ffwd{} outperforms \notnuma{} implementations in the small-sized priority queues (e.g., 100K elements). In large-sized priority queues, \insrt{} operations have a larger impact on the total execution time (due to a longer traversal), and thus \nuddle{} and \notnuma{} implementations perform better than \ffwd, since they provide \emph{higher} thread-level parallelism. Note that \ffwd{} has \emph{single-threaded} performance, since at any point in time only \emph{one} (server) thread performs operations in the data structure. Moreover, as it is expected, the performance of both \ffwd{} and \nuddle{} saturates at the number of server threads used (e.g., 8 server threads for \nuddle) to perform operations in the data structure. Finally, we note that the communication between server and client threads used in \numa{} implementations has negligible overhead; when the number of client threads increases, even though the communication traffic over the interconnect increases, there is \emph{no} performance drop. Overall, we conclude that \nuddle{} achieves the highest throughput in \emph{all} \delete-dominated workloads, and is the most efficient \numa{} approach, since it provides high thread-level parallelism.

On the other hand, \notnuma{} implementations incur high performance degradation in \emph{high-contention} scenarios, such as \delete-dominated workloads, when using more than one NUMA node (i.e., after 8 threads).  As already discussed in prior works~\cite{David2013Everything,Boyd2010AnAnalysis,Lozi2012Remote,Zhang2016Scalable,Molka2009Memory,Giannoula2021SynCron}, the non-uniformity in memory accesses and cache line invalidation traffic significantly affects performance in high-contention scenarios. In \insrt-dominated workloads, which incur lower contention, even though \emph{lotan\_shavit} priority queue still incurs performance degradation when using more than one NUMA nodes of the system, the \emph{relaxed} \notnuma{} implementations, i.e., \emph{alistarh\_fraser} and \emph{alistarh\_herlihy} priority queues, achieve high scalability. This is because relaxed priority queues decrease both (i) the contention among threads, and (ii) the cache line invalidation traffic: the \delete{} operation returns (with a high probability) an element among the \emph{first few} (high-priority) elements of the queue, and thus, threads do not frequently compete for the same elements. Finally, we observe that \emph{alistarh\_herlihy} priority queue achieves higher performance benefits over \emph{alistarh\_fraser} priority queue, when we oversubscribe software threads to the available hardware contexts of our system. Overall, we find that in \insrt-dominated workloads, the \emph{relaxed} \notnuma{} implementations significantly outperform the \numa{} ones.

To sum up, we conclude that there is no one-size-fits-all solution, since none of the priority queues performs best across \emph{all} contention workloads. \nuddle{} achieves the highest throughput in high contention scenarios, while \emph{alistarh\_herlihy} performs best in low and medium contention scenarios. It is thus desirable to design a new approach for a concurrent priority queue to perform best under \emph{all} various contention scenarios.

\subsection{Throughput of \smartpq}

\subsubsection{Classifier Accuracy}\label{sec:accuracy}
We evaluate the efficiency of our proposed classifier (Section~\ref{sec:classifier}) using two metrics: (i) accuracy, and (ii) misprediction cost. First, we define the accuracy of the classifier as the percentage of \emph{correct} predictions, where a prediction is considered correct, if the classifier predicts the \algomode{} mode (either the \numa{} \nuddle{} or the \notnuma{} \emph{alistarh\_herlihy}) that achieves the best performance between the two. We use a test set of 10780 different contention workloads, where we randomly select the values of the features in each workload. In the above test set, our classifier has 87.9\% accuracy, i.e., it mispredicts 1300 times in 10780 different contention workloads. Second, we define the misprediction cost as the performance difference between the correct (best-performing) \algomode{} mode and the wrong predicted mode, normalized to the performance of the wrong predicted mode. Specifically, assuming the throughput of the wrong predicted and correct (best-performing) \algomode{} mode is $Y$ and $X$ respectively, the misprediction cost is defined as $((X-Y)/Y) * 100\%$. In 1300 mispredicted workloads, the geometric mean of misprediction cost for our classifier is 30.2\%. We conclude that the proposed classifier has \emph{high} accuracy, and in case of misprediction, incurs low performance degradation.

\subsubsection{Varying the Contention Workload}

\begin{figure}[!ht]
%\centering
\captionsetup[subfigure]{justification=centering}
    \begin{subfigure}[t]{0.32\textwidth}\centering
        %\hspace{-60pt}
        \includegraphics[scale=0.30]{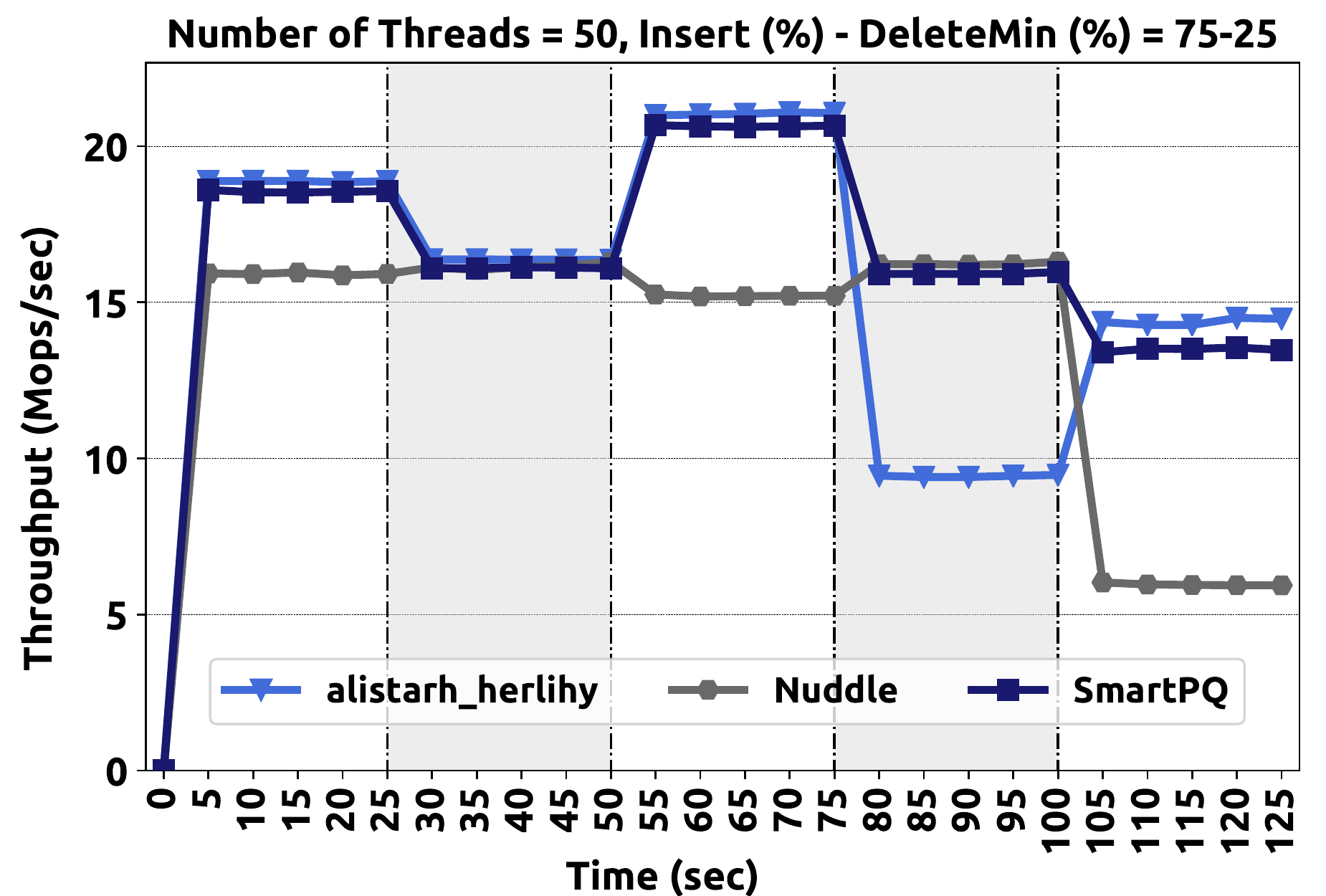}
        \vspace{-5pt}
        \caption{Varying the key range.}
	\label{fig:dynamic_range}
    \end{subfigure}%
    ~
    \begin{subfigure}[t]{0.32\textwidth}\centering
       % \hspace{-22pt}
        \includegraphics[scale=0.30]{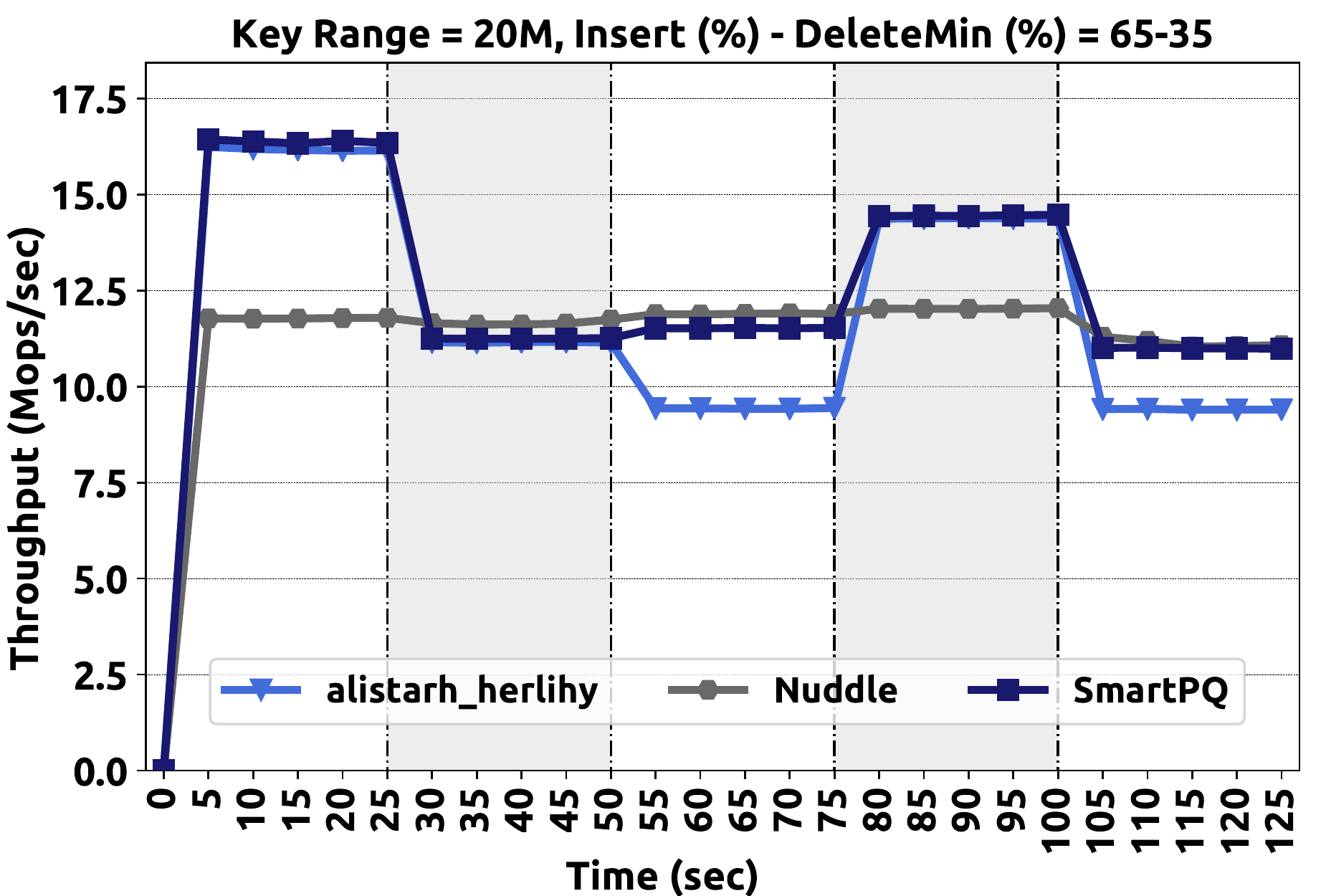}
        \vspace{-5pt}
        \caption{Varying the number of threads.}
	\label{fig:dynamic_threads}
    \end{subfigure}
    ~ 
    \begin{subfigure}[t]{0.32\textwidth}\centering
        %\hspace{1pt}
        \includegraphics[scale=0.30]{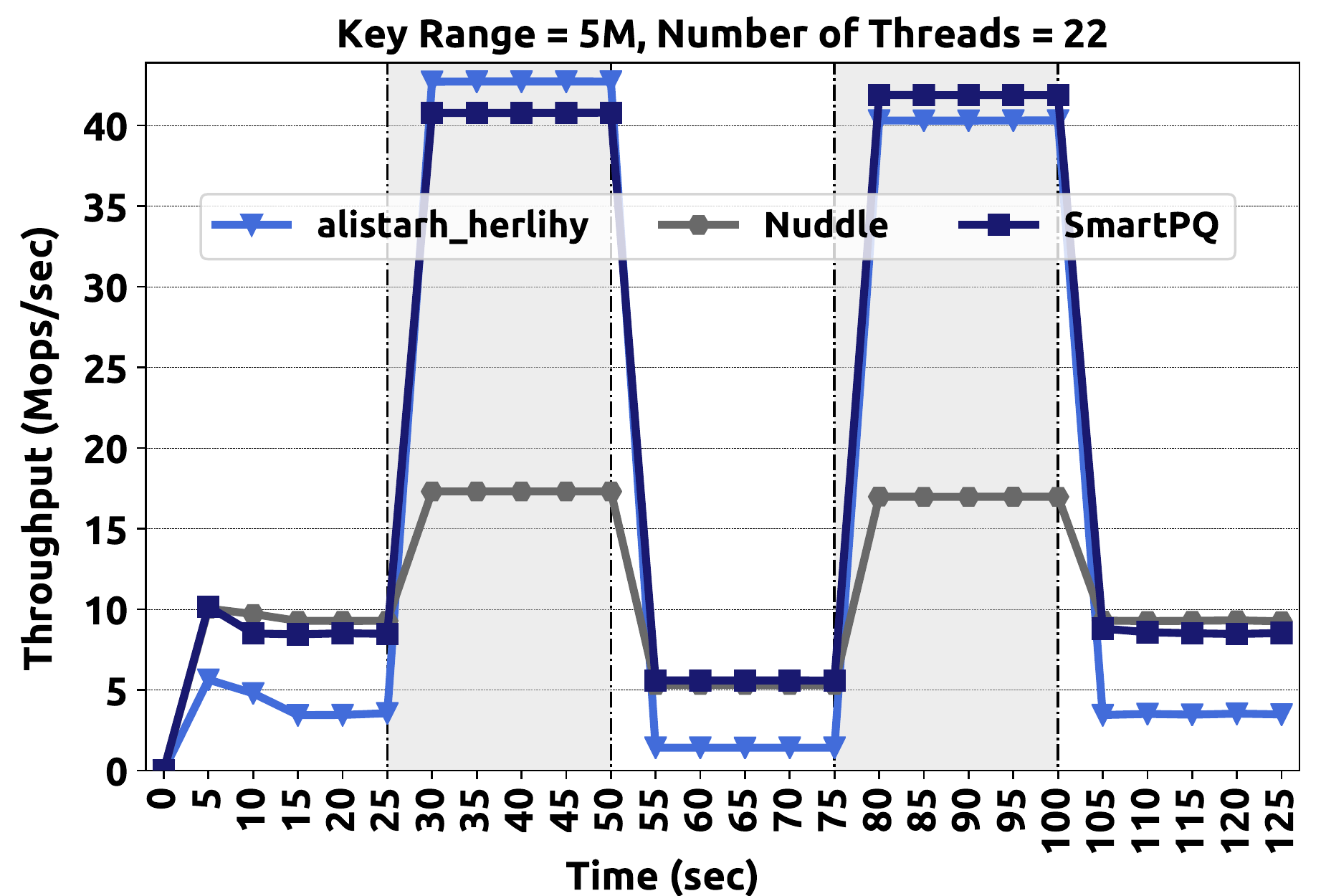}
        \vspace{-5pt}
        \caption{Varying the operation workload.}
	\label{fig:dynamic_operation}
    \end{subfigure}
    \caption{Throughput achieved by \smartpq{}, \nuddle{} and its underlying \emph{base algorithm} (\emph{alistarh\_herlihy}), in synthetic benchmarks, in which we vary a) the key range, b) the number of threads that perform operations in the data structure, and c) the percentage of \insrt/\delete{} operations in the workload.}
    \label{fig:dynamic_bench}
    \vspace{-10pt}
\end{figure}

\begin{figure}[ht]
\centering
\includegraphics[width=\textwidth]{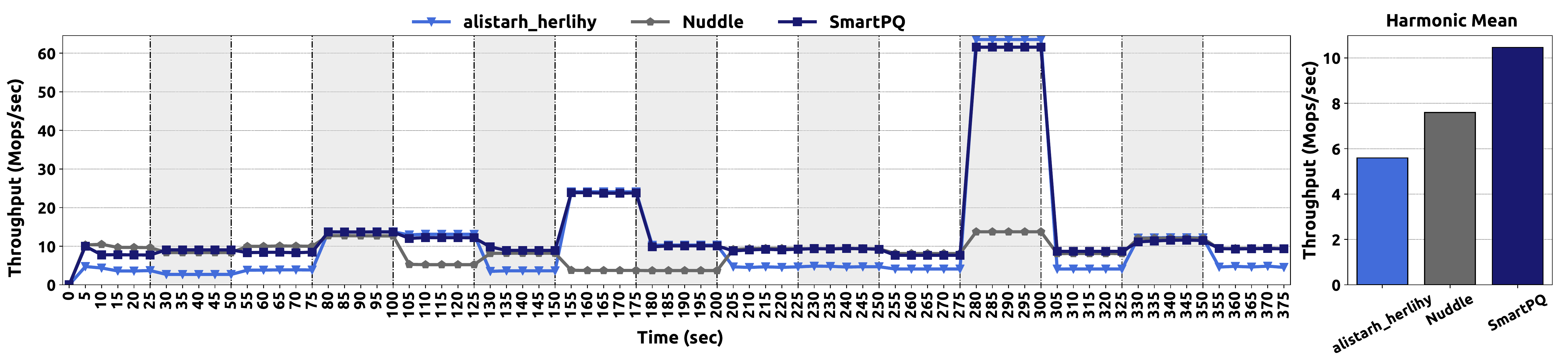}
\caption{Throughput achieved by \smartpq{}, \nuddle{} and its underlying \emph{base algorithm} (\emph{alistarh\_herlihy}), in synthetic benchmarks, in which we vary multiple features in the contention workload.}
\label{fig:dynamic_total}
%\vspace{-8pt}
\end{figure}

We present the performance benefit of \smartpq{} in synthetic benchmarks, in which we vary the contention workload over time, and compare it with \nuddle{} and its underlying \emph{base algorithm}, i.e., \emph{alistarh\_herlihy} priority queue. In all benchmarks, we change the contention workload every 25 seconds. In \smartpq, we set one dedicated sever thread to call the decision tree classifier \emph{every} second, in order to check if a transition to another \algomode{} mode is needed. Figure~\ref{fig:dynamic_bench} and Figure~\ref{fig:dynamic_total} show the throughput achieved by all three schemes, when we vary one and multiple features in the contention workload, respectively. Table~\ref{tab:features_dynamic_bench} and Table~\ref{tab:features_dynamic_total} show the features of the workload as they vary during the execution for the benchmarks evaluated in Figure~\ref{fig:dynamic_bench} and Figure~\ref{fig:dynamic_total}, respectively. Note that the current size of the priority queue changes during the execution due to successful \insrt{} and \delete{} operations.

\begin{table}[t]
\centering
\begin{subtable}{0.94\linewidth}\centering
\resizebox{\textwidth}{!}{
{\begin{tabular}{r r r c c}
\toprule
\textbf{Time (sec)} & \textbf{Current Size} & \textbf{Key Range} & \textbf{Number of Threads} & \textbf{Insert - DeleteMin (\%)} \\
\midrule
0 & \textbf{1149} & \textbf{100K} & 50 & 75-25 \\
25 & \textbf{812} & \textbf{2K} & 50 & 75-25 \\
50 & \textbf{485} & \textbf{1M} & 50 & 75-25 \\
75 & \textbf{2860} & \textbf{10K} & 50 & 75-25 \\
100 & \textbf{2256} & \textbf{50M} & 50 & 75-25 \\
\bottomrule
\end{tabular}}
}
\caption{Varying the key range in the workload.}\label{tab:1a}
\end{subtable}%

\begin{subtable}{0.94\linewidth}\centering
\resizebox{\textwidth}{!}{
{\begin{tabular}{r r c c c}
\toprule
\textbf{Time (sec)} & \textbf{Current Size} & \textbf{Key Range} & \textbf{Number of Threads} & \textbf{Insert - DeleteMin (\%)} \\
\midrule
\midrule
0 & \textbf{1166} & 20M & \textbf{57} & 65-35 \\
25 & \textbf{15567} & 20M & \textbf{29} & 65-35 \\
50 & \textbf{15417} & 20M & \textbf{15} & 65-35 \\
75 & \textbf{15297} & 20M & \textbf{43} & 65-35 \\
100 & \textbf{15346} & 20M & \textbf{15} & 65-35 \\
\bottomrule
\end{tabular}}
}
\caption{Varying the number of threads that perform operations in the data structure.}
\end{subtable}

\begin{subtable}{0.94\linewidth}\centering
\resizebox{\textwidth}{!}{
{\begin{tabular}{r r c c c}
\toprule
\textbf{Time (sec)} & \textbf{Current Size} & \textbf{Key Range} & \textbf{Number of Threads} & \textbf{Insert - DeleteMin (\%)} \\
\midrule
\midrule
0 & \textbf{1M} & 5M & 22 & \textbf{50-50} \\
25 & \textbf{140} & 5M & 22 & \textbf{100-0} \\
50 & \textbf{7403} & 5M & 22 & \textbf{30-70} \\
75 & \textbf{962} & 5M & 22 & \textbf{100-0} \\
100 & \textbf{8236} & 5M & 22 & \textbf{0-100} \\
\bottomrule
\end{tabular}}
}
\caption{Varying the percentage of \insrt/\delete{} operations.}
\end{subtable}
\caption{Features of the contention workload for benchmarks evaluated in Figure~\ref{fig:dynamic_bench}. We use bold font on the features that change in each execution phase.}\label{tab:features_dynamic_bench}
\vspace{-4pt}
\end{table}

We make three observations. First, as already shown in Section~\ref{sec:simbple_bench}, there is no one-size-fits-all solution, since neither \nuddle{} nor \emph{alistarh\_herlihy} performs best across all various contention workloads. For instance, in Figure~\ref{fig:dynamic_threads}, even though the performance of \nuddle{} remains constant, it outperforms \emph{alistarh\_herlihy}, when having 15 running threads, i.e., using 2 NUMA nodes of the system. Second, we observe that \smartpq{} successfully adapts to the best-performing \algomode{} mode, and performs best in \emph{all} contention scenarios. In Figure~\ref{fig:dynamic_total}, even when multiple features in the contention workload vary during the execution, \smartpq{} outperforms \emph{alistarh\_herlihy} and \nuddle{} by 1.87$\times$ and 1.38$\times$ on average, respectively. Note that any of the contention workloads evaluated in Figures~\ref{fig:dynamic_bench} and ~\ref{fig:dynamic_total} belongs in the training data set used for training our classifier. Third, we note that the decision-making mechanism of \smartpq{} has very low performance overheads. Across all evaluated benchmarks, \smartpq{} achieves \emph{only up} to 5.3\% performance slowdown (i.e., when using a range of 50M keys in Figure~\ref{fig:dynamic_range}) over the corresponding baseline implementation (\emph{alistarh\_herlihy} priority queue). Note that since the proposed decision tree classifier has very low traversal cost (Section~\ref{sec:classifier}), we intentionally set a frequent time interval (i.e., one second) for calling the classifier, such that \smartpq{} detects the contention workload change \emph{on time}, and quickly adapts itself to the best-performing \algomode{} mode.  We also tried large time intervals, and observed that \smartpq{} slightly delays to detect the contention workload change, thus achieving lower throughput in the transition points. 

Overall, we conclude that \smartpq{} performs best across \emph{all} contention workloads and at \emph{any} point in time, and incurs negligible performance overheads over the corresponding baseline implementation.

\begin{table}[!ht]
\vspace{-8pt}
\centering
\begin{subtable}{0.96\linewidth}\hspace{-8pt}
\resizebox{\textwidth}{!}{
{\begin{tabular}{r r c c c}
\toprule
\textbf{Time (sec)} & \textbf{Current Size} & \textbf{Key Range} & \textbf{Number of Threads} & \textbf{Insert - DeleteMin (\%)} \\
\midrule
\midrule
0 & \textbf{1M} & 10M & 57 & 50-50 \\
25 & \textbf{26} & 10M & \textbf{36} & \textbf{70-30} \\
50 & \textbf{12} & \textbf{20M} & 36 & \textbf{50-50} \\
75 & \textbf{79} & 20M & 36 & \textbf{80-20} \\
100 & \textbf{29K} & 20M & \textbf{50} & 80-20 \\
125 & \textbf{319K} & \textbf{100M} & 50 & \textbf{50-50} \\
150 & \textbf{13} & 100M & \textbf{57} & 50-50 \\
175 & \textbf{524K} & 100M & \textbf{22} & \textbf{100-0} \\
200 & \textbf{524K} & 100M & 22 & \textbf{50-50} \\
225 & \textbf{1142} & 100M & 22 & 50-50 \\
250 & \textbf{463} & \textbf{200M} & \textbf{57} & \textbf{0-100} \\
275 & \textbf{253} & 200M & 57 & \textbf{100-0} \\
300 & \textbf{33K} & \textbf{20M} & 57 & \textbf{0-100} \\
325 & \textbf{142} & 20M & \textbf{29} & \textbf{80-20} \\
350 & \textbf{25K} & 20M & 29 & \textbf{50-50} \\
\bottomrule
\end{tabular}}
}
\end{subtable}
\caption{Features of the contention workload for benchmarks evaluated in Figure~\ref{fig:dynamic_total}. We use bold font on the features that change in each execution phase.}\label{tab:features_dynamic_total}
\vspace{-12pt}
\end{table}

\section{Discussion and Future Work}\label{sec:discussion}

In Section~\ref{sec:classifier}, we assume that the contention workload is known a priori to extract the features needed for classification. To \emph{on-the-fly} extract these features, and \emph{dynamically} detect when contention changes, the main structure of \smartpq, i.e., {\fontfamily{lmtt}\selectfont \textit{struct smartpq}}, needs to be enriched with additional fields to keep track of workload statistics (e.g., the number of completed \insrt/\delete{} operations, the number of active threads that perform operations on the data structure, the minimum and/or maximum key that has been requested so far). Active threads that perform operations on the data structure could atomically update these statistics. In frequent time lapses, either a background thread or an active thread could extract the features needed for classification based on the workload statistics, and call the classifier to predict if a transition to another \algomode{} mode is needed. Finally, an additional parameter could be also added in \smartpq{} to configure how often to collect workload statistics.

In our experimental evaluation, we pin server threads on a single NUMA node and client threads on all nodes. We have chosen to do so (i) for simplicity, given that this approach fits well with our microbenchmark-based evaluation, and (ii) because this is par with prior works on concurrent data structures~\cite{ascylib,ffwd,spraylist,blackbox,sagonas,Sagonas2016TheCA,numask,pim_cds,choe2019concurrent,Winblad2018Lock,adaptivepq,Siakavaras2017Combining,Guerraoui2016Optimistic,Howley2012ANon,Bronson2010APractical,Ellen-bst}. In a real-world scenario, where \smartpq{} is used as a part of a high-level application, client threads do \emph{not} need to be pinned in hardware contexts, and they can be allowed to run in any core of the system. However, for our approach to be meaningful server threads need to be limited on a single NUMA node. This can easily be done by creating the server threads when \smartpq{} is initialized, and pinning them to hardware contexts that are located at the same NUMA node. In this case, server threads are background threads that only accept and serve requests from various client threads, which are part of the high-level application.

Finally, even though we focus on a microbenchmark-based evaluation to cover a \emph{wide variety} of contention scenarios, it is one of our future directions to explore the efficiency of \smartpq{} in real-world applications, such as web servers~\cite{swift, web_servers}, graph traversal applications~\cite{sagonas,CLRS} and scheduling in operating systems~\cite{scheduler}. As future work, we also aim to investigate the applicability of our approach in other data structures, that may have similar behavior with priority queues (e.g., skip lists, search trees), and extend our proposed classifier (e.g., adding more features) to cover a variety of NUMA CPU-centric systems with different architectural characteristics.

\section{Recommendations}

\noindent\textbf{Recommendation.}
\textit{Design adaptive parallel algorithms and concurrent data structures that on-the-fly adjust their parallelization approach and synchronization scheme depending on the dynamic workload demands and contention.} \\
Our work demonstrates (Figures~\ref{fig:motivation} and ~\ref{fig:nuddle_spraylist}) that there is no one-size-fits-all \algomode{} mode (between \notnuma{} and \numa{}) for a concurrent priority queue in modern computing systems: the best-performing \algomode{} mode depends on multiple characteristics, including the contention/operation workload, the size of the data structure and the underlying hardware platform~\cite{Strati2019AnAdaptive}. Such characteristics can \emph{dynamically} change during runtime, when performing various operations (e.g., \insrt{}, \delete{}) in the data structures used. Therefore, we conclude that to achieve high system performance in real-world scenarios, we need to \emph{dynamically} tune the configuration of parallel kernels based on the characteristics of the current load at each time. To this end, we recommend that software designers propose adaptive parallel algorithms and concurrent data structures that \emph{dynamically} adjust their parallelization technique and synchronization approach depending on the dynamic contention and workload demands. For example, machine learning, dynamic profiling and statistical approaches~\cite{Strati2019AnAdaptive,Singh2022Sibyl,Antic2016Locking} could be integrated in parallel kernels to improve performance.

\section{Related Work}
To our knowledge, this is the first work to propose an adaptive priority queue for NUMA systems, which performs best under \emph{all} various contention workloads, and even when contention varies over time. We briefly discuss prior work.

\textbf{Concurrent Priority Queues.}
A large corpus of work proposes concurrent algorithms for priority queues~\cite{lotan_shavit, linden_jonsson,sagonas,sundell,spraylist,adaptivepq,Wimmer_Martin,rihani,Brodal,Zhang,Sanders1998Randomized,Sagonas2016TheCA,Rab2020NUMA}, or generally for skip lists~\cite{hotspot,fraser,fomitchev,herlihy,herlihy_art,dick,pim_cds,Pugh1990SkipLists,choe2019concurrent}. Recent works~\cite{lotan_shavit, linden_jonsson} designed lock-free priority queues that separate the logical and the physical deletion of an element to increase parallelism. Alistarh et al.~\cite{spraylist} design a relaxed priority queue, called \textit{SprayList}, in which \delete{} operation returns with a high probability, an element among the \emph{first} $\mathcal{O}(p\log{}3p)$ elements of the priority queue, where $p$ is the number of threads. Sagonas et al~\cite{Sagonas2016TheCA} design a contention avoiding technique, in which \delete{} operation returns the highest-priority element of the priority queue under \emph{low} contention, while it enables relaxed semantics when high contention is detected. Specifically, under high-contention a few \delete{} operations are queued, and later several elements are deleted from the head of the queue \emph{at once} via a combined deletion operation. Heidarshenas et al.~\cite{Heidarshenas2020Snug} design a novel architecture for relaxed priority queues. These prior approaches are \notnuma{} implementations. Thus, in NUMA systems, they incur significant performance degradation in high-contention scenarios (e.g., \delete{}-dominated workloads in Section~\ref{sec:simbple_bench}). In contrast, Calciu et al.~\cite{adaptivepq} propose a \emph{NUMA-friendly} priority queue employing the combining and elimination techniques. Elimination allows the complementary operations, i.e., \insrt{} with \delete{}, to complete \emph{without} updating the data structure, while combining is a technique similar to the delegation technique~\cite{Calciu2013Message,Klaftenegger2014Delegation,Lozi2012Remote,Petrovic2015Performance,Suleman2009Accelerating} of \nuddle{} and \ffwd{}~\cite{ffwd}. Finally, Daly et al.~\cite{numask} propose an efficient technique to obtain \numa{} skip lists, which however, can only be applied to skip list-based data structures. In contrast, \nuddle{} is a \emph{generic} technique to obtain \numa{} data structures.

\textbf{Black-Box Approaches.}
Researchers have also proposed black-box approaches: any data structure can be made wait-free or \numa{} without effort or knowledge on parallel programming or NUMA architectures. Herlihy \cite{wait_free} provides a universal method to design wait-free implementations of any sequential object. However, this method remains impractical due to high overheads. Hendler et al.~\cite{flatcombining1} propose flat combining; a technique to reduce synchronization overheads by executing multiple client threads' requests \emph{at once}. Despite significant improvements~\cite{flatcombining2}, this technique provides high performance \emph{only} for a few data structures (e.g., synchronous queues). \ffwd{}~\cite{ffwd} is black-box approach, which uses the delegation technique~\cite{Calciu2013Message,Klaftenegger2014Delegation,Lozi2012Remote,Petrovic2015Performance,Suleman2009Accelerating} to eliminate cache line invalidation traffic over the interconnect. However, \ffwd{} is limited to single threaded performance. Calciu et al.~\cite{blackbox} propose a black-box technique, named \textit{Node Replication}, to obtain concurrent \numa{} data structures. In \textit{Node Replication}, every NUMA node has replicas of the shared data structure, which are synchronized via a shared log. Even though \ffwd{} and \textit{Node Replication} are generic techniques to obtain \numa{} data structures, similarly to \nuddle, both of them use a \emph{serial asynchronized} implementation as the underlying \emph{base algorithm}. Therefore, if they are used as the \numa{} \algomode{} mode in an adaptive data structure, which dynamically tunes itself between a \notnuma{} and a \numa{} mode, both \ffwd{} and \emph{Node Replication} need a synchronization point between transitions to ensure correctness. Consequently, they would incur high performance overheads, when transitions between \algomode{} modes happen at a non-negligible frequency.

\textbf{Machine Learning in Data Structures.}
Even though machine learning is widely used to improve performance in many emerging applications~\cite{athena, ppopp2018, cluster2018, benatia,Gronquist2021Deep,Memeti2019Using,KusumNG2016Safe,Michie1968Memo,Meng2019APattern,Dhulipala2020APattern,Sedaghati2015Automatic,Benatia2016SparseMF,Pengfei2020LISA}, there are a handful of works~\cite{smart,google} that leverage machine learning to design \emph{highly-efficient} concurrent data structures. Recently, Eastep et al.~\cite{smart} use reinforcement learning to on-the-fly tune a parameter in the flat combining technique~\cite{flatcombining1,flatcombining2}, which is used in skip lists and priority queues. Kraska et al.~\cite{google} demonstrate that machine learning models can be trained to predict the position or existence of elements in key-value lookup sets, and discuss under which conditions learned index models can outperform the traditional indexed data structures (e.g., B-trees).
\section{Summary}

We propose \smartpq, an adaptive concurrent priority queue for NUMA architectures, which performs best under \emph{all} various contention scenarios, and even when contention varies over time. \smartpq{} has two key components. First, it is built on top of \nuddle{}; a generic low-overhead technique to obtain efficient \numa{} data structures using \emph{any} concurrent \notnuma{} implementation as its backbone. Second, \smartpq{} integrates a lightweight decision-making mechanism, which is based on a simple decision tree classifier, to decide when to switch between \nuddle{}, i.e., a \numa{} \algomode{} mode, and its underlying \emph{base algorithm}, i.e., a \notnuma{} \algomode{} mode. Our evaluation over a wide range of contention scenarios demonstrates that \smartpq{} switches between the two \algomode{} modes with negligible overheads, and significantly outperforms prior schemes, even when contention varies over time. We conclude that \smartpq{} is an efficient concurrent priority queue for NUMA systems, and hope that this work encourages further study on adaptive concurrent data structures for NUMA architectures.

%SynCron
\chapter{\SynCron}\label{SynCronChapter}
\section{Overview}
\label{Introductionbl}

Recent advances in 3D-stacked memories~\cite{HBM,HMC,kim2015ramulator,HMC_old,HBM_old,Lee2016Simultaneous} have renewed interest in Near-Data Processing (NDP)~\cite{Mutlu2020AMP,ahn2015scalable,Ahn2015PIMenabled,Balasubramonian2014Near}. NDP involves performing computation close to where the application data resides. This alleviates the expensive data movement between processors and memory, yielding significant performance improvements and energy savings in parallel applications. Placing low-power cores or special-purpose accelerators (hereafter called NDP cores) close to the memory dies of high-bandwidth 3D-stacked memories is a commonly-proposed design for NDP systems~\cite{Mutlu2020AMP,Mutlu2019Processing,Ghose2019Workload,Ahn2015PIMenabled,Nair2015Active,ahn2015scalable,Hsieh2016accelerating,pugsley2014ndc,Boroumand2018Google,Gokhale2015Near,Gao2016HRL,Hsieh2016TOM,Drumond2017mondrian,Liu2018Processing,Boroumand2019Conda,Gao2015Practical,gao2017tetris,Kim2016Neurocube,Kim2017GrimFilter,Survive2016Survive,Nai2017GraphPIM,Youwei2019GraphQ,Pattnaik2016Scheduling,fernandez2020natsa,Singh2019Napel,Singh2020NERO,Cali2020GenASM,Zhang2018GraphP,Kim2013memory,Tsai2018Adaptive,boroumand2017lazypim}. Typical NDP architectures support several NDP units connected to each other, with each unit comprising multiple NDP cores close to memory~\cite{Kim2013memory,Youwei2019GraphQ,Tsai2018Adaptive,ahn2015scalable,Hsieh2016TOM,Boroumand2018Google,Zhang2018GraphP}. Therefore, NDP architectures provide high levels of parallelism, low memory access latency, and large aggregate memory bandwidth.

Recent research demonstrates the benefits of NDP for parallel applications, e.g., for genome analysis~\cite{Kim2017GrimFilter,Cali2020GenASM}, graph processing~\cite{ahn2015scalable,Nai2017GraphPIM,Zhang2018GraphP,Youwei2019GraphQ,Ahn2015PIMenabled,Boroumand2019Conda,boroumand2017lazypim}, databases~\cite{Drumond2017mondrian,Boroumand2019Conda}, security~\cite{Gu2016Leveraging},  pointer-chasing workloads~\cite{liu2017concurrent,choe2019concurrent,Hsieh2016accelerating,hashemi2016accelerating}, and neural networks~\cite{Boroumand2018Google,gao2017tetris,Kim2016Neurocube,Liu2018Processing}. In general, these applications exhibit high parallelism, low operational intensity, and relatively low cache locality~\cite{Gagandeep2019Near,Awan2015Performance,Awan2016Node,oliveira2021pimbench,Gomez2021Analysis}, which make them suitable for NDP.

Prior works discuss the need for efficient synchronization primitives in NDP systems, such as locks~\cite{liu2017concurrent,choe2019concurrent} and barriers~\cite{ahn2015scalable,Gao2015Practical,Youwei2019GraphQ,Zhang2018GraphP}. Synchronization primitives are widely used by multithreaded applications~\cite{LeBeane2015Data,Tallent2010Analyzing,Strati2019AnAdaptive,Giannoula2018Combining,Elafrou2019Conflict,Suleman2009Accelerating,Joao2012Bottleneck,Joao2013Utility,Ebrahimi2011Parallel,Suleman2010Data}, and must be carefully designed to fit the underlying hardware requirements to achieve high performance. Therefore, to fully leverage the benefits of NDP for parallel applications, an effective synchronization solution for NDP systems is necessary.

Approaches to support synchronization are typically of two types~\cite{herlihy2008art,Hoefler2004survey}.
First, synchronization primitives can be built through \emph{shared memory}, most commonly using the atomic read-modify-write (\emph{rmw}) operations provided by hardware. In CPU systems, atomic \emph{rmw} operations are typically implemented upon the underlying hardware cache coherence protocols, but many NDP systems do \emph{not} support hardware cache coherence (e.g.,~\cite{Tsai2018Adaptive,Ghose2019Workload,ahn2015scalable,Zhang2018GraphP,Youwei2019GraphQ}). In GPUs and Massively Parallel Processing systems (MPPs), atomic \emph{rmw} operations can be implemented in dedicated hardware atomic units, known as \emph{remote atomics}. However, synchronization using remote atomics has been shown to be inefficient, since sending every update to a fixed location creates high global traffic and hotspots~\cite{Wang2019Fast,li2015fine,yilmazer2013hql,eltantawy2018warp,Mukkara2019PHI}. Second, synchronization can be implemented via a \emph{\mpsync{}} scheme, where cores exchange messages to reach an agreement. Some recent NDP works (e.g.,~\cite{ahn2015scalable,Gao2015Practical,Youwei2019GraphQ,Gu2020iPIM}) propose \mpsync{} barrier primitives among NDP cores of the system. However, these synchronization schemes are still inefficient, as we demonstrate in Section~\ref{Evaluationbl}, and also lack support for lock, semaphore and condition variable synchronization primitives.

Hardware synchronization techniques that do not rely on hardware coherence protocols and atomic \emph{rmw} operations have been proposed for multicore systems~\cite{abell2011glocks,abellan2010g,oh2011tlsync,Zhu2007SSB,Vallejo2010Architectural,Liang2015MISAR,Leiserson1992CM5,Sergi2016WiSync}. However, such synchronization schemes are tailored for the specific architecture of each system, and are not efficient or suitable for NDP systems (Section~\ref{Relatedbl}). For instance, CM5~\cite{Leiserson1992CM5} provides a barrier primitive via a dedicated physical network, which would incur high hardware cost to be supported in large-scale NDP systems. LCU~\cite{Vallejo2010Architectural} adds a control unit to \emph{each} CPU core and a buffer to each memory controller, which would also incur high cost to implement in \emph{area-constrained} NDP cores and controllers. SSB~\cite{Zhu2007SSB} includes a small buffer attached to each controller of the last level cache (LLC) and MiSAR~\cite{Liang2015MISAR} introduces an accelerator distributed at the LLC. Both schemes are built on the shared cache level in CPU systems, which most NDP systems do \emph{not} have. Moreover, in NDP systems with \emph{non-uniform} memory access times, most of these prior schemes would incur significant performance overheads under high-contention scenarios. This is because they are oblivious to the non-uniformity of NDP, and thus would cause excessive traffic across NDP units of the system upon contention (Section~\ref{Flat}).

Overall, NDP architectures have several important characteristics that necessitate a new approach to support efficient synchronization. First, most NDP architectures~\cite{Nair2015Active,ahn2015scalable,Hsieh2016accelerating,pugsley2014ndc,Youwei2019GraphQ,Boroumand2018Google,Gao2016HRL,Drumond2017mondrian,Liu2018Processing,Gao2015Practical,gao2017tetris,choe2019concurrent,Gokhale2015Near,Zhang2018GraphP,fernandez2020natsa,Mutlu2020AMP,Mutlu2019Processing,Ghose2019Workload,Gu2020iPIM} lack shared caches that can enable low-cost communication and synchronization among NDP cores of the system. Second, hardware cache coherence protocols are typically not supported in NDP systems~\cite{ahn2015scalable,Hsieh2016accelerating,pugsley2014ndc,Youwei2019GraphQ,Boroumand2018Google,Gokhale2015Near,Gao2016HRL,Drumond2017mondrian,Liu2018Processing,Gao2015Practical,gao2017tetris,choe2019concurrent,Kim2016Neurocube,Zhang2018GraphP,fernandez2020natsa,Mutlu2019Processing,Gu2020iPIM}, due to high area and traffic
overheads associated with such protocols~\cite{Tsai2018Adaptive,Ghose2019Workload}. Third, NDP systems are non-uniform, distributed architectures, in which inter-unit communication is more expensive (both in performance and energy) than intra-unit communication~\cite{Zhang2018GraphP,Boroumand2019Conda,boroumand2017lazypim,Drumond2017mondrian,Youwei2019GraphQ,ahn2015scalable,Gao2015Practical,Kim2013memory}.

In this work, we present \SynCron{}, an efficient synchronization mechanism for NDP architectures. \SynCron{} is designed to achieve the goals of performance, cost, programming ease, and generality to cover a wide range of synchronization primitives through four key techniques. First, we offload synchronization among NDP cores to dedicated low-cost hardware units, called \myEngine{}s (\myEngineShort{}s). This approach avoids the need for complex coherence protocols and expensive \emph{rmw} operations, at low hardware cost. Second, we directly buffer the synchronization variables in a specialized cache memory structure to avoid costly memory accesses for synchronization. Third, \SynCron{} coordinates synchronization with a hierarchical \mpsync{} scheme: NDP cores only communicate with their local \myEngineShort{} that is located in the same NDP unit. At the next level of communication, all local \myEngineShort{}s of the system's NDP units communicate with each other to coordinate synchronization at a global level. Via its hierarchical communication protocol, \SynCron{} significantly reduces synchronization traffic across NDP units under high-contention scenarios. Fourth, when applications with frequent synchronization oversubscribe the hardware synchronization resources, \SynCron{} uses an efficient and programmer-transparent overflow management scheme that avoids costly fallback solutions and minimizes overheads.

We evaluate \SynCron{} using a wide range of parallel workloads including pointer-chasing, graph applications, and time series analysis. Over prior approaches (similar to~\cite{ahn2015scalable,Gao2015Practical}), \SynCron{} improves performance by 1.27$\times$ on average (up to 1.78$\times$) under high-contention scenarios, and by 1.35$\times$ on average (up to 2.29$\times$) under low-contention scenarios. In real applications with fine-grained synchronization, \SynCron{} comes within 9.5\% of the performance and 6.2\% of the energy of an ideal zero-overhead synchronization mechanism. Our proposed hardware unit incurs very modest area and power overheads (Section~\ref{Areabl}) when integrated into the compute die of an NDP unit.

\vspace{-1pt}
The main \textbf{contributions} of this work are:
\begin{itemize}
\vspace{-8pt}
\setlength\itemsep{-2pt}
    \item We investigate the challenges of providing efficient synchronization in Near-Data-Processing architectures, and propose an end-to-end mechanism, \SynCron{}, for such systems.
    \item We design low-cost synchronization units that coordinate synchronization across NDP cores, and directly buffer synchronization variables to avoid costly memory accesses to them. We propose an efficient \mpsync{} synchronization approach that organizes the process hierarchically, and provide a hardware-only programmer-transparent overflow management scheme to alleviate performance overheads when hardware synchronization resources are exceeded.
    \item We evaluate \SynCron{} using a wide range of parallel workloads and demonstrate that it significantly outperforms prior approaches both in performance and energy consumption. \SynCron{} also has low hardware area and power overheads.
\end{itemize}
\section{Background and Motivation}
\label{Motivationbl}

\subsection{Baseline Architecture}

\begin{figure}[H]
\vspace{-1pt}
  \centering
   \includegraphics[width=0.9\linewidth]{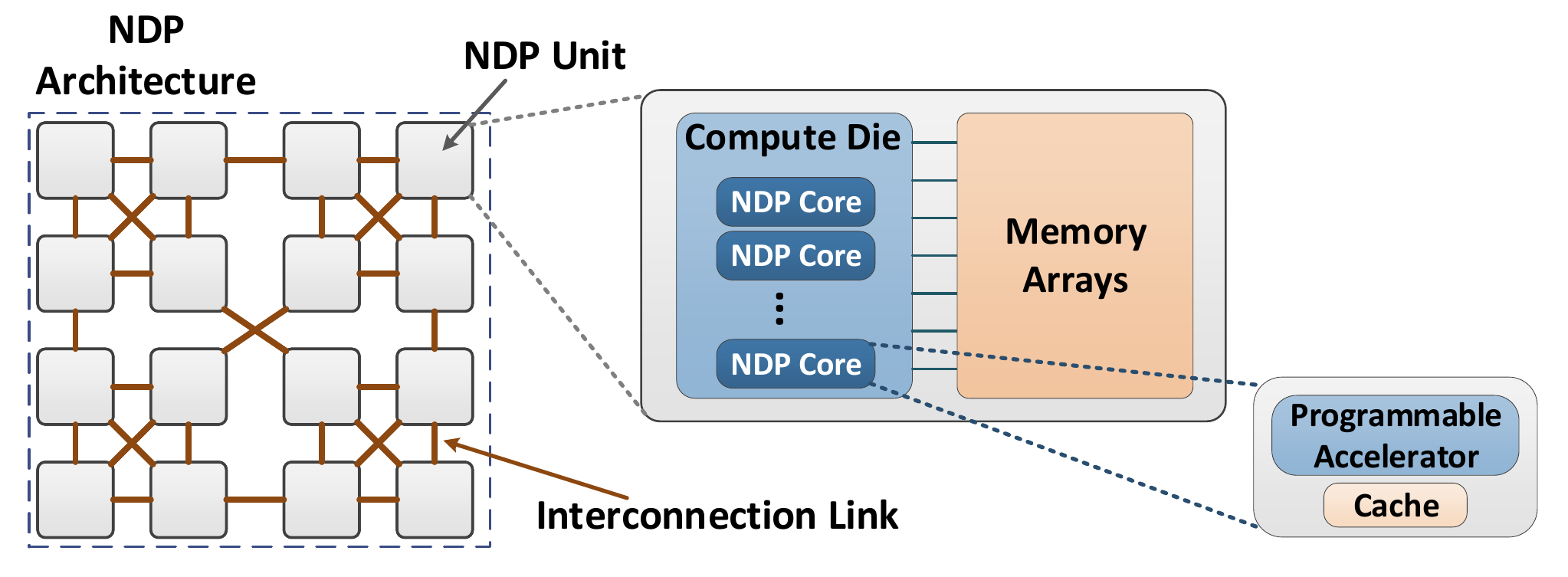}
  \caption{High-level organization of an NDP architecture.}
  \label{fig:PIMArch}
  \vspace{-12pt}
\end{figure}

Numerous works~\cite{Hsieh2016accelerating,ahn2015scalable,Gao2015Practical,Nai2017GraphPIM,Youwei2019GraphQ,Zhang2018GraphP,gao2017tetris,Kim2016Neurocube,Gu2016Leveraging,Drumond2017mondrian,Boroumand2018Google,Ahn2015PIMenabled,Boroumand2019Conda,boroumand2017lazypim,Gu2020iPIM,Tsai2018Adaptive,liu2017concurrent,choe2019concurrent,Seshadri2017Ambit,Kanellopoulos2019SMASH} show the potential benefit of NDP for parallel, irregular applications. These proposals focus on the design of the compute logic that is placed close to or within memory, and in many cases provide special-purpose near-data accelerators for specific applications. Figure~\ref{fig:PIMArch} shows the baseline organization of the NDP architecture we assume in this work, which includes several NDP units connected with each other via serial interconnection links to share the same physical address space. Each NDP unit includes the memory arrays and a compute die with multiple low-power programmable cores or fixed-function accelerators, which we henceforth refer to as NDP cores. NDP cores execute the offloaded NDP kernel and access the various memory locations across NDP units with non-uniform access times~\cite{Zhang2018GraphP,Tsai2018Adaptive,ahn2015scalable,Youwei2019GraphQ,Boroumand2019Conda,boroumand2017lazypim,Drumond2017mondrian}. We assume that there is no OS running in the NDP system. In our evaluation, we use programmable in-order NDP cores, each including small private L1 I/D caches. However, \SynCron{} can be used with any programmable, fixed-function or reconfigurable NDP accelerator. We assume software-assisted cache-coherence (provided by the operating system or the programmer), similar to~\cite{Tsai2018Adaptive,Gao2015Practical}: data can be either thread-private, shared read-only, or shared read-write. Thread-private and shared read-only data can be cached by NDP cores, while shared read-write data is uncacheable.

We focus on three characteristics of NDP architectures that are of particular importance in the synchronization context. First, NDP architectures typically do not have a shared level of cache memory~\cite{Nair2015Active,ahn2015scalable,Hsieh2016accelerating,pugsley2014ndc,Youwei2019GraphQ,Boroumand2018Google,Gao2016HRL,Drumond2017mondrian,Liu2018Processing,Gao2015Practical,gao2017tetris,choe2019concurrent,Gokhale2015Near,Zhang2018GraphP,fernandez2020natsa,Mutlu2020AMP,Mutlu2019Processing,Ghose2019Workload,Gu2020iPIM}, since the NDP-suited workloads usually do not benefit from deep cache hierarchies due to their poor locality~\cite{Gao2015Practical,Gagandeep2019Near,Tsai2018Adaptive,oliveira2021pimbench}. Second, NDP architectures do not typically support conventional hardware cache coherence protocols~\cite{ahn2015scalable,Hsieh2016accelerating,pugsley2014ndc,Youwei2019GraphQ,Boroumand2018Google,Gokhale2015Near,Gao2016HRL,Drumond2017mondrian,Liu2018Processing,Gao2015Practical,gao2017tetris,choe2019concurrent,Kim2016Neurocube,Zhang2018GraphP,fernandez2020natsa,Mutlu2019Processing,Gu2020iPIM}, because they would add area and traffic overheads~\cite{Tsai2018Adaptive,Ghose2019Workload}, and would incur high complexity and latency~\cite{Abeydeera2020Chronos}, limiting the benefits of NDP. Third, communication across NDP units is expensive, because NDP systems are non-uniform distributed architectures. The energy and performance costs of inter-unit communication are typically orders of magnitude greater than the costs of intra-unit communication~\cite{Zhang2018GraphP,Boroumand2019Conda,boroumand2017lazypim,Drumond2017mondrian,Youwei2019GraphQ,ahn2015scalable,Gao2015Practical,Kim2013memory}, and thus inter-unit communication may slow down the execution of NDP cores~\cite{Zhang2018GraphP}.

%\camfive{slow down}

\subsection{The Solution Space for Synchronization}
Approaches to support synchronization are typically either via shared memory or \mpsync{} schemes.

\subsubsection{Synchronization via Shared Memory} In this case, cores coordinate via a consistent view of shared memory locations, 
using atomic read/write operations or atomic read-modify-write (\emph{rmw}) operations. If \emph{rmw} operations are \emph{not} supported by hardware, Lamport's bakery algorithm~\cite{Lamport1974New} can provide synchronization to $N$ participating cores, assuming sequential consistency~\cite{Lamport1979How}. However, this scheme scales poorly, as a core accesses $O(N)$ memory locations at \emph{each} synchronization retry.
In contrast, commodity systems (CPUs, GPUs, MPPs) typically support \emph{rmw} operations in hardware.

GPUs and MPPs support \emph{rmw} operations in specialized hardware units (known as \emph{remote atomics}), located in each bank of the shared cache~\cite{Wittenbrink2011Fermi,Luna2013Performance}, or the memory controllers~\cite{kessler1993crayTA,Laudon1997SGI}. Remote atomics are also supported by an NDP work~\cite{Gao2015Practical} at the vault controllers of Hybrid Memory Cube (HMC)~\cite{HMC,HMC_old}. Implementing synchronization primitives using remote atomics requires a spin-wait scheme, i.e., executing consecutive \emph{rmw} retries. However, performing and sending every \emph{rmw} operation to a shared, fixed location can cause high global traffic and create hotspots~\cite{Wang2019Fast,Mukkara2019PHI,li2015fine,yilmazer2013hql,eltantawy2018warp}. In NDP systems, consecutive \emph{rmw} operations to a remote NDP unit would incur high traffic \emph{across} NDP units, with high performance and energy overheads.

Commodity CPU architectures support \emph{rmw} operations either by locking the bus (or equivalent link), or by relying on the hardware cache coherence protocol~\cite{Sorin2011Primer,intelsys}, which many NDP architectures do not support. Therefore, coherence-based synchronization~\cite{guiroux2016multicore,rudolph1984dynamic,anderson1989performance,mellor1991algorithms,scott2002non,Dice2015Lock,magnusson1994queue,craig1993building,luchangco2006hierarchical,dice2011flat,chabbi2015high,Zhang2016Scalable} cannot be directly implemented in NDP architectures. Moreover, based on prior works on synchronization~\cite{David2013Everything,Boyd2010AnAnalysis,Kaxiras2015Turning,Mellor1991Synchronization,Tallent2010Analyzing,Molka2009Memory}, coherence-based synchronization would exhibit low scalability on NDP systems for two reasons. First, it performs poorly with a \emph{large} number of cores, due to low scalability of conventional hardware coherence protocols~\cite{Heinrich1999Aquantitative,Sorin2011Primer,Kelm2010Cohesion,Kelm2010Waypoint}. Most NDP systems include several NDP units~\cite{Kim2013memory,Zhang2018GraphP,Youwei2019GraphQ,ahn2015scalable}, each typically supporting hundreds of small, area-constrained cores~\cite{ahn2015scalable, Boroumand2018Google,Youwei2019GraphQ,Zhang2018GraphP}. Second, the non-uniformity in memory accesses significantly affects the scalability of coherence-based synchronization~\cite{David2013Everything,Boyd2010AnAnalysis,Zhang2016Scalable,Molka2009Memory}. Prior work on coherence-based synchronization~\cite{David2013Everything} observes that the latency of a lock acquisition that needs to transfer the lock \emph{across} NUMA sockets can be up to 12.5$\times$ higher than that \emph{within} a socket. We expect such effects to be aggravated in NDP systems, since they are by nature \emph{non-uniform} and \emph{distributed}~\cite{Kim2013memory,Zhang2018GraphP,Youwei2019GraphQ,ahn2015scalable,Gao2015Practical,Boroumand2019Conda,boroumand2017lazypim,Drumond2017mondrian} with very low memory access latency within an NDP unit.

We validate these observations on both a real CPU and our simulated NDP system. On an Intel Xeon Gold server, we evaluate the operation throughput achieved by two coherence-based lock algorithms (Table~\ref{Tab:motivation}), i.e., TTAS~\cite{rudolph1984dynamic} and Hierarchical Ticket Lock (HTL)~\cite{mellor1991algorithms}, using a microbenchmark taken from the \emph{libslock} library~\cite{David2013Everything}. When increasing the number of threads from 1 to 14 within a single socket, throughput drops by 3.91$\times$ and 2.77$\times$ for TTAS and HTL, respectively. Moreover, when pinning two threads on different NUMA sockets, throughput drops by up to 2.29$\times$ over when pinning them on the same socket, due to non-uniform memory access times of lock variables.

\vspace{3pt}
\begin{table}[t]
\centering
\resizebox{0.84\columnwidth}{!}{%
\begin{tabular}{c||c c||c c}
\toprule
Million Operations & 1 thread & 14 threads & 2 threads & 2 threads\\
per Second & single-socket & single-socket & same-socket & different-socket \\
\midrule
TTAS lock~\cite{rudolph1984dynamic}                & 8.92  & 2.28 & 9.91 & 4.32                     \\
Hierarchical Ticket lock~\cite{mellor1991algorithms} & 8.06 & 2.91 & 9.01 & 6.79 \\
\bottomrule

\end{tabular}%
}
\caption{Throughput of two coherence-based lock algorithms on an Intel Xeon Gold server using the libslock library~\cite{David2013Everything}.}
\label{Tab:motivation}
\vspace{-2pt}
\end{table}

\begin{figure}[H]
%\centering
    \centering\captionsetup[subfloat]{labelfont=bf}
  \begin{subfigure}[h]{0.48\columnwidth}
    \hspace{-8pt}
    \includegraphics[scale=0.26]{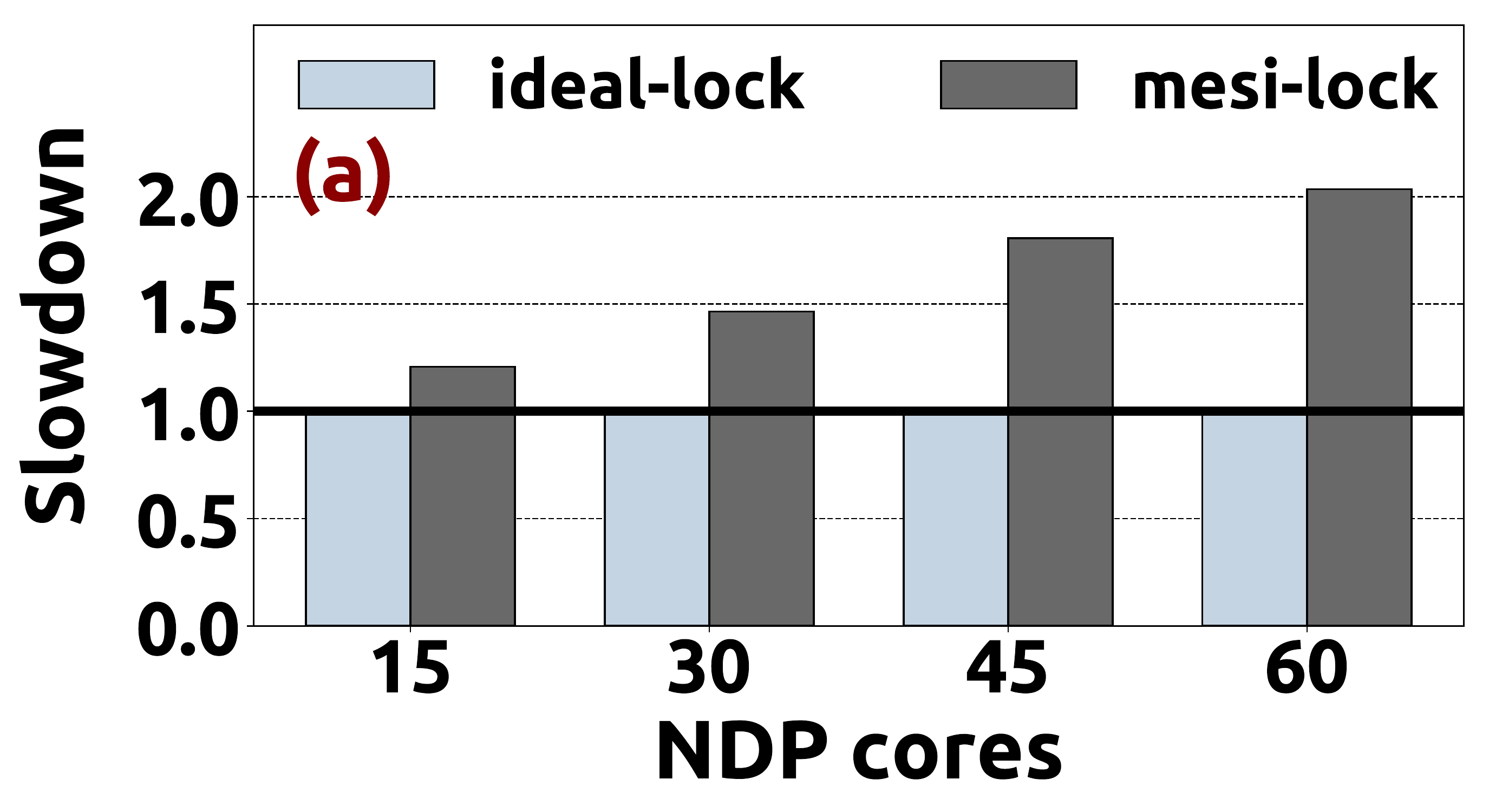}
    \vspace{-12pt}
    %\caption{} 
    \label{fig:coherence} 
 \end{subfigure}
  ~
  \begin{subfigure}[h]{0.48\columnwidth}
   \centering
    \includegraphics[scale=0.26]{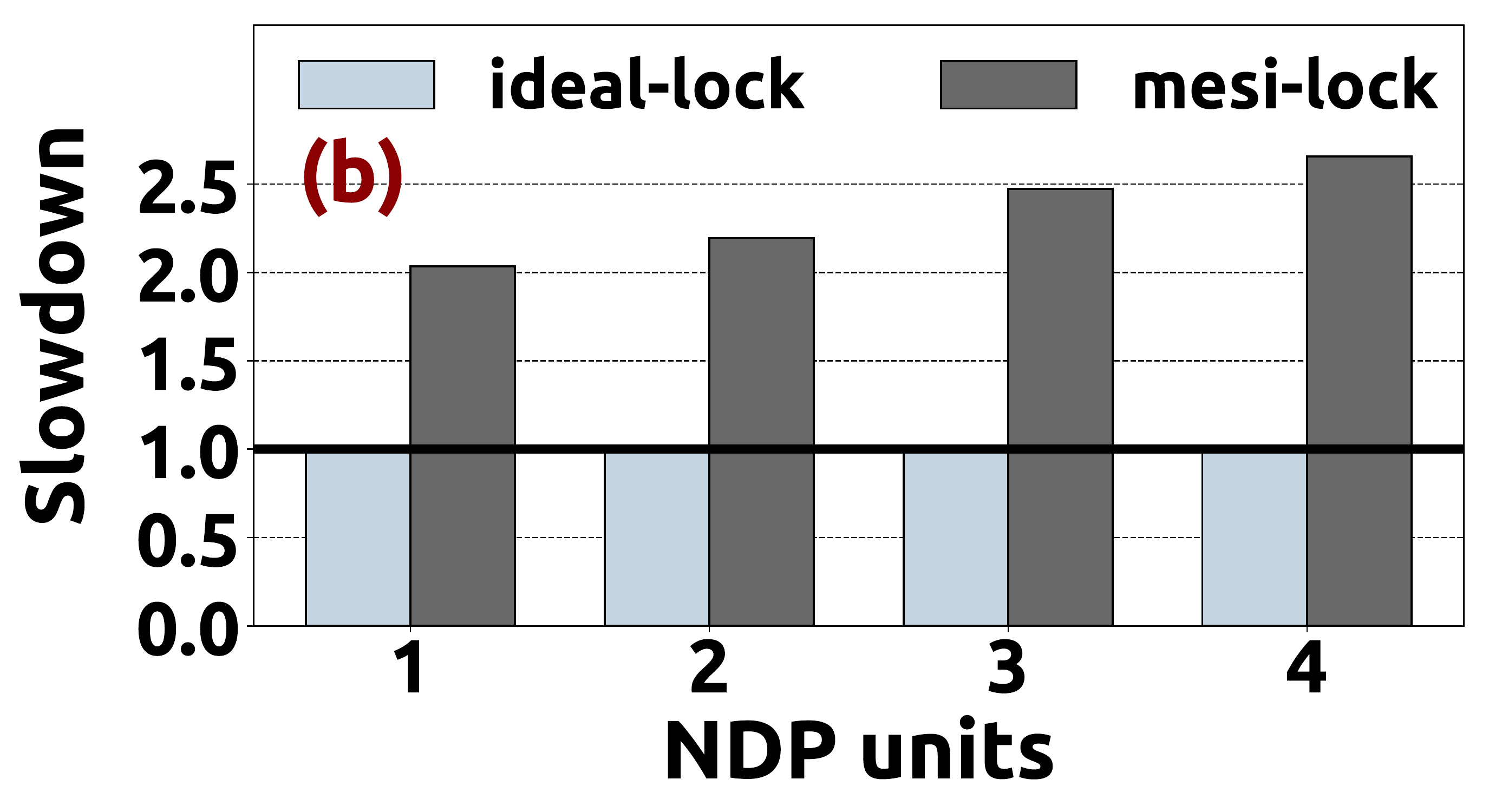}
    \vspace{-4pt}
    %\caption{}
    \label{fig:numaness}
  \end{subfigure} 
  %\vspace{-6pt}
  \caption{Slowdown of a stack data structure using a coherence-based lock over using an \emph{ideal} zero-cost lock, when varying (a) the NDP cores within a single NDP unit and (b) the number of NDP units while keeping core count constant at 60.}
  \label{fig:motcoh}
  \vspace{-4pt}
\end{figure}

In our simulated NDP system, we evaluate the performance achieved by a stack data structure protected with a coarse-grained lock. Figure~\ref{fig:motcoh} shows the slowdown of the stack when using a coherence-based lock~\cite{herlihy2008art} (\emph{mesi-lock}), implemented upon a MESI directory coherence protocol, over using an ideal lock with zero cost for synchronization (\emph{ideal-lock}). First, we observe that the high contention for the cache line containing the \emph{mesi-lock} and the resulting coherence traffic inside the network significantly limit scalability of the stack as the number of cores increases. With 60 NDP cores within a single NDP unit (Figure~\ref{fig:motcoh}a), the stack with \emph{mesi-lock} incurs 2.03$\times$ slowdown over \emph{ideal-lock}. Second, we notice that the non-uniform memory accesses to the cache line containing the \emph{mesi-lock} also impact the scalability of the stack. When increasing the number of NDP units while keeping total core count constant at 60 (Figure~\ref{fig:motcoh}b), the slowdown of the stack with \emph{mesi-lock} increases to 2.66$\times$ (using 4 NDP units) over \emph{ideal-lock}. In \emph{non-uniform} NDP systems, the scalability of coherence-based synchronization is severely limited by the long transfer latency and low bandwidth of the interconnect used between the NDP units.

\subsubsection{\Mpsync{} Synchronization}
In this approach, cores coordinate with each other by exchanging messages (either in software or hardware) in order to reach an agreement. For instance, a recent NDP work~\cite{ahn2015scalable} implements a barrier primitive via hardware message-passing communication among NDP cores, i.e., one core of the system works as a \emph{master} core to collect the synchronization status of the rest.
To improve system performance in \emph{non-uniform} HMC-based NDP systems, Gao et al.~\cite{Gao2015Practical} propose a \emph{tree-style} barrier primitive, where cores exchange messages to first synchronize within a vault, then across the vaults of an HMC cube, and finally across HMC cubes. In general, optimized \mpsync{} synchronization schemes proposed in the literature~\cite{abellan2010g, Tang2019plock,Hoefler2004survey, hensgen1988two,grunwald1994efficient,Gao2015Practical} aim to minimize (i) the number of messages sent among cores, and (ii) expensive network traffic. To avoid the major issues of synchronization via shared memory described above, we design our approach building on the \mpsync{} synchronization concept.

%\vspace{8pt}

\section{\SynCron{}: Overview}\label{Overviewbl}

\SynCron{} is an end-to-end solution for synchronization in NDP architectures that improves performance, has low cost, eases programmability, and supports multiple synchronization primitives.
\SynCron{} relies on the following key techniques:

\noindent \textbf{1. Hardware support for synchronization acceleration:}
We design low-cost hardware units, called \myEngine{}s (\myEngineShort{}s), to coordinate the synchronization among NDP cores of the system. \myEngineShort{}s eliminate the need for complex cache coherence protocols and expensive \emph{rmw} operations, and incur modest hardware cost.

\noindent \textbf{2. Direct buffering of synchronization variables:}
We add a specialized cache structure, the Synchronization Table (\myTableShort{}), inside an \myEngineShort{} to keep synchronization information. Such direct buffering avoids costly memory accesses for synchronization, and enables high performance under low-contention scenarios.

\noindent \textbf{3. Hierarchical message-passing communication:}
We organize the communication hierarchically, with each NDP unit including an \myEngineShort{}. NDP cores communicate with their local \myEngineShort{} that is located in the same NDP unit. \myEngineShort{}s communicate with each other to coordinate synchronization at a global level. Hierarchical communication minimizes expensive communication \emph{across} NDP units, and achieves high performance under high-contention scenarios.

\noindent \textbf{4. Integrated hardware-only overflow management:} 
We incorporate a hardware-only overflow management scheme to efficiently handle scenarios when \myTableShort{} is fully occupied. This programmer-transparent technique effectively limits performance degradation under overflow scenarios.

\subsection{Overview of \SynCron{}}

\begin{figure}[t]
  \centering
  \includegraphics[scale=0.7]{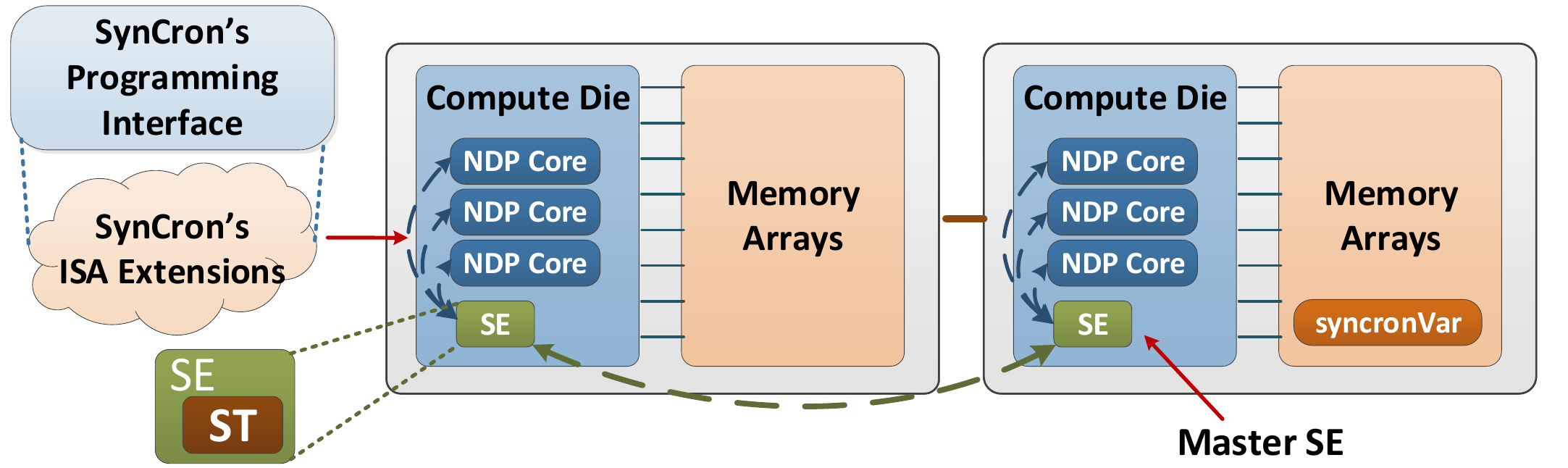}
  \caption{High-level overview of \SynCron{}.}
  \label{fig:overview}
  \vspace{-12pt}
\end{figure}

Figure~\ref{fig:overview} provides an overview of our approach. \SynCron{} exposes a simple programming interface such that programmers can easily use a variety of synchronization primitives in their multithreaded applications when writing them for NDP systems. The interface is implemented using two new instructions that are used by NDP cores to communicate synchronization requests to \myEngineShort{}s. These are general enough to cover all semantics for the most widely used synchronization primitives.

We add one \myEngineShort{} in the compute die of each NDP unit. For a particular synchronization variable allocated in an NDP unit, the \myEngineShort{} that is physically located in the same NDP unit is considered the \masterSE{}. In other words, the \masterSE{} is defined by the address of the synchronization variable. It is responsible for the global coordination of synchronization on that variable, i.e., among all \myEngineShort{}s of the system. All other \myEngineShort{}s are responsible only for the local coordination of synchronization among the cores in the same NDP unit with them.

NDP cores act as clients that send requests to \myEngineShort{}s via hardware message-passing. \myEngineShort{}s act as servers that process synchronization requests. In the proposed hierarchical communication, NDP cores send requests to their local \myEngineShort{}s, while \myEngineShort{}s of different NDP units communicate with the \masterSE{} of the specific variable, to coordinate the process at a global level, i.e., among all NDP units.

When an \myEngineShort{} receives a request from an NDP core for a synchronization variable, it directly buffers the variable in its \myTableShort{}, keeping all the information needed for synchronization in the \myTableShort{}. If the \myTableShort{} is full, we use the main memory as a fallback solution. To hierarchically coordinate synchronization via main memory in \myTableShort{} overflow cases, we design (i) a generic structure, called \emph{\mySyncVar{}}, to keep track of required synchronization information, and (ii) specialized \emph{overflow} messages to be sent among \myEngineShort{}s. The hierarchical communication among \myEngineShort{}s is implemented via corresponding support in message encoding, the \myTableShort{}, and \emph{\mySyncVar{}} structure.

\subsection{\SynCron{}'s Operation} 

\SynCron{} supports locks, barriers, semaphores, and condition variables.
Here, we present \SynCron{}'s operation for locks. \SynCron{} has similar behavior for the other three primitives.

\begin{figure}[t]
  \centering
  \includegraphics[scale=0.68]{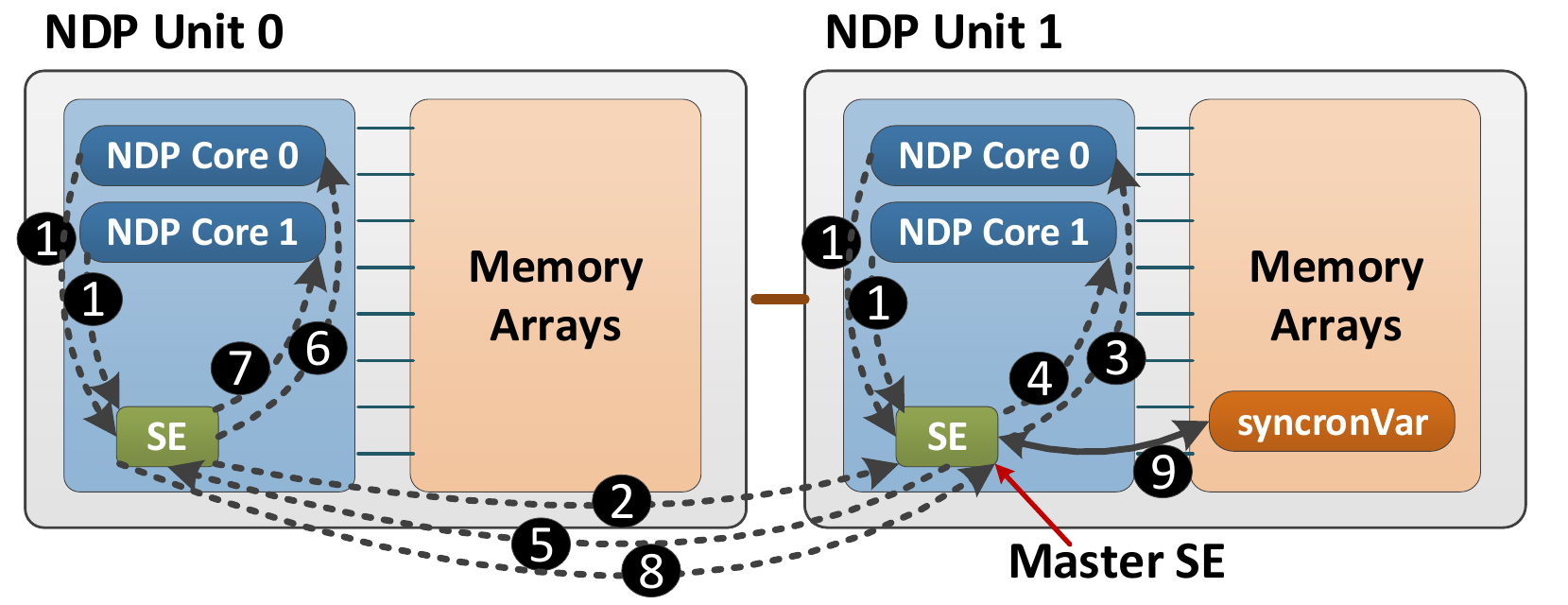}
  \caption{An example execution scenario for a lock requested by \emph{all} NDP cores.}
  \label{fig:lock}
  \vspace{-12pt}
\end{figure}

\noindent
\textbf{Lock Synchronization Primitive:}
Figure~\ref{fig:lock} shows a system 
composed of two NDP units with two NDP cores each. In this example, all cores request and compete for the same lock. First, all NDP cores send \emph{local} lock acquire messages to their \myEngineShort{}s \circled{1}. After receiving these messages, each \myEngineShort{} keeps track of its requesting cores by reserving one new entry in its \myTableShort{}, i.e., directly buffering the lock variable in \myTableShort{}. Each \myTableShort{} entry includes a local waiting list (i.e., a hardware bit queue with one bit for each local NDP core), and a global waiting list (i.e., a bit queue with one bit for each \myEngineShort{} of the system). To keep track of the requesting cores, each \myEngineShort{} sets the bits corresponding to the requesting cores in the local waiting list of the \myTableShort{} entry. When the local \myEngineShort{} receives a request for a synchronization variable \emph{for the first time}, it sends a \emph{global} lock acquire message to the \masterSE{} \circled{2}, which in turn sets the corresponding bit in the global waiting list in its \myTableShort{}. This way, the \masterSE{} keeps track of all requests to a particular variable coming from an \myEngineShort{}, and can arbitrate between different \myEngineShort{}s. The local \myEngineShort{} can then serve successive local requests to the same variable until there are no other local requests. By using the proposed hierarchical communication protocol, the cores send local messages to their local \myEngineShort{}, and the \myEngineShort{} needs to send \emph{only one aggregated} message, on behalf of all its local waiting cores, to the \masterSE{}. As a result, we reduce the need for communication through the narrow, expensive links that connect different NDP units.

The \masterSE{} first prioritizes the local waiting list, granting the lock to its own local NDP cores in sequence (e.g., to NDP Core 0 first \circled{3}, and to NDP Core 1 next \circled{4} in Figure~\ref{fig:lock}). At the end of the critical section, each local lock owner sends a lock release message to its \myEngineShort{} in order to release the lock. When there are no other local requests, the \masterSE{} transfers the control of the lock to the \myEngineShort{} of another NDP unit based on its global waiting list \circled{5}. Then, the local \myEngineShort{} grants the lock to its local NDP cores in sequence (e.g., \circled{6}, \circled{7}). After all local cores release the lock, the \myEngineShort{} sends an \emph{aggregated} global lock release message to the \masterSE{} \circled{8} and releases its \myTableShort{} entry. When the message arrives at the \masterSE{}, if there are no other pending requests to the same variable, the \masterSE{} releases its \myTableShort{} entry. In this example, \myEngineShort{}s directly buffer the lock variable in their \myTableShort{}s. If an \myTableShort{} is \emph{full}, the \masterSE{} globally coordinates synchronization by keeping track of all required information in main memory \circled{9}, via our proposed overflow management scheme (Section~\ref{Overflowbl}).

\section{\SynCron{}: Detailed Design}
\label{Mechanismbl}

\SynCron{} leverages the key observation that all synchronization primitives fundamentally communicate the same information, i.e., a waiting list of cores that participate in the synchronization 
process, and a condition to be met to notify one or more cores. Based on this observation, we design \SynCron{} to cover the four most widely used synchronization primitives.
Without loss of generality, we assume that each NDP core represents a hardware thread context with a unique ID. To support multiple hardware thread contexts per NDP core, the corresponding hardware structures of \SynCron{} need to be augmented to include 1-bit per hardware thread context.

\subsection{Programming Interface and ISA Extensions}\label{SynCronISAbl}

\SynCron{} provides lock, barrier, semaphore and condition variable synchronization primitives, supporting two types of barriers: within cores of the \emph{same} NDP unit and within cores across different NDP units of the system. \SynCron{}'s programming interface (Table~\ref{tab:interface}) implements the synchronization semantics with two new ISA instructions, which are \emph{rich} and \emph{general} enough to express all supported primitives. NDP cores use these instructions to assemble messages for synchronization requests, which are issued through the network to \myEngineShort{}s.

\begin{table}[H]
\vspace{-6pt}
\centering
  \begin{minipage}{.76\textwidth}
  \centering
  \resizebox{\textwidth}{!}{
    \begin{tabular}{l} 
    \toprule
    \textbf{\SynCron{} Programming Interface} \\ %[0.1ex] 
    \midrule
    \textcolor{blue(pigment)}{\mySyncVar} *create\_syncvar (); \\
    \textcolor{blue(pigment)}{void} destroy\_syncvar (\textcolor{blue(pigment)}{\mySyncVar} *svar); \\
    \textcolor{blue(pigment)}{void} lock\_acquire (\textcolor{blue(pigment)}{\mySyncVar} *lock); \\
    \textcolor{blue(pigment)}{void} lock\_release (\textcolor{blue(pigment)}{\mySyncVar} *lock); \\
    \textcolor{blue(pigment)}{void} barrier\_wait\_within\_unit (\textcolor{blue(pigment)}{\mySyncVar} *bar, \textcolor{blue(pigment)}{int} initialCores); \\
    \textcolor{blue(pigment)}{void} barrier\_wait\_across\_units (\textcolor{blue(pigment)}{\mySyncVar} *bar, \textcolor{blue(pigment)}{int} initialCores); \\
    \textcolor{blue(pigment)}{void} sem\_wait (\textcolor{blue(pigment)}{\mySyncVar} *sem, \textcolor{blue(pigment)}{int} initialResources); \\
    \textcolor{blue(pigment)}{void} sem\_post (\textcolor{blue(pigment)}{\mySyncVar} *sem);\\
    \textcolor{blue(pigment)}{void} cond\_wait (\textcolor{blue(pigment)}{\mySyncVar} *cond, \textcolor{blue(pigment)}{\mySyncVar} *lock); \\
    \textcolor{blue(pigment)}{void} cond\_signal (\textcolor{blue(pigment)}{\mySyncVar} *cond);\\ 
    \textcolor{blue(pigment)}{void} cond\_broadcast (\textcolor{blue(pigment)}{\mySyncVar} *cond);\\
    \bottomrule
    \end{tabular}}
  \end{minipage}
   \caption{\label{tab:interface}\SynCron{}'s Programming Interface (i.e., API).}
   \vspace{-14pt}
\end{table}

\textbf{\textit{req\_sync addr, opcode, info}}: This instruction creates a message and commits when a response message is received back. The \emph{addr} register has the address of a synchronization variable, the \emph{opcode} register has the message opcode of a particular semantic of a synchronization primitive (Table~\ref{tab:opcodes}), and the \emph{info} register has specific information needed for the primitive (\emph{MessageInfo} in message encoding of Fig.~\ref{fig:msgencoding}).

\textbf{\textit{req\_async addr, opcode}}: This instruction creates a message and after the message is issued to the network, the instruction commits. The registers \emph{addr}, \emph{opcode} have the same semantics as in \emph{req\_sync} instruction.

\subsubsection{Memory Consistency} We design \SynCron{} assuming a relaxed consistency memory model. The proposed ISA extensions act as memory fences. First, \emph{req\_sync}, commits once a message (ACK) is received (from the local \myEngineShort{} to the core), which ensures that all following instructions will be issued after \emph{req\_sync} has been completed. Its semantics is similar to those of the SYNC and ACQUIRE operations of Weak Ordering (WO)~\cite{Culler1999Parallel} and Release Consistency (RC)~\cite{Culler1999Parallel} models, respectively. Second, \emph{req\_async}, does not require a return message (ACK). It is issued once all previous instructions are completed. Its semantics is similar to that of the RELEASE operation of RC~\cite{Culler1999Parallel}. In the case of WO, \emph{req\_sync} is sufficient. In the case of RC, the \emph{req\_sync} instruction is used for acquire-type semantics, i.e.,  lock\_acquire, barrier\_wait, semaphore\_wait and condition\_variable\_wait, while the \emph{req\_async} instruction is used for release-type semantics, i.e., lock\_release, semaphore\_post, condition\_variable\_signal, and condition\_variable\_broadcast.

\subsubsection{Message Encoding}\label{Encodingbl}

\begin{figure}[H]
  \vspace{-4pt}
  \centering
  \includegraphics[width=0.74\columnwidth]{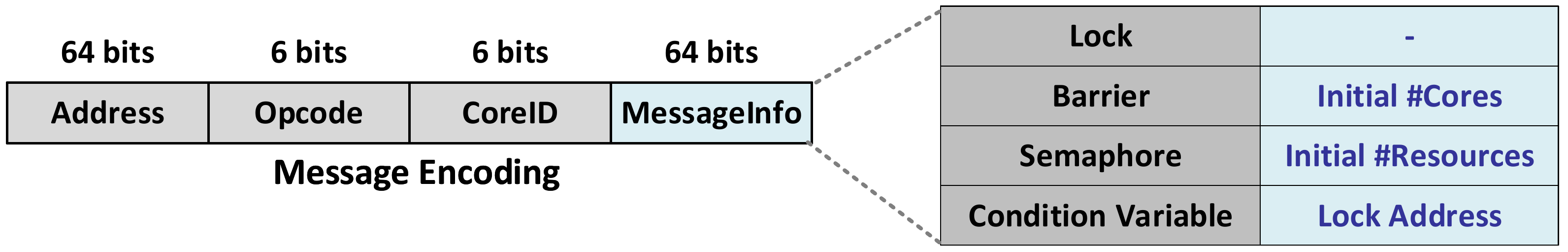}
  \caption{Message encoding of \SynCron{}.} 
  \label{fig:msgencoding}
  \vspace{-12pt}
\end{figure}

Figure~\ref{fig:msgencoding} describes the encoding of the message used for communication between NDP cores and the \myEngineShort{}. Each message includes: (i) the 64-bit address of the synchronization variable, (ii) the message opcode that implements the semantics of the different synchronization primitives (6 bits cover all message opcodes), (iii) the unique ID number of the NDP core (6 bits are sufficient for our simulated NDP system in Section~\ref{Methodologybl}), 
and (iv) a 64-bit field (\emph{MessageInfo}) that communicates specific information needed for each different synchronization primitive, i.e., the number of the cores that participate in a barrier, the initial value of a semaphore, the address of the lock associated with a condition variable.

\noindent\textbf{Hierarchical Message Opcodes.}
\SynCron{} enables a hierarchical scheme, where the \myEngineShort{}s of NDP units communicate with each other to coordinate synchronization at a global level. Therefore, we support two types of messages (Table~\ref{tab:opcodes}): (i) \emph{local}, which are used by NDP cores to communicate with their local \myEngineShort{}, and (ii) \emph{global}, which are used by \myEngineShort{}s to communicate with the \masterSE{}, and vice versa. Since we support two types of barriers (Table~\ref{tab:interface}), we design two message opcodes for a \emph{local} barrier\_wait message sent by an NDP
core to its local \myEngineShort{}: (i) \emph{barrier\_wait\_local\_within\_unit} is used when cores of a single NDP unit participate in the barrier, and (ii) \emph{barrier\_wait\_local\_across\_units} is used when cores from different NDP units participate in the barrier. In the latter case, if a \emph{smaller} number of cores than the total \emph{available} cores of the NDP system participate in the barrier, \SynCron{} supports one-level communication: local \myEngineShort{}s re-direct all messages (received from their local NDP cores) to the \masterSE{}, which globally coordinates the barrier among \emph{all} participating cores. This design choice is a trade-off between performance (\emph{more remote messages}) and hardware/ISA complexity, since the number of participating cores of \emph{each} NDP unit would need to be communicated to the hardware through additional registers in ISA, and message opcodes (\emph{higher complexity}).

\begin{table}[H]
\vspace{-4pt}
\centering
  \begin{minipage}{.70\textwidth}
  \hspace{-6pt}
  \resizebox{\textwidth}{!}{
    \begin{tabular}{c c} 
    \toprule
    \textbf{Primitives} & \textbf{\SynCron{}} Message Opcodes \\ [0.1ex] 
    \midrule
    \midrule
    \multirow{3}{*}{\textbf{Locks}} & lock\_acquire\_global, lock\_acquire\_local, lock\_release\_global \\ 
    & lock\_release\_local,
      lock\_grant\_global,
     lock\_grant\_local \\ 
     & lock\_acquire\_overflow,  
     lock\_release\_overflow, lock\_grant\_overflow \\ 
    \hline
     \multirow{3}{*}{\textbf{Barriers}} & barrier\_wait\_global, barrier\_wait\_local\_within\_unit \\ 
     & barrier\_wait\_local\_across\_units,
    barrier\_depart\_global, barrier\_depart\_local \\ 
    & barrier\_wait\_overflow, barrier\_departure\_overflow \\
    \hline
     \multirow{3}{*}{\textbf{Semaphores}} & sem\_wait\_global, sem\_wait\_local, sem\_grant\_global \\ 
     & sem\_grant\_local,
     sem\_post\_global, sem\_post\_local \\ 
     & sem\_wait\_overflow,  
     sem\_grant\_overflow, sem\_post\_overflow \\ 
    \hline
    \multirow{4}{*}{\shortstack[l]{\textbf{Condition }\\ \textbf{Variables}}} & cond\_wait\_global, cond\_wait\_local, cond\_signal\_global \\ 
    & cond\_signal\_local,
    cond\_broad\_global, cond\_broad\_local \\ 
    & cond\_grant\_global, cond\_grant\_local,
    cond\_wait\_overflow \\ 
    & cond\_signal\_overflow,
    cond\_broad\_overflow,
    cond\_grant\_overflow  \\
    \hline
    \textbf{Other} & decrease\_indexing\_counter \\
    \bottomrule
    \end{tabular}}
  \end{minipage}
  \caption{\label{tab:opcodes}Message opcodes of \SynCron{}.}
  %\vspace{-4pt}
\end{table}

\subsection{Synchronization Engine (\myEngineShort{})} 
\begin{figure}[H]
   \centering
  \includegraphics[scale=0.86]{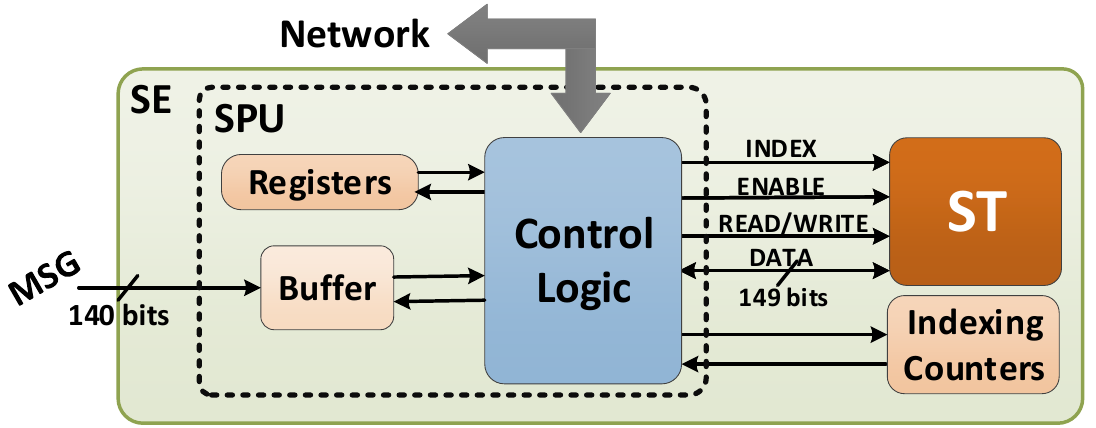}
  \caption{The Synchronization Engine (\myEngineShort{}).}
  \label{fig:syncEng}
  \vspace{-8pt}
\end{figure}

Each \myEngineShort{} module (Figure~\ref{fig:syncEng}) is integrated into the compute die of each NDP unit. An \myEngineShort{} consists of \emph{three} components:

\subsubsection{Synchronization Processing Unit (SPU)}
The SPU is the logic that handles the messages, updates the \myTableShort{}, and issues requests to memory as needed. The SPU includes the control unit, a buffer, and a few registers. The buffer is a small SRAM queue for temporarily storing messages that arrive at the \myEngineShort{}. The control unit implements custom logic with simple logical bitwise operators (and, or, xor, zero) and multiplexers. 

\vspace{-4pt}

\subsubsection{Synchronization Table (\myTableShort{})}
\begin{figure}[H]
  \vspace{2pt}
  \centering
  \includegraphics[width=0.82\columnwidth]{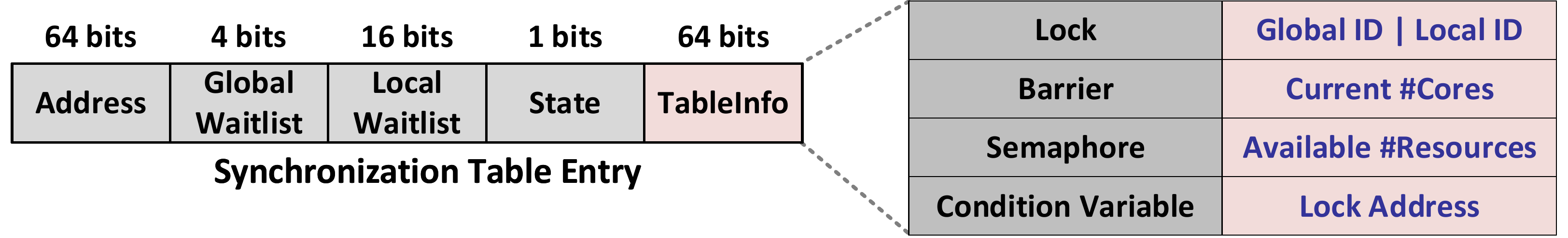}
  \caption{Synchronization Table (\myTableShort{}) entry.}
  \label{fig:stentry}
  %\vspace{-4pt}
\end{figure}

\myTableShort{} keeps track of all the information needed to coordinate synchronization. Each \myTableShort{} has 64 entries. Figure~\ref{fig:stentry} shows an \myTableShort{} entry, which includes: (i) the 64-bit address of a synchronization variable, (ii) the global waiting list used by the \masterSE{} for global synchronization among \myEngineShort{}s, i.e., a hardware bit queue including one bit for each \myEngineShort{} of the system, (iii) the local waiting list used by all \myEngineShort{}s for synchronization among the NDP cores of an NDP unit, i.e., a hardware bit queue including one bit for  each NDP core within the unit, (iv) the state of the \myTableShort{} entry, which can be either \emph{free} or \emph{occupied}, and (v) a 64-bit field (\emph{TableInfo}) to track specific information needed for each synchronization primitive. For the lock primitive, the \emph{TableInfo} field is used to indicate the lock owner that is either an \myEngineShort{} of an NDP unit (\emph{Global ID} represented by the most significant bits)  or a \emph{local} NDP core (\emph{Local ID} represented by the least significant bits). We assume that all NDP cores of an NDP unit have a unique \emph{local ID} within the NDP unit, while all \myEngineShort{}s of the system have a unique \emph{global ID} within the system. The number of bits in the global and local waiting lists of Figure~\ref{fig:stentry} is specific for the configuration of our evaluated system (Section~\ref{Methodologybl}), which includes 16 NDP cores per NDP unit and 4 \myEngineShort{}s (one per NDP unit), and has to be extended accordingly, if the system supports more NDP cores or \myEngineShort{}s.

\subsubsection{Indexing Counters}
If an \myTableShort{} is full, i.e., all its entries are in \emph{occupied} state, \SynCron{} cannot keep track of information for a new synchronization variable in \myTableShort{}. We use the main memory as a fallback solution for such \myTableShort{} overflow (Section~\ref{Overflowbl}). The \myEngineShort{} keeps track of \emph{which} synchronization variables are currently serviced via main memory: similar to MiSAR~\cite{Liang2015MISAR}, we include a small set of counters (\emph{indexing counters}), 256 in current implementation, indexed by the least significant bits of the address of a synchronization variable, as extracted from the message that arrives at an \myEngineShort{}. When an \myEngineShort{} receives a message with acquire-type semantics for a synchronization variable and there is no corresponding entry in the \emph{fully-occupied} \myTableShort{}, the indexing counter for that synchronization variable increases. When an \myEngineShort{} receives a message with release-type semantics for a synchronization variable that is currently serviced using main memory, the corresponding indexing counter decreases. A synchronization variable is currently serviced via main memory, when the corresponding indexing counter is larger than zero. Note that different variables may alias to the same indexing counter. This aliasing does not affect correctness, but it does affect performance, since a variable may unnecessarily be serviced via main memory, while the \myTableShort{} is \emph{not} full.

\subsubsection{Control Flow in \myEngineShort{}} 
\begin{figure}[H]
  \vspace{2pt}
  \centering
  \includegraphics[width=1\columnwidth]{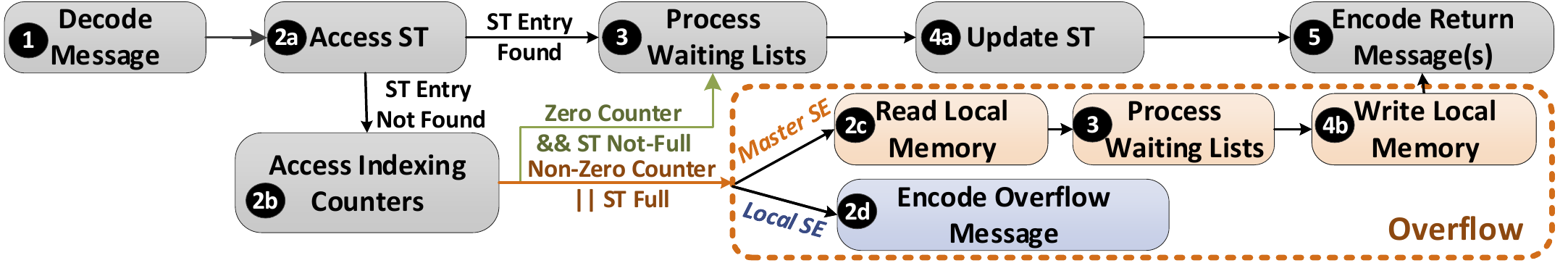}
  \caption{Control flow in \myEngineShort{}.}
  \label{fig:control_flow}
\end{figure}

Figure~\ref{fig:control_flow} describes the control flow in \myEngineShort{}. When an \myEngineShort{} receives a message, it decodes the message \rectangled{1} and accesses the \myTableShort{} \rectangled{\hspace{1pt}2a\hspace{1.2pt}}. If there is an \myTableShort{} entry for the specific variable (depending on its address), the \myEngineShort{} processes the waiting lists \rectangled{3}, updates the \myTableShort{} \rectangled{\hspace{1pt}4a\hspace{1.2pt}}, and encodes return message(s) \rectangled{5}, if needed. If there is \emph{not} an \myTableShort{} entry for the specific variable, the \myEngineShort{} checks the value of the corresponding indexing counter \rectangled{\hspace{1pt}2b\hspace{1.2pt}}: (i) if the indexing counter is zero \emph{and} the \myTableShort{} is not full, the \myEngineShort{} reserves a \emph{new} \myTableShort{} entry and continues with step \rectangled{3}, otherwise (ii) if the indexing counter is larger than zero \emph{or} the \myTableShort{} is full, there is an overflow. In that case, if the \myEngineShort{} is the \masterSE{} for the specific variable, it reads the synchronization variable from \emph{local} memory arrays \rectangled{\hspace{1pt}2c\hspace{1.2pt}}, processes the waiting lists \rectangled{3}, updates the variable in main memory \rectangled{\hspace{1pt}4b\hspace{1.2pt}}, and encodes return message(s) \rectangled{5}, if needed. If the \myEngineShort{} is \emph{not} the \masterSE{} for the specific variable, it encodes an \emph{overflow} message to the \masterSE{} \rectangled{\hspace{1pt}2d\hspace{1.2pt}} to handle overflow.

\subsection{Overflow Management}\label{Overflowbl} 

\SynCron{} integrates a hardware-only overflow management scheme that provides very modest performance degradation (Section~\ref{OverflowEvalbl}) and is programmer-transparent. To handle \myTableShort{} overflow cases, we need to address two issues: (i) where to keep track of required information to coordinate synchronization, and (ii) how to coordinate \myTableShort{} overflow cases between \myEngineShort{}s. For the former issue, we design a generic structure allocated in main memory. For the latter issue, we propose a hierarchical \emph{overflow} communication protocol between \myEngineShort{}s.

\subsubsection{\SynCron{}'s Synchronization Variable}

\begin{figure}[H]
  %\vspace{2pt}
  \centering
  \includegraphics[scale=0.34]{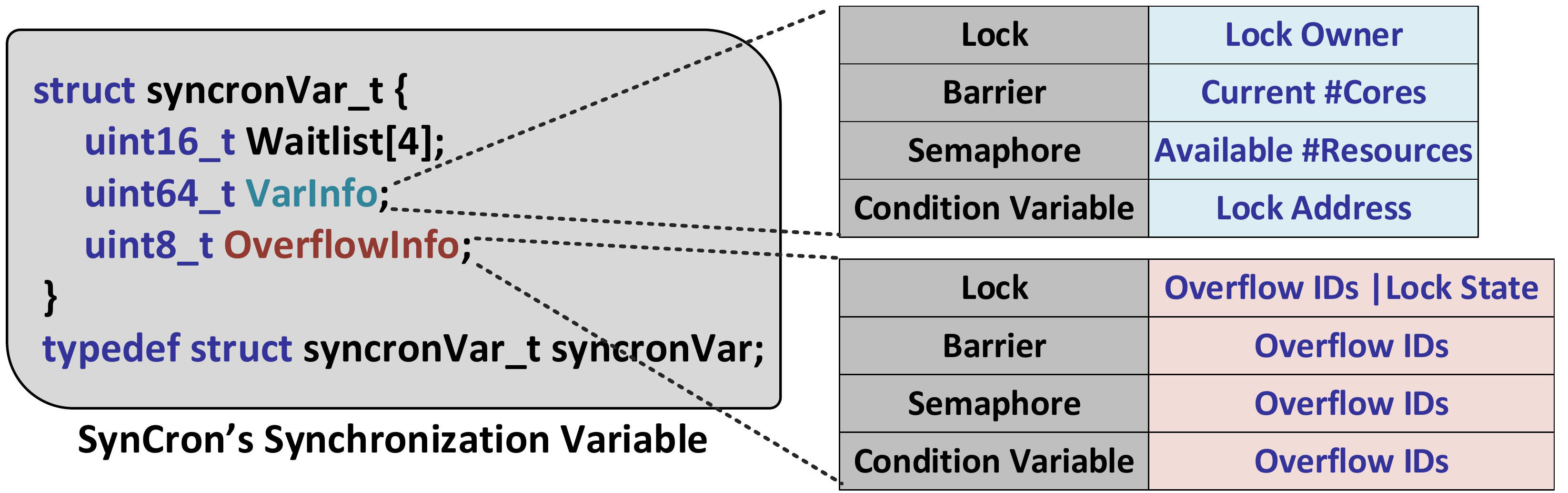}
  \caption{Synchronization variable of \SynCron{} (\emph{\mySyncVar{}}).}
  \label{fig:syncVar}
  \vspace{-7pt}
\end{figure}

We design a generic structure (Figure~\ref{fig:syncVar}), called \emph{\mySyncVar{}}, which is used to coordinate synchronization for all supported primitives in \myTableShort{} overflow cases. \emph{\mySyncVar{}} is defined in the driver of the NDP system, which handles the allocation of the synchronization variables: programmers use \emph{create\_syncvar()} %interface
(Table~\ref{tab:interface}) to create a \emph{new} synchronization variable, the driver allocates the bytes needed for \emph{\mySyncVar{}} in main memory, and returns an opaque pointer that points to the address of the variable. Programmers should not de-reference the opaque pointer and its content can \emph{only} be accessed via \SynCron{}'s API (Table~\ref{tab:interface}).

\emph{\mySyncVar{}} structure includes one waiting list for each %NDP unit
\myEngineShort{} of the system, which has one bit for each NDP core within the NDP unit, and two additional fields (\emph{VarInfo, OverflowInfo}) needed to hierarchically handle \myTableShort{} overflows for all %the supported 
primitives.

\subsubsection{Communication Protocol between \myEngineShort{}s} 
To ensure correctness, \emph{only} the \masterSE{} updates the \emph{\mySyncVar{}} variable: in \myTableShort{} overflow, the SPU of the \masterSE{} issues read or write requests to its local memory to \emph{globally} coordinate synchronization via the \emph{\mySyncVar{}} variable. In our proposed hierarchical design, there are two overflow scenarios: (i) the \myTableShort{} of the \masterSE{} overflows, and (ii) the \myTableShort{} of a local \myEngineShort{} overflows or \myTableShort{}s of multiple local \myEngineShort{}s overflow.

\noindent\textbf{The \myTableShort{} of the \masterSE{} overflows.} 
The other \myEngineShort{}s of the system have \emph{not} overflowed for a specific synchronization variable. Thus, they can still directly buffer this  variable in their local \myTableShort{}s, and serve their local cores %by 
themselves, implementing a hierarchical (two-level) communication with \masterSE{}. %In contrast
The \masterSE{} receives \emph{global} messages from \myEngineShort{}s, and serves a local \myEngineShort{} of an NDP unit using \emph{all} bits in the waiting list of the \emph{\mySyncVar{}} variable associated with that local \myEngineShort{}. Specifically, when it receives a \emph{global} acquire-type message from a local \myEngineShort{}, it sets \emph{all} bits in the corresponding waiting list of the \emph{\mySyncVar{}} variable. When it receives a \emph{global} release-type message from a local \myEngineShort{}, it resets \emph{all} bits in the corresponding waiting list of the \emph{\mySyncVar{}} variable.

\noindent\textbf{The \myTableShort{} of a local \myEngineShort{} overflows.}
%\textbf{In the latter case},
In this scenario,
there are local \myEngineShort{}s that have overflowed for a specific variable, and local \myEngineShort{}s that have \emph{not} overflowed. Without loss of generality, we assume that only one \myEngineShort{} of the system has overflowed. \textbf{The local \myEngineShort{}s that have \emph{not} overflowed} serve their local cores %by 
themselves via their \myTableShort{}s, implementing a hierarchical (two-level) communication with \masterSE{}. When the \masterSE{} receives a \emph{global} message from a local \myEngineShort{} (that has \emph{not} overflowed), it (i) sets (or resets) \emph{all bits} in the waiting list of the \emph{\mySyncVar{}} variable associated with that \myEngineShort{}, and (ii) responds with a \emph{global} message to the local \myEngineShort{}, if needed.

\textbf{The overflowed \myEngineShort{}} needs to notify the \masterSE{} to handle \emph{local} synchronization requests of NDP cores located at \emph{another} NDP unit via main memory. We design \emph{overflow} message opcodes (Table~\ref{tab:opcodes}) to be sent from the local overflowed \myEngineShort{} to the \masterSE{} and back. The overflowed \myEngineShort{} %that overflows 
re-directs all messages (sent from its local NDP cores) for a specific variable to the \masterSE{} using the \emph{overflow} message opcodes, and both the overflowed \myEngineShort{} and the \masterSE{} increase their corresponding indexing counters to indicate that this variable is currently serviced via memory. When the \masterSE{} receives an \emph{overflow} message, it (i) sets (or resets) in the waiting list (associated with the overflowed \myEngineShort{}) of the \emph{\mySyncVar{}} variable, the bit that corresponds to the \emph{local ID} of the NDP core within the NDP unit, (ii) sets (or resets) in the \emph{OverflowInfo} field of the \emph{\mySyncVar{}} variable the bit that corresponds to the \emph{global ID} of the overflowed \myEngineShort{} to keep track of \emph{which} \myEngineShort{} (or \myEngineShort{}s) of the system has overflowed, and (iii) responds with an \emph{overflow} message to that \myEngineShort{}, if needed. The \emph{local ID} of the NDP core, and the \emph{global ID} of the overflowed \myEngineShort{} are encoded in the \emph{CoreID} field of the message (Figure~\ref{fig:msgencoding}). When all bits in the waiting lists of the \emph{\mySyncVar{}} variable become zero (upon receiving a  release-type message), the \masterSE{} decrements the corresponding indexing counter. Then, it sends a \emph{decrease\_index\_counter} message (Table~\ref{tab:opcodes}) to the overflowed \myEngineShort{} (based on the set bit that is tracked in the \emph{OverflowInfo} field), which decrements its corresponding indexing counter.

\subsection{\SynCron{} Enhancements}

\subsubsection{\emph{RMW} Operations}
It is straightforward to extend \SynCron{} to support simple atomic \emph{rmw} operations inside the \myEngineShort{} (by adding a lightweight ALU). The \masterSE{} could be responsible for executing atomic \emph{rmw} operations on a variable depending on its address. We leave that for future work.

\subsubsection{Lock Fairness}
When local cores of an NDP unit repeatedly request a lock from their local \myEngineShort{}, the \myEngineShort{} repeatedly grants the lock within its unit, potentially causing unfairness and delay to other NDP units. To prevent this, an extra field of a local grant counter could be added to the \myTableShort{} entry. The counter increases every time the \myEngineShort{} grants the lock to a local core. If the counter exceeds a predefined threshold, then when the \myEngineShort{} receives a lock release, it transfers the lock to another \myEngineShort{} (assuming other \myEngineShort{}s request the lock). The host OS or the user could dynamically set this threshold via a dedicated register. We leave the exploration of such fairness mechanisms to future work.

\subsection{Comparison with Prior Work}
\SynCron{}'s design shares some of its design concepts 
with SSB~\cite{Zhu2007SSB}, LCU~\cite{Vallejo2010Architectural}, and MiSAR~\cite{Liang2015MISAR}. However, \SynCron{} is more general, supporting the four most widely used synchronization primitives, and easy-to-use thanks to its high-level programming interface.

Table~\ref{tab:comparison} qualitatively compares \SynCron{} with these schemes. SSB and LCU support only lock semantics, thus they introduce two \emph{ISA extensions} for a simple lock. MiSAR introduces seven ISA extensions to support three primitives and handle overflow scenarios. \SynCron{} includes two ISA extensions for four \emph{supported primitives}. A \emph{spin-wait approach} performs consecutive synchronization retries, typically incurring high energy consumption. A \emph{direct notification} scheme sends a direct message to only one waiting core when the synchronization variable becomes available, minimizing the traffic involved upon a release operation. SSB, LCU and MiSAR are tailored for \emph{uniform} memory systems. In contrast, \SynCron{} is the \emph{only} hardware synchronization mechanism that targets NDP systems as well as \emph{non-uniform} memory systems.

SSB and LCU handle \emph{overflow} in hardware synchronization resources using a pre-allocated table in main memory, and if it overflows, they switch to software exception handlers (handled by the programmer), which typically incur large overheads (due to OS intervention) when overflows happen at a non-negligible frequency. To avoid falling back to main memory, which has high latency, and using expensive software exception handlers, MiSAR requires the programmer to handle overflow scenarios using alternative software synchronization libraries (e.g., pthread library provided by the OS). This approach can provide performance benefits in CPU systems, since alternative synchronization solutions can exploit low-cost accesses to caches and hardware cache coherence. However, in NDP systems alternative solutions would by default use main memory due to the absence of shared caches and hardware cache coherence support. Moreover, when overflow occurs, MiSAR's accelerator sends abort messages to all participating CPU cores notifying them to use the alternative solution, and when the cores finish synchronizing via the alternative solution, they notify MiSAR's accelerator to switch back to hardware synchronization. This scheme introduces additional hardware/ISA complexity, and communication between the cores and the accelerator, thus incurring high network traffic and communication costs, as we show in Section~\ref{OverflowEvalbl}. In contrast, \SynCron{} directly falls back to memory via a fully-integrated hardware-only overflow scheme, which provides graceful performance degradation (Section~\ref{OverflowEvalbl}), and is completely transparent to the programmer: programmers \emph{only} use \SynCron{}'s high-level API, similarly to how software libraries are in charge of synchronization.

\vspace{2pt}
\renewcommand{\arraystretch}{1.2}
\begin{table}[t]
\centering
  \begin{minipage}{.78\textwidth}
  \resizebox{\textwidth}{!}{
   %\tiny
    \begin{tabular}{ l c c c c} 
\toprule
 %\noalign{\global\arrayrulewidth=0.52mm}
  %\arrayrulecolor{black}\hline
%\hline
   & SSB~\cite{Zhu2007SSB} &
  LCU~\cite{Vallejo2010Architectural} &
  MiSAR~\cite{Liang2015MISAR} &
  \textbf{SynCron} \\
  \midrule
  \midrule
  %\noalign{\global\arrayrulewidth=0.52mm}
  %\arrayrulecolor{black}\hline
   
 Supported Primitives & 
  1 &
  1 &
  3 &
  \textbf{4} \\   \hline

  ISA Extensions &
  2 &
  2 &
  7 &
  \textbf{2} \\ \hline

  Spin-Wait Approach &
  yes &
  yes &
  no & 
  \textbf{no} \\ \hline
  
  Direct Notification &
  no &
  yes &
  yes &
  \textbf{yes} \\ \hline

  Target System &
  uniform &
  uniform &
  uniform &
  \textbf{non-uniform} \\ \hline
  
  Overflow &
  partially  &
  partially  &
  handled by  &
  %%\textbf{programmer} \\ 
  \textbf{fully} \\ 
  
  Management &
  integrated &
  integrated &
  programmer &
  %%\textbf{transparent} \\ \hline
  \textbf{integrated} \\ %\hline
  
%  Code Complexity &
%  low &
%  low &
%  low &
%  \textbf{very low} \\ \hline

  \bottomrule
  %\noalign{\global\arrayrulewidth=0.52mm}
  %\arrayrulecolor{black}\hline
    \end{tabular}
    }
  \end{minipage}
  \caption{\label{tab:comparison}Comparison of \SynCron{} with prior mechanisms.}
  \vspace{-10pt}
\end{table}
\renewcommand{\arraystretch}{1}

\subsection{Use of \SynCron{} in Conventional Systems}
The baseline NDP architecture ~\cite{ahn2015scalable,Zhang2018GraphP,Youwei2019GraphQ,Tsai2018Adaptive,Gao2015Practical} we assume in this work shares key design principles with conventional NUMA systems. However, unlike NDP systems, NUMA CPU systems (i) have a shared level of cache (within a NUMA socket and/or across NUMA sockets), (ii) run multiple multi-threaded applications, i.e., a high number of software threads executed in hardware thread contexts, and (iii) the OS migrates software threads between hardware thread contexts to improve system performance. Therefore, although \SynCron{} could be implemented in such commodity systems, our proposed hardware design would need extensions. First, \SynCron{} could exploit the low-cost accesses to \emph{shared} caches in conventional CPUs, e.g., including an additional level in \SynCron{}'s hierarchical design to use the shared cache for efficient synchronization within a NUMA socket, and/or handling overflow scenarios by falling back to the low-latency cache instead of main memory. Second, \SynCron{} needs to support use cases (ii) and (iii) listed above in such systems, i.e., including larger \myTableShort{}s and waiting lists to satisfy the needs of multiple multithreaded applications, handling the OS thread migration scenarios across hardware thread contexts, and handling multiple synchronization requests sent from different software threads with the same hardware ID to \myEngineShort{}s, when different software threads are executed on the same hardware thread context. We leave the optimization of \SynCron{}'s design for conventional systems to future work.

\section{{Methodology}}\label{Methodologybl}

\noindent\textbf{Simulation Methodology.} We use an in-house simulator that integrates ZSim~\cite{Sanchez2013Zsim} and Ramulator~\cite{kim2015ramulator}. We model 4 NDP units (Table ~\ref{table:zsim_parameters}), each with 16 in-order cores. The cores issue a memory operation after the previous one has completed, i.e., there are no overlapping operations issued by the same core. Any write operation {is} completed (and the latency is accounted for in our simulations) before executing the next instruction. To ensure memory consistency, compiler support~\cite{Rutgers2013Portable} guarantees that there is no reordering around the \emph{sync} instructions and a read is inserted after a write inside a critical section.

\begin{table}[t]
    \vspace{2pt}
    \centering
    \resizebox{0.88\columnwidth}{!}{%
    %\begin{threeparttable}
    \begin{tabular}{l l}
    \toprule
    %\textbf{NDP Cores}\tnote{*} & 16 in-order cores @2.5~GHz per NDP unit
    \textbf{NDP Cores} & 16 in-order cores @2.5~GHz per NDP unit
    \\
    \midrule
    \textbf{L1 Data + Inst. Cache} & private, 16KB, 2-way, 4-cycle;  64 B line; 23/47 pJ per hit/miss~\cite{Muralimanohar2007Optimizing}  \\
    \midrule
    \textbf{NDP Unit} & buffered crossbar network with packet flow control; 1-cycle arbiter; \\
    {\textbf{Local Network}}  & 1-cycle per hop~\cite{Agarwal2009garnet}; 0.4 pJ/bit per hop~\cite{Wolkotte2005Energy}; \\
    & M/D/1 model~\cite{Narayan2015} for queueing latency; \\
    \midrule
    \multirow{2}{*}{\textbf{DRAM HBM}} & 4 stacks; 4GB HBM 1.0~\cite{HBM_old,Lee2016Simultaneous}; 500MHz with 8 channels; \\
    & nRCDR/nRCDW/nRAS/nWR 7/6/17/8 {ns}~\cite{kim2015ramulator,Ghose2019Demystifying}; 7 pJ/bit~\cite{Mingyu2019Alleviating} \\ 
    \midrule
    \multirow{2}{*}{\textbf{DRAM HMC}} & 4 stacks; 4GB HMC 2.1; 1250MHz; 32 vaults per stack;  \\
    & nRCD/nRAS/nWR 17/34/19 {ns}~\cite{Ghose2019Demystifying,kim2015ramulator} \\
    \midrule
    \multirow{2}{*}{\textbf{DRAM DDR4}} & 4 DIMMs; 4GB {each DIMM} DDR4 2400MHz; \\
    & nRCD/nRAS/nWR 16/39/18 {ns}~\cite{Ghose2019Demystifying,kim2015ramulator} \\
    \midrule
    \textbf{Interconnection Links} & 12.8GB/s per direction; 40 ns per cache line;  \\
    {\textbf{Across NDP Units}} & 20-cycle; 4 pJ/bit\\
    \midrule
    \multirow{1}{*}{\textbf{Synchronization}} & SPU @1GHz clock frequency~\cite{shao2016aladdin}; {8$\times$ 64-bit registers;} \\
    \textbf{Engine} & {{b}uffer: 280B;} \myTableShort{}: 1192B, 64 entries, 1-cycle~\cite{Muralimanohar2007Optimizing}; \\
    & {i}ndexing {c}ounters: 2304B, 256 entries (8 LSB of the address), 2-cycle~\cite{Muralimanohar2007Optimizing} \\
    \bottomrule
    \end{tabular}
 }
 \caption{Configuration of our simulated system.}
\label{table:zsim_parameters}
\vspace{-10pt}
\end{table}

We evaluate three NDP configurations for different memory technologies, namely 2D, 2.5D, 3D NDP. The 2D NDP configuration {uses a} DDR4 memory model and resembles recent 2D NDP systems~\cite{Lavenier2016DNA,upmem,devaux2019,Gomez2021Benchmarking}. In the 2.5D NDP configuration, each compute die of NDP units (16 NDP cores) is connected to an HBM stack via an interposer, similar to current GPUs~\cite{mojumder2018profiling,designontap} and FPGAs~\cite{ultrascale,Singh2020NERO}. For the 3D NDP configuration, we use {the} HMC memory model, {where} the compute die of {the} NDP unit {is} located in the logic layer of the memory stack, as in prior works~\cite{Youwei2019GraphQ,Zhang2018GraphP,ahn2015scalable,Boroumand2018Google}. Due to space limitations, we present detailed evaluation results for the 2.5D NDP configuration, and provide a sensitivity study for the different NDP configurations in Section~\ref{MemoryTechnologies}.

We model a crossbar network within each NDP unit, simulating queuing latency using {the} M/D/1 model~\cite{Narayan2015}. We count in ZSim-Ramulator all events for caches, i.e., number of hits/misses, network, i.e., number of bits transferred inside/across NDP units, and memory, i.e., number of total memory accesses, and use CACTI~\cite{Muralimanohar2007Optimizing} 
and parameters reported in prior {works}~\cite{Mingyu2019Alleviating,Wolkotte2005Energy,Tsai2018Adaptive} to calculate energy. To estimate the latency in \myEngineShort{}, we use CACTI for \myTableShort{} and indexing counters, and Aladdin~\cite{shao2016aladdin} for the SPU with 1GHz at 40nm. Each message is served in 12 cycles, corresponding to the message (barrier\_depart\_global) that takes the longest time.

\noindent\textbf{{Workloads.}} We evaluate workloads with both (i) coarse-grained synchronization, i.e., including only a few synchronization variables to protect shared data, {leading to} cores highly {contending for} them (\emph{high-contention}), and (ii) fine-{grained} synchronization, i.e., including a large number of synchronization variables, each of them protecting a small granularity of shared data, leading to cores not frequently contending for the same variables at the same time (\emph{low-contention}). We use the term \emph{synchronization intensity} to refer to the ratio of synchronization {operations} over {other} computation {in} the workload. As this ratio increases, synchronization latency affects the total execution time of the workload more.

We study three classes of applications (Table~\ref{tab:workloads}), all well suited for NDP. First, we evaluate pointer-chasing workloads, i.e., lock-based concurrent data structures from the ASCYLIB library~\cite{David2015ASCYLIB}, used as key-value sets. In ASCYLIB's Binary Search Tree (BST)~\cite{Drachsler2014Practical}, the 
lock memory requests are only 0.1\% of the total memory requests, so we also evaluate an external fine-grained locking BST from~\cite{Siakavaras2017Combining}. Data structures are initialized with a fixed size and statically partitioned across NDP units, except for BSTs, which are distributed randomly. In these benchmarks, each core performs a fixed number of operations. We use lookup operations for data structures that support it, deletion for the rest, and push and pop operations for stack and queue. 
Second, we evaluate graph applications with fine-grained synchronization from Crono~\cite{Ahmad2015CRONO,hong2014simplifying} (push version), where the output array has read-write data. All real-world graphs~\cite{davis2011florida} used are undirected and statically partitioned across NDP units, where the vertex data is equally distributed across cores. Third, we evaluate time series analysis~\cite{TL17}, using SCRIMP, and \emph{real} data sets from Matrix Profile~\cite{MPROFILEI}. We replicate the input data in each NDP unit and partition the output array (read-write data) across NDP units.

\begin{table}[!htb]
\vspace{-6pt}
\centering
\begin{minipage}{.82\textwidth}
   \resizebox{\textwidth}{!}{
    \begin{tabular}{l l} 
    \toprule
    \textbf{Data Structure} &  \textbf{Configuration} \\ %[0.5ex] 
    \midrule
    \midrule
    Stack~\cite{David2015ASCYLIB} & 100K - 100\% push \\
    %\hline
    Queue~\cite{Michael1996Simple,David2015ASCYLIB} & 100K - 100\% pop \\
    %\hline
    Array Map~\cite{David2015ASCYLIB,Guerraoui2016Optimistic} & 10 - 100\% lookup \\ 
    %\hline
    Priority Queue~\cite{David2015ASCYLIB,Pugh1990Concurrent,Alistarh2015Spraylist} & 20K - 100\% deleteMin \\ 
    %\hline
    Skip List~\cite{David2015ASCYLIB,Pugh1990Concurrent} & 5K - 100\% deletion\\ 
    %\hline
    Hash Table~\cite{David2015ASCYLIB,herlihy2008art} & 1K - 100\% lookup\\
    %\hline
    Linked List~\cite{David2015ASCYLIB,herlihy2008art} & 20K - 100\% lookup \\
    %\hline
    Binary Search Tree Fine-Grained (BST\_FG)~\cite{Siakavaras2017Combining}%~\cite{Siakavaras2017Combining,GithubConcMap} 
    & 20K - 100\% lookup \\ 
    %\hline
    Binary Search Tree Drachsler (BST\_Drachsler)~\cite{David2015ASCYLIB,Drachsler2014Practical} & 10K - 100\% deletion \\
    \bottomrule
    \end{tabular}}
  \end{minipage}\vspace{10pt}
  \begin{minipage}{.98\textwidth}
  \vspace{1pt}
  \resizebox{\textwidth}{!}{
    \begin{tabular}{l c c} 
    \toprule
    \textbf{Real Application} & \textbf{Locks} & \textbf{Barriers} \\ %[0.5ex] 
    \midrule
    \midrule
    Breadth First Search (\textbf{bfs})~\cite{Ahmad2015CRONO} & \checkmark & \checkmark \\ 
    %\hline
    Connected Components (\textbf{cc})~\cite{Ahmad2015CRONO} & \checkmark & \checkmark \\
    %\hline
    Single Source Shortest Paths (\textbf{sssp})~\cite{Ahmad2015CRONO} & \checkmark & \checkmark \\
    %\hline
    Pagerank (\textbf{pr})~\cite{Ahmad2015CRONO} & \checkmark & \checkmark \\
    %\hline
    Teenage Followers (\textbf{tf})~\cite{hong2014simplifying} & \checkmark  & - \\ 
    %\hline
    Triangle Counting (\textbf{{tc}})~\cite{Ahmad2015CRONO}  & \checkmark  & \checkmark \\ 
    \midrule 
    Time Series Analysis (\textbf{ts})~\cite{MPROFILEI} & \checkmark & \checkmark \\
    \bottomrule
    \end{tabular}
    \hspace{1pt}
    \begin{tabular}{l c}
    \toprule
    \textbf{Real Application} & \textbf{Input Data Set} \\ %[0.5ex] 
    \midrule
    \midrule
     & wikipedia \\ 
     & -20051105 (\textbf{wk}) \\
   \textbf{bfs, cc, sssp,} & soc-LiveJournal1 (\textbf{sl})\\  
   \textbf{pr, tf, tc} & sx-stackoverflow (\textbf{sx}) \\ 
    & com-Orkut (\textbf{co}) \\ 
    \midrule 
    \multirow{2}{*}{\shortstack[l]{\textbf{ts}}} & air quality (\textbf{air})\\
    & energy consumption (\textbf{pow}) \\ 
    \bottomrule
    \end{tabular}
    }
  \end{minipage}
  \caption{\label{tab:workloads} Summary of all workloads used in our evaluation.}
  \vspace{-14pt}
\end{table}

\noindent\textbf{{Comparison Points.}} We compare \SynCron{} with three schemes: (i) \emph{Central}: a \mpsync{} scheme that supports all primitives by extending the barrier primitive of Tesseract~\cite{ahn2015scalable}, i.e., one dedicated NDP core {in the entire NDP system} acts as server and coordinates synchronization among all NDP cores of the system by issuing memory requests to synchronization variables {via} its memory hierarchy, while the {remaining} client cores communicate with it via hardware message-passing; (ii) \emph{Hier}: a hierarchical \mpsync{} scheme that supports all {primitives, similar} to the barrier primitive of~\cite{Gao2015Practical} (or hierarchical lock of~\cite{Tang2019plock}), i.e., one NDP core per NDP unit acts as server and coordinates synchronization by issuing memory requests to synchronization variables via its memory hierarchy (including caches), and communicates with other servers and local client cores (located at the same NDP unit with it) via hardware message-passing; (iii) \emph{Ideal}: an ideal scheme with zero performance overhead for synchronization. In our evaluation, each NDP core runs one thread. {For fair comparison, we use the same number of client cores, i.e., 15 per NDP unit, that execute the main workload {for all schemes}. For synchronization, we add one server core for the entire system in \naive{}, one {server core} per NDP unit for \hier{}, and one \myEngineShort{} per NDP unit for \SynCron. For \SynCron{}, we disable one core per NDP unit} to match the same number of client cores as the previous schemes. Maintaining the same thread-level parallelism for executing the main kernel is consistent with prior works on message-passing synchronization~\cite{Tang2019plock,Liang2015MISAR}. 

\section{Evaluation}\label{Evaluationbl}
\vspace{-4pt}
\subsection{Performance}\label{Performancebl}
\vspace{-2pt}

\subsubsection{Synchronization Primitives} 
\begin{figure}[H]
\centering
  \includegraphics[width=0.96\columnwidth]{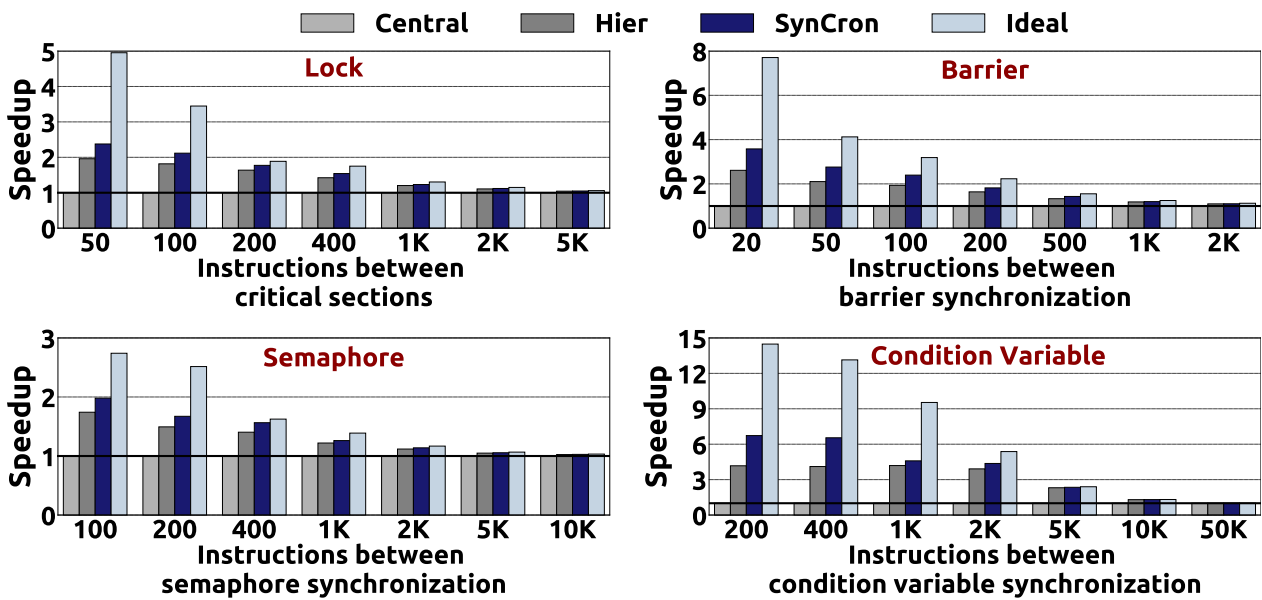}
  \caption{Speedup of different synchronization primitives.}
   \label{fig:mbench}
   \vspace{-13pt}
\end{figure}

Figure~\ref{fig:mbench} evaluates all supported primitives using 60 cores, varying the interval (in terms of instructions) between two synchronization points. We devise simple benchmarks, where cores repeatedly request a single synchronization variable. For lock, the critical section is empty, i.e., it does not include any instruction. For semaphore and condition variable, half of the cores execute sem\_wait/cond\_wait, while the rest execute sem\_post/cond\_signal, respectively. As the interval between synchronization points becomes smaller, \SynCron{}'s performance benefit increases. For an interval of 200 instructions, \SynCron{} outperforms \naive{} and \hier{} by 3.05$\times$ and 1.40$\times$ respectively, averaged across all primitives. \SynCron{} outperforms \hier{} due to directly buffering synchronization variables in low-latency \myTableShort{}s, and achieves the highest benefits in the condition variable primitive (by 1.61$\times$), since this benchmark has higher synchronization intensity compared to the rest: cores coordinate for both the condition variable and the lock associated with it. When the interval between synchronization operations becomes larger, synchronization requests become less dominant in the main workload, and thus all schemes perform similarly. Overall, \SynCron{} outperforms prior schemes for all different synchronization primitives.

\subsubsection{Pointer-Chasing Data Structures} 
\begin{figure}[t]
%\vspace{-8pt}
\hspace{-5pt}
  \begin{subfigure}[h]{\textwidth}
   \includegraphics[scale=0.28]{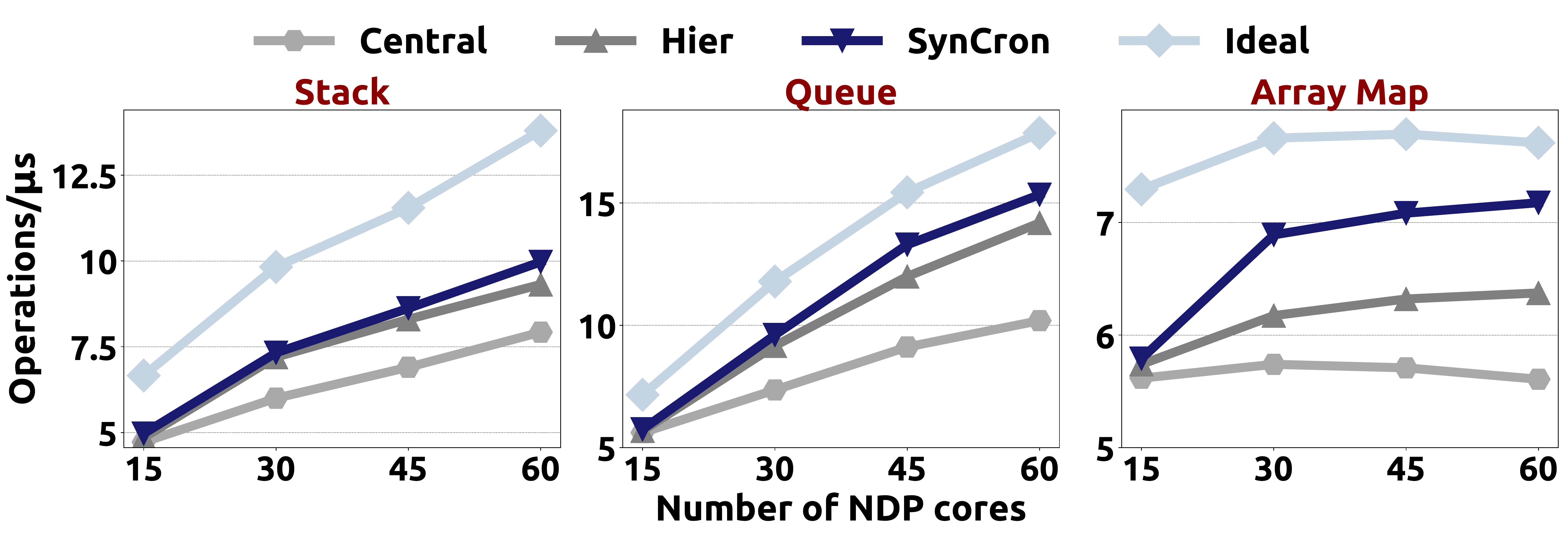}
    %\caption{} 
    %\label{fig:scalability}
  \end{subfigure}
  \begin{subfigure}[h]{\textwidth}
  \hspace{-4pt}
   \includegraphics[scale=0.28]{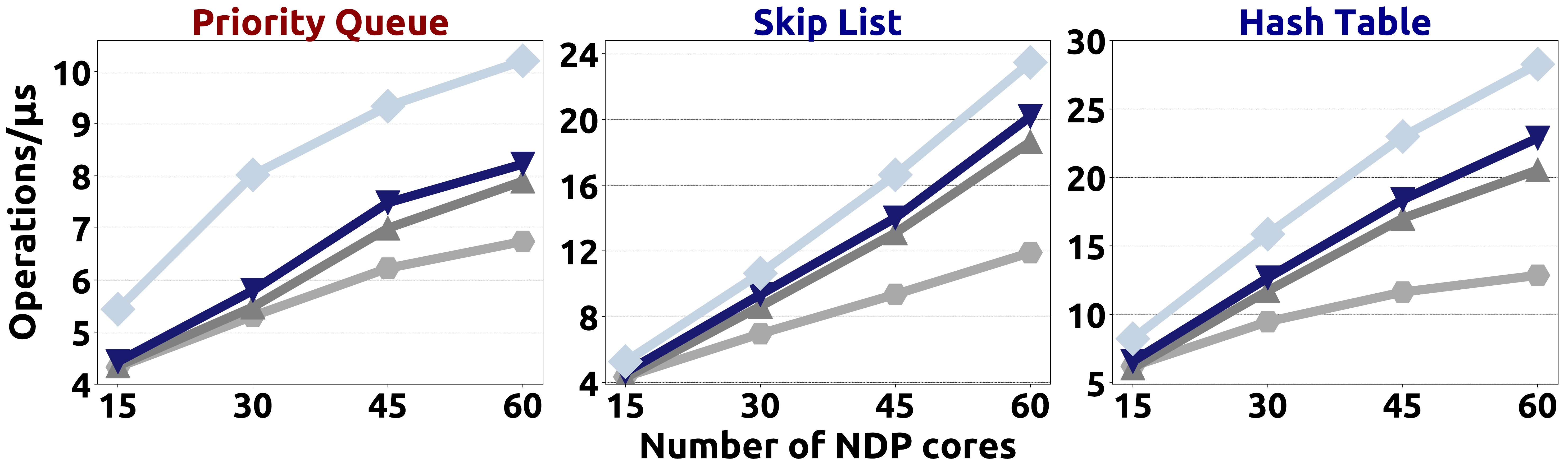}
    %\caption{}
    %\label{fig:numaness}
  \end{subfigure} 
  \begin{subfigure}[h]{\textwidth}
  \hspace{-4pt}
   \includegraphics[scale=0.28]{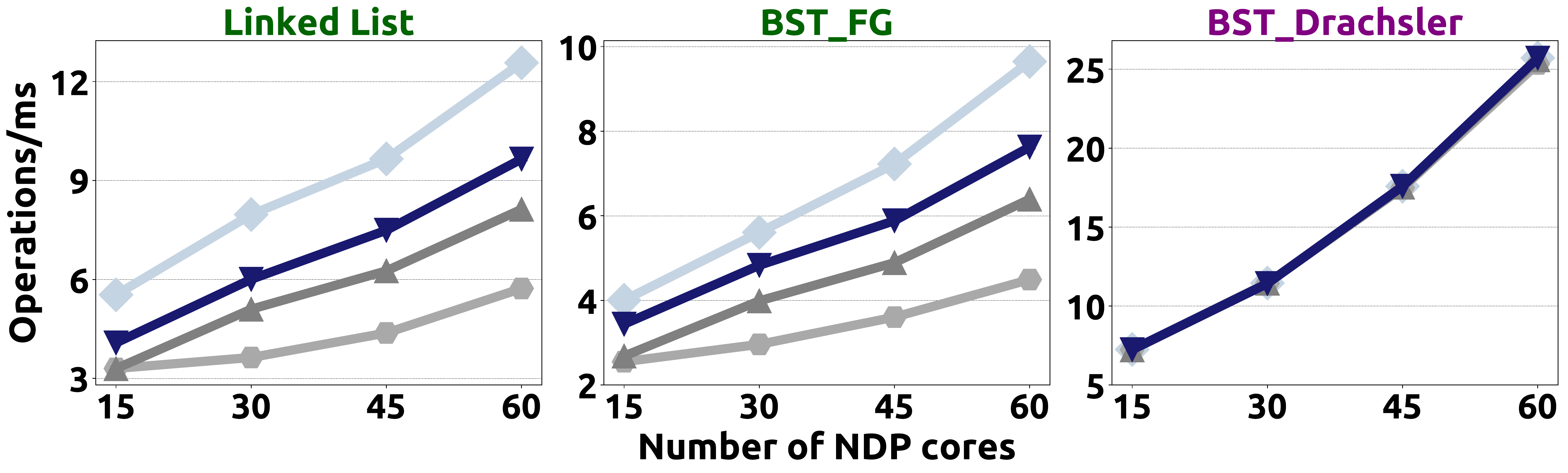}
    %\caption{}
    %\label{fig:numaness}
  \end{subfigure}
  \caption{Throughput of pointer-chasing using data structures.}
  \label{fig:ds_thr}
  %\vspace{-10pt}
  \vspace{-16pt}
\end{figure}

Figure~\ref{fig:ds_thr} shows the throughput for all schemes in pointer-chasing varying the NDP cores in steps of 15, each time adding one NDP unit.

We observe four different patterns. First, \emph{stack}, \emph{queue}, \emph{array map}, and \emph{priority queue} incur high contention, as all cores heavily contend for a few variables. \emph{Array map} has the lowest scalability due to a larger critical section. In high-contention scenarios, hierarchical schemes (\hier{}, \SynCron{}) perform better by reducing the expensive traffic across NDP units. \SynCron{} outperforms \hier{}, since the latency cost of using \myEngineShort{}s that update small \myTableShort{}s is lower than using NDP cores as servers that update larger caches. Second, \emph{skip list} and \emph{hash table} incur medium contention, as different cores may work on different parts of the data structure. For these data structures, hierarchical schemes perform better, as they minimize the expensive traffic, and multiple server cores concurrently serve requests to their local memory. \SynCron{} retains most of the performance benefits of \ideal{}, incurring only 19.9\% overhead with 60 cores, and outperforms \hier{} by 9.8\%. Third, \emph{linked list} and \emph{BST\_FG} exhibit low contention and high synchronization demand, as each core requests multiple locks concurrently. These data structures cause higher synchronization-related traffic inside the network compared to \emph{skip list} and \emph{hash table}, and thus \SynCron{} further outperforms \hier{} by 1.19$\times$ due to directly buffering synchronization variables in \myTableShort{}s. Fourth, in \emph{BST\_Drachsler} lock requests constitute only 0.1\% of the total requests, and all schemes perform similarly.
Overall, we conclude that \SynCron{} achieves higher throughput \emph{than prior mechanisms} under different scenarios with diverse conditions.

\subsubsection{Real Applications} 
\begin{figure}[H]
\vspace{-2pt}
\centering
  \hspace{-5pt}
  \includegraphics[width=\columnwidth]{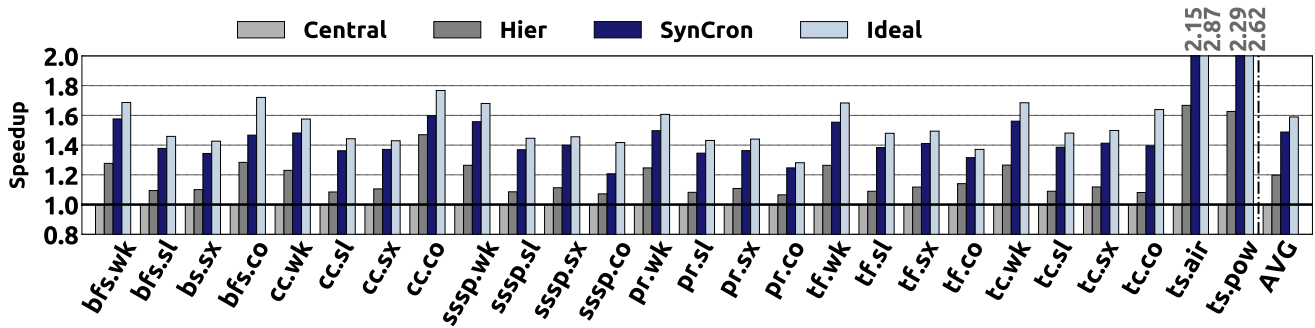}
  \caption{Speedup in real applications normalized to \naive{}.}
  \label{fig:speedup_real}
  \vspace{-14pt}
\end{figure}

Figure~\ref{fig:speedup_real} shows the performance of all schemes with real applications using all NDP units, normalized to \naive{}. Averaged across 26 application-input combinations, \SynCron{} outperforms \naive{} by 1.47$\times$ and \hier{} by 1.23$\times$, and performs within 9.5\% of \ideal{}.

Our real applications exhibit low contention, as two cores rarely contend for the same synchronization variable, and high synchronization demand, as several synchronization variables are active during execution. We observe that \hier{} and \SynCron{} increase parallelism, because the per-NDP-unit servers service different synchronization requests concurrently, and avoid remote synchronization messages across NDP units. Even though \hier{} performs 1.19$\times$ better than \naive{}, on average, its performance is still 1.33$\times$ worse than \ideal{}. \SynCron{} provides most of the performance benefits of \ideal{} (with only 9.5\% overhead on average), and outperforms \hier{} due to directly buffering the synchronization variables in \myTableShort{}s, thereby completely avoiding the memory accesses for synchronization requests. 
Specifically, we find that \emph{time series analysis} has high synchronization intensity, since the ratio of synchronization over other computation of the workload is higher compared to graph workloads. For this application, \hier{} and \SynCron{} outperform \naive{} by 1.64$\times$ and 2.22$\times$, as they serve multiple synchronization requests concurrently. \SynCron{} further outperforms \hier{} by 1.35$\times$ due to directly buffering the synchronization variables in \myTableShort{}s. We conclude that \SynCron{} performs best across \emph{all} real application-input combinations and approaches the \ideal{} scheme with no synchronization overhead.

\begin{figure}[H]
\vspace{-8pt}
\centering
  \includegraphics[width=0.96\columnwidth]{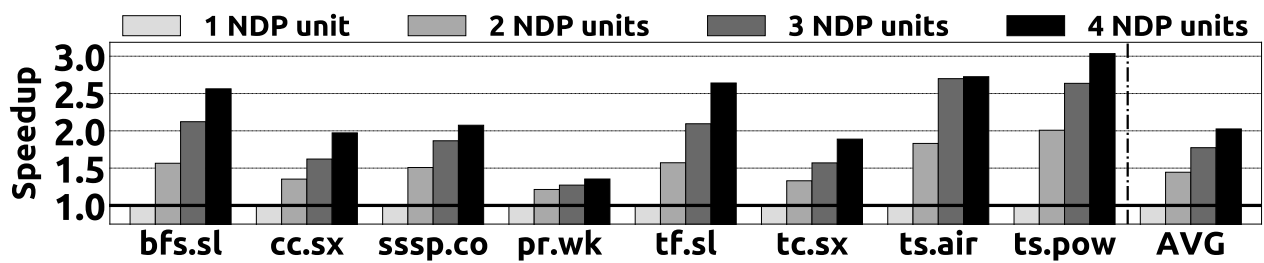}
  \caption{Scalability of real applications using \SynCron{}.}
  \label{fig:scalability}
\end{figure}

\noindent\textbf{Scalability.} 
Figure~\ref{fig:scalability} shows the scalability of real applications using \SynCron{} from 1 to 4 NDP units. Due to space limitations, we present a subset of our workloads, but we report average values for all 26 application-input combinations. This also applies for all figures presented henceforth. Across all workloads, \SynCron{} enables performance scaling by at least 1.32$\times$, on average 2.03$\times$, and up to 3.03$\times$, when using 4 NDP units (60 NDP cores) over 1 NDP unit (15 NDP cores).

\subsection{Energy Consumption}
Figure~\ref{fig:energy} shows the energy breakdown for cache, network, and memory in our real applications when using all cores. \SynCron{} reduces the network and memory energy thanks to its hierarchical design and direct buffering. On average, \SynCron{} reduces energy consumption by 2.22$\times$ over \naive{} and 1.94$\times$ over \hier{}, and incurs only 6.2\% energy overhead over \ideal{}.

\begin{figure}[H]
\vspace{-4pt}
 \centering
  \includegraphics[width=1.0\columnwidth]{sections/SynCron/graphs/energy.pdf}
  \caption{Energy breakdown in real applications for C: \naive{}, H: \hier{}, SC: \SynCron{} and I: \ideal{}.}
  \label{fig:energy}
  \vspace{-8pt}
\end{figure}

We observe that 1) cache energy consumption constitutes a small portion of the total energy, since these applications have irregular access patterns. NDP cores that act as servers for \naive{} and \hier{} increase the cache energy only by 5.1\% and 4.8\% over \ideal{}. 2) \naive{} generates a larger amount of expensive traffic across NDP units compared to hierarchical schemes, resulting in 2.68$\times$ higher network energy over \SynCron{}. \SynCron{} also has less network energy (1.21$\times$) than \hier{}, because it avoids transferring synchronization variables from memory to \myEngineShort{}s due to directly buffering them. 3) \hier{} and \naive{} have approximately the same memory energy consumption, because they issue a similar number of requests to memory. In contrast, \SynCron{}'s memory energy consumption is similar to that of \ideal{}. We note that \SynCron{} provides \emph{higher} energy reductions in applications with high synchronization intensity, such as time series analysis, since it avoids a \emph{higher} number of memory accesses for synchronization due to its direct buffering capability.

\subsection{Data Movement}
Figure~\ref{fig:traffic} shows normalized data movement, i.e., bytes transferred between NDP cores and memory, for all schemes using four NDP units. \SynCron{} reduces data movement across all workloads by 2.08$\times$ and 2.04$\times$ over \naive{} and \hier{}, respectively, on average, and incurs only 13.8\% more data movement than \ideal{}. \naive{} generates high data movement across NDP units, particularly when running time series analysis that has high synchronization intensity. \hier{} reduces the traffic across NDP units; however, it may increase the traffic inside an NDP unit, occasionally leading to slightly higher total data movement (e.g., \emph{ts.air}). This is because when an NDP core requests a synchronization variable that is physically located in another NDP unit, it first sends a message inside the NDP unit to its local server, which in turns sends a message to the global server. In contrast, \SynCron{} reduces the traffic inside an NDP unit due to directly buffering synchronization variables, and across NDP units due to its hierarchical design.

\begin{figure}[H]
  \centering
  \includegraphics[width=1.\columnwidth]{sections/SynCron/graphs/traffic.pdf}
  \caption{Data movement in real applications for C: \naive{}, H: \hier{}, SC: \SynCron{} and I: \ideal{}.}
  \label{fig:traffic}
  \vspace{-14pt}
\end{figure}

\subsection{Non-Uniformity of NDP Systems}
\subsubsection{High Contention} 
\begin{figure}[H]
  \vspace{-6pt}
  \centering
  \includegraphics[width=0.94\columnwidth]{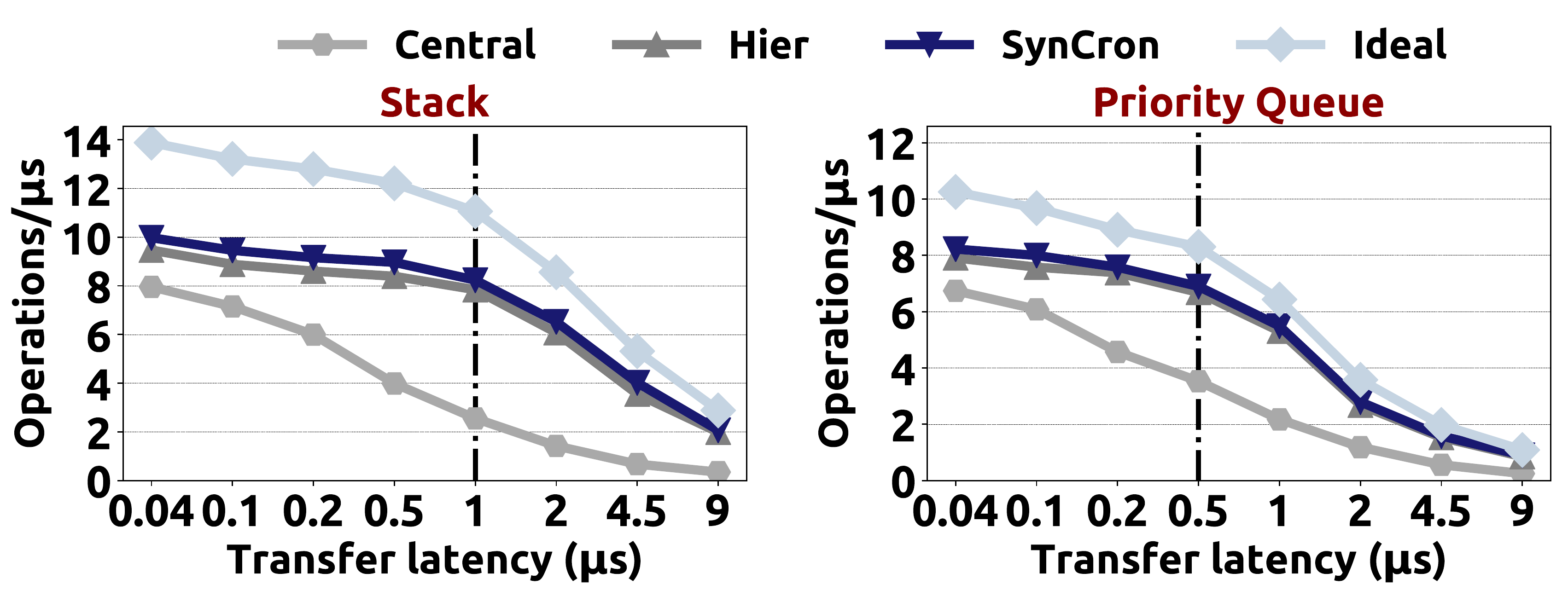}
  \vspace{-1pt}
  \caption{Performance sensitivity to the transfer latency of the interconnection links used to connect the NDP units.}
  \label{fig:cds_numa}
  \vspace{-12pt}
\end{figure}

Hierarchical schemes provide high benefit under high contention, as they prioritize local requests inside each NDP unit. We study their performance benefit in stack and priority queue (Figure~\ref{fig:cds_numa}) when varying the transfer latency of the interconnection links used across four NDP units. \naive{} is significantly affected by the interconnect latency across NDP units, as it is oblivious to the non-uniform nature of the NDP system. Observing \ideal{}, which reflects the actual behavior of the main workload, we notice that after a certain point (vertical line), the cost of remote memory accesses across NDP units become high enough to dominate performance. \SynCron{} and \hier{} tend to follow the actual behavior of the workload, as local synchronization messages within NDP units are much less expensive than remote messages of \naive{}. \SynCron{} outperforms \hier{} by 1.06$\times$ and 1.04$\times$ for stack and priority queue. We conclude that \SynCron{} is the best at hiding the latency of slow links across NDP units.

\subsubsection{Low Contention} 
\begin{figure}[H]
  \centering
  \includegraphics[width=0.93\columnwidth]{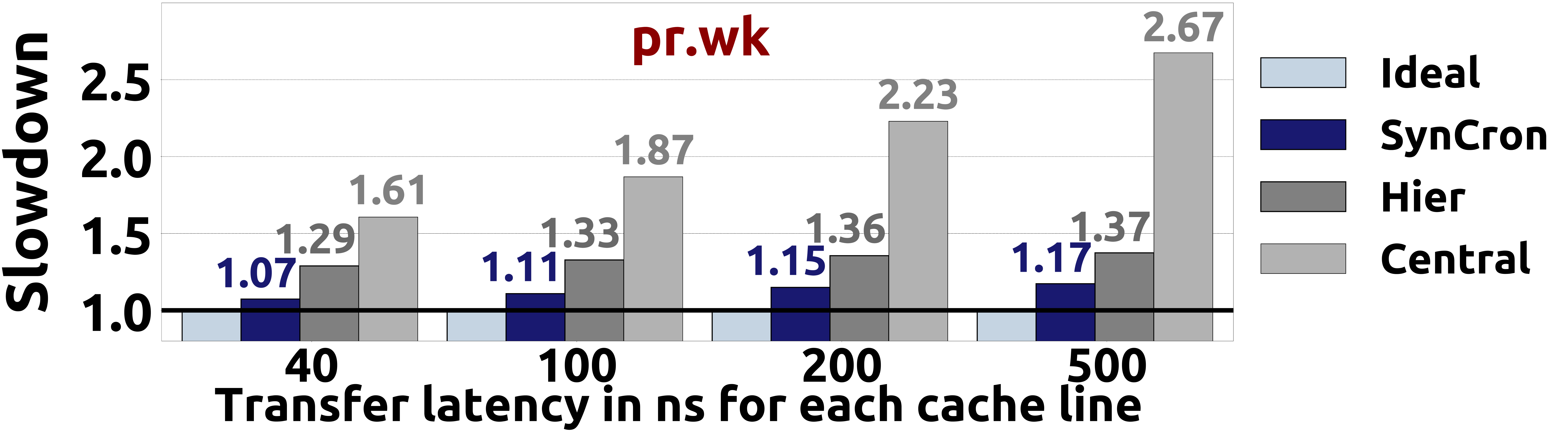}
  \vspace{-1pt}
  \caption{Performance sensitivity to the transfer latency of the interconnection links used to connect the NDP units. All data is normalized to \ideal{} (\emph{lower is better}).}
  \label{fig:numa_pr}
  \vspace{-12pt}
\end{figure}

We also study the effect of interconnection links used across the NDP units in a low-contention graph application (Figure~\ref{fig:numa_pr}). Observing \ideal{}, with 500 ns transfer latency per cache line, we note that the workload experiences 2.46$\times$ slowdown over the default latency of 40 ns, as 24.1\% of its memory accesses are to remote NDP units. As the transfer latency increases, \naive{} incurs significant slowdown over \ideal{}, since all NDP cores of the system communicate with one single server, generating expensive traffic across NDP units. In contrast, the slowdown of hierarchical schemes over \ideal{} is smaller, as these schemes generate less remote traffic by distributing the synchronization requests across multiple local servers. \SynCron{} outperforms \hier{} due to its direct buffering capabilities. Overall, \SynCron{} outperforms prior high-performance schemes even when the network delay across NDP units is large.

\subsection{Memory Technologies}\label{MemoryTechnologies}
\begin{figure}[H]
\centering
  \includegraphics[width=1.0\columnwidth]{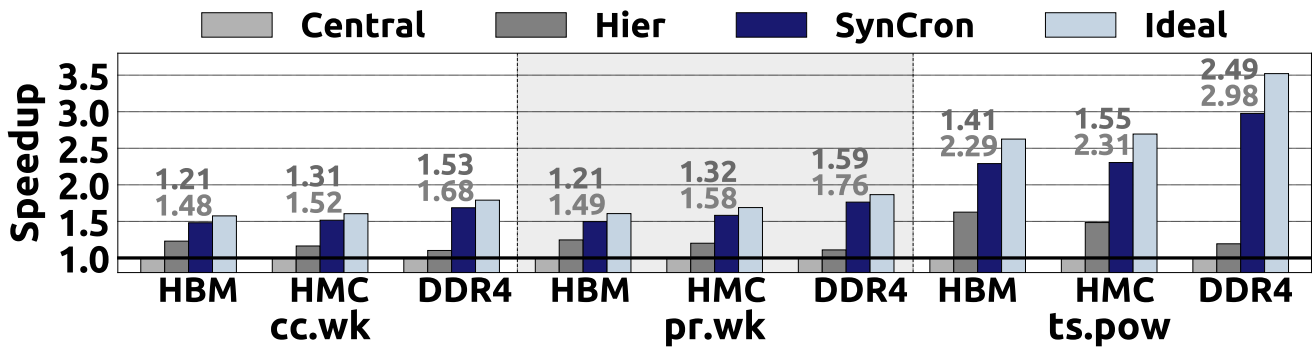}
  \vspace{-8pt}
  \caption{Speedup with different memory technologies.}
  \label{fig:3D_memories}
\end{figure}

We study three memory technologies, which provide different memory access latencies and bandwidth. We evaluate (i) 2.5D NDP using HBM, (ii) 3D NDP using HMC, and (iii) 2D NDP using DDR4. Figure~\ref{fig:3D_memories} shows the performance of all schemes normalized to \naive{} of each memory. The reported values show the speedup of \SynCron{} over \naive{} and \hier{}. \SynCron{}'s benefit is independent of the memory used: its performance versus \ideal{} only slightly varies ($\pm$1.4\%) across different memory technologies, since \myTableShort{}s never overflow. Moreover, \SynCron{}'s performance improvement over prior schemes increases as the memory access latency becomes higher thanks to direct buffering, which avoids expensive memory accesses for synchronization. For example, in \emph{ts.pow}, \SynCron{} outperforms \hier{} by 1.41$\times$ and 2.49$\times$ with HBM and DDR4, respectively, as the latter incurs higher access latency. Overall, \SynCron{} is orthogonal to the memory technology used.

\subsection{Effect of Data Placement}
\begin{figure}[H]
   \vspace{-4pt}
   \begin{minipage}{\columnwidth}
  \includegraphics[width=\columnwidth]{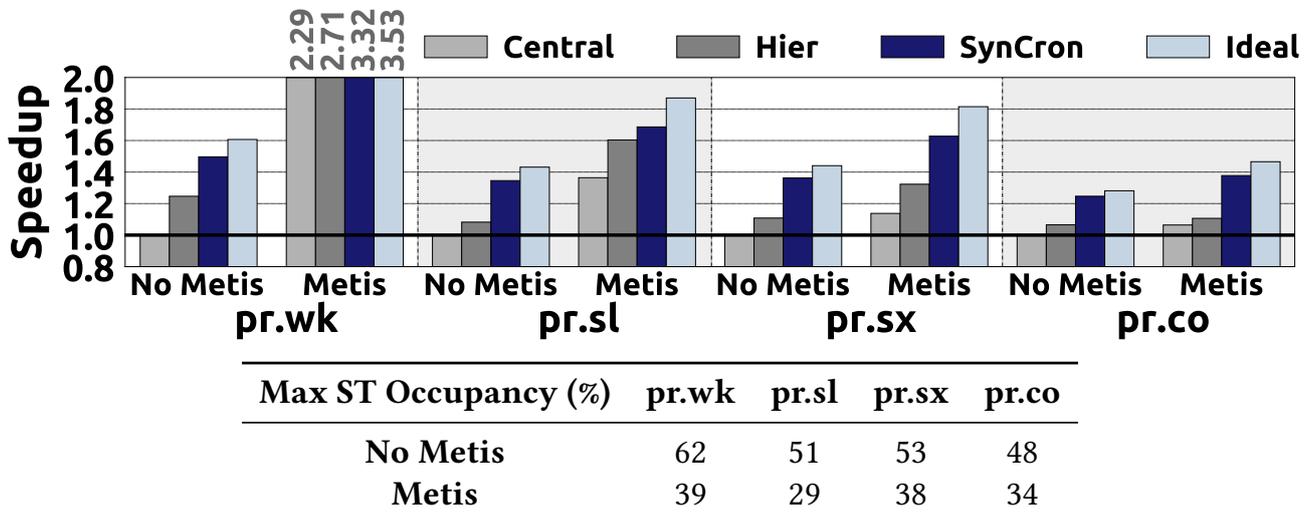}
  \end{minipage}\vspace{6pt}
  \begin{minipage}{\columnwidth} \vspace{1pt}
  \centering
  \resizebox{0.64\columnwidth}{!}{%
    \begin{tabular}{c c c c c} 
    \toprule
    \textbf{Max \myTableShort{} Occupancy (\%)} & \textbf{pr.wk} & \textbf{pr.sl} & \textbf{pr.sx} & \textbf{pr.co}  \\ 
    \midrule
    \textbf{No Metis} & 62 & 51 & 53 & 48 \\
    \textbf{Metis} & 39 & 29 & 38 & 34 \\
    \bottomrule
    \end{tabular}}
  \end{minipage}%
  \caption{Performance sensitivity to a better graph partitioning and maximum \myTableShort{} occupancy of \SynCron{}.}
  \label{fig:metis}
  \vspace{-14pt}
\end{figure}

Figure~\ref{fig:metis} evaluates the effect of better data placement on \SynCron{}'s benefits. We use Metis~\cite{Karypis1998Metis} to obtain a 4-way graph partitioning to minimize the crossing edges between the 4 NDP units. All data values are normalized to \naive{} without Metis. For \SynCron{}, we define \myTableShort{} occupancy as the average fraction of ST entries that are occupied in each cycle.

We make three observations. First, \ideal{}, which reflects the actual behavior of the main kernel (i.e., with zero synchronization overhead), improves performance by 1.47$\times$ across the four graphs. Second, with a better graph partitioning, \SynCron{} still outperforms both \naive{} and \hier{}. Third, we find that \myTableShort{} occupancy is lower with a better graph partitioning. When a local \myEngineShort{} receives a request for a synchronization variable of another NDP unit, \emph{both} the local \myEngineShort{} and the \masterSE{} reserve a new entry in their \myTableShort{}s. With a better graph partitioning, NDP cores send requests to their local \myEngineShort{}, which is also the \masterSE{} for the requested variable. Thus, \emph{only one} \myEngineShort{} of the system reserves a new entry, resulting in a lower \myTableShort{} occupancy. We conclude that, with better data placement \SynCron{} still performs the best while achieving even lower \myTableShort{} occupancy.

\subsection{\SynCron{}'s Design Choices}\label{DesignChoices}
\vspace{1pt}

\subsubsection{Hierarchical Design}\label{Flat}
To demonstrate the effectiveness of \SynCron{}'s hierarchical design in non-uniform NDP systems, we compare it with \SynCron{}'s \emph{flat} variant. Each core in 
\emph{flat} \emph{directly} sends all its synchronization requests to the \masterSE{} of each variable. In contrast, each core in \SynCron{} sends all its synchronization requests to the local \myEngineShort{}. If the local \myEngineShort{} is \emph{not} the \masterSE{} for the requested variable, the local \myEngineShort{} sends a message across NDP units to the \masterSE{}.

We evaluate three synchronization scenarios: (i) low-contention and synchronization non-intensive (e.g., graph applications), (ii) low-contention and synchronization-intensive (e.g., time series analysis), and (iii) high-contention (e.g., queue data structure). 

\begin{figure}[H]
  \vspace{1pt}
  \hspace{-4pt}
  \includegraphics[width=1.\columnwidth]{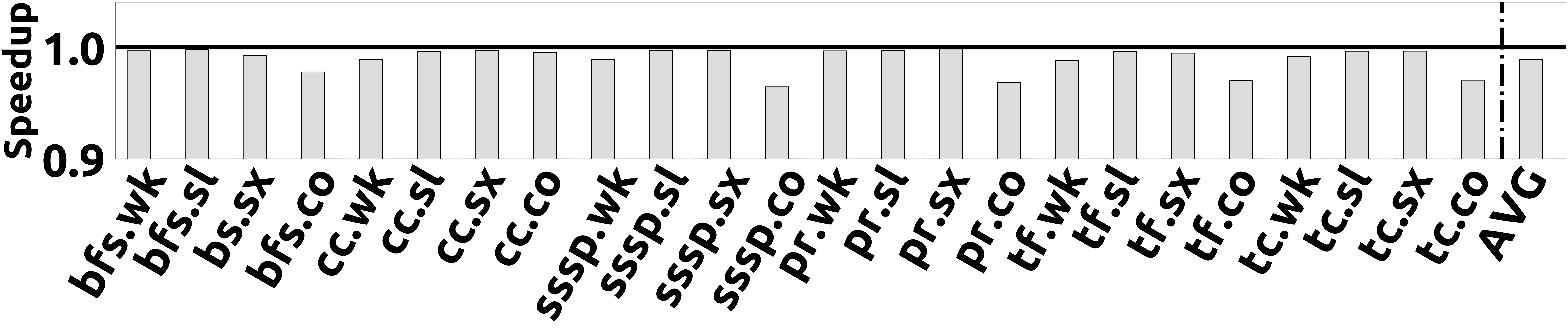}
  \caption{Speedup of \SynCron{} normalized to \emph{flat} with 40 ns link latency between NDP units, under a low-contention and synchronization non-intensive scenario.}
  \label{fig:hier_benefit_graphs}
 \vspace{-8pt}
\end{figure}

\noindent\textbf{Low-contention and synchronization non-intensive.} 
Figure~\ref{fig:hier_benefit_graphs} evaluates this scenario using several graph processing workloads with 40 ns link latency between NDP units. \SynCron{} is 1.1\% worse than \emph{flat}, on average. We conclude that \SynCron{} performs only \emph{slightly} worse than \emph{flat} for low-contention and synchronization non-intensive scenarios.

\begin{figure}[t]
  \centering
  \includegraphics[scale=0.054]{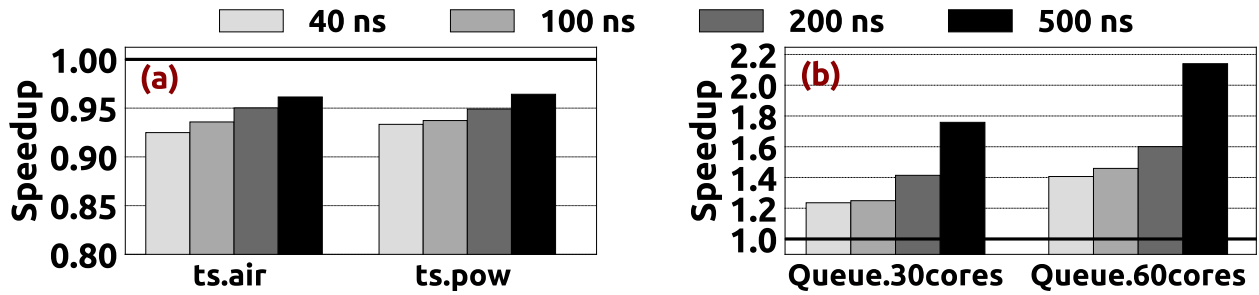}
  \caption{Speedup of \SynCron{} normalized to \emph{flat}, as we vary the transfer latency of the interconnection links used to connect NDP units, under (a) a low-contention and synchronization-intensive scenario using 4 NDP units, and (b) a high-contention scenario using 2 and 4 NDP units.}
  \label{fig:hier_benefit}
  \vspace{-8pt}
\end{figure}

\noindent\textbf{Low-contention and synchronization-intensive.} 
Figure~\ref{fig:hier_benefit}a evaluates this scenario using time series analysis with four different link latency values between NDP units. \SynCron{} performs 7.3\% worse than \emph{flat} with a 40 ns inter-NDP-unit latency. With a 500 ns inter-NDP-unit latency, \SynCron{} performs \emph{only} 3.6\% worse than \emph{flat}, since remote traffic has a larger impact on the total execution time. We conclude that \SynCron{} performs modestly worse than \emph{flat}, and \SynCron{}'s slowdown decreases as non-uniformity, i.e., the latency between NDP units, increases.

\noindent\textbf{High-contention.} 
Figure~\ref{fig:hier_benefit}b evaluates this scenario using a queue data structure with four different link latency values between NDP units, for 30 and 60 NDP cores. \SynCron{} with 30 NDP cores outperforms \emph{flat} from 1.23$\times$ to 1.76$\times$, as the inter-NDP-unit latency increases from 40 ns to 500 ns (i.e., with increasing non-uniformity in the system). In a scenario with high non-uniformity in the system and large number of contended cores, e.g., using a 500 ns inter-NDP-unit latency and 60 NDP cores, \SynCron{}'s benefit increases to a 2.14$\times$ speedup over \emph{flat}. We conclude that \SynCron{} performs \emph{significantly} better than \emph{flat} under high-contention.

\vspace{1pt}

Overall, we conclude that in \emph{non-uniform}, \emph{distributed} NDP systems, \emph{only} a \emph{hierarchical} hardware synchronization design can achieve high performance under \emph{all} various scenarios.

\vspace{1pt}

\subsubsection{\myTableShort{} Size}
\begin{figure}[H]
\vspace{-4pt}
  \centering
  \includegraphics[scale=0.05]{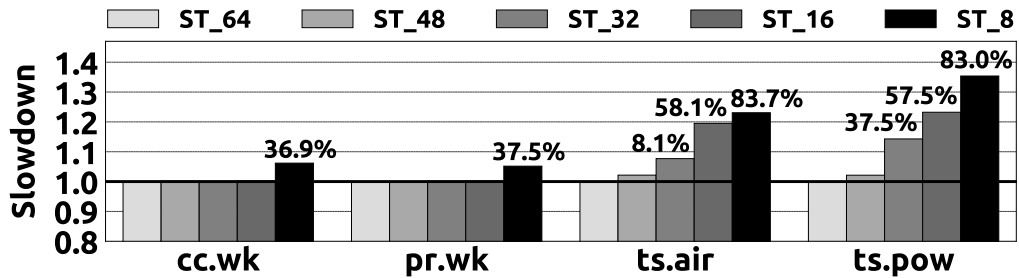}
  \vspace{-1pt}
  \caption{Slowdown with varying \myTableShort{} size (normalized to 64-entry \myTableShort{}). Numbers on top of bars show the percentage of overflowed requests.}
  \label{fig:sens_table}
  \vspace{-10pt}
\end{figure}

We show the effectiveness of the proposed 64-entry \myTableShort{} (per NDP unit) using real applications. Table~\ref{tab:graphs_occ} shows the measured occupancy across all \myTableShort{}s. Figure~\ref{fig:sens_table} shows the performance sensitivity to \myTableShort{} size. In graph applications, the average \myTableShort{} occupancy is low (2.8\%), and the 64-entry \myTableShort{} never overflows: maximum occupancy is 63\% (\emph{cc.wk}). In contrast, time series analysis has higher \myTableShort{} occupancy (reaching up to 89\% in \emph{ts.pow}) due to the high synchronization intensity, but there are no \myTableShort{} overflows. Even a 48-entry \myTableShort{} overflows for only 0.01\% of synchronization requests, and incurs 2.1\% slowdown over a 64-entry \myTableShort{}. 
We conclude that the proposed 64-entry \myTableShort{} meets the needs of applications that have high synchronization intensity.

\subsubsection{Overflow Management}\label{OverflowEvalbl}
The linked list and BST\_FG data structures are the \emph{only} cases where the proposed 64-entry \myTableShort{} overflows, when using 60 cores, for 3.1\% and 30.5\% of the requests, respectively. This is because each core requests at least two locks \emph{at the same time} during the execution. Note that these synthetic benchmarks represent extreme scenarios, where all cores repeatedly perform key-value operations.

\vspace{-2pt}
\begin{center}
%\hspace{-16pt}
\resizebox{0.83\columnwidth}{!}{%
 \begin{tabular}{l c c} 
 \toprule
 \textbf{\myTableShort{} Occupancy} & \textbf{Max (\%)} & \textbf{Avg (\%)} \\ 
 \midrule
 \midrule
  \textbf{bfs.wk} & 51 & 1.33 \\
 %\hline
 \textbf{bfs.sl} & 59 & 1.49 \\
 %\hline
 \textbf{bfs.sx} & 51 & 3.24 \\
 %\hline
 \textbf{bfs.co} & 55 & 6.09 \\
 %\hline
 \textbf{cc.wk} & 63 & 1.27 \\
 %\hline
 \textbf{cc.sl} & 61 & 2.16 \\
 %\hline
 \textbf{cc.sx} & 48 & 2.43 \\
 %\hline 
 \textbf{cc.co} & 46 & 4.53 \\
 %\hline
 \textbf{sssp.wk} & 62 & 1.18 \\
 %\hline 
 \textbf{sssp.sl} & 54 & 2.08 \\
 %\hline
 \textbf{sssp.sx} & 50 & 2.20 \\
 %\hline 
 \textbf{sssp.co} & 48 & 5.23\\
 %\hline
 \textbf{pr.wk} & 62 & 4.27 \\
 \bottomrule
 \end{tabular}%}
 ~ \hspace{18pt}
 \begin{tabular}{l c c} 
 \toprule
 \textbf{\myTableShort{} Occupancy} & \textbf{Max (\%)} & \textbf{Avg (\%)} \\ 
 \midrule
 \midrule
 \textbf{pr.sl} & 51 & 2.27 \\
 %\hline
 \textbf{pr.sx} & 53 & 2.46 \\
 %\hline
 \textbf{pr.co} & 48 & 4.72 \\
 %\hline 
 \textbf{tf.wk} & 62 & 1.44 \\
 %\hline
 \textbf{tf.sl} & 53 & 2.21 \\
 %\hline 
 \textbf{tf.sx} & 50 & 2.99 \\
 %\hline 
 \textbf{tf.co} & 48 & 4.61 \\
 %\hline 
 \textbf{tc.wk} & 62 & 1.26 \\
 %\hline 
 \textbf{tc.sl} & 48 & 2.08 \\
 %\hline
 \textbf{tc.sx} & 50 & 2.77 \\
 %\hline
 \textbf{tc.co} & 51 & 4.52 \\
 \hline 
 \textbf{ts.air} & 84 & 44.20 \\
 %\hline
 \textbf{ts.pow} & 89 & 43.51 \\
 \bottomrule
 \end{tabular}} 
 \vspace{-1pt}
 \captionof{table}{\label{tab:graphs_occ}\myTableShort{} occupancy in real applications.}
\end{center}

\begin{figure}[t]
  \centering
  \includegraphics[scale=0.52]{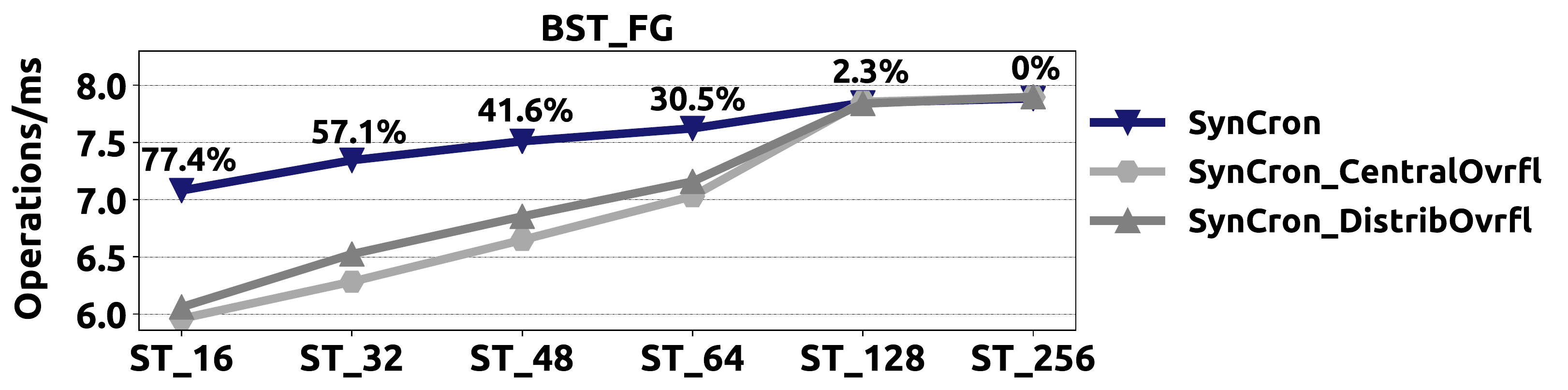}
  \caption{Throughput achieved by BST\_FG using different overflow schemes and varying the \myTableShort{} size. The reported numbers show to the percentage of overflowed requests.}
  \label{fig:overflow}
  \vspace{-4pt}
\end{figure}

Figure~\ref{fig:overflow} compares BST\_FG's performance with \SynCron{}'s integrated overflow scheme versus with a non-integrated scheme as in MiSAR. When overflow occurs, MiSAR's accelerator aborts all participating cores notifying them to use an alternative synchronization library, and when the cores finish synchronizing
via an alternative solution, they notify MiSAR’s accelerator to switch back to hardware synchronization. We adapt this scheme to \SynCron{} for comparison purposes: when an \myTableShort{} overflows, \myEngineShort{}s send abort messages to NDP cores with a hierarchical protocol, notifying them to use an alternative synchronization solution, and after finishing synchronization they notify \myEngineShort{}s to decrease their indexing counters and switch to hardware. We evaluate two alternative solutions: (i) \emph{SynCron\_CentralOvrfl}, where one dedicated NDP core handles all synchronization variables, and (ii) \emph{SynCron\_DistribOvrfl}, where one NDP core per NDP unit handles variables located in the same NDP unit. With 30.5\% overflowed requests (i.e., with a 64-entry \myTableShort{}),
\emph{SynCron\_CentralOvrfl} and \emph{SynCron\_DistribOvrfl} incur 12.3\% and 10.4\% performance slowdown compared to with \emph{no} \myTableShort{} overflow, due to high network traffic and communication costs between NDP cores and \myEngineShort{}s. In contrast, \SynCron{} affects performance by only 3.2\% compared to with \emph{no} \myTableShort{} overflow. We conclude that \SynCron{}'s integrated hardware-only overflow scheme enables very small performance overhead.

\subsection{\SynCron{}'s Area and Power Overhead} \label{Areabl}

Table~\ref{tab:area} compares an \myEngineShort{} with the ARM Cortex A7 core~\cite{ARM_CORTEX}. We estimate the SPU using Aladdin~\cite{shao2016aladdin}, and the \myTableShort{} and indexing counters using CACTI~\cite{Muralimanohar2007Optimizing}. We conclude that our proposed hardware unit incurs very modest area and power costs to be integrated into the compute die of an NDP unit.

\begin{center} 
%\hspace{-10pt}
\resizebox{0.76\columnwidth}{!}{%
 \begin{tabular}{l c c} 
 \toprule
 \textbf{} & \textbf{\myEngineShort{} (Synchronization Engine)} & \textbf{ARM Cortex A7~\cite{ARM_CORTEX}} \\ 
 \midrule
 \midrule
 \textbf{Technology} & 40nm & 28nm \\
 \hline
 \multirow{3}{*}{\shortstack[l]{\textbf{Area}}} & SPU: 0.0141mm\textsuperscript{2}, \myTableShort{}: 0.0112mm\textsuperscript{2} & \multirow{2}{*}{\shortstack[l]{32KB L1 Cache}} \\
  & Indexing Counters: 0.0208mm\textsuperscript{2} & \\
  & \textbf{Total:} 0.0461mm\textsuperscript{2} & \textbf{Total:} 0.45mm\textsuperscript{2} \\
 \hline
 \textbf{Power} & 2.7 mW & ~100mW \\
 \bottomrule
 \end{tabular}
 } 
 \captionof{table}{\label{tab:area}Comparison of \myEngineShort{} with a simple general-purpose in-order core, ARM Cortex A7.}
 %\vspace{2pt}
\end{center}

\section{Recommendations}

This section presents our key takeaways in the form of recommendations for software and hardware designers.

\noindent\textbf{Recommendation \#1.} \textit{Provide hardware synchronization support for NDP architectures.}
Figures~\ref{fig:mbench}, ~\ref{fig:ds_thr} and ~\ref{fig:speedup_real} demonstrate that \SynCron{} significantly outperforms software-based synchronization schemes, e.g., \naive{} and \hier{}, across various contention scenarios and workload demands. In addition, Tables~\ref{tab:graphs_occ} and ~\ref{tab:area} show that \SynCron{} has modest area and power costs for NDP architectures. In contrast to commodity CPU and GPU systems that run multiple software threads executed at each hardware thread context, NDP architectures~\cite{Mutlu2020AMP,Mutlu2019Processing,Ghose2019Workload,Ahn2015PIMenabled,Nair2015Active,ahn2015scalable,Hsieh2016accelerating,pugsley2014ndc,Boroumand2018Google,Gokhale2015Near,Gao2016HRL,Hsieh2016TOM,Drumond2017mondrian,Liu2018Processing,Boroumand2019Conda,Gao2015Practical,gao2017tetris,Kim2016Neurocube,Kim2017GrimFilter,Survive2016Survive,Nai2017GraphPIM,Youwei2019GraphQ,Pattnaik2016Scheduling,fernandez2020natsa,Singh2019Napel,Singh2020NERO,Cali2020GenASM,Zhang2018GraphP,Kim2013memory,Tsai2018Adaptive,boroumand2017lazypim,upmem} typically support a only fixed number of hardware thread contexts, and thus in such computing platforms synchronization can be effectively implemented in hardware with low cost. Therefore, we suggest that hardware designers of NDP architectures provide low-cost synchronization mechanisms implemented in \emph{hardware}.

\noindent\textbf{Recommendation \#2.} \textit{Design hierarchical, non-uniform aware synchronization schemes for non-uniform NDP systems.} NDP systems are typically non-uniform, distributed architectures, in which inter-unit communication is more expensive (both in performance and energy) than intra-unit communication~\cite{Zhang2018GraphP,Boroumand2019Conda,boroumand2017lazypim,Drumond2017mondrian,Youwei2019GraphQ,ahn2015scalable,Gao2015Practical,Kim2013memory}. Our evaluations presented in Figures~\ref{fig:cds_numa} and ~\ref{fig:numa_pr} show that the hierarchical schemes, i.e., \hier{} and \SynCron{}, provide significant performance benefits over \naive{}, since \naive{} is oblivious to the non-uniform nature of NDP systems. Under high-contention scenarios (Figure~\ref{fig:cds_numa}), the hierarchical (non-uniform aware) schemes achieve high system performance by minimizing the expensive traffic across NDP units of the system. Under low-contention scenarios (Figure~\ref{fig:numa_pr}), the hierarchical schemes provide high system performance, because they (i) generate less remote traffic by distributing the synchronization requests across multiple local synchronization units, and (ii) increase parallelism, since the per-NDP-unit synchronization units service different synchronization requests concurrently. To this end, we recommend that hardware architects design non-uniform aware synchronization mechanisms for NDP systems.

\noindent\textbf{Recommendation \#3.} \textit{Design effective data placement schemes of the input data and the associated synchronization variables across multiple NDP units of the NDP system.} In many real-world applications (e.g., graph processing applications), the large input data set given (e.g., real-world graphs with a large number of vertices and edges) is shared among multiple cores, and thus a fine-grained synchronization scheme (i.e., including a large number of synchronization variables, each of them protects a small granularity of shared data) is typically used.  Figure~\ref{fig:metis} demonstrates that a better graph partitioning in graph processing workloads significantly improves performance of the main kernel and reduces the synchronization costs among NDP cores. Specifically, with a better graph partitioning \SynCron{} (i) reduces the remote synchronization messages sent across the NDP units of the system through the expensive interconnection links, and (ii) has lower \myTableShort{} occupancy, thus having lower \myTableShort{} sizes (with lower area and power costs) can be sufficient to meet the synchronization needs of real-world applications without \emph{never} overflowing. Therefore, we suggest that software engineers of real-world applications with fine-grained synchronization schemes design intelligent data placement schemes of the input data and the associated synchronization variables across multiple NDP units of NDP architectures to achieve high system performance and minimize synchronization costs.
\section{Related Work}
\label{Relatedbl}

To our knowledge, our work is the first one to (i) comprehensively analyze and evaluate synchronization primitives in NDP systems, and (ii) propose an end-to-end hardware-based synchronization mechanism for efficient execution of such primitives. We briefly discuss prior work.

\textbf{Synchronization on NDP.} 
Ahn et al.~\cite{ahn2015scalable} include a \mpsync{} barrier similar to our \naive{} baseline. Gao et al.~\cite{Gao2015Practical} implement a {hierarchical} tree-based barrier for HMC~\cite{HMC}, where cores first synchronize inside the vault, then across vaults, and finally across HMC {stacks}. Section~\ref{Performancebl} shows that \SynCron{} outperforms such schemes. Gao et al.~\cite{Gao2015Practical} also provide remote atomics at the vault controllers of HMC. However, synchronization {using} remote atomics creates high global traffic and hotspots~\cite{Wang2019Fast,Mukkara2019PHI,li2015fine,yilmazer2013hql,eltantawy2018warp}.

\textbf{Synchronization on CPUs.}
A range of hardware synchronization mechanisms have been proposed for commodity CPU systems~\cite{abell2011glocks,sampson2006exploiting,abellan2010g,oh2011tlsync,Sergi2016WiSync,akgul2001system}. These are not suitable for NDP systems because they either (i) rely on the underlying cache coherence system~\cite{sampson2006exploiting,akgul2001system}, (ii) are tailored for the 2D-mesh network topology to connect all cores~\cite{abell2011glocks,abellan2010g}, or (iii) use transmission-line technology~\cite{oh2011tlsync} or on-chip wireless technology~\cite{Sergi2016WiSync}. Callbacks~\cite{Ros2015Callback} includes a directory cache structure close to the LLC of a CPU system {built on} self-invalidation coherence protocols~\cite{Kaxiras2010SARC,Choi2011DeNovo,Kaxiras2013ANew,Sung2014DeNovoND,Lebeck1995Dynamic,Ros2012Complexity}. Although it has low area cost, it would be oblivious to the non-uniformity of NDP, thereby incurring high performance overheads {{under} high} contention (Section~\ref{Flat}). Callbacks 
improves performance of spin-wait in hardware, on top of which high-level primitives (locks/barriers) are implemented in software. In contrast, \SynCron{} directly supports high-level primitives {in hardware}, and is tailored {to} all {salient} characteristics of NDP systems.

The closest works to ours are SSB~\cite{Zhu2007SSB}, LCU~\cite{Vallejo2010Architectural}, and MiSAR~\cite{Liang2015MISAR}. SSB, a shared memory scheme, includes a small buffer attached to each controller of LLC to provide lock semantics for a given data address. LCU, a \mpsync{} scheme, incorporates a control unit into each core and a reservation table into each memory controller to provide reader-writer locks. MiSAR is a \mpsync{} synchronization accelerator distributed at each LLC slice of tile-based many-core chips. These schemes provide efficient synchronization for CPU systems \emph{without} relying on hardware coherence protocols. As shown in Table~\ref{tab:comparison}, compared to these works, \SynCron{} is a more effective, general and easy-to-use solution for NDP systems. These works have two major shortcomings. First, they are designed for \emph{uniform} architectures, and would incur high performance overheads in \emph{non-uniform, distributed} NDP systems under high-contetion scenarios, similarly to \emph{flat} in  Figure~\ref{fig:hier_benefit}b. Second, SSB and LCU handle overflow cases using software exception handlers that typically incur large performance overheads, while MiSAR's overflow scheme would incur high performance degradation due to high network traffic and communication costs between the cores and the synchronization accelerator (Section~\ref{OverflowEvalbl}). In contrast, \SynCron{} is a {non-uniformity} aware, hardware-only, end-to-end solution designed to handle key characteristics of NDP systems.

\textbf{Synchronization on GPUs.} 
GPUs support remote atomic units at the shared cache and hardware barriers among threads of the same block~\cite{teslav100}, while inter-block barrier synchronization is inefficiently implemented via the host CPU~\cite{teslav100}. The closest work to ours is HQL~\cite{yilmazer2013hql}, which modifies the tag arrays of L1 and L2 caches to support {the} lock primitive. This scheme incurs high area {cost}~\cite{eltantawy2018warp}, and is tailored to the GPU architecture that includes a {shared} L2 cache, while most NDP systems do \emph{not} have shared caches.

\textbf{Synchronization on MPPs.}
The Cray T3D/T3E~\cite{kessler1993crayTA,scott1996synchronization}, SGI Origin~\cite{Laudon1997SGI}, and AMOs~\cite{zhang2004highly} include remote atomics at the memory controller, while NYU {Ultracomputer}~\cite{gottlieb1998NYU} {provides} \emph{fetch\&and} remote atomics in each network switch. As discussed in Section~\ref{Motivationbl}, synchronization via remote atomics {incurs} high performance overheads due to high global traffic~\cite{Wang2019Fast,Mukkara2019PHI,yilmazer2013hql,eltantawy2018warp}. Cray T3E supports a barrier {using} physical wires, {but it is} designed {specifically} for 3D torus interconnect. Tera {MTA}~\cite{Alverson1990Tera}, HEP~\cite{Jordan1983Performance,Smith1978Pipelined}, J- and M-machines~\cite{Dally1992TheMessage,Keckler1998Exploiting}, and Alewife~\cite{Agarwal1995Alewife} provide synchronization using hardware bits (\emph{full/empty} bits) as tags in \emph{each memory word}. This scheme can incur high area cost~\cite{Vallejo2010Architectural}. QOLB~\cite{Goodman1989EfficientSP} associates one cache line for every lock to track a pointer to the next waiting core, and one cache line for local spinning using bits (\emph{syncbits}). QOLB is {built on} the underlying cache coherence protocol. Similarly, DASH~\cite{Lenoski1992Dash} keeps a queue of waiting cores for a lock in the directory used for coherence to notify caches when the lock is released. CM5~\cite{Leiserson1992CM5} supports remote atomics and a barrier among cores via a dedicated physical control network (organized as a binary tree), which would incur high hardware cost to be supported in NDP systems.

%\vspace{-3pt}
%\vspace{-1pt}
\section{Summary}
\label{Conclusionbl}

\SynCron{} is the first end-to-end synchronization solution for NDP systems. \SynCron{} avoids the need for complex coherence protocols and expensive \emph{rmw} operations, incurs very modest hardware cost, generally supports many synchronization primitives and is easy-to-use. Our evaluations show that it outperforms prior designs under various conditions, providing high performance both under high-contention (due to reduction of expensive traffic across NDP units) and low-contention scenarios (due to direct buffering of synchronization variables and high execution parallelism). We conclude that \SynCron{} is an efficient synchronization mechanism for NDP systems, and hope that this work encourages further comprehensive studies of the synchronization problem in heterogeneous systems, including NDP systems.

%SparseP
\chapter{\SparseP{}}\label{SparsePChapter}

\section{Overview} 

Sparse Matrix Vector Multiplication (\spmv) is a fundamental linear algebra kernel for important applications from the scientific computing, machine learning, and graph analytics domains. In commodity systems, it has been repeatedly reported to achieve only a small fraction of the peak performance~\cite{Elafrou2018SparseX,Shengen2014YaSpMV,Liu2018Towards,Elafrou2017PerformanceAA,Goumas2009Performance,Karakasis2009Performance,Vuduc2005Fast,im2004sparsity,Vuduc2003PhD,Vuduc2002Performance,Goumas2008Understanding,Elafrou2017PerformanceXeon} due to its algorithmic nature, the employed compressed matrix storage format, and the sparsity pattern of the input matrix. \spmv{} performs indirect memory references as a result of storing the matrix in a compressed format, and irregular memory accesses to the input vector due to sparsity. The matrices involved are very sparse, i.e., the vast majority of elements are zeros~\cite{Kanellopoulos2019SMASH,Elafrou2018SparseX,Elafrou2017PerformanceAA,YouTubeGraph,FacebookGraph,Goumas2008Understanding,White97Improving,Helal2021ALTO,Pelt2014Medium}. For example, the matrices that represent Facebook’s and YouTube’s network connectivity contain 0.0003\%~\cite{YouTubeGraph,Kanellopoulos2019SMASH} and 2.31\%~\cite{FacebookGraph,Kanellopoulos2019SMASH} non-zero elements, respectively. Therefore, in processor-centric systems, \spmv{} is a memory-bandwidth-bound kernel for the majority of real sparse matrices, and is bottlenecked by data movement between memory and processors~\cite{Gomez2021Benchmarking,Elafrou2018SparseX,Elafrou2017PerformanceAA,Elafrou2019Conflict,Xie2021SpaceA,Goumas2009Performance,Pal2018OuterSpace,Gomez2021Analysis,Vuduc2002Performance,Vuduc2003PhD,Vuduc2005Fast,Karakasis2009Performance,dongarra1996sparse,im2004sparsity,Liu2018Towards,Kourtis2011CSX,Goumas2008Understanding,Kourtis2008Optimizing,Elafrou2017PerformanceXeon}.

One promising way to alleviate the data movement bottleneck is the Processing-In-Memory (PIM) paradigm~\cite{Gomez2021Benchmarking,ahn2015scalable,Kwon2019TensorDIMM,gao2017tetris,Lee2021HardwareAA,fernandez2020natsa,devaux2019,Gao2015Practical,ke2019recnmp,Kwon2021Function,Hadi2016Chameleon,Gomez2021Analysis,Giannoula2021SynCron,mutlu2020modern,Mutlu2019Processing,Ghose2019Workload,Mutlu2019Enabling,Gagandeep2019Near,Oliveira2021Damov,Boroumand2018Google,Zhang2018GraphP,Lockerman2020Livia,bostanci2022drstrange,Kim2019DRange,Alser2020Accelerating,Cali2020GenASM,Kim2017GrimFilter,Gao2016HRL,Farmahini2015NDA,Dai2018GraphH,Nair2015Active,choe2019concurrent,liu2017concurrent,Boroumand2021Google,Nag2021OrderLight,Gu2021DLUX,Aga2019coml,Shin2018MCDRAM,Cho2020MCDRAM,Yazdanbakhsh2018InDRAM,Farmahini2015DRAMA,Alian2019NetDIMM,Kautz1969Cellular,Stone1970Logic,Ahn2015PIMenabled,Hsieh2016TOM,hashemi2016accelerating,Singh2019Napel,Singh2020NEROAN,seshadri2020indram,Olgun2021QuacTrng,Alves2015Opportunities,hashemi2016continuous,huangfu2019medal,Seshadri2017Ambit,Aga2017Compute,Eckert2018Neural,Fujiki2019Duality,Kang2014Energy,Li2016Pinatubo,Seshadri2013RowClone,Angizi2019GraphiDe,Chang2016LISA,Gao2019ComputeDRAM,Xin2020ELP2IM,li2017drisa,Deng2018DrAcc,Hajinazar2021SIMDRAM,Rezaei2020NoM,Wang2020Figaro,Ali2020InMemory,levi2014Loci,Kvatinsky2014Magic,Shafiee2016ISAAC,Kvatinsky2011Memristor,Gaillardon2016Programmable,Bhattacharjee2017ReVAMP,Hamdioui2015Memristor,Xie2015FastBL,Song2018GraphR,Ankit2020Panther,Ankit2019PUMA,Chi2016PRIME,Xi2021Memory,Zheng2016RRAM,Hamdioui2017Memristor,Yu2018Memristive,Kim2018PUF,orosa2021codic,ferreira2021pluto,Sun2021ABCDIMMAT,olgun2021pidram,Wu2021Sieve,Yuan2021FORMS,Khan2020Survey}. PIM moves computation close to application data by equipping memory chips with processing capabilities~\cite{mutlu2020modern,Mutlu2019Enabling}. Prior works~\cite{ahn2015scalable,fernandez2020natsa,Gao2015Practical,gao2017tetris,Giannoula2021SynCron,Boroumand2019Conda,Dai2018GraphH,Nai2017GraphPIM,Youwei2019GraphQ,Lee2008Adaptive,Lockerman2020Livia,choe2019concurrent,liu2017concurrent,boroumand2017lazypim,Drumond2017mondrian,Hsieh2016accelerating,Kim2021Functionality,Huang2019active,Boroumand2018Google,Zhang2018GraphP,Lockerman2020Livia,Gao2016HRL,pugsley2014ndc,Zhang2014TOPPIM,Nair2015Active,Farmahini2015DRAMA} propose PIM architectures wherein a processor logic layer is tightly integrated with DRAM memory layers using 2.5D/3D-stacking technologies~\cite{HMC,HBM,Lee2016Simultaneous}. Nonetheless, the 2.5D/3D integration itself might not always be able to provide significantly higher memory bandwidth for processors than standard DRAM~\cite{Lee2021HardwareAA,Hadi2016Chameleon}. To provide even higher bandwidth for the in-memory processors, \textit{near-bank} PIM designs have been explored~\cite{Lee2021HardwareAA,upmem,Hadi2016Chameleon,devaux2019,Kwon2021Function,Gomez2021Analysis,Gomez2021Benchmarking,Gu2020iPIM,Cho2020Near,Cho2021Accelerating,Kumar2020Parallel,Nag2021OrderLight,Park2021Trim,Sadredini2021Sunder,Gu2021DLUX,Aga2019coml,Shin2018MCDRAM,Cho2020MCDRAM,Yazdanbakhsh2018InDRAM,Alves2015Opportunities,li2017drisa}. \textit{Near-bank} PIM designs tightly couple a PIM core with each DRAM bank, exploiting bank-level parallelism to expose high on-chip memory bandwidth of standard DRAM to processors. Moreover, manufacturers of near-bank PIM architectures avoid disturbing the key components (i.e., subarray and bank) of commodity DRAM to provide a cost-efficient and practical way for silicon materialization. Two \textit{real} near-bank PIM architectures are Samsung's FIMDRAM~\cite{Lee2021HardwareAA,Kwon2021Function} and the UPMEM PIM system~\cite{upmem2018,devaux2019,Gomez2021Analysis,Gomez2021Benchmarking}.

Most near-bank PIM architectures~\cite{Lee2021HardwareAA,upmem,Hadi2016Chameleon,devaux2019,Kwon2021Function,Gomez2021Analysis,Gomez2021Benchmarking,Gu2020iPIM,Cho2020Near,Cho2021Accelerating,Kumar2020Parallel,Nag2021OrderLight,Park2021Trim} support several PIM-enabled memory chips connected to a host CPU via memory channels. Each memory chip comprises multiple PIM cores, which are low-area and low-power cores with relatively low computation capability~\cite{Gomez2021Benchmarking,Gomez2021Analysis}, and each of them is located close to a DRAM bank~\cite{Lee2021HardwareAA,upmem,Hadi2016Chameleon,devaux2019,Kwon2021Function,Gomez2021Analysis,Gomez2021Benchmarking,Gu2020iPIM,Cho2020Near,Cho2021Accelerating,Kumar2020Parallel,Nag2021OrderLight,Park2021Trim}. Each PIM core can access data located on their local DRAM banks, and typically there is no direct communication channel among PIM cores. Overall, near-bank PIM architectures provide high levels of parallelism and very large memory bandwidth, thereby being a very promising computing platform to accelerate memory-bound kernels. Recent works leverage near-bank PIM architectures to provide high performance and energy benefits on bioinformatics~\cite{lavenier2020Variant,Gomez2021Benchmarking,Gomez2021Analysis,Lavenier2016DNA}, skyline
computation~\cite{Zois2018Massively}, compression~\cite{Nider2020Processing} and neural network~\cite{Lee2021HardwareAA,Gomez2021Benchmarking,Gomez2021Analysis,Gu2020iPIM,Cho2021Accelerating} kernels. A recent study~\cite{Gomez2021Analysis,Gomez2021Benchmarking} provides PrIM benchmarks~\cite{PrIMLibrary}, which are a collection of 16 kernels for evaluating near-bank PIM architectures, like the UPMEM PIM system. However, there is \emph{no} prior work to thoroughly study the widely used, memory-bound \spmv{} kernel on a real PIM system.

Our work is the first to efficiently map the \spmv{} execution kernel on near-bank PIM systems, and understand its performance implications on a real PIM system. Specifically, our \textbf{goal} in this work is twofold: (i) design efficient \spmv{} algorithms to accelerate this kernel in current and future PIM systems, while covering a wide variety of sparse matrices with diverse sparsity patterns, and (ii) provide an extensive characterization analysis of the widely used \spmv{} kernel on a real PIM architecture. To this end, we provide a wide variety of \spmv{} implementations for real PIM architectures, and conduct a rigorous experimental analysis of \spmv{} kernels in the UPMEM PIM system, the first publicly-available real-world PIM architecture.

We present the \SparseP{} library~\cite{SparsePLibrary} that includes 25 \spmv{} kernels for real PIM systems, supporting various (1) data types, (2) data partitioning techniques of the sparse matrix to PIM-enabled memory, (3) compressed matrix formats, (4) load balancing schemes across PIM cores, (5) load balancing schemes across threads of a multithreaded PIM core, and (6) synchronization approaches among threads within PIM core. We support a wide range of data types, i.e., 8-bit integer, 16-bit integer, 32-bit integer, 64-bit integer, 32-bit float and 64-bit float data types to cover a wide variety of real-world applications that employ \spmv{} as their underlying kernel. We design two types of well-crafted data partitioning techniques: (i) the 1D partitioning technique to perform the complete \spmv{} computation only using PIM cores, and (ii) the 2D partitioning technique to strive a balance between computation and data transfer costs to PIM-enabled memory. In the 1D partitioning technique, the matrix is horizontally partitioned across PIM cores, and the \textit{whole} input vector is copied into the DRAM bank of \textit{each} PIM core, while PIM cores directly compute the elements of the final output vector. In the 2D partitioning technique, the matrix is split in 2D tiles, the number of which is equal to the number of PIM cores, and a \textit{subset} of the elements of the input vector is copied into the DRAM bank of each PIM core. However, in the 2D partitioning technique, PIM cores create a large number of partial results for the elements of the output vector which are gathered and merged by the host CPU cores to assemble the final output vector. We support the most popular compressed matrix formats, i.e., CSR~\cite{bjorck1996numerical,Pooch1973Survey},  COO~\cite{Pooch1973Survey,Shubhabrata2007Scan}, BCSR~\cite{Im1999Optimizing}, BCOO~\cite{Pooch1973Survey}, and for each compressed format we implement various load balancing schemes across PIM cores to provide efficient \spmv{} execution for a wide variety of sparse matrices with diverse sparsity patterns. Finally, we design several load balancing schemes and synchronization approaches among parallel threads within a PIM core to cover a variety of real PIM systems that provide multithreaded PIM cores.

We conduct an extensive characterization analysis of \SparseP{} kernels on the UPMEM PIM system~\cite{upmem,Gomez2021Analysis,Gomez2021Benchmarking,devaux2019} analyzing the \spmv{} execution using (1) one single multithreaded PIM core, (2) thousands of PIM cores, and (3) comparing it with that achieved on conventional processor-centric CPU and GPU systems. First, we characterize the limits of a single multithreaded PIM core, and show that (i) high operation imbalance across threads of a PIM core can impose high overhead in the core pipeline, and (ii) fine-grained synchronization approaches to increase parallelism cannot outperform a coarse-grained approach, if PIM hardware serializes accesses to the local DRAM bank. Second, we analyze the end-to-end \spmv{} execution of 1D and 2D partitioning techniques using thousands of PIM cores. Our study indicates that the performance (i) of the 1D partitioning technique is limited by data transfer costs to \textit{broadcast} the whole input vector into \textit{each} DRAM bank of PIM cores, and (ii) of the 2D partitioning technique is limited by data transfer costs to \textit{gather} partial results for the elements of the output vector from PIM-enabled memory to the host CPU. Such data transfers incur high overheads, because they take place via the narrow memory bus. In addition, our detailed study across a wide variety of compressed matrix formats and sparse matrices with diverse sparsity patterns demonstrates that (i) the compressed matrix format determines the data partitioning strategy across DRAM banks of PIM-enabled memory, thereby affecting the computation balance across PIM cores with corresponding performance implications, and (ii) there is \textit{no one-size-fits-all} solution. The load balancing scheme across PIM cores (and across threads within a PIM core) and data partitioning technique that provides the best-performing \spmv{} execution depends on the characteristics of the input matrix and the underlying PIM hardware. Finally, we compare the \spmv{} execution on a state-of-the-art UPMEM PIM system with 2528 PIM cores to state-of-the-art CPU and GPU systems, and observe that \spmv{} on the UPMEM PIM system achieves a much higher fraction of the machine's peak performance compared to that on the state-of-the-art CPU and GPU systems. Our extensive evaluation provides programming recommendations for software designers, and suggestions and hints for hardware and system designers of future PIM systems.

Our most significant recommendations for PIM software designers are:
\begin{enumerate}
\vspace{-8pt}
\setlength\itemsep{-3pt}
    \item Design algorithms that provide high load balance across threads of PIM core in terms of computations, loop control iterations, synchronization points and memory accesses.
    \item Design compressed data structures that can be effectively partitioned across DRAM banks, with the goal of providing high computation balance across PIM cores.
    \item Design \textit{adaptive} algorithms that trade off computation balance across PIM cores for lower data transfer costs to PIM-enabled memory, and adapt their configuration to the particular patterns of each input given, as well as the characteristics of the PIM hardware.
\end{enumerate}

Our most significant suggestions for PIM hardware and system designers are:
\begin{enumerate}
\vspace{-8pt}
\setlength\itemsep{-3pt}
    \item Provide low-cost synchronization support and hardware support to enable concurrent memory accesses by multiple threads to the local DRAM bank to increase parallelism in a multithreaded PIM core.
    \item Optimize the broadcast collective operation in data transfers from main memory to PIM-enabled memory to minimize overheads of copying the input data into all DRAM banks in the PIM system.
    \item Optimize the gather collective operation \textit{at DRAM bank granularity} for data transfers from PIM-enabled memory to the host CPU to minimize overheads of retrieving the output results.
    \item Design high-speed communication channels and optimized libraries for data transfers to/from thousands of DRAM banks of PIM-enabled memory.
\end{enumerate}

Our \SparseP{} software package is freely and publicly available~\cite{SparsePLibrary} to enable further research on \spmv{} in current and future PIM systems. The main contributions of this work are as follows:
\begin{itemize}
\vspace{-8pt}
\setlength\itemsep{-3pt}
    \item We present \SparseP{}, the first open-source \spmv{} software package for real PIM architectures. \SparseP{} includes 25 \spmv{} kernels, supporting the four most widely used compressed matrix formats and a wide range of data types. \SparseP{} is publicly available at~\cite{SparsePLibrary}, and can be useful for researchers to improve multiple aspects of future PIM hardware and software. 
    \item We perform the first comprehensive study of the widely used \spmv{} kernel on the UPMEM PIM architecture, the first real commercial PIM architecture. We analyze performance implications of \spmv{} PIM execution using a wide variety of (1) compressed matrix formats, (2) data types, (3) data partitioning and load balancing techniques, and (4) 26 sparse matrices with diverse sparsity patterns.
    \item We compare the performance and energy of \spmv{} on the state-of-the-art UPMEM PIM system with 2528 PIM cores to state-of-the-art CPU and GPU systems. \spmv{} execution achieves less than 1\% of the peak performance on processor-centric CPU and GPU systems, while it achieves on average 51.7\% of the peak performance on the UPMEM PIM system, thus better leveraging the computation capabilities of underlying hardware. The UPMEM PIM system also provides high energy efficiency on the \spmv{} kernel.
\end{itemize}

\section{Background and Motivation}
\vspace{-2pt}
\subsection{Sparse Matrix Vector Multiplication (\spmv{})}

\begin{figure}[H]
    \centering
    \includegraphics[width=0.9\linewidth]{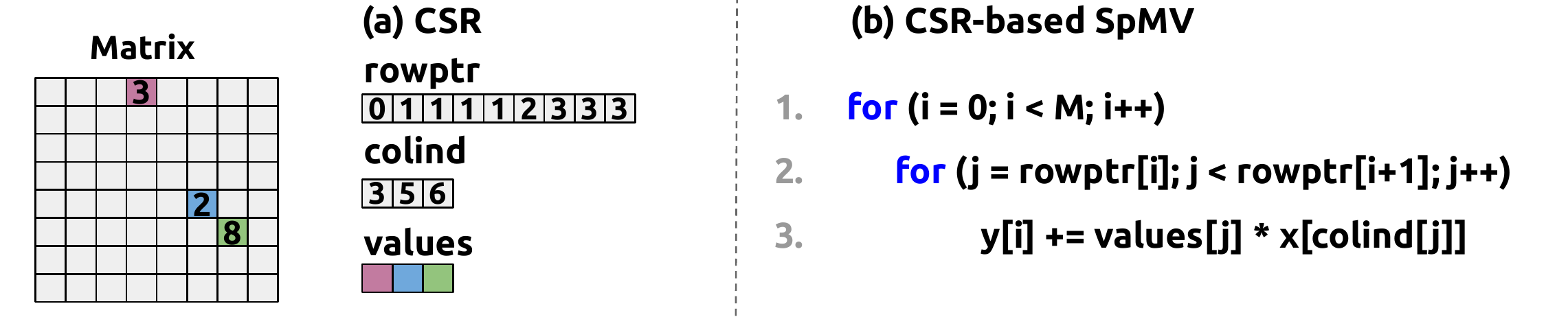}
    \vspace{-5pt}
    \caption{(a) CSR representation of a sparse matrix. (b) CSR-based \spmv{} implementation. }
    \label{fig:csr-spmv}
    \vspace{-14pt}
\end{figure}

The \spmv{} kernel multiples a sparse matrix of size $M\times N$ with a dense input vector of size $1\times N$ to compute an output vector of size $M\times 1$. The \spmv{} kernel is widely used in a variety of applications including graph processing~\cite{Brin1998the,besta2017slimsell,Kanellopoulos2019SMASH,Giannoula2018Combining}, neural networks~\cite{liu2015sparse,Zhou2018CambriconS,Han2016EIE,Han2015Learning}, machine learning~\cite{dnn2018,linden2003amazon,recommenderFB2019,recommendfb2,Zhang2016CambriconX,Gupta2020DeepRecSys}, and high performance computing~\cite{solversGPU, Falgout2006an,dongarra1996sparse, falgout2002hypre, henon2002pastix,Cho2020Near,Elafrou2017PerformanceXeon}. These applications involve matrices with
very high sparsity~\cite{Kanellopoulos2019SMASH,Elafrou2018SparseX,Elafrou2017PerformanceAA,YouTubeGraph,FacebookGraph,Goumas2008Understanding,White97Improving,Helal2021ALTO,Pelt2014Medium}, i.e., a large fraction of zero elements. Thus, using a compression scheme is a straightforward approach to avoid unnecessarily storing zero elements and performing computations on them. For general sparse matrices, the most widely used storage format is the Compressed Sparse Row (CSR) format~\cite{bjorck1996numerical,Pooch1973Survey}. Figure~\ref{fig:csr-spmv} presents an example of a compressed matrix using the CSR format (left), and the CSR-based \spmv{} execution (right), assuming an input vector $x$ and an output vector $y$.

\subsubsection{Compressed Matrix Storage Formats}

\begin{figure}[H]
    \vspace{-4pt}
    \centering
    \includegraphics[width=0.99\linewidth]{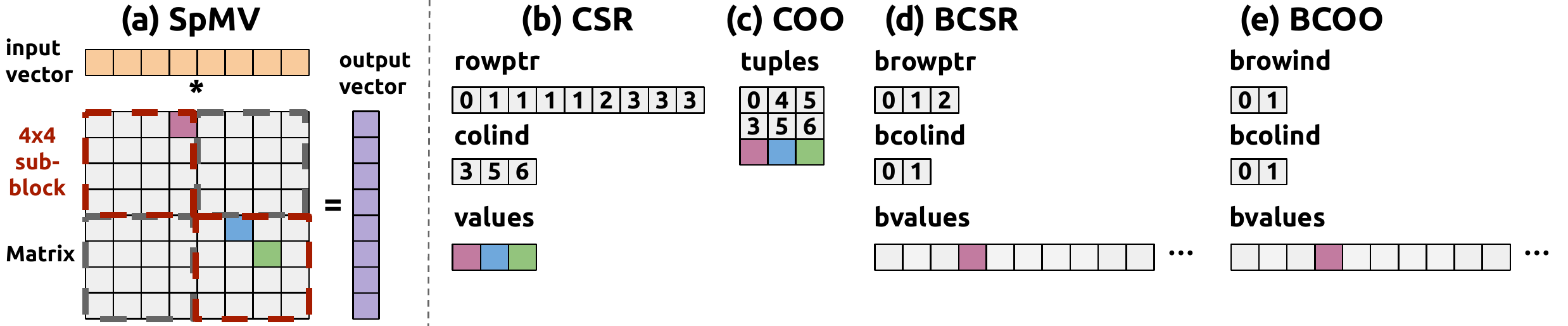}
    \vspace{-2pt}
    \caption{(a) \spmv{} with a dense matrix representation, and (b) CSR, (c) COO, (d) BCSR, (e) BCOO formats.}
    \label{fig:sparse_formats}
\end{figure}

Several prior works~\cite{Im1999Optimizing,Langr2016Evaluation,Liu2015CSR5,Pinar1999Improving,Vuduc2005Fast,Yang2014Optimization,Shengen2014YaSpMV,Kourtis2011CSX,Kourtis2008Optimizing,Belgin2009Pattern,LIL,ELL,bjorck1996numerical,Pooch1973Survey,Shubhabrata2007Scan,Changwan2018Efficient,Liu2013Efficient,Monakov2010Automatically,Saad1989Krylov,Buluc2009Parallel,Martone2014251,Martone2010Blas,Kreutzer2012Sparse,Benatia2018BestSF} propose compressed storage formats for sparse matrices, which are typically of two types~\cite{Kanellopoulos2019SMASH}. The first approach is to design general purpose compressed formats, such as CSR~\cite{Pooch1973Survey,bjorck1996numerical}, CSR5~\cite{Liu2015CSR5}, COO~\cite{Pooch1973Survey,Shubhabrata2007Scan}, BCSR~\cite{Im1999Optimizing}, and BCOO~\cite{Pooch1973Survey}. Such encodings are general in applicability and are highly-efficient in storage. The second approach is to leverage a certain known structure in a given type of sparse matrix. For example, the DIA format~\cite{Belgin2009Pattern} is effective in matrices where the non-zero elements are concentrated along the diagonals of the matrix. Such encodings aim to improve performance of sparse matrix computations by specializing to particular matrix patterns, but they sacrifice generality. In this work, we explore with the four most widely used \textit{general} compressed formats (Figure~\ref{fig:sparse_formats}), which we describe in more detail next.

\noindent\textbf{Compressed Sparse Row (CSR)~\cite{bjorck1996numerical,Pooch1973Survey}.} The CSR format (Figure~\ref{fig:sparse_formats}b) sequentially stores values in a row-wise order. A column index array (\texttt{colind[]}) and a value array (\texttt{values[]}) store the column index and value of each non-zero element, respectively. An array, named \texttt{rowptr[]}, stores the location of the first non-zero element of each row within the \texttt{values[]} array. The values of an adjacent pair of the \texttt{rowptr[]} array, i.e., \texttt{rowptr[i, i+1]}, represent a slice of the \texttt{colind[]} and \texttt{values[]} arrays. The corresponding slice of the \texttt{colind[]} and \texttt{values[]} arrays stores the column indices and the values of the non-zero elements, respectively, for the i-th row of the original matrix.

\noindent\textbf{Coordinate Format (COO)~\cite{Pooch1973Survey,Shubhabrata2007Scan}.} The COO format (Figure~\ref{fig:sparse_formats}c) stores the non-zero elements as a series of tuples (\texttt{tuples[]} array). Each tuple includes the row index, column index, and value of the non-zero element.

\noindent\textbf{Block Compressed Sparse Row (BCSR)~\cite{Im1999Optimizing}.} The BCSR format (Figure~\ref{fig:sparse_formats}d) is a block representation of CSR. Instead of storing and indexing single non-zero elements, BCSR stores and indexes $r\times c$ sub-blocks with at least one non-zero element. The original matrix is split into $r\times c$ sub-blocks. Figure~\ref{fig:sparse_formats}d shows an example of BCSR assuming $4\times 4$ sub-blocks. The original matrix of Figure~\ref{fig:sparse_formats}a is split into four sub-blocks, and two of them (highlighted with red color) contain at least one non-zero element. The \texttt{bvalues[]} array stores the values of all the \emph{non-zero sub-blocks} of the original matrix. Each non-zero sub-block is stored in the \texttt{bvalues[]} array with a dense representation, i.e., padding with zero values when needed. The \texttt{bcolind[]} array stores the block-column index of each non-zero sub-block. The \texttt{browptr[]} array stores the location of the first non-zero sub-block of each block row within the \texttt{bcolind[]} array, assuming a block row represents $r$ consecutive rows of the original matrix, where $r$ is the vertical dimension of the sub-block.

\noindent\textbf{Block Coordinate Format (BCOO)~\cite{Pooch1973Survey}.} The BCOO format is the block counterpart of COO. The \texttt{browind[]}, \texttt{bcolind[]} and  \texttt{bvalues[]} arrays store the row indices, column indices and values of the non-zero sub-blocks, respectively. Figure~\ref{fig:sparse_formats}e shows an example of BCOO, assuming $4\times 4$ sub-blocks.

\subsubsection{\spmv{} in Processor-Centric Systems}

Many prior works~\cite{Elafrou2018SparseX,Shengen2014YaSpMV,Elafrou2017PerformanceAA,Goumas2009Performance,Karakasis2009Performance,Vuduc2005Fast,im2004sparsity,Vuduc2003PhD,Vuduc2002Performance,White97Improving,Elafrou2017PerformanceXeon,Elafrou2019BASMAT} generally show that \spmv{} performs poorly on commodity CPU and GPU systems, and achieves a small fraction of the peak performance (e.g., 10\% of the peak performance~\cite{Vuduc2003PhD}) due to its algorithmic nature,  the employed compressed matrix storage format and the sparsity pattern of the matrix.

The \spmv{} kernel is highly bottlenecked by the memory subsystem in processor-centric CPU and GPU systems due to three reasons. First, due to its algorithmic nature there is \textit{no} temporal locality in the input matrix. Unlike  traditional algebra kernels like Matrix Matrix Multiplication or LU decomposition, the elements of the matrix in \spmv{} are used only \textit{once}~\cite{Goumas2009Performance,Goumas2008Understanding}. Second, due to the sparsity of the matrix, the matrix is stored in a compressed format (e.g., CSR) to avoid unnecessary computations and data accesses. Specifically, the non-zero elements of the matrix are stored contiguously in memory, while additional data structures assist in the proper traversal of the matrix, i.e., to discover the positions of the non-zero elements. For example, CSR uses the \texttt{rowptr[]} and \texttt{colind[]} arrays to discover the positions of the non-zero elements of the matrix. These additional data structures cause additional memory access operations, memory bandwidth pressure and contention with other requests in the memory subsystem. Third, due to the sparsity of the input matrix, \spmv{} causes irregular memory accesses to the elements of the input vector $x$. The memory accesses to the elements of the input vector are input driven, i.e., they follow the sparsity pattern of the input matrix. This irregularity results to poor data locality on the elements of the input vector and expensive data accesses, because it increases the average access latency due to a high number of cache misses on commodity systems with deep cache hierarchies~\cite{Goumas2009Performance,Goumas2008Understanding}.
As a result, memory-centric near-bank PIM systems constitute a better fit for the widely used \spmv{} kernel, because they provide high levels of parallelism, large aggregate memory bandwidth and low memory access latency~\cite{Gomez2021Benchmarking,Gomez2021Analysis, upmem, Hadi2016Chameleon, Lee2021HardwareAA}.

\vspace{-2pt}
\subsection{Near-Bank PIM Systems}

\begin{figure}[H]
\vspace{-2pt}
    \centering
    \includegraphics[width=0.99\linewidth]{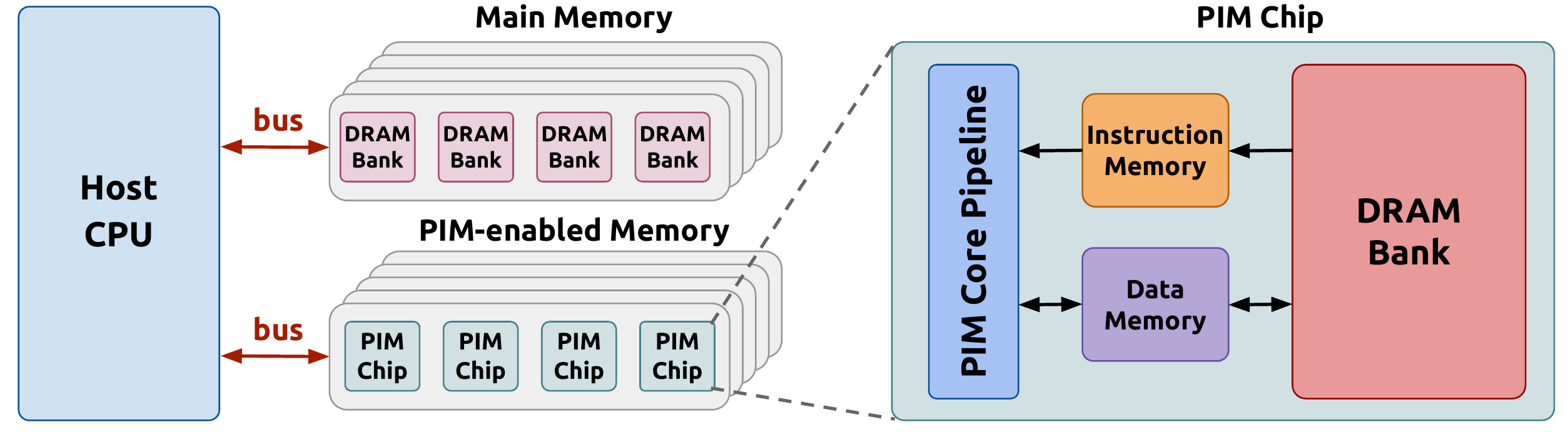}
    \vspace{-4pt}
    \caption{High-level organization of a near-bank PIM architecture.}
    \label{fig:near-bank-pim}
    \vspace{-12pt}
\end{figure}

Figure~\ref{fig:near-bank-pim} shows the baseline organization of a near-bank PIM system that we assume in this work. The PIM system consists of a host CPU, standard DRAM memory modules, and PIM-enabled memory modules. PIM-enabled modules are connected to the host CPU using one or more memory channels, and include multiple PIM chips. A PIM chip (Figure~\ref{fig:near-bank-pim} right) tightly integrates a low-area PIM core with a DRAM bank. We assume that each PIM core can additionally include a small private instruction memory and a small data (scratchpad or cache) memory. PIM cores can access data located on their local DRAM bank, and typically there is no direct communication channel among PIM cores. The DRAM banks of PIM chips are accessible by the host CPU for copying input data and retrieving results via the memory bus.

\subsubsection{The UPMEM PIM Architecture} 

The UPMEM PIM system~\cite{Gomez2021Analysis,Gomez2021Benchmarking,devaux2019} includes the host CPU with standard main memory, and UPMEM PIM modules. An UPMEM PIM module is a standard DDR4-2400 DIMM~\cite{ddr4jedec} with 2 ranks. Each rank contains 64 PIM cores, which are called \underline{D}RAM \underline{P}rocessing \underline{U}nits (DPUs). In the current UPMEM PIM system, there are 20 double-rank PIM DIMMs with 2560 DPUs.\footnote{There are thirty two faulty DPUs in the system where we run our experiments. They cannot be used and do not affect the correctness of our results, but take away from the system’s full computational power of 2560 DPUs.}

\noindent\textbf{DPU Architecture and Interface.} Each DPU has exclusive access to a 24-KB instruction memory, called \textbf{IRAM}, a 64-KB scratchpad memory, called \textbf{WRAM}, and a 64-MB DRAM bank, called \textbf{MRAM}. A DPU is a multithreaded in-order 32-bit RISC core that can potentially reach 500 MHz~\cite{upmem}. The DPU has 24 hardware threads, each of which has 24 32-bit general purpose registers. The DPU pipeline has 14 stages, and supports a single cycle 8x8-bit multiplier. Multiplications on 64-bit integers, 32-bit floats and 64-bit floats are not supported in hardware, and require longer routines with a large number of operations~\cite{Gomez2021Benchmarking,Gomez2021Analysis,upmem}. Threads share the IRAM and WRAM, and can access the MRAM by executing transactions at 64-bit granularity via a DMA engine, i.e., data can be accessed from/to MRAM as a multiple of 8 bytes, up to 2048 bytes. MRAM transactions are serialized in the DMA engine. The ISA provides DMA instructions to move instructions from MRAM to IRAM, or data between MRAM and WRAM. The DPU accesses the WRAM through 8-, 16-, 32- and 64-bit load/store instructions. DPUs use the \textit{Single Program Multiple Data} programming model, where software threads, called \textbf{tasklets}, execute the same code, but operate in different pieces of data, and can execute different control-flow paths during runtime. Tasklets can synchronize using mutexes, barriers, handshakes and semaphores provided by the UPMEM runtime library.

\noindent\textbf{CPU-DPU Data Transfers.} Standard main memory and PIM-enabled memory have different data layouts. The UPMEM SDK~\cite{upmem-guide} has a transposition library to execute necessary data shuffling when moving data between main memory and MRAM banks of PIM-enabled memory modules via a programmer-transparent way. The CPU-DPU and DPU-CPU data transfers can be performed in parallel, i.e., concurrently across multiple MRAM banks, with the limitation that \textit{the transfer sizes from/to all MRAM banks need to be the same}. The UPMEM SDK provides two options: (i) perform parallel transfers to all MRAM banks of all ranks, or (ii) iterate over each rank to perform parallel transfers to MRAM banks of the same rank, and serialize data transfers across ranks.

\section{The \SparseP{} Library}

This section describes the parallelization techniques that we explore for \spmv{} on real PIM architectures, and presents the \spmv{} implementations of our \SparseP{} package. Section~\ref{sec:pim_exec} describes \spmv{} execution on a real PIM system. Section~\ref{sec:lib_overview} presents an overview of the data partitioning techniques that we explore.
Section~\ref{sec:lib_1d-2d} and Section~\ref{sec:lib_1dpu} describe in detail the parallelization techniques across PIM cores, and across threads within a PIM core, respectively. Section~\ref{sec:kernel_impl} describes the kernel implementation for all compressed matrix storage formats.

\subsection{\spmv{} Execution on a PIM System}\label{sec:pim_exec}
\begin{figure}[H]
    \centering
    \includegraphics[width=0.99\linewidth]{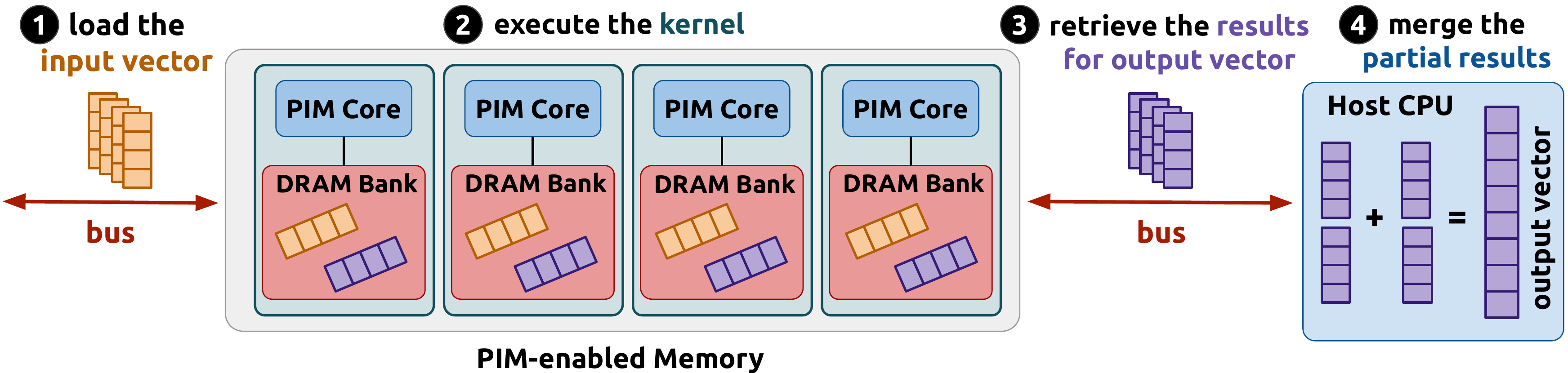}
    \vspace{-2pt}
    \caption{Execution of the \spmv{} kernel on a real PIM system.}
    \label{fig:spmv-pim-execution}
    \vspace{-8pt}
\end{figure}

Figure~\ref{fig:spmv-pim-execution} shows the \spmv{} execution on a real PIM system, which is broken down in four steps: (1) the time to load the input vector into DRAM banks of PIM-enabled memory (\texttt{\textbf{load}}), (2) the time to execute the \spmv{} kernel on PIM cores (\texttt{\textbf{kernel}}), (3) the time to retrieve from DRAM banks to the host CPU results for the output vector (\texttt{\textbf{retrieve}}), and (4) the time to merge partial results and assemble the final output vector on the host CPU (\texttt{\textbf{merge}}). In our analysis, we omit the time to load the matrix into PIM-enabled memory, since this step can typically be hidden in real-world applications (it can be overlapped with other computation performed by the application or amortized if the application performs multiple \spmv{} iterations on the same matrix).

\subsection{Overview of Data Partitioning Techniques}\label{sec:lib_overview}
\begin{figure}[H]
    \centering
    \includegraphics[width=0.78\linewidth]{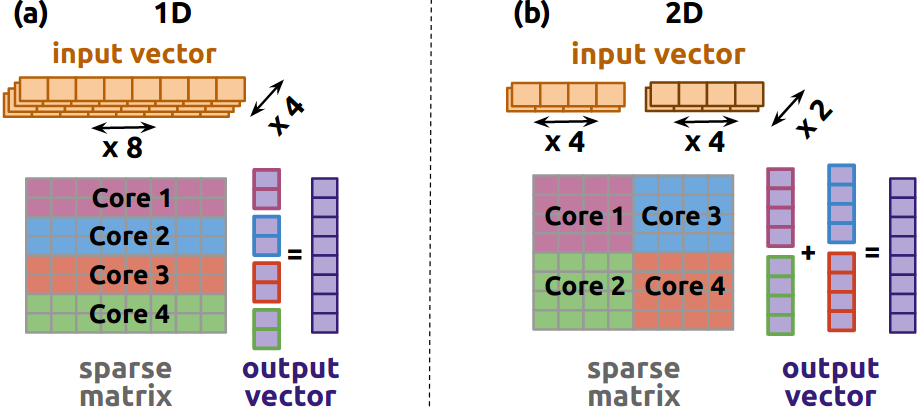}
    \vspace{-2pt}
    \caption{Data partitioning techniques of the \SparseP{} package.}
    \label{fig:partitioning_overview}
    \vspace{-8pt}
\end{figure}

To parallelize the \spmv{} kernel, we implement well-crafted data partitioning schemes to split the matrix across multiple DRAM banks of PIM cores. \SparseP{} supports two general types of data partitioning techniques, shown in Figure~\ref{fig:partitioning_overview}.

First, we provide an 1D partitioning technique (Figure~\ref{fig:partitioning_overview}a), where the matrix is horizontally partitioned across PIM cores, and the whole input vector is copied into the DRAM bank of each PIM core. With the 1D partitioning technique, almost the entire \spmv{} computation is performed using only PIM cores, since the \texttt{merge} step in the host CPU is negligible: a very small number of partial results is created, i.e., only for a few rows that are split across neighboring PIM cores. Thus, the number of partial elements of the output vector is at most equal to the number of PIM cores used. Second, we provide a 2D partitioning technique (Figure~\ref{fig:partitioning_overview}b), where the matrix is partitioned into 2D tiles, the number of which is equal to the number of PIM cores. With the 2D partitioning technique, we aim to strive a balance between computation and data transfer costs, since only a subset of the elements of the input vector is copied into the DRAM bank of each PIM core. However, PIM cores assigned to tiles that horizontally overlap, i.e., tiles that share the same rows of the original matrix (rows that are split across multiple tiles), produce \emph{many} partial results for the elements of the output vector. These partial results are transferred to the host CPU, and merged by CPU cores, which assemble the final output vector. In the \SparseP{} library, the \texttt{merge} step performed by the CPU cores is parallelized using the OpenMP API~\cite{Dagum98OpenMP}.

In both data partitioning schemes, matrices are stored in a row-sorted way, i.e., the non-zero elements are sorted in increasing order of their row indices. Therefore, each PIM core computes results for a \textit{continuous} subset of elements of the output vector. This way we minimize data transfer costs, since we only transfer necessary data to the host CPU, i.e., \textit{the values} of the elements of the output vector produced at PIM cores. If each PIM core instead computed results for a \textit{non-continuous} subset of elements of the output vector, an additional array \textit{per core}, which would store \textit{the indices} of the \textit{non-continuous} elements within the output vector, would need to be transferred to the host CPU, causing additional data transfer overheads.

\subsection{Parallelization Techniques Across PIM Cores}\label{sec:lib_1d-2d}
To parallelize \spmv{} across multiple PIM cores \SparseP{} supports various parallelization schemes for both 1D and 2D partitioning techniques.

\subsubsection{1D Partitioning Technique}

\begin{figure}[t]
    
    \centering
    \includegraphics[width=1\linewidth]{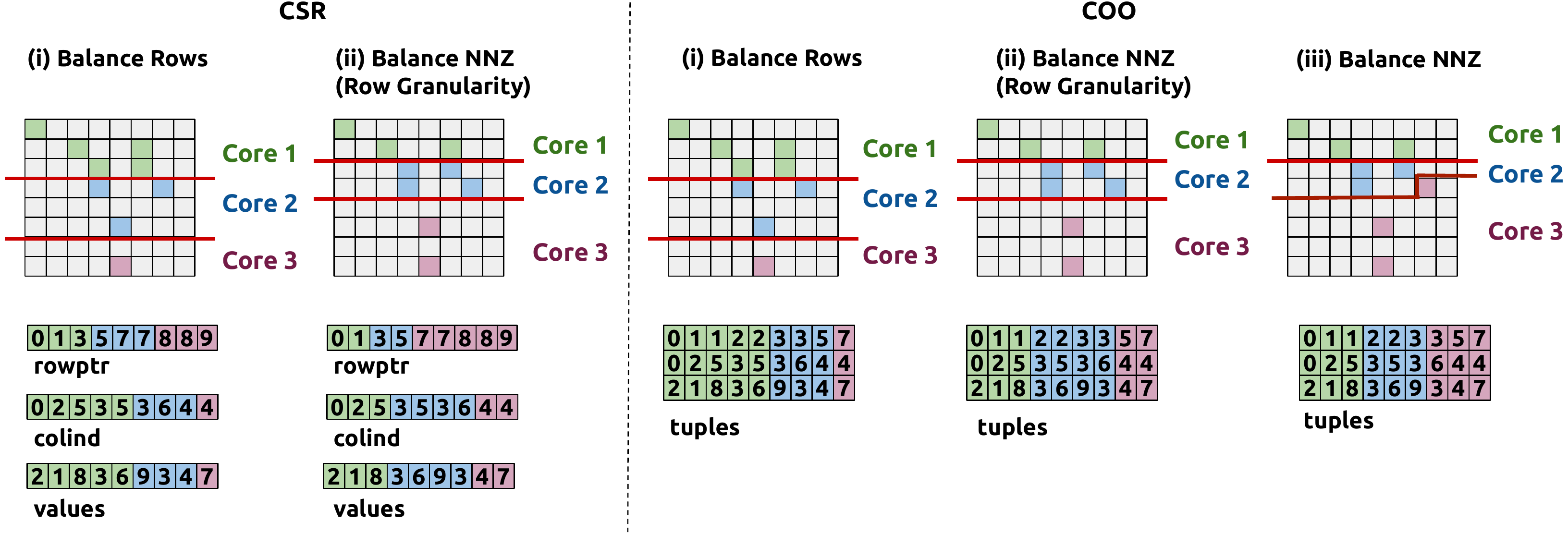}
    \vspace{-18pt}
    \caption{Load balancing schemes across PIM cores for the CSR (left) and COO (right) formats with the 1D partitioning technique. The colored cells of the matrix represent non-zero elements.}
    \label{fig:1d_partitioning}
\end{figure}

To efficiently parallelize \spmv{} across multiple PIM cores via the 1D partitioning technique, \SparseP{} provides various load balancing schemes for each supported compressed matrix format. Figure~\ref{fig:1d_partitioning} presents an example of parallelizing \spmv{} across multiple PIM cores using load balancing schemes for the CSR and COO formats. For the CSR and COO formats, we balance either the rows, such that each PIM core processes almost the same number of rows, or the non-zero elements, such that each PIM core processes almost the same number of non-zero elements. In the CSR format, since the matrix is stored in row-order, i.e., the \texttt{rowptr[]} array stores the index pointers of the non-zero elements of \textit{each row}, and thus balancing the non-zero elements across PIM cores is performed at row granularity. In the COO format, the matrix is stored in non-zero order using the \texttt{tuples[]} array, and thus balancing the non-zero elements can be performed either at row granularity, or by splitting a row across two neighboring PIM cores to provide a near-perfect non-zero element balance across cores. In the latter case, as mentioned, a small number of partial results for the output vector is merged by the host CPU: if the row is split between two neighboring PIM cores at most one element needs to be accumulated at the host CPU cores.

\begin{figure}[t]
    %\vspace{-4pt}
    \centering
    \includegraphics[width=1\linewidth]{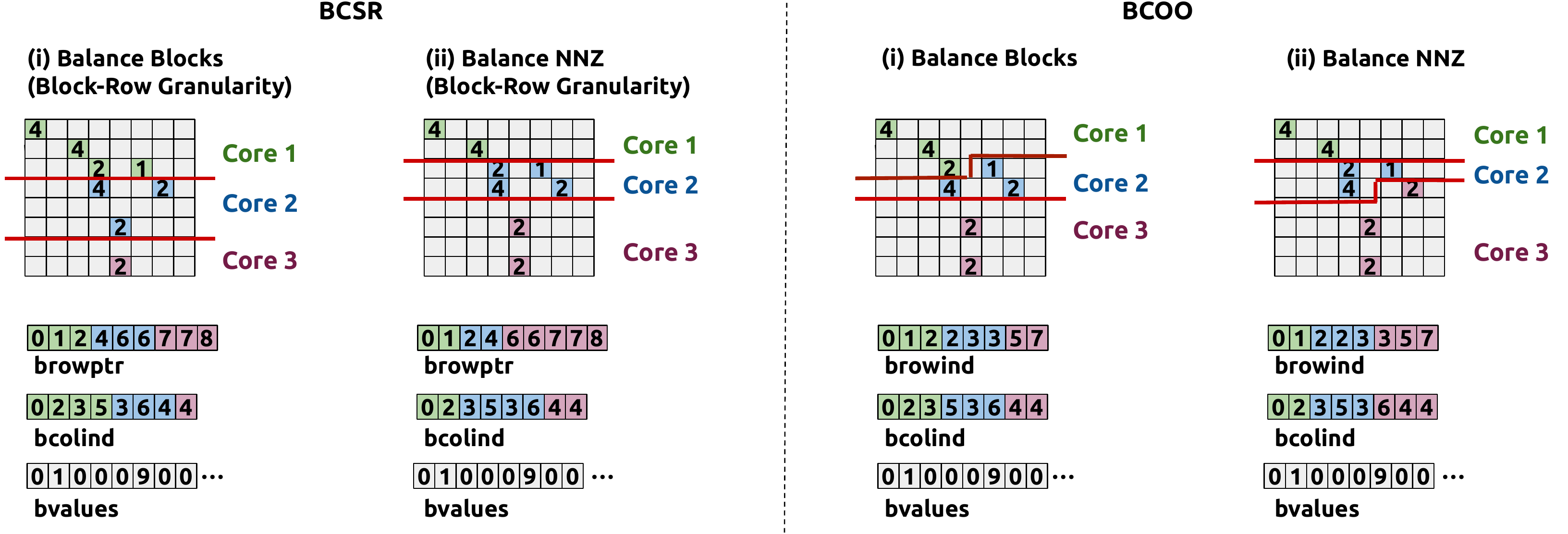}
    \vspace{-18pt}
    \caption{Load balancing schemes across PIM cores for the BCSR (left) and BCOO (left) formats with the 1D partitioning technique. The cells of the matrix represent sub-blocks of size 4x4. The colored cells of the matrix represent non-zero sub-blocks, and the number inside a colored cell describes the number of non-zero elements of the corresponding sub-block.}
    \label{fig:1d_partitioning2}
    \vspace{-8pt}
\end{figure}

Figure~\ref{fig:1d_partitioning2} presents an example of parallelizing \spmv{} across multiple PIM cores using load balancing schemes of the BCSR and BCOO formats. In Figure~\ref{fig:1d_partitioning2}, the cells of the matrix represent sub-blocks of size 4x4: the \textit{grey} cells represent sub-blocks that do not have \textit{any} non-zero element, and the \textit{colored} cells represent sub-blocks that have \textit{$k$} non-zero elements, where $k$ is the number shown inside the colored cell. In the BCSR and BCOO formats, since the matrix is stored in sub-blocks of non-zero elements, we balance either the blocks, such that each PIM core processes almost the same number of blocks, or the non-zero elements, such that each PIM core processes almost the same number of non-zero elements. Similarly to CSR, in the BCSR format, the matrix is stored in block-row-order,  i.e., the \texttt{browptr[]} array stores the index pointers of the non-zero blocks of \textit{each block row} (recall that a block row represents $r$ consecutive rows of the original matrix, where $r$ is the vertical dimension of the sub-block), and thus balancing the blocks or the non-zero elements across cores is limited to be performed at block-row granularity. In the BCOO format, given that a block-row might be split across two PIM cores, a small number of partial results for the output vector is merged by the host CPU: between two neighboring PIM cores at most block size $r$ elements ($r$ is the vertical dimension of the block size) might need to be accumulated at the host CPU cores.

\subsubsection{2D Partitioning Technique}

\begin{figure}[H]
    \vspace{-4pt}
    \centering
    \includegraphics[width=1\linewidth]{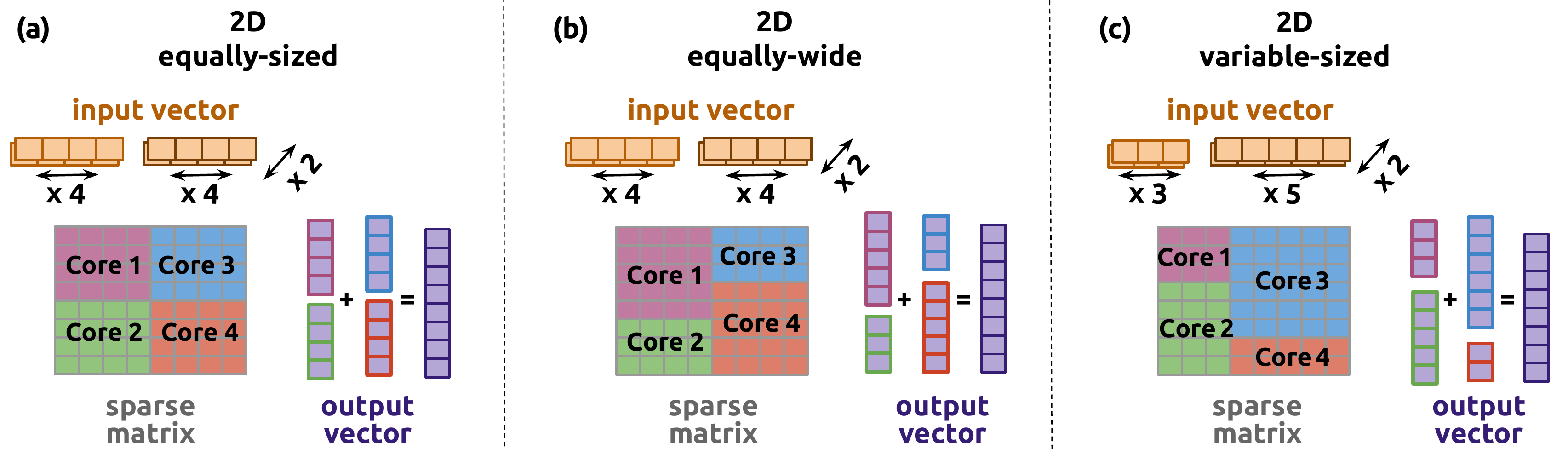}
    \vspace{-16pt}
    \caption{The 2D partitioning techniques of \SparseP{} package assuming
    4 PIM cores and 2 vertical partitions.}
    \label{fig:2d_partitioning}
    \vspace{-12pt}
\end{figure}

\SparseP{} includes three 2D partitioning techniques, shown in Figure~\ref{fig:2d_partitioning}:
\begin{enumerate}
\vspace{-8pt}
\setlength\itemsep{-4pt}
    \item \textbf{\equallySized} (Figure~\ref{fig:2d_partitioning}a): The 2D tiles are statically created to have the same height and width. This way the subsets of the elements for the input and output vectors have the same sizes across all PIM cores.
    \item \textbf{\equallyWidth} (Figure~\ref{fig:2d_partitioning}b): The 2D tiles have the same width and variable height. This way the subset of the elements for the input vector has the same size across PIM cores, while the subset of the elements for the output vector varies across PIM cores. We balance the non-zero elements across the tiles of the \textit{same} vertical partition, such that we can provide high non-zero element balance across PIM cores assigned to the same vertical partition.
    \item \textbf{\variableSized} (Figure~\ref{fig:2d_partitioning}c): The 2D tiles have both variable width and height. We balance the non-zero elements both across the vertical partitions and across the tiles of the \textit{same} vertical partition. This way we can provide high non-zero element balance across all PIM cores.
\end{enumerate}

\SparseP{} provides various load balancing schemes across PIM cores in the \equallyWidth{} and \variableSized{} techniques. In the \equallyWidth{} technique, for the CSR and COO formats, we balance the non-zero elements across the tiles of the same vertical partition. Load balancing in the CSR format is performed at row-granularity, i.e., splitting the \texttt{rowptr[]} array across PIM cores. For the BCSR and BCOO formats, we balance either the blocks or the non-zero elements across the tiles of the same vertical partition. Load balancing in the BCSR format is performed at block-row granularity, i.e., splitting the \texttt{browptr[]} array across PIM cores. In the \variableSized{} technique, we first balance the non-zero elements across the vertical partitions, such that the vertical partitions include the same number of non-zero elements. Then, across the tiles of the same vertical partition, we balance the non-zero elements for the CSR (at row-granularity) and COO formats, and either the blocks or the non-zero elements for the BCSR (at block-row granularity) and BCOO formats.

Table~\ref{table:library} summarizes the parallelization approaches across PIM cores. Please also see Appendix~\ref{sec:appendix-sparseP} for all \spmv{} kernels provided by the \SparseP{} software package.  All kernels support a wide range of data types, i.e., 8-bit integer (\textbf{int8}), 16-bit integer (\textbf{int16}), 32-bit integer (\textbf{int32}), 64-bit integer (\textbf{int64}), 32-bit float (\textbf{fp32}), and 64-bit float (\textbf{fp64}) data types.

\begin{table}[t]
\centering
\begin{minipage}{0.66\linewidth}
%\ra{0.9}
\resizebox{1.0\linewidth}{!}{
\begin{tabular}{| l | l | l | l |} 
 \hline 
 \cellcolor{gray!15}\raisebox{-0.30\height}{\textbf{Partitioning}}& \cellcolor{gray!15}\raisebox{-0.30\height}{\textbf{Compressed}} & \cellcolor{gray!15}\raisebox{-0.30\height}{\textbf{Load Balancing}} \\ 
 \cellcolor{gray!15}\textbf{Technique} & \cellcolor{gray!15}\textbf{Format} & \cellcolor{gray!15}\textbf{Across PIM Cores}  \\ [0.5ex] \hline \hline
% \hline\hline
    \multirow{9}{*}{\hspace*{2.1em}\turnbox{0}{1D}} & \multirow{2}{*}{CSR} &  rows (\textbf{CSR.row})  \\
    & & nnz$^{\star}$ (\textbf{CSR.nnz}) \\ \cline{2-3}
    & \multirow{3}{*}{COO} &  rows (\textbf{COO.row}) \\
    & & nnz$^{\star}$ (\textbf{COO.nnz-rgrn})  \\
    & & nnz (\textbf{COO.nnz}) \\  \cline{2-3}
    & \multirow{2}{*}{BCSR} &  blocks$^{\dagger}$ (\textbf{BCSR.block})   \\
    & & nnz$^{\dagger}$ (\textbf{BCSR.nnz}) \\  \cline{2-3}
    & \multirow{2}{*}{BCOO} &  blocks (\textbf{BCOO.block}) \\
    & & nnz (\textbf{BCOO.nnz}) \\ \hline
    \multirow{4}{*}{\shortstack{2D \\ \equallySized}} & CSR (\textbf{DCSR}) & - \\  \cline{2-3}
    &  COO (\textbf{DCOO}) & -  \\  \cline{2-3}
    &  BCSR (\textbf{DBCSR}) & -  \\  \cline{2-3}
    &  BCOO (\textbf{DBCOO}) & - \\ \hline
  \multirow{6}{*}{\shortstack{2D \\ \equallyWidth}} &  CSR (\textbf{RBDCSR}) & nnz$^{\star}$ \\  \cline{2-3}
    &  COO (\textbf{RBDCOO}) & nnz  \\  \cline{2-3}
    & \multirow{2}{*}{BCSR} & blocks$^{\dagger}$ (\textbf{RBDBCSR}) \\
    &  & nnz$^{\dagger}$ \\ \cline{2-3}
    & \multirow{2}{*}{BCOO} & blocks (\textbf{RBDBCOO})  \\ 
    & & nnz  \\ \hline    
  \multirow{6}{*}{\shortstack{2D \\ \variableSized}} &  CSR (\textbf{BDCSR}) & nnz$^{\star}$  \\ \cline{2-3}
    &  COO (\textbf{BDCOO}) & nnz   \\  \cline{2-3}
    &  \multirow{2}{*}{BCSR} & blocks$^{\dagger}$ (\textbf{BDBCSR})  \\
    &  & nnz$^{\dagger}$ \\ \cline{2-3}
    &  \multirow{2}{*}{BCOO } & blocks (\textbf{BDBCOO}) \\ 
    &  &  nnz  \\ \hline     
 %\bottomrule
\end{tabular}
}
\end{minipage} % 
\vspace{4pt}
\caption{Parallelization techniques across PIM cores of the \SparseP{} library. $^{\star}$: row-granularity, $^{\dagger}$: block-row-granularity}
\label{table:library}
\vspace{-12pt}
\end{table}

\subsection{Parallelization Techniques Across Threads within a PIM Core}\label{sec:lib_1dpu}
PIM cores can support multiple hardware threads to exploit high memory bank bandwidth~\cite{Gomez2021Analysis,Gomez2021Benchmarking}. To parallelize \spmv{} across multiple threads within a multithreaded PIM core \SparseP{} supports various load balancing schemes for each compressed matrix format, and three synchronization approaches to ensure correctness among threads of a PIM core.

\subsubsection{Load Balancing Approaches} 
\vspace{-12pt}

In a similar way as explained in Figure~\ref{fig:1d_partitioning}, for the CSR and COO formats, we balance either the rows, such that each thread processes almost the same number of rows, or the non-zero elements, such that each thread processes almost the same number of non-zero elements. In the CSR format, matrix is stored in row-order, and thus load balancing across threads is performed at row granularity. In the UPMEM PIM system, elements of the output vector are accessed at 64-bit granularity in DRAM memory. Thus, when balancing is performed at row granularity, we assign rows to threads in chunks of $8 / sizeof(data\_type)$ to ensure 8-byte alignment on the elements of the output vector. In the COO format, balancing the non-zero elements can be performed either at row granularity or by splitting the row between threads, i.e., providing an almost perfect non-zero balance  across threads. In the latter case, synchronization among threads for write accesses on the elements of the output vector can be implemented with three synchronization approaches described in Section~\ref{sec:1dpu_sync}.

For the BCSR and BCOO formats, we balance either the blocks, such that each thread processes almost the same number of blocks, or the non-zero elements, such that each thread processes almost the same number of non-zero elements. In the BCSR format, the matrix is stored in block-row order, and thus load balancing across threads is performed at block row granularity. For both formats, the block sizes are \textit{configurable} in \SparseP{}. In our evaluation, we use block sizes of 4x4, since these are the most common dimensions to cover various sparse matrices~\cite{asgari2020copernicus,Elafrou2018SparseX,Karakasis2009Performance}. In the UPMEM PIM architecture, elements of the output vector are accessed at 64-bit granularity. Therefore, for the BCSR format, with an 8-bit integer data type and small block sizes (4x4 or smaller), threads use synchronization primitives to ensure correctness when writing the elements of the output vector. This is because different threads may write to the same 64-bit-aligned DRAM memory location. Synchronization among threads for writes to the elements of the output vector is necessary for all configurations of the BCOO format, and can be implemented with three approaches described next.

\subsubsection{Synchronization Approaches}\label{sec:1dpu_sync}

\SparseP{} provides three synchronization approaches.
\begin{enumerate}
\vspace{-8pt}
\setlength\itemsep{-4pt}
\item \textbf{Coarse-Grained Locking (\textcolor{darkred}{lb-cg}).} One global mutex protects the elements of the entire output vector.
\item\textbf{Fine-Grained Locking (\textcolor{darkred}{lb-fg}).} Multiple mutexes protect the elements of the output vector. \SparseP{} associates mutexes to the elements of the output vector in a round-robin manner. The UPMEM API supports up to 56 mutexes~\cite{upmem-guide}. In our evaluation, we use 32 mutexes such that we can find the corresponding mutex for a particular element of the output vector only with a shift operation on the MRAM address, avoiding costly division operations.
\item\textbf{Lock-Free (\textcolor{darkred}{lf}).} Since the formats are row-sorted or block-row-sorted, race conditions in the elements of the output vector arise \textit{only in a few elements}, i.e., either when a row (or a block row for BCSR/BCOO) is split across threads, or when continuous elements of the output vector processed by different threads belong to the same 64-bit-aligned DRAM location in the UPMEM PIM system. In our proposed lock-free approach, threads temporarily store partial results for these few elements in the data (scratchpad) memory (i.e., WRAM in the UPMEM PIM system), and later one single thread merges the partial results, and writes the final result for the corresponding element of the output vector to the DRAM bank.
\end{enumerate}

Table~\ref{table:1DPU-impl} summarizes the parallelization techniques across threads of a PIM core. All kernels support a wide range of data types, i.e., 8-bit integer (\textbf{int8}), 16-bit integer (\textbf{int16}), 32-bit integer (\textbf{int32}), 64-bit integer (\textbf{int64}), 32-bit float (\textbf{fp32}), and 64-bit float (\textbf{fp64}) data types.

\begin{table}[t]
\centering
%\ra{0.9}
\begin{minipage}{0.9\linewidth}
\resizebox{1.0\linewidth}{!}{
\begin{tabular}{| l | l | l |} 
 \hline
  \cellcolor{gray!15}\raisebox{-0.30\height}{\textbf{Compressed}} & \cellcolor{gray!15}\raisebox{-0.30\height}{\textbf{Load Balancing}} & \cellcolor{gray!15}\raisebox{-0.30\height}{\textbf{Synchronization}}  \\ 
  \cellcolor{gray!15}\textbf{Format} & \cellcolor{gray!15}\textbf{Across Threads} & \cellcolor{gray!15}\textbf{Approach}  \\
  [0.5ex] \hline \hline
% \hline\hline
   \multirow{2}{*}{CSR} & rows (\textbf{CSR.row}) & -  \\
   & nnz$^{\star}$ (\textbf{CSR.nnz}) & -   \\ \hline
   \multirow{3}{*}{COO} & rows (\textbf{COO.row}) & -   \\
   & nnz$^{\star}$ (\textbf{COO.nnz-rgrn}) & -   \\
   & nnz (\textbf{COO.nnz}) & lb-cg / lb-fg / lf   \\ \hline
   \multirow{2}{*}{BCSR} & blocks$^{\dagger}$ (\textbf{BCSR.block}) & lb-cg / lb-fg (only for int8 and small block sizes)   \\
   & nnz$^{\dagger}$ (\textbf{BCSR.nnz}) & lb-cg / lb-fg (only for int8 and small block sizes)  \\  \hline
   \multirow{2}{*}{BCOO} & blocks (\textbf{BCOO.block}) & lb-cg / lb-fg / lf \\
   & nnz (\textbf{BCOO.nnz}) & lb-cg / lb-fg / lf  \\
 \hline
\end{tabular}
}
\end{minipage}
\vspace{4pt}
\caption{Parallelization schemes across threads of a PIM core. 
$^{\star}$: row-granularity, $^{\dagger}$: block-row-granularity}
\label{table:1DPU-impl}
\vspace{-16pt}
\end{table}

\subsection{Kernel Implementation}\label{sec:kernel_impl}
We briefly describe the \SparseP{} implementations for all compressed matrix formats, i.e., the way that threads access data involved in the kernel from/to the local DRAM bank. The \spmv{} kernels include three types of data structures: (i) the arrays that store the non-zero elements, i.e., the values (\texttt{values[]}) and the positions of the non-zero elements (\texttt{rowptr[]}, \texttt{colind[]} for CSR, \texttt{tuples[]} for COO, \texttt{browptr[]}, \texttt{bcolind[]} for BCSR, \texttt{browind[]}, \texttt{bcolind[]} for BCOO), (ii) the array that stores the elements of the input vector, and (iii) the array that stores the partial results created for the elements of the output vector.

First, \spmv{} performs streaming memory accesses to the arrays that store the non-zero elements and their positions. Therefore, to exploit spatial locality and immense bandwidth in data (scratchpad or cache) memory, each thread reads the non-zero elements by fetching large chunks of bytes in a coarse-grained manner from DRAM to data memory (i.e., WRAM in the UPMEM PIM system). Then, it accesses elements through data memory in a fine-grained manner. In the UPMEM PIM system, we fetch chunks of 256-byte data to discover the non-zero elements, as suggested by the UPMEM API~\cite{upmem-guide}, since 256-byte transfer sizes highly exploit the available local bandwidth of DRAM bank~\cite{Gomez2021Benchmarking,Gomez2021Analysis}. For the BCSR and BCOO formats, only for the array that stores the values of the non-zero elements (i.e., \texttt{bvalues[]}), we fetch from DRAM to data memory block size chunks, i.e., chunks of $r\times c \times sizeof(data\_type)$ bytes, assuming that the matrix is stored in blocks of size $r\times c$.

Second, \spmv{} causes irregular memory accesses to the elements of the input vector. Specifically, the accesses to the elements of the input vector are input-driven, i.e.,  they are determined by the column positions (column indexes) of the non-zero elements of each particular matrix. Given that matrices involved in \spmv{} are very sparse~\cite{Kanellopoulos2019SMASH,Elafrou2018SparseX,Elafrou2017PerformanceAA,YouTubeGraph,FacebookGraph,Goumas2008Understanding,White97Improving,Helal2021ALTO,Pelt2014Medium}, i.e., the column indexes of the non-zero elements significantly vary, memory accesses to the input vector incur poor data locality. Thus, in our \spmv{} implementations, threads of a PIM core directly access elements of the input vector through DRAM bank at fine-granularity~\cite{Gomez2021Benchmarking,upmem-guide,Gomez2021Analysis}, i.e., using the smallest possible granularity: for the CSR and COO formats at 64-bit granularity, and for the BCSR and BCOO formats at the granularity of $c\times sizeof(data\_type)$ bytes, where $c$ is the horizontal dimension of the block size.

Third, regarding the output vector, threads temporarily store partial results for the same elements of the output vector in data (scratchpad or cache) memory to exploit data locality, until all the non-zero elements of the \textit{same} row or the \textit{same} block row have been traversed (recall matrices are stored in a row-sorted way). Then, the produced results are written to DRAM bank at fine-granularity~\cite{Gomez2021Benchmarking,upmem-guide,Gomez2021Analysis}: for the CSR and COO formats at 64-bit granularity, and for the BCSR and BCOO formats at the granularity of $r\times sizeof(data\_type)$ bytes, where $r$ is the vertical dimension of the block size.

\section{Evaluation Methodology}\label{methodology}
We conduct our evaluation on an UPMEM PIM system that includes a 2-socket Intel Xeon Silver 4110 CPU~\cite{intel4110} at 2.10 GHz (host CPU), standard main memory (DDR4-2400)~\cite{ddr4jedec} of 128 GB, and 20 UPMEM PIM DIMMs with 160 GB PIM-capable memory and 2560 DPUs.\footnote{There are thirty two faulty DPUs in the system where we run our experiments. They cannot be used and do not affect the correctness of our results, but take away from the system’s full computational power of 2560 DPUs.}

First, we evaluate \spmv{} execution using one single DPU and multiple tasklets (Section~\ref{1DPU}). Table~\ref{tab:small-matrices} shows our evaluated small matrices that fit in the 64 MB DRAM memory of a single DPU. The evaluated matrices vary in sparsity (i.e., NNZ / (rows x columns)), standard deviation of non-zero elements among rows (NNZ-r-std) and columns (NNZ-c-std). The highlighted matrices in Table~\ref{tab:small-matrices} with red color exhibit block pattern~\cite{Kourtis2011CSX,Elafrou2018SparseX}, i.e., they include \emph{a lot} of dense sub-blocks (almost all their non-zero elements fit in dense sub-blocks).

\begin{table}[H]
\vspace{-4pt}
\begin{center}
\centering
%\begin{minipage}{.48\linewidth}
%\resizebox{1.0\linewidth}{0.13\textheight}{
\resizebox{0.9\linewidth}{!}{
\begin{tabular}{|l||r|r|r|}
    \hline
    \cellcolor{gray!15}\raisebox{-0.10\height}{\textbf{Matrix Name}} & \cellcolor{gray!15}\raisebox{-0.10\height}{\textbf{Sparsity}} & \cellcolor{gray!15}\raisebox{-0.10\height}{\textbf{NNZ-r-std}} & \cellcolor{gray!15}\raisebox{-0.10\height}{\textbf{NNZ-c-std}} \\
    \hline \hline
    delaunay\_n13  & 	7.32e-04 &  1.343  & 1.343 \\ \hline 
    wing\_nodal &	1.26e-03 &  2.861 & 2.861 \\ 
    \hline
    
    \textcolor{darkred}{raefsky4} & 3.396e-03  & 15.956 & 15.956 \\ \hline
    \textcolor{darkred}{pkustk08} & 0.006542 &  61.537 &  61.537 \\     
    \hline
\end{tabular}
}
\end{center}
\vspace{-4pt}
\caption{Small Matrix Dataset.}
\label{tab:small-matrices}
\vspace{-20pt}
\end{table}

Second, we evaluate \spmv{} execution using \emph{multiple} DPUs of the UPMEM PIM system (Section~\ref{MultipleDPUs}). We evaluate \spmv{} execution using both 1D (Section~\ref{1D}) and 2D (Section~\ref{2D}) partitioning techniques, and compare them (Section~\ref{1D-2D}) using a wide variety of sparse matrices with diverse sparsity patterns. We select 22 representative sparse matrices from the Sparse Suite Collection~\cite{davis2011florida}, the characteristics of which are shown in Table~\ref{tab:large-matrices}. As the values of the last two metrics increase (i.e., NNZ-r-std and NNZ-c-std), the matrix becomes very irregular~\cite{Namashivayam2021Variable,Tang2015Optimizing}, and is referred to as \textit{scale-free} matrix. In our evaluation, we refer to all matrices between \texttt{hgc} to \texttt{bns} matrices of Table~\ref{tab:large-matrices} as \textit{regular} matrices. The  matrices in which NNZ-r-std is larger than 25, i.e., all matrices between \texttt{wbs} to \texttt{ask} in Table~\ref{tab:large-matrices}, we refer to as \textit{scale-free} matrices. Please see Appendix~\ref{sec:appendix-matrix-dataset} for a complete description of our dataset of large sparse matrices.

% Flops / byte = (2 * nnz) / (8 * nnz + 12 * N), N : matrix dimension
% source: https://github.com/Sable/fait-maison-spmv
% Flops / byte = (2 * nnz) / (sizeof(datatype) * nnz + 12 * N), N : matrix dimension
\begin{table}[t]
\begin{center}
\centering
%\begin{minipage}{.7\linewidth}
%\resizebox{1.0\linewidth}{0.13\textheight}{
\resizebox{.68\linewidth}{!}{
\begin{tabular}{|l||r|r|r|}
    \hline
    \cellcolor{gray!15}\raisebox{-0.10\height}{\textbf{Matrix Name}} & \cellcolor{gray!15}\raisebox{-0.10\height}{\textbf{Sparsity}} & \cellcolor{gray!15}\raisebox{-0.10\height}{\textbf{NNZ-r-std}} & \cellcolor{gray!15}\raisebox{-0.10\height}{\textbf{NNZ-c-std}} \\
    \hline
    \hline
    hugetric-00020 (\textbf{hgc}) & 4.21e-07 & 0.031 & 0.031 \\ \hline
    
    mc2depi (\textbf{mc2}) & 7.59e-06 & 0.076 & 0.076 \\ \hline
    
    parabolic\_fem (\textbf{pfm}) & 	1.33e-05 & 0.153 &  0.153 \\ \hline 
    
    roadNet-TX (\textbf{rtn}) & 	1.98e-06 &  1.037 & 1.037 \\ \hline
    
    rajat31 (\textbf{rjt})  &  9.24e-07  & 1.106  & 1.106 \\ \hline
    
    \textcolor{darkred}{af\_shell1 (\textbf{ash})} & 	6.90e-05 & 1.275 & 1.275 \\ \hline  
    
    delaunay\_n19 (\textbf{del}) & 	1.14e-05 &  1.338 & 1.338 \\ \hline
    
    thermomech\_dK  (\textbf{tdk}) & 6.81e-05 & 1.431 &  1.431 \\ \hline
    
    memchip	(\textbf{mem}) & 	2.02e-06 &  2.062 & 1.173 \\ \hline
        
    amazon0601	(\textbf{amz}) & 	2.08e-05 &  2.79 & 15.29  \\ \hline
    
    FEM\_3D\_thermal2 (\textbf{fth}) & 	1.59e-04 & 4.481  & 4.481 \\ \hline

    web-Google (\textbf{wbg}) & 	6.08e-06 & 6.557 &  38.366 \\ \hline
    
    \textcolor{darkred}{ldoor (\textbf{ldr})} & 	5.13e-05  & 11.951 & 11.951 \\ \hline 
    
    poisson3Db (\textbf{psb}) & 	3.24e-04  & 14.712  & 14.712 \\ \hline
    
    \textcolor{darkred}{boneS10 (\textbf{bns})}  & 	6.63e-05 &  20.374 &  20.374 \\ \hline \hline  
    
    webbase-1M (\textbf{wbs}) & 	3.106e-06 & 25.345  & 36.890 \\ \hline  
    
    in-2004	(\textbf{in}) & 	8.846e-06  & 37.230 & 144.062 \\ \hline 
    
    \textcolor{darkred}{pkustk14 (\textbf{pks})} & 	6.428e-04 &  46.508  & 46.508 \\ \hline
    
    com-Youtube (\textbf{cmb}) & 4.639e-06  & 50.754  & 50.754 \\ \hline

    as-Skitter (\textbf{skt})  & 	7.71e-06  & 136.861 &  136.861 \\ \hline
        
    sx-stackoverflow (\textbf{sxw}) & 	5.352e-06 &  137.849 & 65.367 \\ \hline
    
    ASIC\_680k (\textbf{ask})  & 	8.303e-06  & 659.807 & 659.807 \\ \hline 
    %\hline
\end{tabular}
}
%\end{minipage} \hspace{2pt}% 

\end{center}
\vspace{-4pt}
\caption{Large Matrix Dataset. Matrices are sorted by NNZ-r-std, i.e., based on their irregular pattern. The highlighted matrices with red color exhibit block pattern~\cite{Kourtis2011CSX,Elafrou2018SparseX}.}
\label{tab:large-matrices}
\vspace{-8pt}
\end{table}

Third, we compare the performance and energy consumption
of \spmv{} execution on the UPMEM PIM system to those
on the Intel Xeon Silver 4110 CPU~\cite{intel4110} and the NVIDIA Tesla V100 GPU~\cite{nvidiaTeslaV100} (Section~\ref{cpu-gpu}).

In Section~\ref{recommendations}, we summarize our key takeaways and provide programming recommendations for software designers, and suggestions and hints for hardware and system designers of future PIM systems.

\section{Analysis of \spmv{} Execution on One DPU}\label{1DPU}
This section characterizes \spmv{} performance with various load balancing schemes and compressed matrix formats using multiple tasklets in a single DPU. Section~\ref{1DPU-MulTskl} compares load balancing schemes of each compressed matrix format, and Section~\ref{1DPU-Formats} compares the scalability of various compressed matrix formats.

\subsection{Load Balancing Schemes Across Tasklets of One DPU}\label{1DPU-MulTskl}

\begin{figure}[t]
\centering
\begin{minipage}{\textwidth}
\includegraphics[width=.245\textwidth]{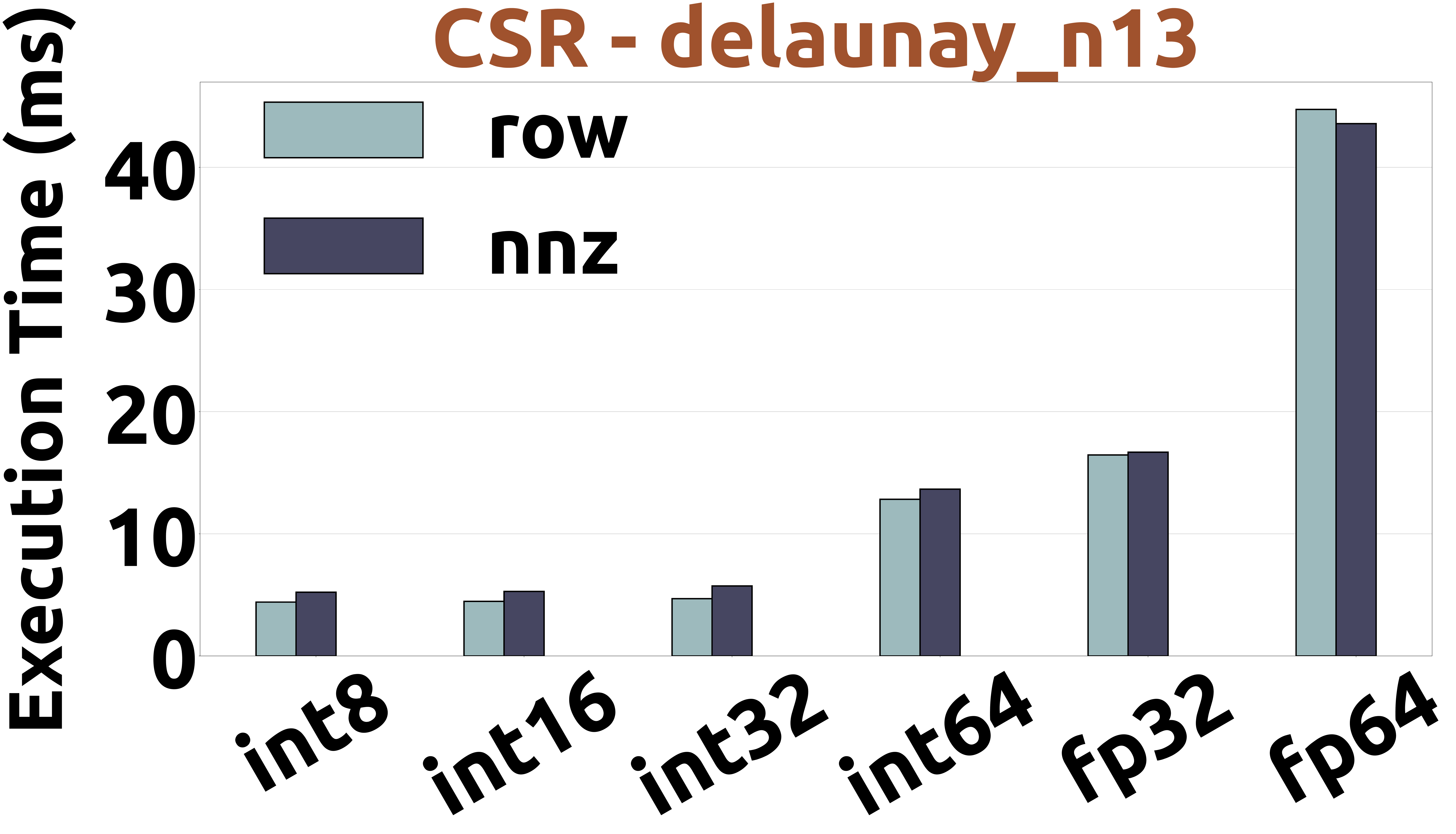}
\includegraphics[width=.245\textwidth]{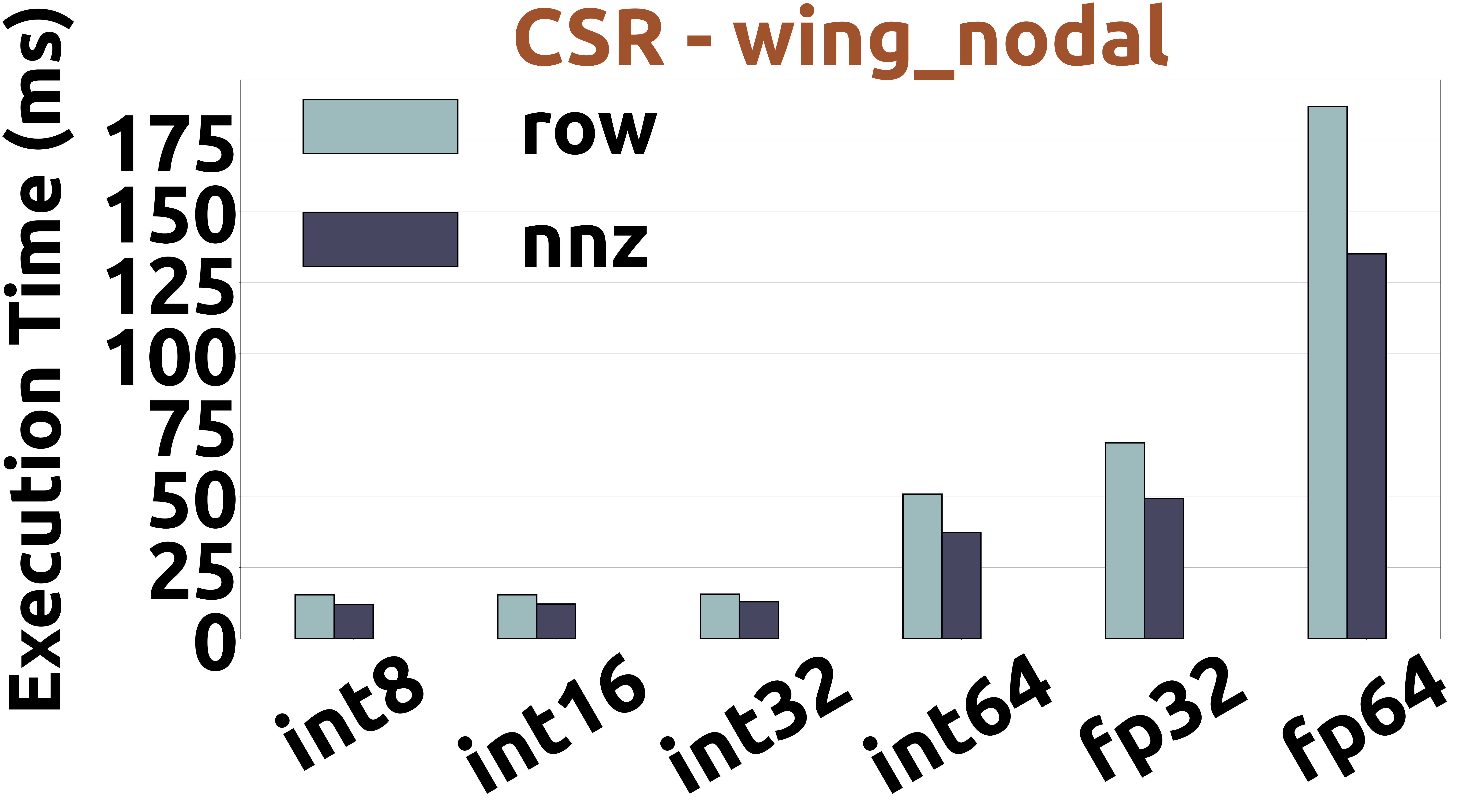}
\includegraphics[width=.245\textwidth]{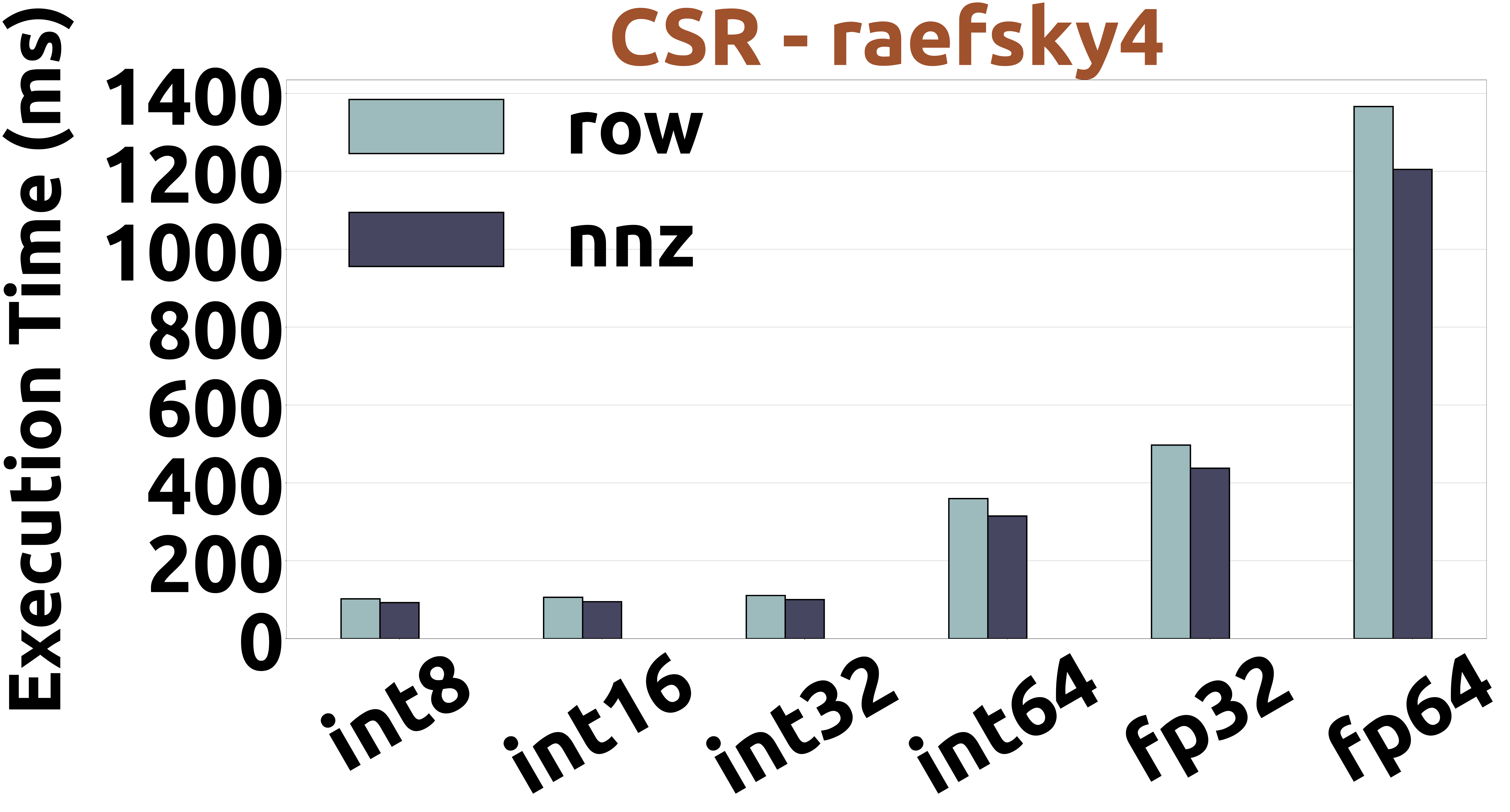}
\includegraphics[width=.245\textwidth]{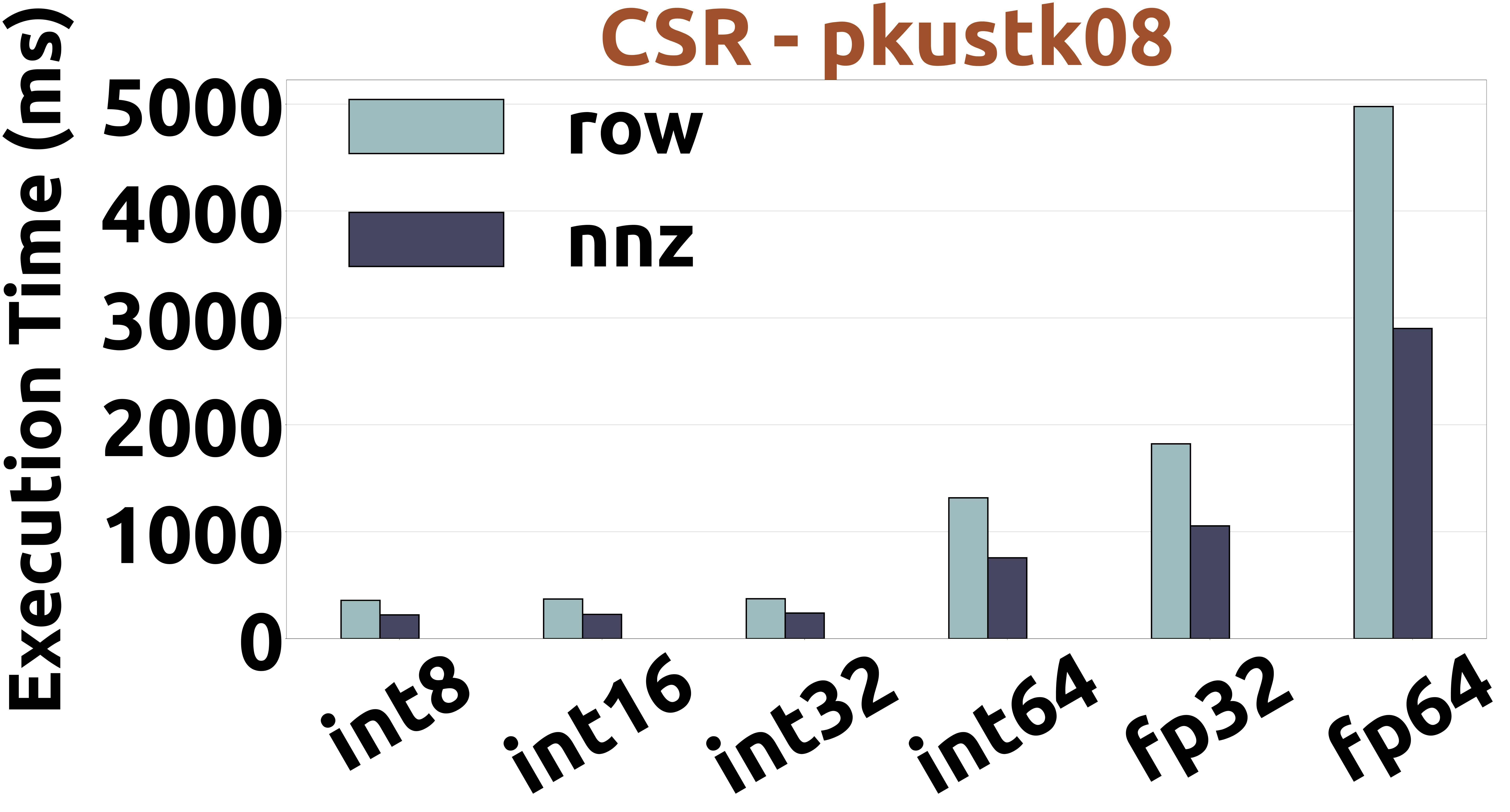}
\end{minipage}\hspace{2pt}%
\begin{minipage}{\textwidth}
\includegraphics[width=.245\textwidth]{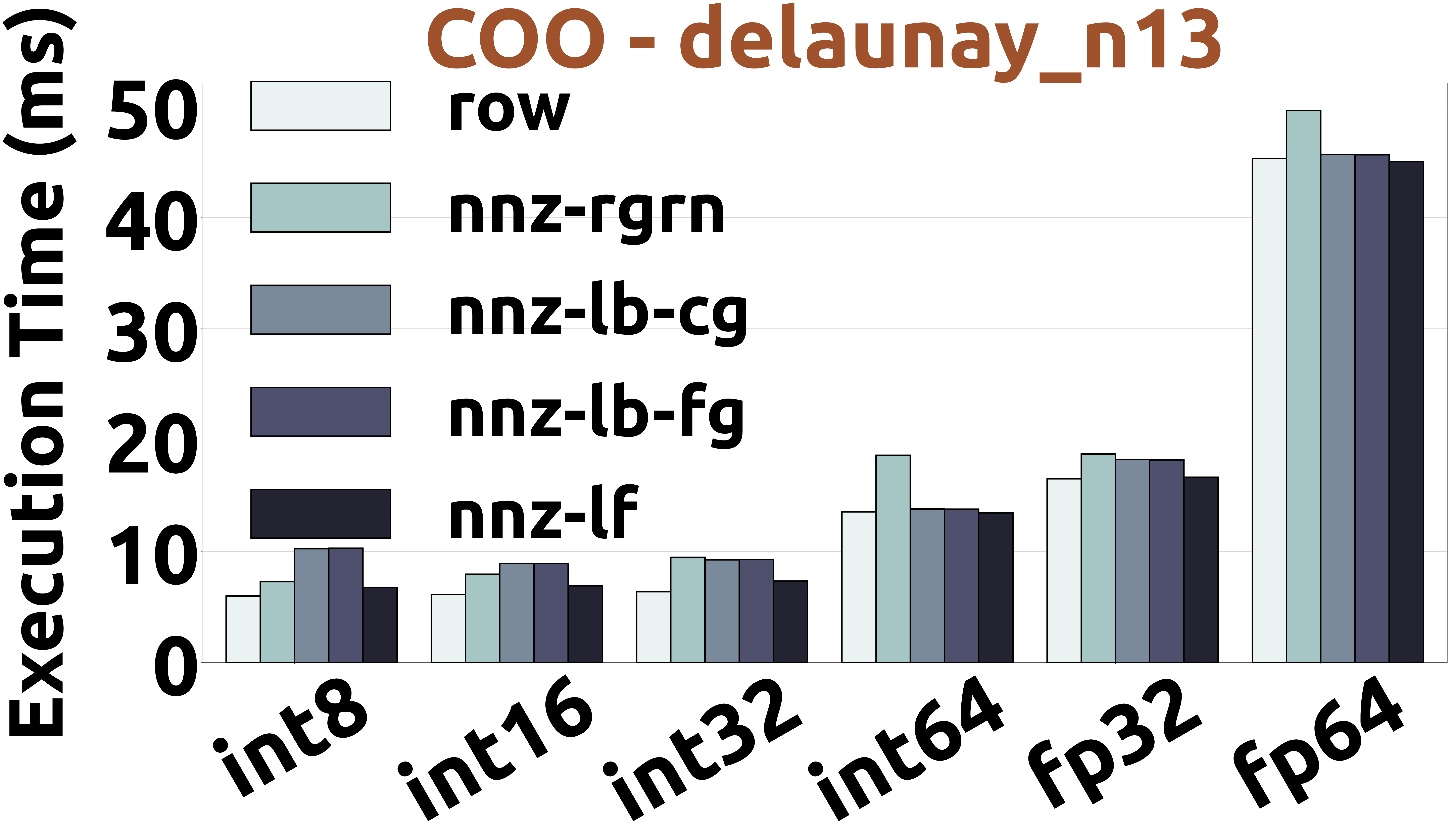}
\includegraphics[width=.245\textwidth]{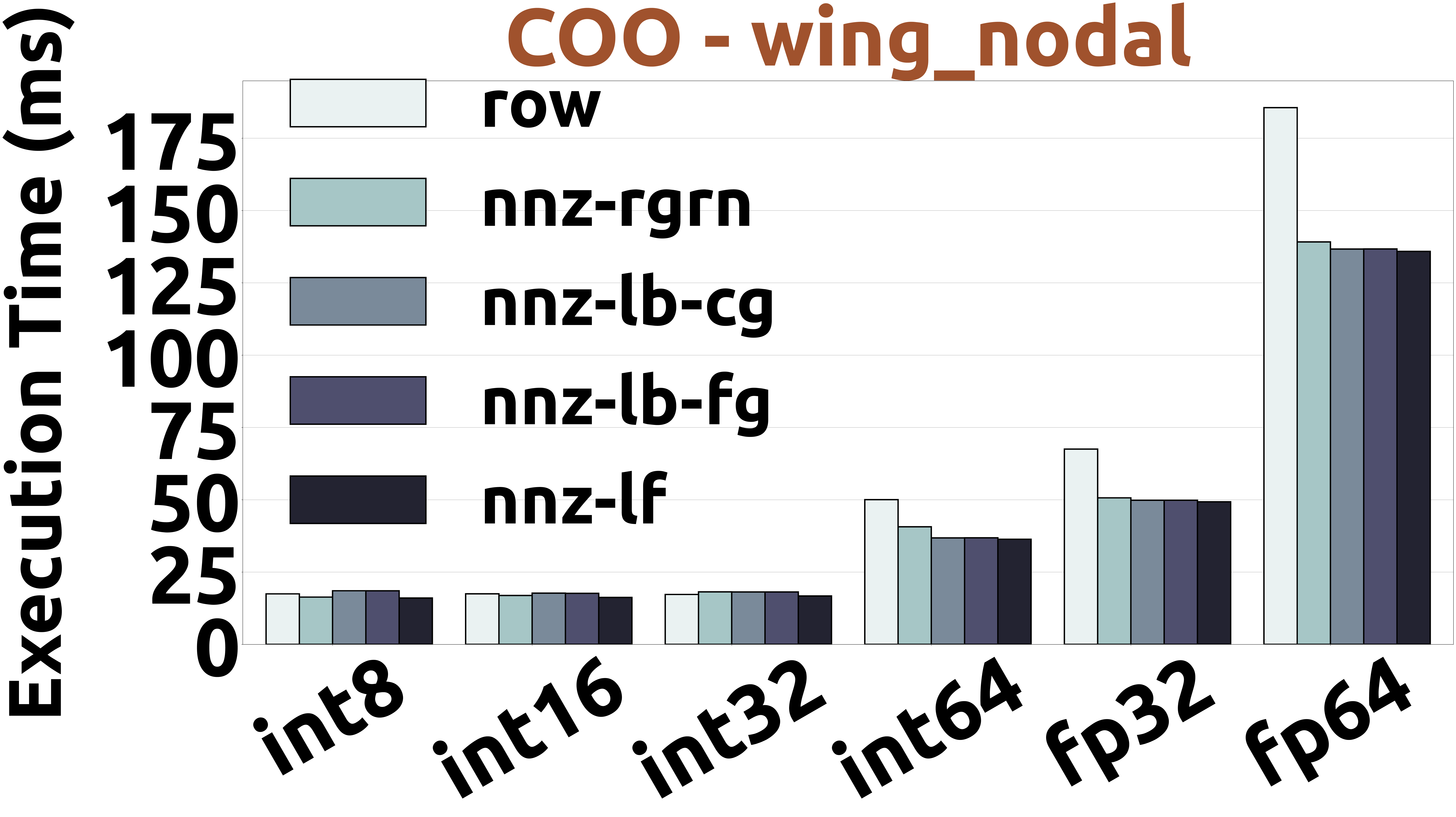}
\includegraphics[width=.245\textwidth]{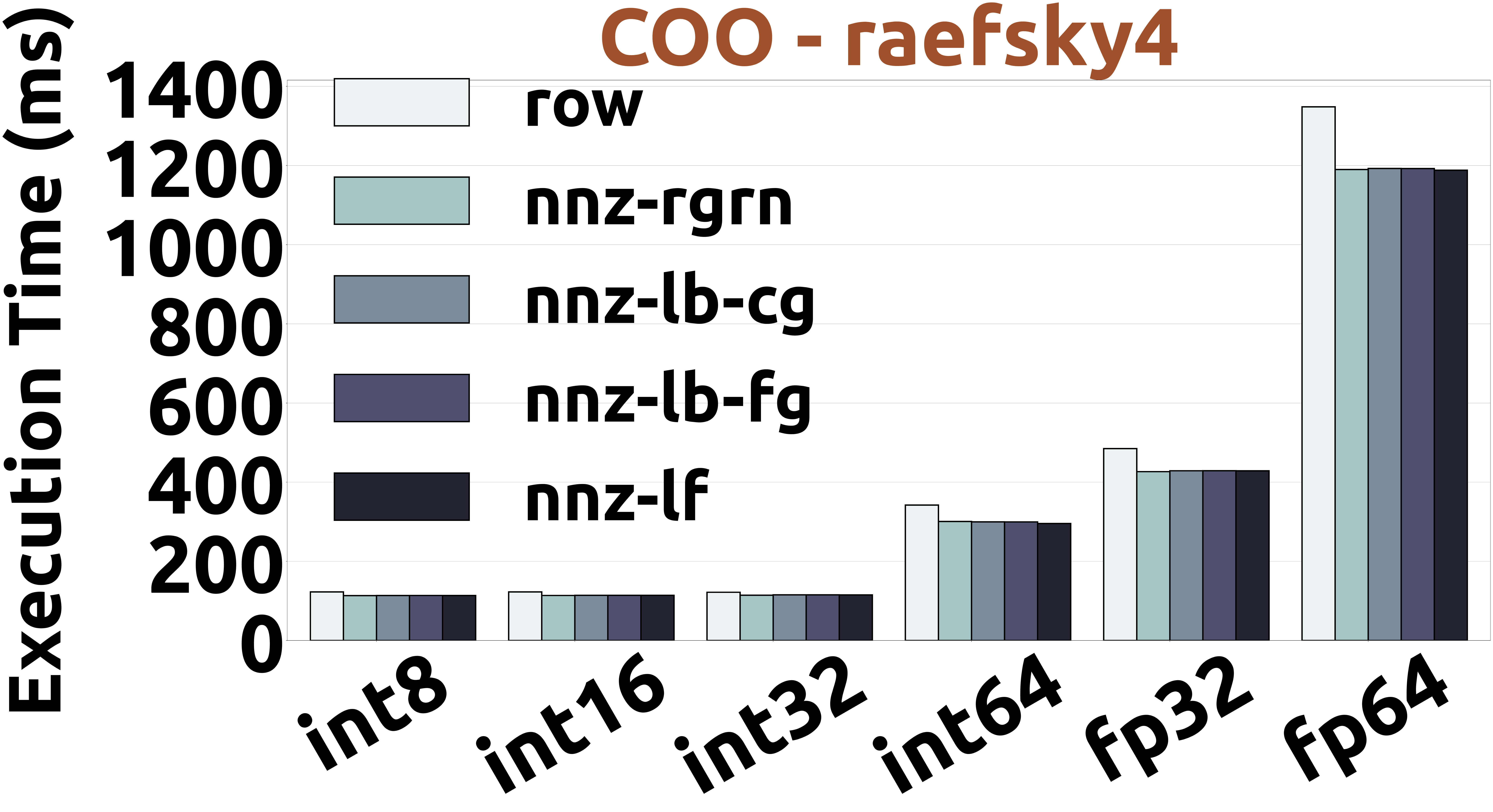}
\includegraphics[width=.245\textwidth]{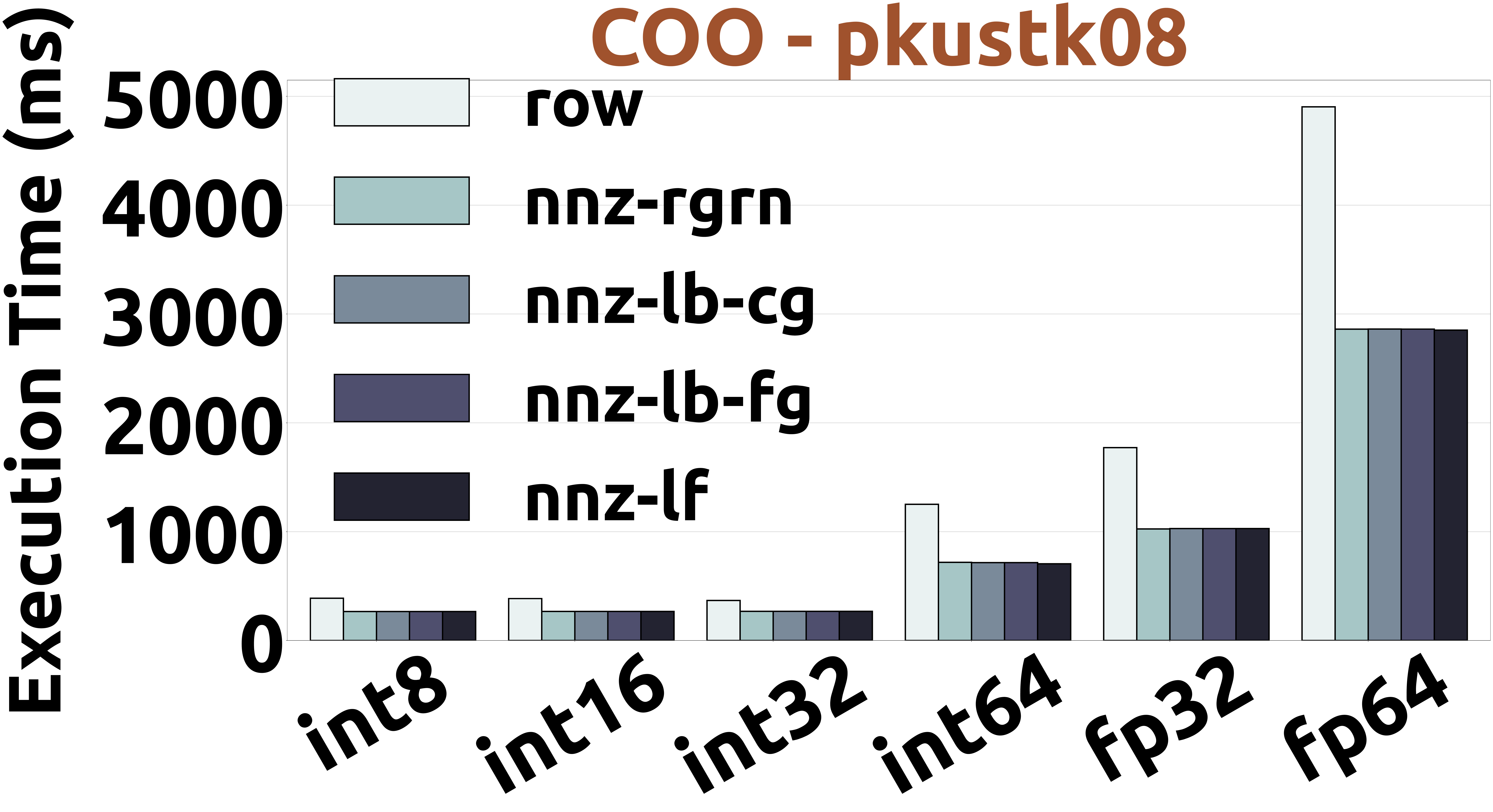}
\end{minipage}\hspace{2pt}%
\begin{minipage}{\textwidth}
\includegraphics[width=.245\textwidth]{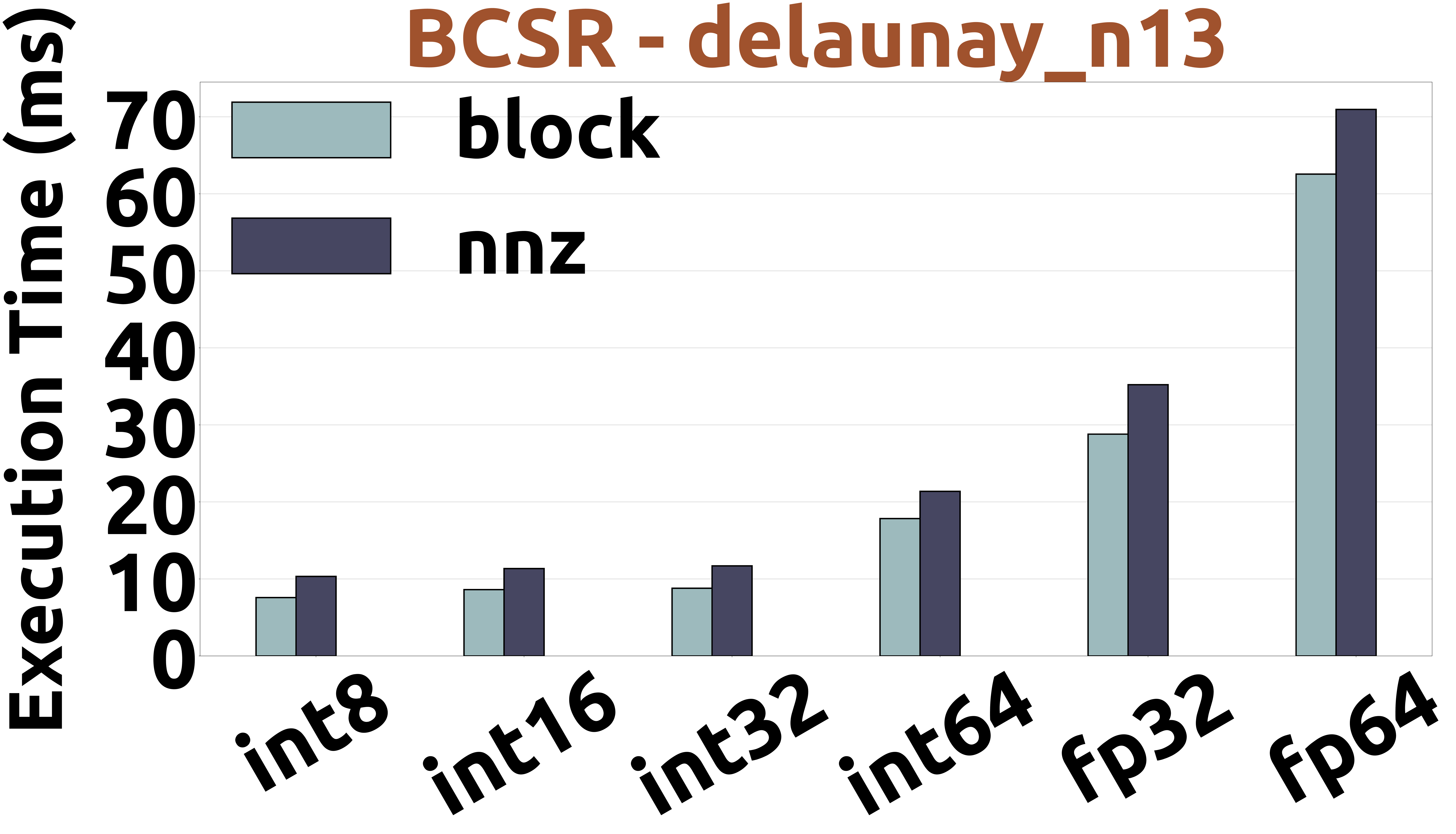}
\includegraphics[width=.245\textwidth]{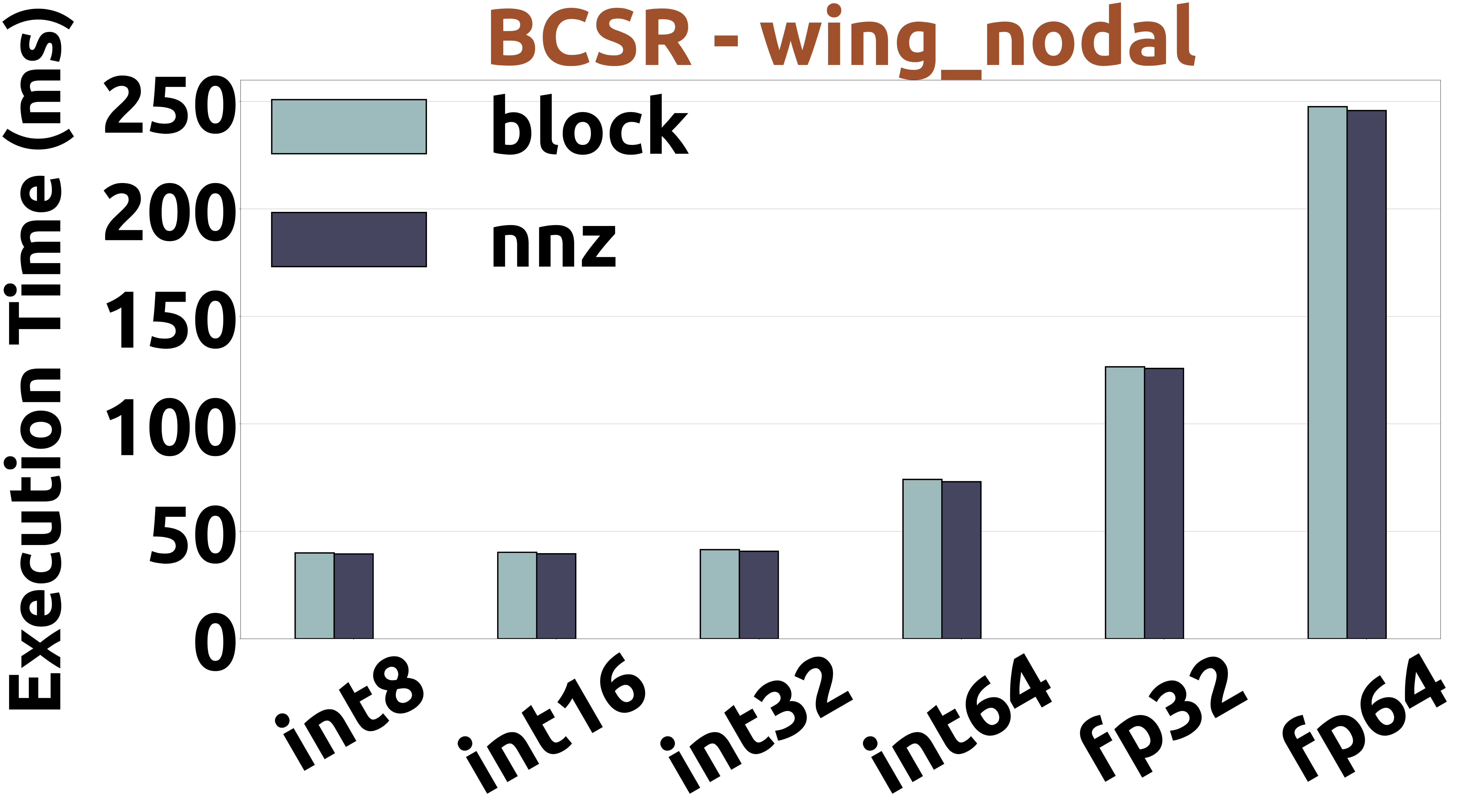}
\includegraphics[width=.245\textwidth]{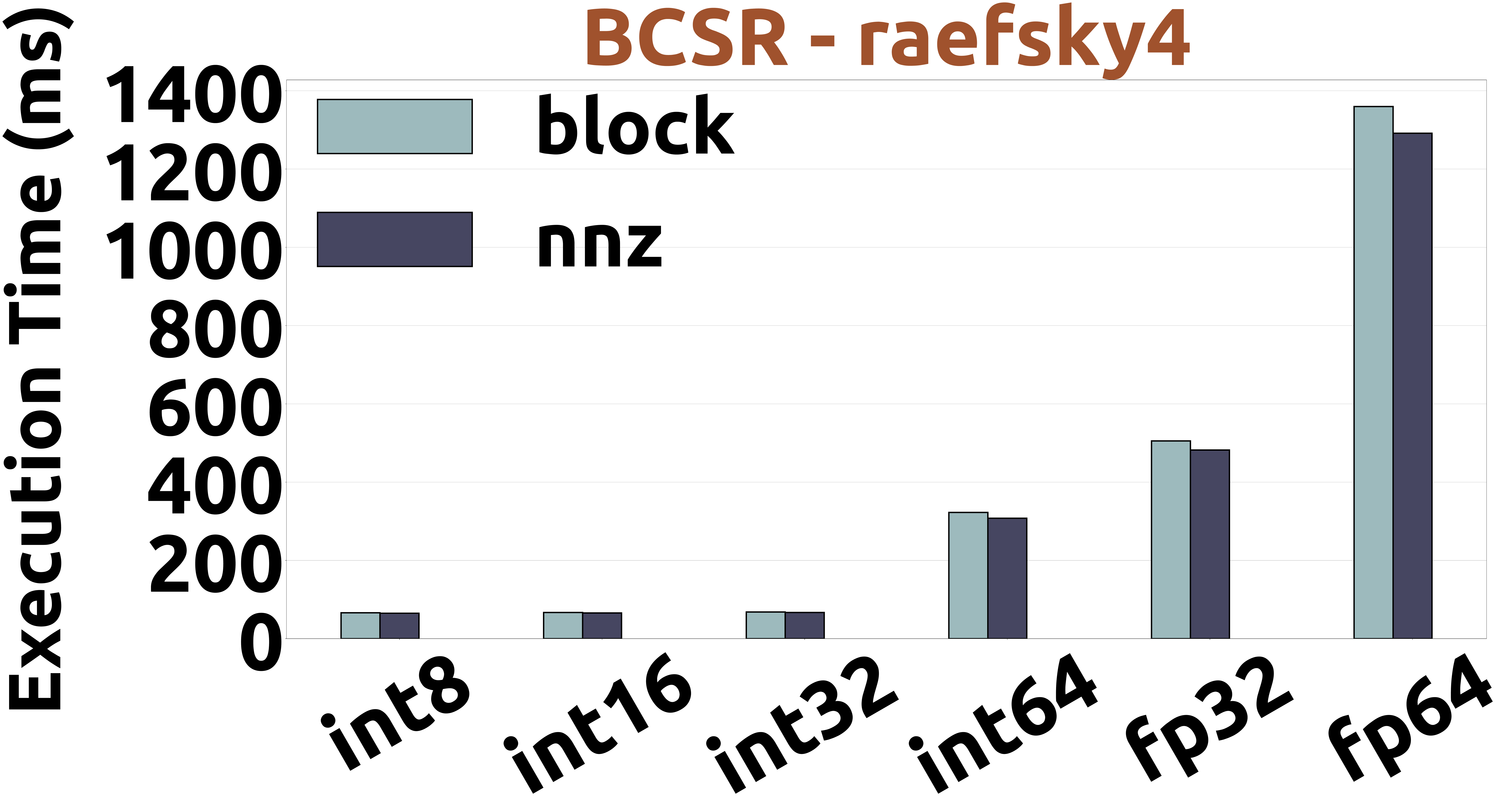}
\includegraphics[width=.245\textwidth]{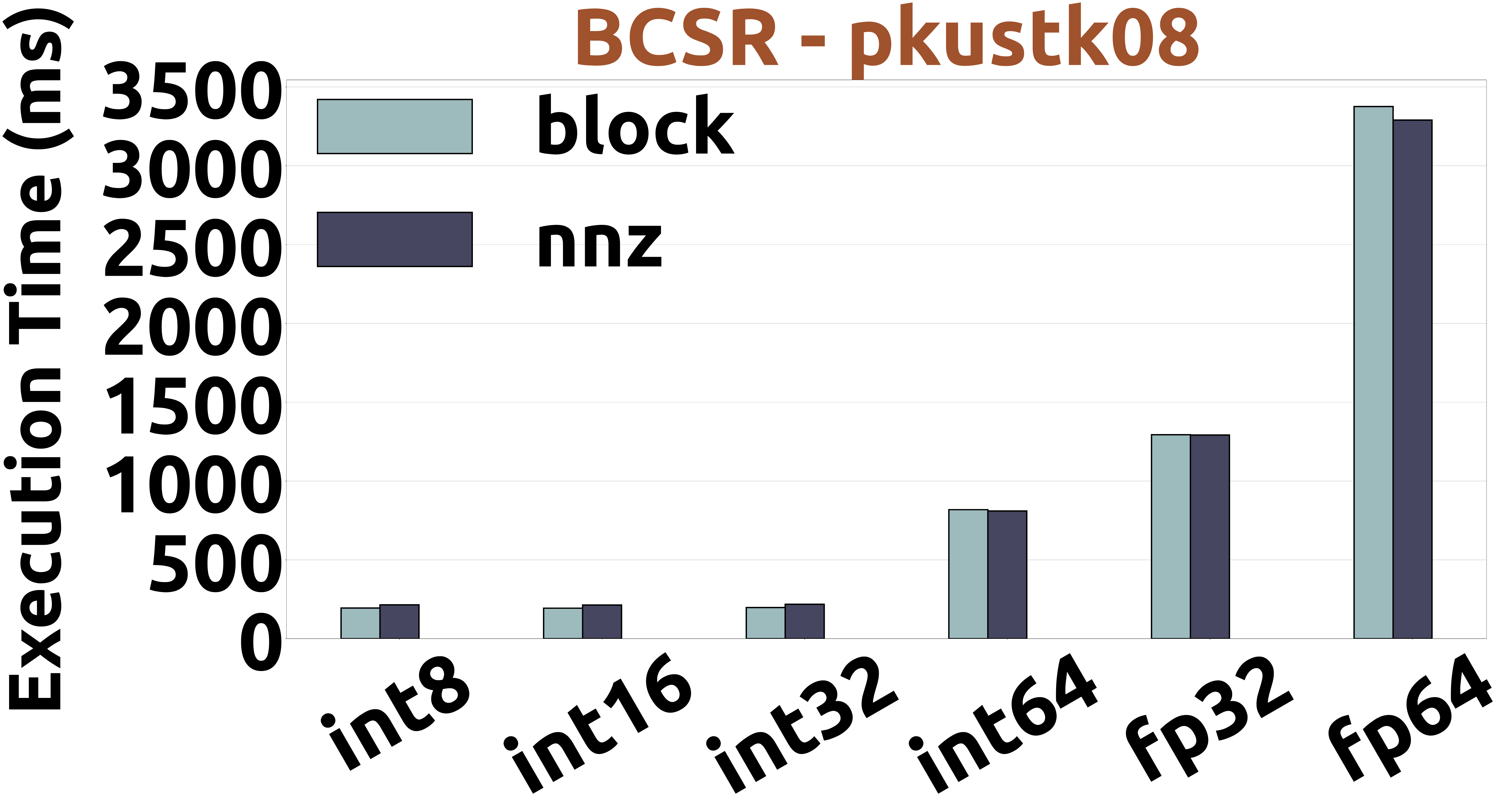}
\end{minipage}\hspace{2pt}%
\begin{minipage}{\textwidth}
\includegraphics[width=.245\textwidth]{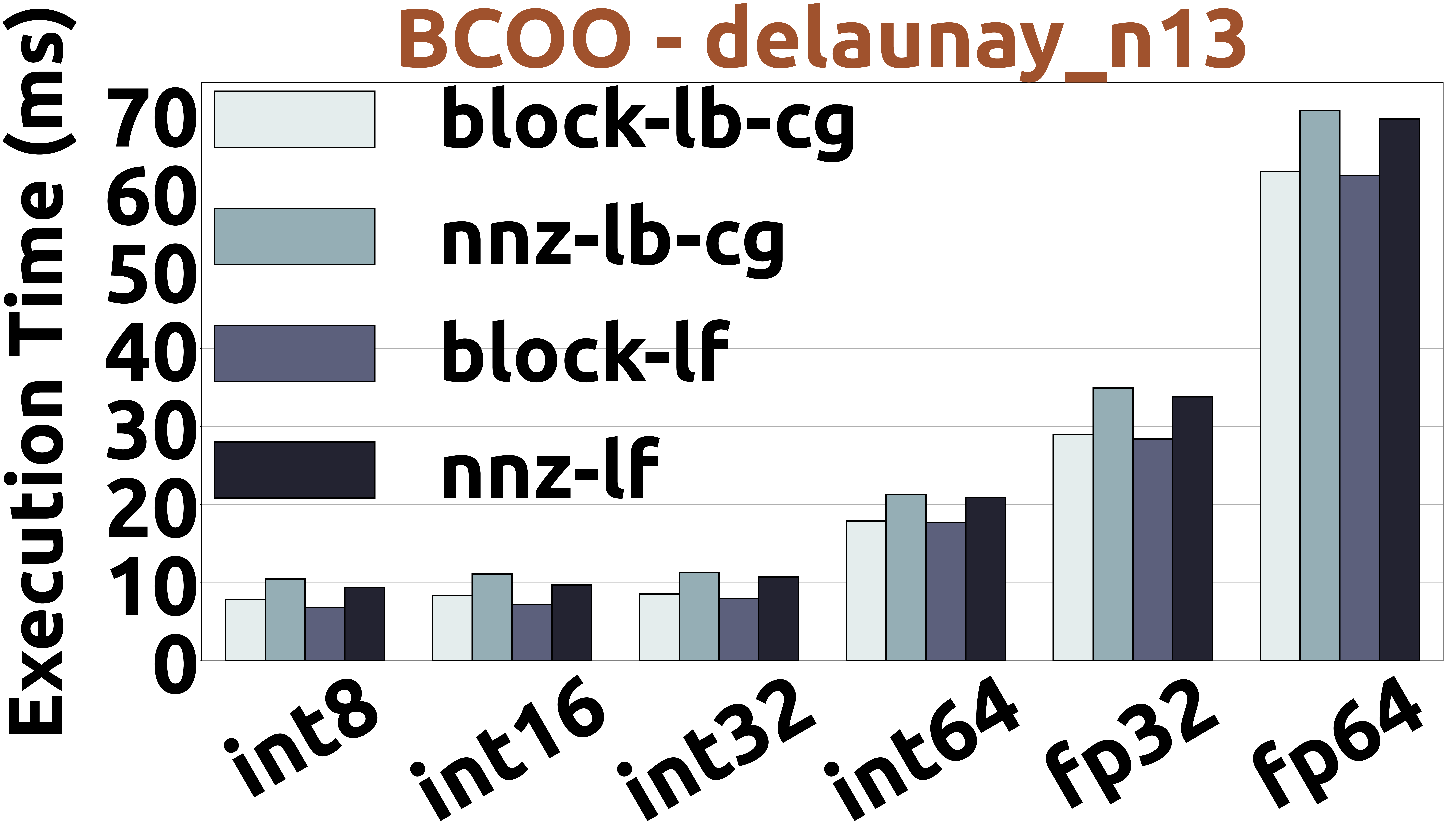}
\includegraphics[width=.245\textwidth]{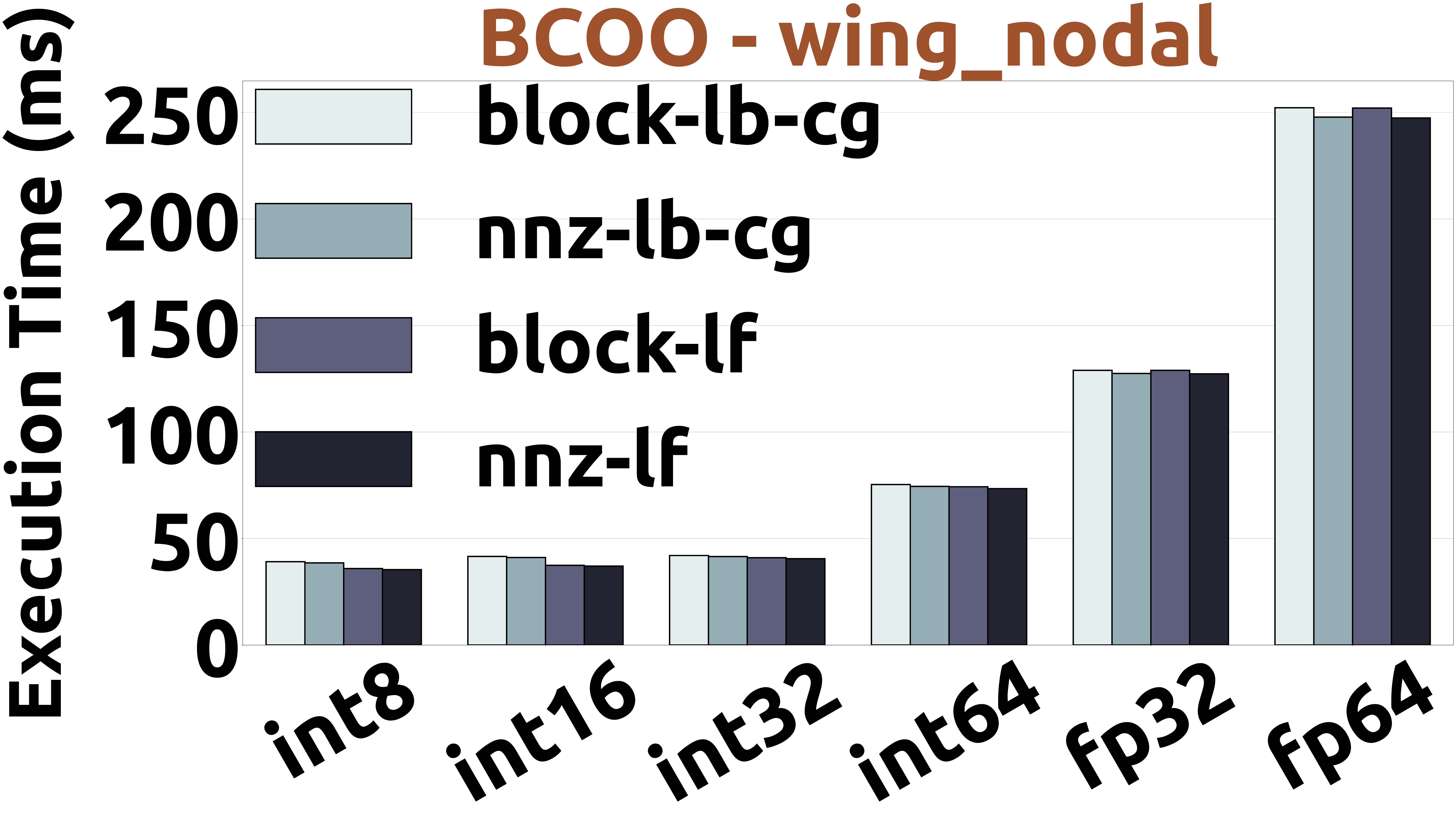}
\includegraphics[width=.245\textwidth]{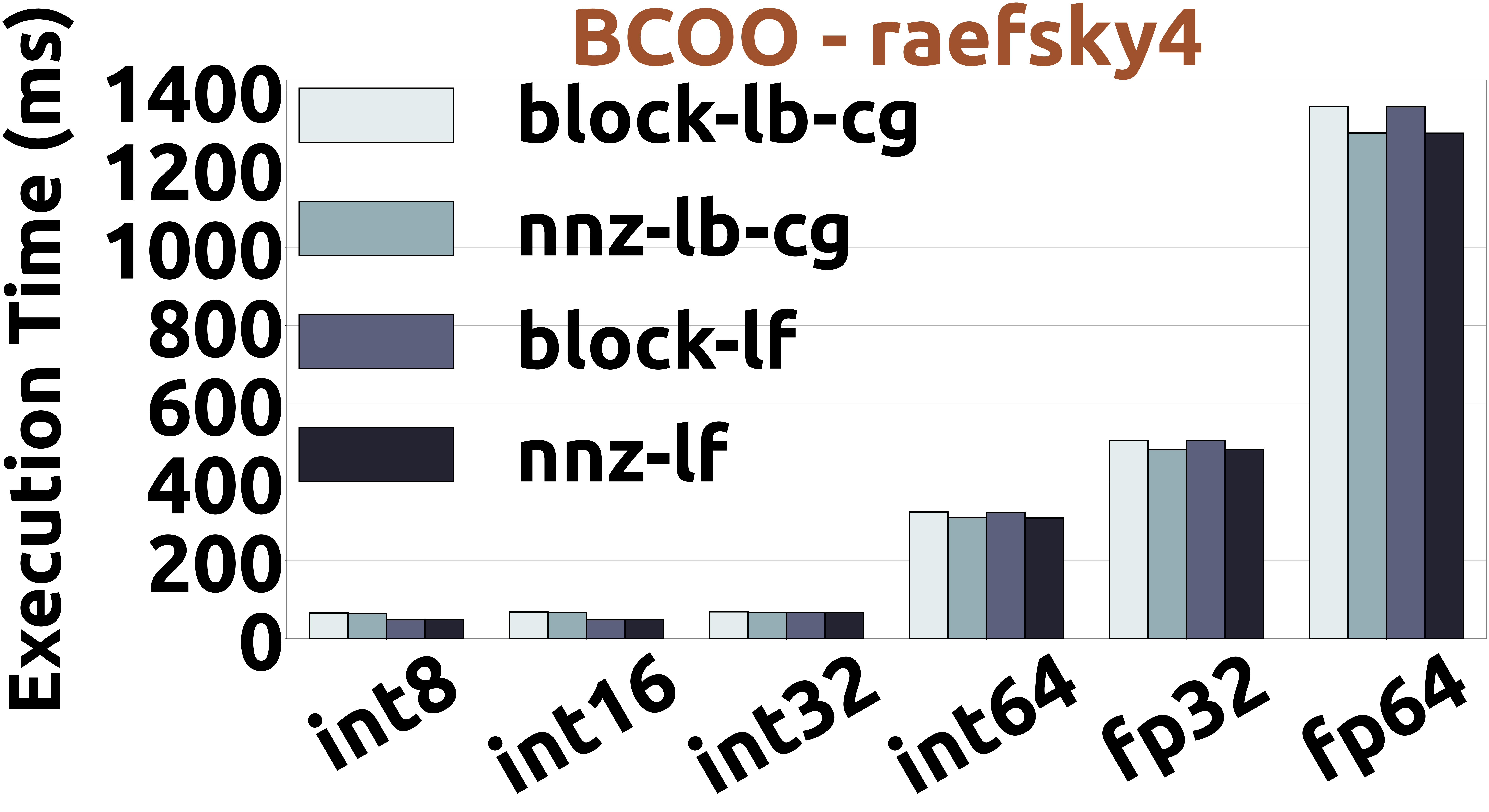}
\includegraphics[width=.245\textwidth]{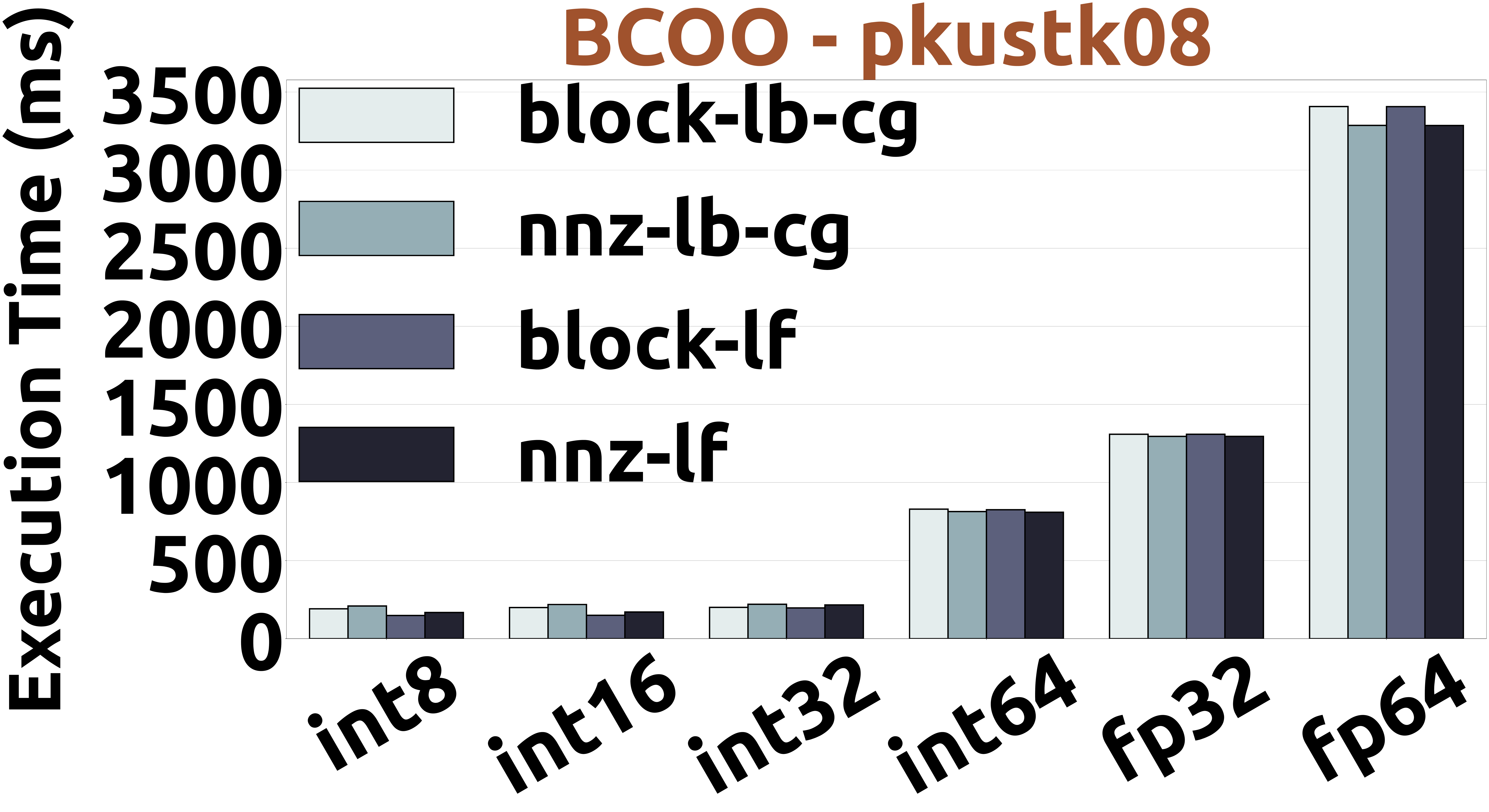}
\end{minipage}
\vspace{-6pt}
\caption{Execution time achieved by various load balancing schemes of each compressed matrix format using 16 tasklets of a single DPU.}
\label{fig:1DPU-datatypes}
\vspace{-14pt}
\end{figure}

We compare the parallelization schemes of each compressed matrix format supported by \SparseP{} library (presented in Table~\ref{table:1DPU-impl}) across multiple threads of a multithreaded PIM core.
Figure~\ref{fig:1DPU-datatypes} compares the load balancing schemes of each compressed matrix format using 16 tasklets in a single DPU. For the BCSR and BCOO formats, we omit results for the fine-grained locking approach, since it performs similarly with the coarse-grained locking approach: as we explain in Appendix ~\ref{sec:appendix-1DPU-BCOO}, fine-grained locking does not increase parallelism over coarse-grained, since in the UPMEM PIM hardware, DRAM memory accesses of the critical section are serialized in the DMA engine of the DPU~\cite{Gomez2021Analysis,upmem-guide,Gomez2021Benchmarking}.

We draw four findings from Figure~\ref{fig:1DPU-datatypes}. First, we find that \spmv{} execution using int8, int16, and int32 data types achieves similar execution times across them. This is because the multiplication operation of these data types is sufficiently supported by hardware~\cite{Gomez2021Benchmarking}. In contrast, execution time sharply increases when using more heavyweight data types, i.e., int64 and floating point data types, in which multiplication is emulated in software using the 8x8-bit multiplier of the DPU~\cite{upmem-guide,Gomez2021Benchmarking,Gomez2021Analysis}.

Second, we observe that balancing the non-zero elements across tasklets typically outperforms balancing the rows for the CSR/COO formats or blocks for the BCSR/BCOO formats, since the non-zero element multiplications are computationally very expensive and can significantly affect load balance across tasklets. However, in \texttt{delaunay\_n13 matrix}, balancing the non-zero elements causes high row/block imbalance across tasklets, since one tasklet processes a significantly higher number of rows/blocks over the rest, thereby causing high operation imbalance across tasklets within the DPU core pipeline. As a result, balancing the rows/blocks outperforms balancing the non-zero elements due to the particular pattern of \texttt{delaunay\_n13 matrix}. In addition, performance benefits of balancing the blocks over balancing the non-zero elements are significant in the BCSR/BCOO formats, because they operate at block granularity and incur high loop control costs.

Third, we observe that the lock-free approach (\texttt{COO.nnz-lf}) outperforms the lock-based approaches (\texttt{COO.nnz-lb-cg} and \texttt{COO.nnz-lb-fg}) in \texttt{delaunay\_n13 matrix}, especially in data types where the multiplication operation is supported directly in hardware. In \texttt{delaunay\_n13 matrix}, one tasklet processes a much larger number of rows than the rest, i.e., it performs a much larger number of critical sections than the rest. In other words, one tasklet performs a much larger number of lock acquisitions/releases and memory instructions than the rest. Thus, lock-based approaches cause high operation imbalance in the DPU core pipeline with significant performance costs. Instead, lock-free and lock-based approaches in the BCOO format perform similarly, since lock acquisition/release costs can be hidden due to BCOO's higher loop control costs and larger critical sections. Overall, based on the second and the third findings, we conclude that in matrices and formats, where the load balancing and/or the synchronization scheme used cause \emph{high} disparity in the number of non-zero elements/blocks/rows processed across tasklets or the number of
lock acquisitions/lock releases/memory accesses performed across tasklets, the DPU core pipeline can incur significant performance overheads.

\begin{tcolorbox}
\noindent\textbf{OBSERVATION 1:} \\
\textit{High operation imbalance} in computation, control, synchronization, or memory instructions executed by multiple threads of a PIM core can cause \textit{high performance overheads} in the compute-bound and area-limited PIM cores.
\end{tcolorbox}

Fourth, we find that the fine-grained locking approach (\texttt{COO.nnz-lb-fg}) performs similarly with the coarse-grained locking approach (\texttt{COO.nnz-lb-cg}). This is because the critical section includes memory accesses to the local DRAM bank, which, in the UPMEM PIM hardware, are serialized in the DMA engine of the DPU. Therefore, fine-grained locking does not increase execution parallelism over coarse-grained locking, since concurrent accesses to MRAM bank are not supported in the UPMEM PIM hardware. Fine-grained locking does not improve performance over coarse-grained locking, also when using block-based formats (e.g., BCSR/BCOO formats), as we demonstrate in Appendix~\ref{sec:appendix-1DPU-BCOO}. Therefore, we recommend PIM hardware designers to provide lightweight synchronization mechanisms~\cite{Giannoula2021SynCron} for PIM cores, and/or enable concurrent accesses to local DRAM memory, e.g., supporting sub-array level parallelism~\cite{Kim2012Case,Seshadri2013RowClone,Seshadri2017Ambit,Hajinazar2021SIMDRAM,Seshadri2017Simple,Chang2016LISA,seshadri2020indram,Chang2014Improving} or multiple DRAM banks per PIM core.

\begin{tcolorbox}
\noindent\textbf{OBSERVATION 2:} \\
\textit{Fine-grained} locking approaches to parallelizing critical sections that perform memory accesses to different DRAM memory locations cannot improve performance over \textit{coarse-grained} locking, when the PIM hardware does not support \textit{concurrent accesses to a DRAM bank}. 
\end{tcolorbox}

\subsection{Analysis of Compressed Matrix Formats on One DPU}\label{1DPU-Formats}
 
\begin{figure}[t]
\centering
\includegraphics[width=.92\textwidth]{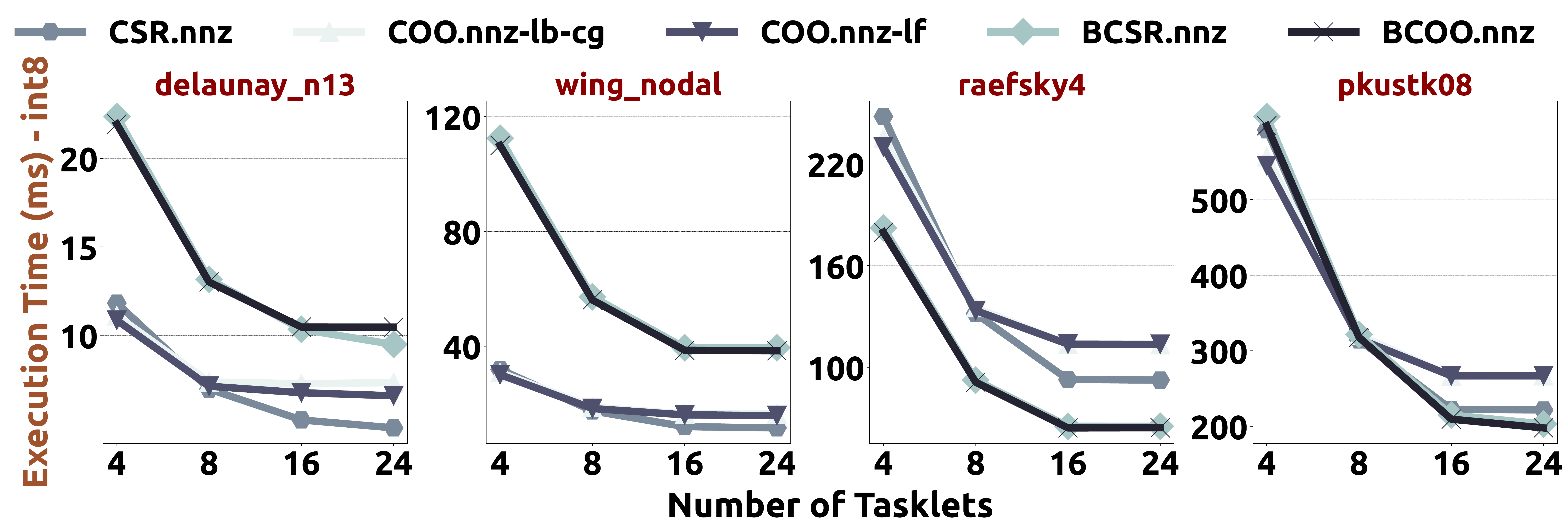}
\includegraphics[width=.9\textwidth]{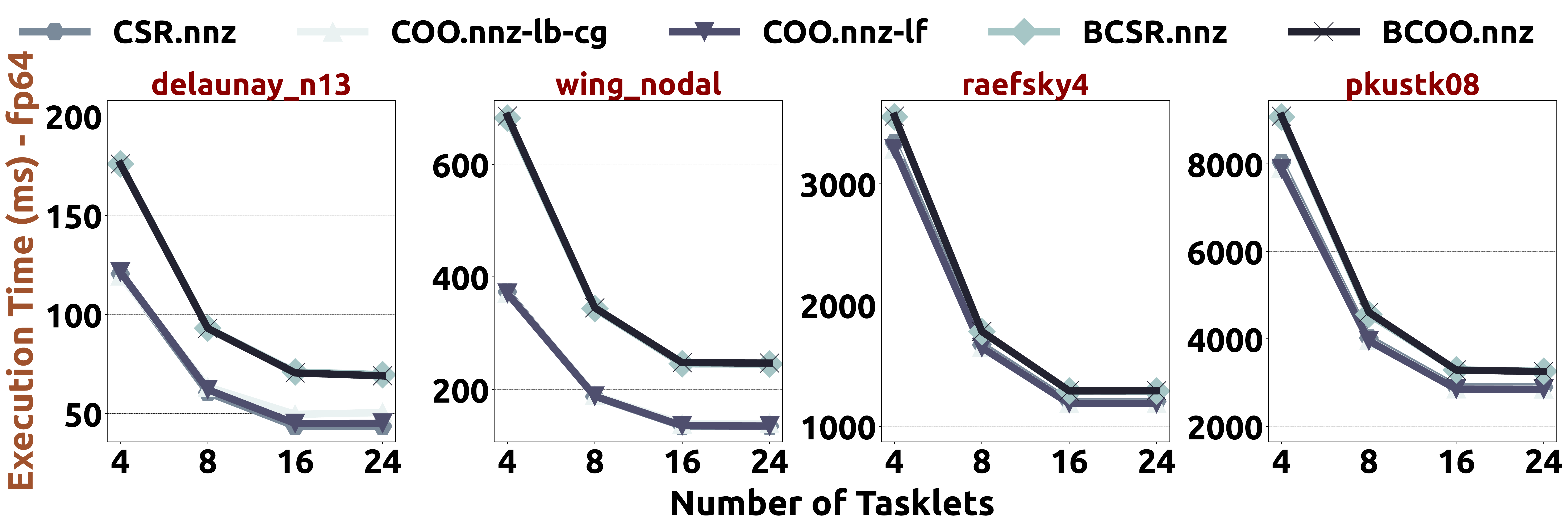}
\vspace{-5pt}
\caption{Scalability of all compressed formats for the int8 (top graphs) and fp64 (bottom graphs) data types as the number of tasklets of a single DPU increases.}
\label{fig:1DPU-scalability}
\vspace{-8pt}
\end{figure}

We compare the scalability and the performance achieved by various compressed matrix formats. Figure~\ref{fig:1DPU-scalability} compares the supported compressed formats for the int8 (top graphs) and fp64 (bottom graphs) data types when balancing the non-zero elementsacross tasklets of a DPU.

We draw three findings. First, we find that even though a DPU supports 24 tasklets, \spmv{} execution typically scales up to 16 tasklets, since the DPU pipeline is fully utilized. In \texttt{delaunay\_n13} matrix, \texttt{CSR.nnz} scales up to 24 tasklets. In this matrix, when using 16 tasklets, performance of the \texttt{CSR.nnz} scheme is limited by memory accesses:  \textit{only} one tasklet processes 6 $\times$ more rows than the rest, i.e., it performs 6 $\times$ more memory accesses to fetch elements from the \texttt{rowptr[]} array. Thus, as we increase the number of tasklets from 16 to 24, the disparity in the number of rows across tasklets decreases, and the performance of the \texttt{CSR.nnz} scheme improves.Second, we observe that for the data types with hardware-supported multiplication operation (e.g., int8 data type), CSR achieves the highest scalability, since it provides a better balance between memory access and computation. In contrast, in the floating point data types (e.g., fp64 data type), the DPU is significantly bottlenecked by the expensive software-emulated multiplication operations, and thus all formats scale similarly. Third, we observe that the BCSR and BCOO formats outperform the CSR and COO formats in matrices that exhibit block pattern (i.e., \texttt{raefsky4} and \texttt{pkustk08} matrices), only when multiplication is supported by hardware (e.g., int8 data type). This is  because they exploit spatial and temporal locality in data memory (i.e., WRAM) in the accesses of the elements of the input vector. Instead, in the fp64 data type, performance is severely bottlenecked by computation, thus the BCSR/BCOO formats perform worse than the CSR/COO formats, since they incur higher indexing costs to discover the positions of the non-zero elements~\cite{asgari2020copernicus,Kanellopoulos2019SMASH}.

\begin{tcolorbox}
\noindent\textbf{OBSERVATION 3:} \\
Block-based formats (e.g., BCSR/BCOO) and can provide high performance gains over non-block-based formats (e.g., CSR/COO) in matrices that exhibit block pattern, if the multiplication operation is supported by hardware. Otherwise, the state-of-the-art CSR and COO formats can provide high performance and scalability.
\end{tcolorbox}

\section{Analysis of \spmv{} Execution on Multiple DPUs}\label{MultipleDPUs}

This section analyzes \spmv{} execution using multiple DPUs in the UPMEM PIM system using the large matrix data set of Table~\ref{tab:large-matrices}. 

Section~\ref{1D} evaluates the 1D partitioning schemes. Section~\ref{1D-Kernel} evaluates the actual kernel time of \spmv{} by comparing (a) all load balancing schemes of each compressed matrix format, and (b) the performance of all compressed matrix formats. Section~\ref{1D-EndToEnd} characterizes end-to-end \spmv{} execution time of the 1D partitioning technique including the data transfer costs for the input and output vectors. 

Section~\ref{2D} evaluates the 2D partitioning techniques. Section~\ref{2D-Studies} presents three characterization studies on (a) performing fine-grained data transfers to transfer the elements of the input and output vectors to/from PIM-enabled memory, (b) the scalability of 2D partitioning techniques to thousands of DPUs, and (c) the number of vertical partitions to perform on the matrix. Section~\ref{2D-Formats} compares the end-to-end performance of all compressed matrix formats for each of the three types of 2D partitioning techniques. Section~\ref{2D-Comparison} compares the best-performing \spmv{} implementations of all three types of 2D partitioning techniques. 

Section~\ref{1D-2D} compares the best-performing (on average across all matrices and data types) \spmv{} implementations of the 1D and 2D partitioning techniques.

\subsection{Analysis of \spmv{} Execution Using 1D Partitioning Techniques}\label{1D}
We evaluate the 1D partitioning schemes highlighted in bold in Table~\ref{table:library}. Specifically, for \texttt{COO.nnz}, we present the coarse-grained locking (\texttt{COO.nnz-lb}) and lock-free (\texttt{COO.nnz-lf}) approaches, since the fine-grained locking approach performs similarly with the coarse-grained locking approach, as shown in the previous section (Section~\ref{1DPU-MulTskl}). Similarly, for the BCSR (int8 data type) and BCOO formats, we present only the coarse-grained locking approach, since all synchronization approaches perform similarly (Section~\ref{1DPU-MulTskl}). Finally, in all experiments presented henceforth, we use 16 tasklets and load-balance the non-zero elements across tasklets within the DPU, since this load balancing scheme provides the highest performance benefits on average across all matrices and data types, according to our evaluation shown in Section~\ref{1DPU}.

\subsubsection{Analysis of \texttt{Kernel} Time}\label{1D-Kernel}  

We compare the \texttt{kernel} time of \spmv{} achieved by various load balancing schemes for each particular compressed matrix format, and then we compare the \texttt{kernel} time of the compressed matrix formats.

\begin{figure}[t]
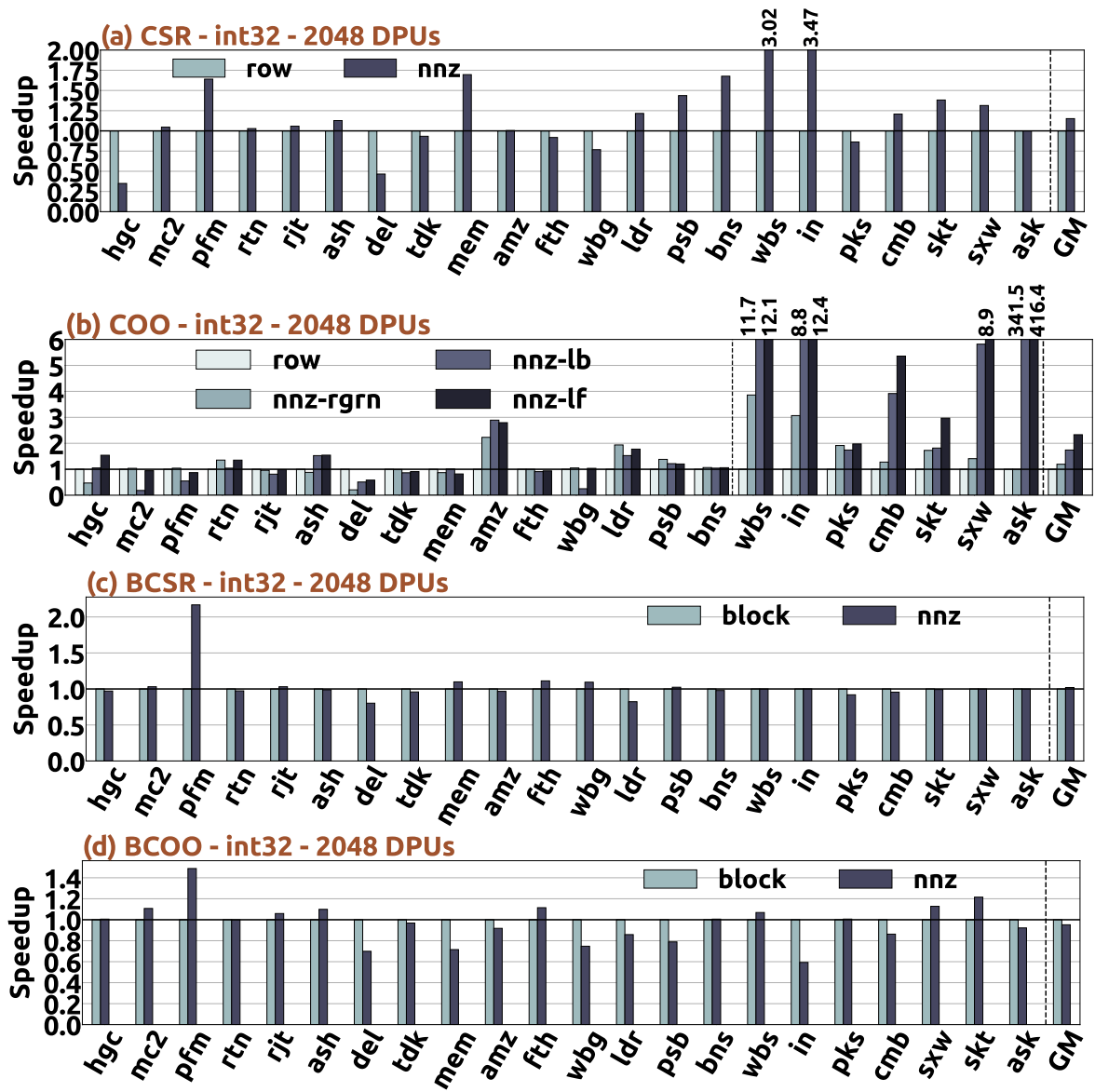

\centering
\includegraphics[width=.89\textwidth]{sections/SparseP/1D-partitioning/csr_balance_dpu_2048_int32.pdf}
\includegraphics[width=.89\textwidth]{sections/SparseP/1D-partitioning/coo_balance_dpu_2048_int32.pdf}
\includegraphics[width=.89\textwidth]{sections/SparseP/1D-partitioning/bcsr_balance_dpu_2048_int32.pdf}
\includegraphics[width=.89\textwidth]{sections/SparseP/1D-partitioning/bcoo_balance_dpu_2048_int32.pdf}
\vspace{-7pt}
\caption{Performance comparison of load balancing techniques for each particular compressed format using 2048 DPUs and the int32 data type.}
\label{fig:1D_balance}
\vspace{-12pt}
\end{figure}

\noindent\textbf{Analysis of Load Balancing Schemes Across DPUs.}
Figure~\ref{fig:1D_balance} compares load balancing techniques for each compressed matrix format using 2048 DPUs and the int32 data type. 

We draw four findings. First, we observe that \texttt{CSR.nnz} and \texttt{COO.nnz-rgrn}, i.e., balancing the non-zero elements across DPUs (at row granularity), either outperform or perform similarly to \texttt{CSR.row} and \texttt{COO.row}, respectively, i.e., balancing the rows across DPUs, except for \texttt{hgc} and \texttt{del} matrices. In these two matrices, \texttt{CSR.nnz} and \texttt{COO.nnz-rowgrn} incur a high disparity in rows assigned to DPUs, i.e., only one DPU processes $4\times$ and $11\times$ more rows than the rest, for \texttt{hgc} and \texttt{del} matrices, respectively. This in turn creates a high disparity in the elements of the output vector processed across DPUs, causing performance to be limited by the DPU that processes the largest number of rows. Thus, we find that adaptive load balancing approaches and selection methods based on the characteristics of each input matrix need to be developed to achieve high performance across all matrices.

\begin{tcolorbox}
\noindent\textbf{OBSERVATION 4:} \\
\textit{Adaptive} load balancing schemes and selection methods for the balancing scheme on rows/blocks/non-zero elements based on the characteristics of each input matrix need to be developed to provide best performance across all matrices.
\end{tcolorbox}

Second, we find that \texttt{COO.nnz-lb} and \texttt{COO.nnz-lf}, which provide an almost perfect non-zero element balance across DPUs, significantly outperform \texttt{COO.row} and \texttt{COO.nnz-rgrn} in \textit{scale-free} matrices (i.e., from \texttt{wbs} to \texttt{ask} matrices) by on average 6.73$\times$. Scale-free matrices have only a few rows, that include a much larger number of non-zero elements compared to the remaining rows of the matrix. Therefore, perfectly balancing the non-zero elements across DPUs provides high performance gains. 

\begin{tcolorbox}
\noindent\textbf{OBSERVATION 5:} \\
\textit{Perfectly balancing the non-zero elements} across PIM cores can provide significant performance benefits in \textit{highly irregular, scale-free matrices.}
\end{tcolorbox}

Third, we find that the lock-free \texttt{COO.nnz-lf} scheme outperforms the lock-based \texttt{COO.nnz-lb} scheme by 1.34$\times$ on average, and provides high performance benefits when there is a high row imbalance across tasklets within the DPU. When one tasklet processes a much larger number of rows versus the rest, it executes a much larger number of critical sections. As a result, the core pipeline incurs high imbalance in lock acquisitions/releases, causing the lock-based approach to incur high performance overheads in relatively compute-bound DPUs~\cite{Gomez2021Benchmarking,Gomez2021Analysis}.

\begin{tcolorbox}
\noindent\textbf{OBSERVATION 6:} \\
\textit{Lock-free} approaches can provide high performance benefits over \textit{lock-based} approaches in PIM architectures, because they minimize synchronization overheads in PIM cores.
\end{tcolorbox}

Finally, in the BCSR and BCOO formats, balancing the blocks across DPUs performs similarly (on average across all matrices) to balancing the non-zero elements across DPUs.

\begin{figure}[t]
\begin{minipage}{\textwidth}
\centering
\includegraphics[width=.48\textwidth]{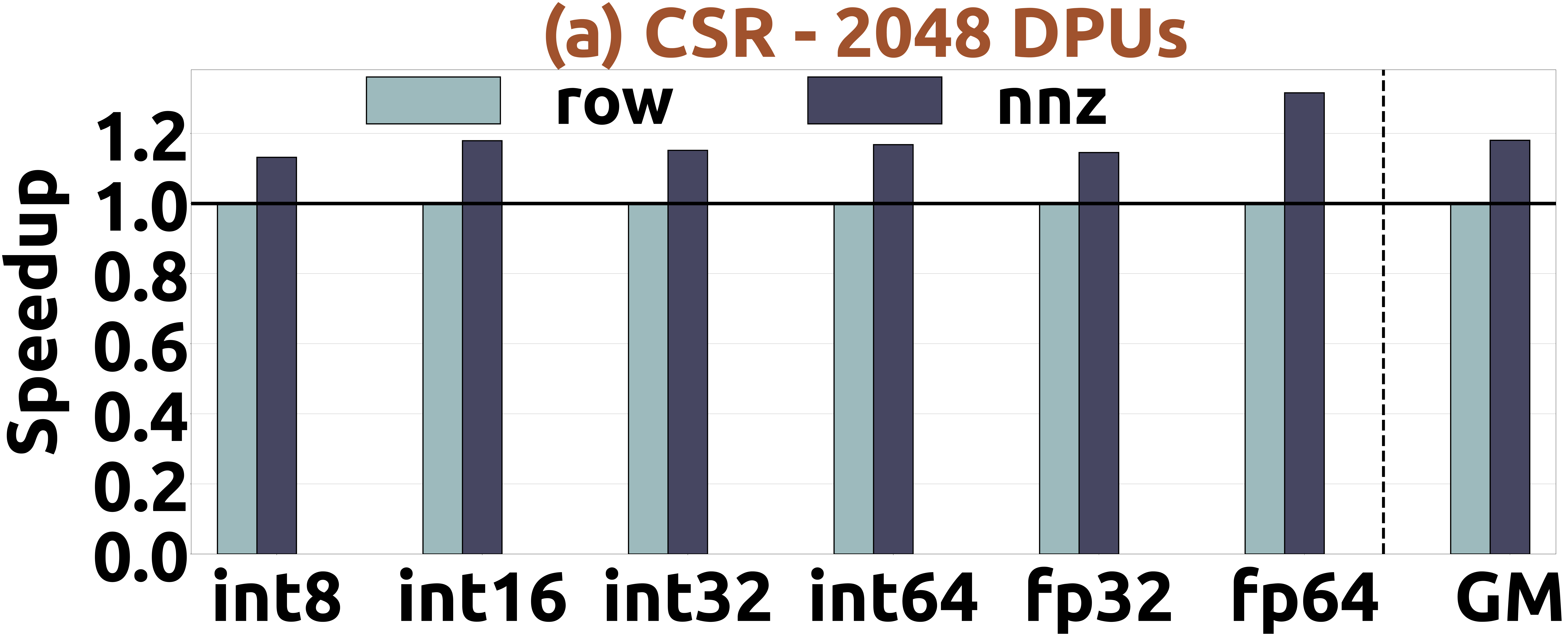} \hspace{10pt}
\includegraphics[width=.48\textwidth]{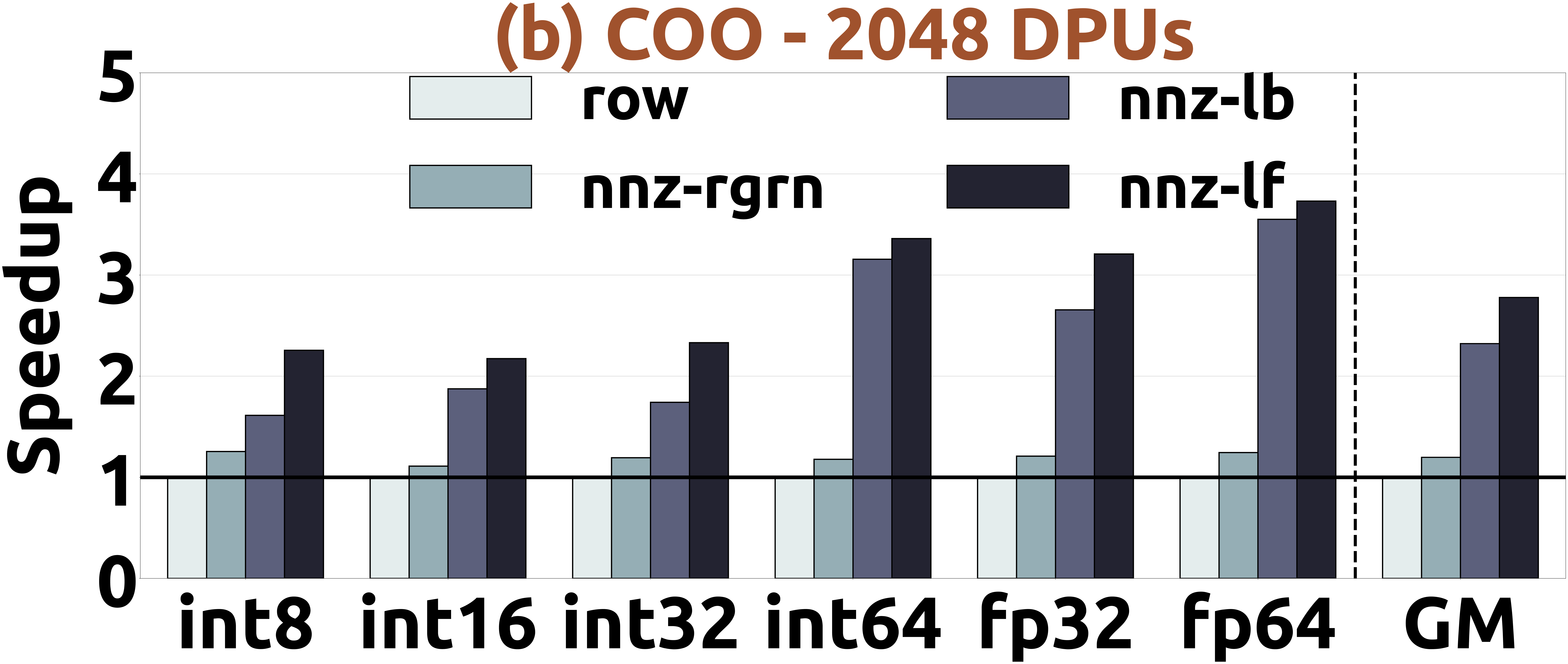} 
\vspace{8pt}
\end{minipage} 
\begin{minipage}{\textwidth}
\centering
\includegraphics[width=.48\textwidth]{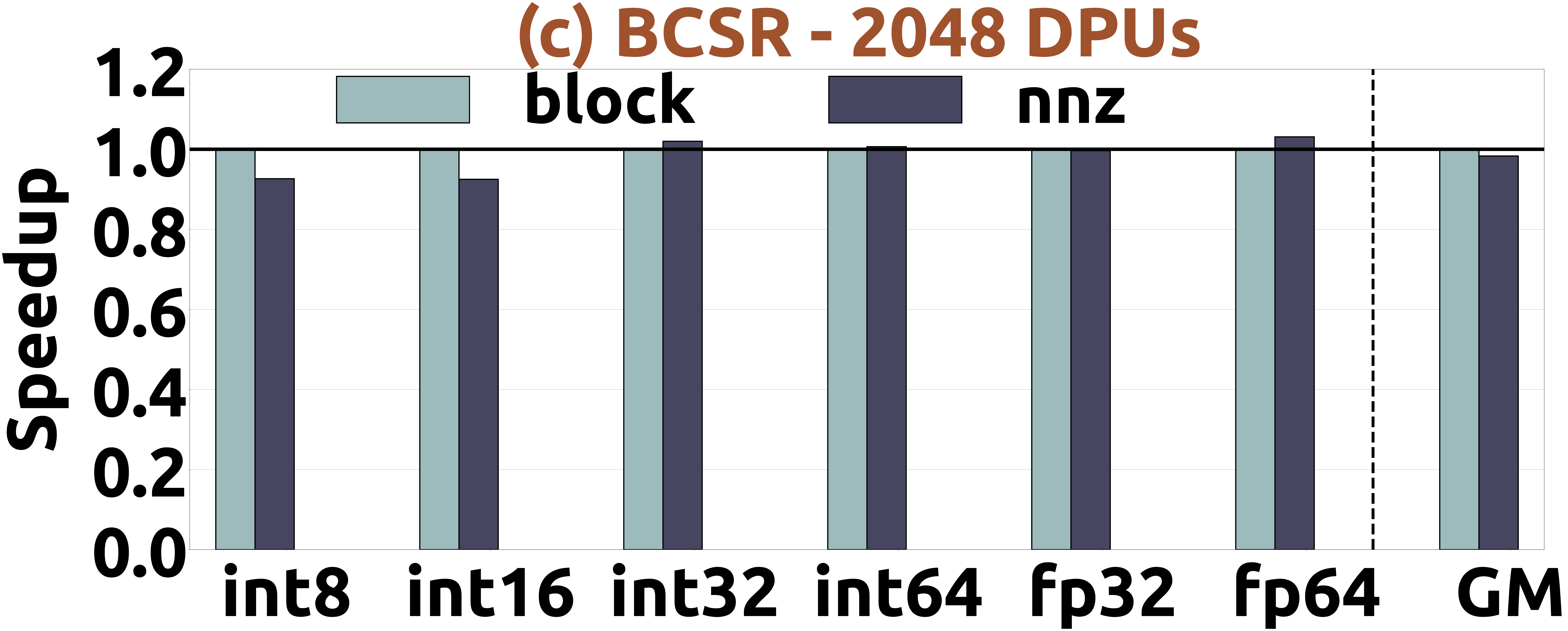} \hspace{10pt}
\includegraphics[width=.48\textwidth]{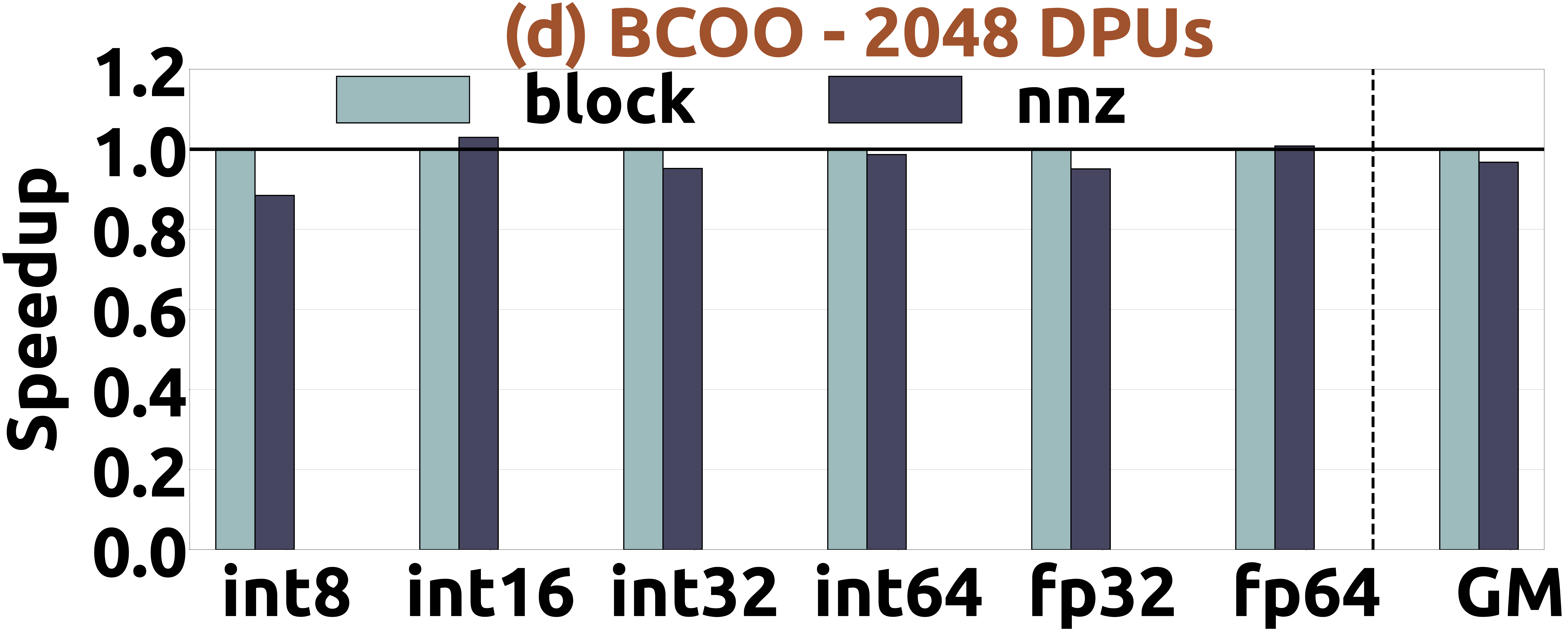}
\end{minipage}  
\vspace{-6pt}
\caption{Performance comparison of load balancing techniques for each data type using 2048 DPUs.}
\label{fig:1D_datatypes}
\vspace{-14pt}
\end{figure}

To further investigate the performance of the various load balancing schemes, Figure~\ref{fig:1D_datatypes} compares them using all the data types. We present the geometric mean of all matrices using 2048 DPUs. In the CSR and COO formats, balancing the non-zero elements across DPUs on average outperforms balancing the rows across DPUs by 1.18$\times$ and 1.20$\times$, respectively. We observe that in the COO format almost perfectly balancing the non-zero elements across DPUs provides significant performance benefits (2.55$\times$, averaged across all the data types), compared to balancing the rows, especially when multiplication is not supported by hardware (e.g., for the floating point data types). In contrast, in the BCSR and BCOO formats, balancing the blocks across DPUs performs only slightly better (on average 2.7\% across all the data types) than balancing the non-zero elements.

\begin{figure}[H]
    %\vspace{-1pt}
    \centering
    \includegraphics[width=\textwidth]{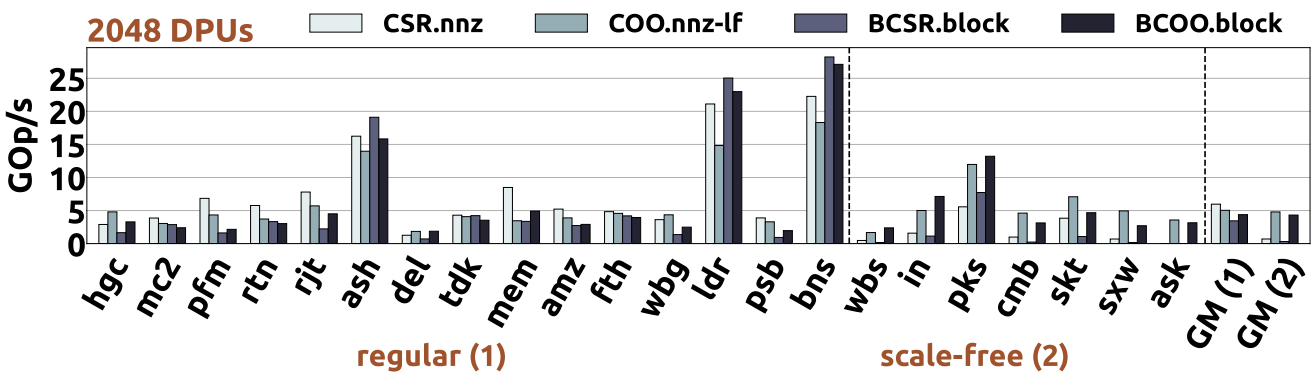}
    \vspace{-20pt}
    \caption{Throughput of various compressed formats using 2048 DPUs and the int32 data type.}
    \label{fig:1D_kernel}
    \vspace{-8pt}
\end{figure}

\begin{figure}[H]
    %\vspace{-1pt}
    \centering
    \includegraphics[width=\textwidth]{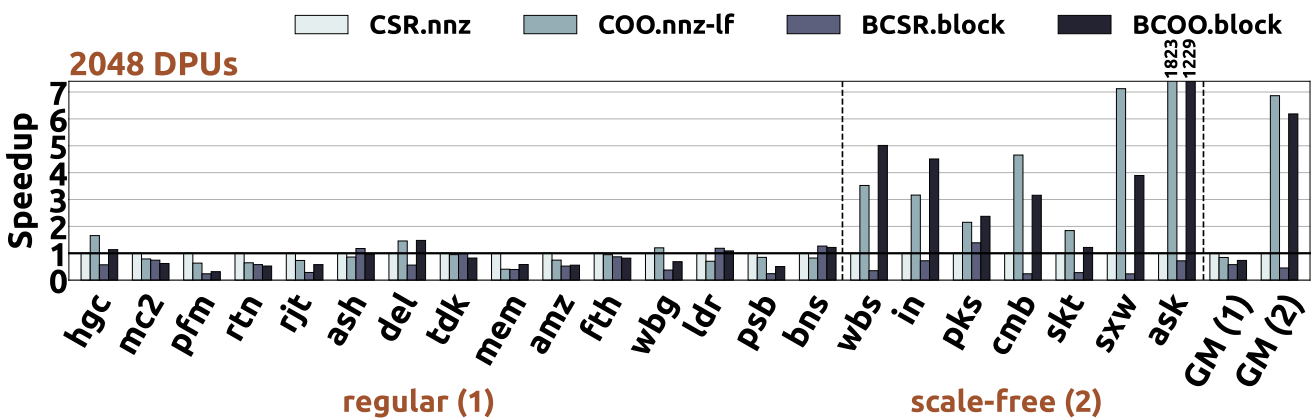}
    \vspace{-20pt}
    \caption{Performance comparison of various compressed formats using 2048 DPUs and the int32 data type. Performance is normalized to that of \texttt{CSR.nnz}.}
    \label{fig:1D_kernel_perf}
    \vspace{-8pt}
\end{figure}

\noindent{\textbf{Comparison of Compressed Matrix Formats.}}
Figures~\ref{fig:1D_kernel} and ~\ref{fig:1D_kernel_perf} compare the throughput (in GOperations per second) and the performance, respectively, achieved by various compressed formats using 2048 DPUs and the int32 data type. For the CSR and COO formats, we select balancing the non-zero elements across DPUs, and for the BCSR and BCOO formats, we select balancing the blocks across DPUs, since these are the best-performing schemes for each format averaged across all matrices and data types (Figure~\ref{fig:1D_datatypes}).

We draw four findings. First, matrices that exhibit block pattern (almost all non-zero elements of the matrix fit in dense sub-blocks), i.e., \texttt{ash}, \texttt{ldr}, \texttt{bns}, \texttt{pks} matrices, have the highest throughput, since they leverage higher data locality compared to matrices with non-block pattern. Second, in scale-free matrices, the COO and BCOO formats significantly outperform the CSR and BCSR formats by 6.94$\times$ and 13.90$\times$, respectively. This is because they provide better non-zero element balance across DPUs. In the CSR and BCSR formats, the non-zero element balance is limited to be performed at row and block-row granularity, respectively, causing performance to be limited by the DPU that processes the largest number of non-zero elements. Third, we observe that the BCOO format can outperform the CSR format even in \textit{non-blocked} scale-free matrices. Fourth, we find that when the CSR and BCSR formats provide sufficient non-zero element balance across DPUs, i.e., in many regular matrices such as \texttt{rtn}, \texttt{tdk}, \texttt{amz}, and \texttt{fth}, they can outperform the COO and BCOO formats, respectively.

\begin{tcolorbox}
\noindent\textbf{OBSERVATION 7:} \\ 
In \textit{scale-free} matrices, the COO and BCOO formats significantly outperform the CSR and BCSR formats, because they provide higher non-zero element balance across PIM cores.
\end{tcolorbox}

\subsubsection{Analysis of End-To-End \spmv{} Execution}\label{1D-EndToEnd}

\begin{figure}[H]
    \centering
    \includegraphics[width=\textwidth]{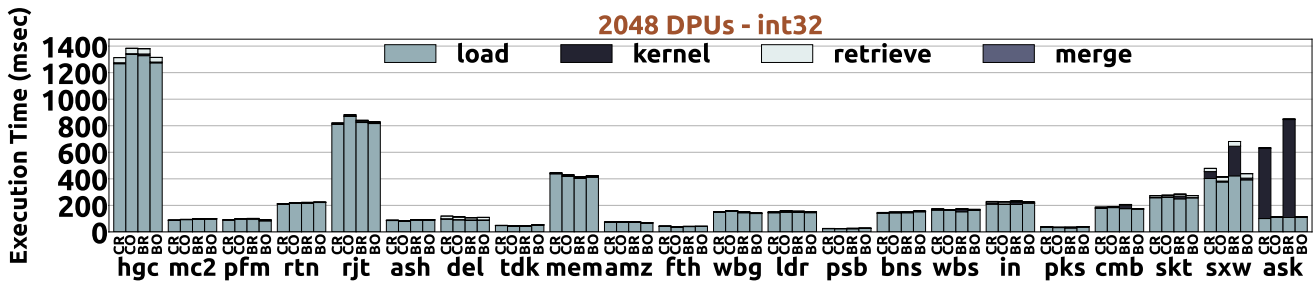}
    \vspace{-12pt}
    \caption{Total execution time when using 2048 DPUs and the int32 data type for CR: \texttt{CSR.nnz}, CO:  \texttt{COO.nnz-lf}, BR: \texttt{BCSR.block} and BO: \texttt{BCOO.block} kernels.}
    \label{fig:1D_transfers}
    \vspace{-6pt}
\end{figure}

Figure~\ref{fig:1D_transfers} shows the end-to-end execution time of 1D-partitioned kernels using 2048 DPUs and the int32 data type. The times are broken down into (i) the time for CPU to DPU transfer to load the input vector into DRAM banks (\texttt{load}), (ii) the kernel time on DPUs (\texttt{kernel}), (iii) the time for DPU to CPU transfer to retrieve the results for the output vector (\texttt{retrieve}), and (iv) the time to merge partial results on the host CPU cores (\texttt{merge}).

We draw four findings. First, the \texttt{load} data transfers constitute more than 90\% of the total execution time, because the input vector is replicated and broadcast into each DPU, causing a large number of bytes to be transferred through the narrow off-chip memory bus. An exception is in the CSR and BCSR formats for \texttt{sxw}, \texttt{ask} matrices, which include one very dense row, and thus \texttt{kernel} time is highly bottlenecked by one DPU that processes a significantly larger number of non-zero elements than the rest. Second, the \texttt{kernel} time constitutes on average only 4.3\% of the total execution time, since \spmv{} is effectively parallelized to thousands of DPUs. Third, the \texttt{retrieve} data transfers constitute on average 3.4\% of the total execution time, because the output vector is split across DPUs. Fourth, the \texttt{merge} time on the host CPU is negligible (less than 1\% of the total execution time), since only a few partial results for the elements of the output vector are merged by the host CPU cores in the 1D partitioning techniques.

\begin{tcolorbox}
\noindent\textbf{OBSERVATION 8:} \\
The end-to-end performance of the 1D partitioning techniques is severely bottlenecked by the data transfer costs to replicate and broadcast the whole input vector into \textit{each} DRAM bank of PIM cores, which takes place through the narrow off-chip memory bus.
\end{tcolorbox}

\begin{figure}[H]
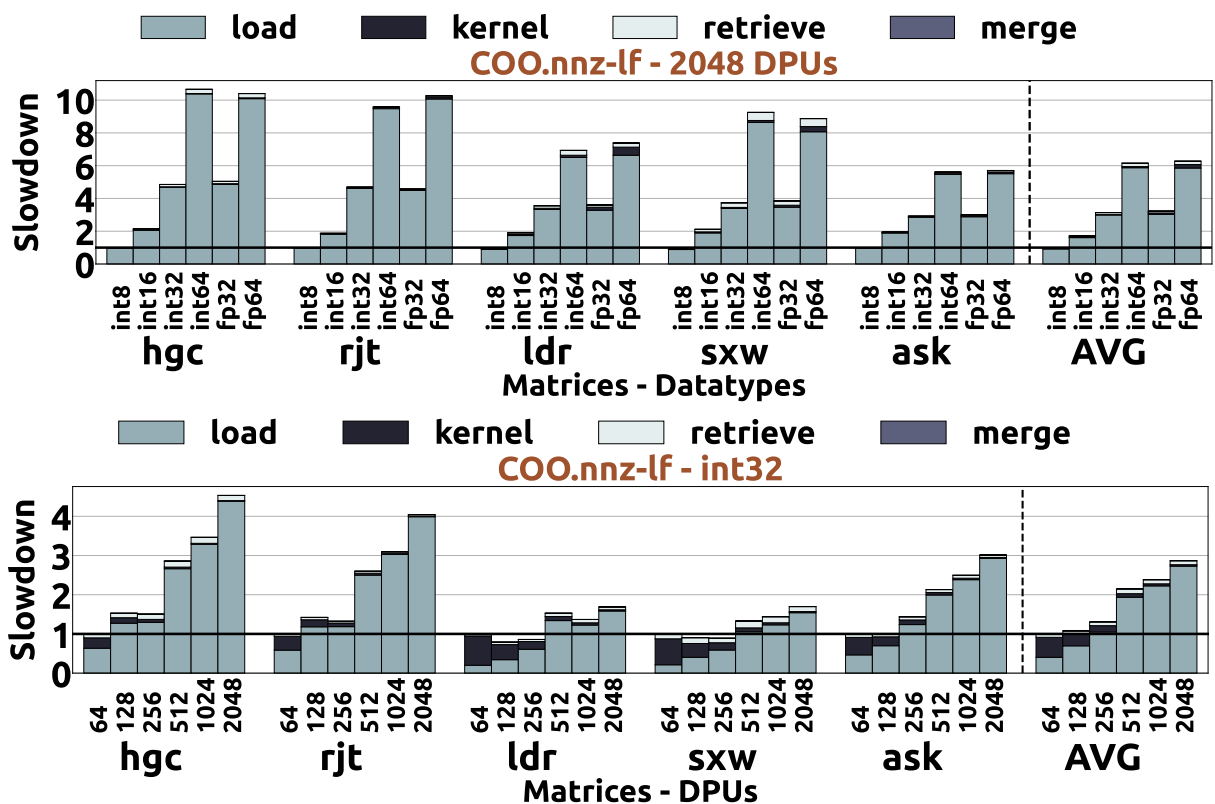

    \centering
    \includegraphics[width=.92\textwidth]{sections/SparseP/1D-partitioning/transfers_datatypes_dpus2048.pdf}
    \includegraphics[width=.92\textwidth]{sections/SparseP/1D-partitioning/transfers_scalability_int32.pdf}
    \vspace{-4pt}
    \caption{End-to-end execution time breakdown achieved by \texttt{COO.nnz-lf} when varying (a) the data type using 2048 DPUs (normalized to the experiment for the int8 data type), and (b) the number of DPUs for the int32 data type (normalized to 64 DPUs).}
    \label{fig:1D_transfers_scalability}
    \vspace{-8pt}
\end{figure}

To further investigate on the costs to the load input vector into all DRAM banks of PIM-enabled memory, we present in 
Figure~\ref{fig:1D_transfers_scalability} the total execution time achieved by \texttt{COO.nnz-lf} when varying (a) the data type using 2048 DPUs (normalized to the experiment for the int8 data type), and (b) the number of DPUs for the int32 data type (normalized to 64 DPUs).

We draw two conclusions. First, the \texttt{load} data transfer costs increase proportionally to the number of bytes of the data type, and still dominate performance even for the data type with the smallest memory footprint (int8). Second, the \texttt{load} data transfer costs and the associated memory footprint for the input vector increase proportionally to the number of DPUs used, and thus the best end-to-end performance is achieved using only a small portion of the available DPUs on the system.

\begin{tcolorbox}
\noindent\textbf{OBSERVATION 9:} \\
\spmv{} execution of the 1D-partitioned schemes cannot scale up to a large number of PIM cores due to high data transfer overheads to copy the input vector into \textit{each} DRAM bank of PIM-enabled memory.
\end{tcolorbox}

\subsection{Analysis of \spmv{} Execution Using 2D Partitioning Techniques}\label{2D}
We evaluate the 2D-partitioned kernels highlighted in bold in Table~\ref{table:library}. Specifically, for the COO format we use the lock-free approach, and for the BCSR (in the int8 data type) and BCOO formats we use the coarse-grained locking approach. In the \equallyWidth{} and \variableSized{} techniques, for the BCSR and BCOO formats we balance the blocks across DPUs of the same vertical partition, since doing so performs slightly better than balancing the non-zero elements, as explained in Section~\ref{1D-Kernel}. In all experiments, we balance the non-zero elements across 16 tasklets within a single DPU.

\subsubsection{Sensitivity Studies on 2D Partitioning Techniques}\label{2D-Studies}

We present three characterization studies on the 2D partitioning techniques. First, we evaluate the performance of fine-grained data transfers from/to PIM-enabled memory for the input and output vectors. Second, we evaluate the scalability of the 2D partitioning techniques to thousands of DPUs. Finally, we explore performance implications on the number of vertical partitions used in the 2D-partitioned kernels.

\noindent{\textbf{Analysis of Fine-Grained Data Transfers.}}
The UPMEM API~\cite{upmem-guide} has the limitation that \textit{the transfer sizes from/to all DRAM banks involved in the same parallel transfer need to be the same}. The UPMEM API provides \textit{parallel data transfers} either to all DPUs of all ranks (henceforth referred to as \textit{coarse-grained} transfers), or at rank granularity, i.e., to 64 DPUs of the same rank (henceforth referred to as \textit{fine-grained} transfers). In the first case, parallel data transfers are performed to all DPUs used at once, padding with empty bytes at the granularity of \textit{all} DPUs used, e.g., 2048 DPUs in Figure~\ref{fig:2D_fgtransfers}. In the latter case, programmers iterate over the ranks of PIM-enabled DIMMs, and for \textit{each} rank perform parallel data transfers to the 64 DPUs of the same rank padding with empty bytes at the granularity of 64 DPUs.

In \spmv{} execution, for the \equallyWidth{} and \variableSized{} techniques the heights and widths of 2D tiles vary, and thus padding with empty bytes is necessary for the \texttt{load} and \texttt{retrieve} data transfers of the elements of the input and output vector, respectively. Figure~\ref{fig:2D_fgtransfers} compares coarse-grained data transfers, i.e., performing parallel transfers to all 2048 DPUs at once, with fine-grained data transfers, i.e., iterating over the ranks and for each rank performing parallel transfers to the 64 DPUs of the same rank. We evaluate both the \equallyWidth{} and \variableSized{} techniques using the COO format and with 2 and 32 vertical partitions. Please see Appendix~\ref{sec:appendix-2D-fgtrans} for all matrices.

\begin{figure}[t]
    \centering
    \includegraphics[width=.98\textwidth]{sections/SparseP/2D-partitioning/trans_reduced_dpus2048_int32_cps2.pdf}
    \includegraphics[width=.98\textwidth]{sections/SparseP/2D-partitioning/trans_reduced_dpus2048_int32_cps32.pdf}
    \vspace{-8pt}
    \caption{Performance comparison of \texttt{RC}: \texttt{RBDCOO} with coarse-grained transfers, \texttt{RY}: \texttt{RBDCOO} with fine-grained transfers in the output vector, \texttt{BC}: \texttt{BDCOO} with coarse-grained transfers, \texttt{BY}: \texttt{BDCOO} with fine-grained transfers only in the output vector, and \texttt{BT}: \texttt{BDCOO} with fine-grained transfers in both the input and the output vector using the int32 data type, 2048 DPUs and having 2 (left) and 32 (right) vertical partitions. Performance is normalized to that of the \texttt{RC} scheme.}
    \label{fig:2D_fgtransfers}
    \vspace{-10pt}
\end{figure}

We draw two findings. First, when the number of vertical partitions is small, e.g., 2 vertical partitions, the disparity in widths across tiles in the \variableSized{} scheme is low. Thus, \texttt{BT} only slightly outperforms \texttt{BY} by 1\% on average, since in \texttt{BY} \textit{only} a small amount of padding is added on the \texttt{load} data transfers of the input vector. In contrast, the disparity in heights across tiles in the \equallyWidth{} and \variableSized{} schemes is high. Thus, \texttt{RY} and \texttt{BY} significantly outperform \texttt{RC} and \texttt{BC} by an average of 1.68$\times$ and 1.60$\times$, respectively. This is because fine-grained transfers to retrieve the elements of the output vector significantly decrease the amount of bytes transferred from PIM-enabled memory to host CPU over coarse-grained transfers. Second, when the number of vertical partitions is large, e.g., 32 vertical partitions, the disparity in heights across tiles in the \equallyWidth{} and \variableSized{} schemes is lower compared to when the number of vertical partitions is small. Thus, \texttt{RY} and \texttt{BY} provide smaller performance benefits over \texttt{RC} and \texttt{BC} (on average 1.24$\times$ and 1.22$\times$, respectively), respectively, compared to a small number of vertical partitions. In contrast, the disparity in heights across tiles in the \equallyWidth{} and \variableSized{} schemes is higher compared to when the number of vertical partitions is small. Thus, \texttt{BT} outperforms \texttt{BY} by 4.7\% on average. Overall, we conclude that fine-grained data transfers (i.e., at rank granularity in the UPMEM PIM system) can significantly improve performance in the \equallyWidth{} and \variableSized{} schemes.

\begin{tcolorbox}
\noindent\textbf{OBSERVATION 10:} \\ 
\textit{Fine-grained} parallel transfers in the \equallyWidth{} and \variableSized{} 2D partitioning techniques, i.e., minimizing the amount of padding with empty bytes in parallel data transfers to/from PIM-enabled memory, can provide large performance gains. 
\end{tcolorbox}

\begin{figure}[t]
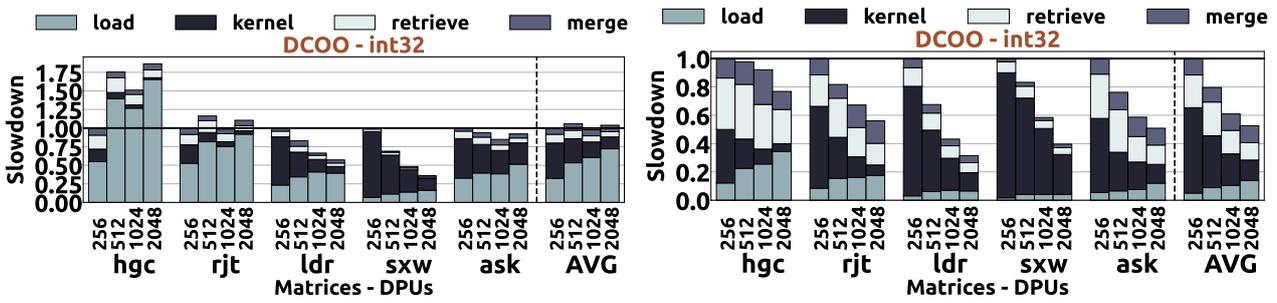

    \centering
    \begin{minipage}{\textwidth}
    \includegraphics[width=.484\textwidth]{sections/SparseP/2D-partitioning/scalability_fix_int32_col4.pdf}
    \includegraphics[width=.484\textwidth]{sections/SparseP/2D-partitioning/scalability_fix_int32_col16.pdf}
    \end{minipage}
    \vspace{-8pt}
    \caption{Execution time breakdown of \equallySized{} partitioning technique of the COO format using 4 (left) and 16 (right) vertical partitions when varying the number of DPUs used for the int32 data type. Performance is normalized to that with 256 DPUs.}
    \label{fig:2D_scale_equally_sized}
    \vspace{-8pt}
\end{figure}

\begin{figure}[!ht]
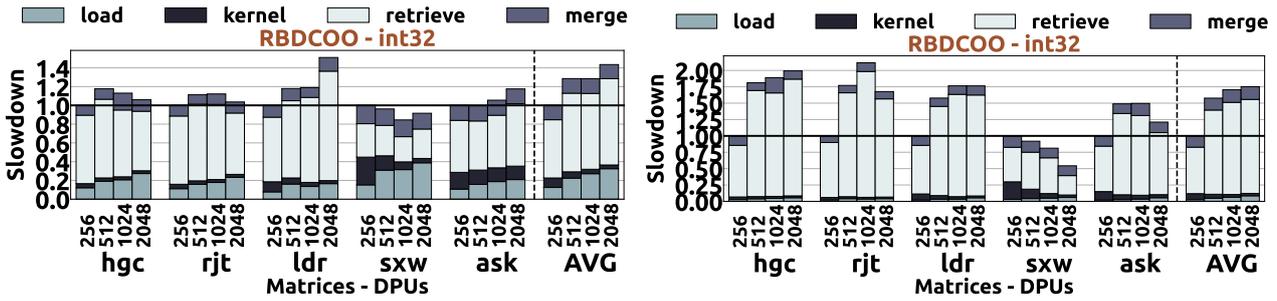

    \centering
    \begin{minipage}{\textwidth}
    \includegraphics[width=.484\textwidth]{sections/SparseP/2D-partitioning/scalability_bal_int32_col4.pdf}
    \includegraphics[width=.484\textwidth]{sections/SparseP/2D-partitioning/scalability_bal_int32_col16.pdf}
    \end{minipage}
    \vspace{-8pt}
    \caption{Execution time breakdown of \variableSized{} partitioning technique of the COO format using 4 (left) and 16 (right) vertical partitions when varying the number of DPUs used for the int32 data type. Performance is normalized to that with 256 DPUs.}
    \label{fig:2D_scale_variable_sized}
    \vspace{-10pt}
\end{figure}

\noindent\textbf{Scalability of the 2D Partitioning Techniques.}
We analyze scalability with the number of DPUs for the 2D partitioning techniques. Figures~\ref{fig:2D_scale_equally_sized}, ~\ref{fig:2D_scale_equally_wide} and ~\ref{fig:2D_scale_variable_sized} compare the performance of the \equallySized{}, \equallyWidth{} and \variableSized{} schemes, respectively, using the COO format and the int32 data type, as the number of DPUs increases.

\begin{figure}[t]
    \centering
    \begin{minipage}{\textwidth}
    \includegraphics[width=.484\textwidth]{sections/SparseP/2D-partitioning/scalability_rbal_int32_col4.pdf}
    \includegraphics[width=.484\textwidth]{sections/SparseP/2D-partitioning/scalability_rbal_int32_col16.pdf}
    \end{minipage}
    \vspace{-8pt}
    \caption{Execution time breakdown of \equallyWidth{} partitioning technique of the COO format using 4 (left) and 16 (right) vertical partitions when varying the number of DPUs used for the int32 data type. Performance is normalized to that with 256 DPUs.}
    \label{fig:2D_scale_equally_wide}
    \vspace{-8pt}
\end{figure}

We draw two findings. First, the \equallySized{} scheme (i.e., \texttt{DCOO}) achieves high scalability with a large number of vertical partitions. The \texttt{kernel} time of \equallySized{} scheme is mainly limited by the DPU (or a few DPUs) that processes the largest number of non-zero elements. With a large number of \textit{static} vertical partitions, the non-zero element disparity across DPUs is high, i.e., the \texttt{kernel} time is highly bottlenecked by the DPU that processes the largest number of non-zero elements. As a result, increasing the number of DPUs improves performance by decreasing the \texttt{kernel} time via better non-zero element balance across DPUs.

\begin{tcolorbox}
\noindent\textbf{OBSERVATION 11:} \\
The \texttt{kernel} time in the \equallySized{} schemes is limited by the PIM core (or a few PIM cores) assigned to the 2D tile with the largest number of non-zero elements. 
\end{tcolorbox}

Second, we observe that the \equallyWidth{} and \variableSized{} schemes (i.e., \texttt{RBDCOO} and \texttt{BDCOO}) are severely bottlenecked by \texttt{retrieve} data transfer costs (a large number of partial results is created on PIM cores), and thus they are difficult to scale up to thousands of DPUs. Moreover, when the number of vertical partitions is high, the disparity in heights of the tiles is high. Thus, as the number of DPUs increases, the amount of padding needed in \texttt{retrieve} data transfers becomes very large, causing significant performance degradation.

\begin{tcolorbox}
\noindent\textbf{OBSERVATION 12:} \\
The scalability of the \equallyWidth{} and \variableSized{} schemes to a large number of PIM cores is severely limited by large data transfer overheads to retrieve partial results for the elements of the output vector from the DRAM banks of PIM-enabled memory to the host CPU via the narrow memory bus.
\end{tcolorbox}

\begin{figure}[H]
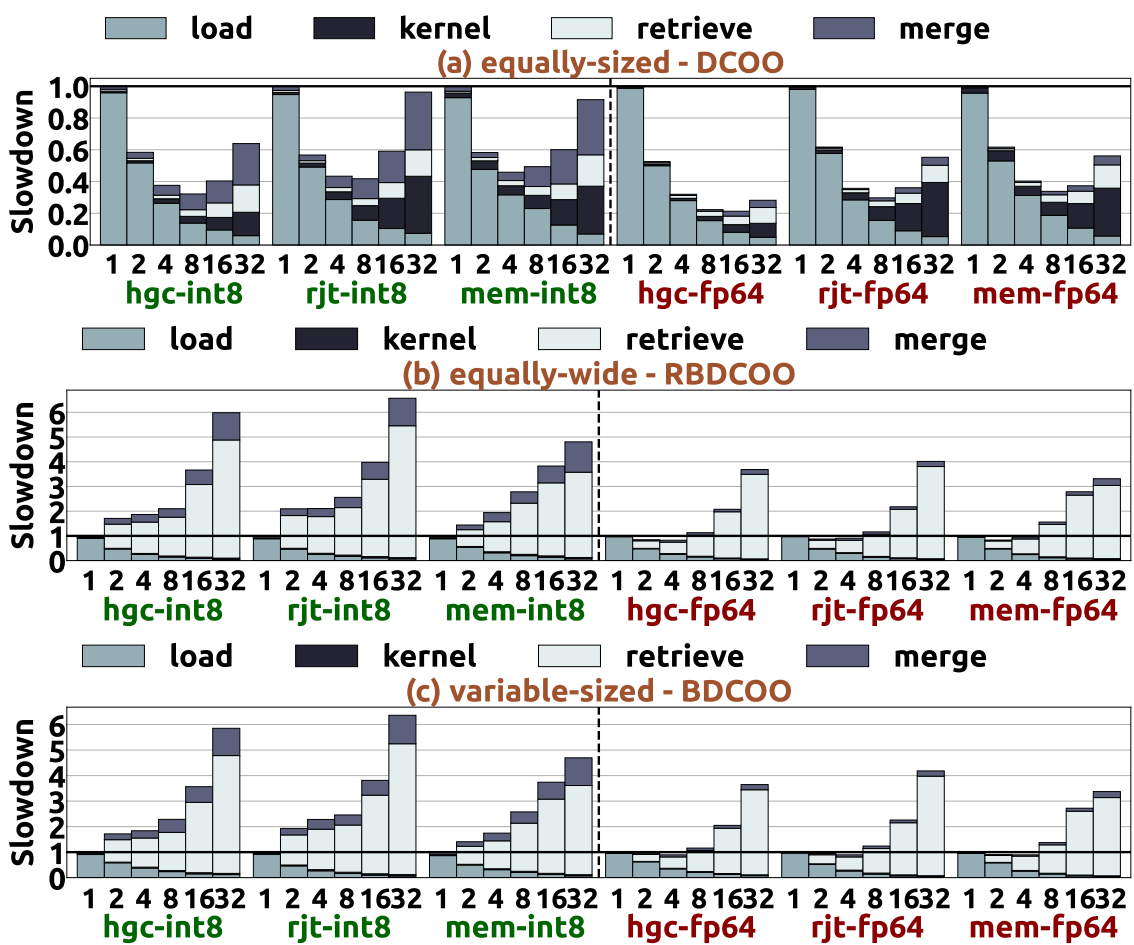

\vspace{-6pt}
    \centering
    \begin{minipage}{\textwidth}
    \centering
    \includegraphics[width=0.86\textwidth]{sections/SparseP/2D-partitioning/fixed_vertical_dpus2048.pdf}
    \includegraphics[width=0.86\textwidth]{sections/SparseP/2D-partitioning/rbal_vertical_dpus2048.pdf}
    \includegraphics[width=0.86\textwidth]{sections/SparseP/2D-partitioning/bal_vertical_dpus2048.pdf}
    \end{minipage}
    \vspace{-4pt}
    \caption{Execution time breakdown of 2D partitioning schemes using the COO format and 2048 DPUs when varying the number of vertical partitions from 1 to 32 for the int8 and fp64 data types. Performance is normalized to the performance of the experiment with 1 vertical partition.}
    \label{fig:2D_vertpartitions}
    \vspace{-10pt}
\end{figure}

\noindent\textbf{Effect of the Number of Vertical Partitions.}
In all experiments presented henceforth, we perform fine-grained data transfers (at rank granularity, i.e., 64 DPUs in the UPMEM PIM system) in the 2D partitioning schemes. Figure~\ref{fig:2D_vertpartitions} evaluates performance implications on the number of vertical partitions performed in 2D-partitioned kernels. We use the COO format and vary the number of vertical partitions from 1 to 32, in steps of multiple of 2. We draw four findings.

First, in the \equallySized{} scheme, as the number of vertical partitions increases, \texttt{kernel} time increases, if there is \textit{no} dense row in the matrix. This is because the disparity in the non-zero elements across 2D tiles increases as the number of vertical partitions increases. Thus, performance is limited by one DPU or a few DPUs that process the largest number of non-zero elements.

\begin{tcolorbox}
\noindent\textbf{OBSERVATION 13:} \\
As the number of vertical partitions increases, the \equallySized{} 2D partitioning scheme typically increases the non-zero element disparity across PIM cores (unless there is one dense row on the matrix), thereby increasing the \texttt{kernel} time. 
\end{tcolorbox}

Second, as the number of vertical partitions increases, \texttt{retrieve} data transfer costs and \texttt{merge} time increase. This is because the partial results created for the output vector increase proportionally with the number of vertical partitions. The performance overheads of \texttt{retrieve} data transfer costs are highly affected by the characteristics of the underlying hardware (e.g., the bandwidth provided on I/O channels of the memory bus between host CPU and PIM-enabled DIMMs). Similarly, the performance cost of the \texttt{merge} step depends on the hardware characteristics of the host CPU (e.g., the number of the CPU cores, the available hardware threads, microarchitecture of CPU cores). We refer the reader to Appendix~\ref{sec:appendix-2D-vertpartitions} for a comparison of \spmv{} execution using two different UPMEM PIM systems with different hardware characteristics (Table ~\ref{tab:pim-systems}).

Third, we find that in the \equallyWidth{} and \variableSized{} schemes, there is high disparity in heights of 2D tiles, and as a result on the number of partial results created across DPUs. Even with fine-grained parallel \texttt{retrieve} data transfers at rank granularity, the amount of padding needed in the \equallyWidth{} and \variableSized{} schemes is at 88.6\% and 88.0\%, respectively, causing high bottlenecks in the narrow memory bus. Therefore, in PIM systems that do not support very fine-grained parallel transfers to gather results from PIM-enabled memory to the host CPU \textit{at DRAM bank granularity}, execution is highly limited by the amount of padding performed in \texttt{retrieve} data transfers, which can be very large in irregular workloads~\cite{Gomez2021Analysis,Gomez2021Benchmarking,Oliveira2021Damov,Lockerman2020Livia,Kanellopoulos2019SMASH,Giannoula2018Combining,besta2017slimsell,dongarra1996sparse,Elafrou2018SparseX,Elafrou2017PerformanceAA,YouTubeGraph,FacebookGraph,Goumas2008Understanding,White97Improving,Helal2021ALTO,Pelt2014Medium,Strati2019AnAdaptive} like the \spmv{} kernel.

\begin{tcolorbox}
\noindent\textbf{OBSERVATION 14:} \\
The \equallyWidth{} and \variableSized{} 2D partitioning schemes require fine-grained parallel transfers \textit{at DRAM bank granularity} to be supported by the PIM system, i.e., \textit{zero} padding in \textit{parallel} \texttt{retrieve} data transfers from PIM-enabled memory to the host CPU, to achieve high performance.
\end{tcolorbox}

Fourth, we find that the number of vertical partitions that provides the best performance depends on the sparsity pattern of the input matrix, the data type, and the underlying hardware parameters (e.g., number of PIM cores, off-chip memory bus bandwidth, transfer latency costs between main memory and PIM-enabled memory, characteristics and microarchitecture of the host CPU cores that perform the \texttt{merge} step). For example, with the int8 data type, \texttt{DCOO} performs best for \texttt{hgc} and \texttt{mem} matrices with 8 and 4 vertical partitions, respectively. Instead, with the fp64 data type, \texttt{DCOO} performs best for \texttt{hgc} and \texttt{mem} matrices with 16 and 8 vertical partitions, respectively. We refer the reader to Appendix~\ref{sec:appendix-2D-vertpartitions} for a characterization study on the number of vertical partitions to perform in the 2D-partitioned kernels using two UPMEM PIM systems with different hardware characteristics. As we demonstrate in Appendix~\ref{sec:appendix-2D-vertpartitions}, the number of vertical partitions that provides best performance on \spmv{} varies across the two different UPMEM PIM platforms. In this work, we leave for future work the exploration of selection methods for the number of vertical partitions that provide best \spmv{} execution. Overall, based on our analysis we conclude that the parallelization scheme that achieves the best performance in \spmv{} depends on both the input sparse matrix and the hardware characteristics of the PIM system.

\begin{tcolorbox}
\noindent\textbf{OBSERVATION 15:} \\
There is \textit{no one-size-fits-all} parallelization approach for \spmv{} in PIM systems, since the performance of each parallelization scheme depends on the characteristics of the input matrix and the underlying PIM hardware.
\end{tcolorbox}

\subsubsection{Analysis of Compressed Formats}\label{2D-Formats}

\begin{figure}[b]
\vspace{-4pt}
    \centering
    \includegraphics[width=0.9\textwidth]{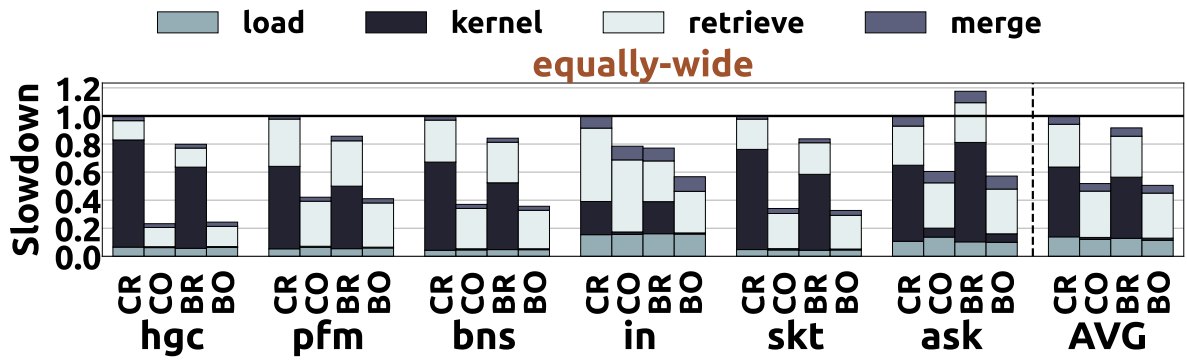}
    \vspace{-4pt}
    \caption{End-to-end execution time breakdown of the \equallySized{} 2D partitioning technique for CR: \texttt{DCSR}, CO: \texttt{DCOO}, BR: \texttt{DBCSR} and BO: \texttt{DBCOO} schemes using 4 vertical partitions and the int32 data type. Performance is normalized to that of \texttt{DCSR}.}
    \label{fig:2D_balance_fixed}
\end{figure}

\begin{figure}[t]
    \centering
    \includegraphics[width=0.9\textwidth]{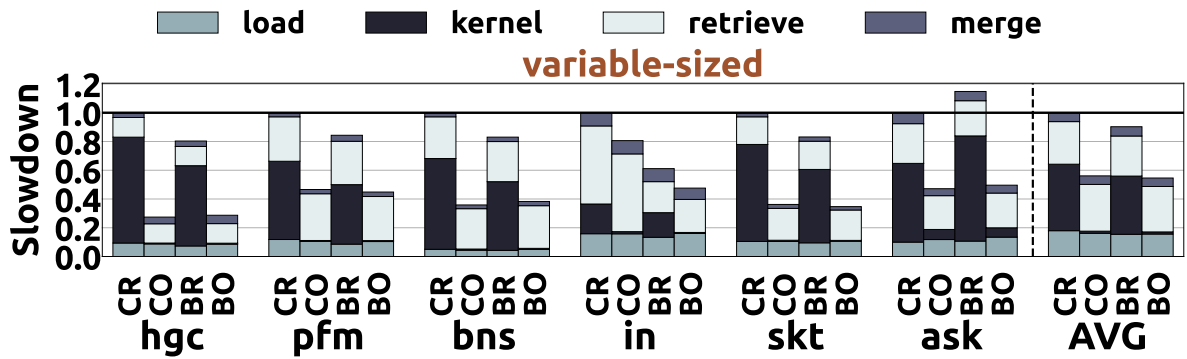}
    \vspace{-4pt}
    \caption{End-to-end execution time breakdown of the \variableSized{} 2D partitioning technique for CR: \texttt{BDCSR}, CO: \texttt{BDCOO}, BR: \texttt{BDBCSR} and BO: \texttt{BDBCOO} schemes using 4 vertical partitions and the int32 data type. Performance is normalized to that of \texttt{BDCSR}.}
    \label{fig:2D_balance_bal}
    \vspace{-10pt}
\end{figure}

We compare the performance achieved by various compressed matrix formats for each of the three types of the 2D partitioning technique. The goal of this experiment is to find the best-performing compressed format for each 2D partitioning technique. Figures~\ref{fig:2D_balance_fixed}, ~\ref{fig:2D_balance_rbal}, and ~\ref{fig:2D_balance_bal} compare the performance of compressed matrix formats for the \equallySized{}, \equallyWidth{} and \variableSized{} 2D partitioning techniques, respectively. We use 2048 DPUs and the int32 data type having 4 vertical partitions. See Appendix~\ref{sec:appendix-2D-formats} for the complete evaluation on all large sparse matrices.

\begin{figure}[t]
    \centering
    \includegraphics[width=0.9\textwidth]{sections/SparseP/2D-partitioning/rbal_time_reduced_norm_dpus2048_int32_cps4.pdf}
    \vspace{-4pt}
    \caption{End-to-end execution time breakdown of the \equallyWidth{} 2D partitioning technique for CR: \texttt{RBDCSR}, CO: \texttt{RBDCOO}, BR: \texttt{RBDBCSR} and BO: \texttt{RBDBCOO} schemes using 4 vertical partitions and the int32 data type. Performance is normalized to that of \texttt{RBDCSR}.}
    \label{fig:2D_balance_rbal}
    \vspace{-10pt}
\end{figure}

We draw two findings. First, as already explained, \texttt{kernel} time of the \equallySized{} scheme is limited by the DPU (or a few DPUs) assigned to the 2D tile with the largest number of non-zero elements. In scale-free matrices (e.g., \texttt{in} and \texttt{ask}), the disparity in the non-zero elements across 2D tiles is higher than in regular matrices (e.g., \texttt{pfg} and \texttt{bns}), causing \texttt{kernel} time to be a larger portion of the total execution time. Second, we find that the CSR and BCSR formats perform worse than the COO and BCOO formats, especially in the \equallyWidth{} and \variableSized{} schemes, due to higher \texttt{kernel} times. In the CSR and BCSR formats, data partitioning across DPUs and/or across tasklets within a DPU is performed at row and block-row granularity, respectively. Thus, the CSR and BCSR formats can cause higher non-zero element imbalance across processing units compared to the COO and BCOO formats. Overall, the COO and BCOO formats outperform the CSR and BCSR formats by 1.59 $\times$ and 1.53 $\times$ (averaged across all three types of 2D partitioning techniques), respectively.

\begin{tcolorbox}
\noindent\textbf{OBSERVATION 16:} \\
The compressed matrix format used to store the input matrix determines the data partitioning across DRAM banks of PIM-enabled memory. Thus, it affects the load balance across PIM cores with corresponding performance implications. Overall, the COO and BCOO formats outperform the CSR and BCSR formats, because they provide higher non-zero element balance across PIM cores.
\end{tcolorbox}

\subsubsection{Comparison of 2D Partitioning Techniques}\label{2D-Comparison}

\begin{figure}[H]
    \centering
    \includegraphics[width=\textwidth]{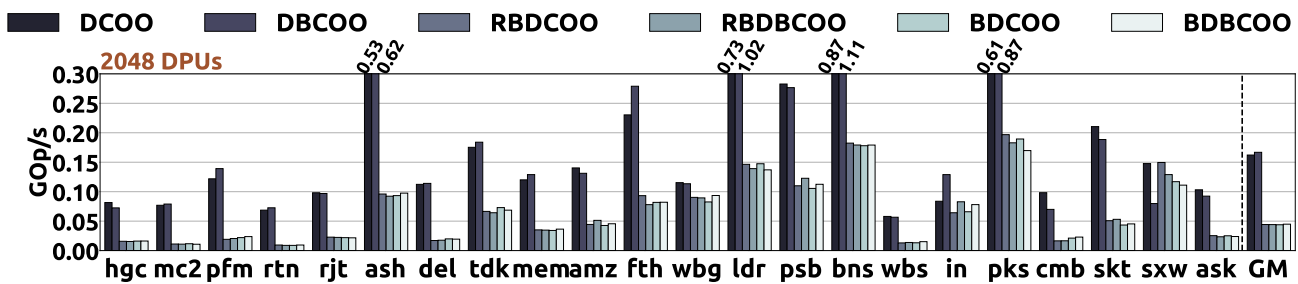}
    \vspace{-17pt}
    \caption{Throughput of 2D partitioning techniques using the COO and BCOO formats, 2048 DPUs and the int32 type.}
    \label{fig:2D_best_partition}
    \vspace{-4pt}
\end{figure}

\begin{figure}[H]
    \vspace{-6pt}
    \centering
    \includegraphics[width=\textwidth]{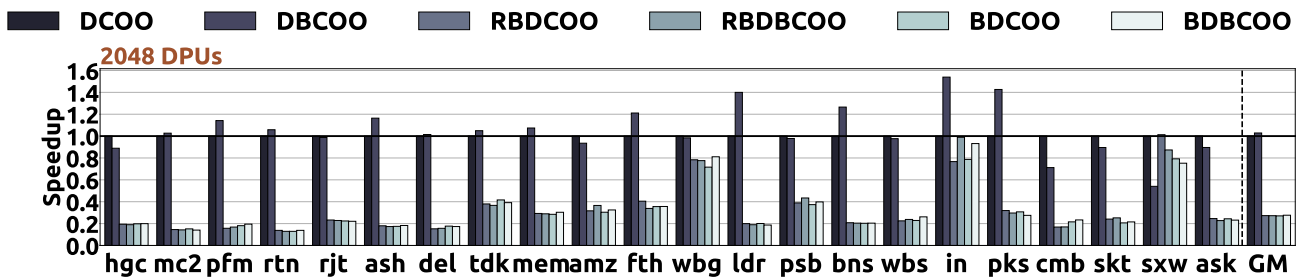}
    \vspace{-17pt}
    \caption{Performance comparison of 2D partitioning techniques using the COO and BCOO formats, 2048 DPUs and the int32 type. Performance is normalized to that of \texttt{DCOO}.}
    \label{fig:2D_best_partition_perf}
    \vspace{-4pt}
\end{figure}

We compare the best-performing \spmv{} implementations of all 2D partitioning schemes, i.e., using the COO and BCOO formats. Figures~\ref{fig:2D_best_partition} and ~\ref{fig:2D_best_partition_perf} compare the throughput (in GOperations per second) and the performance, respectively, of \texttt{DCOO}, \texttt{DBCOO}, \texttt{RBDCOO}, \texttt{RBDBCOO}, \texttt{BDCOO}, \texttt{BDBCOO} schemes using 2048 DPUs and the int32 data type. For each implementation, we vary the number of vertical partitions from 2 to 32, in steps of multiple of 2, and select the best-performing execution throughput.

We draw two conclusions. First, similarly to 1D-partitioned kernels, matrices that exhibit block pattern (e.g., \texttt{ash}, \texttt{ldr}, \texttt{bns}, \texttt{pks}) have the highest throughput (Figure~\ref{fig:2D_best_partition}). Second, the \equallyWidth{} and \variableSized{} schemes perform similarly, i.e., their performance varies only by $\pm$1.1\% on average. Even though the \variableSized{} technique can improve the non-zero element balance across DPUs, and thus \texttt{kernel} time, compared to the \equallyWidth{} technique, the total execution time does not improve. In the UPMEM PIM system, performance of both techniques is severely bottlenecked by data transfer overheads due to a large amount of padding needed to retrieve results from PIM-enabled memory to the host CPU. Third, we find that the \equallySized{} technique outperforms the \equallyWidth{} and \variableSized{} techniques by 3.71$\times$ on average, because it achieves lower data transfer overheads. The \equallyWidth{} and \variableSized{} techniques provide near-perfect non-zero element balance across DPUs, but they significantly increase the \texttt{retrieve} data transfer costs due to the large amount  of padding with empty bytes performed. As a result, we recommend software designers to explore \textit{relaxed} load balancing schemes, i.e., schemes that trade off computation balance across PIM cores for lower amounts of data transfer.

\subsection{Comparison of 1D and 2D Partitioning Techniques}\label{1D-2D}
\begin{figure}[H]
\vspace{-2pt}
    \centering
    \includegraphics[width=\textwidth]{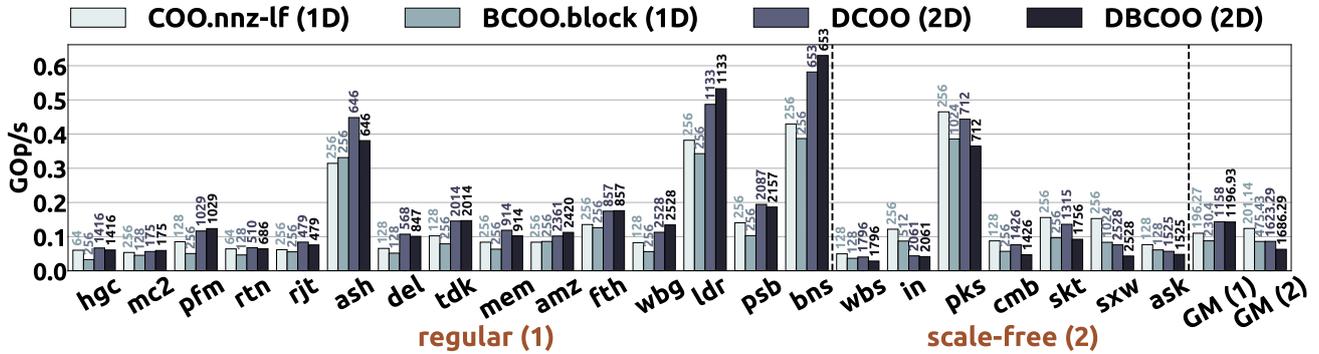}
    \vspace{-20pt}
    \caption{Throughput of the best-performing 1D- and 2D-partitioned kernels for the fp32 data type.}
    \label{fig:2D_1D-2D}
    \vspace{-6pt}
\end{figure}

\begin{figure}[H]
    \vspace{-4pt}
    \centering
    \includegraphics[width=\textwidth]{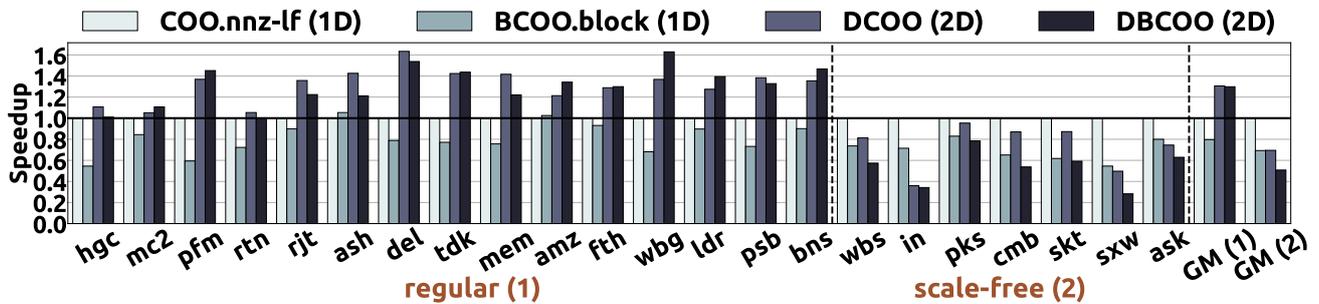}
    \vspace{-18pt}
    \caption{Performance comparison of the best-performing 1D- and 2D-partitioned kernels for the fp32 data type. Performance is normalized to that of \texttt{COO.nnz-lf}.}
    \label{fig:2D_1D-2D-perf}
    \vspace{-10pt}
\end{figure}

We compare the throughput (in GOperations per second) and the performance of the best-performing 1D- and 2D-partitioned kernels in Figures~\ref{fig:2D_1D-2D} and ~\ref{fig:2D_1D-2D-perf}, respectively. For 1D partitioning, we use the lock-free COO (\texttt{COO.nnz-lf}) and coarse-grained locking BCOO (\texttt{BCOO.block}) kernels. For each matrix, we vary the number of DPUs from 64 to 2528, and select the best-performing end-to-end execution throughput. For 2D partitioning, we use the \equallySized{} COO (\texttt{DCOO}) and BCOO  (\texttt{DBCOO}) kernels with \dpuActive{}. For each matrix, we vary the number of vertical partitions from 2 to 32 (in steps of multiple of 2), and select the best-performing end-to-end execution throughput. The numbers shown over each bar of Figure~\ref{fig:2D_1D-2D} present the number of DPUs that provide the best-performing end-to-end execution throughput for each input-scheme combination. Please see Appendix~\ref{sec:appendix-1D_2D} for a performance comparison of the best-performing \spmv{} kernels on two UPMEM PIM systems with different hardware characteristics.

We draw two findings. First, we find that best performance is achieved using a much smaller number of DPUs than the available DPUs on the system. In the 1D-partitioned kernels (i.e., \texttt{COO.nnz-lf} and \texttt{BCOO.block}), replicating the input vector into a large number of DPUs significantly increases the \texttt{load} data transfer costs. Thus, best performance is achieved using 253 DPUs on average across all matrices. In the 2D-partitioned kernels (i.e.,  \texttt{DCOO} and  \texttt{DBCOO}), creating \equallySized{} 2D tiles leads to a large disparity in non-zero element count across tiles, causing many tiles to be empty, i.e., without \textit{any} non-zero element. Thus, best performance is achieved using 1329 DPUs on average across all matrices, since DPUs associated with empty tiles are idle.

\begin{tcolorbox}
\noindent\textbf{OBSERVATION 17:} \\
Expensive data transfers to PIM-enabled memory performed via the narrow memory bus impose significant performance overhead to end-to-end \spmv{} execution. Thus, it is hard to fully exploit all available PIM cores of the system.
\end{tcolorbox}

Second, we observe that in regular matrices, the 2D-partitioned kernels outperform the 1D-partitioned kernels by 1.45$\times$ on average. This is because the 2D-partitioned kernels use a larger number of DPUs, and thus their \texttt{kernel} times are lower. In contrast, in scale-free matrices, the 1D-partitioned kernels outperform the 2D-partitioned kernels by 1.41$\times$ on average. This because the \equallySized{} 2D technique significantly increases the non-zero element disparity across DPUs, i.e., \texttt{kernel} time is bottlenecked by only one DPU or a few DPUs that process a much larger number of non-zero elements compared to the rest.

\begin{tcolorbox}
\noindent\textbf{OBSERVATION 18:} \\
In \textit{regular} matrices, 2D-partitioned kernels outperform 1D-partitioned kernels, since the former provide a better trade-off between computation and data transfer overheads. In contrast, in \textit{scale-free} matrices, 2D-partitioned kernels perform worse than 1D-partitioned kernels, since the former's performance is limited by one DPU or a few DPUs that process the largest number of non-zero elements.
\end{tcolorbox}

\section{Comparison with CPUs and GPUs}\label{cpu-gpu}

We compare \spmv{} execution on the UPMEM PIM architecture to a state-of-the-art CPU and a state-of-the-art GPU in terms of performance and energy consumption. Our goal is to quantify the potential of the UPMEM PIM architecture on the widely used memory-bound \spmv{} kernel.

We compare the UPMEM PIM system with \dpuActive{} to an Intel Xeon CPU~\cite{intel4110} and an NVIDIA Tesla V100 GPU~\cite{nvidiaTeslaV100}, the characteristics of which are shown in Table~\ref{tab:cpu-gpu}. We use peakperf~\cite{peak-perf} and stream~\cite{stream} for CPU and GPU systems to calculate the peak performance, memory bandwidth, and Thermal Design Power (TDP). For the UPMEM PIM system, we estimate the peak performance as $Total\_DPUs * AT$, where the arithmetic throughput (AT) is calculated for the multiplication operation in Appendix~\ref{sec:appendix-1DPU-AT} (Figure~\ref{fig:1DPU-ai-cloud4}), the total bandwidth as $Total\_DPUs * Bandwidth\_DPU$, where the $Bandwidth\_DPU$ is 700 MB/s~\cite{Gomez2021Benchmarking,Gomez2021Analysis,devaux2019}, and TDP as $(Total\_DPUs / DPUs\_per\_chip) * 1.2W/chip$ from prior work~\cite{Gomez2021Benchmarking,Gomez2021Analysis,devaux2019}.

\begin{table}[H]
\begin{center}
\centering
\resizebox{1.0\linewidth}{!}{
\begin{tabular}{|l||c|c|c|c|c|c|c|}
    \hline
    \cellcolor{gray!15} & \cellcolor{gray!15}\raisebox{-0.20\height}{\textbf{Process}} & \cellcolor{gray!15} & \cellcolor{gray!15} & \cellcolor{gray!15}\raisebox{-0.20\height}{\textbf{Peak}} & \cellcolor{gray!15}\raisebox{-0.20\height}{\textbf{Memory}} & \cellcolor{gray!15}\raisebox{-0.20\height}{\textbf{Total}} & \cellcolor{gray!15} \\
     \multirow{-2}{*}{\cellcolor{gray!15}\textbf{System}} & \cellcolor{gray!15}\textbf{Node} & \multirow{-2}{*}{\cellcolor{gray!15}\textbf{Total Cores}} &  \multirow{-2}{*}{\cellcolor{gray!15}\textbf{Frequency}} & \cellcolor{gray!15}\textbf{Performance} & \cellcolor{gray!15}\textbf{Capacity} & \cellcolor{gray!15}\textbf{Bandwidth} & \multirow{-2}{*}{\cellcolor{gray!15}\textbf{TDP}} \\
    \hline \hline
    Intel Xeon 4110 CPU~\cite{intel4110} & 14 nm & 2x8 x86 cores (2x16 threads) & 2.1 GHz & 660 GFLOPS & 128 GB & 23.1 GB/s  & 2x85 W \\ \hline 
    NVIDIA Tesla V100~\cite{nvidiaTeslaV100} & 12 nm & 5120 CUDA cores & 1.25 GHz & 14.13 TFLOPS & 32 GB & 897 GB/s & 300 W \\ \hline 
    PIM System & 2x nm & \dpuActive{} & 350 MHz & 4.66 GFLOPS & 159 GB & 1.77 TB/s & 379 W \\ \hline 

   % \bottomrule
\end{tabular}
}
\end{center}
\vspace{-6pt}
\caption{Evaluated CPU, GPU, and UPMEM PIM Systems.}
\label{tab:cpu-gpu}
\vspace{-8pt}
\end{table}

\subsection{Performance Comparison}

For the CPU system, we use the optimized CSR kernel from the TACO library~\cite{Kjolstad2017Taco}. For the GPU system, we use the CSR5 CUDA~\cite{Weifeng2015CSR5,CSR5-cuda} for the int32 data type and cuSparse~\cite{cuSparse} for the other data types. For the UPMEM PIM system, we use the lock-free COO 1D-partitioned kernel (\texttt{\textbf{COO}.nnz-lf}) and the \equallySized{} COO 2D-partitioned kernel (\texttt{\textbf{DCOO}}). In the former, we run experiments from 64 to \dpuActive{}, and in the latter, we use \dpuActive{}, and vary the number of vertical partitions from 2 to 32, in steps of multiple of 2. In both schemes, we select the best-performing end-to-end execution throughput. We also include the lock-free COO 1D-partitioned kernel using \dpuActive{}, named \texttt{\textbf{COO.kl}}, to evaluate \spmv{} execution using \textit{all} available DPUs of the system.

\begin{figure}[t]
    \begin{minipage}{1.0\textwidth}
    \centering
    \includegraphics[width=.44\textwidth]{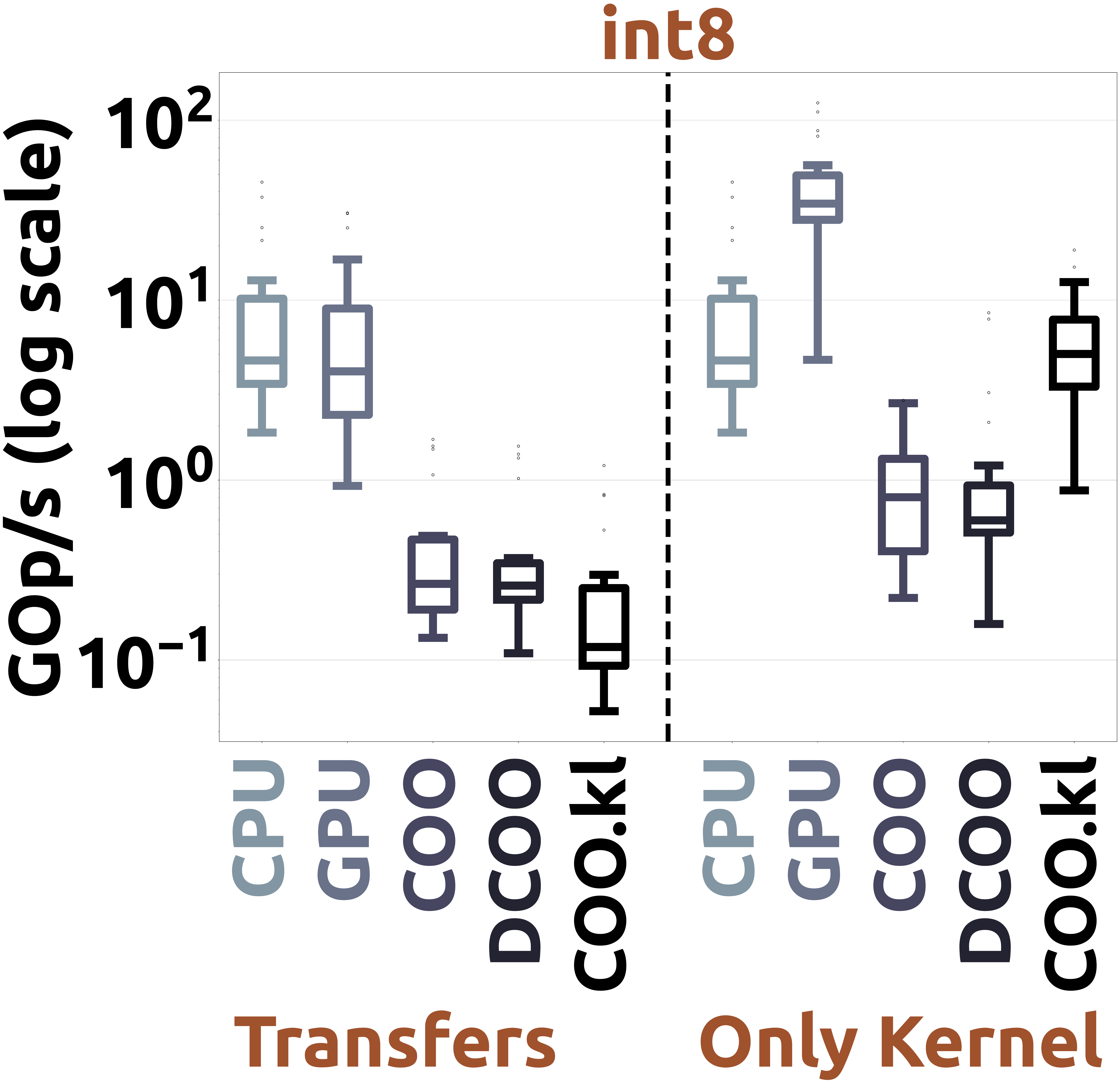}\hspace{16pt}
    \includegraphics[width=.44\textwidth]{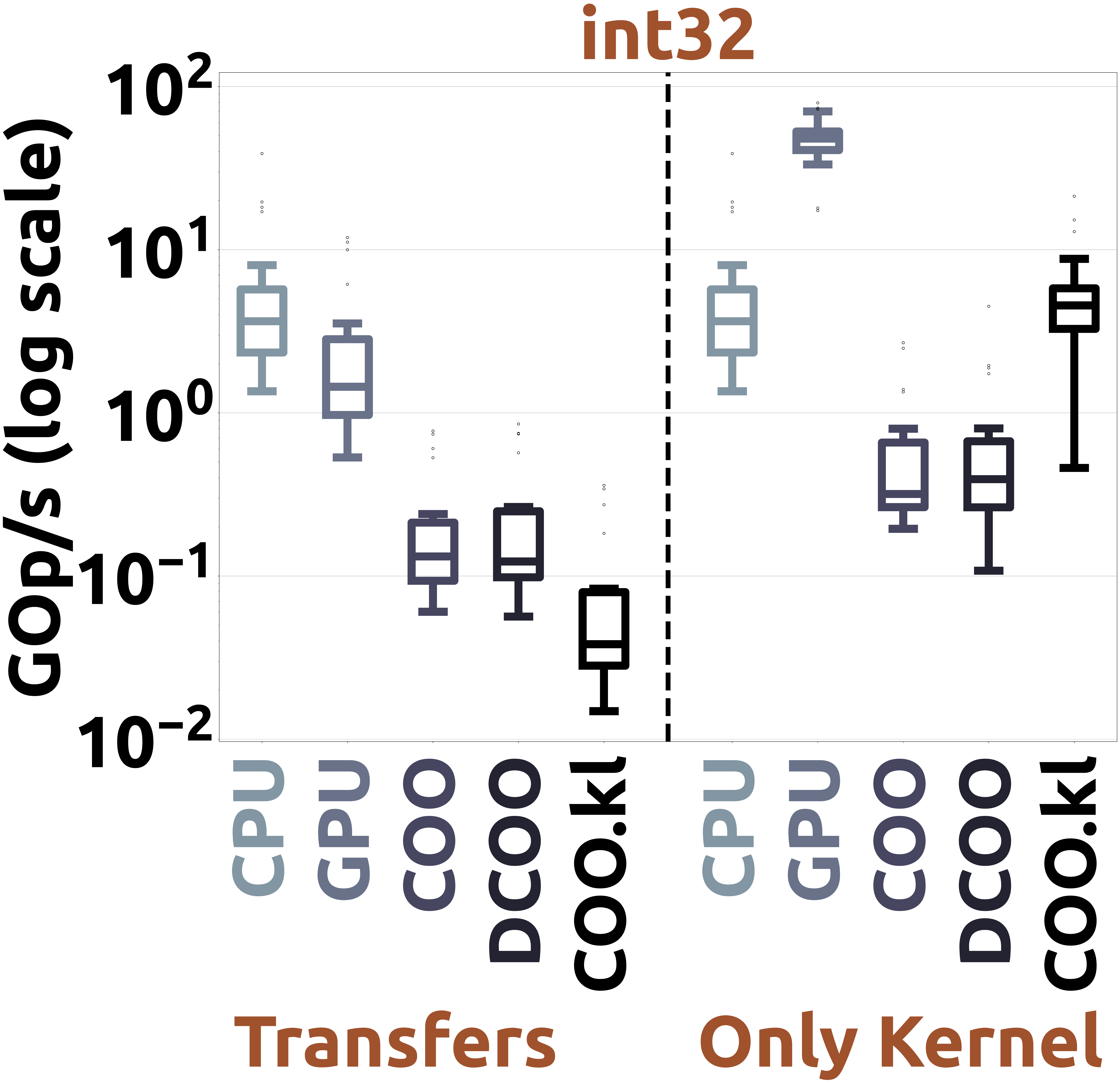}
    \end{minipage}\vspace{18pt}
    \begin{minipage}{1.0\textwidth}
    \centering
    \includegraphics[width=.44\textwidth]{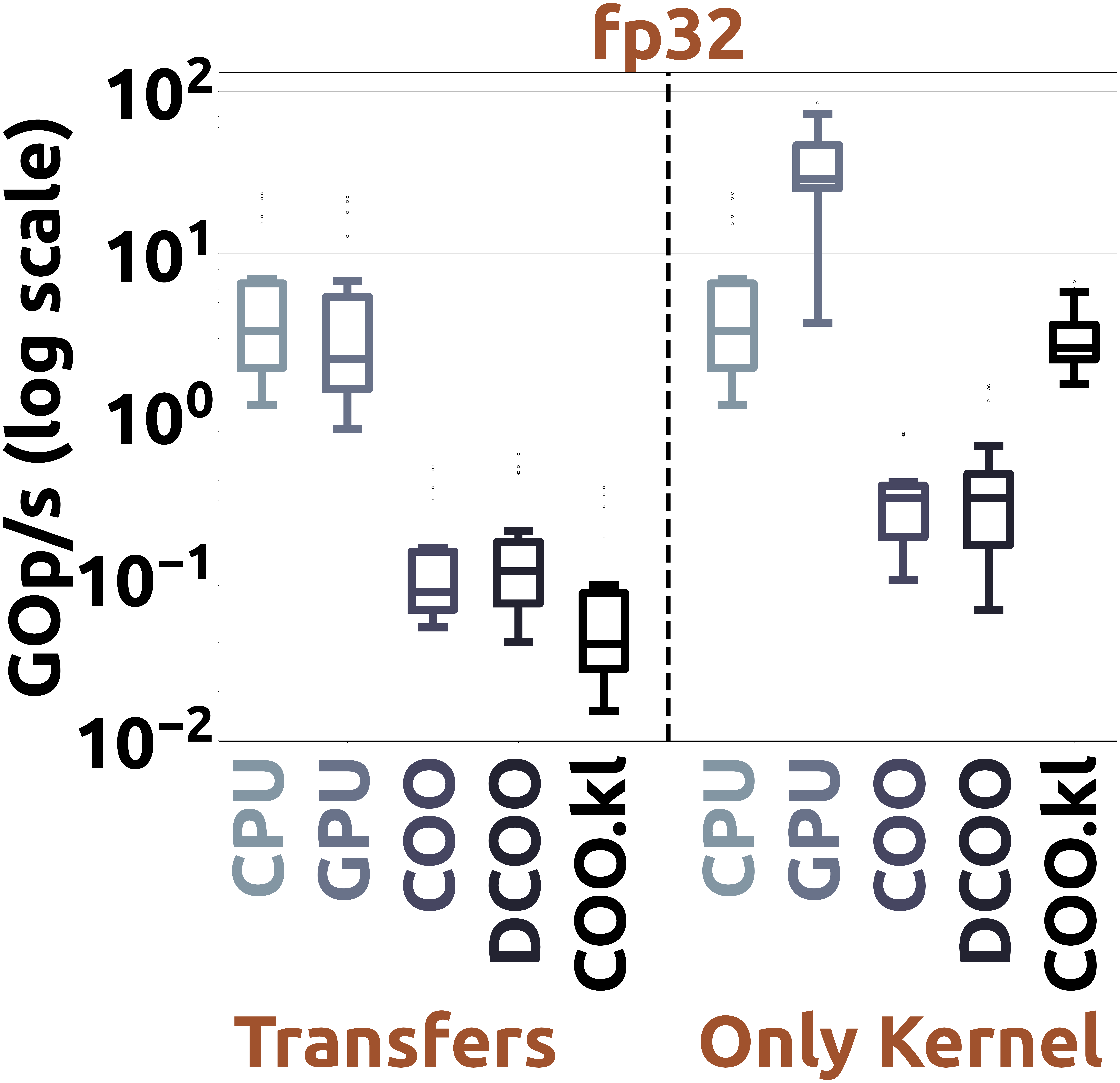}\hspace{16pt}
    \includegraphics[width=.44\textwidth]{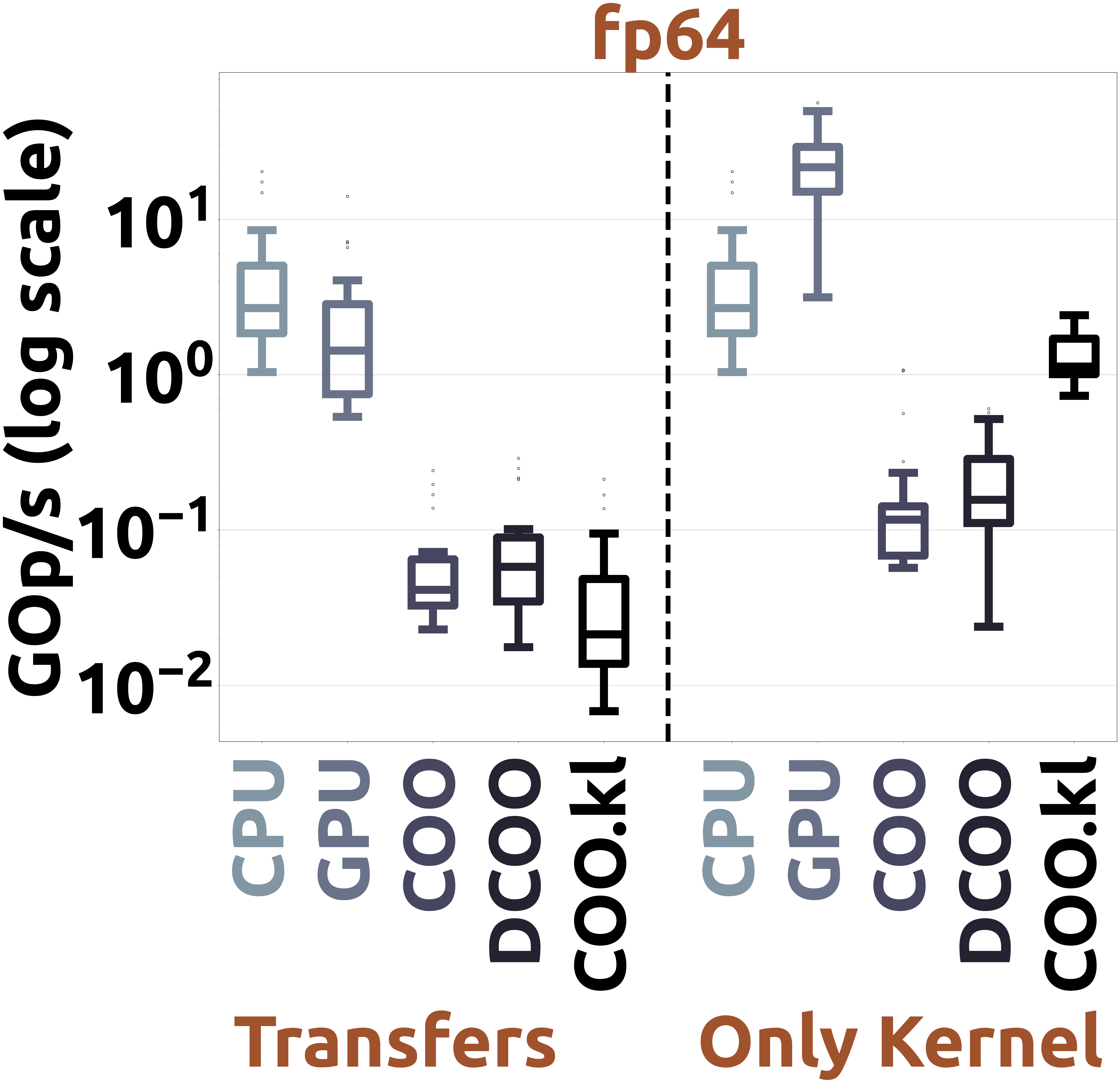}
    \end{minipage}
    \vspace{-6pt}
    \caption{Performance comparison between the UPMEM PIM system, Intel Xeon CPU and Tesla V100 GPU on \spmv{} execution.}
    \label{fig:cpu-gpu-end_to_end}
\end{figure}

Figure~\ref{fig:cpu-gpu-end_to_end} shows the throughput of \spmv{} (in GOperations per second) in all systems, comparing both the end-to-end execution throughput (i.e., including the \texttt{load} and \texttt{retrieve} data transfer costs for the input and output vectors in case of the UPMEM PIM and GPU systems), and only the actual kernel throughput (i.e., including the \texttt{kernel} time in DPUs and the \texttt{merge} time in host CPU for the UPMEM PIM system).

We draw three conclusions. First, when data transfer costs to/from host CPU are included, CPU outperforms both the GPU and UPMEM PIM systems, since data transfers impose high overhead. When only the actual kernel time is considered, GPU performs best, since it is the system that provides the highest computation throughput, e.g., 14.13 TFlops for the fp32 data type. Second, we evaluate the portion of the machine's peak performance achieved on \spmv{} in all systems, and observe that \spmv{} execution on the UPMEM PIM system achieves a much higher fraction of the peak performance compared to CPU and GPU systems. For the fp32 data type, \spmv{} achieves on average 0.51\% and 0.21\% of the peak performance in CPU and GPU, respectively, while it achieves 51.7\% of the peak performance in the UPMEM PIM system using the \texttt{COO.kl} scheme. Achieving a high portion of machine's peak performance is highly desirable, since the software highly exploits the computation capabilities of the underlying hardware. This way, it improves the processor/resource utilization, and the cost of ownership of the underlying hardware. Third, we observe that when all DPUs are used, as in \texttt{COO.kl}, \spmv{} execution on the UPMEM PIM outperforms \spmv{} execution on the CPU by 1.09$\times$ and 1.25$\times$ for the int8 and int32 data types, respectively, the multiplication of which is supported by hardware. In contrast, \spmv{} execution on the UPMEM PIM performs 1.27$\times$ and 2.39$\times$ worse than \spmv{} execution on the CPU for the fp32 and fp64 data types, the multiplication of which is software emulated in the DPUs of the UPMEM PIM system.

\begin{tcolorbox}
\noindent\textbf{OBSERVATION 19:} \\
\spmv{} execution can achieve a \textit{significantly higher} fraction of the peak performance on real memory-centric PIM architectures compared to that on processor-centric CPU and GPU systems, since PIM architectures greatly mitigate data movement costs.
\end{tcolorbox}

\subsection{Energy Comparison}

\begin{figure}[t]
    %\vspace{-2pt}
    \begin{minipage}{1.0\textwidth}
    \centering
    \includegraphics[width=.45\textwidth]{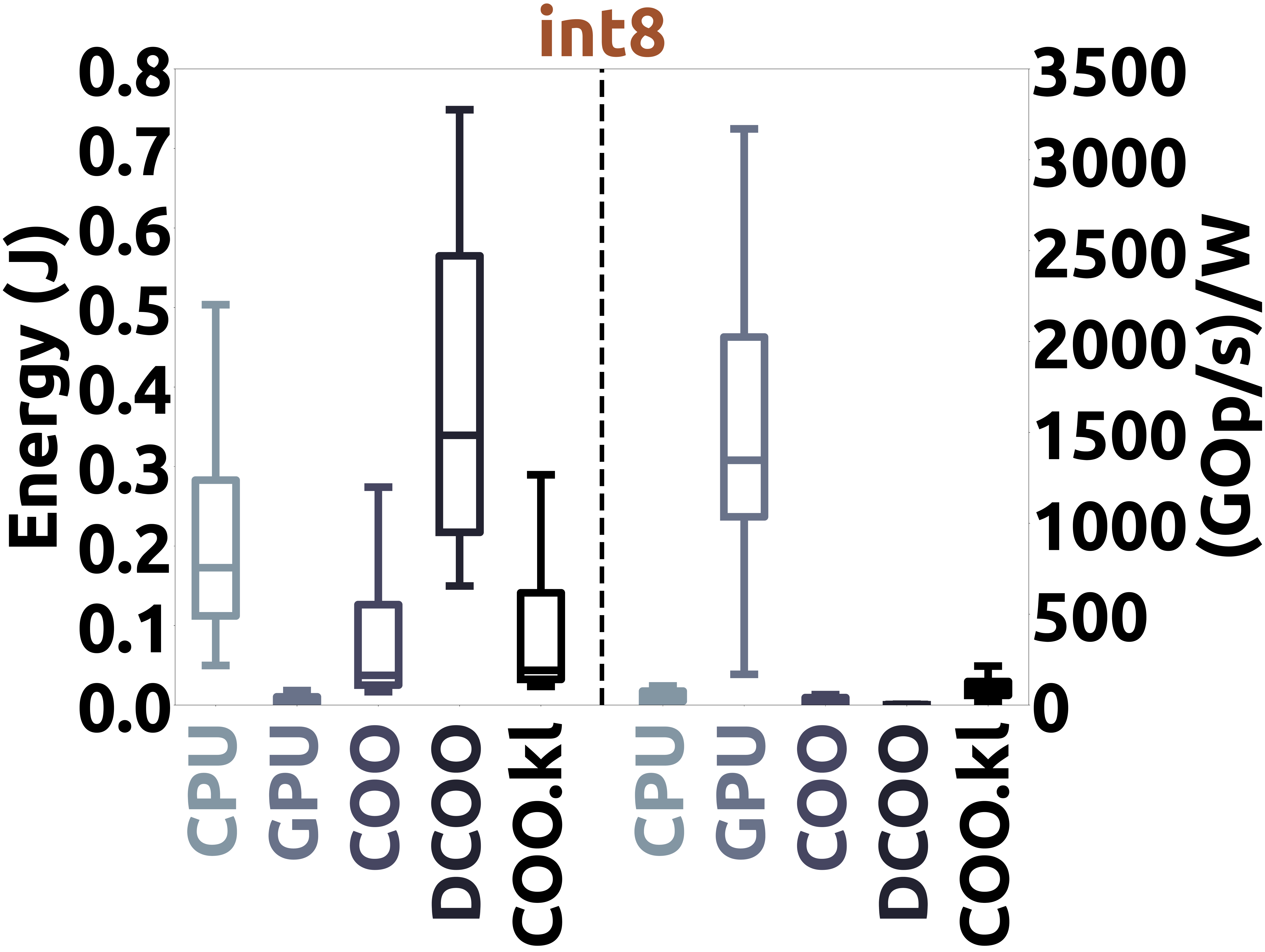}\hspace{16pt}
    \includegraphics[width=.45\textwidth]{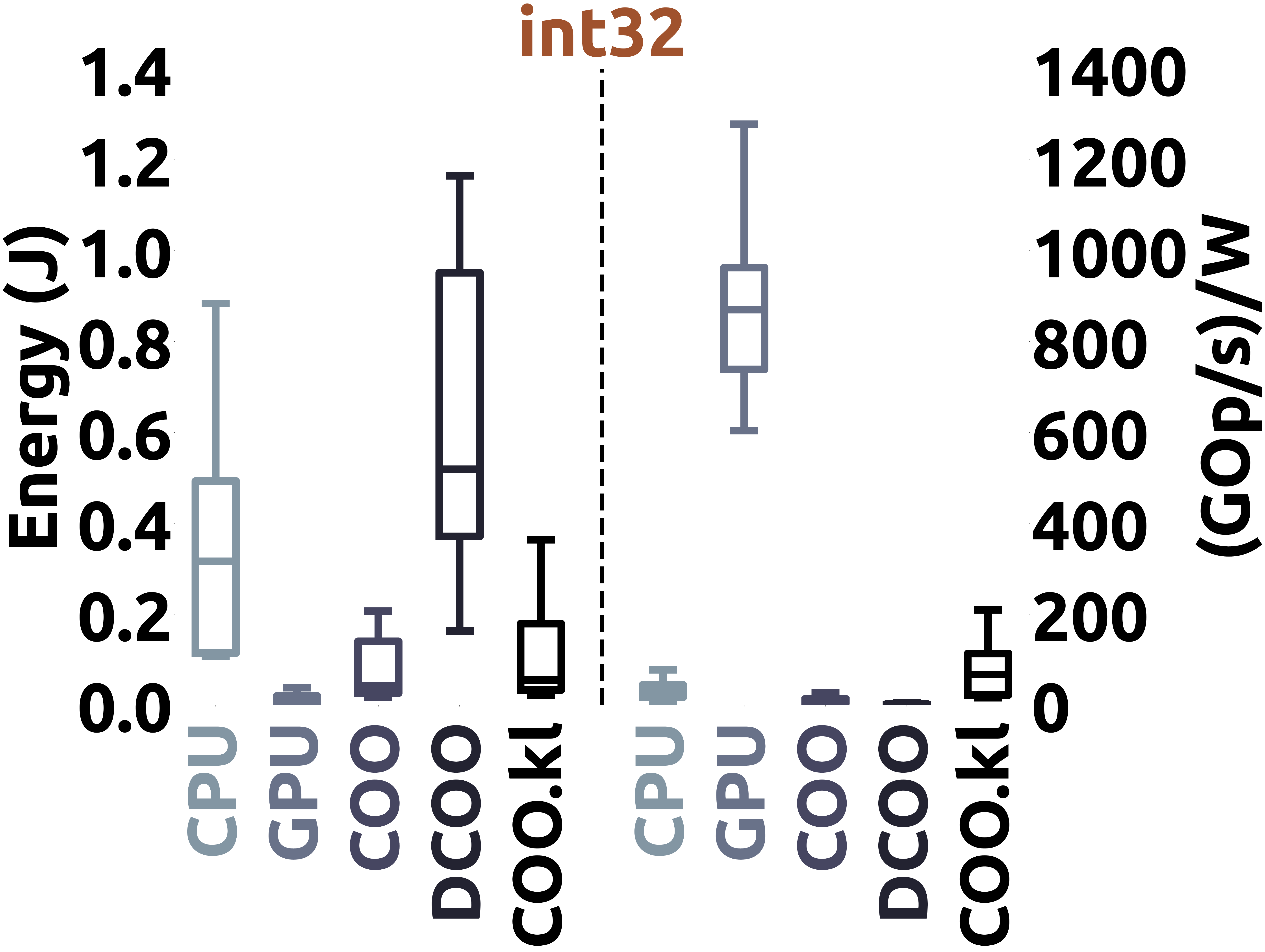}
    \end{minipage} 
    \begin{minipage}{1.0\textwidth}
    \vspace{10pt}
    \centering
    \includegraphics[width=.45\textwidth]{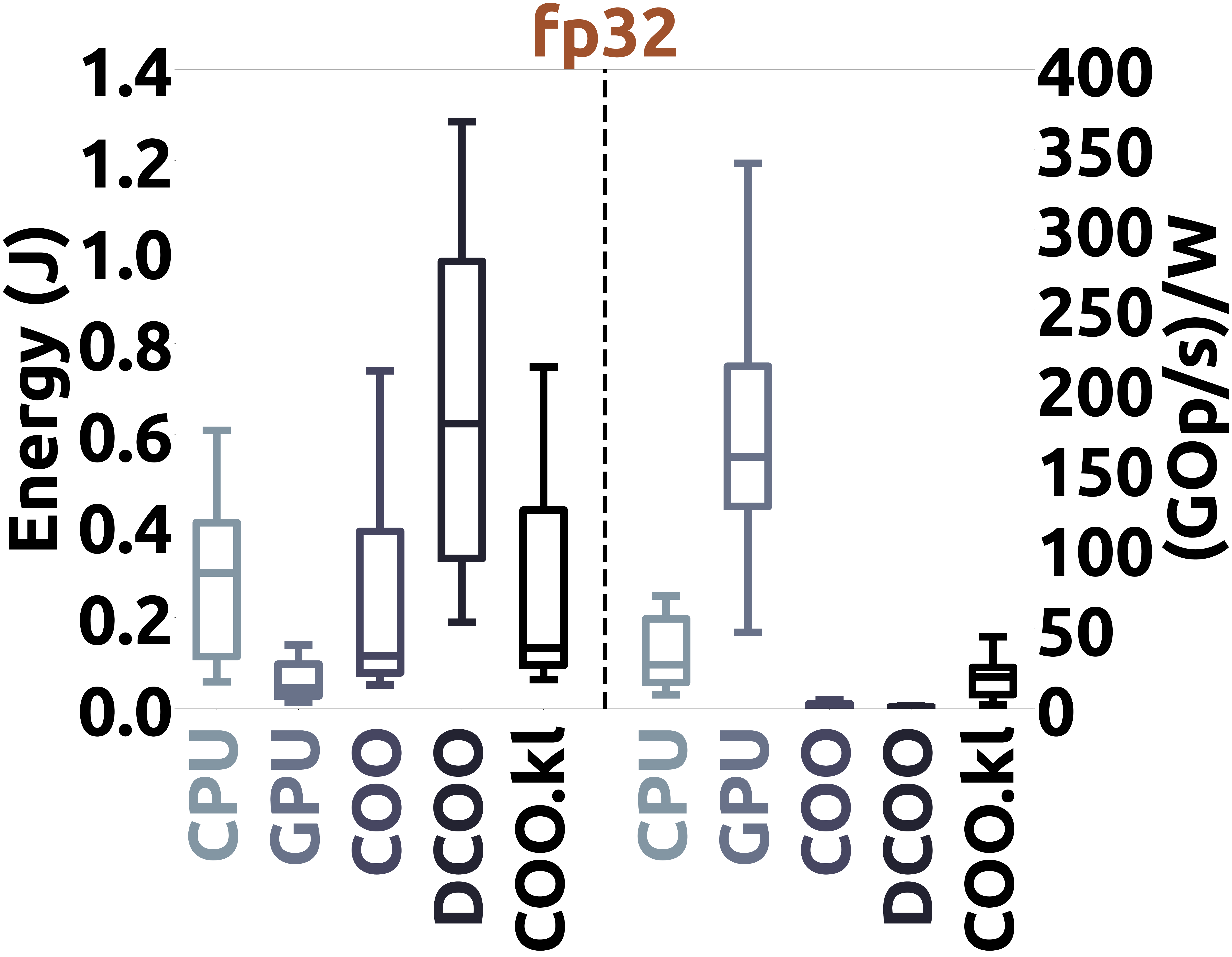}\hspace{16pt}
    \includegraphics[width=.45\textwidth]{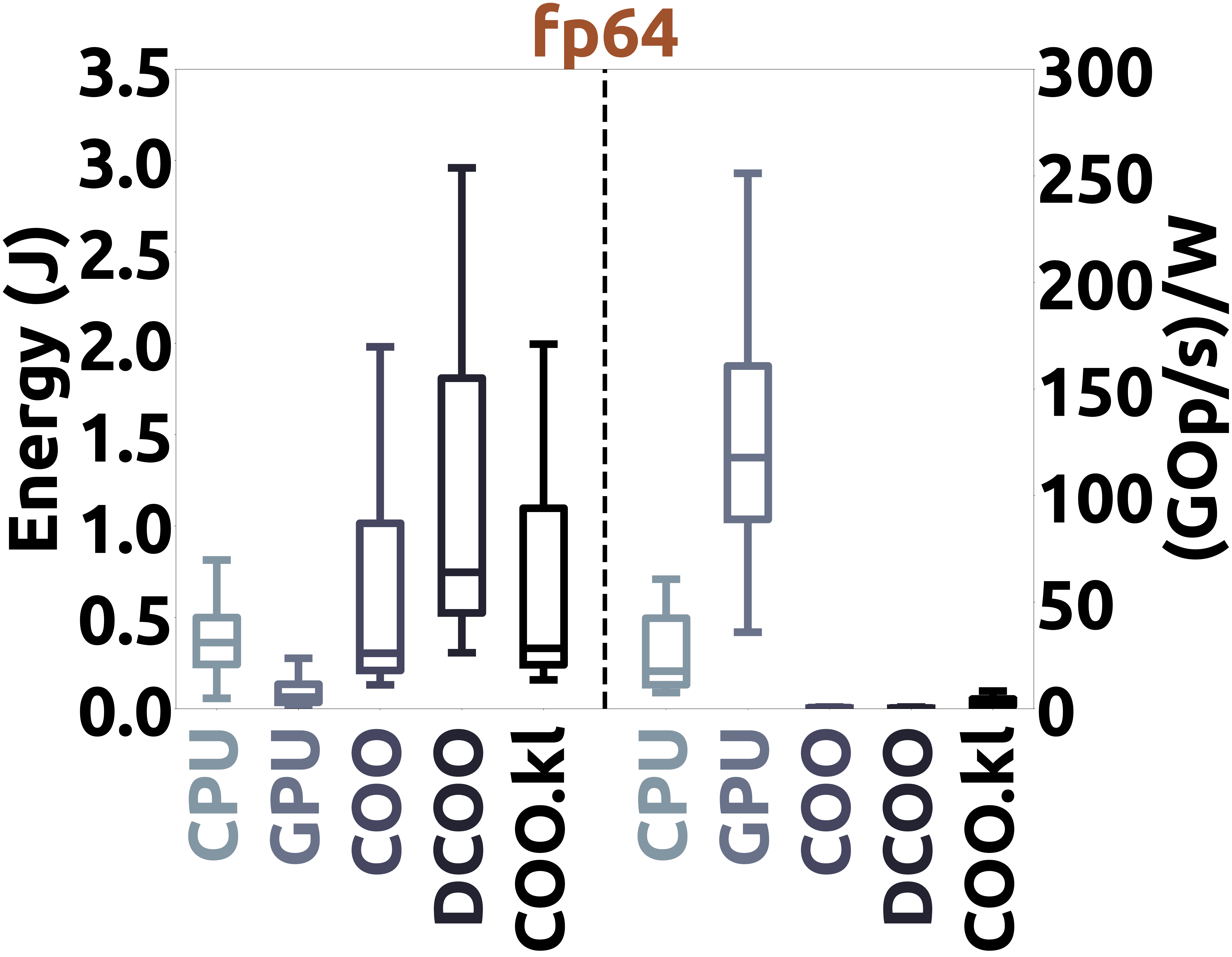}
    \end{minipage}
    \vspace{-4pt}
    \caption{Energy comparison between the UPMEM PIM system, Intel Xeon CPU and Tesla V100 GPU on \spmv{} execution.}
    \label{fig:cpu-gpu-energy}
    \vspace{-14pt}
\end{figure}

For energy measurements, we consider only the actual kernel time in all systems (in the UPMEM PIM we consider the \texttt{kernel} and \texttt{merge} steps of \spmv{} execution). We use Intel RAPL~\cite{rapl} on the CPU, and NVIDIA SMI~\cite{smi} on the GPU. For the UPMEM PIM system, we measure the number of cycles, instructions, WRAM accesses and MRAM accesses of each DPU, and estimate energy with energy weights provided by the UPMEM company~\cite{upmem}. Figure~\ref{fig:cpu-gpu-energy} shows the energy consumption (in Joules) and performance per energy (in (GOp/s)/W) for all systems.

We draw three findings. First, GPU provides the lowest energy on \spmv{} over the other two systems, since the energy results typically follow the performance results. Second, we find that the 2D-partitioned kernel, i.e., \texttt{DCOO}, consumes more energy than the 1D-partitioned kernels, i.e., \texttt{COO} and \texttt{COO.kl}, due to the energy consumed in the host CPU cores. CPU cores merge a large number of partial results in the 2D-partitioned kernels to assemble the final output vector, thereby increasing the energy consumption. Finally, we find that the 1D-partitioned kernels provide better energy efficiency on \spmv{} over the CPU system, when the multiplication operation is supported by hardware. Specifically, 1D-partitioned kernels provide 3.16$\times$ and 4.52$\times$ less energy consumption, and 1.74$\times$ and 1.14$\times$ better performance per energy over the CPU system for the int8 and int32 data types, respectively.

\begin{tcolorbox}
\noindent\textbf{OBSERVATION 20:} \\
Real PIM architectures can provide high energy efficiency on \spmv{} execution.
\end{tcolorbox}

\vspace{-8pt}
\subsection{Discussion}
\vspace{-2pt}
These evaluations are useful for programmers to anticipate how much performance and energy savings memory-centric PIM systems can provide on \spmv{} over commodity processor-centric CPU and GPU systems. However, our evaluated \spmv{} kernels do not constitute the best-performing approaches for \textit{all} matrices. Designing methods to select the best-performing \spmv{} parallelization scheme depending on the particular characteristics of the input matrix would further improve performance and energy savings of \spmv{} execution on memory-centric PIM systems. Moreover, the UPMEM PIM hardware is still maturing and is expected to run at a higher frequency in the near future (500 MHz instead of 350 MHz)~\cite{upmem,Gomez2021Benchmarking}. Hence, \spmv{} execution on the UPMEM PIM architecture might achieve even higher performance and energy benefits over the results we report in this comparison. Finally, note that our proposed \SparseP{} kernels can be adapted and evaluated on other current and future real PIM systems with potentially higher computation capabilities and energy efficiency than the UPMEM PIM system.

\vspace{-8pt}
\section{Key Takeaways and Recommendations}\label{recommendations}

This section summarizes our key takeaways in the form of recommendations to improve multiple aspects of PIM hardware and software.

\noindent\textbf{\textit{Recommendation \#1.}} \textit{Design algorithms that provide high load balance across threads of a PIM core in terms of computations, loop control iterations, synchronization points and memory accesses.} 
Section~\ref{1DPU} shows that in matrices and formats where the parallelization scheme used causes \textit{high disparity} in the non-zero elements/blocks/rows processed across threads of a PIM core, or the number of lock acquisitions/lock releases/DRAM memory accesses performed across threads,  \spmv{} performance severely degrades in compute-bound DPUs~\cite{Gomez2021Analysis,Gomez2021Benchmarking}. Therefore, from a programmer’s perspective, providing high operation balance across parallel threads is of vital importance in low-area and low-power PIM cores with relatively low computation capabilities~\cite{Gomez2021Benchmarking,Gomez2021Analysis}.

\noindent\textbf{\textit{Recommendation \#2.}} \textit{Design compressed data structures that can be effectively partitioned across DRAM banks, with the goal of providing high computation balance across} PIM cores. Sections~\ref{1D-Kernel} and ~\ref{2D-Formats} demonstrate that (i) the compressed matrix format used to store the input matrix determines the data partitioning across DRAM banks of PIM-enabled memory, and (ii) \spmv{} execution using the CSR and BCSR formats performs significantly worse than \spmv{} execution using the COO and BCOO formats. This is because the matrix is stored in row- or block-row-order for the CSR and BCSR formats, respectively, and thus data partitioning across DRAM banks is limited to be performed at row or block-row granularity, respectively, leading to high non-zero element imbalance across PIM cores. Therefore, we recommend that programmers design compressed data structures that can provide effective data partitioning schemes with high computation balance across thousands of PIM cores.

\noindent\textbf{\textit{Recommendation \#3.}} \textit{Design \textit{adaptive} algorithms that (i) trade off computation balance across PIM cores for lower data transfer costs to PIM-enabled memory, and (ii) adapt their configuration to the particular patterns of each input given, as well as the characteristics of the PIM hardware.} Our analysis in Sections~\ref{1D-Kernel}, ~\ref{2D-Studies} and ~\ref{2D-Comparison} demonstrates that the best-performing \spmv{} execution on the UPMEM PIM system can be achieved using algorithms that (i) trade off computation for lower data transfer costs, and (ii) select the load balancing strategy and data partitioning policy based on the particular sparsity pattern of the input matrix. In addition, the performance of each balancing scheme and data partitioning technique for \spmv{} execution highly depends on the characteristics of the underlying PIM hardware, as we explain in Section~\ref{2D-Studies} and Appendix~\ref{sec:appendix-2D-vertpartitions}. To this end, we recommend that software designers implement heuristics and selection methods for their algorithms to adapt their configuration to the underlying hardware characteristics of the PIM system and the input data given.

\noindent\textbf{\textit{Recommendation \#4.}} \textit{Provide low-cost synchronization support and hardware support to enable concurrent memory accesses by multiple threads to the local DRAM bank to increase parallelism in a multithreaded PIM core.} Section~\ref{1DPU} shows that (i) lock acquisitions/releases can cause high overheads in the DPU pipeline, and (ii) fine-grained locking approaches to increase parallelism in critical sections do not improve performance over coarse-grained approaches in the UPMEM PIM hardware. This is because the DMA engine of the DPU serializes DRAM memory accesses included in the critical sections. Based on these key takeaways, we recommend that hardware designers provide lightweight synchronization mechanisms for multithreaded PIM cores~\cite{Giannoula2021SynCron}, and enable concurrent access to local DRAM memory arrays to increase execution parallelism. For example, sub-array level parallelism~\cite{Kim2012Case,Chang2014Improving} or multiple DRAM banks per PIM core could be supported in the PIM hardware to improve parallelism.

\noindent\textbf{\textit{Recommendation \#5.}} \textit{Optimize the broadcast collective operation in data transfers from main memory to PIM-enabled memory to minimize overheads of copying the input data into all DRAM banks in the PIM} system. Figures~\ref{fig:1D_transfers} and ~\ref{fig:1D_transfers_scalability} show that \spmv{} execution using the 1D partitioning technique cannot scale up to a large number of PIM cores. This is because it is severely limited by data transfer costs to broadcast the input vector into \textit{each} DRAM bank of PIM-enabled DIMMs via the narrow off-chip memory bus. To this end, we suggest that hardware and system designers provide a fast broadcast collective primitive to DRAM banks of PIM-enabled memory modules~\cite{Sun2021ABCDIMMAT}.

\noindent\textbf{\textit{Recommendation \#6.}} \textit{Optimize the gather collective operation \textit{at DRAM bank granularity} for data transfers from PIM-enabled memory to the host CPU to minimize overheads of retrieving the output results.} Figures~\ref{fig:2D_scale_equally_wide}, ~\ref{fig:2D_scale_variable_sized} and ~\ref{fig:2D_vertpartitions} demonstrate that \spmv{} execution using the \equallyWidth{} and \variableSized{} 2D partitioning schemes is severely limited by data transfers to retrieve results for the output vector from DRAM banks of PIM-enabled DIMMs. This is due to two reasons: (i) 2D-partitioned kernels create a large number of partial results that need to be transferred from PIM-enabled memory to the host CPU via the narrow memory bus in order to assemble the final output vector, and (ii) the UPMEM PIM system has the limitation that the transfer sizes from/to all DRAM banks involved in the same parallel transfer need to be the same, and therefore a large amount of padding with empty bytes is performed in the \equallyWidth{} and \variableSized{} schemes. To this end, we suggest that hardware and system designers provide an optimized \textit{gather} primitive to efficiently collect results from multiple DRAM banks to host CPU~\cite{Sun2021ABCDIMMAT}, and support parallel fine-grained data transfers from PIM-enabled memory to host CPU \textit{at DRAM bank granularity} to avoid padding with empty bytes.

\noindent\textbf{\textit{Recommendation \#7.}} \textit{Design high-speed communication channels and optimized libraries for data transfers to/from thousands of DRAM banks of PIM-enabled memory.} Section~\ref{cpu-gpu} demonstrates that \spmv{} execution on the memory-centric UPMEM PIM system achieves a much higher fraction of the machine's peak performance (on average 51.7\% for the 32-bit float data type), compared to that on processor-centric CPU and GPU systems. However, the end-to-end performance of both 1D- and 2D-partitioned kernels is significantly limited by data transfer overheads on the narrow memory bus. To this end, we recommend that the hardware architecture and the software stack of real PIM systems be enhanced with low-cost and fast data transfers to/from PIM-enabled memory modules, and/or with support for efficient direct communication among PIM cores~\cite{Seshadri2013RowClone,Chang2016LISA,Rezaei2020NoM,Wang2020Figaro,Seshadri2017Ambit,Seshadri2017Simple}.

\section{Related Work}

To our knowledge, this is the first work that (i) extensively characterizes the Sparse Matrix Vector Multiplication (\spmv{}) kernel in a real PIM system, and (ii) presents an open-source \spmv{} library for real-world PIM systems. We briefly discuss closely related prior work.

\noindent\textbf{Processing-In-Memory (\emph{PIM}).}
A large body of prior work examines Processing-Near-Memory (\emph{PNM})~\cite{Mutlu2019Enabling,Mutlu2019Processing,boroumand2017lazypim,Boroumand2019Conda,Boroumand2018Google,Giannoula2021SynCron,fernandez2020natsa,Alser2020Accelerating,Kim2017GrimFilter,Ahn2015PIMenabled,Cali2020GenASM,Singh2019Napel,Singh2020NEROAN,ahn2015scalable,Hadi2016Chameleon,hashemi2016accelerating,Farmahini2015NDA,Gao2015Practical,Gao2016HRL,hashemi2016continuous,Hsieh2016accelerating,Hsieh2016TOM,choe2019concurrent,liu2017concurrent,Gu2020iPIM,ke2019recnmp,Kwon2019TensorDIMM,huangfu2019medal,Kim2016Neurocube,Nai2017GraphPIM,Zhang2018GraphP,Youwei2019GraphQ,Drumond2017mondrian,pugsley2014ndc,Zhang2014TOPPIM,Zhu2013Accelerating,upmem,Lee2021HardwareAA,gao2017tetris,Dai2018GraphH,Nair2015Active,Alves2015Opportunities,Nag2021OrderLight,Park2021Trim,Sadredini2021Sunder,Gu2021DLUX,Oliveira2021Damov,Lockerman2020Livia,Sun2021ABCDIMMAT,olgun2021pidram}. PNM integrates processing units near or inside the memory via a 3D PNM configuration (i.e., processing units are located at the logic layer of 3D-stacked memories)~\cite{boroumand2017lazypim,Boroumand2019Conda,Boroumand2018Google,Nai2017GraphPIM,Zhang2018GraphP,ahn2015scalable,Gao2015Practical,liu2017concurrent,choe2019concurrent,Youwei2019GraphQ,Drumond2017mondrian,Zhang2014TOPPIM,Nair2015Active,pugsley2014ndc}, a 2.5D PNM configuration (i.e., processing units are located in the same package as the CPU connected via silicon interposers)~\cite{Singh2020NEROAN,fernandez2020natsa,Giannoula2021SynCron}, a 2D PNM configuration (i.e., processing units are placed inside DDRX DIMMs)~\cite{Hadi2016Chameleon,Gu2020iPIM,Alves2015Opportunities,Nag2021OrderLight,Park2021Trim,Sadredini2021Sunder,Gu2021DLUX,ke2019recnmp,Kwon2019TensorDIMM,Nider2020Processing,Cho2021Accelerating,Zois2018Massively,Lavenier2016DNA,lavenier2020Variant}, or at the memory controller of CPU systems~\cite{hashemi2016accelerating,hashemi2016continuous,Lockerman2020Livia}. These works propose hardware designs for irregular applications like graph processing~\cite{Dai2018GraphH,Nai2017GraphPIM,Youwei2019GraphQ,ahn2015scalable,Ahn2015PIMenabled,boroumand2017lazypim,Boroumand2019Conda}, bioinformatics~\cite{Cali2020GenASM,Kim2017GrimFilter,Lavenier2016DNA,lavenier2020Variant,Giannoula2021SynCron}, neural networks~\cite{Choi2010Model,Gu2020iPIM,Boroumand2018Google,gao2017tetris,Kim2016Neurocube,Boroumand2021Google,fernandez2020natsa,Singh2020NEROAN}, pointer-chasing workloads~\cite{liu2017concurrent,choe2019concurrent,Hsieh2016accelerating,Giannoula2021SynCron}, and databases~\cite{Drumond2017mondrian}. However, \emph{none} of these works examines the \spmv{} kernel in such systems.

Several prior works enable Processing-Using-Memory (\emph{PUM})~\cite{Seshadri2017Ambit,Aga2017Compute,Eckert2018Neural,Fujiki2019Duality,Kang2014Energy,Li2016Pinatubo,Seshadri2013RowClone,Angizi2019GraphiDe,Chang2016LISA,Gao2019ComputeDRAM,Xin2020ELP2IM,li2017drisa,Deng2018DrAcc,Hajinazar2021SIMDRAM,Rezaei2020NoM,Wang2020Figaro,Ali2020InMemory,levi2014Loci,Kvatinsky2014Magic,Shafiee2016ISAAC,Kvatinsky2011Memristor,Gaillardon2016Programmable,Bhattacharjee2017ReVAMP,Hamdioui2015Memristor,Xie2015FastBL,Song2018GraphR,Ankit2020Panther,Ankit2019PUMA,Chi2016PRIME,Xi2021Memory,Zheng2016RRAM,Hamdioui2017Memristor,Yu2018Memristive,Kim2018PUF,ferreira2021pluto,Wu2021Sieve,Yuan2021FORMS}. PUM exploits the operational principles of memory cells to perform computation within the memory chip. Prior works propose PUM designs using SRAM~\cite{Aga2017Compute,Eckert2018Neural,Fujiki2019Duality,Kang2014Energy}, DRAM~\cite{Seshadri2017Ambit,Seshadri2013RowClone,Angizi2019GraphiDe,Chang2016LISA,Gao2019ComputeDRAM,Xin2020ELP2IM,li2017drisa,Deng2018DrAcc,Hajinazar2021SIMDRAM,Rezaei2020NoM,Wang2020Figaro,Ali2020InMemory,Kim2018PUF,ferreira2021pluto,Wu2021Sieve}, PCM~\cite{Li2016Pinatubo} or RRAM/memristive memory technologies~\cite{levi2014Loci,Kvatinsky2014Magic,Shafiee2016ISAAC,Kvatinsky2011Memristor,Gaillardon2016Programmable,Bhattacharjee2017ReVAMP,Hamdioui2015Memristor,Xie2015FastBL,Song2018GraphR,Ankit2020Panther,Ankit2019PUMA,Chi2016PRIME,Xi2021Memory,Zheng2016RRAM,Hamdioui2017Memristor,Yu2018Memristive,Yuan2021FORMS}. A few PUM works~\cite{Shafiee2016ISAAC,Chi2016PRIME,Aga2017Compute,Eckert2018Neural,li2017drisa,Deng2018DrAcc,Hajinazar2021SIMDRAM} enable the multiplication operation inside memory cells with the goal of performing efficient matrix vector multiplication at low cost within the memory chip. These works design hardware-based solutions to accelerate the \emph{dense} matrix vector multiplication (GEMV) kernel via PUM. However, there is \emph{no} prior work that leverages PUM to accelerate the \emph{Sparse} Matrix Vector Multiplication (\spmv{}) kernel using state-of-the-art compressed matrix storage formats.

\noindent\textbf{Sparse Matrix Kernels in PIM Systems.} 
Xie et al.~\cite{Xie2021SpaceA} design heterogenous PIM units to accelerate \spmv{} via a 3D PNM configuration, i.e., in HMC-based PIM systems. Sun et al.~\cite{Sun2021ABCDIMMAT} leverage the buffer device space of DIMM modules to add one processing unit per each DIMM module, and design low-cost inter-DIMM broadcast collectives to minimize data transfer overheads on irregular workloads, like \spmv{} and graph processing, executed in 2D PNM configurations. Zhu et al.~\cite{Zhu2013Accelerating} propose a PIM accelerator for Sparse Matrix Matrix Multiplication via a 3D PNM configuration. Fujiki et al.~\cite{Fujiki2019Near} enhance the memory controllers of GPUs with PIM cores to transform the matrix from the CSR to the DCSR format~\cite{Changwan2018Efficient} on the fly to minimize memory traffic on \spmv{} execution. These works propose hardware designs for sparse matrix kernels. In contrast, our work studies software optimizations and strategies to efficiently map compressed matrix storage formats on real near-bank PIM systems, and accelerate \spmv{} execution on such systems.

%Xie et al.~\cite{Xie2021SpaceA} design heterogenous PIM units to accelerate \spmv{} in HMC-based PIM systems. Fujiki et al.~\cite{Fujiki2019Near} enhance the memory controllers of GPUs with PIM cores to transform the matrix from CSR to DCSR format~\cite{Changwan2018Efficient} on the fly to minimize memory traffic. Zhu et al.~\cite{Zhu2013Accelerating} propose a PIM accelerator for Sparse Matrix Matrix Multiplication. These works propose hardware designs for sparse matrix kernels, while our work studies software optimizations to accelerate \spmv{} execution on real near-bank PIM systems. 

\noindent\textbf{\spmv{} in Commodity Systems.} Numerous prior works propose optimized \spmv{} algorithms for CPUs~\cite{Elafrou2018SparseX,Buluc2011Reduced,Elafrou2017PerformanceAA,Kjolstad2017Taco,Merrill2016Merge,Willcock2006Accelerating,Williams2007Optimization,Namashivayam2021Variable,Tang2015Optimizing,Elafrou2019Conflict,Vuduc2005oski,Elafrou2017PerformanceXeon,Rong2016Sparso,Xie2018CVR,Xiao2021CASpMV,Hou2017Auto,Pinar1999Improving,Liu2013Efficient,Mellor2004Optimizing,Oliker2002Effects,Vuduc2005Fast,Toledo1997Improving,Temam1992Characterizing,Aktemur2018ASM,Zhao2020Exploring}, GPUs~\cite{Bolz2003Sparse,Hong2018Efficient,Liu2014AnEfficient,Wu2010Efficient,Guo2014APerformance,su2012ClSpMV,Steinberger2017Globally,Shengen2014YaSpMV,Bell99Implementing,Choi2010Model,Pichel2012Optimization,Sun2011Optimizintg,Vazquez2011New,Yang2011Fast,Elafrou2019BASMAT,Filippone2017Sparse}, heterogeneous CPU-GPU systems~\cite{Yang2017Hybrid,Indarapu2014Architecture,Yang2015Performance,Benatia2020Sparse,Boyer2010Exact,Pichel2013Sparse,Indarapu2013Architecture,Anastasiadis2021CoCoPeLia}, and distributed CPU systems~\cite{Lee2008Adaptive,Bisseling2005Communication,Bylina2014Performance,Page2018Scalability,Kayaaslan2015Semi,Liu2018Towards,Catalyurek1999Hypergraph,Vastenhouw2005Two,Nastea1996Load,Pelt2014Medium,Grandjean2012Optimal,Boman2013Scalable}. Optimized \spmv{} kernels for processor-centric CPU and GPU systems exploit the shared memory model of these systems and data locality in deep cache hierarchies. However, these kernels  cannot be directly mapped to most near-bank PIM systems, which have a distributed memory model and a shallow cache hierarchy. Most well-tuned \spmv{} kernels for distributed CPU and CPU-GPU systems improve performance by overlapping computation with communication among processing units, and exploiting data locality in large cache memories. In contrast, real near-bank PIM architectures are fundamentally different from CPU-GPU systems, since they are \textit{highly distributed}, i.e., there is no direct communication among PIM cores, and include a shallow memory hierarchy. Therefore, \spmv{} kernels designed for common processor-centric systems cannot be directly used in near-bank PIM systems.

\noindent\textbf{Hardware Accelerators for \spmv{}.} Recent works design accelerators for \spmv{}~\cite{Sadi2019Efficient,Fowers2014AHigh,Grigoras2015Accelerating,Lin2010Design,Umuroglu2014Anenergy,Kanellopoulos2019SMASH,Mukkara2018Exploiting,Nurvitadhi2015Sparse} or other sparse kernels~\cite{Asgari2020Alrescha,Hegde2019ExTensor,Zhang2016CambriconX,Pal2018OuterSpace,Nurvitadhi2016Hardware,Eric2020sigma,Zhou2018CambriconS,Mishra2017Fine,zhang2021asplos,zhang2020sparch,parashar2017scnn,Hwang2020Centaur,qin2020sigma}. In contrast, our work proposes software optimizations and provides the first characterization study of \spmv{} on a real PIM system.

\noindent\textbf{Compressed Matrix Storage Formats.} Prior works propose a range of compressed matrix storage formats~\cite{Im1999Optimizing,Langr2016Evaluation,Liu2015CSR5,Pinar1999Improving,Vuduc2005Fast,Yang2014Optimization,Shengen2014YaSpMV,Kourtis2011CSX,Kourtis2008Optimizing,Belgin2009Pattern,LIL,ELL,bjorck1996numerical,Pooch1973Survey,Shubhabrata2007Scan,Changwan2018Efficient,Liu2013Efficient,Monakov2010Automatically,Saad1989Krylov,Buluc2009Parallel,Martone2014251,Martone2010Blas,Kreutzer2012Sparse} and selection methods to find the most efficient compressed format~\cite{su2012ClSpMV,Niu2021TileSpMV,asgari2020copernicus,Zhao2018Overhead,Sedaghati2015Automatic,Benatia2016Sparse,Zhao2018Bridging,Li2013SMAT,Maggioni2013AdELL,Tan2018Design,Li2015Performance,Benatia2018BestSF}. In this work, we extensively explore the four most widely used \emph{general} compressed matrix formats, and observe that the compressed format (i) needs to provide good balance between computation and memory accesses inside the core pipeline, and (ii) affects load balancing across PIM cores, with corresponding performance implications. Therefore, some compressed formats designed for commodity processor-centric systems might not be suitable or efficient for real PIM systems. We leave the exploration of other PIM-suitable compressed matrix storage formats for future work.

\section{Summary}

We  present \SparseP{}, the first open-source \spmv{} library for real Processing-In-Memory (PIM) systems, and conduct the first comprehensive characterization analysis of the widely used \spmv{} kernel on a real-world PIM architecture. 

First, we design efficient \spmv{} kernels for real PIM systems. Our proposed \SparseP{} software package supports (1) a wide range of data types, (2) two types of well-crafted data partitioning techniques of the sparse matrix to DRAM banks of PIM-enabled memory, (3) the most popular compressed matrix formats, (4) a wide variety of load balancing schemes across PIM cores, (5) several load balancing schemes across threads of a multithreaded PIM core, and (6) three synchronization approaches among threads within PIM core. 

Second, we conduct an extensive characterization study of \SparseP{} kernels on the state-of-the-art UPMEM PIM system. We analyze \spmv{} execution on one single multithreaded PIM core and thousands of PIM cores using 26 sparse matrices with diverse sparsity patterns. We also compare the performance and energy consumption of \spmv{} on the UPMEM PIM system with those of state-of-the-art CPU and GPU systems to quantify the potential of a real memory-centric PIM architecture on the widely used \spmv{} kernel over conventional processor-centric architectures. Our analysis of \SparseP{} kernels provides programming recommendations for software designers, as well as suggestions and hints for hardware and system designers of future PIM systems. 

We believe and hope that our work will provide valuable insights to programmers in the development of efficient sparse linear algebra kernels and other irregular kernels from different application domains tailored for real PIM systems, as well as to architects and system designers in the development of future memory-centric computing systems.

\chapter{Conclusions and Future Directions}\label{FutureWorkChapter}

The goal of this dissertation is to significantly improve performance and efficiency of important irregular applications in modern processor-centric CPU and memory-centric NDP/PIM systems. To this end, we develop low-overhead synchronization and well-crafted data access approaches for emerging irregular applications including graph processing kernels, pointer-chasing, data analytics, and sparse linear algebra.

First, we comprehensively analyze prior state-of-the-art algorithms for the widely used graph coloring kernel, and we find that they are still inefficient, since they access application data from the last levels of the memory hierarchy (e.g., main memory) of commodity CPU architectures. Therefore, we introduce the \ColorTM{} parallel algorithm, which provides highly efficient execution of the graph coloring kernel. \ColorTM{} (i) accesses application data by leveraging the low-cost on-chip cache memories of CPU systems to minimize data access costs, and (ii) executes short and small critical sections by performing many computations and data accesses \emph{outside} the critical section to minimize synchronization overheads and increase the levels of parallelism among parallel threads. We also extend our proposed design to introduce a highly efficient \emph{balanced} graph coloring algorithm (\BalColorTM{}) that can provide high load balance and high resource utilization in the real-world end-applications of graph coloring. Our evaluations show that \ColorTM{} and \BalColorTM{} can provide significant performance improvements over prior state-of-the-art parallel graph coloring algorithms. We hope that \ColorTM{} and \BalColorTM{} will encourage further studies on the graph coloring kernel in modern multicore computing systems.

Second, we extensively characterize prior state-of-the-art \notnuma{} and \numa{} concurrent priority queues in a NUMA CPU architecture using a wide variety of contention scenarios, and find that none of them %prior state-of-the-art concurrent priority queues 
performs best across all various contention scenarios. Based on this observation, we introduce \smartpq{}, an adaptive concurrent priority queue for NUMA architectures that achieves the highest performance in all different contention scenarios. We design \smartpq{} that integrates (i) \nuddle{}, a generic framework that wraps \emph{any} arbitrary \notnuma{} concurrent data structure and transforms it to its \numa{} counterpart, and (ii) a simple decision tree classifier which predicts the best-performing \algomode{} mode between a \notnuma{} and a \numa{} \algomode{} mode. Therefore, \smartpq{} can dynamically switch during runtime between the \numa{} \nuddle{} and its underlying \notnuma{} implementation with negligible transition overheads. We demonstrate that \smartpq{} outperforms prior state-of-the-art \notnuma{} and \numa{} concurrent priority queues under various contention scenarios, and when the contention of the workload varies over time. We hope that our study will inspire future work on designing \emph{adaptive} algorithmic designs and/or adaptive runtime frameworks for concurrent data structures for modern computing systems.

Third, we rigorously examine the applicability of synchronization mechanisms tailored for processor-centric systems, including CPU, GPU and Massively Parallel Processing systems, to memory-centric NDP architectures, and find that such synchronization approaches are \emph{not} efficient or suitable for NDP systems. To this end, we introduce \SynCron{}, the first end-to-end hardware synchronization mechanism for NDP architectures. \SynCron{} achieves the goals of high performance, low cost, high programming ease and generality to cover a wide range of synchronization primitives by (1) adding low-cost hardware support near memory for synchronization acceleration, (2) including a specialized cache memory structure to store synchronization information and minimize latency overheads, (3) implementing a hierarchical message-passing communication protocol to minimize expensive network traffic, and (4) integrating a programmer-transparent hardware-only overflow management scheme to minimize performance degradation when hardware resources for synchronization tracking are exceeded. Our evaluations show that \SynCron{} can significantly improve system performance and system energy in NDP systems across a wide variety of emerging irregular applications and under various contention scenarios. We hope that \SynCron{} will encourage further studies of the synchronization problem in NDP systems and other unconventional computing systems.

Finally, we examine and efficiently map the fundamental memory-bound \spmv{} kernel on near-bank PIM systems. Specifically, we design \SparseP{}, the first open-source \spmv{} library for real PIM systems that includes 25 efficient \spmv{} kernels to cover a wide variety of sparse matrices and real-world applications of \spmv{}. \SparseP{} supports various (1) data types, (2) compressed matrix storage formats, (3) data partitioning techniques of the sparse matrix to PIM-enabled memory modules, (4) load balancing schemes across PIM cores of the system, (5) load balancing schemes across parallel threads of a multithreaded PIM core, and (6) synchronization approaches among parallel threads within PIM core. We comprehensively evaluate the \SparseP{} kernels on a real PIM system with 2528 PIM cores using 26 sparse matrices with diverse sparsity patterns. Our extensive evaluations provide new recommendations for software, system and hardware designers of real PIM systems. We also demonstrate that the \spmv{} execution on a memory-centric PIM system achieves a much higher fraction of the machine's peak performance compared to that on processor-centric CPU and GPU systems, while also having high energy efficiency. We hope that our \SparseP{} analysis on a real PIM system will provide valuable insights to software engineers in the development of efficient irregular kernels for real PIM systems, as well as to system designers and hardware architects in the development of future memory-centric computing platforms.

\section{Future Research Directions}

The concepts and methods proposed in this dissertation can potentially enable and open up several new research directions. This section describes some promising directions for future work.

\subsection{Accelerating Irregular Applications in Unconventional Systems}

Traditional data centers comprise monolithic servers that use DRAM as the main memory of the system, and tightly integrate it with the compute units, e.g., processors or accelerators. However, the increasing demand and growing size of data in modern applications in combination with the device scaling problems of DRAM memory technology~\cite{Mandelman2002Challenges} have enabled the commercialization of new unconventional systems that consist of heterogeneous memory technologies (e.g., combine DRAM with alternative memory technologies such as 3D-DRAM~\cite{HBM,HMC}, Phase Change Memory~\cite{Qureshi2011PCM}, STT-RAM~\cite{kultursay.ispass13},  NAND flash-based SSD~\cite{luo2018heatwatch}) or physically separate compute and memory devices as independent network-attached hardware components (e.g., disaggregated memory systems~\cite{Shan2018LegoOS,Lee2021MIND,guo2021clio,Gao2016Network}). These unconventional computing systems can satisfy the increasing memory capacity demands of emerging applications by providing a large pool of main memory either as a second-tier main memory tightly integrated within the server~\cite{Singh2022Sibyl,Doudali2019Kleio,Kokolis2019PageSeer,Agarwal2017Thermostat,Mitesh2015Heterogeneous,Kotra2018Chameleon} or as remote disaggregated memory
components accessed over a high-bandwidth network~\cite{Shan2018LegoOS,Gao2016Network}. Therefore, future work can take inspiration from the techniques proposed in this dissertation to accelerate irregular applications in other unconventional computing systems.

\textbf{In Heterogeneous Memory Systems} 

Hybrid or heterogeneous memory systems typically include two (or even three) tiers of memory, e.g., integrating a die-stacked DRAM~\cite{HMC,HBM} organized as a cache of a larger main memory. Therefore, the key challenge to fully leverage the heterogeneity of such systems is to accurately identify the performance-criticality of application data and place the corresponding memory pages in the "best-fit" tier of main memory.

At the same time, memory pages corresponding to application data of irregular applications exhibit high variability in their memory access patterns. For example, in \spmv{}, the memory pages that store the compressed sparse matrix exhibit high spatial locality~\cite{Kanellopoulos2019SMASH}, since the values and the positions of non-zero elements of the compressed matrix are accessed and traversed with a streaming manner in the \spmv{} execution. Instead, the memory pages that store the input vector typically exhibit low spatial locality~\cite{Kanellopoulos2019SMASH,Giannoula2022SparsePPomacs}, since \spmv{} causes irregular/random memory accesses to the elements of the input vector. However, the accesses on the input vector are \emph{input driven}, i.e., they follow the sparsity pattern of the particular input matrix given: e.g., in sparse matrices with power-law distribution, a small subset of the rows of the matrix has a very large number of non-zero elements (accounting for the majority of the matrices' non-zero elements)~\cite{Giannoula2022SparsePPomacs,Barabasi2009Scale}, and thus processing these few rows can lead to high spatial locality in the memory pages that store the input vector. Therefore, irregular applications have dynamic access patterns, e.g., memory pages might exhibit either low or high spatial locality during runtime, a fact that also depends on the particular characteristics of the input data given.

Future work could investigate intelligent hot memory page placement approaches and selection methods tailored for irregular applications executed in heterogeneous memory systems. Even though past works~\cite{Singh2022Sibyl,Kokolis2019PageSeer,Doudali2019Kleio,Doudali2021Cori} propose many different memory page placement techniques, these works do not handle variability in memory access patterns of irregular applications, and do not consider the \emph{dynamic} access patterns exhibited at memory pages for each particular input data given. Therefore, the first steps would involve to investigate the memory access patterns and page hotness/coldness across a wide variety of irregular applications (e.g., graph analytics, pointer-chasing, sparse matrix kernels) executed in modern heterogeneous memory systems, and understand the variability on the memory access patterns exhibited across memory pages. The long-term research goal is to design (i) intelligent data placement approaches for irregular applications that take into consideration the characteristics of the particular input data given, (ii) easy-to-use programming interfaces that communicate information for the characteristics of the application data to the underlying system and hardware in order to leverage data properties, and (iii) cost-effective frameworks and runtime systems that are general to support various types of memory/storage devices and more than two tiers of main memory.

\textbf{In Disaggregated Memory Systems}

Disaggregated memory systems propose to physically separate compute (e.g., processors, accelerators), memory (e.g., DRAM) and storage (e.g., disk) devices as independent and failure-isolated components connected over a high-bandwidth network\cite{Shan2018LegoOS,Lee2021MIND,guo2021clio,Gao2016Network}. This way they can provide a cost-effective solution to improve resource utilization, resource scaling and failure handling in data centers, thus decreasing data center costs. In disaggregated systems, almost all the memory in the data center is separated as network-attached disaggregated memory components, and the majority of the application working sets are accessed from the remote disaggregated memory components over the network. Moreover, disaggregated memory systems are not monolithic: each component in the system implements its own resource allocation and management policy in a completely transparent way from the remaining components in the system.

Achieving high system performance for irregular applications in disaggregated memory systems is challenging for three reasons. First, accesses across the network can be significantly slower than these within the server, and data is typically migrated at a page granularity (e.g., 4KB)\cite{Yan2019Nimble,Aguilera2017Remote,Zhang2020RethinkingDM,Lim2012System,Shan2018LegoOS,Angel2020Disaggregation,Gu2017Infiniswap,Aguilera2018Remote,Lee2021MIND}, thus incurring high data movement overheads. Second, there is high variability in data access latencies as they depend on the location of the remote disaggregated memory components and the contention with other compute components that share the same remote memory components and network. Third, prior runtime systems and hot page selection/placement schemes for heterogeneous systems~\cite{Singh2022Sibyl,Kokolis2019PageSeer,Doudali2019Kleio,Doudali2021Cori} are not suitable for fully disaggregated memory systems: prior approaches for heterogeneous systems assume that the management of memory pages is handled by the compute component itself and the OS running on it. Instead, this is not the case with fully disaggregated systems, in which remote disaggregated memory components have their own kernel modules and hardware controllers to manage their resources and memory pages (transparently to compute components)~\cite{Shan2018LegoOS,guo2021clio,Lee2021MIND}. Therefore, to efficiently execute irregular applications in such systems new software and hardware solutions are necessary.

Future work would investigate the following new challenges in the execution of irregular applications in fully disaggregated memory systems: (i) the high data movement overheads imposed by remotely accessing data over the network, (ii) the high variability in data access costs during runtime due to network and memory sharing, (iii) the unconventional distributed approach of managing the data on multiple components in the system with a completely transparent way to each other, and (iv) the high memory sharing and the memory protection issues for pages located in remote disaggregated memory components, which can be accessed by multiple processes that concurrently run at different compute components of the system.

The first step is to develop a cost model, a software-based simulator, or a hardware-based emulator for fully disaggregated memory systems, which can support various configurations for the network characteristics (e.g., network topology, network bandwidth/latency), and evaluate, analyze and understand critical performance overheads in the execution of a wide variety of irregular applications with diverse access patterns. Rigorously and comprehensively understanding performance implications of irregular applications in fully disaggregated memory systems can provide valuable insights to software engineers, system designers and hardware architects of this architecture. The next steps are to propose new address translation approaches and kernel modules to minimize system-level overheads (e.g., page faults), flexible (asynchronous and synchronous) programming interfaces and abstractions to easily access remote data over the network, fast network technologies to mitigate network-related bottlenecks, low-overhead synchronization and memory sharing/coherence mechanisms for multiple memory components in the system to ensure correctness at low cost, as well as to leverage the NDP paradigm~\cite{Mutlu2019Processing} in disaggregated memory components to reduce access costs to remote data. The long-term goal is to perform research on designing fundamentally new approaches for all key components of the computing stack, which need to be distributed, \emph{disaggregated} and scale elastically, in keeping with the promise of resource disaggregation.

\subsection{Adaptive Algorithmic, System-Level and Hardware-Based Approaches for Irregular Applications}
%\subsection{Adaptive Algorithmic Designs, Runtime Systems and Hardware Mechanisms for Irregular Workloads}

Emerging irregular applications exhibit dynamic workload demands and contention, i.e., their memory access patterns, bandwidth, latency and parallelization demands vary over time. For instance, irregular key-value stores such as binary search trees~\cite{Ellen-bst,Howley-bst,Natarajan-bst,Siakavaras2017Combining,Siakavaras2021RCUHTM}, linked lists~\cite{Harris-ll,Michael-ll,fraser,spraylist}, priority queues~\cite{spraylist,ffwd,lotan_shavit,Strati2019AnAdaptive}, hash tables~\cite{Guerraoui2016Optimistic,CHEN2018Concurrent,Michael2002High} are used in database management systems, and multiple users perform lookup and update operations (e.g., insert or delete) on them with various frequencies over runtime: concurrent key-value store data structures exhibit high variability during time in the levels of contention and their memory access patterns as they depend on the amount and types of operations (lookup, insert, delete) that users perform on them during runtime. Similarly, modern computing systems and large-scale architectures exhibit high variability into network characteristics (e.g., memory bandwidth, latency, network topology), runtime contention (e.g., co-running applications), and available hardware resources (e.g., memory devices, accelerators). For example, in disaggregated memory systems, the architectures, component placements and network characteristics can highly vary over time, since multiple hardware components can be dynamically added, removed or upgraded, and network technologies or topologies can also flexibly change over time~\cite{Shan2018LegoOS,guo2021clio,Lee2021MIND}. Similarly, virtualized environments support dynamic sets of resources, in which virtual machines can be dynamically added, removed or change their hardware characteristics/configuration over time~\cite{Athlur2022Varuna}. The dynamic variability on the (i) runtime workload demands of irregular applications, and (ii) architecture and network characteristics of modern computing systems results in significant variations on data access latencies and data movement overheads in the execution of emerging irregular applications, which might thus significantly degrade system performance and resource utilization. To this end, future work would involve (i) designing \emph{adaptive} algorithmic approaches for irregular applications depending on the runtime contention, and application, hardware and network characteristics, and (ii) enabling the system and hardware to \emph{dynamically} change their configurations depending on the availability of resources and runtime application behavior.

\textbf{Adaptive Algorithmic Designs}

The research goal is to design adaptive algorithms that \emph{on-the-fly} change their parallelization strategy, synchronization approach and data management policy over time to significantly improve system performance, energy efficiency and data access costs in irregular applications. The key idea is to enable parallel threads or background/monitor threads to track properties of application data and/or runtime statistics, and employ low-overhead decision-making mechanisms to select between multiple configurations (e.g., different parallelization strategies, synchronization approaches, data management policies) during the execution. For example, in Chapter~\ref{SmartPQChapter}, we propose an adaptive priority queue for NUMA CPU architectures that dynamically changes its data access policy and synchronization scheme by tracking the levels of contention during runtime and integrating a lightweight decision tree classifier that predicts the optimal parallelization strategy based on runtime statistics. Other examples include to change on-the-fly the graph traversal strategy on graph processing kernels depending on the number of the edges of the current vertex that is being processed (e.g., in real-world graphs with power-law distribution~\cite{Barabasi2009Scale} a few vertices have a significantly larger number of edges compared the remaining vertices) or alternate the data access policy in sparse matrix kernels when processing rows with a small/large number of non-zero elements. The challenge in designing adaptive algorithms for irregular applications is to minimize the performance overheads between transitions on different configurations.

\textbf{Adaptive Runtime Systems}

The research goal is to develop adaptive runtime systems that dynamically adjust the task assignments, task scheduling and data distribution policies during runtime to significantly improve system performance, resource utilization and financial costs. The key idea is to integrate in the runtime systems dedicated managers that monitor tasks and jobs running across the computing nodes of the system, and decide on the optimal configuration, e.g., optimal task scheduling across the computing nodes of the system, when architecture, hardware and/or network characteristics change. For instance, a recent work~\cite{Athlur2022Varuna} proposes a novel runtime system for distributed machine learning training, that \emph{dynamically} tunes the number of pipeline stages, depending on the network load/contention at any given time and the number of available computing nodes (i.e., GPUs) in the system. Therefore, future work could investigate designing adaptive runtime systems for distributed training of sparse neural networks, that include dedicated monitoring managers which track the execution of running tasks and on-the-fly tune the parallelization approach and data distribution policy across multiple computing nodes (e.g., GPUs, TPUs, NPUs) of large-scale clusters and in cloud environments (e.g., when using virtual machines), when new computing nodes are added or removed in the system and when network load/contention changes. Similarly to adaptive algorithmic designs, the key challenge in such intelligent runtime systems is to achieve low synchronization overheads between transitions from one configuration to another.

\textbf{Adaptive Hardware Mechanisms}

The key research goal is to propose adaptive hardware mechanisms that on-the-fly adjust their performance optimization strategies depending on availability of resources and runtime application characteristics. The key idea is to integrate  hardware controllers in the computing system that decide between different optimization policies by leveraging system-level metadata (e.g., page tables/TLBs), simple prediction heuristics, or statistics collected during runtime at low cost. For instance, a few recent works~\cite{Alameldeen2004Adaptive,Alameldeen2007Interactions,Tuduce2005Adaptive,Arelakis2015HyComp,Kim2017Transparent} propose hardware compression mechanisms for cache and main memory of CPU systems that dynamically enable/disable compression~\cite{Alameldeen2004Adaptive,Alameldeen2007Interactions,Tuduce2005Adaptive} or on-the-fly select the best-performing compression algorithm~\cite{Arelakis2015HyComp,Kim2017Transparent} based on properties of application data or the runtime application behavior. Other examples include to design (i) intelligent hardware prefetchers that on-the-fly enable/disable fetching application data from main memory to cache memory depending on the current bandwidth utilization and locality of application data, or (ii) effective selection granularity mechanisms that on-the-fly decide the granularity at which data migrations should be served (e.g., in heterogeneous systems choosing if a data migration from the second tier main memory to the cache-based main memory should be served by a page or a smaller granularity, e.g., cache line granularity) depending on the runtime network and application characteristics, and the available memory resources. The key challenge in designing adaptive hardware mechanisms is to implement intelligent prediction heuristics and/or to enable keeping metadata for the runtime system and application behavior at low hardware- and system-level cost.

\section{Concluding Remarks}

In this dissertation we extensively characterize the execution of irregular applications in modern processor-centric (e.g., CPUs) and memory-centric (e.g., NDP/PIM) systems, and provide directions to bridge the gap between processor-centric systems and memory-centric systems in the context of important yet difficult irregular applications. We observe that excessive synchronization and high memory intensity of irregular applications can significantly degrade system performance. Therefore, we propose low-overhead synchronization and well-crafted data access techniques for irregular applications, and demonstrate that they can significantly increase parallelism, improve energy efficiency, minimize data access costs, and accelerate performance of emerging irregular applications in CPU and NDP/PIM systems. Specifically, we introduce four new designs that enable efficient execution of irregular applications in modern computing systems: (1) \ColorTM{}, a speculative synchronization scheme co-designed with an effective data access policy that accelerates the graph coloring kernel in modern CPU systems, (2) \smartpq{}, an adaptive algorithm design that improves performance of priority queue data structure in NUMA CPU architectures, (3) \SynCron{}, a practical and low-overhead hardware synchronization mechanism that effectively leverages the benefits of NDP for a wide range of irregular applications, and (4) \SparseP{}, a wide collection of parallel algorithms to easily attain high  performance of the \spmv{} kernel on real PIM systems. We hope that the ideas, analysis, methods and techniques presented in this dissertation will enable new studies and research directions to accelerate the execution of important data-intensive irregular applications in current and future computing platforms.

\chapter{\textbf{Other Works of the Author}}\label{OtherWorksChapter}
In addition to the works presented in this dissertation, the author of this dissertation has also contributed to several other research works done in collaboration with SAFARI Research Group members at ETH Zürich. This chapter briefly overviews these works.

\textbf{PrIM~\cite{Gomez2021Analysis,Gomez2021Benchmarking,Gomez2022Benchmarking,PrIMLibrary}}: In modern computing systems like CPU and GPU systems, a large fraction of the execution time and energy consumption of modern data-intensive irregular workloads is spent on moving data between memory and processor cores. Recent research explores different PIM configurations~\cite{Gomez2022Benchmarking,Giannoula2021SynCron,devaux2019,Mutlu2019Processing,Ghose2019Workload,mutlu2020modern,Gomez2021Analysis,Gomez2021Benchmarking,Stone1970Logic,Kautz1969Cellular,shaw1981non,kogge1994,gokhale1995processing,patterson1997case,oskin1998active,kang1999flexram,Mai:2000:SMM:339647.339673,Draper:2002:ADP:514191.514197,aga.hpca17,Eckert2018Neural,Fujiki2019Duality,kang.icassp14,seshadri.micro17,seshadri.arxiv16,Seshadri:2015:ANDOR,Seshadri2013RowClone,Angizi2019GraphiDe,kim.hpca18,kim.hpca19,gao2020computedram,chang.hpca16,Xin2020ELP2IM,li.micro17,deng.dac2018,hajinazarsimdram,Rezaei2020NoM,Wang2020Figaro,Ali2020InMemory,Li2016Pinatubo,angizi2018pima,angizi2018cmp,angizi2019dna,levy.microelec14,kvatinsky.tcasii14,Shafiee2016ISAAC,kvatinsky.iccd11,kvatinsky.tvlsi14,gaillardon2016plim,Bhattacharjee2017ReVAMP,Hamdioui2015Memristor,xie2015fast,hamdioui2017myth,Yu2018Memristive,fernandez2020natsa,Cali2020GenASM,kim.bmc18,Ahn2015PIMenabled,boroumand2017lazypim,Boroumand2019Conda,Chi2016PRIME,Farmahini2015NDA,Gao2015Practical,Gao2016HRL,Gu2016Leveraging,Guo2014APerformance,hashemi2016accelerating,Hsieh2016accelerating,Kim2017GrimFilter,LeBeane2015Data,Nai2017GraphPIM,Drumond2017mondrian,Dai2018GraphH,Zhang2018GraphP,Youwei2019GraphQ,ferreira2021pluto,Olgun2021QuacTrng,Singh2020NEROAN,asghari-moghaddam.micro16,BabarinsaI15,kim2016bounding,morad.taco15,pattnaik.pact16,zhang.hpdc14,Zhu2013Accelerating,denzler2021casper,boroumand2021polynesia,boroumand2021icde,singh2021fpga,singh2021accelerating,herruzo2021enabling,yavits2021giraf,asgarifafnir,seshadri.bookchapter17,diab2022high,diab2022hicomb,fujiki2018memory,zha2020hyper,Saugata2018Enabling,Mutlu2013Memory,Mutlu2014Research,Mutlu2019InDram,SESHADRI2017107,impica,DBLP:conf/isca/AkinFH15,huang2020heterogeneous,santos2017operand,wen2017rebooting,besta2021sisa,lloyd2015memory,elliott1999computational,zheng2016tcam,landgraf2021combining,rodrigues2016scattergather,lloyd2018dse,lloyd2017keyvalue,gokhale2015rearr,jacob2016compiling,sura2015data,nair2015evolution,xi2020memory,Singh2019Napel,pugsley2014ndc,gao2017tetris,Nair2015Active,Balasubramonian2014Near}, since the PIM paradigm provides a promising way to alleviate the data movement bottleneck between memory and processors. The UPMEM company~\cite{upmem,upmem2018,devaux2019} has designed and fabricated the first commercially-available near-bank PIM architecture. In this work, we conduct an experimental characterization of the UPMEM-based PIM system using microbenchmarks to assess various architecture limits such as compute throughput and memory bandwidth, and we present PrIM (Processing-In-Memory benchmarks), a benchmark suite of 16 irregular workloads from different application domains (e.g., dense/sparse linear algebra, databases, data analytics, graph processing, neural networks, bioinformatics, image processing), which we identify as memory-bound. We evaluate the performance and scaling characteristics of PrIM benchmarks on the UPMEM PIM architecture, and compare their performance and energy consumption to their state-of-the-art CPU and GPU counterparts. Our extensive evaluation conducted on two real UPMEM-based PIM systems provides new insights about suitability of different irregular workloads to the PIM system, programming recommendations for software designers, and suggestions and hints for hardware and architecture designers of future PIM systems.

\textbf{NATSA~\cite{fernandez2020natsa,NatsaCode}}: Time series analysis is an irregular computation kernel that processes a chronologically ordered set of samples of a real-valued variable that can contain millions of observations, and is used to analyze information in a wide variety of domains including epidemiology, genomics, neuroscience, medicine and environmental sciences. Matrix profile is the state-of-the-art algorithm to perform time series analysis, by computing the most similar subsequence for a given query subsequence within a sliced time series. In this work, we evaluate the state-of-the-art CPU implementation of the matrix profile algorithm on a real multi-core machine, i.e., Intel Xeon
Phi KNL, and observe that its performance is heavily bottlenecked by data movement between the off-chip memory units and the on-chip computation units that execute matrix profile. To reduce the data movement overheads, we design a near-data processing accelerator for time series analysis, called NATSA. NATSA exploits the low-latency, high-bandwidth, and energy-efficient memory access provided by modern 3D-stacked High Bandwidth Memory (HBM), and integrates specialized custom processing units in the logic layer of HBM. This way NATSA enables energy-efficient and fast matrix profile computation near memory, i.e., where time series data resides, and reduces the data movement costs between the computation units and the memory units. NATSA provides generality and flexibility supporting a wide range of time series applications, and significantly improves system performance and energy efficiency over state-of-the-art CPU, GPU and NDP systems.

\textbf{SMASH~\cite{Kanellopoulos2019SMASH,SmashCode}}: The matrices involved in irregular sparse linear algebra computation kernels are very large in size and highly sparse, i.e., the vast majority of the elements are zeros. Prior research works~\cite{Im1999Optimizing,Langr2016Evaluation,Liu2015CSR5,Pinar1999Improving,Vuduc2005Fast,Yang2014Optimization,Shengen2014YaSpMV,Kourtis2011CSX,Kourtis2008Optimizing,Belgin2009Pattern,LIL,ELL,bjorck1996numerical,Pooch1973Survey,Shubhabrata2007Scan,Changwan2018Efficient,Liu2013Efficient,Monakov2010Automatically,Saad1989Krylov,Buluc2009Parallel,Martone2014251,Martone2010Blas,Kreutzer2012Sparse} design compressed storage formats for sparse matrices: the non-zero elements and their positions within the matrix are stored using additional data structures and different encodings. However, determining the positions of the non-zero elements in the compressed encoding (i.e., \emph{indexing}) requires a series of pointer-chasing operations in memory, that are highly inefficient in modern processors and memory hierarchies, and incur high data access costs. The key idea of SMASH is to explicitly enable the hardware to recognize and exploit the compression encoding used in software for any sparse matrix. On the software side, SMASH efficiently compresses any sparse matrix via a novel software encoding that is based on a hierarchy of bitmaps. On the hardware side, SMASH includes a lightweight hardware unit, named Bitmap Management Unit, that is used to perform highly-efficient scans of the hierarchy of bitmaps, and thus enabling highly efficient indexing in sparse matrices and minimizing data access costs in sparse linear algebra computation kernels. SMASH provides significant speedups in sparse matrix computations by eliminating the expensive pointer-chasing operations required in state-of-the-art compressed matrix storage formats.

\newpage

\chapter{Appendix A}\label{AppendixChapter}

\section{Extended Results for \SparseP{}}

\subsection{Synchronization Approaches in Block-Based Compressed Matrix Formats}\label{sec:appendix-1DPU-BCOO}

\begin{figure}[t]
    \centering
    \includegraphics[width=\textwidth]{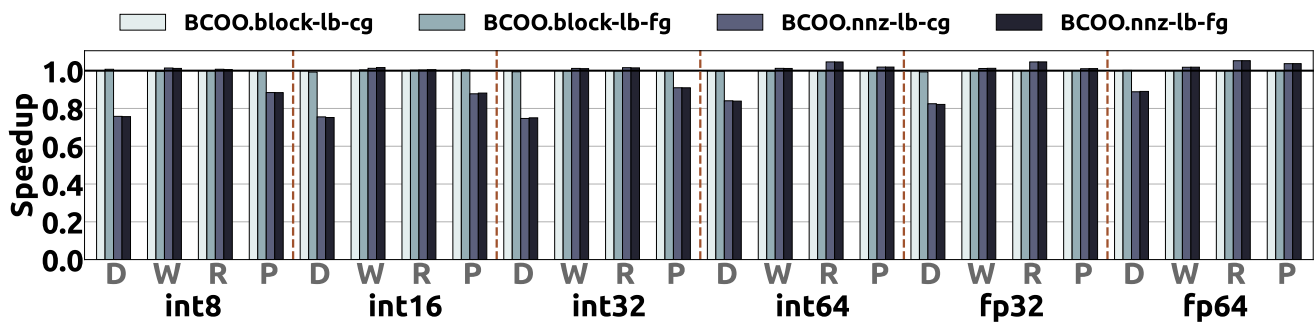}
    \vspace{-15pt}
    \caption{Performance of the BCOO format with various load balancing schemes and synchronization approaches for all the data types and small matrices using 16 tasklets of one DPU.}
    \label{fig:1DPU-bcoo-balance}
    \vspace{-8pt}
\end{figure}

We compare the coarse-grained locking (\textit{lb-cg}) and the fine-grained locking (\textit{lb-fg}) approaches in the BCOO format. Figure~\ref{fig:1DPU-bcoo-balance} shows the performance achieved by the BCOO format for all the data types when balancing the blocks or the non-zero elements across 16 tasklets of one DPU. We evaluate all small matrices of Table~\ref{tab:small-matrices}, i.e., delaunay\_n13 (\textbf{D}), wing\_nodal (\textbf{W}), raefsky4 (\textbf{R}) and pkustk08 (\textbf{P}) matrices.

Our key finding is that the fine-grained locking approach performs similarly with the coarse-grained locking approach. The fine-grained locking approach does not increase parallelism in the UPMEM PIM architecture, since memory accesses executed by multiple tasklets to the local DRAM bank are serialized in the DMA engine of the DPU. The same key finding holds independently of the compressed matrix format used.

\subsection{Fine-Grained Data Transfers in 2D Partitioning Techniques}\label{sec:appendix-2D-fgtrans}
\begin{figure}[H]
    \centering
    \includegraphics[width=1\textwidth]{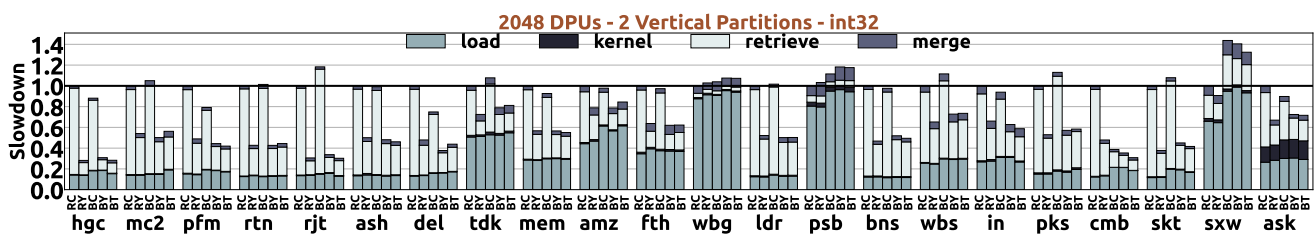}
    \vspace{-15pt}
    \caption{Performance comparison of \texttt{RC}: \texttt{RBDCOO} with coarse-grained transfers, \texttt{RY}: \texttt{RBDCOO} with fine-grained transfers in the output vector, \texttt{BC}: \texttt{BDCOO} with coarse-grained transfers, \texttt{BY}: \texttt{BDCOO} with fine-grained transfers only in the output vector, and \texttt{BT}: \texttt{BDCOO} with fine-grained transfers in both the input and the output vector using the int32 data type, 2048 DPUs and having 2 vertical partitions. Performance is normalized to that of the \texttt{RC} scheme.}
    \label{fig:2D_vp2_fgtransfers}
\end{figure}

\begin{figure}[H]
    \centering
    \includegraphics[width=1\textwidth]{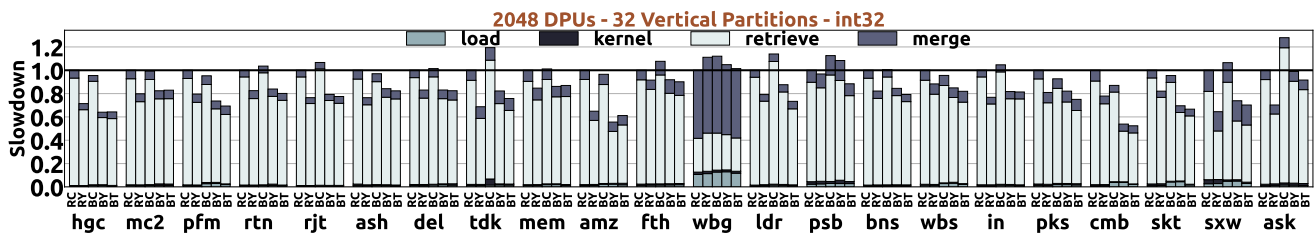}
    \vspace{-15pt}
    \caption{Performance comparison of \texttt{RC}: \texttt{RBDCOO} with coarse-grained transfers, \texttt{RY}: \texttt{RBDCOO} with fine-grained transfers in the output vector, \texttt{BC}: \texttt{BDCOO} with coarse-grained transfers, \texttt{BY}: \texttt{BDCOO} with fine-grained transfers only in the output vector, and \texttt{BT}: \texttt{BDCOO} with fine-grained transfers in both the input and the output vector using the int32 data type, 2048 DPUs and having 32 vertical partitions. Performance is normalized to that of the \texttt{RC} scheme.}
    \label{fig:2D_vp32_fgtransfers}
    \vspace{-4pt}
\end{figure}

Figures~\ref{fig:2D_vp2_fgtransfers} and ~\ref{fig:2D_vp32_fgtransfers} compare coarse-grained data transfers (i.e., performing parallel data transfers to all 2048 DPUs at once, padding with empty bytes at the granularity of 2048 DPUs) with fine-grained data transfers (i.e., iterating over the ranks and for each rank performing parallel data transfers to the 64 DPUs of the same rank, padding with empty bytes at the granularity of 64 DPUs) for all matrices of our large matrix dataset in the \equallyWidth{} and \variableSized{} schemes, respectively. The reported key findings of Figure~\ref{fig:2D_fgtransfers} (Section~\ref{2D-Studies}) apply to all matrices with diverse sparsity patterns.

\subsection{Effect of the Number of Vertical Partitions Using Two Different UPMEM PIM Systems}\label{sec:appendix-2D-vertpartitions}

We compare \spmv{} execution in the two different UPMEM PIM sytems using 2048 DPUs and 16 tasklets for each DPU. Table~\ref{tab:pim-systems} shows the characteristics of two different UPMEM PIM systems. We calculate the available PIM peak performance and PIM bandwidth assuming 2048 DPUs for both PIM systems\footnote{Both UPMEM PIM systems support 20 UPMEM PIM DIMMs with 2560 DPUs in total. However, both UPMEM-based PIM systems include multiple faulty DPUs. Thus, for a fair comparison between two systems we conduct our experiments using 2048 DPUs in both systems.}. We estimate the PIM peak performance as $Total\_DPUs * AT$, where the arithmetic throughput (AT) is calculated for the multiplication operation by running the arithmetic throughput microbenchmark of the PrIM benchmark suite~\cite{Gomez2021Analysis,Gomez2021Benchmarking} in each of the two UPMEM PIM systems (See Appendix~\ref{sec:appendix-1DPU-AT}). We estimate the PIM bandwidth as $Total\_DPUs * Bandwidth\_DPU$, where the $Bandwidth\_DPU$ is calculated according to prior work~\cite{Gomez2021Analysis, Gomez2021Benchmarking}. Specifically, the theoretical maximum MRAM bandwidth (i.e., $Bandwidth\_DPU$) is 700 MB/s and 850 MB/s at a DPU frequency of 350 MHz (PIM system A) and 425 MHz (PIM system B), respectively.

\begin{table}[H]
\vspace{-2pt}
\begin{center}
\centering
\resizebox{1.0\linewidth}{!}{
\begin{tabular}{|l||c|c|c|c|c|c|c|c|}
    \hline
      \cellcolor{gray!15} & \cellcolor{gray!15}\raisebox{-0.20\height}{\textbf{Avail.}} & \cellcolor{gray!15} & \cellcolor{gray!15}\raisebox{-0.20\height}{\textbf{PIM Peak}} & \cellcolor{gray!15}\raisebox{-0.20\height}{\textbf{PIM}} &  \cellcolor{gray!15}  & \cellcolor{gray!15}\raisebox{-0.20\height}{\textbf{CPU Peak}} & \cellcolor{gray!15}\raisebox{-0.20\height}{\textbf{Bus}}  \\
       \multirow{-2}{*}{\cellcolor{gray!15}\textbf{System}}   & \cellcolor{gray!15} \textbf{DPUs} & \multirow{-2}{*}{\cellcolor{gray!15}\textbf{Frequency}}  & \cellcolor{gray!15}\textbf{Performance} & \cellcolor{gray!15}\textbf{Bandwidth} & \multirow{-2}{*}{\cellcolor{gray!15} \textbf{Host CPU}} & \cellcolor{gray!15}\textbf{Performance} & \cellcolor{gray!15}\textbf{Bandwidth}  \\
    
    \hline \hline
    PIM System A & 2048 DPUs & 350 MHz & 3.78 GFLOPS & 1.43 TB/s &  Intel Xeon Silver 4110 @2.1 GHz & 660 GFLOPS & 23.1 GB/s\\ \hline 
    PIM System B & 2048 DPUs & 425 MHz & 4.63 GFLOPS & 1.74 TB/s &  Intel Xeon Silver 4215 @2.5 GHz & 1016 GFLOPS & 21.8 GB/s \\ \hline 
\end{tabular}
}
\end{center}
\vspace{-4pt}
\caption{Evaluated UPMEM PIM Systems.}
\label{tab:pim-systems}
\vspace{-12pt}
\end{table}

\begin{figure}[!ht]
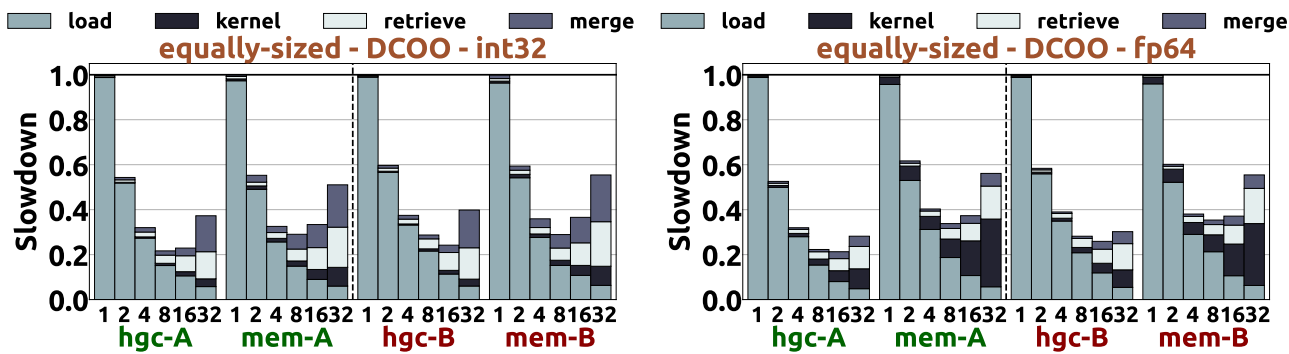

    \vspace{-6pt}
    \centering
    \begin{minipage}{\textwidth}
    \centering
    \includegraphics[width=.494\textwidth]{sections/SparseP/2D-partitioning/cloud4_cloud7_bal_vertical_dpus2048_int32.pdf}
    \includegraphics[width=.494\textwidth]{sections/SparseP/2D-partitioning/cloud4_cloud7_bal_vertical_dpus2048_dbl64.pdf}
    \end{minipage}
    \vspace{-3pt}
    \caption{Execution time breakdown of \texttt{BDCOO} using 2048 DPUs when varying the number of vertical partitions from 1 to 32 for the int32 (left) and fp64 (right) data types on two different UPMEM PIM systems.}
    \label{fig:cloud4-cloud7_bal_vertpartitions}
    %\vspace{-10pt}
\end{figure}

Figures~\ref{fig:cloud4-cloud7_fixed_vertpartitions}, ~\ref{fig:cloud4-cloud7_rbal_vertpartitions} and ~\ref{fig:cloud4-cloud7_bal_vertpartitions} compare \spmv{} execution in the two different UPMEM PIM systems (2048 DPUs) using 2D-partitioned kernels with the COO format, when varying the number of vertical partitions from 1 to 32 (in steps of multiple of 2) for the int32 (left) and fp64 (right) data types.

We observe that the number of vertical partitions that provides the best performance on \spmv{} execution varies depending on the input matrix and the PIM system. For example, in PIM system B with the int32 data type, \texttt{DCOO} performs best for the \texttt{hgc} matrix with 16 vertical partitions, while in PIM system A, \texttt{DCOO} performs best for the same matrix with 8 vertical partitions. Similarly, in PIM system A with the fp64 data type, \texttt{BDCOO} performs best for the \texttt{rjt} matrix with 4 vertical partitions. Instead, in PIM system B with the fp64 data type, \texttt{BDCOO}'s performance does not improves for the \texttt{rjt} matrix when having more than 1 vertical partition (i.e., compared to when using the 1D partitioning technique). We conclude that the best-performing parallelization scheme that achieves the best performance in \spmv{} depends on the characteristics of both the input sparse matrix and the underlying PIM system.

%\vspace{6pt}
\begin{figure}[t]
    \centering
    \begin{minipage}{\textwidth}
    \centering
    \includegraphics[width=.494\textwidth]{sections/SparseP/2D-partitioning/cloud4_cloud7_fixed_vertical_dpus2048_int32.pdf}
    \includegraphics[width=.494\textwidth]{sections/SparseP/2D-partitioning/cloud4_cloud7_fixed_vertical_dpus2048_dbl64.pdf}
    \end{minipage}
    \vspace{-3pt}
    \caption{Execution time breakdown of \texttt{DCOO} using 2048 DPUs when varying the number of vertical partitions from 1 to 32 for the int32 (left) and fp64 (right) data types on two different UPMEM PIM systems.}
    \label{fig:cloud4-cloud7_fixed_vertpartitions}
    \vspace{-8pt}
\end{figure}

\begin{figure}[t]
    \centering
    \begin{minipage}{\textwidth}
    \centering
    \includegraphics[width=.494\textwidth]{sections/SparseP/2D-partitioning/cloud4_cloud7_rbal_vertical_dpus2048_int32.pdf}
    \includegraphics[width=.494\textwidth]{sections/SparseP/2D-partitioning/cloud4_cloud7_rbal_vertical_dpus2048_dbl64.pdf}
    \end{minipage}
    \vspace{-3pt}
    \caption{Execution time breakdown of \texttt{RBDCOO} using 2048 DPUs when varying the number of vertical partitions from 1 to 32 for the int32 (left) and fp64 (right) data types on two different UPMEM PIM systems.}
    \label{fig:cloud4-cloud7_rbal_vertpartitions}
    \vspace{-2pt}
\end{figure}

%\newpage

%\vspace{400pt}
\vspace{-8pt}
\subsection{Performance of Compressed Matrix Formats Using 2D Partitioning Techniques}\label{sec:appendix-2D-formats}
\begin{figure}[H]
    \centering
    \includegraphics[width=1\textwidth]{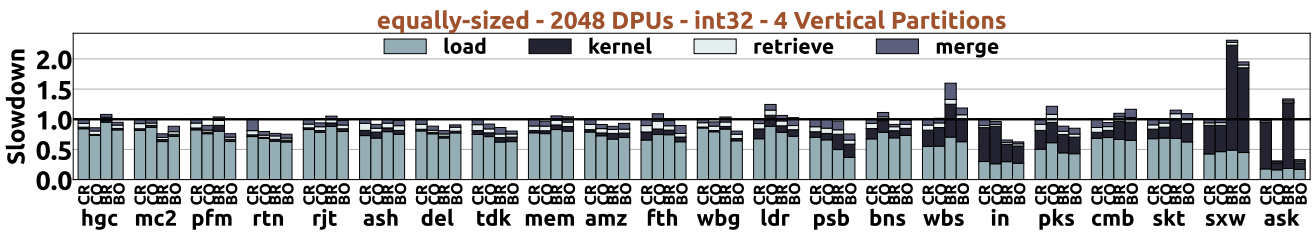}
    \vspace{-16pt}
    \caption{End-to-end execution time breakdown of the \equallySized{} 2D partitioning technique for CR: \texttt{DCSR}, CO: \texttt{DCOO}, BR: \texttt{DBCSR} and BO: \texttt{DBCOO} schemes using 4 vertical partitions and the int32 data type. Performance is normalized to that of \texttt{DCSR}.}
    \label{fig:2D_formats_fixed_all}
\end{figure}

\begin{figure}[H]
    \centering
    \includegraphics[width=1\textwidth]{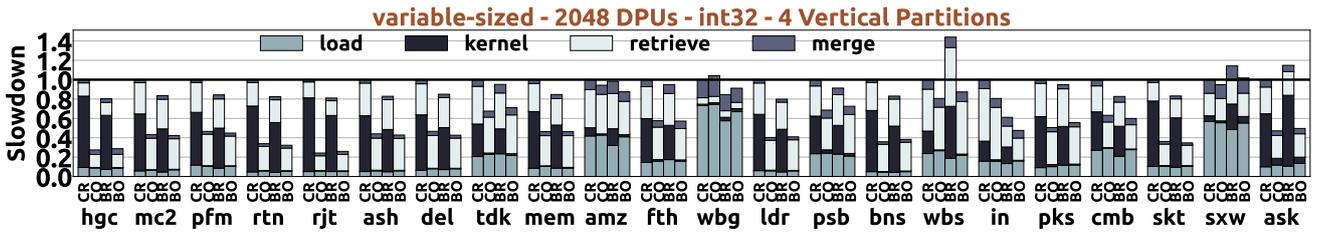}
    \vspace{-16pt}
    \caption{End-to-end execution time breakdown of the \variableSized{} 2D partitioning technique for CR: \texttt{BDCOO}, CO: \texttt{BDCOO}, BR: \texttt{BDBCSR} and BO: \texttt{BDBCOO} schemes using 4 vertical partitions and the int32 data type. Performance is normalized to that of \texttt{BDCSR}.}
    \label{fig:2D_formats_bal_all}
    \vspace{-14pt}
\end{figure}

Figures~\ref{fig:2D_formats_fixed_all},~\ref{fig:2D_formats_rbal_all},~\ref{fig:2D_formats_bal_all} compare the performance achieved by various
compressed matrix formats for each of the three types of the 2D partitioning technique for all matrices of our large matrix dataset. The reported key findings explained in Section~\ref{2D-Formats} apply to all matrices with diverse sparsity patterns.

\begin{figure}[H]
\vspace{2pt}
    \centering
    \includegraphics[width=1\textwidth]{sections/SparseP/2D-partitioning/rbal_time_dpus2048_int32_cps4.pdf}
    \vspace{-16pt}
    \caption{End-to-end execution time breakdown of  the \equallyWidth{} 2D partitioning technique for CR: \texttt{RBDCSR}, CO: \texttt{RBDCOO}, BR: \texttt{RBDBCSR} and BO: \texttt{RBDBCOO} schemes using 4 vertical partitions and the int32 data type. Performance is normalized to that of \texttt{RBDCSR}.}
    \label{fig:2D_formats_rbal_all}
    \vspace{-14pt}
\end{figure}

\subsection{Analysis of 1D- and 2D-Partitioned Kernels in Two  UPMEM PIM Systems}\label{sec:appendix-1D_2D}

\begin{figure}[H]
    \vspace{-4pt}
    \centering
    \includegraphics[width=\textwidth]{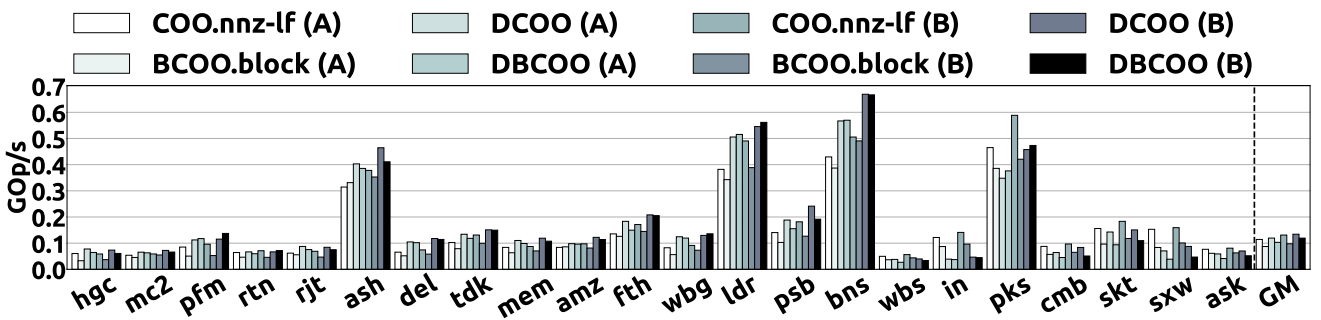}
    \vspace{-18pt}
    \caption{Throughput of 1D- and 2D-partitioned kernels for the fp32 data type using two different UPMEM PIM systems.}
    \label{fig:cloud4-cloud7_ops}
\end{figure}

\begin{figure}[H]
    %\vspace{-1pt}
    \centering
    \includegraphics[width=\textwidth]{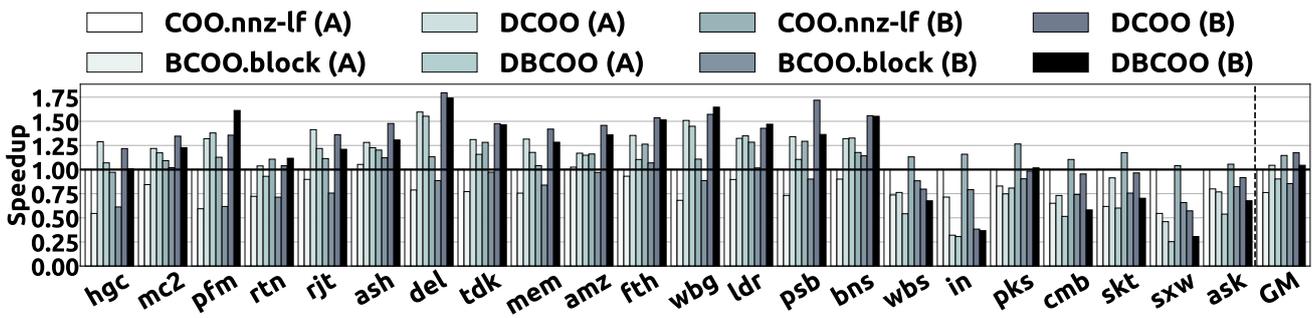}
    \vspace{-16pt}
    \caption{Performance comparison of 1D- and 2D-partitioned kernels for the fp32 data type using two different UPMEM PIM systems. Performance is normalized to that of \texttt{COO.nnz-lf (A)}.}
    \label{fig:cloud4-cloud7_perf}
    \vspace{-12pt}
\end{figure}

Figures~\ref{fig:cloud4-cloud7_ops} and ~\ref{fig:cloud4-cloud7_perf} compare the throughput and the performance, respectively, achieved by the best-performing 1D- and 2D-partitioned kernels in two different UPMEM PIM systems (Table~\ref{tab:pim-systems} presents the characteristics of the two UPMEM PIM systems). For  1D partitioning, we use the lock-free COO (\texttt{COO.nnz-lf}) and coarse-grained locking BCOO (\texttt{BCOO.block}) kernels. For each matrix, we vary the number of DPUs from 64 to 2048 DPUs, and select the best-performing end-to-end execution throughput. For 2D partitioning, we use the \equallySized{} COO (\texttt{DCOO}) and BCOO (\texttt{BCOO}) kernels with 2048 DPUs for both systems. For each matrix, we vary the number of vertical partitions from 2 to 32 (in steps of multiple of 2), and select the best-performing end-to-end execution throughput.

We draw three findings. First, we observe that in both systems the best performance is achieved using a smaller number of DPUs than 2048 DPUs. This is because \spmv{} execution in both UPMEM PIM systems is significantly bottlenecked by expensive data transfers performed via the narrow memory bus. As a result, the best-performing 1D- and 2D-partitioned kernels trade off computation with lower data transfer costs, thus causing many DPUs to be \textit{idle}. Second, we find that in both systems the 2D-partitioned kernels outperform the 1D-partitioned kernels in regular matrices (i.e., from \texttt{hgc} to \texttt{bns} matrices on x axis), while the 1D-partitioned kernels outperform the 2D-partitioned kernels in scale-free matrices, i.e., in matrices that have high non-zero element disparity among rows and columns (i.e., from \texttt{wbs} to \texttt{ask} matrices on x axis). Third, we observe that PIM system B improves performance over PIM system A by 1.14$\times$ (averaged across all matrices). This is because the DPUs of the PIM system B run at a higher frequency than that of PIM system A (425 MHz vs 350 MHz), providing higher peak performance on the system. Specifically, with 2048 DPUs, peak performance of the PIM system A and PIM system B is 3.78 GFlops and 4.63 GFlops, respectively, i.e., PIM system B provides 1.22 $\times$ higher computation throughput than PIM system A.

\newpage

\section{Arithmetic Throughput of One DPU for the Multiplication Operation}\label{sec:appendix-1DPU-AT}
%\subsubsection{Microbenchmark Description.}
We evaluate the arithmetic throughput of the DPU for the multiplication (MUL) operation. We use the arithmetic throughput microbenchmark of the PrIM benchmark suite~\cite{Gomez2021Benchmarking,Gomez2021Analysis} and configure it for the all data types. 

\begin{figure}[H]
\centering
\begin{minipage}{1.0\textwidth}
\centering
\includegraphics[width=.48\textwidth]{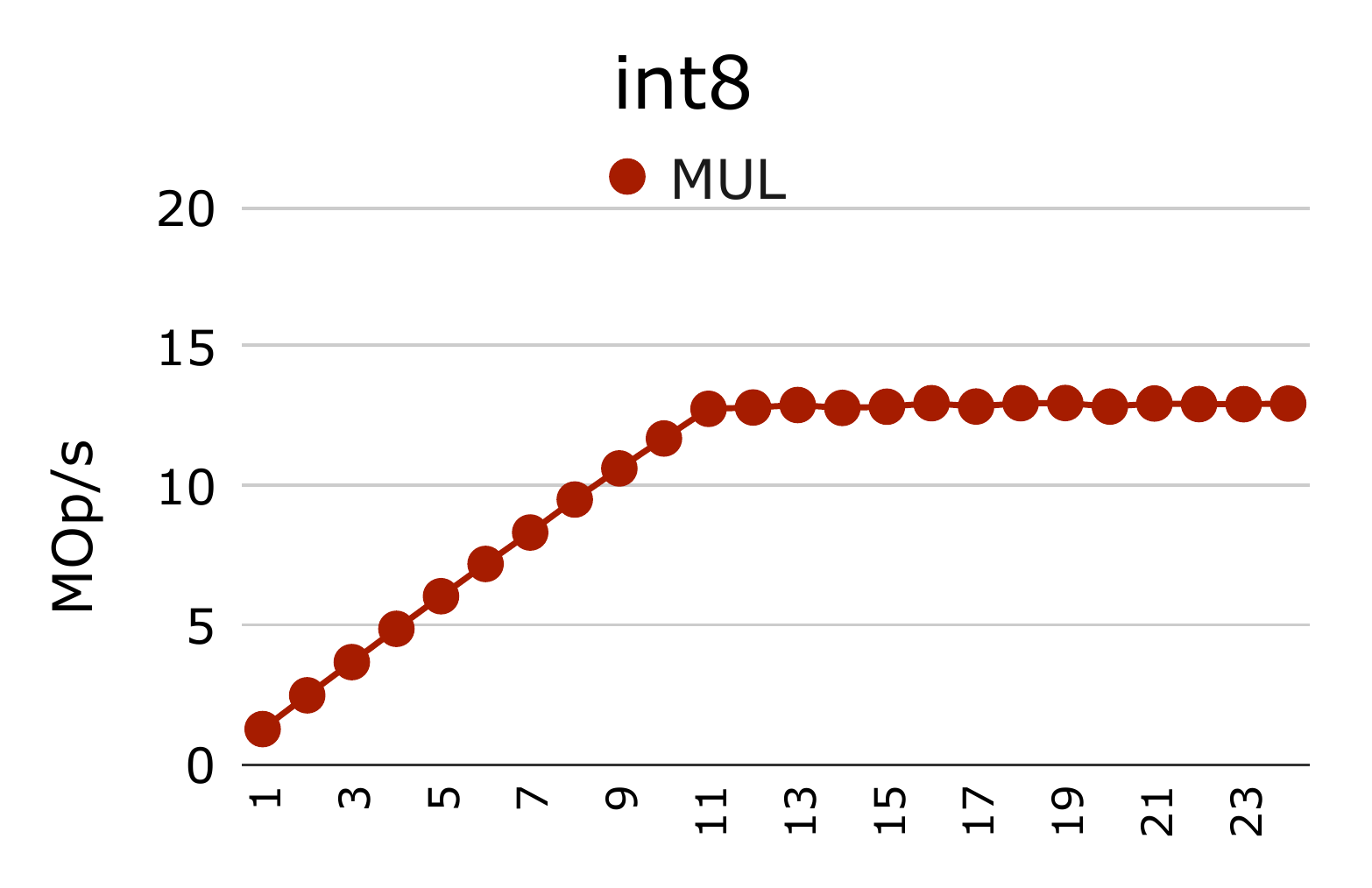}\hspace{12pt}
\includegraphics[width=.48\textwidth]{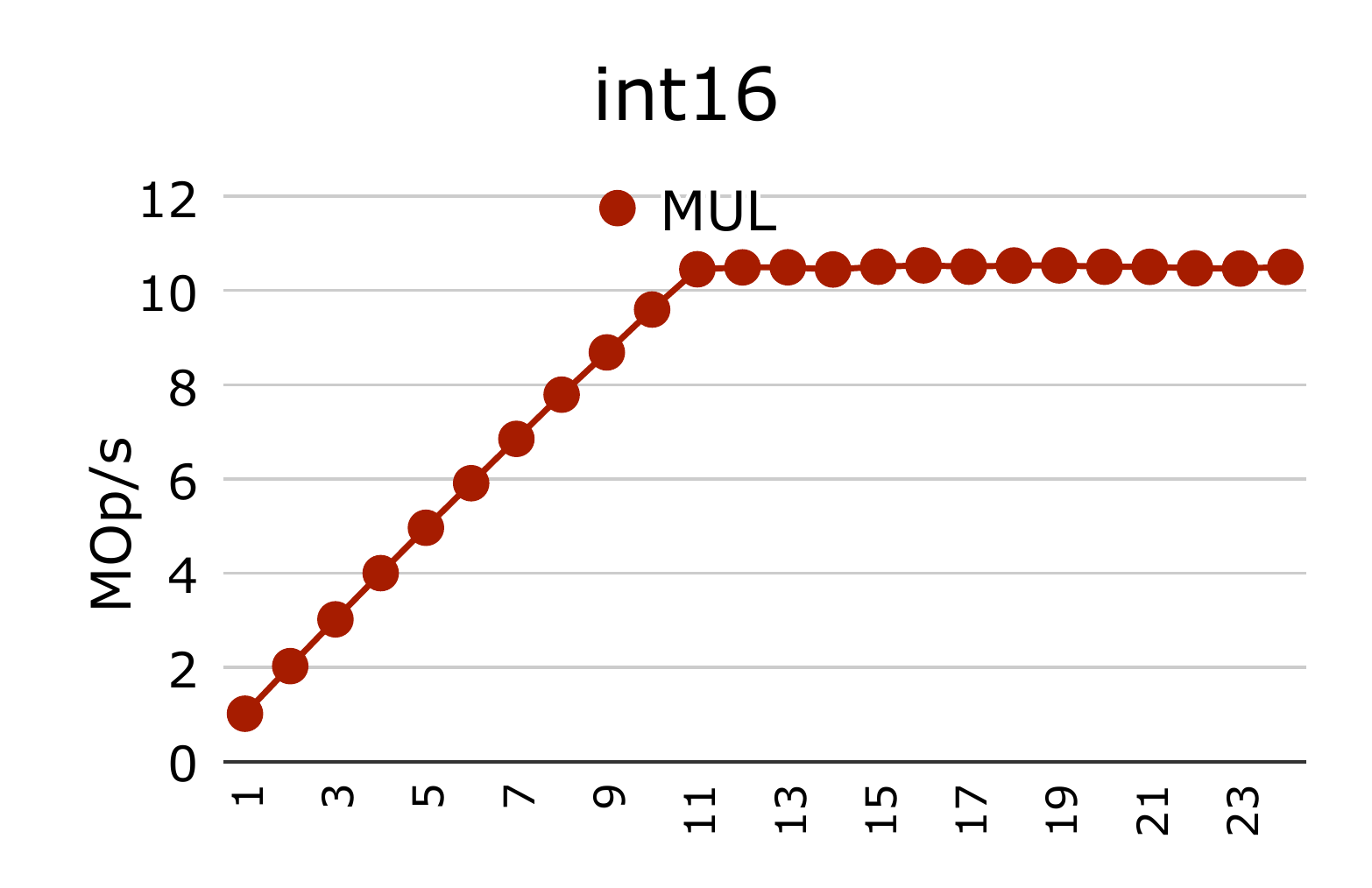}
\end{minipage} % 
\begin{minipage}{1.0\textwidth}
\centering
\includegraphics[width=.48\textwidth]{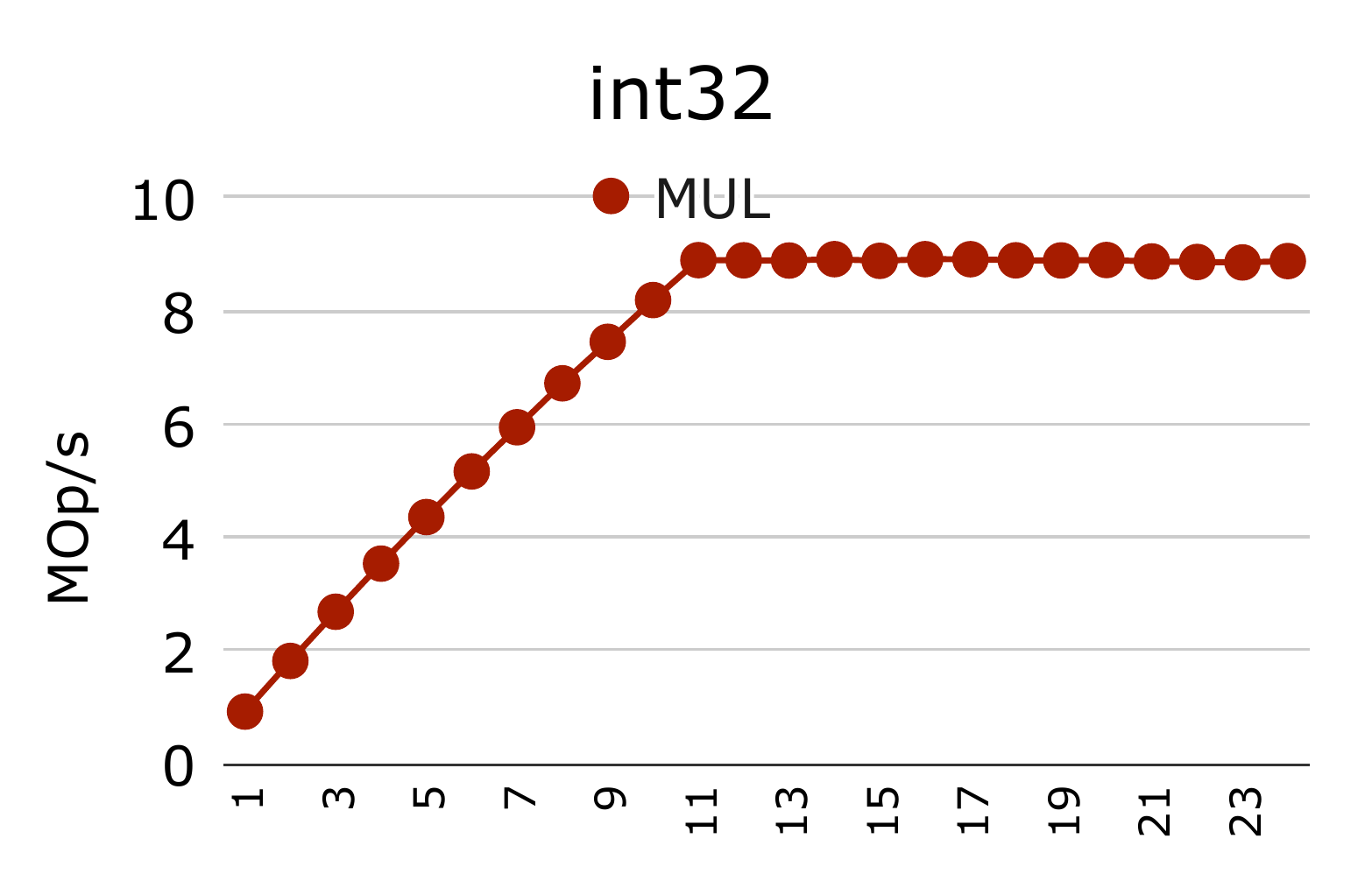}\hspace{12pt}
\includegraphics[width=.48\textwidth]{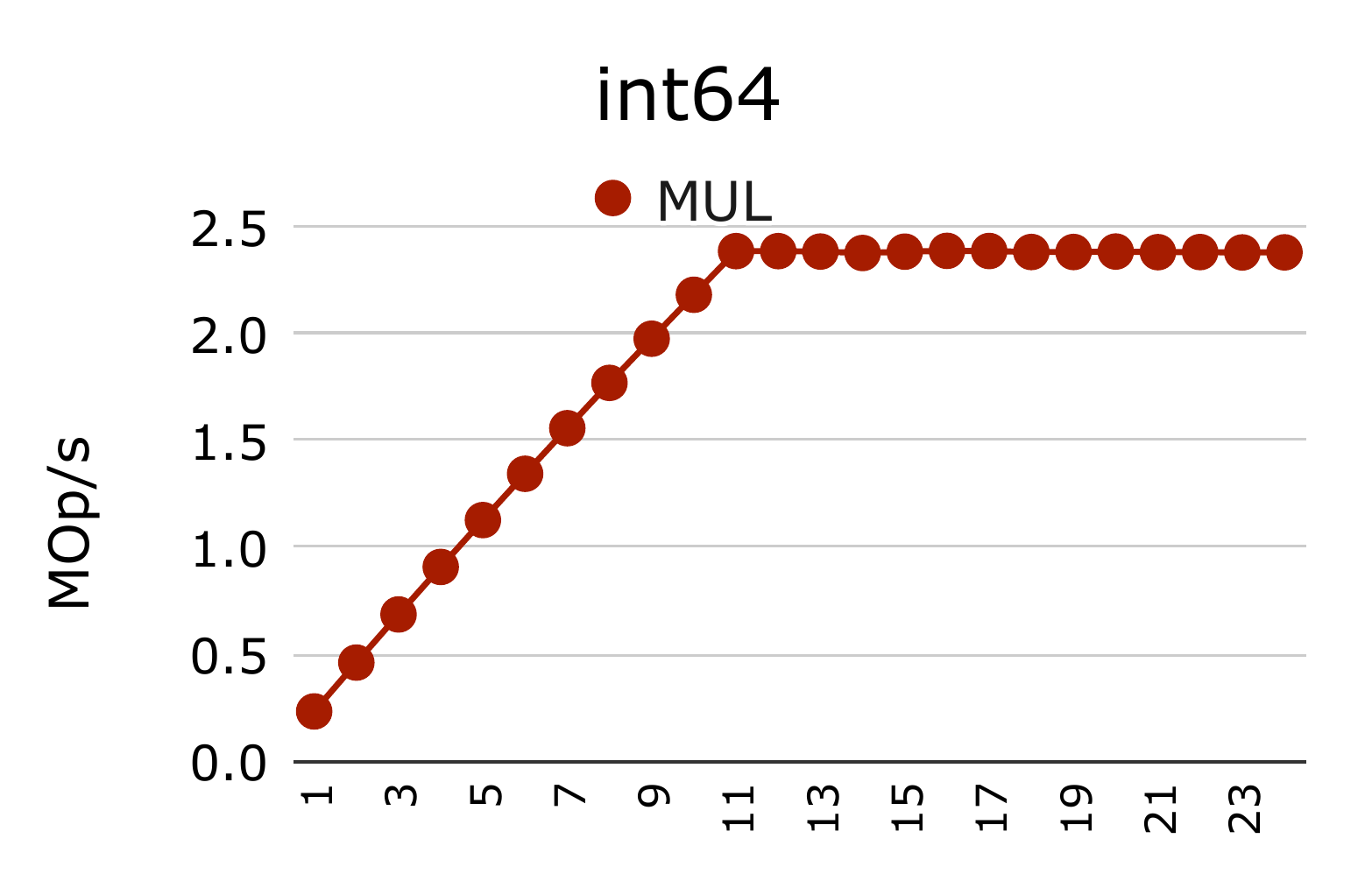}
\end{minipage} % 
\begin{minipage}{1.0\textwidth}
\centering
\includegraphics[width=.48\textwidth]{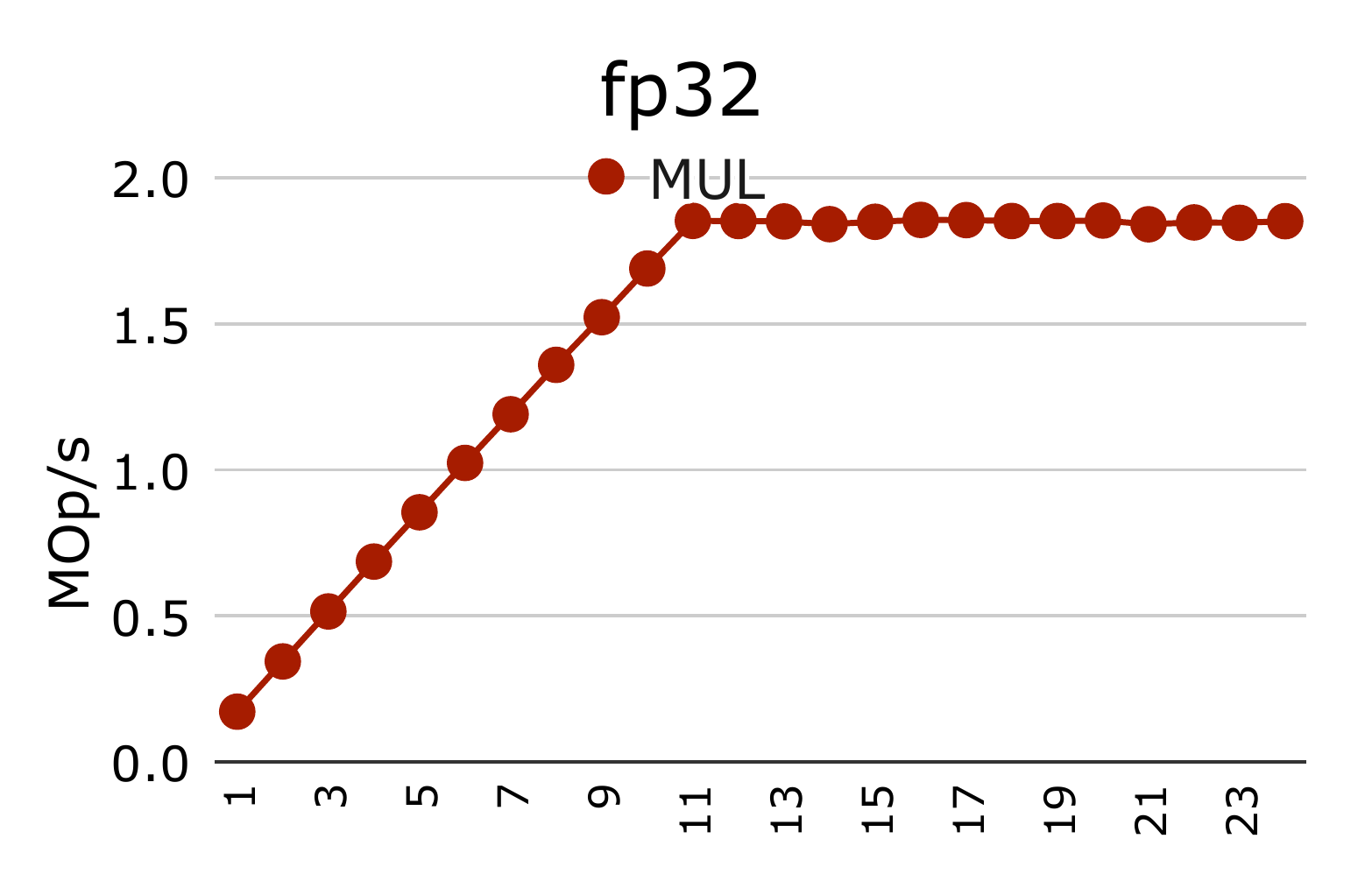}\hspace{12pt}
\includegraphics[width=.48\textwidth]{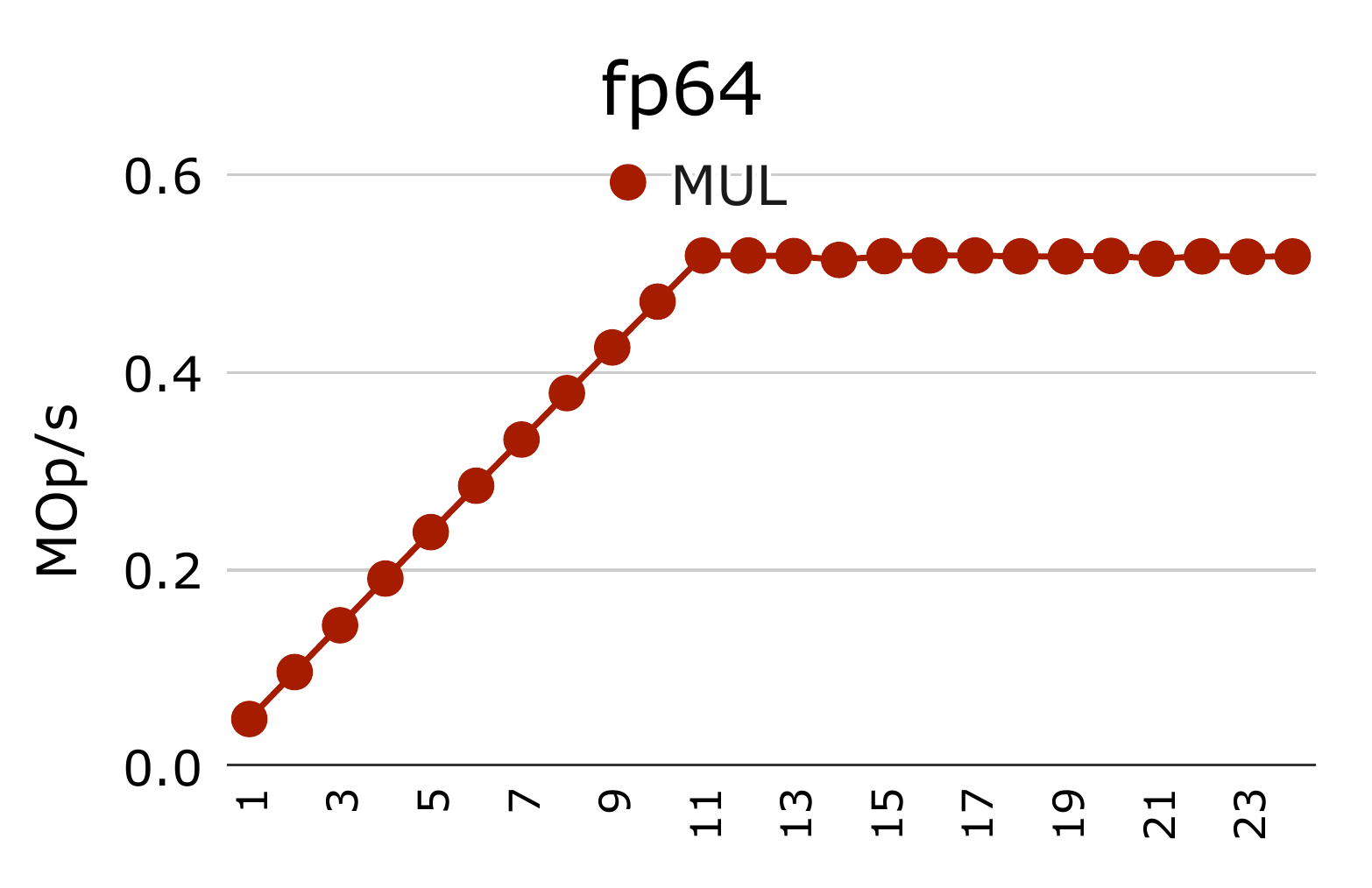}
\end{minipage}
\vspace{-4pt}
\caption{Throughput of the MUL operation on one DPU at 350 MHz for all the data types.}
\label{fig:1DPU-ai-cloud4}
%\vspace{-8pt}
\end{figure}

Figure~\ref{fig:1DPU-ai-cloud4} shows the measured arithmetic throughput (in MOperations per second) for the MUL operation varying the number of tasklets of one DPU at 350 MHz (PIM system A in Table~\ref{tab:pim-systems}) for all the data types. The arithmetic throughput for the MUL operation is 12.941 MOps, 10.524 MOps, 8.861 MOps, 2.381 MOps, 1.847 MOps, and 0.517 MOps for the int8, int16, int32, int64, fp32 and fp64 data types, respectively.

\newpage

\begin{figure}[H]
\centering
\begin{minipage}{1.0\textwidth}
\centering
\includegraphics[width=.48\textwidth]{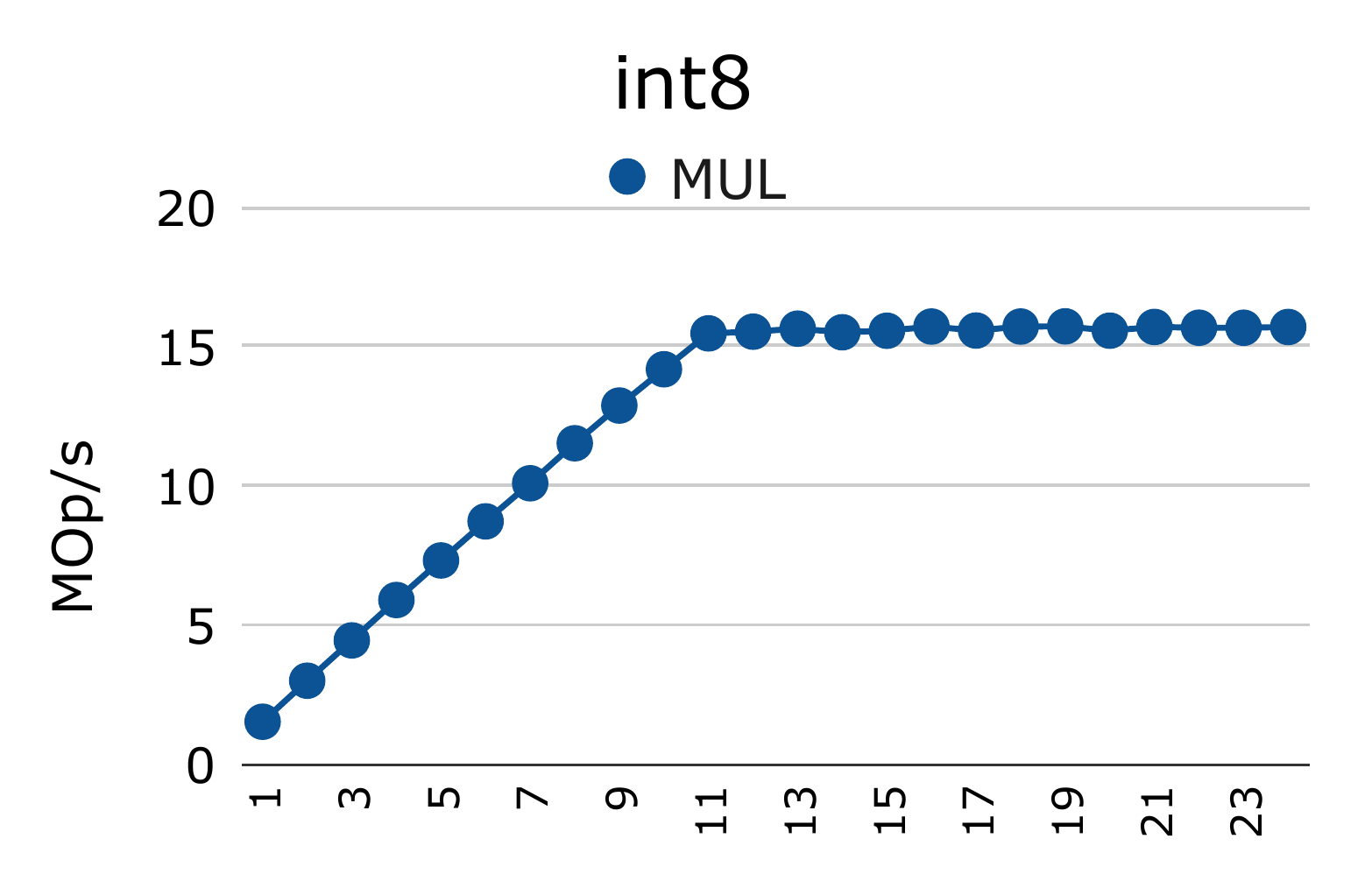}\hspace{12pt}
\includegraphics[width=.48\textwidth]{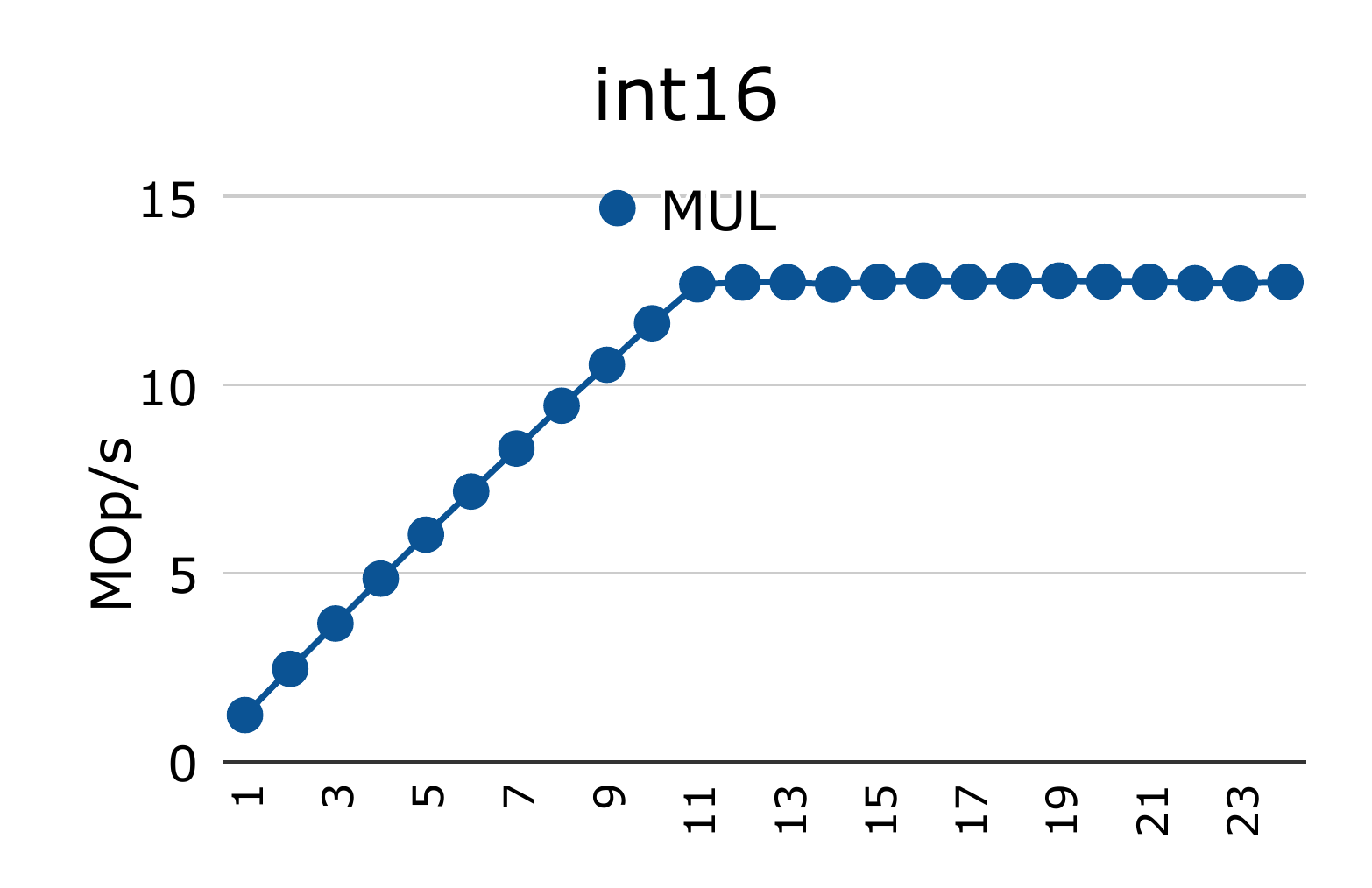}
\end{minipage} % 
\begin{minipage}{1.0\textwidth}
\centering
\includegraphics[width=.48\textwidth]{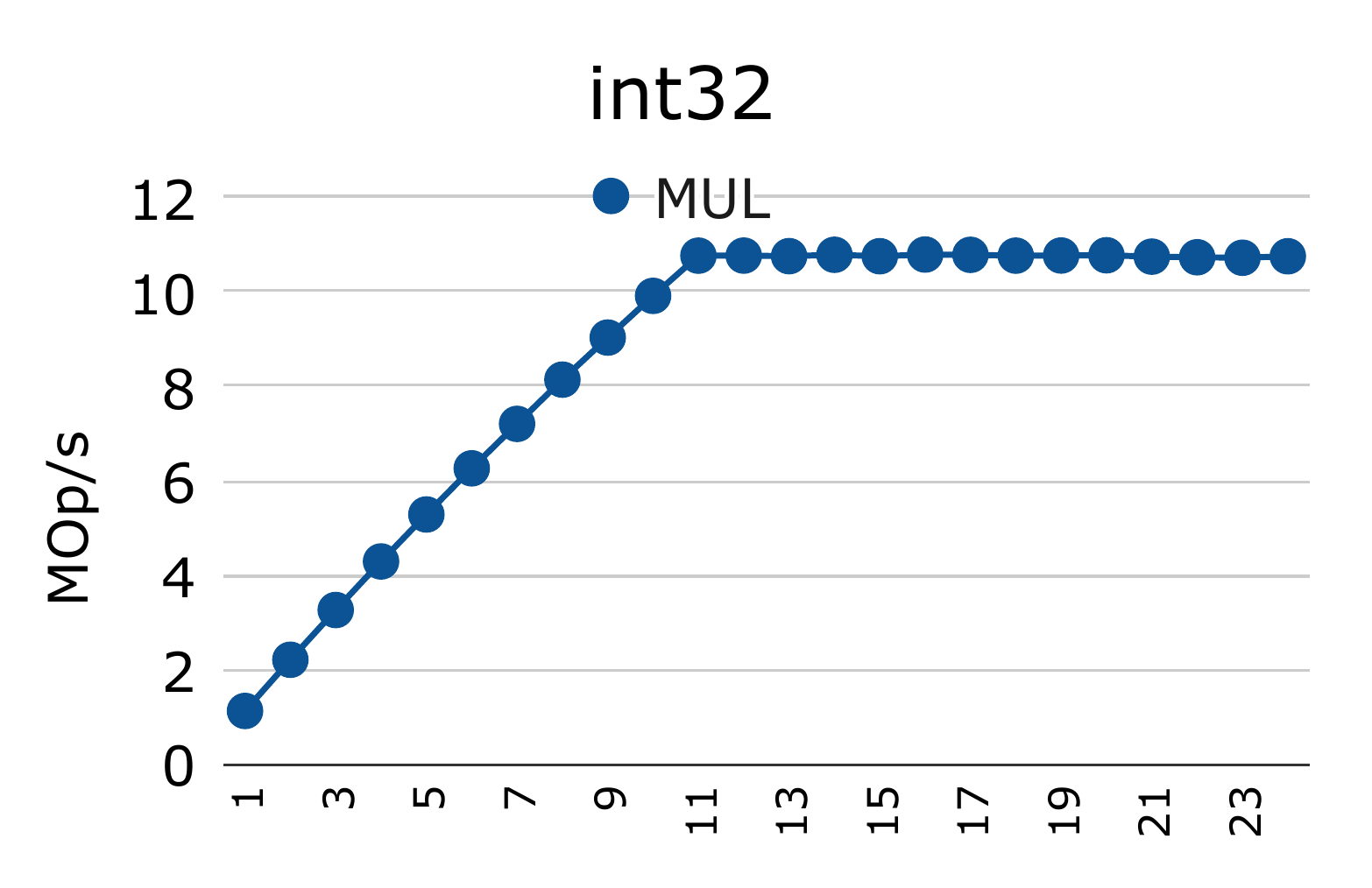}\hspace{12pt}
\includegraphics[width=.48\textwidth]{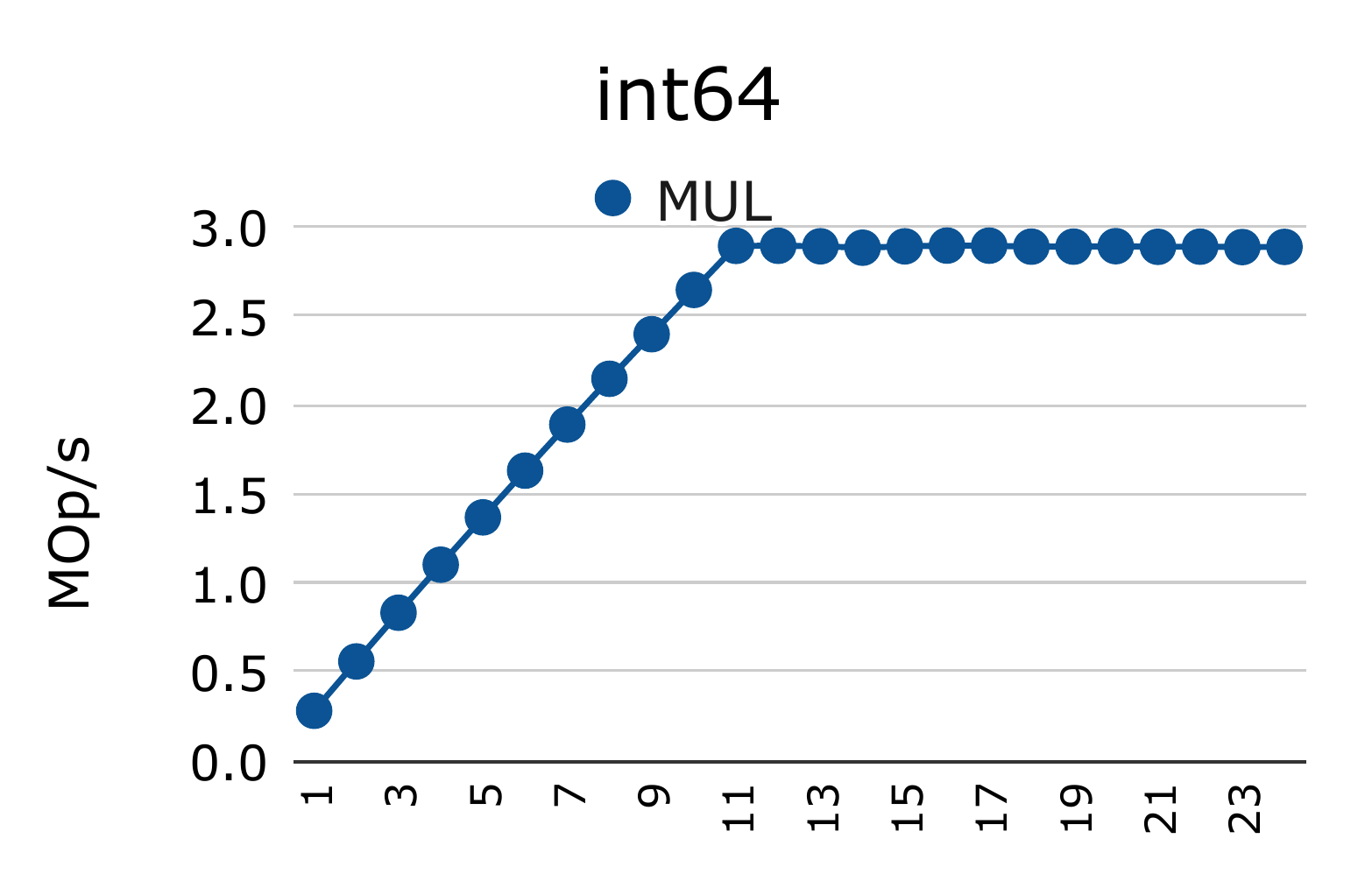}
\end{minipage} % 
\begin{minipage}{1.0\textwidth}
\centering
\includegraphics[width=.48\textwidth]{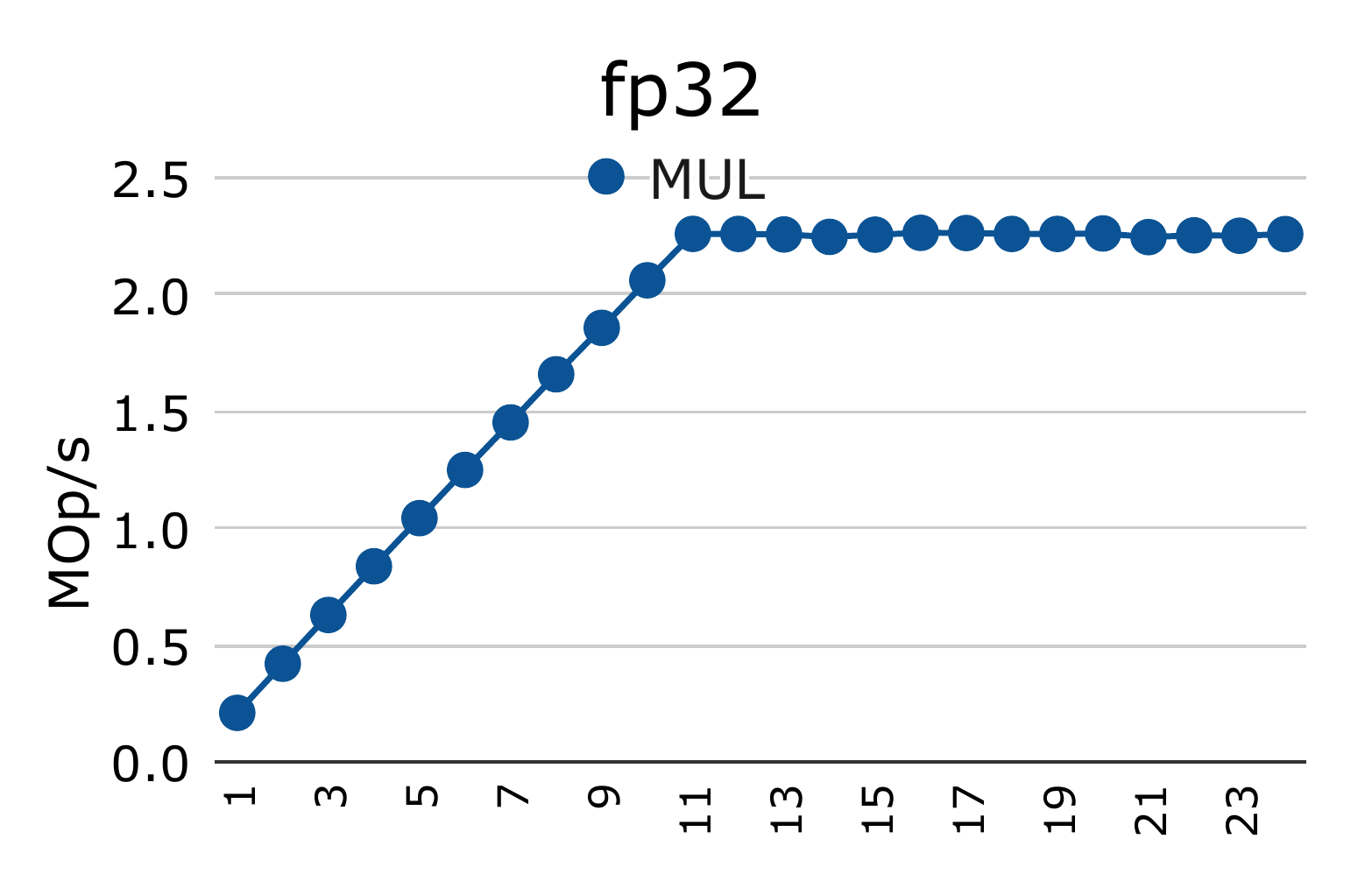}\hspace{12pt}
\includegraphics[width=.48\textwidth]{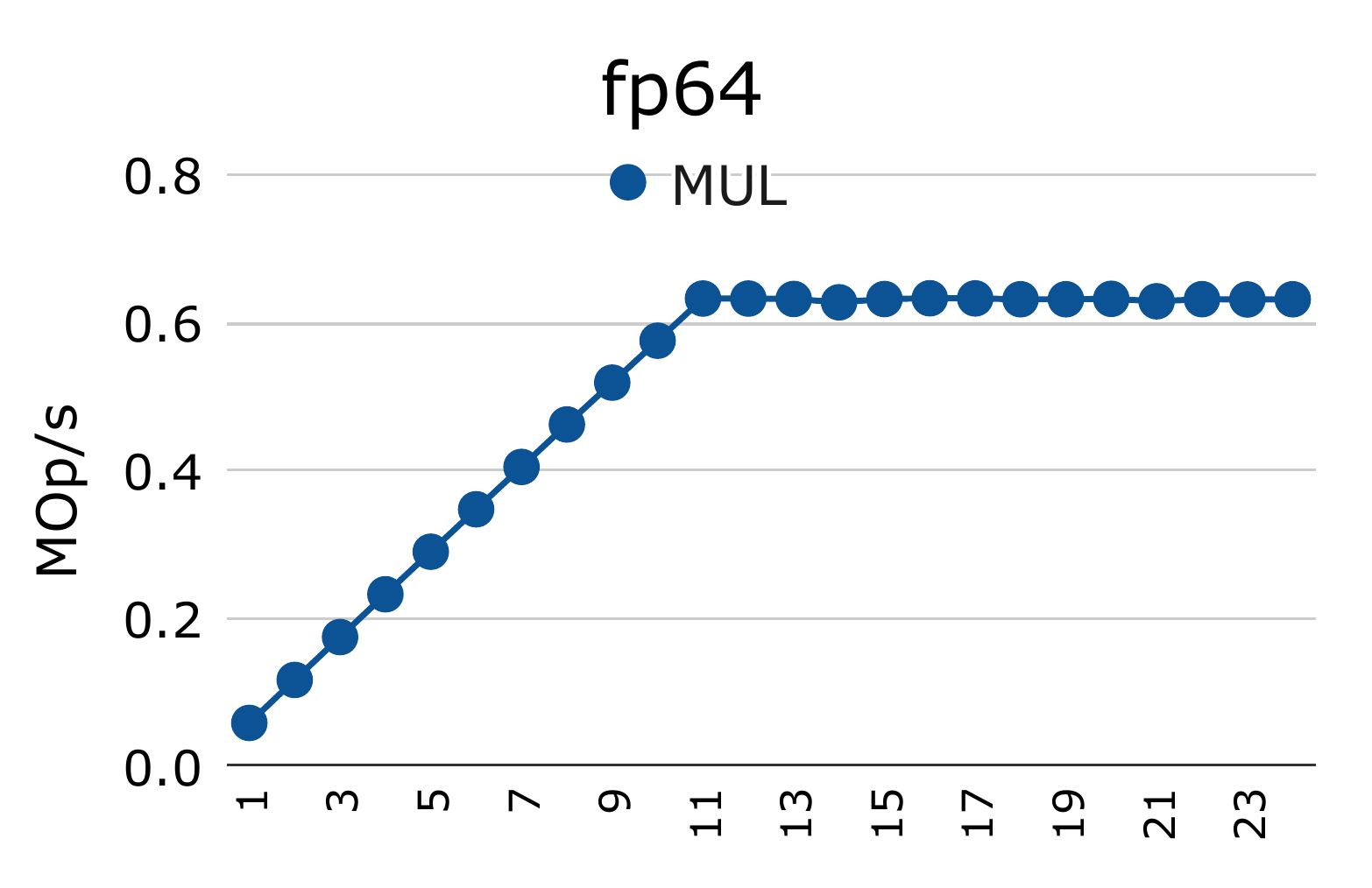}
\end{minipage}
\vspace{-4pt}
\caption{Throughput of the MUL operation on one DPU at 425 MHz for all the data types.}
\label{fig:1DPU-ai-cloud7}
%\vspace{-8pt}
\end{figure}

Figure~\ref{fig:1DPU-ai-cloud7} shows the measured arithmetic throughput (in MOperations per second) for the MUL operation varying the number of tasklets of one DPU at 425 MHz (PIM system B in Table~\ref{tab:pim-systems}) for all the data types. The arithmetic throughput for the MUL operation is 15.656 MOps, 12.721 MOps, 10.732 MOps, 2.888 MOps, 2.259 MOps, and 0.631 MOps for the int8, int16, int32, int64, fp32 and fp64 data types, respectively.

\newpage

\section{The \SparseP{} Software Package}\label{sec:appendix-sparseP}
Table~\ref{table:appendix-library} summarizes the \spmv{} PIM kernels provided by the \SparseP{} library. All kernels support a wide range of data types, i.e., 8-bit integer, 16-bit integer, 32-bit integer, 64-bit integer, 32-bit float, and 64-bit float data types.

\begin{table}[H]
\vspace{10pt}
\centering
\begin{minipage}{1\linewidth}
%\ra{0.9}
\resizebox{1.0\linewidth}{!}{
\begin{tabular}{| l | l | l | l | l | l |} 
 \hline
\cellcolor{gray!15}\raisebox{-0.20\height}{\textbf{Partitioning}} & \cellcolor{gray!15}\raisebox{-0.20\height}{\textbf{Compressed}} & \cellcolor{gray!15}\raisebox{-0.20\height}{\textbf{Balancing}} & \cellcolor{gray!15}\raisebox{-0.20\height}{\textbf{Balancing}}  & \cellcolor{gray!15}\raisebox{-0.20\height}{\textbf{Synchronization}} \\ 
 \cellcolor{gray!15}\textbf{Technique} & \cellcolor{gray!15}\textbf{Format} & \cellcolor{gray!15}\textbf{Across PIM Cores} & \cellcolor{gray!15}\textbf{Across Threads}  & \cellcolor{gray!15}\textbf{Approach}  \\ [0.5ex] \hline \hline
% \hline\hline
    \multirow{9}{*}{\hspace*{2.1em}\turnbox{0}{1D}} & \multirow{2}{*}{CSR} &  rows  &  rows, nnz$^{\star}$ & -   \\
    & & nnz$^{\star}$ &  rows, nnz$^{\star}$ & -   \\ \cline{2-5}
    & \multirow{3}{*}{COO} &  rows  &  rows, nnz$^{\star}$ & -   \\
    & & nnz$^{\star}$  &  rows, nnz$^{\star}$ & -   \\
    & & nnz  &  nnz & lb-cg / lb-fg / lf \\  \cline{2-5}
    & \multirow{2}{*}{BCSR} &  blocks$^{\dagger}$  &  blocks$^{\dagger}$, nnz$^{\dagger}$ & lb-cg$^{\ddagger}$ / lb-fg$^{\ddagger}$   \\
    & & nnz$^{\dagger}$  &  blocks$^{\dagger}$, nnz$^{\dagger}$ & lb-cg$^{\ddagger}$ / lb-fg$^{\ddagger}$ \\  \cline{2-5}
    & \multirow{2}{*}{BCOO} &  blocks  &  blocks, nnz & lb-cg / lb-fg / lf \\
    & & nnz &  blocks, nnz & lb-cg / lb-fg / lf  \\ \hline
    \multirow{4}{*}{\shortstack{2D \\ \equallySized}} & CSR & - & rows, nnz$^{\star}$ & -  \\  \cline{2-5}
    &  COO  & - & nnz & lb-cg / lb-fg / lf \\  \cline{2-5}
    &  BCSR  & - & blocks$^{\dagger}$, nnz$^{\dagger}$ & lb-cg$^{\ddagger}$ / lb-fg$^{\ddagger}$  \\  \cline{2-5}
    &  BCOO & - & blocks, nnz & lb-cg / lb-fg  \\ \hline
  \multirow{6}{*}{\shortstack{2D \\ \equallyWidth}} &  CSR & nnz$^{\star}$ & rows, nnz$^{\star}$ & -  \\  \cline{2-5}
    &  COO & nnz & nnz & lb-cg / lb-fg / lf   \\  \cline{2-5}
    & \multirow{2}{*}{BCSR} & blocks$^{\dagger}$ & blocks$^{\dagger}$, nnz$^{\dagger}$ & lb-cg$^{\ddagger}$ / lb-fg$^{\ddagger}$ \\
    &  & nnz$^{\dagger}$ & blocks$^{\dagger}$, nnz$^{\dagger}$ & lb-cg$^{\ddagger}$ / lb-fg$^{\ddagger}$ \\ \cline{2-5}
    & \multirow{2}{*}{BCOO} & blocks  & blocks, nnz & lb-cg / lb-fg   \\ 
    & & nnz  & blocks, nnz & lb-cg / lb-fg   \\ \hline     
  \multirow{6}{*}{\shortstack{2D \\ \variableSized}} &  CSR & nnz$^{\star}$ & rows, nnz$^{\star}$& -  \\  \cline{2-5}
    &  COO & nnz  & nnz & lb-cg / lb-fg / lf \\  \cline{2-5}
    &  \multirow{2}{*}{BCSR} & blocks$^{\dagger}$  & blocks$^{\dagger}$, nnz$^{\dagger}$ & lb-cg$^{\ddagger}$ / lb-fg$^{\ddagger}$   \\
    &  & nnz$^{\dagger}$ & blocks$^{\dagger}$, nnz$^{\dagger}$ & lb-cg$^{\ddagger}$ / lb-fg$^{\ddagger}$   \\ \cline{2-5}
    &  \multirow{2}{*}{BCOO } & blocks  & blocks, nnz & lb-cg / lb-fg  \\ 
    &  &  nnz & blocks, nnz &  lb-cg / lb-fg  \\   \hline   
 %\bottomrule
\end{tabular}
}
\end{minipage} \hspace{2pt}% 
\vspace{6pt}
\caption{The \SparseP{} library. $^{\star}$: row-granularity, $^{\dagger}$: block-row-granularity, $^{\ddagger}$: (only for 8-bit integer and small block sizes)}
\label{table:appendix-library}
\end{table}

\newpage

\section{Large Matrix Dataset}\label{sec:appendix-matrix-dataset}
We present the characteristics of the sparse matrices of our large matrix data set. Table~\ref{tab:appendix-large-matrices} presents the sparsity of the matrix (i.e., NNZ / (rows x columns)), the standard deviation of non-zero elements among rows (NNZ-r-std) and columns (NNZ-c-std). %Table~\ref{tab:appendix-large-matrices-plot} visualizes the sparsity patterns of each sparse matrix of our large matrix data set.

% Flops / byte = (2 * nnz) / (8 * nnz + 12 * N), N : matrix dimension
% source: https://github.com/Sable/fait-maison-spmv
% Flops / byte = (2 * nnz) / (sizeof(datatype) * nnz + 12 * N), N : matrix dimension
\begin{table}[H]
\begin{center}
\vspace{10pt}
\centering
\begin{minipage}{1\linewidth}
%\resizebox{1.0\linewidth}{0.13\textheight}{
%\resizebox{1.0\linewidth}{!}{
\begin{tabular}{|l||c|r|r|r|r|}
    \hline
    \cellcolor{gray!15}\raisebox{-0.10\height}{\textbf{Matrix Name}} & \cellcolor{gray!15}\raisebox{-0.10\height}{\textbf{Rows x Columns}} & \cellcolor{gray!15}\raisebox{-0.10\height}{\textbf{NNZs}} & \cellcolor{gray!15}\raisebox{-0.10\height}{\textbf{Sparsity}} & \cellcolor{gray!15}\raisebox{-0.10\height}{\textbf{NNZ-r-std}} & \cellcolor{gray!15}\raisebox{-0.10\height}{\textbf{NNZ-c-std}} \\
    \hline \hline
    hugetric-00020 & 7122792 x 7122792 & 21361554 & 4.21e-07 & 0.031 & 0.031 \\ \hline 
    
    mc2depi & 525825 x 525825	 & 2100225 & 7.59e-06 & 0.076 & 0.076 \\ \hline
    
    parabolic\_fem &	525825	 x 525825 & 	3674625 & 	1.33e-05 & 0.153 &  0.153 \\ \hline
    
    roadNet-TX & 	1393383	x 1393383	& 3843320 & 	1.98e-06 &  1.037 & 1.037 \\ \hline
    
    rajat31 & 	4690002	 x 4690002 & 	20316253 &  9.24e-07  & 1.106  & 1.106 \\ \hline
    
    af\_shell1 & 	504855 x 	504855	& 17588875 & 	6.90e-05 & 1.275 & 1.275 \\ \hline  
    
    delaunay\_n19 & 	524288 x 	524288 &	3145646 & 	1.14e-05 &  1.338 & 1.338 \\ \hline
    
    thermomech\_dK  &	204316 x 	204316 & 	2846228 & 6.81e-05 & 1.431 &  1.431 \\ \hline
    
    memchip & 2707524 x 2707524	& 14810202 & 	2.02e-06 &  2.062 & 1.173 \\ \hline
        
    amazon0601 & 403394 x 403394	& 3387388 & 	2.08e-05 &  2.79 & 15.29  \\ \hline
    
    FEM\_3D\_thermal2 & 147900 x 	147900 & 	3489300 & 	1.59e-04 & 4.481  & 4.481 \\ \hline

    %\bottomrule
%\end{tabular}
%}
%\end{minipage} \hspace{2pt}% 
%\begin{minipage}{.49\linewidth}
%\resizebox{1.0\linewidth}{!}{
%\begin{tabular}{|l|c|r|r|r|r|}
%    \toprule
%    \textbf{Matrix Name} & \textbf{Rows x Columns} & \textbf{NNZs} & \textbf{Sparsity} & \textbf{NNZ-r-std} & \textbf{NNZ-c-std} \\
    %\midrule
     
    web-Google	& 916428 x 	916428 & 	5105039 & 	6.08e-06 & 6.557 &  38.366 \\ \hline
    
    ldoor	 & 952203 x 952203	& 46522475 & 	5.13e-05  & 11.951 & 11.951 \\ \hline
    
    poisson3Db & 	85623 x 85623	 & 2374949 & 	3.24e-04  & 14.712  & 14.712 \\ \hline
    
    boneS10 & 	914898  x 	914898 & 	55468422 & 	6.63e-05 &  20.374 &  20.374 \\\hline
    
    webbase-1M	& 1000005 x 1000005	& 3105536 & 	3.106e-06 & 25.345  & 36.890 \\ \hline
    
    in-2004 & 1382908  x 	1382908	 & 16917053 & 	8.846e-06  & 37.230 & 144.062 \\ \hline 
    
    pkustk14 & 	151926  x	151926	 & 14836504 & 	6.428e-04 &  46.508  & 46.508 \\ \hline
    
    com-Youtube & 	1134890	 x  1134890	 & 5975248 & 4.639e-06  & 50.754  & 50.754 \\ \hline

    as-Skitter & 	1696415 x	1696415	& 22190596 & 	7.71e-06  & 136.861 &  136.861 \\ \hline
    
    sx-stackoverflow &	2601977	 x  2601977 & 	36233450 & 	5.352e-06 &  137.849 & 65.367 \\ \hline 

    ASIC\_680 & 	682862 x  682862 & 	3871773 & 	8.303e-06  & 659.807 & 659.807 \\ \hline
 
    %\hline
\end{tabular}
%}
\end{minipage}%
\end{center}
\vspace{6pt}
\caption{Large Matrix Dataset. Matrices are sorted by NNZ-r-std, i.e., based on their irregular pattern. }
\label{tab:appendix-large-matrices}
\end{table}

\newpage
\fancyhead[RE,LO]{Bibliography}
\fancyhead[LE,RO]{\thepage}
%\selectlanguage{english}
\bibliographystyle{unsrt}
\bibliography{sections/bibliography}

\end{document}